\documentclass[a4paper,10pt]{thesis}

\usepackage{a4}
\usepackage[english]{babel}
\usepackage{amsmath}
\usepackage{amsfonts}
\usepackage{amssymb}
\usepackage{theorem}
\usepackage[dvips]{color}
\usepackage{bm}
\usepackage[dvips]{graphicx}
\usepackage{floatflt}
\usepackage{enumerate}
\usepackage{subfigure}
\usepackage{multicol}
\usepackage{multirow}
\usepackage{epsf}
\usepackage{array}
\usepackage[page,title,titletoc,header]{appendix}
\usepackage{isoent, slashed, sidecap}

\def\rpv{$\slashed{R}_p \:$}
\def\rpvm{\slashed{R}_p \:}
\def\vb#1{\vbox to #1 pt{}}
\def\eslash{\slashed{E}}
\newcommand{\eVq}{\text{eV}^2}

\begin{document}

\frontmatter

\thispagestyle{empty}

\begin{center}

  \vspace{2.5cm}

  \begin{figure}[h!]
  \center
  \resizebox{0.25\textwidth}{!}{\includegraphics*{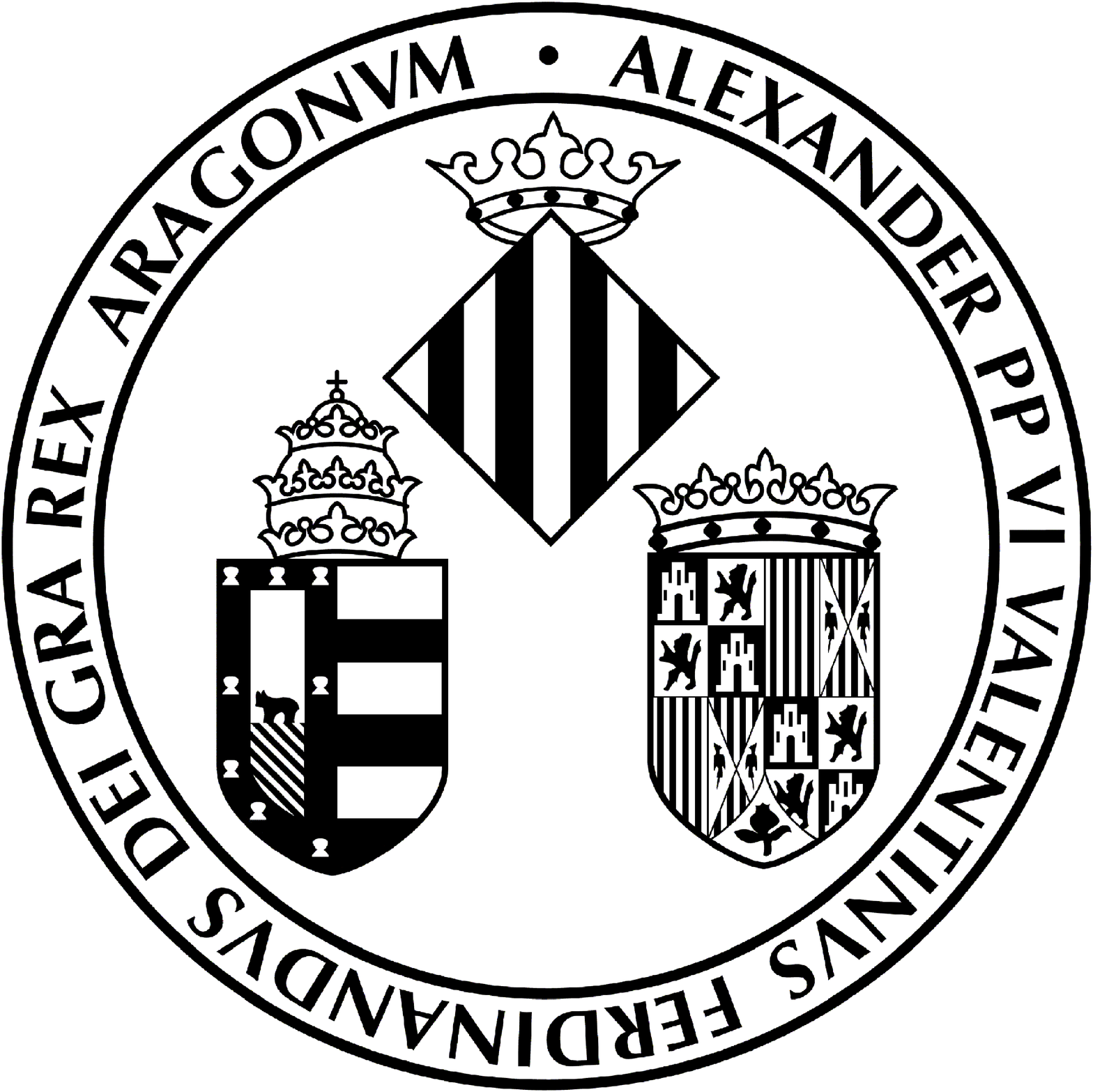}} \\
  \resizebox{0.65\textwidth}{!}{\includegraphics*{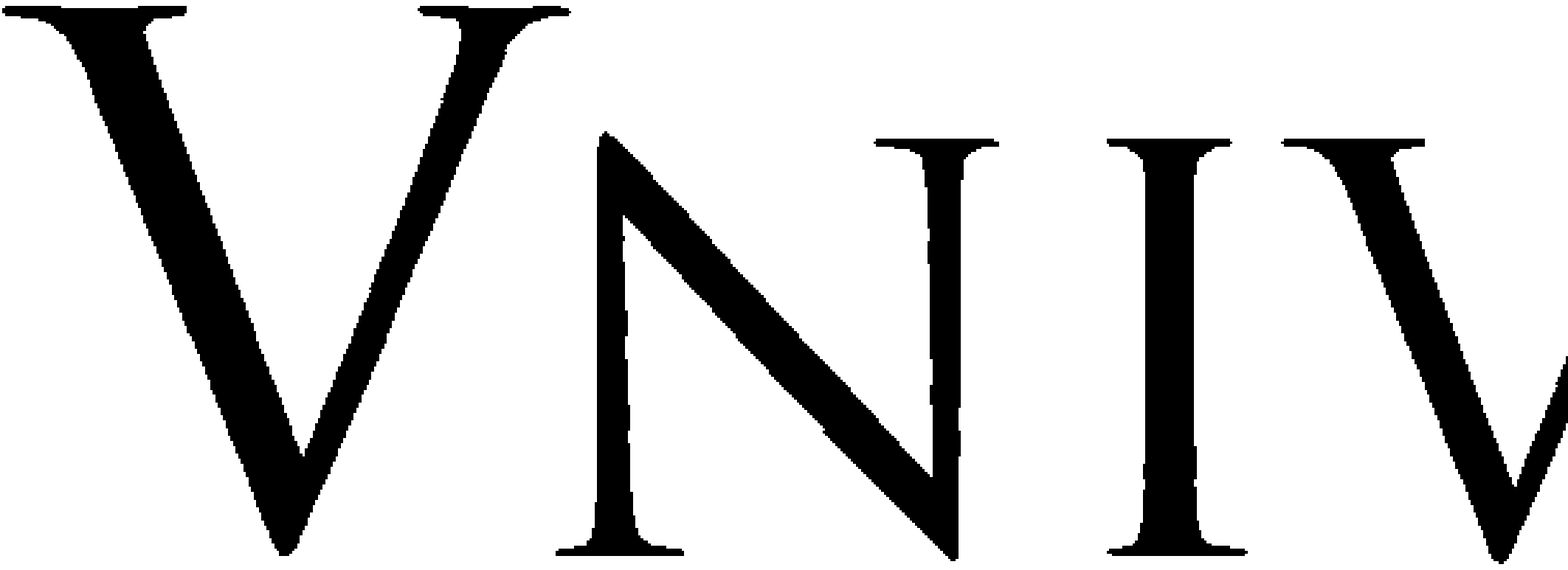}}	 
  \end{figure}

  {\LARGE Departamento de F\'isica Te\'orica}
  \vspace{3 cm}

  {\bf \huge Phenomenology of}\\
  \vspace{0.4 cm}
  {\bf \huge supersymmetric}\\
  \vspace{0.4 cm}
  {\bf \huge neutrino mass models}\\

  \vspace{3.0cm}

  {\bf \large Ph.D. Thesis}
  \vspace{0.2 cm}

  {\bf \large Avelino Vicente Montesinos}
  \vspace{1.0 cm}

  {\bf \large Supervisors: Jos\'e W. F. Valle and Martin Hirsch }
  \vspace{1.0 cm}
  
  {\bf \large Valencia 2011 }

\end{center}


\newpage
\thispagestyle{empty}
\vspace*{18cm}

\newpage
\thispagestyle{empty}

\begin{abstract}

The origin of neutrino masses is currently one of the most intriguing questions of particle physics and many extensions of the Standard Model have been proposed in that direction. This experimental evidence is a very robust indication of new physics, but is not the only reason to go beyond the Standard Model. The existence of some theoretical issues supports the idea of a wider framework, supersymmetry being the most popular one. In this thesis, several supersymmetric neutrino mass models have been studied. In the first part, the phenomenology of models with bilinear-like R-parity violation is discussed in great detail, highlighting the most distinctive signatures at colliders and low energy experiments. In particular, the correlations between the LSP decay and neutrino physics are shown to be a powerful tool to put this family of models under experimental test. Other important signatures are investigated as well, like majoron emission in charged lepton decays for the case of models with spontaneous breaking of R-parity. A very different approach is followed in the second part of the thesis. Supersymmetric models with a Left-Right symmetry have all the ingredients to incorporate a type-I seesaw mechanism for neutrino masses and conserve R-parity at low energies. In this case, which only allows for indirect tests, the generation of neutrino masses at the high seesaw scale is encoded at low energies in the slepton soft masses. Contrary to minimal seesaw models, sizeable flavor violation in the right slepton sector is expected. Its experimental observation would be a clear hint of an underlying Left-Right symmetry, providing valuable information about physics at very high energies.

\end{abstract}


\newpage
\thispagestyle{empty}
\vspace*{18cm}


\newpage
\thispagestyle{myheadings}
\markright{Agradecimientos}
\vspace*{1cm}

{\LARGE \it Agradecimientos}

\vspace*{1cm}

Pasados los a\~nos, todos vemos el comienzo de la tesis de una forma muy particular. Para algunos, los inicios se esconden en un rinc\'on lejano de la memoria, mientras que para otros, parece que fue ayer cuando ojearon su primer \emph{paper}. En mi caso, todav\'ia estoy sorprendido por la rapidez con que todo ha sucedido. Incluso recuerdo esa sensaci\'on extra\~na cuando me preguntaron por primera vez si hab\'ia comenzado a escribir la tesis. Mi reacci\'on inmediata fue un sorprendido ``?`?`ya??'' para luego darme cuenta de que, efectivamente, deb\'ia comenzar a trabajar en la redacci\'on del documento que el lector tiene entre manos.

Al escribir este breve texto, y con ello reflexionar sobre mi \'epoca de doctorado, me ha resultado obvio que si estos cuatro a\~nos han pasado tan r\'apido ha sido gracias a las fant\'asticas personas de las que me he visto rodeado. A algunos los he conocido durante el doctorado y a otros los conoc\'ia de mi vida anterior. En cualquier caso, todos han puesto su granito de arena para hacer de esta etapa de mi vida algo inolvidable. Espero que estas l\'ineas sirvan de agradecimiento a todos aquellos que me han acompa\~nado en este viaje.

Recibir unas buenas directrices es fundamental para el desarrollo de un doctorado. Por ello, me encuentro profundamente en deuda con Jos\'e y Martin, mis directores de tesis. Siempre os he encontrado cercanos, accesibles y dispuestos a darme un consejo o resolver una de mis interminables dudas. Echar\'e de menos escuchar un ``Avelinooooo'' por el pasillo o comentar el \'ultimo cap\'itulo de The Big Bang Theory. Gracias Martin por todo lo que me has ense\~nado estos a\~nos y, tambi\'en, por haberte le\'ido esta tesis una y otra vez (!`siento que haya salido algo larga!). Por supuesto, hago extensibles estas palabras a todos los miembros pasados y presentes del grupo AHEP. El ambiente de trabajo todos estos a\~nos ha sido realmente genial. Por cierto, lamento haberos hecho esperar cada d\'ia a que terminara de comer, aunque estoy seguro de que alguno disfrutaba de ese descanso a\~nadido antes de volver al despacho.

Asimismo, mando un afectuoso agradecimiento a todos aquellos que han compartido la f\'isica conmigo. Eso incluye, por supuesto, a mis colaboradores en los trabajos presentados en esta tesis. De todos ellos he podido aprender much\'isimo y sus valiosas aportaciones me han enriquecido notablemente. En particular, me gustar\'ia dirigir unas palabras a Jorge, por su hospitalidad, atenci\'on y trato amable, y a Werner, por la confianza depositada en m\'i, especialmente en este periodo de transici\'on en el que tanto me est\'a ayudando. Del mismo modo, no podr\'ia olvidar a los compa\~neros y compa\~neras que hicieron que los cinco a\~nos de licenciatura sean un c\'umulo de gratos recuerdos. Juntos aprendimos a amar (y en algunos momentos tambi\'en a odiar) la f\'isica, lo que sirvi\'o como pretexto para establecer unos lazos de amistad que aprecio enormemente y que con toda seguridad se mantendr\'an por el resto de nuestras vidas.

Trabajar estos a\~nos en el IFIC no podr\'ia haber sido una tarea m\'as agradable y eso se lo debo a quienes han alegrado mi d\'ia a d\'ia. Esas breves (y no tan breves) paradas en el pasillo para contarnos alegr\'ias y preocupaciones o esas visitas a la cafeter\'ia a media tarde han formado parte de mi mundo durante estos a\~nos. Sin lugar a dudas se echar\'an de menos, as\'i como todo lo que las ha acompa\~nado: el f\'utbol de los jueves (!`el f\'utbol es mi vida!), las cenas en el Ambit, los pedestrians, las peleas a muerte por la \'ultima plaza de aparcamiento, tantos y tantos saludos amistosos en la cafeter\'ia. Hab\'eis sido fant\'asticos. Lo mismo puedo decir de todo el grupo de amigos que ha nutrido el Campus de Burjassot con su alegr\'ia y buen humor. Astrof\'isicos, te\'oricos, nucleares, \'opticos, bi\'ologos, qu\'imicos \dots he encontrado gente maravillosa de muchos tipos diferentes. Aprovecho tambi\'en para mandar un saludo a todos los amigos que hice en mis estancias en Jap\'on y Portugal, dos lugares de los que guardo recuerdos muy singulares. Sin duda nuestros caminos futuros volver\'an a cruzarse y podremos compartir de nuevo un bol de ramen o un bacalhau com natas. 

Igualmente, y de un modo muy especial, no puedo olvidarme de la gente que ha visto nacer y crecer mis sue\~nos desde el principio, mis amigos de siempre. Seguro que ahora pod\'eis perdonar que os haya dado la tabarra tantas veces con esas locuras a las que me dedicaba. Much\'isimas gracias por vuestra ayuda, tanto en los peque\~nos favores que os he pedido (Manolo, tranquilo, espero no volver a quedarme tirado con el coche) como en las grandes reservas de apoyo que me hab\'eis brindado. Espero que estas breves l\'ineas sean capaces de mostraros mi enorme gratitud por las numerosas veces en las que hab\'eis arrimado el hombro por m\'i.

Ignoro donde estar\'ia hoy en d\'ia sin mi familia, que siempre me ha apoyado en las decisiones que he tomado. Agradezco a mis padres la educaci\'on que me han dado, siempre facilitando mi camino por este mundo. Gracias, mam\'a, porque nunca me ha faltado tu ayuda constante. Ojal\'a el pap\'a estuviera con nosotros en este momento tan importante.

S\'olo puedo tener palabras de cari\~no hacia la persona m\'as importante en mi vida. Gracias Isa por tu inestimable apoyo en tantos momentos. Por escucharme y ofrecerme tus consejos siempre que los necesit\'e. Por alegrarte con cada peque\~no paso que he dado y deleitarme con tu sonrisa con cada buena noticia que ha llegado. Sabes que sin ti esto no ser\'ia igual. No lo disfrutar\'ia del mismo modo. De todo coraz\'on, gracias.

A todos, !`os espero al otro lado del doctorado!

\setcounter{secnumdepth}{-1}

\chapter{Publications}
The articles published during this thesis are the following ones:\\

M.~Hirsch, W.~Porod and A.~Vicente, {\it Phys. Rev. D}, {\bf 77}, 075005 (2008).\\

M.~Hirsch, J.~Meyer, W.~Porod and A.~Vicente, {\it Phys. Rev. D}, {\bf 79}, 055023 (2009).\\

A.~Bartl, M.~Hirsch, S.~Liebler, W.~Porod and A.~Vicente, {\it JHEP}, {\bf 05}, 120 (2009).\\

J.~N.~Esteves, M.~Hirsch, W.~Porod, J.~C.~Rom\~{a}o, F.~Staub and A.~Vicente, {\it JHEP}, {\bf 12}, 077 (2010).

\setcounter{secnumdepth}{2}

\tableofcontents

\mainmatter

\pagebreak

\chapter{Introduction}

The research developed in this doctoral thesis aims to characterize neutrino masses in supersymmetric theories, both in models at low and high energies, and the study of the resulting phenomenology at colliders (such as the LHC or a future ILC), astroparticle physics and cosmology.

The historic discovery of neutrino oscillations made a major breakthough in particle physics and implies that neutrinos have masses. This is nowadays an active field of research that deals with the interdependencies between particle and nuclear physics, astrophysics and cosmology, areas in which the neutrino plays a central role.

Thanks to the precision reached by modern experiments it has been possible to determine the values of the parameters involved in the flavor oscillations observed in different experiments, obtaining a total agreement and leading to the acceptance of the oscillation mechanism as the cause of the observed phenomena. Therefore, due to the lack of neutrino masses in the Standard Model of particle physics, it has become necessary to go beyond the established theoretical framework in order to explain their origin.

Furthermore, the Standard Model has also some theoretical problems that make us think that it is an incomplete theory. In particular, the sensitivity of the Higgs boson mass to the existence of heavy particles requires a strong fine-tuning of the parameters if the mass of this particle lies at the electroweak scale. This naturalness problem is known as the hierarchy problem. Many ideas have been developed in order to solve this flaw of the Standard Model. The most popular one, Supersymmetry, provides a technical solution to the hierarchy problem, while at the same time it has the necessary ingredients to accomodate new physics. In the case of neutrinos, a discrete symmetry, known as R-parity and defined as $R_p = (-1)^{3(B-L)+2s}$, with $B$ and $L$ the baryon and lepton number and $s$ the spin of the particle, plays a determinant role. Both the origin of this symmetry and its connection to the generation of neutrino masses are topics that deserve investigation in depth.

The present thesis can be divided into two research lines closely linked. On the one hand, the phenomenology of supersymmetric neutrino mass models with R-parity violation has been studied in great detail. This type of theories offers a very rich phenomenology in present and near future experiments. Therefore, the study of the experimental signals that are predicted is of fundamental relevance to understand the connection between the underlying theory and the data obtained by the experimental collaborations. On the other hand, the opposite situation has been studied as well. If R-parity is conserved, neutrino masses must be generated in a different way. This has led us into the investigation on the origin of R-parity and how this can be related to neutrino masses. In particular, supersymmetric theories with a left-right symmetry represent an ideal framework for this purpose, since they can lead to R-parity conservation at low energies and incorporate the seesaw mechanism to generate neutrino masses.

Finally, we should keep in mind that we are living a moment of high expectation in particle physics, due to the recent startup of the LHC. This will hopefully soon imply the arrival of large amounts of new experimental data with very valuable information about the fundamental components of matter. Therefore, this thesis mainly focuses on phenomenological issues, concentrating on the LHC and on related experiments that will attempt to unravel the misteries that nature is still hidding from us.

\section{Organization of the manuscript}

I start with a list of publications, and the Chapters based on them. The outline is briefly described.

Chapter 1 contains the introduction, an overview which summarizes the main ideas and the conclusions of the thesis in Spanish.

Chapter 2 contains the introduction and the organization of the thesis.

In Chapter 3, an introduction to supersymmetry is presented. Starting with the Standard Model and the well-known hierarchy problem, supersymmetry is motivated and briefly reviewed. Special attention is paid at the end of the chapter to the role of R-parity.

Chapter 4 is a review of neutrino mass models. First, the current experimental situation is discussed and the need for neutrino masses is highlighted. Then, after a brief discussion on Dirac neutrinos, the chapter focuses on Majorana neutrinos, describing several models that can generate them.

Chapters 5, 6, 7 and 8 contain the work developed in this thesis on R-parity violation. After a general introduction in chapter 5, chapter 6 concentrates on Spontaneous R-parity violation, defining the model and discussing its phenomenology at colliders and low energy experiments. Chapter 7 focuses on a different model, the so-called $\mu \nu$SSM, and studies its phenomenology in great detail. Finally, chapter 8 summarizes this part of the thesis by comparing the different \rpv models under investigation.

Chapter 9 represents the second part of the thesis, focused on supersymmetric left-right models and lepton flavor violation. The current status in the theory of LR models is reviewed and a particular model is chosen for numerical study. After a detailed discussion on the basic properties of the model our results and conclusions are presented.

Finally, chapter 10 summarizes the thesis.

\chapter{Supersymmetry}
\label{chap:susy}

Supersymmetry is one of the most popular extensions of the Standard Model. Widely studied over the last decades, it addresses many of the problems, both theoretical and phenomenological, present in the Standard Model. In addition, it leads to a plethora of new phenomena, expected to appear at current and future experiments. A general review is presented here in order to introduce the setup for the following chapters.

\section{The Standard Model and the hierarchy problem}

An impressive number of measurements have established that the Standard Model can describe the world of fundamental particles to a very high precision. Colliders like SPS, LEP or Tevatron have explored its predictions and studied each one of its fundamental pieces, finding a stunning agreement between theory and experiment. As a result of this amazing success, the Standard Model has become one of the essential ingredients in our current understanding of physics.

However, more than two decades after its invention, there are some experimental results in contradiction with the Standard Model. One of them is the fundamental motivation for this thesis: neutrino masses. As we will see in the next chapter, the Standard Model was built under the assumption of massless neutrinos, something that neutrino oscillation experiments have shown to be wrong.

Furthermore, there are still many theoretical questions that the Standard Model is not able to answer. Just to mention a few, the unknown origin of the flavor structure or the uncontrolled radiative contributions to the Higgs boson mass are unsolved problems in the Standard Model.

These reasons, both experimental and theoretical, point to the need of a new paradigm beyond the Standard Model. Among the different possibilities, Supersymmetry (SUSY) is the most popular choice, due to its capability to account for many of the problems of the Standard Model.

\subsection{Standard Model basic}

The Standard Model (SM) was born as several brilliant ideas in particle physics gathered together to become a coherent framework\footnote{Reference \cite{Weinberg:2004kv} presents the historical development of the Standard Model from the experience of one of its fathers.}. The consistent combination of the different pieces led to a global picture that has been shown to describe very accurately the world of subatomic physics. In fact, after its establishment in the early 70s the SM has been put to constant experimental test, and only a few anomalies have been recently found\footnote{Neutrino masses is the most important of these experimental anomalies. The existence of dark matter and the baryon asymmetry of the universe are other issues, in this case based on cosmological observations, not explained by the Standard Model.}. In that sense, the SM can be considered as a very good description of particle physics up to the energies explored, but also as the starting point for model builders who want to extend it to higher energies.

The overwhelming success of Quantum ElectroDynamics (QED) in the late 40s was taken as proof in favor of quantum field theory as a good theoretical framework to describe particle physics interactions. The most controversial issue was the interpretation of the infinities that appeared in computations beyond the leading order in perturbation theory \cite{Oppenheimer:1930zz}. However, the technique of renormalization, whose purpose is to eliminate these infinities, was shown to give results in good agreement with the experimental data and after many years was finally regarded as consistent. The first application of this technique was done by Hans Bethe, who calculated the famous Lamb shift in 1947 \cite{Bethe:1947id}, opening a new way to handle the infinities. This way, the theory developed by Tomonaga, Feynman, Schwinger and Dyson\footnote{See \cite{SchwingerBook} for selected collection of papers on QED.} became totally consistent, and even today it remains the most accurate description of a physical phenomena.

On the other hand, the progress in our understanding of the weak interactions was a step behind. The four fermion interaction theory developed by Fermi \cite{Fermi:1934hr,Fermi:1934sk} was known to be plagued with infinities that were impossible to remove. It was clear that a high energy completion was required and the possibility of an intermediate vector boson was being discussed as a potential solution to these problems.

Finally, the zoo of baryons and mesons that were discovered along the years was understood after the establishment of the quark model proposed by Gell-Mann and Zweig \cite{GellMann:1964nj,Zweig:1981pd}. The idea that hadrons (baryons and mesons) were not fundamental particles but bound states of quarks was used to classify the discovered states according to their quantum numbers, in analogy to what Mendeleev did with the periodic table of the chemical elements. Results from deep inelastic scattering experiments and the discovery of predicted new states provided strong experimental evidence in favor of internal structure for hadrons, and Gell-Mann was awarded the Nobel Prize in 1969. However, the underlying dynamics responsible for the quark interactions inside hadrons was still to be found.

One of the major breakthroughs in the development of the Standard Model was the work by Yang and Mills \cite{Yang:1954ek}. The extension of gauge symmetries to non-abelian Lie groups was the fundamental piece of the puzzle that allowed theorists to build intermediate vector boson models with a symmetry basis. However, the masslessness of the gauge bosons was still a problem, since it was already clear that the weak interactions are mediated by a massive particle. The missing ingredient, key for the consistency of the full theory, was the mechanism of spontaneous symmetry breaking. The works by Englert, Brout, Guralnik, Hagen, Kibble and Higgs \cite{Higgs:1964ia,Higgs:1964pj,Higgs:1966ev,Guralnik:1964eu,Englert:1964et}, preceded by Goldstone and its famous theorem \cite{Goldstone:1961eq}, showed that the gauge bosons get masses when the vacuum structure of the theory leads to the spontaneous breaking of the gauge symmetry. Nowadays, the so-called Higgs mechanism is vastly employed as a way to generate masses for particles and its applications beyond fundamental particle physics are well known.

With all these pieces, Weinberg, Salam, Glashow and other leading theorists in the late 60s gave birth to the Standard Model \cite{Glashow:1961tr,Weinberg:1967tq,SalamBook}. Soon after its invention, a remarkable theoretical development gave strong support to the model. The work by Veltman and 't Hooft, who showed in 1971 that gauge theories are renormalizable, even after spontaneous symmetry breaking \cite{'tHooft:1971fh,'tHooft:1971rn,'tHooft:1972fi,'tHooft:1972ue}, was fundamental to ensure the validity of the model beyond the tree-level approximation. This consistency check was prior to the experimental support given by the discovery of weak neutral currents mediated by the $Z$ boson at CERN. Many other tests were done in the subsequent years and the SM passed all the challenges, becoming one of the most robust foundations of current physics.

A full description of the Standard Model and its most important features is far from the scope of this thesis. However, in order to introduce properly the hierarchy problem a brief review will be presented. For more detailed and complete introductions the reader is referred to the well-known book \cite{Quigg:1983gw} or more recent reviews \cite{Novaes:1999yn,Quigg:2007dt}.

The Standard Model is a gauge theory based on the group $SU(3)_c \times SU(2)_L \times U(1)_Y$. This already sets the spin 1 particle content, this is, the gauge bosons of the model. Since $SU(N)$ has $N^2-1$ generators, 8 gauge bosons are required for $SU(3)_c$, the gluons, and 3 for $SU(2)_L$, the W bosons. Finally, the gauge boson for the $U(1)_Y$ subgroup is the B boson.

Fermions must be assigned to irreducible representations of the gauge group. On the one hand, the $SU(2)_L \times U(1)_Y$ piece describes the weak and electromagnetic interactions. The left-handed fermions are given doublet representations under $SU(2)_L$, whereas the right-handed ones are singlets. The observed violation of parity is thus explicitly introduced in the theory by making left- and right-handed fermions different from the beginning. Note however that, for practical purposes, it is convenient to work with the conjugates of the left-handed fermions, which are right-handed, instead of the right-handed components themselves. On the other hand, the $SU(3)_c$ piece is the one responsible for the the strong interactions that only quarks feel, and so they are assigned to triplets of this subgroup.

Finally, it is a well known fact in Yang-Mills theories that mass terms for non-singlets representations are forbidden by gauge invariance. Therefore, the symmetry must be broken in order to generate masses for fermions and gauge bosons. For this purpose a scalar $SU(2)_L$ doublet is added. When the so-called Higgs doublet gets a vacuum expectation value (VEV) the gauge symmetry is broken to the group $SU(3)_c \times U(1)_Q$, where $Q$ stands for electric charge. This is the observed symmetry at low energies.

In table \ref{sm-particle-content} the full particle content of the SM is presented.

\begin{table}

\begin{center}
{
\renewcommand\arraystretch{1.5} 
\begin{tabular}{|c|c|c|c|c|}
\hline
Name & Representation & $SU(3)_c$, $SU(2)_L$, $U(1)_Y$ \\
\hline
\hline
\multirow{3}{*}{quarks} &  ($u_L$ $d_L$) & ( 3 , 2 , $\frac{1}{6}$ )\\
& $u_R^c$ & ( 3 , 1 , $-\frac{2}{3}$ )\\
& $d_R^c$ & ( 3 , 1 , $\frac{1}{3}$ )\\
\hline
\multirow{2}{*}{leptons} &  ($\nu_L$ $l_L$) & ( 1 , 2 , $-\frac{1}{2}$ )\\
& $l_R^c$ & ( 1 , 1 , 1 )\\
\hline
\hline
Higgs & ($\phi^+$ $\phi^0$) & ( 1 , 2 , $\frac{1}{2}$ )\\
\hline
\hline
B boson & $B^0$ & ( 1 , 1 , 0 )\\
\hline
W bosons & $W_1$, $W_2$, $W_3$ & ( 1 , 3 , 0 )\\
\hline
gluons & $g$ & ( 8 , 1 , 0 )\\
\hline
\end{tabular}
}
\end{center}
\caption[SM particle content]{SM particle content. Hypercharge is defined following the convention $Q = I_{3L}+Y$. Only the first quark and lepton families are presented.} \label{sm-particle-content}
\end{table}

The most general lagrangian for $H$, the Higgs doublet, invariant under the gauge symmetry is

\begin{equation}
\mathcal{L}_{Higgs} = \partial_\mu H^\dagger \partial^\mu H - V(H^\dagger H)
\end{equation}

where the scalar potential $V(H^\dagger H)$ is

\begin{equation}
V(H^\dagger H) = \mu^2 H^\dagger H + \lambda (H^\dagger H)^2
\end{equation}

\begin{figure}
\centering
\includegraphics[width=0.8\textwidth]{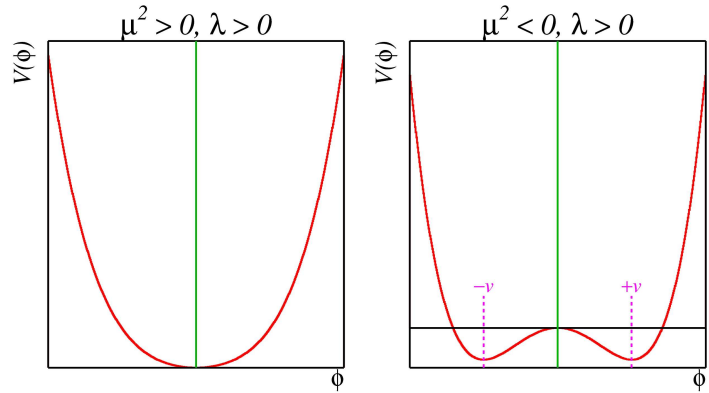}
\caption[SM Higgs potential.]{The well-known SM Higgs potential for two parameter configurations. If $\lambda > 0$ and $\mu^2 < 0$ the minimum of the potential lies at $\langle H \rangle \neq 0$, leading to the spontaneous breaking of the gauge symmetry. Picture thanks to Ian C. Brock and taken from his website http://www-zeus.physik.uni-bonn.de/~brock/index.php.}
\label{mexhat}
\end{figure}

The parameters $\mu$ and $\lambda$ determine the vacuum structure and their values are crucial for the validity of the model. For example, for the potential to be bounded from below one needs the condition $\lambda > 0$. Moreover, as shown in figure \ref{mexhat}, if $\mu^2 < 0$ the minimum of the potential is not at $\langle H \rangle = 0$, but at

\begin{equation}
\langle H \rangle = \left( \begin{array}{c}
0 \\
\frac{v}{\sqrt{2}}
\end{array} \right)
\end{equation}

where

\begin{equation}
v = \sqrt{- \frac{\mu^2}{\lambda}}
\end{equation}

The vacuum expectation value of the Higgs doublet breaks the gauge symmetry $SU(3)_c \times SU(2)_L \times U(1)_Y$ to $SU(3)_c \times U(1)_Q$, giving masses to the gauge bosons and leading to the observed symmetries at low energies. In addition, Yukawa couplings between the Higgs doublet and the fermions

\begin{equation}
Y H \bar{f}_R f_L + h.c.
\end{equation}

induce Dirac masses for all of them except for the neutrino, due to the lack of a right-handed neutrino component\footnote{In principle, one can generate Dirac masses for the neutrinos just by adding a right-handed neutrino field to the SM. However, this proposal would require tiny neutrino Yukawa couplings ($Y_\nu \lesssim 10^{-11}$) as needed to explain the smallness of neutrino masses. Such Yukawa parameters, much smaller than the other fermion Yukawas, would be theoretically unmotivated, making the idea of Dirac neutrinos in the SM unappealing. That is the reason why neutrino masses are usually supposed to come from a different origin, as explained in chapter \ref{chap:numass}.}.

This is the Higgs mechanism. The spontaneous breaking of the gauge symmetry gives masses to gauge bosons and fermions. By writing the neutral component of the Higgs doublet as $\phi^0 = (h + v)/\sqrt{2}$, this is, relative to the minimum of the potential, one finds that the interactions of the Higgs doublet with the other particles generate masses

\begin{eqnarray}
M_{\text{gauge boson}}^2 &\sim& v^2 \\
M_{\text{fermion}} &\sim& v
\end{eqnarray}

In addition, there is one scalar degree of freedom left in the spectrum. The neutral scalar $h$ is a physical state, the so-called Higgs boson, with mass

\begin{equation}\label{hmass-tree}
m_h = \sqrt{-2 \mu^2}
\end{equation}

However, no numerical prediction for this mass can be made, since the value of $\mu^2$ is unknown. In that sense, the Standard Model predicts the existence of a new particle, the Higgs boson, but cannot determine its mass.

So far only the tree-level mass of the Higgs boson has been discussed. To the result in equation \eqref{hmass-tree} one has to add radiative corrections coming from the interactions of the Higgs boson with the rest of particles.

\begin{figure}
\centering
\subfigure[Scalar contribution]{
\includegraphics[width=0.45\textwidth]{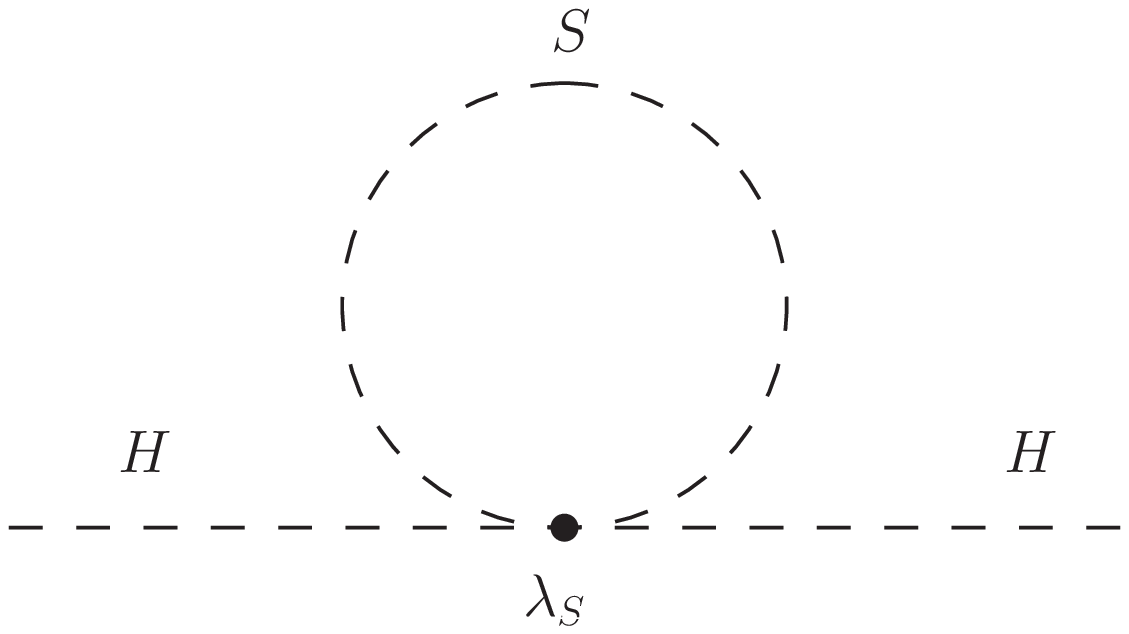}
\label{mHloopS4}
}
\subfigure[Fermion contribution]{
\includegraphics[width=0.45\textwidth]{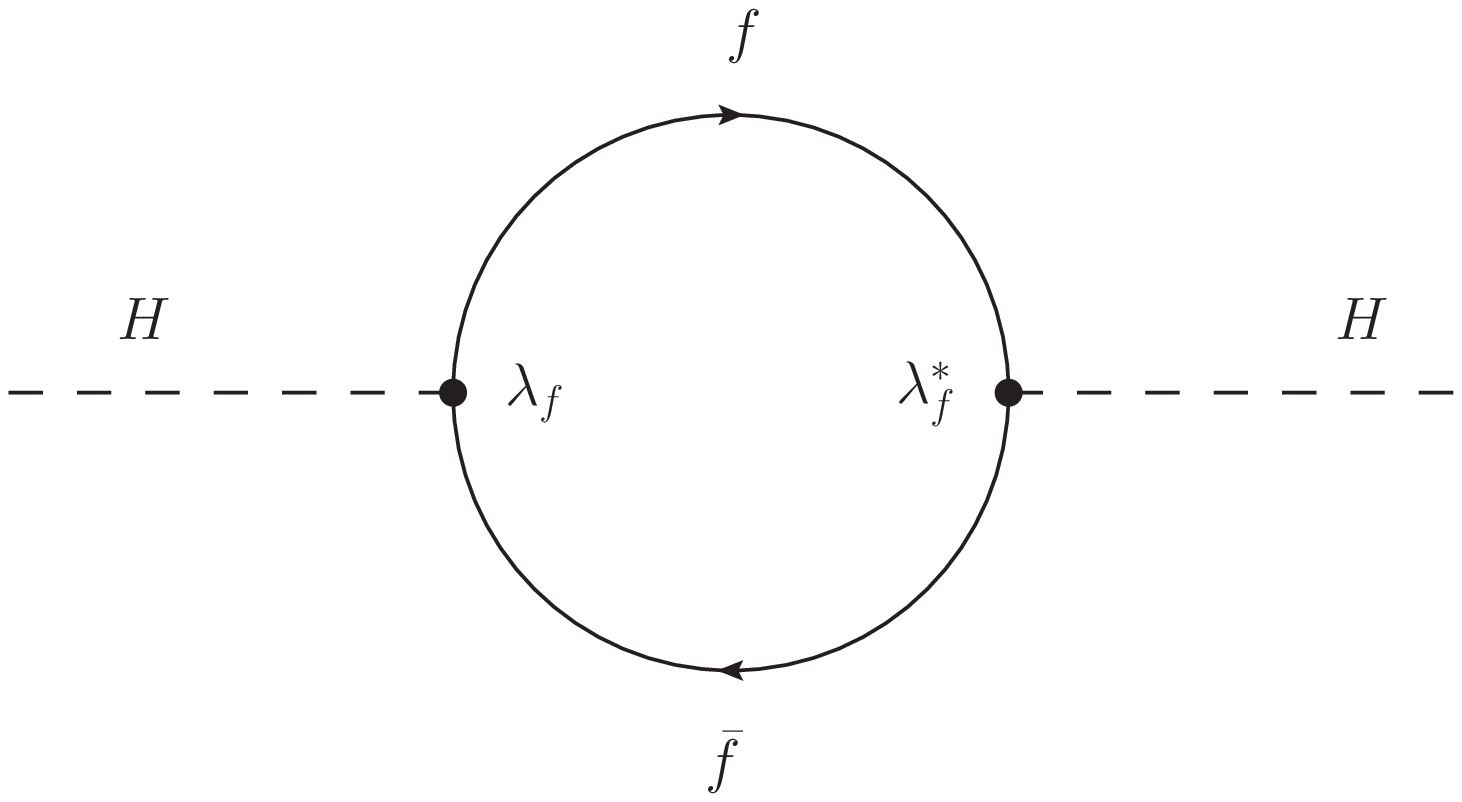}
\label{mHloopF}
}
\label{mHloop}
\caption[1-loop corrections to the Higgs boson mass]{Feynman diagrams leading to 1-loop corrections to the Higgs boson mass.}
\end{figure}

Let us consider a scalar $S$ with mass $m_S$ that couples to the Higgs boson with an interaction term of the form $- \lambda_S |H|^2 |S|^2$. Then the Feynman diagram in figure \ref{mHloopS4} gives a contribution

\begin{equation} \label{hie-1}
\left( \Delta m_h^2 \right)_S \sim \lambda_S \int \frac{d^4p}{(2 \pi)^4} \frac{1}{p^2-m_S^2}
\end{equation}

By dimensional analysis, this contribution is proportional to $m_S^2$. If the scalar $S$ is a heavy particle, with a mass much above the electroweak scale, such quadratic corrections will be much larger than the Higgs boson tree-level mass.

Let us now consider a Dirac fermion $f$ with mass $m_f$ that couples to the Higgs boson with a Yukawa interaction term $- \lambda_f H \bar{f} P_L f + h.c.$, where $P_L = \frac{1}{2}(1-\gamma_5)$ is the left chirality projector. Then, its contribution to the Higgs boson mass is given by diagram \ref{mHloopF} and turns out to be

\begin{equation} \label{hie-2}
\left( \Delta m_h^2 \right)_f \propto - |\lambda_f|^2 \int \frac{d^4p}{(2 \pi)^4} \frac{Tr}{(p^2-m_f^2)^2}
\end{equation}

where $Tr = Tr[(\slashed{p}+m_f)P_L(\slashed{p}+m_f)P_R] = 2 p^2$. Using this result for the fermionic trace equation \eqref{hie-2} splits into

\begin{equation}
\left( \Delta m_h^2 \right)_f \propto - |\lambda_f|^2 \int \frac{d^4p}{(2 \pi)^4} \frac{1}{p^2-m_f^2} + \text{log}
\end{equation}

Where log corresponds to a logarithmic integral that can be absorbed by choosing the right renormalization scale. Again, there is a quadratically divergent integral that is proportional to $m_f^2$. As we saw for the case of the scalar diagram, if the fermion $f$ has a large mass $m_f \gg m_h$ the correction $\left( \Delta m_h^2 \right)_f$ will be also much larger than the tree-level mass.

The problem appears when we think of the SM as an effective theory, obtained at low energies from an extended model that describes physics at higher energies. For example, if the SM is to be embedded in a Grand Unification Theory (GUT), the corrections to the Higgs boson mass given by the particles at the GUT scale are, according to equations \eqref{hie-1} and \eqref{hie-2}, proportional to the square of their masses. Moreover, new physics is also expected at the Planck scale $M_p$, at which gravity must be described in terms of a new quantum theory. Therefore, when this is taken into account one obtains corrections to the Higgs boson mass proportional to $M_p^2 = (10^{18}$ GeV$)^2$, much larger than its natural value at the electroweak scale.

This shows that the Higgs boson, and any scalar particle in general, is extremely sensitive to the existence of physics at higher energies. With such large quadratic corrections it is hard to understand how the mass of the Higgs boson could be at the electroweak scale. Unless a very precise conspiracy among the different contributions makes them cancel, there is no reason to think that the Higgs boson would remain as light as to be part of the spectrum of the SM. This is the famous hierarchy problem\footnote{For a very pedagogical description of the hierarchy problem the reader is referred to the introductory review \cite{Drees:1996ca}.}.

Nevertheless, note that if

\begin{equation} \label{hie-3}
m_S^2 = m_f^2
\end{equation}

and

\begin{equation}  \label{hie-4}
\lambda_S = | \lambda_f |^2
\end{equation}

the scalar and fermion contributions cancel exactly. However, for that to happen there must be a reason, a symmetry that relates fermions and bosons.

That symmetry is Supersymmetry.

In the following sections the basic concepts of Supersymmetry and the MSSM will be introduced. However, a very brief review will be presented, leaving many technical details for more complete references. For a rigorous superfield treatment it is highly recommended to use the references \cite{Nilles:1983ge,Haber:1984rc,Sohnius:1985qm}. On the other hand, the well-known reference \cite{Martin:1997ns} has a more intuitive approach, highlighting the most important phenomenological details of the MSSM and its minimal extensions. Finally, the recent text \cite{Baer:2006rs} gives a modern review of our current understanding and perspectives.

\section{Basic concepts of Supersymmetry}

Supersymmetry is a symmetry that relates bosons and fermions. Although its origin is far from the hierarchy problem, the most important consequence of its introduction is the technical solution that it offers to this theoretical drawback of the SM.

Supersymmetry has many advantages with respect to non-supersymmetric theories. Both from the phenomenological and the purely theoretical sides, motivations for Supersymmetry can be found:

\begin{itemize}

\item It solves the hierarchy problem.

\item It helps to understand the mass problem (radiative symmetry breaking).

\item It extends the particle spectrum, introducing new states which can potentially be dark matter candidates.

\item Minimal models without additional states lead naturally to gauge coupling unification at high scales.

\item It extends the symmetries of space-time.

\item Local supersymmetry opens a window to a quantum theory for gravity: supergravity.

\item Supersymmetric theories have better ultraviolet behavior than non-super\-symmetric ones.

\end{itemize}

For brevity only the gauge hierarchy problem and how the cancellation of quadratic divergencies occurs in SUSY will be discussed in some detail below. For the rest the reader can see the well-known literature \cite{Nilles:1983ge,Haber:1984rc,Sohnius:1985qm,Martin:1997ns,Baer:2006rs}.

These advantages have made supersymmetric theories the object of detailed study for the last years, as can be clearly seen in the scientific literature. Many proposals for supersymmetric models have been made in order to solve questions unanswered in the standard model. From the simplest SUSY extension, the MSSM, to more complicated constructions, issues like electroweak symmetry breaking, neutrino masses, flavor structure and grand unification have been addressed with the help of Supersymmetry.

\subsection{Foundations of Supersymmetry}

Symmetries play a central role in our current understanding of physics. Not only do they have the power to explain the links between different phenomena, but also they have been shown to be a fundamental tool to describe the dynamics of a physical system. Many examples support this idea and this thesis cannot list them all.

In particular, symmetries have been used for a long time in the field of particle physics. From the pioneering works of Yang and Mills \cite{Yang:1954ek} and their subsequent transformation into our present gauge theories, to their constant use in flavor physics, with Gell-Mann's Eightfold way \cite{GellMann:1964nj} as prominent example, one can find many cases that clearly show how important symmetries are in the development of our current theories. It is therefore a good strategy to look for higher symmetry principles that can provide a connection for phenomena that seem to be unconnected.

Nowadays, particle physics is built on the theoretical framework of Quantum Field Theory (QFT). Having special relativity as one of its constituent pieces, the symmetries of spacetime in QFT are encoded in the Poincar\'e group. Under this group, coordinates transform as

\begin{equation}
x_\mu \: \to \: x_\mu' = \Lambda_\mu^\nu x_\nu + a_\mu
\end{equation}

where $\Lambda$ is a Lorentz transformation and $a_\mu$ a spacetime translation. The generators of these transformations are $M^{\mu \nu}$, a tensor that includes rotations and boosts, and the four-momentum $P^\mu$. They follow the algebra

\begin{eqnarray}
\left[ P^\mu,P^\nu \right] &=& 0 \\
\left[ M^{\mu \nu},P^\lambda \right] &=& i (P^\mu g^{\nu \lambda} - P^\nu g^{\mu \lambda}) \\
\left[ M^{\mu \nu},M^{\lambda \sigma} \right] &=& i (M^{\mu \sigma} g^{\nu \lambda} + M^{\nu \lambda} g^{\mu \sigma} - M^{\mu \lambda} g^{\nu \sigma} - M^{\nu \sigma} g^{\mu \lambda})
\end{eqnarray}

where $g^{\mu \nu} = diag(1,-1,-1,-1)$ is the Minkowski metric tensor.

In order to characterize the irreducible representations one has to find the Casimir operators, this is, the operators that commute with all the generators of the algebra. A representation will be labeled by the value that these invariant operators take when they act on them. In the case of the Poincar\'e group the Casimir operators are

\begin{eqnarray}
P^2 &=& P_{\mu}P^{\mu} \\
W^2 &=& W_{\mu}W^{\mu}
\end{eqnarray}

The eigenvalues of the $P^2$ operator are known to be the masses of the particles, whereas

\begin{equation}
W_{\mu}=-\frac{i}{2}\epsilon_{\mu \nu \lambda \sigma} M^{\nu \lambda} P^{\sigma}
\end{equation}

is the Pauli-Lubanski operator, whose eigenvalues are related to the spin of the particle or, in the case of a massless particle, to its helicity. Therefore, an irreducible representation of the Poincar\'e group, what we call \emph{particle}, will be characterized by its mass and its spin or helicity.

In the sixties the question arose whether it was possible to extend the Poincar\'e group in order to combine it with internal symmetries of the particles in a non-trivial way. By internal symmetries we understand symmetry groups that act on internal properties of the particles and whose generators are Lorentz scalars. Examples of such symmetries are electric charge and isospin.

An answer to this issue was given by the famous no-go theorem by Coleman and Mandula \cite{Coleman:1967ad}, who showed that the most general symmetry of the S matrix is the direct product Poincar\'e $\otimes$ internal, where these two groups act independently in a trivial way. However, the proof contained a loophole, since it only considered commutating generators. When one allows for the existence of anticommutating generators new possibilities open up, as shown by Haag, Lopuszanski and Sohnius, who extended the Coleman-Mandula theorem in 1975 \cite{Haag:1974qh}. Instead of following commutation relations, the fermionic generators (as they are also known in opposition to commutating or bosonic generators) follow anticommutation relations. Including such operators it is possible to unify the Poincar\'e group with the internal symmetries.

A few years earlier the first works that incorporated fermionic generators \cite{Golfand:1971iw} and supersymmetric field theory \cite{Volkov:1972jx,Volkov:1973ix,Wess:1974tw} had already appeared in what can be considered the birth of Supersymmetry. The work by Wess and Zumino \cite{Wess:1974tw} is usually considered the starting point for the study of supersymmetric field theory and most of our current formalism is based on their early developments.

Let us consider fermionic generators $Q_{\alpha}$ and $\bar{Q}_{\dot{\alpha}}$, two components Weyl spinors ($\alpha, \dot{\alpha} = 1, 2$). These operators satisfy the algebra

\begin{gather}
\{Q_{\alpha},Q_{\beta}\}=0 \\
\{\bar{Q}_{\dot{\alpha}},\bar{Q}_{\dot{\beta}}\}=0\\
\{Q_{\alpha},\bar{Q}_{\dot{\beta}}\}=2(\sigma^{\mu})_{\alpha \dot{\beta}} P_{\mu}
\end{gather}

where $\sigma^{\mu}=(1,\sigma^i)$ and $\sigma^i$ are the Pauli matrices. Their commutation relations with the usual generators of the Poincar\'e group are

\begin{gather}
[P^{\mu},Q_{\alpha}]=0\\
[M^{\mu \nu},Q_{\alpha}]=-i{(\sigma^{\mu \nu})_{\alpha}}^{\beta} Q_{\beta}
\end{gather}

The resulting symmetry group is the Super-Poincar\'e group and the algebra followed by its generators is the Super-Poincar\'e algebra. An algebra that includes fermionic generators is called graded algebra or super-algebra.

Analogously to the Poincar\'e group one must find the Casimir operators in order to characterize the irreducible representations of the group. It is easy to show that, after the addition of the new fermionic generators, the vanishing conmutator

\begin{equation}
[Q_{\alpha},P^2]=0
\end{equation}

is still obtained. However

\begin{equation}
[Q_{\alpha},W^2]\ne0
\end{equation}

which shows that the irreducible representations of the SUSY algebra, the so-called supermultiplets, must include particles with the same mass, but different spins/helicities. The fermionic generators $Q_{\alpha}$ and $\bar{Q}_{\dot{\alpha}}$ transform bosons into fermions and vice versa

\begin{equation}
Q_{\alpha} |\text{fermion} \rangle = |\text{boson} \rangle \quad , \quad \bar{Q}_{\dot{\alpha}} |\text{boson} \rangle = | \text{fermion} \rangle
\end{equation}

Particles in the same supermultiplets are called superpartners. It is easy to show that in any supermultiplet the number of bosonic degrees of freedom equals the number of fermionic degrees of freedom.

There are several methods to build supersymmetric lagrangians, this is, lagrangians which are invariant under SUSY transformations. The most elegant one is based on the superfield formalism \cite{Nilles:1983ge,Haber:1984rc,Sohnius:1985qm} where all the fields belonging to the same supermultiplet are described using one single function defined in superspace. However, it is also possible to work with component fields \cite{Martin:1997ns} and get the same conclusions. In either case, these are the basic ingredients that one has to specify in order to build a supersymmetric theory:

\begin{enumerate}

\item \textbf{Particle content}

\end{enumerate}

The first step is to specify the particles in our model. One has to take into account that SUSY implies that for every particle in the spectrum one must add another one with the same mass but different spin. Therefore, instead of particles one usually speaks of superfields that contain both. Two types of superfields will be considered in the following: (1) chiral superfields, which contain a scalar and a fermion, and (2) vector superfields, which contain a fermion and a vector boson.

\begin{enumerate}

\setcounter{enumi}{1}

\item \textbf{Gauge group}

\end{enumerate}

Next, one has to choose a gauge group and assign gauge charges to the particles in the model.

\begin{enumerate}

\setcounter{enumi}{2}

\item \textbf{Superpotential}

\end{enumerate}

The non-gauge interactions are introduced in Supersymmetry using a mathematical object called superpotential. This is an holomorphic function of the superfields and it is how Yukawa couplings are included in the theory. Its general form is highly restricted by SUSY.

\begin{enumerate}

\setcounter{enumi}{3}

\item \textbf{Soft lagrangian}

\end{enumerate}

Up to now only unbroken SUSY theories have been discussed. As already explained, SUSY implies that for every fermion there must be a boson with the same mass. That would mean, for example, the existence of a boson with the mass of the electron. However, such particle would have been already discovered. Therefore, if Supersymmetry is realized in nature, it cannot be an exact symmetry. It must be broken. This way, particles in the same supermultiplet would have different masses as needed to account for the non-discovery of the superpartners of the SM fermions.

The way SUSY is broken is unknown and many mechanisms have been proposed, see \cite{Intriligator:2007cp} for a review. In practice, this ignorance is solved by introducing by hand new terms in the lagrangian that break SUSY explicitly but preserve the properties that motivate it. In particular, these new terms, known as soft SUSY breaking terms, do not spoil the solution that SUSY offers for the hierarchy problem. It has been rigorously shown \cite{Girardello:1981wz} that this requirement constrains the possible new terms, that can only contain masses for scalars and gauginos, and couplings for the scalars of the same type as in the superpotential\footnote{In general, a term is said to break SUSY \emph{softly} if it has positive mass dimension. The rest of SUSY breaking terms are called \emph{hard}.}.

With these four ingredients one can build a realistic supersymmetric model. The simplest example is the MSSM, the minimal supersymmetric extension of the standard model.

\subsection{The MSSM}

The Minimal Supersymmetric Standard Model (MSSM) was proposed in the 80s as a solution to the hierarchy problem \cite{Dimopoulos:1981zb} and since then it has been object of intense study.

In the MSSM, and any SUSY model in general, each SM particle is put in a supermultiplet together with its superpartner. This leads to a duplication in the number of particles and makes necessary to create a new way to name particles. The names of the scalar superpartners are made by adding the prefix `s-'. For example, the scalar partners of the leptons are the sleptons. On the other hand, the name of the fermion superpartners is obtained by adding the sufix `-ino' to the name of the particle. This way, the superpartner of the photon is the photino, whereas the superpartner of the Higgs boson is the higgsino. Both for scalars and fermions the standard notation for the SM superpartners is given by a tilde added to the symbol of the SM particle. For example, the selectron would be written as $\tilde{e}$.

The gauge group of the MSSM is the same as in the SM, $SU(3)_c \times SU(2)_L \times U(1)_Y$, and the gauge quantum numbers of the superfields are assigned in the same way. In principle one could think of having twice as particles as in the SM, due to the SUSY duplication. However, there is an additional detail that further extends the spectrum. One Higgs doublet superfield is not sufficient, and a second one has to be added. There are two reasons for this:

\begin{itemize}

\item The superpotential has to be holomorphic. That means that it has to be written in terms of the superfields, and not their complex conjugates. With this restriction, and due to $U(1)_Y$ gauge invariance, one cannot write Yukawa couplings for both up-type and down-type quarks only with a Higgs doublet with hypercharge $Y=+\frac{1}{2}$. Therefore, one needs to add a second Higgs doublet, with hypercharge $Y=-\frac{1}{2}$, in order to build gauge invariant Yukawa couplings for both quark sectors.

\item In the SM the gauge anomalies cancel exactly. However, after extending the spectrum to include the new SM superpartners, the higgsino spoils this cancellation. The solution is again the addition of a second Higgs doublet superfield with the opposite hypercharge, that cancels this new contribution.

\end{itemize}

\begin{table}
\begin{center}
{
\renewcommand\arraystretch{1.5} 
\begin{tabular}{|c|c|c|c|c|}
\hline
\multicolumn{2}{|c|}{Name} & Spin 0 & Spin 1/2 & $SU(3)_c$, $SU(2)_L$, $U(1)_Y$ \\
\hline
\hline
\multirow{3}{*}{squarks, quarks} & $\widehat{Q}$ & ($\tilde{u}_L$ $\tilde{d}_L$) &  ($u_L$ $d_L$) & ( 3 , 2 , $\frac{1}{6}$ )\\
& $\widehat{u}^c$ & $\tilde{u}_R^*$ & $u_R^c$ & ( 3 , 1 , $-\frac{2}{3}$ )\\
& $\widehat{d}^c$ & $\tilde{d}_R^*$ & $d_R^c$ & ( 3 , 1 , $\frac{1}{3}$ )\\
\hline
\multirow{2}{*}{sleptons, leptons} & $\widehat{L}$ & ($\tilde{\nu}_L$ $\tilde{l}_L$) &  ($\nu_L$ $l_L$) & ( 1 , 2 , $-\frac{1}{2}$ )\\
& $\widehat{e}^c$ & $\tilde{l}_R^*$ & $l_R^c$ & ( 1 , 1 , 1 )\\
\hline
\multirow{2}{*}{Higgs, higgsinos} & $\widehat{H}_d$ & ($H_d^0$ $H_d^-$) &  ($\tilde{H}_d^0$ $\tilde{H}_d^-$) & ( 1 , 2 , $-\frac{1}{2}$ )\\
& $\widehat{H}_u$ & ($H_u^+$ $H_u^0$) & ($\tilde{H}_u^+$ $\tilde{H}_u^0$) & ( 1 , 2 , $\frac{1}{2}$ )\\
\hline
\end{tabular}

\vspace{0.4cm}

\begin{tabular}{|c|c|c|c|c|}
\hline
\multicolumn{2}{|c|}{Name} & Spin 1/2 & Spin 1 & $SU(3)_c$, $SU(2)_L$, $U(1)_Y$ \\
\hline \hline
bino, B boson & $\widehat{V}_1$ & $\tilde{B}^0$ & $B^0$ & ( 1 , 1 , 0 )\\
\hline
wino, W bosons & $\widehat{V}_2$ & $\tilde{W}_1$, $\tilde{W}_2$, $\tilde{W}_3$ & $W_1$, $W_2$, $W_3$ & ( 1 , 3 , 0 )\\
\hline
gluino, gluon & $\widehat{V}_3$ & $\tilde{g}$ & $g$ & ( 8 , 1 , 0 )\\
\hline
\end{tabular}
}
\end{center}
\caption[MSSM particle content]{MSSM particle content. Hypercharge is defined following the convention $Q = I_{3L}+Y$. Only the first quark and lepton families are presented.} \label{mssm-particle-content}
\end{table}

The superpotential of the MSSM is

\begin{equation}\label{mssm-superpotencial}
W^{MSSM} = \epsilon_{ab} \big[ Y_u^{ij} \widehat{Q}_i^a
\widehat{u}_j^c \widehat{H}_u^b + Y_d^{ij} \widehat{Q}_i^b
\widehat{d}_j^c \widehat{H}_d^a + Y_e^{ij} \widehat{L}_i^b
\widehat{e}_j^c \widehat{H}_d^a - \mu \widehat{H}_d^a \widehat{H}_u^b
\big]
\end{equation}

where $i,j = 1,2,3$ are family indices, $a,b = 1,2$ are $SU(2)$ indices and $\epsilon_{ab}$ is the totally antisymmetric Levi-Civita tensor. The $Y_u$, $Y_d$ and $Y_e$ matrices are the usual Yukawa couplings while $\mu$ is a parameter with dimensions of mass.

If SUSY was conserved the number of free parameters in the model would be smaller than in the SM. However, the introduction of the soft lagrangian changes this fact since it contains many new free parameters. Moreover, note that without the soft terms it would be impossible to break the electroweak symmetry. In the SUSY limit the scalar potential is constrained to be strictly positive, and thus spontaneous symmetry breaking cannot be realized.

The soft SUSY breaking terms of the MSSM are

\begin{eqnarray}\label{mssm-soft}
- \mathcal{L}_{soft}^{MSSM} = && (m_Q^2)^{ij} \tilde{Q}_i^{a \ast} \tilde{Q}_j^a + (m_{u^c}^2)^{ij} \tilde{u}^c_i \tilde{u}_j^{c \: \ast} + (m_{d^c}^2)^{ij} \tilde{d}^c_i \tilde{d}_j^{c \: \ast} + (m_L^2)^{ij} \tilde{L}_i^{a \ast} \tilde{L}_j^a \nonumber \\
&& + (m_{e^c}^2)^{ij} \tilde{e}^c_i \tilde{e}_j^{c \: \ast} + m_{H_d}^2 H_d^{a \ast} H_d^a +  m_{H_u}^2 H_u^{a \ast} H_u^a \nonumber \\
&& - \big[ \frac{1}{2} M_1 \tilde{B}^0 \tilde{B}^0 + \frac{1}{2} M_2 \tilde{W}^c \tilde{W}^c + \frac{1}{2} M_3 \tilde{g}^d \tilde{g}^d + h.c. \big] \\
&& + \epsilon_{ab} \big[ T_u^{ij} \tilde{Q}_i^a \tilde{u}^c_j
H_u^b + T_d^{ij} \tilde{Q}_i^b \tilde{d}^c_j H_d^a + T_e^{ij}
\tilde{L}_i^b \tilde{e}^c_j H_d^a - B_\mu H_d^a H_u^b \big] \nonumber
\end{eqnarray}

where $c = 1,2,3$ and $d = 1,\dots,8$. Sometimes the trilinear parameters in equation \eqref{mssm-soft} are expanded as $T_\alpha^{ij} = A_\alpha^{ij} Y_\alpha^{ij}$, with $\alpha = u,d,e$. However, we will follow the convention of \cite{Skands:2003cj,Allanach:2008qq} instead. Note that $\mathcal{L}_{soft}$ clearly breaks SUSY, since it introduces interaction terms for some fields but not for their superpartners.

The electroweak symmetry is broken when the scalar Higgs doublets $H_d$ and $H_u$ get VEVs

\begin{eqnarray}
\langle H_d^0 \rangle &=& \frac{v_d}{\sqrt{2}} \\
\langle H_u^0 \rangle &=& \frac{v_u}{\sqrt{2}}
\end{eqnarray}

and, like in the SM, this mechanism gives masses to the gauge bosons $Z^0$ and $W^{\pm}$ and, due to the Yukawa couplings, to the quarks and charged leptons. However, unlike the SM, there are two important differences

\begin{enumerate}

\item Due to the existence of two complex Higgs doublets we have eight degrees of freedom. When the gauge bosons \emph{eat} three of them to give mass to their longitudinal components, five physical states remain in the spectrum. These are the neutral $h^0$, $H^0$, $A^0$ and the charged $H^\pm$.

\item Higgsinos and $SU(2)_L \times U(1)_Y$ gauginos mix after the breaking of the electroweak symmetry. The neutral higgsinos ($\tilde{H}_d^0$ and $\tilde{H}_u^0$) and the neutral gauginos ($\tilde{B}^0$ and $\tilde{W}^0$) lead to four mass eigenstates known as neutralinos, $\tilde{\chi}_{1 \dots 4}^0$. On the other hand, the charged higgsinos ($\tilde{H}_d^-$ and $\tilde{H}_u^+$) and the charged winos $\tilde{W}^\pm$ also mix in two mass eigenstates known as charginos, $\tilde{\chi}_{1 , 2}^\pm$. Squarks, sleptons and gluinos also mix separately.

\end{enumerate}

Apart from these novelties, very interesting from the phenomenological point of view, supersymmetry offers an important improvement in our understanding of electroweak symmetry breaking. As discussed below, the mechanism of radiative symmetry breking is one of its most attractive motivations and is a natural consequence once the MSSM is embedded in a larger framework with universal conditions at the GUT scale\footnote{Another key ingrediente for radiative symmetry breaking is a large top quark mass, which is in fact the case as measured by the Tevatron collider.}.

Models with many parameters are not predictive. In the case of the MSSM, the large number of arbitrary parameters that appear in the soft lagrangian \eqref{mssm-soft} makes it hard to give numerical predictions for observables\footnote{There are some exceptions, however, like the mass of the lightest Higgs boson, $h^0$, which can be predicted in supersymmetry due to the constrained form of the scalar potential.}. From a practical point of view, it would be necessary to reduce the number of parameters by making some assumptions on the way supersymmetry gets broken. In fact, if the SUSY breaking mechanism was known, all the soft parameters would be linked to a few, the ones that describe the dynamics of the sector that breaks SUSY. However, if the soft parameters are totally free, the high dimensional parameter space will be extremely difficult to analyze.

Therefore, it is common to assume some SUSY breaking scenarios. Without specifying the complete theory that breaks SUSY one can describe its effects by imposing some universal conditions to the soft parameters at some high energy scale where SUSY gets broken. These conditions clearly reduce the number of parameters. Then, with the embedding of the MSSM in a more general framework one can use the renormalization group equations (RGEs) of the MSSM to obtain the resulting values of these soft parameters at the electroweak scale.

Moreover, these universal conditions are also motivated by two well-known problems of the MSSM: the SUSY flavor problem and the SUSY CP problem. The values of the soft parameters at the SUSY scale are strongly constrained due to their contributions to flavor-changing and CP violating processes. For example, the off-diagonal element $(m_{e^c}^2)_{12}$ contributes to the lepton flavor violating decay $\mu \to e \gamma$, whose branching ratio is bounded to be below $1.2 \times 10^{-11}$ by the MEGA experiment \cite{Ahmed:2001eh,Amsler:2008zzb}. By assuming flavor and CP blind conditions for the soft parameters at some high scale, the resulting flavor off-diagonal entries and CP violating phases at the SUSY scale are strongly suppressed, leading to low rates for the dangerous processes.

One of the most popular scenarios is the cMSSM (constrained MSSM), also known as mSUGRA (minimal SUperGRAvity)\footnote{In the following we will use the names mSUGRA and cMSSM as synonyms, although this is not totally correct. In fact, minimal supergravity models lead to a more contrained SUSY breaking scenario than what will be considered here. However, due to the usual naming in the literature, both names will be used in this thesis without making a distinction. See \cite{Olive:2010tm} for some comments on this issue.}. The breaking of SUSY is due to gravitational effects. These conditions are imposed at the GUT scale:

\begin{gather}
m_Q^2 = m_{u^c}^2 = m_{d^c}^2 = m_L^2 = m_{e^c}^2 = m_0^2 \mathbb{I}\\
m_{H_d}^2 = m_{H_u}^2 = m_0^2\\
M_1 = M_2 = M_3 = M_{1/2}\\
T_u = A_0 Y_u \, , \, T_d = A_0 Y_d \, , \, T_e = A_0 Y_e
\end{gather}

where $\mathbb{I}$ is the identity in family space. Then one is left with the following free parameters: a universal gaugino mass ($M_{1/2}$), a universal scalar mass ($m_0$), the trilinear coupling $A_0$ and, since they are not embedded in this framework, the $\mu$ and $B_\mu$ parameters.

The $\mu$ and $B_\mu$ parameters belong to the Higgs sector of the MSSM and thus can be linked to the VEVs $v_d$ and $v_u$ through the minimization conditions. The MSSM tadpole equations are

\begin{eqnarray}
m_{H_d}^2 + |\mu|^2 - B_\mu \tan \beta + \frac{m_Z^2}{2} \cos 2\beta &=& 0 \label{mssm-tad1} \\
m_{H_u}^2 + |\mu|^2 - B_\mu \cot \beta - \frac{m_Z^2}{2} \cos 2\beta &=& 0 \label{mssm-tad2}
\end{eqnarray}

where

\begin{equation}
\tan \beta = \frac{v_u}{v_d}
\end{equation}

is the ratio between the VEVs of the Higgs doublets.

A comment concerning the $\mu$ parameter must be made here. As shown in the tadpole equations, its interplay with the rest of the parameters in the Higgs sector determines the vacuum structure of the theory. Therefore, the value of the $\mu$ parameter must lie around the electroweak scale. Otherwise, equations \eqref{mssm-tad1} and \eqref{mssm-tad2} would need to be fine-tuned in order to cancel the different contributions and lead to the correct symmetry breaking, with $v_d$ and $v_u$ at the electroweak scale. On the other hand, the $\mu$ term in the superpotential is a supersymmetry conserving term and thus it is expected to lie at the GUT scale or beyond. In that case one would need an extreme fine-tuning for electroweak symmetry breaking to work. This discrepancy between the phenomenologically required value and the theoretical expectation is known as the $\mu$-problem \cite{Kim:1983dt}. Several solutions to this naturalness problem of the MSSM have been proposed and one of them will be discussed in this thesis.

From equations \eqref{mssm-tad1} and \eqref{mssm-tad2} it is clear that one can exchange $\mu$ and $B_\mu$ for $\tan \beta$, since for any value of this ratio one can solve them to find the appropriate values for $\mu$ and $B_\mu$ that lead to the minimum of the scalar potential. Note however that the sign of $\mu$ is not determined and thus is left as a free parameter. Therefore, the mSUGRA parameter space is described by $m_0$, $M_{1/2}$, $A_0$, $\tan \beta$ and $sign(\mu)$.

In addition to making the model much more predictive, the MSSM embedding in this framework leads to an additional consequence concerning electroweak symmetry breaking. By inspection of the potential one can find some conditions that must be fulfilled by the parameters in the Higgs sector. These come from demanding a potential bounded from below and the non-stability of the trivial minimum with $v_d = v_u = 0$. The resulting conditions are

\begin{eqnarray}
(B_\mu)^2 &>& (m_{H_d}^2 + |\mu|^2)(m_{H_u}^2 + |\mu|^2) \label{mssm-cond1} \\
|B_\mu| &<& 2|\mu|^2 + m_{H_d}^2 + m_{H_u}^2 \label{mssm-cond2}
\end{eqnarray}

In order for these inequalities to make sense the soft SUSY breaking parameters are expected to be close to the EW scale. Otherwise very large cancellations are required. As will be shown below, this is also needed to provide a technical solution to the hierarchy problem.

Note that equations \eqref{mssm-cond1} and \eqref{mssm-cond2} cannot be satisfied simultaneously if $m_{H_d}^2 = m_{H_u}^2$, equality that holds at the GUT scale due to the mSUGRA universal conditions. However, these two parameters evolve differently in their RGE running from the GUT scale. In fact, the large contributions given by the top Yukawa, present for $m_{H_u}^2$ but not for $m_{H_d}^2$, naturally imply a negative value for $m_{H_u}^2$, allowing for the spontaneous breaking of the EW symmetry. In conclusion, the EW symmetry is broken by radiative effects, in the so-called Radiative EW symmetry breaking mechanism, another interesting property of supersymmetric models.

\subsection{Supersymmetry and the hierarchy problem}

\begin{figure}
\centering
\subfigure[]{
\includegraphics[width=0.35\textwidth]{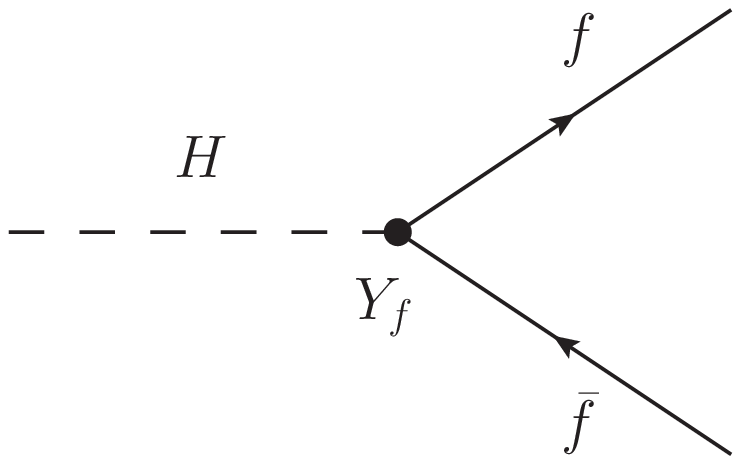}
\label{yukawa-SUSY}
}
\hspace{0.5cm}
\subfigure[]{
\includegraphics[width=0.25\textwidth]{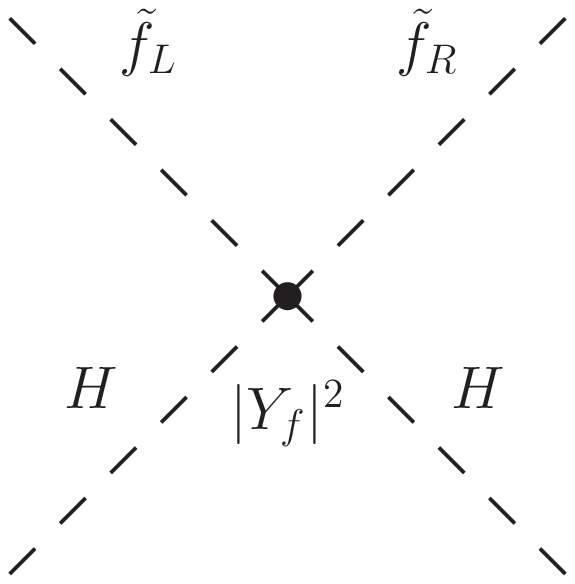}
\label{4point-SUSY}
}
\caption[Yukawa and 4-point scalar interactions obtained from a superpotential term $Y_f \widehat H \widehat f_L \widehat f_R$.]{Yukawa and 4-point scalar interactions obtained from a superpotential term $Y_f \widehat H \widehat f_L \widehat f_R$.}
\label{interactions-SUSY}
\end{figure}

\begin{figure}
\centering
\subfigure[]{
\includegraphics[width=0.45\textwidth]{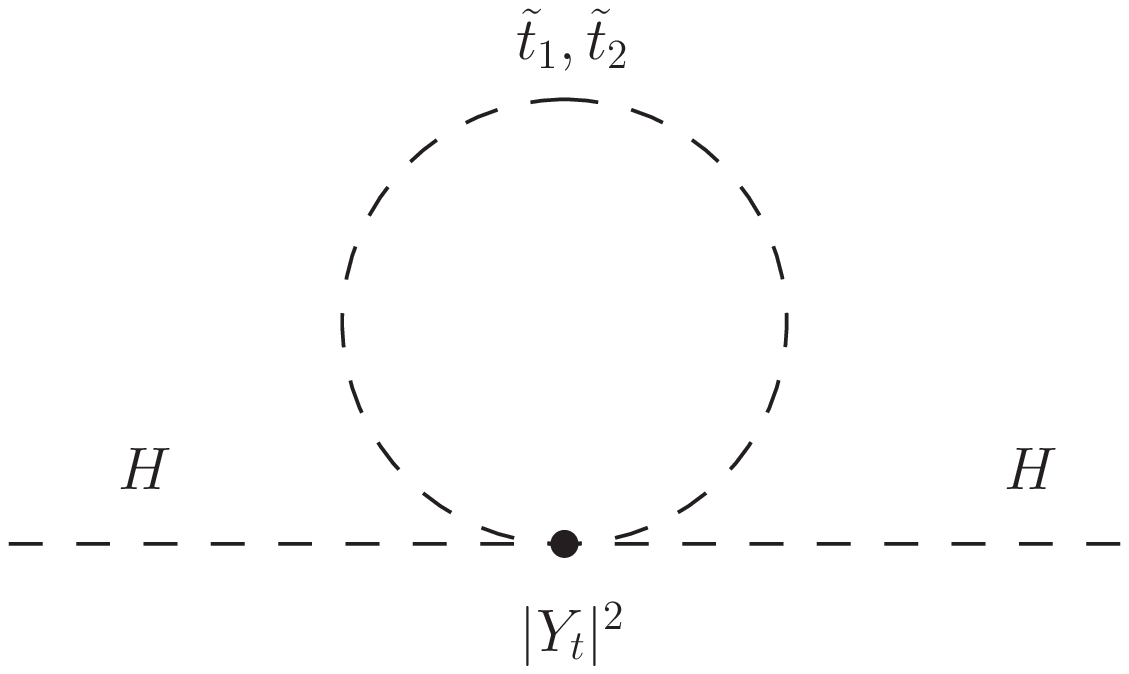}
\label{mHloopS4MSSM}
}
\subfigure[]{
\includegraphics[width=0.45\textwidth]{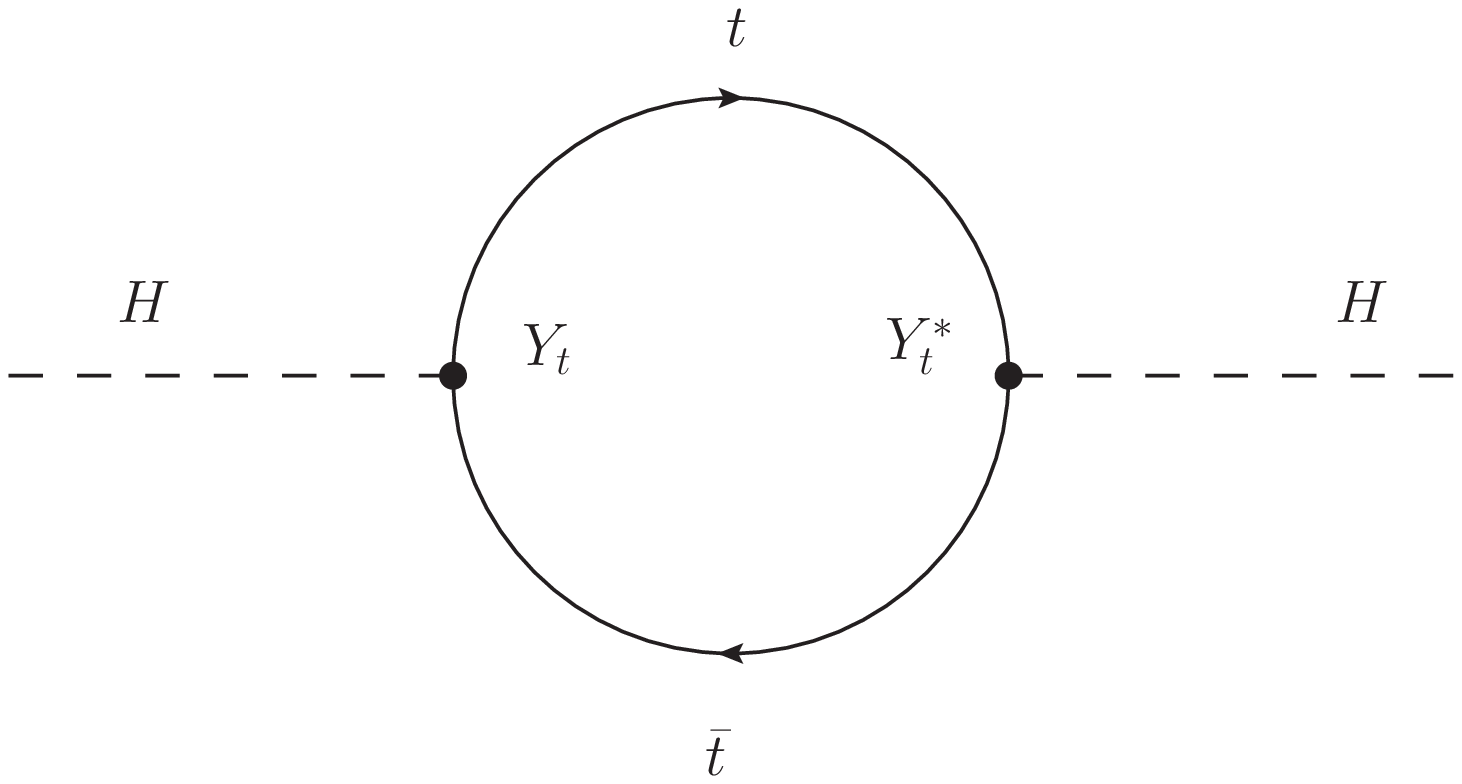}
\label{mHloopFMSSM}
}
\caption[Dominant 1-loop corrections to the Higgs boson mass in the MSSM]{Top-Stop 1-loop corrections to the Higgs boson mass in the MSSM.}
\label{mHloopMSSM}
\end{figure}

After this brief introduction let us consider the main motivation for supersymmetry: the solution to the hierarchy problem \cite{Dimopoulos:1981zb}. The Yukawa interaction in diagram \ref{yukawa-SUSY} and the scalar 4-point interaction in diagram \ref{4point-SUSY} are obtained from the same superpotential term. With this common origin, imposed by SUSY, it is clear that a cancellation is likely to happen. In fact, if we consider the top quark and the corresponding scalar top, the so-called stop, and their contributions to the Higgs boson mass, we get the diagrams in figure \ref{mHloopMSSM}. Here $\tilde{t}_1$ and $\tilde{t}_2$ are the resulting mass eigenstates induced by $\tilde{t}_L$ and $\tilde{t}_R$ mixing. When the electroweak symmetry gets broken by the Higgs VEV these chiral states mix leading to two mass eigenstates $\tilde{t}_1$ and $\tilde{t}_2$.

Note that both diagrams in figure \ref{mHloopMSSM} share the same coupling, following the condition \eqref{hie-4}, necessary for cancellation. This is a direct consequence of SUSY, not spoiled by the fact that it is broken softly.

\begin{figure}
\centering
\includegraphics[width=0.45\textwidth]{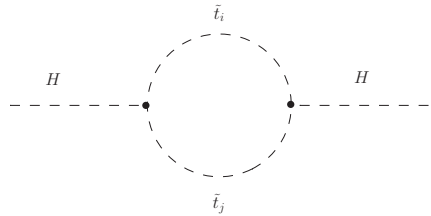}
\caption[Dominant 1-loop corrections to the Higgs boson mass in the MSSM]{Top-Stop 1-loop corrections to the Higgs boson mass in the MSSM. This diagram only appears after electroweak symmetry breaking.}
\label{mHloopS3MSSM}
\end{figure}

In addition, after electroweak symmetry breaking the Feynman diagram in figure \ref{mHloopS3MSSM} appears. However, this contribution gives a correction proportional to

\begin{equation}
\int \frac{d^4 p}{(2 \pi)^4} \frac{1}{(p^2-m_{\tilde{t}_i}^2)(p^2-m_{\tilde{t}_j}^2)} \sim \int \frac{d^4 p}{p^4}
\end{equation}

which is logarithmic and thus does not contribute to the dangerous quadratic divergences.

Summing up all contributions one can calculate the 1-loop correction to the Higgs boson mass. In the SUSY limit, with $\mathcal{L}_{soft} = 0$, one gets a total cancellation. If the soft breaking terms are switched on this cancellation is not exact anymore, but a correction to the mass of the Higgs boson is obtained \cite{Haber:1990aw,Okada:1990vk,Okada:1990gg,Ellis:1990nz,Ellis:1991zd}. An explicit calculation gives the leading order result

\begin{equation}
\Delta m_h^2 \propto Y_t^2 m_t^2 \ln \left( \frac{m_{\tilde{t}_1} m_{\tilde{t}_2}}{m_t^2} \right)
\end{equation}

Three comments are in order:

\begin{itemize}

\item If $m_t = m_{\tilde{t}_1} = m_{\tilde{t}_2}$ the correction vanishes. This is, as explained, what is obtained with unbroken SUSY.

\item The correction is logarithmic. No quadratic divergence appears.

\item Without this correction the MSSM would be ruled out. It can be easily shown that the tree-level scalar potential of the MSSM predicts a very light Higgs boson, $m_h < m_Z \cos(2\beta)$, ruled out by LEP data. This tree-level bound is due to the fact that the quartic coupling of the Higgs boson in the MSSM is not free, but originates from gauge interactions, directly connected to the gauge bosons masses. With the top-stop contribution, the 1-loop corrected Higgs boson mass evades the LEP bound making the MSSM a viable model.

\end{itemize}

From the previous discussion it is clear that the soft SUSY breaking parameters must have values close to the electroweak scale. This was already shown when the tadpole equations of the MSSM were discussed (see eqs. \eqref{mssm-cond1} and \eqref{mssm-cond2}). Now, another indication has been presented. If the SUSY scale, this is, the scale at which the soft parameters lie, is much higher than the electroweak scale, the 1-loop corrections to the Higgs boson mass will be large, and a small fine-tuning will be required. These two arguments point to the same direction: the SUSY scale cannot be very far away from the energy scales under current exploration.

In conclusion, if the soft parameters have values close to the electroweak scale the correction to the Higgs mass are under control. Only logarithmic corrections are obtained and the Higgs boson mass remains naturally at the electroweak scale. This is the supersymmetric solution to the hierarchy problem.

\section{R-parity} \label{sec:rp}

The superpotential \eqref{mssm-superpotencial} is not the most general renormalizable superpotential that is compatible with SUSY and gauge invariance. It is actually possible to add new terms, leading to

\begin{equation}
W = W^{MSSM} + W^{\textnormal{\rpv}}
\end{equation}

where

\begin{equation}\label{rpv-superpotencial}
\begin{split}
W^{\textnormal{\rpv}} = & \: \epsilon_{ab} \big[ \frac{1}{2} \lambda_{ijk} \widehat{L}_i^a \widehat{L}_j^b \widehat{e}^c_k  + \lambda'_{ijk} \widehat{L}_i^a \widehat{Q}_j^b \widehat{d}^c_k + \epsilon_i \widehat{L}_i^a \widehat{H}_u^b \big]\\
& \: + \frac{1}{2} \lambda''_{ijk} \widehat{u}^c_i \widehat{d}^c_j \widehat{d}^c_k
\end{split}
\end{equation}

\begin{figure}
\begin{center}
\includegraphics[width=0.6\textwidth]{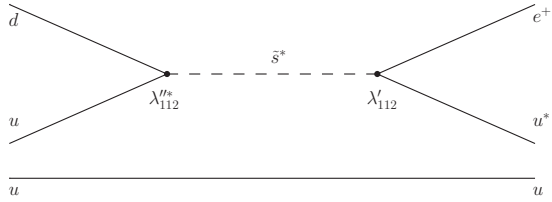}
\end{center}
\caption[Proton decay]{The squarks could mediate a disastrous fast proton decay if both $\lambda'$ and $\lambda''$ are present simultaneously.}
\label{proton-decay}
\end{figure}

The first three terms in $W^{\textnormal{\rpv}}$ break lepton number (L) whereas the last one breaks baryon number (B). The presence of these new couplings is not welcome, since there is no evidence so far for any physical process that breaks B or L. The strongest restriction comes from the non-observation of proton decay, that would break both B and L. If the $\lambda'$ and $\lambda''$ couplings were simultaneously present the life time of the proton would be extremely short unless they are extremely tiny \cite{Hinchliffe:1992ad,Vissani:1995hp,Smirnov:1996bg,Hoang:1997kf,Bhattacharyya:1998bx,Bhattacharyya:1998dt}. For example, Feynman diagramms like the one in figure \ref{proton-decay} contribute to proton decay. A simple estimate based on dimensional analysis gives

\begin{equation}
\Gamma_{p \rightarrow e^+ \pi^0} \sim m_{\textnormal{proton}}^5 \sum_{i=2,3} \frac{| \lambda'_{11i} \lambda''_{11i} |^2}{m_{\tilde{d}_i}^4}
\end{equation}

If $\lambda'$ y $\lambda''$ are of order $1$ and the squarks have masses around the TeV, this equation would be translated into a life time of a fraction of a second. This phenomenological disaster is solved in the MSSM by introducing a new symmetry by hand that forbids the terms in $W^{\textnormal{\rpv}}$. This symmetry is known as R-parity \cite{Fayet:1974pd,Farrar:1978xj}.

The R-parity of a particle is defined as

\begin{equation}\label{Rp}
R_p = (-1)^{3(B-L)+2s}
\end{equation}

where $s$ is the spin of the particle. It is easy to check that all SM particles have $R_p=+1$ while their supartners have $R_p=-1$ (in the following particles with $R_p=-1$ will be called superparticles). It is straightforward to verify that all terms in \eqref{rpv-superpotencial} break R-parity. The MSSM is defined as R-parity conserving.

This phenomenological requirement can be seen as a step back from the SM, where this problem does not exist, since all the allowed renormalizable interactions preserve both B and L. However, as discussed below in this thesis, R-parity can be motivated as a remaining symmetry of a larger group, broken at higher energy scales leading to the MSSM. Nevertheless, the conservation of R-parity has very important implications:

\begin{itemize}

\item The Lightest Supersymmetric Particle (LSP) is stable. If it is electrically neutral and it has no color, it would only interact weakly, being a good dark matter candidate.

\item Superparticles are produced in pairs at colliders.

\item Every superparticle, apart from the LSP, decays into a final state with an odd number of LSPs. Since they are stable and escape detection at colliders, this is seen as a missing energy signal.

\end{itemize}

With these advantages, R-parity is a very practical assumption, made by most of the theoretical studies based on supersymmetry.

\subsection{R-parity violation}\label{subsec:susy-rpv}

R-parity is a central element in the MSSM. Its conservation is imposed to forbid the highly constrained lepton and baryon number violating interactions. In addition, it provides a dark matter candidate, opening the way to the solution to one of the major problems in modern cosmology.

However, no robust theoretical argument supports R-parity and its violation is an interesting alternative that offers many changes with respect to the standard picture\footnote{See \cite{Barbier:2004ez} for a detailed review on R-parity violation.}. In particular, in this thesis R-parity violation \rpv has been explored in connection with neutrino masses, since the violation of lepton number automatically leads to masses for the neutrinos.

This is easy to understand. The \rpv superpotential term $\epsilon_i \widehat{L}_i \widehat{H}_u$ mixes neutrinos with neutral higgsinos, which have Majorana masses coming from the MSSM couplings. This way, by choosing a small $\epsilon$ mixing term neutrinos get masses in the proper range. This is a simple realization of the seesaw mechanism at the EW/SUSY scale.

A comment should be made here about proton decay. As shown a few lines above, both L and B violations are needed for the proton to decay. Therefore, if one relaxes R-parity conservation to B conservation (or any other symmetry that allows for L violation) neutrino masses are generated while the proton is still stable.

If R-parity is broken, collider phenomenology will be completely different. In particular, in \rpv SUSY the LSP is no longer stable and decays to final states containing SM particles. This changes all the standard strategies for SUSY searches, focused on final states with missing energy. There are important implications in cosmology as well. The standard LSP dark matter candidate is lost. Nevertheless, other theoretical possibilities are available and an explanation for the observed amount of dark matter is still possible. Well studied examples are the gravitino \cite{Borgani:1996ag,Takayama:2000uz,Hirsch:2005ag}, the axion \cite{Kim:1986ax,Raffelt:1996wa} and its supersymmetric partner, the axino \cite{Chun:1999cq,Chun:2006ss}.

The main advantage of R-parity violating neutrino mass models with respect to R-parity conserving ones is their clear predictions for colliders. As will be shown along the lines of this thesis, LSP decays are closely linked to the flavor structure in the neutrino sector, allowing for precise predictions at colliders. This makes \rpv neutrino mass models one of the very few examples where colliders can really tell us about the origin of neutrino masses, in opposition to the standard SUSY seesaw framework that can only be tested indirectly in some particular scenarios\footnote{The non-SUSY seesaw mechanism does even worse, and neither direct nor indirect tests are known. The large mass of the right-handed neutrinos, or the heavy fields in other realizations, makes impossible any chance to produce them at colliders. As we will see, SUSY seesaw models have an advantage, with additional phenomenology due to the flavor information hidden in the slepton sector.}.

Finally, the constraints coming from the non-observation of L or B violating processes imply that \rpv couplings must be small \cite{Barbier:2004ez,Kao:2009fg}. Therefore, if one allows for \rpv, model building is required if one wants to explain the smallness of the new couplings. This is another important issue related to \rpv that will be considered in this thesis.\\

In conclusion, two approaches are considered in this thesis concerning the role of R-parity:

\begin{itemize}

\item R-parity violation: If R-parity is not conserved, how is it broken? And how can we probe it? What are the signals at the LHC? What is the connection to neutrino masses?

\item R-parity conservation: If R-parity is a good symmetry at the EW/SUSY scale, what is the mechanism behind? How does the neutrino get a mass and how can we test it?

\end{itemize}

These are the two main lines of research that will be further explained in the following chapters of this thesis.

\chapter{Neutrino mass models}
\label{chap:numass}

Since the invention of the Standard Model and its subsequent development no experimental result challenged its foundations. This agreement between theory and experiment, obtained even in high precision measurements, established the Standard Model as a good description of all the known phenomena in particle physics. However, the discovery of neutrino oscillations in the late 90s was the first clear evidence for physics beyond the Standard Model, and pointed to the necessity of an extension in the leptonic sector that can account for neutrino masses. In this chapter several neutrino mass models will be reviewed, emphasizing their most important properties and how they are related to the present thesis.

\section{Current experimental situation}

Based on the fact that neutrinos were always observed to be left-handed, as opposed to the other fermions that could be found with both chiralities, and because no experimental result pointed to a non-zero neutrino mass, the fathers of the Standard Model decided not to add right-handed neutrinos to the particle spectrum.

Without right-handed neutrinos it is not possible to write down a Yukawa term that can lead to Dirac masses for neutrinos. Therefore, neutrinos are massless in the Standard Model. This statement is valid at any order in perturbation theory due to the underlying $U(1)_{B-L}$ symmetry.

However, this simplistic choice was shown to be wrong after the establishment of neutrino oscillations as the mechanism behind the observed neutrino deficits. For many years, different experiments showed a clear difference between the neutrinos that were detected and the theoretical expectations. This is nowadays understood in terms of neutrino flavor oscillations, a mechanism that implies that neutrinos must be massive.

Let us begin with a short review of the experimental situation.

\subsection{Neutrino deficits}

The Sun produces neutrinos in the nuclear reactions that continuously occur in its interior. These neutrinos escape in all directions, some of them reaching the Earth and our detectors. With a precise knowledge of the solar evolution and structure one can determine the neutrino flux on Earth and the number of neutrinos that are expected to be detected by a given experiment.

For many years, several experiments have been accumulating data on solar neutrinos. To name a few, Homestake \cite{Davis:1994jw,Cleveland:1998nv,Lande:1999cv}, SAGE \cite{Abdurashitov:1999zd,Abdurashitov:2002nt}, GALLEX-GNO \cite{Hampel:1998xg,Altmann:2000ft,Cattadori:2002rd}, Kamiokande \cite{Fukuda:1996sz}, Super-Kamiokande \cite{Fukuda:2001nj,Fukuda:2002pe,Smy:2002fs,Poon:2005qu} and SNO \cite{Ahmad:2002jz,Ahmad:2002ka,Ahmed:2003kj,Aharmim:2005gt} are experiments designed to detect solar neutrinos and measure their properties. These collaborations were able to determine solar neutrino fluxes that one can compare with the theoretical expectation.

However, all the experiments detected less neutrinos than predicted by the Standard Solar Model (SSM) \cite{Bahcall:1995bt,Bahcall:1998wm,Bahcall:2000nu,Bahcall:2004fg,Bahcall:2004pz,Bahcall:2005va}. This theoretical model has successfully passed the observational tests along the years and is considered a robust description of the processes inside the sun. However, it fails to predict the number of neutrinos detected on Earth, which is reduced by $30-60$\%. This discrepancy is known as the solar neutrino problem.

Furthermore, neutrinos are also produced in the atmosphere. When a cosmic ray hits an air molecule in the higher parts of the atmosphere, a particle shower is produced, including some neutrinos that travel towards the Earth, where detectors are placed underground to measure their properties. Theoretical predictions for the atmospheric neutrino fluxes and the expected event rates at different experiments can be obtained by estimating the primary cosmic ray flux, something that has been done with increasing accuracy along the years \cite{Gaisser:1988ar,Barr:1989ru,Honda:2004yz}.

In fact, one can take advantage of the capability of the experiments to distinguish between electron and muon neutrinos. The main neutrino production mechanism is charged pion decay, which has a branching ratio close to $100$\% into the channel \cite{Amsler:2008zzb}

\begin{displaymath}
\pi^\pm \to \mu^\pm + \nu_\mu (\bar{\nu}_\mu)
\end{displaymath}

This is followed by the standard muon decay

\begin{displaymath}
\mu^\pm \to e^\pm + \nu_e (\bar{\nu}_e) + \bar{\nu}_\mu (\nu_\mu)
\end{displaymath}

and thus one expects two muon neutrinos for each electron neutrino

\begin{equation}
R_{e \mu} \equiv \frac{N(\nu_\mu)}{N(\nu_e)} \simeq 2
\end{equation}

As in the case of the solar neutrinos, several experiments have detected the neutrinos produced in the atmosphere. Frejus \cite{Daum:1994bf}, IMB \cite{BeckerSzendy:1992hq}, NUSEX \cite{Aglietta:1988be} and Kamiokande \cite{Hirata:1992ku,Fukuda:1994mc,Fukuda:1996sz} are examples of such experimental setups, sensitive to this type of neutrinos. And again, as for solar neutrinos, a deficit with respect to the theoretical expection has been observed. The measured ratio $R_{e \mu}$ was about $0.6$ times the predicted value, showing a clear deficit of muon neutrinos. This anomaly is known as the atmospheric neutrino problem.

For many years, these two anomalies stimulated the imagination of theoreticians, who invented several mechanisms to explain the observed deficits. Non-standard neutrino interactions \cite{Wolfenstein:1977ue,Valle:1987gv,Guzzo:1991hi,Roulet:1991sm}, decaying neutrinos \cite{Bahcall:1972my,Bahcall:1986gq,Acker:1990zj,Acker:1992eh,Acker:1993sz,Barger:1998xk,Barger:1999bg}, spin flavor precession \cite{Schechter:1981hw,Lim:1987tk,Akhmedov:1988uk}, neutrino decoherence \cite{Grossman:1998jq,Lisi:2000zt} or violation of Lorentz invariance \cite{Gasperini:1988zf,Gasperini:1989rt} were theoretical ideas that appeared in the community to address the neutrino anomalies. However, as the accuracy in the experiments increased all these hypothesis were discarded as the main contribution to the phenomenon, in favor of neutrino oscillations, nowadays established as the main mechanism behind these observations.

\subsection{The neutrino oscillation mechanism}

Neutrino oscillations were first discussed by Bruno Pontecorvo in 1957 \cite{Pontecorvo:1957cp,Pontecorvo:1957qd}. However, these first references concentrate on neutrino-antineutrino oscillations. Ten years later, Pontecorvo was again the first to discuss neutrino flavor oscillations \cite{Pontecorvo:1967fh}, in an important pioneering paper that opened up a rich field in particle physics.

If neutrinos are massive their flavor changes while they propagate. As a consequence, a neutrino which is originally produced as electron neutrino can be detected as muon or tau neutrino. This oscillating effect explains the deficits found in solar and atmospheric neutrino experiments.

Neutrinos are produced by charged current weak interactions as flavor eigenstates $\nu_\alpha = (\nu_e, \nu_\mu, \nu_\tau)$. If they are massive, the mass eigenstates $\nu_i = (\nu_1, \nu_2, \nu_3)$ will be in principle different. These two basis are connected by a unitary matrix $U$, defined as

\begin{equation}
| \nu_\alpha \rangle = U_{\alpha i}^* | \nu_i \rangle
\end{equation}

$U$ is known nowadays as Pontecorvo-Maki-Nakagawa-œôòóSakata (PMNS) matrix and is the analog of the Cabibbo-Kobayashi-Maskawa (CKM) matrix that describes the mixing in the quark sector.

The mismatch between these two basis leads to flavor oscillations in the neutrino propagation. This is because, although the production mechanism is flavor diagonal, the quantum mechanical evolution is mass diagonal. Given the state $| \nu_i (t_0) \rangle$ at time $t_0$, one can compute the resulting state at a later time $t$ as 

\begin{equation}
| \nu_i (t) \rangle = e^{i E_i (t-t_0)} | \nu_i (t_0) \rangle
\end{equation}

where $E_i^2 = p^2 + m_i^2$. Then, by projecting both mass eigenstates into the flavor basis one can easily compute the probability of a flavor eigenstate $\nu_\alpha$ oscillating into the flavor eigenstate $\nu_\beta$:

\begin{equation} \label{transition-prob}
P(\nu_\alpha \to \nu_\beta) = \sum_{j,k=1}^3 U_{\alpha k}^* U_{\beta k} U_{\alpha j} U_{\beta j}^* \exp \left( -i \frac{\Delta m^2_{kj} L}{2 E} \right)
\end{equation}

where $L \simeq t - t_0$ for ultrarelativistic neutrinos and $\Delta m^2_{kj} \equiv m_k^2 - m_j^2$.

Some additional assumptions have been made in this derivation. The interested reader can find good references in the scientific literature, where these details are discussed and neutrino oscillations are treated using more robust theoretical tools \cite{Boehm:1992nn,Kim:1994dy,Bilenky:1998dt,Giunti:2004yg,Strumia:2006db,Akhmedov:2009rb}.

In particular, equation \eqref{transition-prob} is lacking matter effects. If neutrinos oscillate in matter, the picture can change drastically, leading to very different results. In fact, the so-called Mikheyevœôòó-Smirnovœôòó-Wolfenstein (MSW) effect \cite{Mikheev:1986gs,Wolfenstein:1977ue}, a resonant process that enhances the transition probability inside the Sun, is fundamental for our understanding of the solar neutrino fluxes and without it the oscillation solution to the solar neutrino problem would not work. 

Note that equation \eqref{transition-prob} implies that neutrino oscillations are not sensitive to the absolute value of neutrino masses, but only to their squared mass differences $\Delta m^2$. As will be discussed below, a different type of experiments is needed in order to measure the absolute scale of neutrino masses.

As the precision in neutrino physics experiments increased, the oscillation solution to the solar and atmospheric neutrino problems got established. In addition to the discussed experiments, new reactor and accelerator experiments gave strong support for the interpretation in terms of flavor oscillation, ruling out other theoretical explanations \cite{kl:2008ee,Hosaka:2006zd,Adamson:2008zt}.

Neutrino oscillation experiments have demonstrated that at least two 
neutrinos have non-zero mass \cite{Ahmad:2002jz,Fukuda:1998mi,Eguchi:2002dm}. 
Especially remarkable is that data from both atmospheric neutrino 
\cite{Ashie:2004mr} and from reactor neutrino measurements \cite{kl:2008ee} 
now show the characteristic $L/E$ dependence expected from oscillations, see equation \eqref{transition-prob},
ruling out or seriously disfavouring other explanations of the observed 
neutrino deficits.

It is fair to say that with the most recent data by 
the KamLAND \cite{kl:2008ee}, Super-Kamiokande \cite{Hosaka:2006zd} and MINOS 
collaborations \cite{Adamson:2008zt} neutrino physics has finally entered 
the precision era. In fact, global fits to the available experimental data allow to determine the involved parameters with good accuracy. This way, one can set important constraints on the flavor structure of many neutrino mass models.

For the case of three neutrinos the $3 \times 3$ PMNS mixing matrix can be parametrized as \cite{Schechter:1980gr}

\begin{equation} \label{pmns-par}
U = \omega_{23} \omega_{13} \omega_{12}
\end{equation}

where $\omega_{ij}$ are effective $2 \times 2$ unitary matrices characterized by an angle and a CP phase. For example

\begin{equation}
\omega_{12} = \left( \begin{array}{c c c}
\cos \theta_{12} & e^{i \phi_{12}} \sin \theta_{12} & 0 \\
- e^{- i \phi_{12}} \sin \theta_{12} & \cos \theta_{12} & 0 \\
0 & 0 & 1 \end{array} \right)
\end{equation}

Expanding the product in equation \eqref{pmns-par} and neglecting the CP violating phases, one obtains

\begin{equation} \label{pmns-product}
    U=\left(
    \begin{array}{ccc}
        c_{13} c_{12}
        & s_{12} c_{13}
        & s_{13} \\
        -s_{12} c_{23} - s_{23} s_{13} c_{12}
        & c_{23} c_{12} - s_{23} s_{13} s_{12}
        & s_{23} c_{13} \\
        s_{23} s_{12} - s_{13} c_{23} c_{12}
        & -s_{23} c_{12} - s_{13} s_{12} c_{23}
        & c_{23} c_{13}
    \end{array} \right)
\end{equation}

where $c_{ij} \equiv \cos \theta_{ij}$ and $s_{ij} \equiv \sin \theta_{ij}$. In addition to these three mixing angles, neutrino oscillations are sensitive to two squared mass differences, $\Delta m_{21}^2$, responsible for solar neutrino oscillations, and $\Delta m_{31}^2$, responsible for atmospheric neutrino oscillations.

\begin{table}
\centering
{
\renewcommand\arraystretch{1.3} 
\begin{tabular}{|l|c|c|c|}
        \hline
        Parameter & Best fit & 2$\sigma$ & 3$\sigma$ 
        \\
        \hline
        $\Delta m^2_{21}\: [10^{-5}\eVq]$
        & $7.59^{+0.23}_{-0.18}$  & 7.22--8.03 & 7.02--8.27 \\
        $|\Delta m^2_{31}|\: [10^{-3}\eVq]$
        & $2.40^{+0.12}_{-0.11}$  & 2.18--2.64 & 2.07--2.75 \\
        $\sin^2\theta_{12}$
        & $0.318^{+0.019}_{-0.016}$ & 0.29--0.36 & 0.27--0.38\\
        $\sin^2\theta_{23}$
        & $0.50^{+0.07}_{-0.06}$ & 0.39--0.63 & 0.36--0.67\\
        $\sin^2\theta_{13}$
        & $0.013^{+0.013}_{-0.009}$  & $\leq$ 0.039 & $\leq$ 0.053 \\
        \hline
\end{tabular}
}
\caption{Best-fit values with 1 $\sigma$ errors, and 2 $\sigma$ and 3 $\sigma$ intervals for the three-flavor neutrino oscillation parameters from global data. Table taken from reference \cite{Schwetz:2008er}, which is continuosly being updated online with the inclusion of new data.}
\label{tab:nudata}
\end{table}

After parametrizing the PMNS matrix one can compare the resulting mixing angles and squared mass differences with the available experimental data. Table \ref{tab:nudata} shows the best-fit values, and the corresponding 1 $\sigma$ errors, and 2 $\sigma$ and 3 $\sigma$ intervals, for the three-flavor neutrino oscillation parameters from global data \cite{Schwetz:2008er}. Three comments are in order:

\begin{itemize}

\item There is a clear hierarchy between the mass scales responsible for solar and atmospheric oscillations.

\item The atmospheric angle, $\theta_{23}$ is compatible with maximal mixing, whereas the solar angle, $\theta_{12}$ is large as well.

\item The so-called reactor angle, $\theta_{13}$, is very small and compatible with zero. In fact, it was only very recently that a slight preference for $\theta_{13} \neq 0$ was found \cite{Fogli:2008jx,Schwetz:2008er,GonzalezGarcia:2010er}.

\end{itemize}

This data is to be explained by any neutrino mass model\footnote{The discussion has been focused on an interpretation of the data based on 3 flavor neutrino oscillations. However, some recent experimental results might point towards a more complicated picture. The LSND experiment \cite{Aguilar:2001ty} reported a signal for a third mass difference in antineutrino oscillations. This would imply the existence of, at least, a fourth sterile neutrino, which does not have weak interactions but mixes with the active neutrinos. More recently, MiniBooNE \cite{AguilarArevalo:2007it,AguilarArevalo:2008rc,AguilarArevalo:2010wv} also found an anomaly, not compatible with 3 flavor neutrino oscillations, adding supporting evidence in favor of the sterile neutrino hypothesis. Nevertheless, more experimental input from other collaborations is required in order to confirm these results and thus they will not be taken into account in this thesis.}.

\subsection{The absolute scale of neutrino masses}

In the last section we saw that neutrino oscillations are not sensitive to the absolute value of neutrino masses, but only to the squared mass differences. Therefore, although we know that it cannot be zero, we cannot get $m_\nu$ from the experiments discussed above.

There are three main experimental/observational sources of information on the absolute scale of neutrino masses.

\begin{itemize}

\item Tritium beta decay experiments

\end{itemize}

As Fermi pointed out in 1934, the shape of the electron spectrum near the endpoint is very sensitive to the scale of neutrino masses \cite{Fermi:1934hr}. Several experiments have applied this idea using tritium (${}^3H$) as decaying nucleus. This is because the decay

\begin{displaymath}
{}^3H \to {}^3He + e^- + \bar{\nu}
\end{displaymath}

is specially favorable, since it is a super-allowed nuclear transition with a low $Q$-value. Examples of such experimental setups are the Mainz \cite{Barth:1998ra} and Troitsk \cite{Lobashev:1999tp} experiments, which are sensitive to the effective electron neutrino mass \cite{Shrock:1980vy}

\begin{equation}
m_\beta = \sqrt{\sum_{i=1}^3 |U_{ei}|^2 m_i^2}
\end{equation}

and have reported the result \cite{Bonn:2001tw,Weinheimer:2003fj}

\begin{equation}
m_\beta < 2.2 \: \text{eV} (95\% c.l.)
\end{equation}

In the near future, the KATRIN experiment \cite{Bornschein:2003xi} will start its operation, with an expected sensitivity of $0.2$ eV after five years of data taking. See references \cite{Bilenky:2002aw,Weinheimer:2002rs,Giunti:2005qd} for reviews on beta decay experiments designed to measure neutrino masses.

\begin{itemize}

\item Neutrinoless double beta decay experiments

\end{itemize}

Another type of experiment searching for the absolute scale of neutrino masses is neutrinoless double beta decay ($0 \nu 2 \beta$). These are lepton number violating processes of the type

\begin{displaymath}
N (A,Z) \to N(A,Z \pm 2) + e^{\mp} + e^{\mp}
\end{displaymath}

without emission of neutrinos. This signal, only possible if neutrinos are Majorana particles \cite{Schechter:1981bd,Hirsch:2006yk}, would allow to measure the combination

\begin{equation}
m_{\beta \beta} = \sum_{i=1}^3 U_{ei}^2 m_i
\end{equation}

The current experimental situation is controversial. A possible indication of $0 \nu 2 \beta$ in ${}^{76}Ge$ decays was obtained by the authors of reference \cite{KlapdorKleingrothaus:2004ge}, who reported the following limits for the half-life of the process

\begin{equation}
T_{1/2}^{0 \nu}({}^{76}Ge) = (0.69 - 4.18) \times 10^{25} y \quad (3 \sigma)
\end{equation}

However, this result has not been confirmed by other experiments. The most stringent bound on the half-life for ${}^{76}Ge$ comes from the Heidelberg-Moscow experiment \cite{KlapdorKleingrothaus:2000sn}

\begin{equation} \label{h-m-bound}
T_{1/2}^{0 \nu}({}^{76}Ge) > 1.9 \times 10^{25} y \quad (90\% c.l.)
\end{equation}

whereas the IGEX experiment \cite{Aalseth:2002rf} obtained a similar lower bound

\begin{equation}
T_{1/2}^{0 \nu}({}^{76}Ge) > 1.57 \times 10^{25} y \quad (90\% c.l.)
\end{equation}

Hopefully the issue will be clarified in the near future, since the experiments GERDA \cite{Zuzel:2010zz}, CUORE \cite{Guardincerri:2009zz} and EXO \cite{Akimov:2005mq} have expected sensitivities that allow to check the positive signal claimed in \cite{KlapdorKleingrothaus:2004ge}.

The extraction of $m_{\beta \beta}$ from an eventual positive signal would have a large theoretical error due to the uncertanties in the computation of the nuclear matrix elements \cite{Civitarese:2002tu,Elliott:2004hr}. Nevertheless, the detection of neutrinoless double beta decay would represent the discovery of a new type of particle, a Majorana particle, and therefore it would be of great relevance even if the extracted effective neutrino mass cannot be measured with high accuracy.

Applying the bound from the Heidelberg-Moscow collaboration, see equation \eqref{h-m-bound}, and taking into account the uncertanties in the nuclear matrix elements \cite{Elliott:2004hr}, one obtains \cite{Giunti:2005qd}

\begin{equation}
m_{\beta \beta} \lesssim 0.3 - 0.6 \: \text{eV}
\end{equation}

which is slightly better than the bound obtained with tritium beta decay experiments.

\begin{itemize}

\item Cosmology

\end{itemize}

Cosmology sets important constraints on the absolute scale of neutrino masses. In fact, the most stringent bounds can be obtained from cosmological observables. In this case the quantity that is constrained is the sum of neutrino masses which, depending on the cosmological data set that is used, is bounded as \cite{Lesgourgues:2006nd}

\begin{equation} \label{cosmo-bounds}
\sum_{i=1}^3 m_i \lesssim 0.3 - 1.0 \: \text{eV}
\end{equation}

The way neutrino masses are constrained by cosmology is easy to understand. If neutrinos have masses of the order of the eV they would constitute a hot dark matter component of the universe. As it is well known, see for example \cite{Hu:1997mj}, this type of dark matter suppresses the formation of structures at small scales of the order of $1-10$ Mpc. Therefore, by studying density fluctuations in the CMB and the Large Scale Structure distribution of galaxies, one can put strong bounds on the sum of neutrino masses as in equation \eqref{cosmo-bounds}.

For more details on the subject see the review \cite{Lesgourgues:2006nd}. \\

In the following sections several neutrino mass models will be discussed. Very detailed reviews on this subject exist in the literature, see for example \cite{King:2003jb,Mohapatra:2005wg,Mohapatra:2006gs,Valle:2006vb,Nunokawa:2007qh}.

\section{Dirac neutrinos} \label{dirac-nu}

The simplest way to introduce neutrino masses in the Standard Model is to follow the same approach as for the rest of the fermions. The addition of three families of right-handed neutrinos, singlets under the SM gauge group, and the corresponding Yukawa couplings

\begin{equation} \label{nu-yuk}
- \mathcal{L}_Y = Y_\nu H L \nu_R + h.c.
\end{equation}

leads to neutrino Dirac masses $m_D = Y_\nu \langle H \rangle$ after electroweak symmetry breaking. Here $Y_\nu$ is a $3 \times 3$ matrix. Since $\langle H \rangle \sim 100$ GeV and the absolute scale for neutrino masses is known to be below $1$ eV, the entries in $Y_\nu$ must be below $10^{-11}$.

Note that the right-handed neutrinos, being gauge singlets, only couple to the rest of particles through the $Y_\nu$ coupling. They do not have gauge interactions and there is no way to write down another gauge invariant operator involving the $\nu_R$ field.

Therefore, if $Y_\nu < 10^{-11}$ as demanded by current data, the right-handed neutrinos couple very weakly to the rest of matter, suppressing their production cross section at colliders. Moreover, with such small couplings their contributions to other processes are negligible. In conclusion, the introduction of Dirac masses in the standard model does not lead to any phenomenological consequence, due to the smallness of the neutrino Yukawa couplings.

Supersymmetry offers additional ingredients. This is due to the fact that the right-handed neutrino comes together with its scalar partner, the right-handed sneutrino $\tilde{\nu}_R$. In general, the soft term $T_\nu H \tilde{L} \tilde{\nu}_R$ induces $\tilde{\nu}_L - \tilde{\nu}_R$ mixing after electroweak symmetry breaking. However, if this trilinear coupling is small, for example due to the assumption $T_\nu = A_\nu Y_\nu$, the right-handed sneutrino will be a pure state with very weak couplings. With these properties, it is a potential dark matter candidate \cite{Asaka:2006fs,Gopalakrishna:2006kr}.

Concerning collider phenomenology, this scenario has a novelty with respect to the non-supersymmetric case. If the right-handed sneutrino is the LSP and R-parity is conserved, all decay chains at colliders will end up producing a pair of them, even though it couples very weakly to the rest of particles. However, the smallness of the Yukawa couplings implies that the NLSP, whatever character it has, will have a long decay length, possibly measurable at LHC \cite{deGouvea:2006wd}. In conclusion, the MSSM with Dirac neutrinos has some testable signatures, as opposed to the non-SUSY case.

\begin{figure}
\begin{center}
\vspace{5mm}
\includegraphics[width=0.65\textwidth]{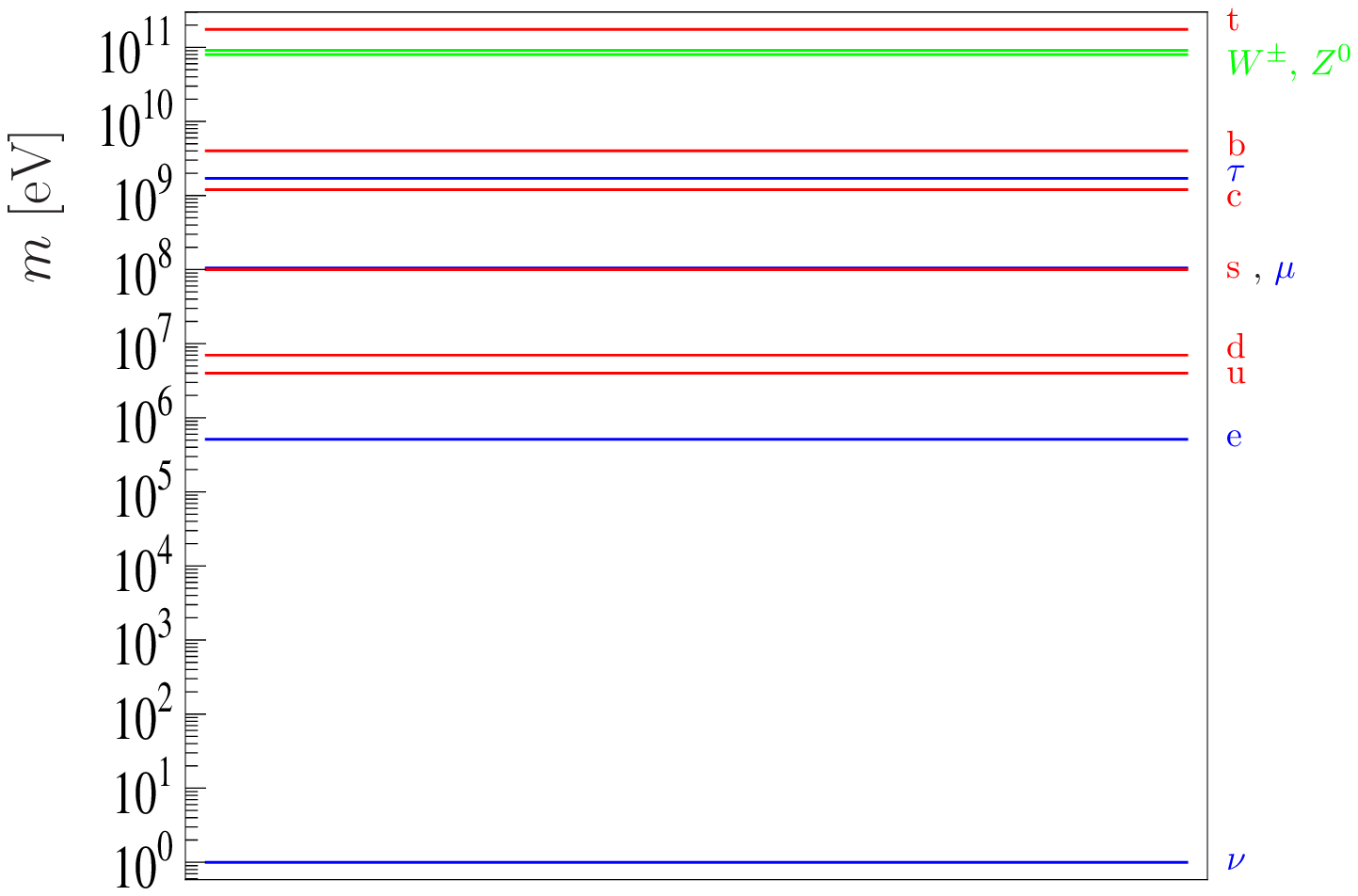}
\end{center}
\vspace{0mm}
\caption{Masses of the known fundamental particles. For the neutrino, the conservative upper bound $m_\nu = 1$ eV is used and only one generation is represented. Leptons are drawn in blue, quarks in red and massive gauge bosons in green. The massless gauge bosons, photon and gluon, are not included in the plot.}
\label{fig:particlemasses}
\end{figure}

The previous discussion assumes that neutrino masses have the same origin as the other fermion masses. This, however, would imply a naturalness problem. Figure \ref{fig:particlemasses} shows the masses of the known fundamental particles \cite{Amsler:2008zzb}. For the neutrino, the conservative upper bound $m_\nu = 1$ eV is used and only one generation is represented. Note the huge difference between the upper bound for the neutrino mass and the masses of the other particles. This can be hardly understood if they share a common source. In fact, this picture suggests that neutrino masses come from a different origin, with an underlying mechanism that can explain why they lie much below the electroweak scale. This will be the subject of the following sections.

\section{Majorana neutrinos} \label{maj-nu}

In the previous section we discussed Dirac neutrinos. They can be easily introduced in the Standard Model, although they lead to very small couplings, with the subsequent difficulties to explore their phenomenology. A different alternative will be discussed in the following: Majorana neutrinos.

\subsection{Theory of Majorana neutrinos} \label{maj-nu-th}

The question whether neutrinos are their own antiparticle has no answer yet. For charged particles there would be no doubt, since charge, electric or other type, can be used to distinguish between the particle and the antiparticle. For example, the trajectory of electrons and positrons in a magnetic field is different. However, for neutrinos, this distintion cannot be applied and the question remains.

From the theoretical point of view, this has to do with the type of spinor used to describe the neutrino. In the previous section we used Dirac spinors, with four independent complex components. A mass term is built as

\begin{equation} \label{nu-yuk-2}
- \mathcal{L}_{D}^{mass} = m_D \bar{\nu}_L \nu_R + h.c. = m_D \bar{\nu} \nu
\end{equation}

with

\begin{equation}
\nu_L = P_L \: \nu \qquad \nu_R = P_R \: \nu
\end{equation}

where $P_{L,R} = \frac{1}{2}(1 \mp \gamma_5)$ are the chirality projectors. In the Weyl basis\footnote{The Weyl basis is also known as chiral basis in some textbooks.} for the Dirac matrices one has

\begin{equation}
\gamma_5 = \left( \begin{array}{c c}
- I_2 & 0 \\
0 & I_2 \end{array} \right)
\end{equation}

$I_2$ being the $2 \times 2$ identity matrix. Then, if the 4-component Dirac spinor $\nu$ is written as

\begin{equation} \label{dirac-nu-1}
\nu = \left( \begin{array}{c}
\chi \\
\sigma_2 \phi^* \end{array} \right)
\end{equation}

where $\chi$ and $\phi$ are 2-component Weyl spinors and $\sigma_2$ is a Pauli matrix, introduced just for convenience, one obtains

\begin{equation} \label{dirac-nu-2}
\nu_L = \left( \begin{array}{c}
\chi \\
0 \end{array} \right) \qquad \nu_R = \left( \begin{array}{c}
0 \\
\sigma_2 \phi^* \end{array} \right)
\end{equation}

Now one can apply charge conjugation to $\nu$. Its charge conjugate state is

\begin{equation} \label{c-conjugate}
\nu^c = C \bar{\nu}^T = - \gamma^2 \gamma^0 \bar{\nu}^T = \left( \begin{array}{c}
\phi \\
\sigma_2 \chi^* \end{array} \right)
\end{equation}

The Weyl spinors $\chi$ and $\phi$ get exchanged after applying charge conjugation. As a consequence of that $\nu \neq \nu^c$ and we conclude that Dirac fermions are not their own antiparticles.

However, the simplest fermionic representation of the Lorentz group is not a Dirac fermion. This has been known for many years since Majorana proposed a new type of neutrinos \cite{Majorana:1937vz}, now called Majorana neutrinos. Equation \eqref{dirac-nu-1} shows that a Dirac fermion is given in terms of four independent complex quantities. However, as will be shown below, one can build Lorentz invariant theories with spinors that only have two independent components.

Let $\rho$ be a 2-component Weyl spinor with the following lagrangian density \cite{Schechter:1980gr}

\begin{equation} \label{maj-lag}
\mathcal{L}_M = - i \rho^\dagger \sigma_\mu \partial^\mu \rho - \frac{m}{2} \rho^T \sigma_2 \rho + h.c.
\end{equation}

Here $\sigma_i$, with $i=1,2,3$, are the Pauli matrices, suplemented with $\sigma_0 = I$. One can show that both terms in the lagrangian \eqref{maj-lag} are invariants under Lorentz transformations. Given a general Lorentz transformation

\begin{equation} \label{trans1}
x \to \Lambda x
\end{equation}

the spinor field $\rho$ transforms as

\begin{equation} \label{trans2}
\rho(x) \to S(\Lambda) \rho(\Lambda^{-1}x)
\end{equation}

where $S$ is a $2 \times 2$ matrix that obeys the relation

\begin{equation} \label{trans3}
S^\dagger \sigma_\mu S = \Lambda_{\mu \nu} \sigma^\nu
\end{equation}

Using equations \eqref{trans1}, \eqref{trans2} and \eqref{trans3} and the unimodular property $\det S = 1$ it is easy to show that both, the kinetic and mass terms, are Lorentz invariant quantities, and thus a spinor field $\rho$ with the lagrangian \eqref{maj-lag} is a consistent physical description for a fermionic field. Note, however, that the mass term is not invariant under the $U(1)$ transformation

\begin{equation}
\rho \to e^{i \delta} \rho
\end{equation}

In conclusion, Majorana fermions cannot have conserved $U(1)$ charges. This is of great relevance for Majorana neutrinos, whose mass terms break lepton number by two units.

The connection between Dirac and Majorana fermions becomes clear if we expand a Dirac fermion into its 2-component pieces. Consider the lagrangian for the Dirac fermion $\Psi$

\begin{equation} \label{dir-lag}
\mathcal{L}_D = i \bar{\Psi} \gamma_\mu \partial^\mu \Psi - m \bar{\Psi} \Psi
\end{equation}

The 4-component spinor $\Psi$ can be split as

\begin{equation} \label{dirac-fermion}
\Psi = \left( \begin{array}{c}
\chi \\
\sigma_2 \phi^* \end{array} \right)
\end{equation}

where $\chi$ and $\phi$ are two 2-component spinors, similar to equation \eqref{dirac-nu-1}. Expanding the lagrangian \eqref{dir-lag} one obtains

\begin{equation} \label{dir-lag-2}
\mathcal{L}_D = -i \sum_{\alpha=1}^2 \rho_\alpha^\dagger \sigma_\mu \partial^\mu \rho_\alpha - \frac{m}{2} \sum_{\alpha=1}^2 \rho_\alpha^T \sigma_2 \rho_\alpha + h.c.
\end{equation}

where the following redefinition has been done

\begin{eqnarray}
\chi &=& \frac{1}{\sqrt{2}}(\rho_2 + i \rho_1) \label{split-1} \\
\phi &=& \frac{1}{\sqrt{2}}(\rho_2 - i \rho_1) \label{split-2}
\end{eqnarray}

Therefore, one Dirac fermion is equivalent to two Majorana fermions of equal mass but opposite CP. This clearly shows that the Majorana fermion is a more fundamental representation of the Lorentz group.

Let us consider now charge conjugation acting on a Majorana fermion. We will show that a Majorana fermion is self-conjugate and therefore its own antiparticle. In order to do that, we can reverse our last computation and build a 4-component spinor from the 2-component fermion $\rho$ in equation \eqref{maj-lag}. Using equations \eqref{split-1} and \eqref{split-2} with $\rho_2 = 0$ one finds that the lagrangian \eqref{maj-lag} can be written as

\begin{equation}
\mathcal{L}_M = i \bar{\Psi}_M \gamma_\mu \partial^\mu \Psi_M - m \bar{\Psi}_M \Psi_M
\end{equation}

where

\begin{equation} \label{maj-fermion-4comp}
\Psi_M = \frac{i}{\sqrt{2}} \left( \begin{array}{c}
\rho \\
\sigma_2 \rho^* \end{array} \right)
\end{equation}

Applying now the operation of charge conjugation, as we did to obtain equation \eqref{c-conjugate}, it is easy to check that $\Psi_M = \Psi_M^c$ and thus it describes both the particle and the antiparticle.

Finally, let us emphasize that the nature of neutrinos is not only a theoretical question, but it also has phenomenological implications. For example, if neutrinoless double beta decay is ever observed, neutrinos will be known to be of Majorana type, since a theory with pure Dirac neutrinos cannot lead to that process \cite{Schechter:1981bd,Hirsch:2006yk}.

\subsection{Weinberg operator}

It has been noted long ago by Weinberg \cite{Weinberg:1979sa} that one can add to the Standard Model a dimension five operator, now called the Weinberg operator, with the following definition

\begin{equation} \label{wop}
W_{op} = \frac{1}{\Lambda} L_i L_j H H
\end{equation}

\begin{figure}
\begin{center}
\vspace{5mm}
\includegraphics[width=0.49\textwidth]{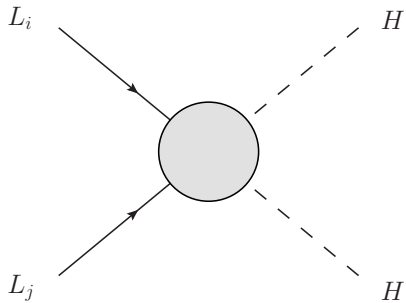}
\end{center}
\vspace{0mm}
\caption{Weinberg dimension five operator responsible for neutrino masses.}
\label{fig:weinberg-op}
\end{figure}

Here $L_i$ and $L_j$ are lepton doublets, whereas $H$ is the standard Higgs doublet.

\begin{equation}
L_i = \left( \begin{array}{c}
\nu_i \\
l_i \end{array} \right) \qquad H = \left( \begin{array}{c}
\phi^+ \\
\phi^0 \end{array} \right)
\end{equation}

Finally, $\Lambda$ is the energy scale at which this operator is generated. Note that, unless one couples $L_i$ and $L_j$ to form an antisymmetric combination, the Weinberg operator includes the piece

\begin{equation} \label{wop-nu}
W_{op}^{\nu \nu} = \frac{1}{\Lambda} \nu_i \nu_j \phi^0 \phi^0
\end{equation}

Note that the Weinberg operator breaks lepton number by two units, as required to obtain a Majorana mass for the neutrinos. In fact, after electroweak symmetry breaking $\langle \phi^0 \rangle = \frac{v}{\sqrt{2}}$ and Majorana neutrino masses are generated. Note the dependence $m_\nu \propto \frac{v^2}{\Lambda}$, which is quadratic in $v$, as opposed to the rest of fermions that get masses linear in $v$.

It can be shown that this dimension five operator is unique in the sense that one must go beyond dimension five to find other lepton number violating effective operators. This observation by Weinberg points out that lepton number conservation looks rather accidental, since non-renormalizable operators like the one in equation \eqref{wop} violate it.

In fact, the Weinberg operator is the effective description of most high-energy models. The details of the mechanism that generates this dimension five operator might be very different and, as we will show below, several realizations are possible. In the rest of this chapter we will follow reference \cite{Ma:1998dn}, pointing out the main distinctions between the different realizations and describing the type of models that generate them.

\subsection{Tree-level models}

Tree-level realizations of the Weinberg operator are well known and the literature is full with examples. In fact, most of the models and mechanisms to generate neutrino masses are based on them and they have been deeply studied over the years.

There are only three ways to build up the Weinberg operator at tree-level. These are

\begin{itemize}

\item $L_i$ and $H$ combine to form a fermion singlet

\item $L_i$ and $L_j$ combine to form a scalar triplet

\item $L_i$ and $H$ combine to form a fermion triplet

\end{itemize}

The proof is based on gauge invariance \cite{Ma:1998dn}. There are four $SU(2)_L$ doublets involved in the Weinberg operator and they must combine in pairs in such a way that the total operator is a gauge invariant. Therefore, two possibilities arise: $H - H$ and $L_i - L_j$ or $H - L_i$ and $H - L_j$. Moreover, in $SU(2)$ one has $2 \otimes 2 = 1 \oplus 3$, this is, two doublets can combine either to a singlet or to a triplet. Then, one obtains four combinations in total. However, note that the one with $L_i - L_j$ combining to a singlet does not lead to the Weinberg operator. This is due to the fact that it is antisymmetric in the indices $i,j$ and then one cannot obtain the piece $\nu_i \nu_j$. Therefore, we are left with three possible realizations.

In fact, although the low-energy effective operator is the same, the different realizations imply very different models in the high-energy regime. In the following we are going to discuss these three tree-level realizations, highlighting their main differences and presenting some high-energy models that generate them at low energies.

\subsubsection{Tree-level: Realization 1}

\begin{figure}
\begin{center}
\vspace{5mm}
\includegraphics[width=0.49\textwidth]{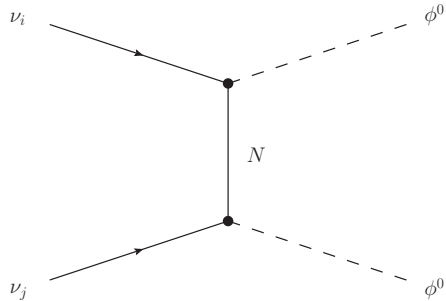}
\end{center}
\vspace{0mm}
\caption{First tree-level realization of the Weinberg operator.}
\label{fig:tree-r1}
\end{figure}

Figure \ref{fig:tree-r1} is the most common realization of the Weinberg operator and it has dominated the literature over the years. Here $H - L_i$ and $H - L_j$ combine to form gauge singlets and thus the intermediary particle is a singlet as well. Being $M_N$ its mass and $y$ the coupling $H L N$, one finds

\begin{equation} \label{lambda-r1}
\frac{1}{\Lambda} = \frac{y^2}{M_N}
\end{equation}

When $\phi^0$ gets a VEV, $\langle \phi^0 \rangle = v/\sqrt{2}$, neutrino masses are generated

\begin{equation}
\mathcal{L}_\nu^{mass} = - \frac{1}{2} \nu^T m_\nu \nu + h.c.
\end{equation}

where

\begin{equation} \label{numass-r1}
m_\nu = \frac{y^2 v^2}{2 M_N} = \frac{m_D^2}{M_N}
\end{equation}

where we have identified $m_D = y \langle \phi^0 \rangle$ as the $\nu - N$ Dirac mass that is generated after electroweak symmetry breaking. Equation \eqref{numass-r1} shows that a large mass $M_N$ for the singlets implies a small mass $m_\nu$ for the neutrinos. This suppression by a high scale is a natural consequence in many models and thus this realization has become a very popular way to generate small neutrino masses.

The most popular high-energy model that leads to this realization is the famous Type-I Seesaw \cite{Minkowski:1977sc,Gell-Mann:1980vs,Yanagida:1979,Mohapatra:1979ia}. In this setup one adds three families of right-handed neutrinos to the Standard Model particle spectrum\footnote{In principle, two families are sufficient to generate neutrino masses in the observed range, but in most of the cases three are assumed.}. If one allows for lepton number violation, in addition to the neutrino Yukawa coupling in equation \eqref{nu-yuk} one can write down a mass for the right-handed neutrinos. With this additional piece the part of the lagrangian that involves the right-handed neutrino becomes

\begin{equation} \label{nu-lag}
- \mathcal{L}_{\nu_R} = Y_\nu \phi^0 \bar{\nu}_R \nu_L + \frac{1}{2} \nu_R^T M_R \nu_R + h.c.
\end{equation}

Here $M_R$ is a $3 \times 3$ symmetric matrix. Note that the Majorana mass $M_R$ is allowed by the gauge symmetry because the right-handed neutrinos are singlets under the SM gauge group:

\begin{center}
\begin{tabular}{c c c c}
\hline
Field & $SU(3)_c$ & $SU(2)_L$ & $U(1)_Y$ \\
\hline
$\nu_R$ & 1 & 1 & 0 \\
\hline
\end{tabular}
\end{center}

This piece was missing in our previous discussion on Dirac neutrinos, see section \ref{dirac-nu}, and it completely changes the picture. In fact, the Majorana nature of the right-handed neutrinos is transferred to the left-handed ones through the Dirac $\nu_L - \nu_R$ mixing. After electroweak symmetry breaking the lagrangian \eqref{nu-lag} leads to

\begin{equation} \label{lagmass-typeI}
- \mathcal{L}_{\nu}^{mass} = m_D \bar{\nu}_L \nu_R + \frac{1}{2} \nu_R^T M_R \nu_R + h.c. = \bar{\chi} M \chi
\end{equation}

where

\begin{equation}
\chi = \left( \begin{array}{c}
\nu_L^c \\
\nu_R \end{array} \right)
\end{equation}

and

\begin{equation} \label{seesaw-matrix}
M = \left( \begin{array}{c c}
0 & m_D \\
m_D^T & M_R \end{array} \right)
\end{equation}

The Majorana mass $M_R$ of the right-handed neutrinos is a free parameter of the model. Since its origin is not tied to electroweak symmetry breaking, $M_R$ can take any value. In section \ref{dirac-nu} we chose $M_R = 0$, which leads to pure Dirac neutrinos. However, theoretical considerations prefer heavy right-handed neutrinos. For example, $M_R$ can be generated at a very high scale by the breaking of a larger gauge group, under which the right-handed neutrinos are not singlets. Moreover, as we will see below, the assumption of heavy right-handed neutrinos helps to understand the smallness of the light neutrino masses.

If we assume $M_R \gg m_D$, the matrix in equation \eqref{seesaw-matrix} can be block-diagona\-lized in good approximation to give

\begin{equation} \label{seesaw-matrix-ap}
\hat M \simeq \left( \begin{array}{c c}
m_{light} & 0 \\
0 & M_{heavy} \end{array} \right)
\end{equation}

with

\begin{eqnarray}
m_{light} &=& -m_D^T \cdot M_R^{-1} \cdot m_D\\
M_{heavy} &=& M_R
\end{eqnarray}

Here we recover the generic result in equation \eqref{numass-r1}. The mass of the light neutrinos is given by $m_\nu \sim m_D^2 / M_R$. This, usually called \emph{the seesaw formula}, provides a natural explanation for the observed lightness of neutrinos \cite{Minkowski:1977sc,Gell-Mann:1980vs,Yanagida:1979,Mohapatra:1979ia}. Let us consider the value $m_\nu \sim 1$ eV. If, for example, we take $M_R = 10^{13}$ GeV, the Dirac mass turns out to be $m_D = Y_\nu \langle \phi^0 \rangle \sim 100$ GeV. This implies Yukawa couplings of order $1$, $Y_\nu \sim 1$, what can be compared to the results in our discussion on Dirac neutrinos, where we showed that the same mass for the light neutrinos implies $Y_\nu \sim 10^{-11}$ in that case.

Moreover, under the same assumption, $M_R \gg m_D$, the mass eigenstates can be approximated as $\chi_{light} \simeq \nu_L$ and $\chi_{heavy} \simeq \nu_R$. This explains why neutrinos have always been observed to be left-handed in all performed experiments.

This popular framework, widely studied and extended in several directions, is one of the most common ways to generate neutrino masses and it can be used with or without supersymmetry. As we will see below, there are other types of Seesaw mechanism. They all share the common feature of the suppression of neutrino masses by the existence of a high energy scale. If this scale is very high, like in the numerical example given above, direct tests of the model become impossible. The energies reached at colliders do not allow to produce such heavy particles, and thus only indirect tests are at best available. As discussed in chapter \ref{chap:susylr}, the supersymmetric version of the seesaw mechanism, in each of its variations, offers some experimental chances due to the existence of the superparticles. In particular, the sleptons carry some information on the high energy regime, which allows to test some scenarios. The non-SUSY case, however, does not provide this possibility, and thus the model cannot be put to experimental test.

\subsubsection{Tree-level: Realization 2}

\begin{figure}
\begin{center}
\vspace{5mm}
\includegraphics[width=0.6\textwidth]{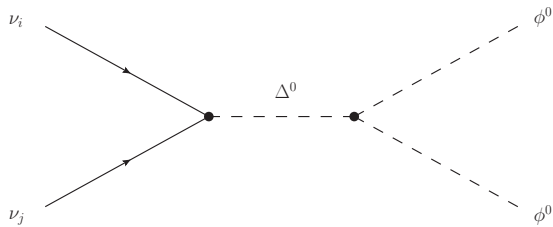}
\end{center}
\vspace{0mm}
\caption{Second tree-level realization of the Weinberg operator.}
\label{fig:tree-r2}
\end{figure}

Figure \ref{fig:tree-r2} shows a different tree-level realization of the Weinberg operator. In this case, $L_i - L_j$ and $H - H$ combine to form a $SU(2)_L$ triplet. Therefore, the intermediary scalar $\Delta^0$ must belong to a $SU(2)_L$ triplet, called $\Delta$ in the following, in order to obtain a gauge invariant Feynman diagram.

Following the same approach as in the discussion for the first tree-level realization, one finds that in this case the $\Lambda$ scale that appears in the non-renormalizable Weinberg operator is

\begin{equation}
\frac{1}{\Lambda} = \frac{f \mu}{M_\Delta^2}
\end{equation}

where $f$ is the $L L \Delta$ coupling, $\mu$ is the $H H \Delta$ coupling and $M_\Delta$ is the mass of the $\Delta$ triplet. Note that triplets are allowed to have invariant masses, since the $SU(2)$ product $3 \otimes 3$ includes the singlet representation. As in the previous case, when $\phi^0$ gets a VEV, $\langle \phi^0 \rangle = v/\sqrt{2}$, neutrino masses are generated

\begin{equation} \label{numass-r2}
m_\nu = \frac{f \mu v^2}{2 M_\Delta^2}
\end{equation}

This realization can be obtained in the so-called Type-II Seesaw mechanism \cite{Schechter:1980gr,Konetschny:1977bn,Cheng:1980qt,Mohapatra:1980yp,Lazarides:1980nt}. In this framework one adds a scalar Higgs triplet to the SM particle spectrum:

\begin{center}
\begin{tabular}{c c c c}
\hline
Field & $SU(3)_c$ & $SU(2)_L$ & $U(1)_Y$ \\
\hline
$\Delta$ & 1 & 3 & 2 \\
\hline
\end{tabular}
\end{center}

The triplet $\Delta$ can be written as

\begin{equation}
\Delta = \left( \begin{array}{c}
\Delta^{++} \\
\Delta^+ \\
\Delta^0 \end{array} \right)
\end{equation}

although it is more common to use the $2 \times 2$ matrix notation

\begin{equation}
\Delta = \left( \begin{array}{c c}
\Delta^+ / \sqrt{2} & \Delta^{++} \\
\Delta^0 & - \Delta^+ / \sqrt{2} \end{array} \right)
\end{equation}

Once the representation for $\Delta$ is chosen, the corresponding $SU(2)_L$ generators have to be chosen accordingly. This way one ensures the gauge invariance of the lagrangian. In particular, the piece that involves the $\Delta$ triplet can be written as

\begin{equation}
\mathcal{L}_\Delta = (D_\mu \Delta)^\dagger (D^\mu \Delta) + \mathcal{L}_Y - V(H,\Delta)
\end{equation}

Here $\mathcal{L}_Y$ is the Yukawa coupling

\begin{equation} \label{seesawII-yuk}
\mathcal{L}_Y = - f L \Delta L + h.c.
\end{equation}

and $V(H,\Delta)$ is the scalar potential, which includes two important terms

\begin{equation}
V(H,\Delta) \supset M_\Delta^2 \Delta^\dagger \Delta + \mu H \Delta H + h.c.
\end{equation}

In the last two equations gauge and flavor indices have been omited for the sake of simplicity. Note that the simultaneous presence of the operators $f L \Delta L$ and $\mu H \Delta H$ necessarily break lepton number by two units. As discussed in section \ref{maj-nu-th}, this is a requirement for Majorana neutrinos, whose mass term also has $\Delta L = 2$.

By studying the scalar potential one finds that its minimum its shifted from the trivial configuration $\langle H \rangle = \langle \Delta \rangle = 0$. In fact, the tadpole equations link both VEVs, leading to the relation

\begin{equation} \label{seesaw-vev}
v_\Delta = \frac{\mu v^2}{\sqrt{2} M_\Delta^2}
\end{equation}

Here $\langle \Delta \rangle = v_\Delta / \sqrt{2}$ and $\langle H \rangle = v / \sqrt{2}$, as usual. Note that the larger the triplet mass is, the smaller its VEV becomes. This unexpected feature of the model has an important implication for neutrino masses and it is sometimes called Seesaw VEV relation. Substituting \eqref{seesaw-vev} into \eqref{seesawII-yuk} one obtains a Majorana mass for the left-handed neutrinos that is just the same result as in equation \eqref{numass-r2}.

Again, the heaviness of a field, the $\Delta$ triplet in this case, is used to explain the lightness of the left-handed neutrinos observed at low energies. This is the seesaw mechanism at work.

It is also worth to mention that the addition of a $SU(2)_L$ triplet modifies some fundamental relations in the SM and, in fact, its properties are highly restricted by SM precision measurements \cite{Gunion:1990dt,Rizzo:1990uu,Blank:1997qa}. Moreover, it also contributes to lepton number violating processes, what can be translated into strong bounds for its mass and couplings \cite{Cuypers:1996ia}. Therefore, the heaviness of the $\Delta$ triplet is also well motivated from the phenomenological point of view, since it helps avoiding the constraints.

As in the case of the type-I seesaw, this idea can be also employed in a supersymmetric context. Although it has attracted less attention than the type-I version of the seesaw mechanism, the type-II seesaw is also a very nice way to generate neutrino masses. In fact, it naturally appears in some high energy constructions, like in left-right symmetric models \cite{Mohapatra:1980yp,Aulakh:1998nn,Ma:2003zy,Akhmedov:2005np,Akhmedov:2006de} and in GUTs based on $SU(5)$ \cite{Ma:1998dn,Rossi:2002zb} or $SO(10)$ \cite{Bajc:2001fe,Goh:2003sy,Goh:2003hf,Bertolini:2004eq}, sometimes in combination with type-I.

\subsubsection{Tree-level: Realization 3}

\begin{figure}
\begin{center}
\vspace{5mm}
\includegraphics[width=0.49\textwidth]{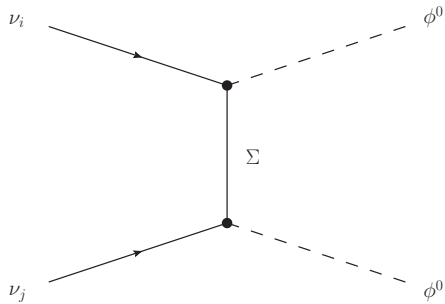}
\end{center}
\vspace{0mm}
\caption{Third tree-level realization of the Weinberg operator. Note that this diagram is equivalent to the one in figure \ref{fig:tree-r1}. However, here $\Sigma$ belongs to a $SU(2)_L$ triplet.}
\label{fig:tree-r3}
\end{figure}

Finally, figure \ref{fig:tree-r3} shows the third tree-level realization of the Weinberg operator. Here, $H - L_i$ and $H - L_j$ combine to form a fermion triplet and thus the intermediary particle must be a fermion triplet as well. Since this diagram is equivalent to the one for the first tree-level realization, equations analogous to \eqref{lambda-r1} and \eqref{numass-r1} are valid in this case.

This case has been less studied than the other two, although the basic idea is just the same. Let us consider a fermionic $SU(2)_L$ triplet $\Sigma$ with hypercharge $Y=0$:

\begin{center}
\begin{tabular}{c c c c}
\hline
Field & $SU(3)_c$ & $SU(2)_L$ & $U(1)_Y$ \\
\hline
$\Sigma$ & 1 & 3 & 0 \\
\hline
\end{tabular}
\end{center}

If we add three $\Sigma$ families to the SM matter content one can write down two additional lagrangian terms

\begin{equation}
- \mathcal{L}_\Sigma = Y_\Sigma H L \Sigma + \frac{1}{2} \Sigma M_\Sigma \Sigma + h.c.
\end{equation}

where $Y_\Sigma$ and $M_\Sigma$ are $3 \times 3$ matrices. Note the similarity between the last expression and equation \eqref{nu-lag}. The formal structure of this model resembles the structure of a type-I seesaw. Although the gauge indices in these two terms contract differently, due to the introduction of triplets instead of singlets, neutrino masses can be shown to follow an analogous expression

\begin{equation} \label{numass-r3}
m_\nu = \frac{Y_\Sigma^2 v^2}{2 M_\Sigma}
\end{equation}

Therefore, if the fermion triplet $\Sigma$ is very heavy, as in the type-I case, the light neutrino masses are naturally obtained in the correct range. This is the so-called Type-III Seesaw mechanism \cite{Foot:1988aq}.

Let us emphasize that, although the common practice is to consider a high seesaw scale in order to fit neutrino masses with order $1$ couplings, there is no reason to limit ourselves to that possibility. In principle, and due to the fact that other Yukawa couplings are known to be much smaller than $1$ (for example, the electron Yukawa, with $Y_e \sim 10^{-6}$), one could also consider much lower seesaw scales. In that case the seesaw formula $m_\nu \sim \frac{y^2 v^2}{M}$ would also require to lower the size of the $y$ coupling. For example, if neutrino masses are generated at the electroweak scale, $M \sim 100$ GeV and the corresponding Yukawa coupling must be around $y \sim 10^{-7}-10^{-6}$.

This scenario sometimes appears in the literature under the name of electroweak scale seesaw. An example is Bilinear R-parity Violation \cite{Hall:1983id}, briefly discussed in this thesis, see section \ref{sec:brpv}. In this supersymmetric model, R-parity is explicitly broken by the superpotential term

\begin{equation}
\mathcal{W} \supset \epsilon \widehat L \widehat H_u
\end{equation}

\begin{figure}
\begin{center}
\vspace{5mm}
\includegraphics[width=0.6\textwidth]{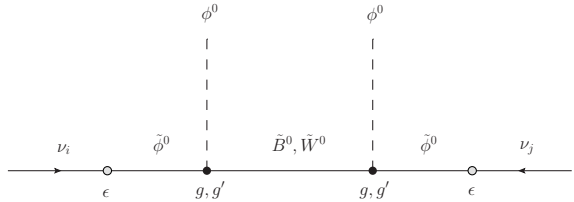}
\end{center}
\vspace{0mm}
\caption{Tree-level generation of neutrino masses in Bilinear R-parity Violation. The mechanism is a combination of type-I and type-III seesaw.}
\label{fig:numass-brpv}
\end{figure}

This lepton number violating term leads to mixing between neutrinos and higgsinos, which in turn mix with the electroweak gauginos after symmetry breaking. Figure \ref{fig:numass-brpv} shows the realization of the Weinberg operator in this model. Note that the intermediary particles are the bino $\tilde{B}^0$, a singlet under $SU(2)_L$, and the neutral wino $\tilde{W}^0$, which belongs to a triplet of $SU(2)_L$. Therefore, neutrino masses are generated as a combination of type-I and type-III seesaw. One can estimate their magnitude as

\begin{equation} \label{brpv-estimate}
m_\nu \sim \frac{(g,g')^2 \epsilon^2 v^2}{m_{SUSY}^3}
\end{equation}

where we have used $M_{gauginos} \sim M_{higgsinos} \equiv m_{SUSY}$. Equation \eqref{brpv-estimate} shows that by choosing $\epsilon \ll v,m_{SUSY}$ one obtains small neutrino masses.

Furthermore, there are extended models that generate the dimensionful $\epsilon$ coupling after electroweak symmetry breaking, see for example references \cite{Masiero:1990uj,LopezFogliani:2005yw}. The superpotential term $Y \widehat S \widehat L \widehat H_u$, where $\widehat S$ is a new singlet superfield, leads to $Y v_S \widehat L \widehat H_u = \epsilon_{eff} \widehat L \widehat H_u$, where $v_S$ is the VEV of the scalar component of the $S$ singlet. For $v_S \sim 100$ GeV a value $Y \sim 10^{-7}-10^{-6}$ is required, as in most electroweak scale seesaw models.

Let us close this section by mentioning that other tree-level models can be found in the literature with the name of seesaw mechanism. The Inverse Seesaw \cite{Mohapatra:1986bd} and the Linear Seesaw \cite{Akhmedov:1995ip,Akhmedov:1995vm,Malinsky:2005bi} are good examples. Nevertheless, they can be understood as extended versions of the three realizations discussed here. Finally, a detailed discussion of the low energy effects of the three versions of the seesaw mechanism was given in reference \cite{Abada:2007ux}.

\subsection{Radiative models}

The Weinberg operator also admits radiative realizations. In fact, models that generate neutrino masses radiatively are very well motivated due to the loop suppression, which allows to lower the scale at which neutrino mass generation takes place. Therefore, the new particles needed to complete the loops can have masses at the electroweak scale. This would imply a very rich phenomenology at the LHC. See \cite{Babu:1989fg,Chang:1999hga} for general reviews on radiative models.

If we concentrate on 1-loop topologies leading to neutrino masses, three types of Feynman diagrams can be drawn. Let us briefly discuss them.

\subsubsection{1-loop: Realization 1}

\begin{figure}
\begin{center}
\vspace{5mm}
\includegraphics[width=0.49\textwidth]{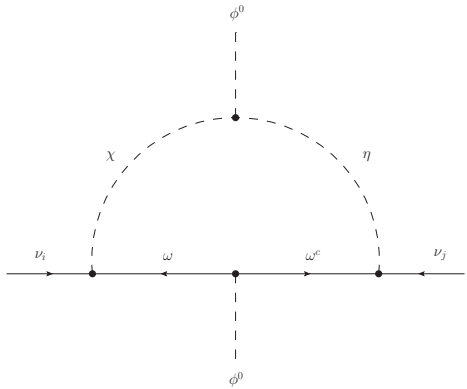}
\end{center}
\vspace{0mm}
\caption{First 1-loop realization of the Weinberg operator.}
\label{fig:1loop-r1}
\end{figure}

The first 1-loop realization of the Weinberg operator is presented in figure \ref{fig:1loop-r1}. When the two $\phi^0$ Higgs fields get a VEV this diagram leads to Majorana neutrino masses for the neutrinos. By imposing gauge invariance in every interaction vertex one can obtain the gauge charges of the involved particles. For example, either $\omega$ or $\omega^c$ must be a $SU(2)_L$ doublet, since they both couple to $\phi^0$. If we choose $\omega(q_3,2,q_1)$ under $SU(3)_c \times SU(2)_L \times U(1)_Y$, one easily obtains that the rest of particles must be $\omega^c(q_3^*,q_2,-q_1+1/2)$, $\chi(q_3^*,q_2',-q_1+1/2)$ and $\eta(q_3,2,q_1)$, where $q_2$ and $q_2'$ are constrained to be either $1$ or $3$.

There are well known models that lead to this particular 1-loop realization. The most famous one is the Zee model \cite{Zee:1980ai}. In this case one has $q_1 = q_2 = q_2' = q_3 = 1$ and $\omega(1,2,-1/2) \equiv L$, $\omega^c(1,1,1) \equiv e^c$ and $\eta(1,2,-1/2) \equiv H$ are identified with the standard left-handed lepton doublet, right-handed lepton singlet and Higgs doublet. The additional $\chi(1,1,1)$ is a new charged scalar, usually called $h^+$. Note that the coupling $H_i H_j h^+$ requires at least two Higgs doublets, since the $H_i H_j$ contraction to form a $SU(2)_L$ singlet is antisymmetric in the $SU(2)_L$ indices. Therefore, the Zee model contains, beside the SM particles, a new charged scalar and a second Higgs doublet. In fact, it shows how a small modification of the scalar sector has dramatic consequences concerning lepton number conservation, which is accidental in the minimalistic Standard Model.

Let us mention that the Minimal Zee Model \cite{Wolfenstein:1980sy} is nowadays ruled out by experimental data \cite{Koide:2001xy,Frampton:2001eu,He:2003ih}. In this version only one of the Higgs doublets couples to the leptons. This automatically implies a very particular structure for the generated neutrino mass matrix. Just by inspection of the diagram in figure \ref{fig:1loop-r1}, adapted for the Zee Model, one finds

\begin{equation} \label{zeemass}
m_{ab}^{\nu \nu} \propto \frac{1}{16 \pi^2} f_{ac} m_{cd} Y_\gamma^{db} v_{\alpha} M_{\alpha \beta} \frac{\left[ \log \left( \frac{m_H^2}{m_{h^+}^2} \right) \right]_{\beta \gamma}}{m_{h^+}^2}
\end{equation}

where $f_{ab}$ is the antisymmetric matrix that appears in the lepton number violating coupling $L_a L_b h^+$, $m_{ab}$ is the mass matrix of the charged leptons, $Y_\gamma^{ab}$ is the Yukawa coupling $H_\gamma L_a (e^c)_b$, $v_\alpha = \langle \phi_\alpha^0 \rangle$ and $M_{\alpha \beta}$ is the mass that appears in the scalar potential term $H_\alpha H_\beta h^+$. In order to avoid confusion, the Higgs mass is denoted here as $m_H$. The last factor in equation \eqref{zeemass} comes from the loop integral, that can be easily approximated in the limit $m_{h^+} \gg m_H$. It can be shown that if only one of the Higgs doublets couples to the leptons, equation \eqref{zeemass} reduces to

\begin{equation}
m_{ab}^{\nu \nu} \propto f_{ab} (m_b^2 - m_a^2)
\end{equation}

where $m_a$ and $m_b$ are charged leptons masses. With this structure one obtains the following texture

\begin{equation}
m^{\nu \nu} \sim \left( \begin{array}{ccc}
0 & a & c \\
a & 0 & b \\
c & b & 0 \end{array} \right)
\end{equation}

which has been shown not to reproduce the observed pattern of neutrino masses and mixing angles \cite{Harrison:2002er}. If one allows both Higgs doublets to couple to the leptons, as in the General Zee Model \cite{Balaji:2001ex}, the previous relation is broken and one can evade this problem. In fact, this generalized version of the model leads to a very rich phenomenology at colliders, where the charged scalar $h^+$ has very clear signatures \cite{AristizabalSierra:2006ri}.

Another good example of this 1-loop contribution to neutrino masses is obtained in \rpv SUSY \cite{Koide:2003mh}. Again, one can choose $q_1 = q_2 = q_2' = q_3 = 1$ and identify $\omega(1,2,-1/2) \equiv L$ and $\omega^c(1,1,1) \equiv e^c$ as the standard left- and right-handed lepton fields. In this case, however, the additional $\eta$ and $\chi$ scalar fields are the corresponding sleptons, $\eta(1,2,-1/2) \equiv \tilde{L}$ and $\chi(1,1,1) \equiv \tilde{e}^c$. Then, the lepton number violating interactions in the external vertices come from the \rpv superpotential trilinear term $\lambda \widehat L \widehat L \widehat e^c$, see equation \eqref{rpv-superpotencial}. Furthermore, if one chooses $q_2 = q_2' = 1$, $q_1 = 1/6$ and $q_3 = 3$ the fields $\eta$ and $\chi$ are identified with left- and right-handed squarks, and thus the involved \rpv term is $\lambda' \widehat L \widehat Q \widehat d^c$.

\subsubsection{1-loop: Realization 2}

\begin{figure}
\begin{center}
\vspace{5mm}
\includegraphics[width=0.49\textwidth]{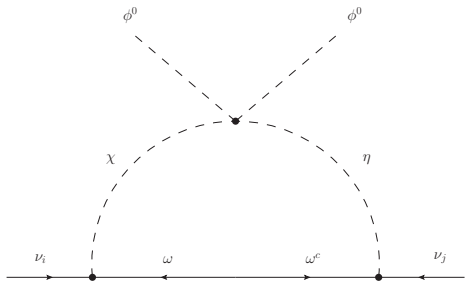}
\end{center}
\vspace{0mm}
\caption{Second 1-loop realization of the Weinberg operator. Note that for this diagram to be possible $\omega - \omega^c$ must have an invariant mass term.}
\label{fig:1loop-r2}
\end{figure}

The second realization, shown in figure \ref{fig:1loop-r2} has attracted less attention among model builders \cite{Branco:1988ex}, although it got some renewed interest in the last years. Note that the fermions $\omega$ and $\omega^c$ must combine to form an invariant mass term. If they are chosen to be a right-handed neutrino $\omega = \omega^c = N$ then the $\eta$ and $\chi$ scalar fields are forced to be $SU(2)_L$ doublets, $\eta(1,2,1/2)$ and $\chi(1,2,1/2)$, and they could be identified with the same extra scalar doublet. This setup has been used recently in the so-called scotogenic neutrino mass models \cite{Ma:2006km,Ma:2008uza}. By adding a new $Z_2$ symmetry, under which only the right-handed neutrinos $N$ and the second scalar doublet $\eta$ are charged, one forbids the Yukawa couplings with the first scalar doublet. Moreover, the conservation of $Z_2$ implies that $\langle \eta^0 \rangle = 0$, and thus no Dirac masses for the neutrinos are generated. Therefore, the loop in figure \ref{fig:1loop-r2} is the dominant contribution to neutrino masses. In addition, the conservation of the $Z_2$ symmetry has an additional consequence: it provides a dark matter candidate, the lightest $Z_2$-odd particle \cite{LopezHonorez:2006gr,Gustafsson:2007pc}. The model may also provide visible signatures at the LHC \cite{Cao:2007rm}.

\subsubsection{1-loop: Realization 3}

\begin{figure}
\begin{center}
\vspace{5mm}
\includegraphics[width=0.49\textwidth]{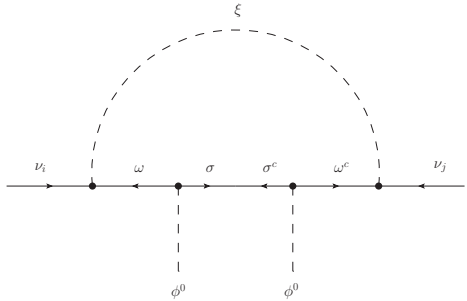}
\end{center}
\vspace{0mm}
\caption{Third 1-loop realization of the Weinberg operator. Note that for this diagram to be possible $\sigma - \sigma^c$ must have an invariant mass term.}
\label{fig:1loop-r3}
\end{figure}

Finally, the third 1-loop realization is shown in figure \ref{fig:1loop-r3}. This has been used for quark and charged lepton masses \cite{Babu:1989fg}, but no examples for neutrino masses are found in the literature.

Note that there are no other 1-loop realizations of the Weinberg operator. In the first realization one Higgs boson line is attached to the scalar part of the loop while the other is attached to the fermionic part, in the second realization both lines are attached to the scalar part, whereas in the third relization they are both attached to the fermionic part. Therefore, this covers all the possibilities.

Other possibilities exist if one goes beyond the 1-loop level. For example, if there is only one right-handed neutrino, only one light neutrino picks up a mass at tree-level. Then, the masses of the other active neutrinos are obtained at the 2-loop level via the exchange of $W$ bosons \cite{Petcov:1984nz,Babu:1988ig,Babu:1989pz}. In this scenario the GIM supression \cite{Glashow:1970gm} naturally leads to very small neutrino masses.

The most famous 2-loop model is the Zee-Babu model \cite{Cheng:1980qt,Zee:1985id,Babu:1988ki}. In this case one adds to the standard model two charged scalars $h^+$ and $k^{++}$, with charges $h(1,1,1)$ and $k(1,1,2)$ under $SU(3)_c \times SU(2)_L \times U(1)_Y$. These two particles allow us to write down new Yukawa couplings:

\begin{equation} \label{yukzeebabu}
\mathcal{L}_Y^{new} = f_{ab} L_a L_b h^+ + h_{ab} e^c_a e^c_b k^{++} + h.c.
\end{equation}

\begin{figure}
\begin{center}
\vspace{5mm}
\includegraphics[width=0.49\textwidth]{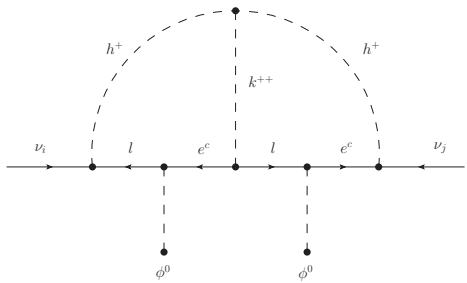}
\end{center}
\vspace{0mm}
\caption{Neutrino mass generation in the Zee-Babu model.}
\label{fig:zeebabu}
\end{figure}

Here $f_{ab}$ is an antisymmetric matrix, whereas $h_{ab}$ is symmetric. Note that the first term was already present in the Zee model but the second one is only possible due to the introduction of the $k^{++}$ field. With this particle content and the Yukawa couplings in equation \eqref{yukzeebabu} neutrino masses are generated at the 2-loop level, as shown in figure \ref{fig:zeebabu}.

The cubic term $\mu (k^{++})^* h^+ h^+$, contained in the scalar potential, is of great relevance. Without this dimensionful coupling it would be possible to find a lepton number assignment that recovers lepton number conservation and thus Majorana neutrino masses would vanish. In fact, note that the diagram in figure \ref{fig:zeebabu} would not be possible without a $(k^{++})^* h^+ h^+$ interaction term.

As in 1-loop radiative models, the loop suppression, higher in this case, lowers the scale at which neutrino masses are generated. This implies a very rich phenomenology at current and future experiments \cite{Babu:2002uu,AristizabalSierra:2006gb}. For example, the rates for lepton flavor violating processes, like $\mu \to e \gamma$ and $\mu \to 3e$, could be in reach of the present experimental searches. Moreover, if the additional charged scalars are light enough to be produced at the LHC one could search for their decays in channels like $h^+ \to l^+_i \nu$, which are correlated with neutrino physics \cite{Babu:2002uu,AristizabalSierra:2006gb}.

In conclusion, radiative models provide an interesting alternative way to generate neutrino masses. The suppression coming from the loop factors allows to lower the scale of new physics, what implies a very rich phenomenology at present and future experiments.

\section{Summary}

This chapter is a review on neutrino mass models. Although the topic has been so heavily investigated that by no means this can give a complete picture on the field, it is useful, at least, to set a common language to be used throughout the thesis.

After a short presentation of the current experimental status, several neutrino mass models have been discussed. Although the possibility of Dirac masses has been briefly mentioned, the discussion has focused on Majorana masses. This case is much more natural and nicely fits wider theoretical frameworks. In particular, the seesaw mechanism, in its different variations, provides a natural explanation for the smallness of neutrino masses and can be easily accommodated in extended versions of the SM or the MSSM.

Nevertheless, many important topics related to neutrino masses have been ignored in this chapter since they are not directly connected to the work done in this thesis.

For example, the observed pattern of neutrino masses and mixing angles has stimulated an intense investigation on flavor symmetries and the literature is full of models that attempt to explain the so-called tribimaximal mixing \cite{Harrison:2002er}. The references \cite{Ishimori:2010au,Altarelli:2010gt} are detailed reviews on the subject.

Another good example is leptogenesis, a subject closely linked to the seesaw mechanism and neutrino masses. This idea, proposed many years ago \cite{Fukugita:1986hr} to explain the observed baryon asymmetry of the universe, is nowadays an active field of research, see references \cite{Giudice:2003jh,Buchmuller:2005eh,Chen:2007fv,Davidson:2008bu} for general reviews.

In the following chapters two types of supersymmetric neutrino mass models will be discussed: models with R-parity violation and a left-right symmetric model that conserves R-parity.

\chapter{Introduction to R-parity violation}
\label{chap:rpv-intro}

Neutrino masses can be easily accommodated in SUSY with R-parity violation. This interesting framework leads to a very rich phenomenology at current and future experiments, being one of the few examples in which the origin of neutrino masses can be directly tested. This chapter provides an illustrative introduction to R-parity violation, emphasizing the most important ideas that will be further explored along the thesis.

\section{General concepts on R-parity violation}

As explained in chapter \ref{chap:susy}, R-parity plays a central role in supersymmetric model building. It forbids the dangerous dimension-4 operators leading to fast proton decay due to lepton and baryon number violation and predicts the existence of a stable particle with the right properties to be a good dark matter candidate.

However, the breaking of R-parity by L violating operators generates non-zero neutrino masses as demanded by the experiments, and thus is a well motivated scenario beyond the standard SUSY models\footnote{Neutrino mass generation in \rpv was briefly described in chapter \ref{chap:susy}, section \ref{sec:rp} and will be further explained in the next chapters. See section \ref{subsec:brpv-numass} for the simplest case.}. 

In fact, several arguments can be raised against R-parity:

\begin{itemize}

\item R-parity is imposed by hand.

\end{itemize}

Unlike the Standard Model, where L and B conservation is automatic, the MSSM requires to introduce an additional ad-hoc symmetry to forbid dangerous operators that contribute to L and B violating processes. However, there is no theoretical explanation for such symmetry\footnote{It is possible to embed the MSSM in a wider picture with a larger symmetry group that leads to R-parity conservation at low-energies. The reference \cite{Martin:1992mq} explores this idea and studies what type of theories leave R-parity as a remnant after symmetry breaking. Examples of such theories are models with a Left-Right gauge group, as will be discussed in chapter \ref{chap:susylr}.}.

\begin{itemize}

\item In fact, R-parity does not solve fast proton decay.

\end{itemize}

It is a well known fact that R-parity does not forbid some dimension-5 operators that lead to proton decay \cite{Ibanez:1991pr,Sakai:1981pk,Weinberg:1981wj}. In particular, the operator

\begin{equation}
\mathcal{O}_5 = \frac{f}{M} QQQL
\end{equation}

has $R_p(\mathcal{O}_5)=+1$ and thus conserves R-parity. However, the bounds on the life time of the proton require $f < 10^{-7}$ even for $M = M_{\text{Planck}}$ \cite{Ellis:1983qm}. Operators like $\mathcal{O}_5$ are typically generated in GUTs, due to the fact that quarks and leptons belong to the same multiplet. Therefore, R-parity does not completely solve fast proton decay when one considers SUSY GUTs.

\begin{itemize}

\item There is no reason to forbid all the L and B violating operators.

\end{itemize}

Proton decay requires both L and B violation. Therefore, for the proton to be stable it is sufficient to impose the conservation of just one of these two symmetries, without any phenomenological reason to impose the conservation of both. This gives rise to a wide variety of possible discrete symmetries, apart from the usual R-parity, that protect the proton while allowing for L or B violation.

One of such symmetries is baryon triality ($Z_3^B$) \cite{Ibanez:1991pr,Dreiner:2005rd}, defined as

\begin{equation}
Z_3^B = \exp \left[ 2 \pi i (B-2Y)/3 \right]
\end{equation}

which exactly conserves baryon number but allows for lepton number violation. Models based on this type of symmetries break R-parity but do not suffer from fast proton decay.

Apart from these arguments against the introduction of R-parity one should not forget the rich phenomenology that it predicts at current and future experiments. This extension of the MSSM includes additional couplings that lead to distinctive signatures at the SUSY scale. This makes SUSY with \rpv a testable framework, possible to rule out at the LHC.

In fact, the \rpv couplings are highly constrained by experiments due to the non-observation of L and B violating processes. Searches for neutrinoless double beta decay \cite{Mohapatra:1986su,Hirsch:1995zi,Hirsch:1995ek,Faessler:1997db,Hirsch:1999ce}, nucleon-antinucleon oscillations \cite{Zwirner:1984is,Goity:1994dq} and proton decay \cite{Hinchliffe:1992ad,Vissani:1995hp,Smirnov:1996bg,Hoang:1997kf,Bhattacharyya:1998bx,Bhattacharyya:1998dt} have led to strong constraints on the size of the \rpv couplings.

Furthermore, if present, the \rpv couplings would also contribute to L and B conserving processes. Some examples are rare leptonic decays of mesons, like $K^0 \to l_i^+ l_j^-$ \cite{Choudhury:1996ia}, $B \bar{B}$ mixing \cite{deCarlos:1996yh,Bhattacharyya:1998be} and $\mu \to e \gamma$ \cite{deGouvea:2000cf}. The agreement between the data and the SM predictions implies that the \rpv couplings must be below the experimental accuracy, which can be used to derive robust indirect bounds on the individual couplings or on combinations of them.

\begin{figure}
\centering
\subfigure[SM]{
\includegraphics[width=0.46\textwidth]{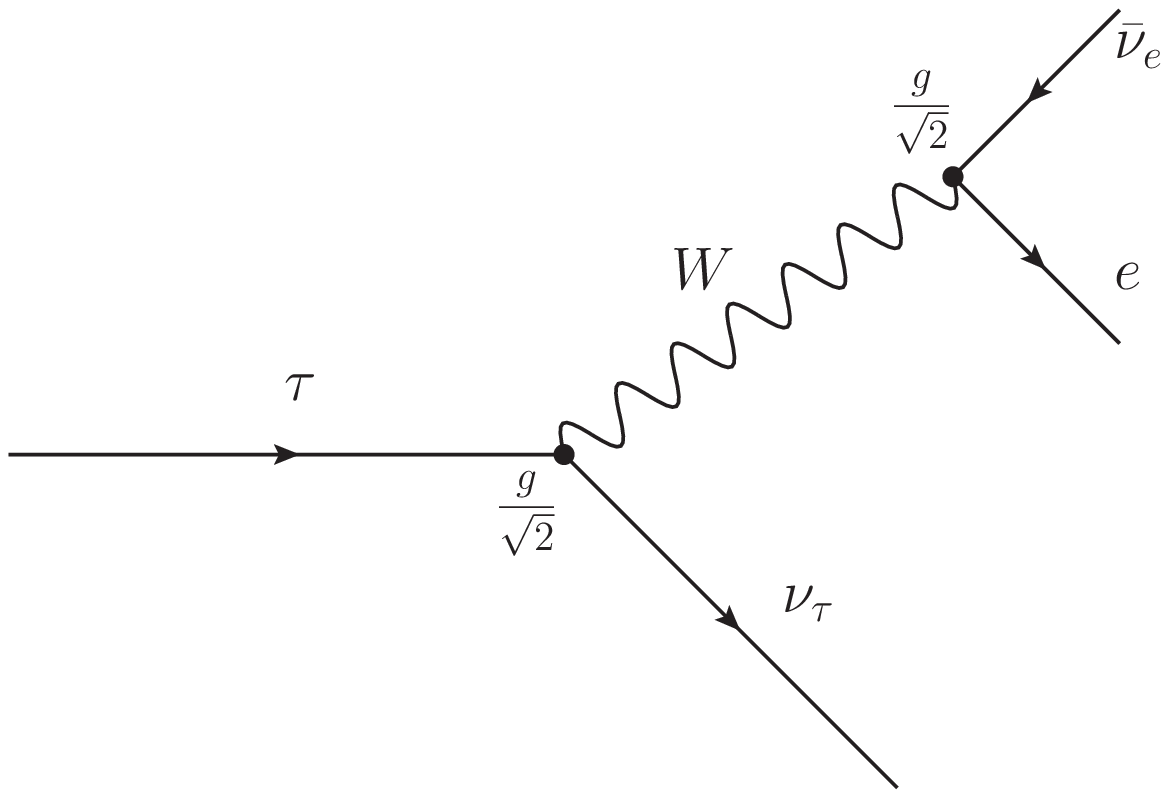}
\label{tau-decay-sm}
}
\subfigure[\rpv]{
\includegraphics[width=0.46\textwidth]{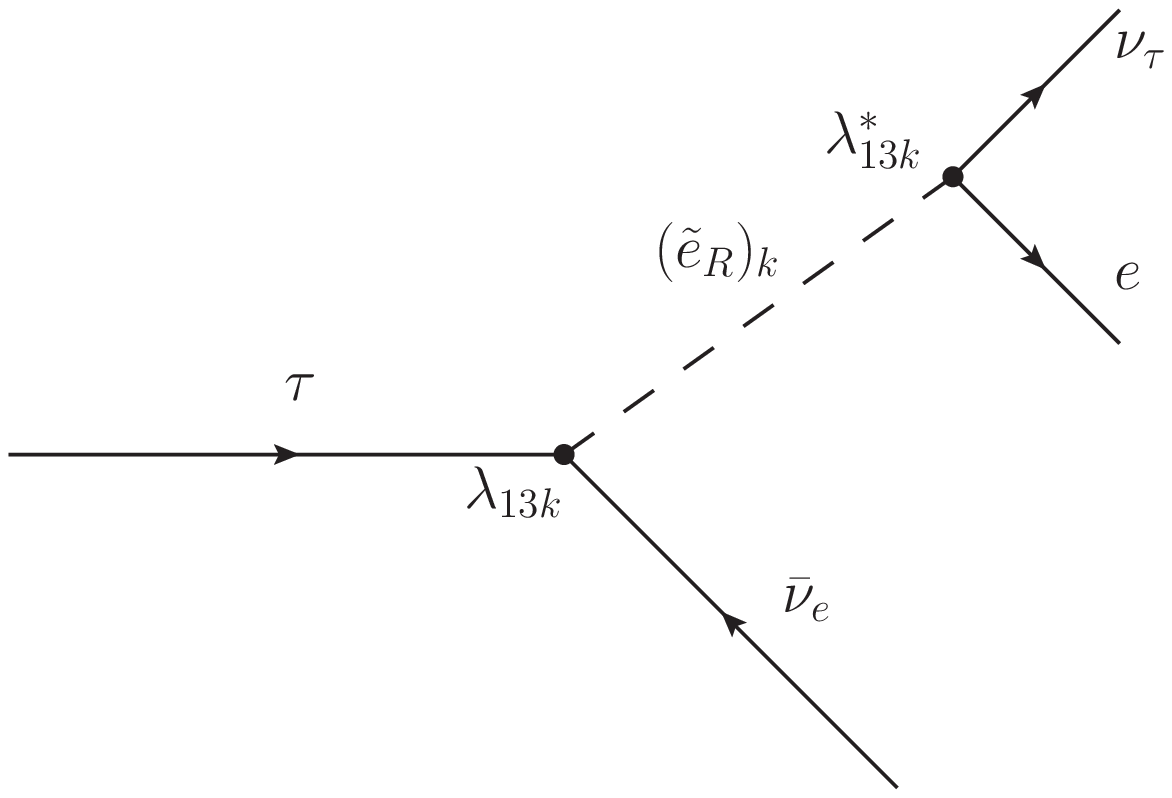}
\label{tau-decay-rpv}
}
\caption{$\tau$ lepton decay: Standard Model and \rpv contributions.}
\end{figure}

Let us consider an example. The operator $\widehat{L}_1 \widehat{L}_3 \widehat{e}^c_k$ contributes to the $\tau$ lepton decay, $\tau \rightarrow e \nu \bar{\nu}$, violating lepton number twice, as shown in figure \ref{tau-decay-rpv}. In the first vertex one has $\Delta L=+1$ while in the second $\Delta L=-1$, and then the global process conserves lepton number. Since the mediator of the decay, the right-handed slepton $(\tilde{e}_R)_k$, is much heavier than the rest of involved particles one can describe the process by an effective 4-fermion lagrangian

\begin{equation} \label{lageffrpv}
\mathcal{L}_{\text{eff}} = \frac{|\lambda_{13k}|^2}{2\tilde{m}_k^2} (\bar{e}_L \gamma^{\mu} \nu_{e L}) (\bar{\nu}_{\tau L} \gamma_{\mu} \tau_L)
\end{equation}

where $\tilde{m}_k \equiv m(\tilde{e}_R)_k$. This decay can be represented in the SM by the Feynman diagram in figure \ref{tau-decay-sm}, which has the Fermi lagrangian as effective theory, with the same structure as in \eqref{lageffrpv}. Therefore, the \rpv contribution generates an apparent shift in the Fermi constant $G_F$. By measuring the ratio

\begin{equation}
R_{\tau} \equiv \frac{\Gamma(\tau \rightarrow e \nu \bar{\nu})}{\Gamma(\tau \rightarrow \mu \nu \bar{\nu})}
\end{equation}

one obtains the following deviation from the SM

\begin{equation}
R_{\tau} = R_{\tau}(SM) \left[ 1 + 2\frac{m_W^2}{g^2} \left( \frac{|\lambda_{13k}|^2}{\tilde{m}_k^2} \right) \right]
\end{equation}

Then, the combination of the experimental value $R_{\tau} = 1.028 \pm 0.004$, published in the final report by the ALEPH collaboration \cite{Schael:2005am}, and the theoretical computation $R_{\tau}(SM) = 1.028$ implies the bound

\begin{equation}
|\lambda_{13k}| < 0.05 \left( \frac{\tilde{m}_k}{100 \: \text{GeV}} \right)
\end{equation}

Let us mention that this bound assumes that this is the only \rpv operator contributing to the process. In a more general case one obtains more complicated contraints that involve combinations of different operators.

It is possible to derive similar bounds on other coupligs following the same approach and using constrains from other physical processes. See the review \cite{Barbier:2004ez} for a collection of bounds on trilinear \rpv couplings. In addition, the reference \cite{Kao:2009fg} gives an update including recent experimental data.

Finally, in \rpv the standard neutralino is lost as a dark matter candidate. Recent WMAP data \cite{Komatsu:2010fb}, however, have confirmed the existence of non-baryonic dark matter and measured its contribution to the energy budget of the universe with unprecedented accuracy. Thus, in \rpv one needs a non-standard explanation of DM. Examples for DM candidates in \rpv include (i) light gravitinos \cite{Borgani:1996ag,Takayama:2000uz,Hirsch:2005ag}, (ii) the axion \cite{Kim:1986ax,Raffelt:1996wa} or (iii) its superpartner, the axino \cite{Chun:1999cq,Chun:2006ss}, to mention a few.

In the following section the simplest \rpv model that generates neutrino masses is discussed. For general reviews see \cite{Barbier:2004ez,Chemtob:2004xr}.

\section{Minimal realization: b-\rpv} \label{sec:brpv}

Bilinear R-parity Violation (b-\rpv) \cite{Hall:1983id} is the minimal extension of the MSSM that incorporates lepton number violation. It is therefore interesting to begin the discussion with this simple case that already contains many of the features present in extended models. For a pedagogical review on b-\rpv see the reference \cite{Hirsch:2004he}.

\subsection{The model}

The b-\rpv superpotential is

\begin{equation}\label{brpv-superpotential}
W=W^{MSSM} + \epsilon_{ab} \, \epsilon_i \widehat{L}_i^a \widehat{H}_u^b
\end{equation}

The three new $\epsilon_i = (\epsilon_e,\epsilon_\mu,\epsilon_\tau)$ parameters have dimensions of mass and break lepton number. As will be shown below, they are constrained by neutrino masses to be much smaller than the EW scale ($\epsilon_i \ll m_W$). Note however that they are SUSY conserving parameters and thus one would naively expect them to be at the GUT or Planck scales. This problem is analogous to the $\mu$-problem of the MSSM and can be addressed using the same approach. In fact, any solution to the $\mu$-problem potentially solves the $\epsilon_i$-problem \cite{Nilles:1996ij}. This will be discussed in detail in the sections devoted to s-\rpv and the $\mu \nu$SSM model.

The introduction of the new superpotential terms implies new soft terms as well

\begin{equation}\label{brpv-soft}
V_{soft}^{b\mbox{-}\rpvm} = - B_i \epsilon_i \, \epsilon_{ab} \, \tilde{L}_i^a H_u^b
\end{equation}

where the $B_i$ parameters have dimensions of mass.

With these new couplings with respect to the MSSM the scalar potential of the theory is modified. At tree-level one obtains the following linear terms

\begin{equation}
V_{linear}^0=t_d^0\sigma^0_d+t_u^0\sigma^0_u+t_1^0\tilde\nu^R_1 +t_2^0\tilde\nu^R_2+t_3^0\tilde\nu^R_3
\end{equation}

where $t^0$ are the tree-level tadpoles, given by

\begin{equation}\label{tadpole}
\begin{split}
t_d^0 = & -B_\mu v_u+\big(m_{H_d}^2+\mu^2\big)v_d+v_dD-\mu v_i\epsilon_i \\
t_u^0 = & -B_\mu v_d+\big(m_{H_u}^2+\mu^2\big)v_u-v_uD+v_iB_i\epsilon_i+v_u\epsilon^2\\
t_1^0 = & v_1D+\epsilon_1\big(-\mu v_d+v_uB_1+v_i\epsilon_i\big)+\frac{1}{2}\left(v_i (m_L^2)_{i1}+ (m_L^2)_{1i} v_i\right)\\
t_2^0 = & v_2D+\epsilon_2\big(-\mu v_d+v_uB_2+v_i\epsilon_i\big)+\frac{1}{2}\left(v_i (m_L^2)_{i2}+ (m_L^2)_{2i} v_i\right)\\
t_3^0 = & v_3D+\epsilon_3\big(-\mu v_d+v_uB_3+v_i\epsilon_i\big)+\frac{1}{2}\left(v_i (m_L^2)_{i3}+ (m_L^2)_{3i} v_i\right)
\end{split}
\end{equation}

Here a repeated index $i$ implies summation over $i=1,2,3$. Moreover, the usual shifts in the neutral fields have been performed

\begin{gather}
H_d^0 = \frac{1}{\sqrt{2}}[\sigma_d^0 + v_d + i\phi_d^0] \nonumber\\
H_u^0 = \frac{1}{\sqrt{2}}[\sigma_u^0 + v_u + i\phi_u^0] \\
\tilde{\nu}_i = \frac{1}{\sqrt{2}}[\tilde{\nu}_i^R + v_i +
i\tilde{\nu}_i^I] \nonumber
\end{gather}

and thus left-handed sneutrino VEVs $v_i = (v_e , v_\mu , v_\tau)$ have been introduced. Finally two useful quantities have been defined

\begin{eqnarray}
D &=& \frac{1}{8}(g^2 + g'^2)(v_e^2+v_\mu^2+v_\tau^2+v_d^2-v_u^2) \\
\epsilon^2 &=& \epsilon_e^2 + \epsilon_\mu^2 + \epsilon_\tau^2
\end{eqnarray}

The minimum of the potential is computed by solving the system of equations obtained when all the tadpoles $t^0$ vanish. Note that the superpotential term $\epsilon \widehat L \widehat H_u$ generates a tadpole for the sneutrino once $H_u^0$ gets a VEV. Therefore, the non-vanishing left-handed sneutrino VEV is a generic feature of $\epsilon$-type models, this is, \rpv models with explicit or effective $\epsilon$ coupling. In conclusion, besides the usual Higgs VEVs

\begin{eqnarray}
\langle H_d^0 \rangle &=& \frac{v_d}{\sqrt{2}} \\
\langle H_u^0 \rangle &=& \frac{v_u}{\sqrt{2}}
\end{eqnarray}

one obtains

\begin{equation}
\langle \tilde{\nu}_i \rangle = \frac{v_i}{\sqrt{2}} \neq 0
\end{equation}

The MSSM limit of the model is obtained when $\epsilon_i \rightarrow 0$. Note that in this limit the $v_i$ VEVs vanish (see tadpole equations $t_1^0$, $t_2^0$ and $t_3^0$) and thus the sneutrinos do not get VEVs, fully recovering the MSSM. Furthermore, the smallness of the $\epsilon_i$ parameters implies that the sneutrino VEVs $v_i$ must be small as well. This is of fundamental importance due to the doublet nature of the sneutrinos, which has an impact on the electroweak sector of the model. For example, after the gauge symmetry is broken the $W$ boson gets a mass

\begin{equation}
m_W^2 = \frac{1}{4} g^2 (v_d^2 + v_u^2 + v_e^2 + v_\mu^2 + v_\tau^3) \simeq \frac{1}{4} g^2 (v_d^2 + v_u^2)
\end{equation}

and the smallness of the sneutrino VEVs guarantees that the well-known SM result is obtained in good approximation.

\subsection{Neutrino masses} \label{subsec:brpv-numass}

The main motivation for \rpv is the generation of neutrino masses \cite{Hirsch:2000ef,Diaz:2003as,Chun:1999bq}. In the basis $(\psi^0)^T= (-i\tilde{B}^0,-i\tilde{W}_3^0,\widetilde{H}_d^0,\widetilde{H}_u^0, \nu_{e}, \nu_{\mu}, \nu_{\tau} )$ the neutral fermion mass matrix $M_N$ is given by

\begin{equation} \label{brpv-massF0}
M_N=\left(
\begin{array}{cc}
{\cal M}_{\chi^0}& m^T \cr \vb{20} m & 0 \cr
\end{array}
\right)
\end{equation}

where

\begin{equation}
{\cal M}_{\chi^0}\hskip -2pt=\hskip -4pt \left( \begin{array}{cccc}
M_1 & 0 & -\frac 12g^{\prime }v_d & \frac 12g^{\prime }v_u \cr
\vb{12} 0 & M_2 & \frac 12gv_d & -\frac 12gv_u \cr \vb{12} -\frac
12g^{\prime }v_d & \frac 12gv_d & 0 & -\mu  \cr \vb{12} \frac
12g^{\prime }v_u & -\frac 12gv_u & -\mu & 0  \cr
\end{array} \right)
\end{equation}

is the usual neutralino mass matrix of the MSSM and

\begin{equation}
m=\left(
\begin{array}{cccc}
-\frac 12g^{\prime }v_e & \frac 12g v_e & 0 & \epsilon_e \cr \vb{18}
-\frac 12g^{\prime }v_\mu & \frac 12g v_\mu & 0 & \epsilon_\mu  \cr
\vb{18} -\frac 12g^{\prime }v_\tau & \frac 12g v_\tau & 0 & \epsilon_\tau
\cr
\end{array}
\right)
\end{equation}

is the matrix that characterizes the breaking of R-parity, mixing the neutrinos with the neutralinos. Note that its elements are suppressed with respect to the elements in ${\cal M}_{\chi^0}$ due to the smallness of the $\epsilon_i$ parameters. In fact, it is useful to define the expansion parameters

\begin{equation}
\xi = m \cdot {\cal M}_{\chi^0}^{-1}
\end{equation}

where $\xi$ denotes a $3\times 4$ matrix given as~\cite{Hirsch:1998kc}

\begin{eqnarray}
\xi_{i1} &=& \frac{g' M_2 \mu}{2 {\rm Det}({\cal M}_{\chi^0})}\Lambda_i \cr
\vb{20}
\xi_{i2} &=& -\frac{g M_1 \mu}{2 {\rm Det}({\cal M}_{\chi^0})}\Lambda_i \cr
\vb{20}
\xi_{i3} &=& - \frac{\epsilon_i}{\mu} + 
          \frac{(g^2 M_1 + {g'}^2 M_2) v_u}
               {4 {\rm Det}({\cal M}_{\chi^0})}\Lambda_i \cr
\vb{20}
\xi_{i4} &=& - \frac{(g^2 M_1 + {g'}^2 M_2) v_d}
               {4 {\rm Det}({\cal M}_{\chi^0})}\Lambda_i
\end{eqnarray}

The resulting $M_N$ matrix has a type-I seesaw structure and thus the effective mass matrix of the light neutrinos can be obtained with the usual formula

\begin{equation} \label{meff-brpv}
m_{eff}^{\nu \nu} = - m \cdot {\cal M}_{\chi^0}^{-1} m^T
\end{equation}

By explicit computation one can show that \eqref{meff-brpv} can be expanded to give

\begin{equation} \label{meff-lambda-brpv}
m_{eff}^{\nu \nu} = \frac{M_1 g^2 \!+\! M_2 {g'}^2}{4\, {\rm Det}({\cal M}_{\chi^0})}\left(\hskip -2mm \begin{array}{ccc}
\Lambda_e^2
\hskip -1pt&\hskip -1pt
\Lambda_e \Lambda_\mu
\hskip -1pt&\hskip -1pt
\Lambda_e \Lambda_\tau \\
\Lambda_e \Lambda_\mu
\hskip -1pt&\hskip -1pt
\Lambda_\mu^2
\hskip -1pt&\hskip -1pt
\Lambda_\mu \Lambda_\tau \\
\Lambda_e \Lambda_\tau
\hskip -1pt&\hskip -1pt
\Lambda_\mu \Lambda_\tau
\hskip -1pt&\hskip -1pt
\Lambda_\tau^2
\end{array}\hskip -3mm \right)
\end{equation}

where

\begin{equation}
\Lambda_i = \mu v_i + v_d \epsilon_i
\end{equation}

It is important to emphasize once again that in the limit $\epsilon_i \to 0$ all the entries in $m_{eff}^{\nu \nu}$ vanish and one recovers the MSSM with massless neutrinos.

By diagonalizing equation \eqref{meff-lambda-brpv} one obtains

\begin{equation}
U_{\nu}^T m_{eff}^{\nu \nu} U_{\nu} = {\rm diag}(0,0,m_{\nu})
\end{equation}

with

\begin{equation}
m_{\nu} = Tr(m_{eff}^{\nu \nu}) = \frac{M_1 g^2 + M_2 {g'}^2}{4\, {\rm Det}({\cal M}_{\chi^0})} |{\vec \Lambda}|^2
\end{equation}

The special form of $m_{eff}^{\nu \nu}$ shown in equation \eqref{meff-lambda-brpv} implies that it is a rank $1$ matrix which only has one non-zero eigenvalue. With this degeneracy between the two massless eigenstates one can rotate away one of the angles in the matrix $U_{\nu}$

\begin{equation}
U_{\nu}= \left(\begin{array}{ccc}
  1 &                0 &               0 \\
  0 &  \cos\theta_{23} & -\sin\theta_{23} \\
  0 &  \sin\theta_{23} & \cos\theta_{23}
\end{array}\right) \times \left(\begin{array}{ccc}
  \cos\theta_{13} & 0 & -\sin\theta_{13} \\
                0 & 1 &               0 \\
  \sin\theta_{13} & 0 & \cos\theta_{13}
\end{array}\right)
\end{equation}

and compute the other two angles as follows

\begin{eqnarray}
\tan\theta_{13} &=& - \frac{\Lambda_e}{(\Lambda_{\mu}^2+\Lambda_{\tau}^2)^{\frac{1}{2}}} \label{brpv-angle13} \\
\tan\theta_{23} &=& - \frac{\Lambda_{\mu}}{\Lambda_{\tau}} \label{brpv-angle23}
\end{eqnarray}

Of course, this is not enough to explain the oscillation data, which at least requires the generation of two mass scales, $\Delta m_{atm}^2$ and $\Delta m_{sol}^2$. However, \eqref{meff-lambda-brpv} is just the tree-level neutrino mass matrix. Its non-vanishing eigenvalue $m_{\nu_3}$ is interpreted as the atmospheric mass scale, whereas for the generation of the solar mass scale, which is much smaller ($\Delta m_{sol}^2 \ll \Delta m_{atm}^2$), one must go beyond the tree-level approximation.

In general, every mass matrix $M_{ij}$ can be written as

\begin{equation}
M_{ij} = M_{ij}^{(0)} + M_{ij}^{(1)} + \dots
\end{equation}

where $M_{ij}^{(0)}$ is the tree-level mass matrix, $M_{ij}^{(1)}$ is the 1-loop correction and the dots stand for higher order corrections. In the lines above only the tree-level neutrino mass matrix has been computed, leading to the generation of the atmospheric mass scale. The addition of loop corrections changes the structure of the resulting mass matrix and generates the solar mass scale.

\begin{figure}
\begin{center}
\begin{tabular}{cc}
\includegraphics[width=0.45\linewidth]{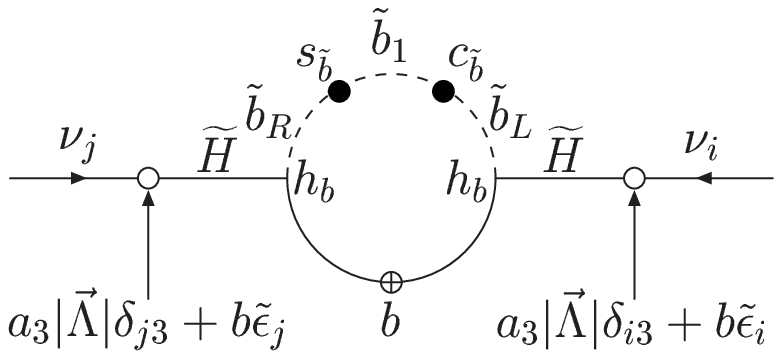}
&\includegraphics[width=0.45\linewidth]{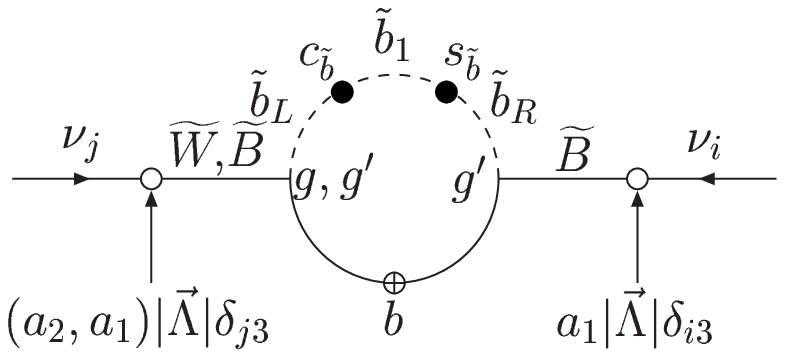}
\end{tabular}
\end{center}
\vspace{-3mm}
\caption{Bottom--Sbottom diagrams for solar neutrino mass in the b-\rpv model. The following conventions have been adopted: a) Open circles correspond to small R-parity violating projections, indicating how much of a weak eigenstate is present in a given mass eigenstate, (b) full circles correspond to R-parity conserving projections and (c) open circles with a cross inside indicate genuine mass insertions which flip chirality. In these figures $h_b \equiv Y_b$ is the bottom quark Yukawa coupling. For more details see \cite{Diaz:2003as}.}
\label{fig:BottomSbottomLoop}
\end{figure}

In particular, it turns out that in mSUGRA like models the most important 1-loop contribution to the neutrino mass matrix comes from the bottom-sbottom loop shown in figure \ref{fig:BottomSbottomLoop} \cite{Hirsch:2000ef,Diaz:2003as}. Neglecting the other contributions one gets

\begin{equation}
m_{\nu_2} \simeq \frac{3}{16 \pi^2} \sin(2\theta_{\tilde b}) m_b Y_b^2 \Delta B_0^{\tilde b_2\tilde b_1}\frac{({\tilde \epsilon}_1^2 + {\tilde \epsilon}_2^2)}{\mu^2}
\end{equation}

In this expression $\theta_{\tilde b}$ is the mixing angle in the sbottom sector, $\Delta B_0^{\tilde b_2\tilde b_1}$ is the difference between two Passarino-Veltman functions \cite{Passarino:1978jh}

\begin{equation}
\Delta B_0^{\tilde b_2\tilde b_1} = B_0(0,m_b^2,m_{\tilde{b}_1}^2) - B_0(0,m_b^2,m_{\tilde{b}_2}^2)
\end{equation}

and the parameters $\tilde\epsilon$ are defined as $\tilde\epsilon_i = (U_\nu^T)^{ij} \epsilon_j$ with the following resulting expressions in terms of the original parameters

\begin{eqnarray}
\tilde\epsilon_1 &=&{{\epsilon_e(\Lambda_{\mu}^2+\Lambda_{\tau}^2)-\Lambda_e(\Lambda_{\mu}\epsilon_{\mu}+\Lambda_{\tau}\epsilon_{\tau})
}\over{\sqrt{\Lambda_{\mu}^2+\Lambda_{\tau}^2}\sqrt{\Lambda_e^2+\Lambda_{\mu}^2+\Lambda_{\tau}^2}}} \\
\tilde\epsilon_2 &=&{{\Lambda_{\tau}\epsilon_{\mu}-\Lambda_{\mu}\epsilon_{\tau}}\over{\sqrt{\Lambda_{\mu}^2+\Lambda_{\tau}^2}}} \\
\tilde\epsilon_3 &=&{{ \vec \Lambda \cdot \vec \epsilon}\over{\sqrt{\Lambda_{e}^2 + \Lambda_{\mu}^2+\Lambda_{\tau}^2}}}
\end{eqnarray}

Complete numerical calculations show that this single contribution reproduces the exact result in good approximation in most parts of parameter space. Other relevant 1-loop corrections are the tau-stau and neutrino-sneutrino contributions, see references \cite{Diaz:2003as,Grossman:2003gq} for detailed studies.

Once $m_{\nu_1}$ and $m_{\nu_2}$ are generated and the degeneracy between the two mass eigenstates is broken, a new mixing angle appears. This mixing angle is identified with the solar mixing angle and it can be expressed as

\begin{equation}
\tan^2\theta_{12}  = \frac{\tilde{\epsilon}_1^2}{\tilde{\epsilon}_2^2}
\end{equation}

In conclusion, the b-\rpv model provides a simple framework to accommodate current data on neutrino masses and mixing angles. The breaking of R-parity mixes neutrinos and neutralinos, giving rise to a EW scale type-I seesaw. Moreover, the hierarchy $\Delta m_{sol}^2 \ll \Delta m_{atm}^2$ is nicely explained in b-\rpv, since the atmospheric scale is obtained at tree-level where as the solar scale has its origin in radiative corrections.

\subsection{Phenomenology}
\label{sec:brpv-pheno}

The phenomenology of b-\rpv has been extensively discussed in the literature, see for example \cite{Mukhopadhyaya:1998xj,Choi:1999tq,Romao:1999up,Porod:2000hv,Restrepo:2001me,Bartl:2003uq,deCampos:2007bn,DeCampos:2010yu}, and is beyond the scope of this thesis. Nevertheless, two important points will be briefly mentioned here due to their relevance also in extended $\epsilon$-type models, discussed in the following chapters.

The breaking of R-parity has an immediate consequence at colliders: the LSP in no longer stable and decays. In fact, this is the main change with respect to the standard MSSM phenomenology, because the \rpv couplings are constrained to be small, and thus they do not affect either the production cross sections or the initial steps of the decay chains.

The most important lesson that one can learn from b-\rpv is that the LSP decay is governed by neutrino physics. The connection is clear. The same $\epsilon$ parameters that break R-parity and lead to LSP decay are the ones which generate neutrino masses and mixing angles. In fact, the simplicity of the model, that only has $6$ \rpv parameters ($3 \: \epsilon_i$ and the corresponding soft parameters), tightens this link and allows to give definite predictions at colliders\footnote{For comparison, we mention that the lepton sector of a type-I seesaw contains 21 physical parameters, see for example \cite{Hirsch:2008dy}.}. 

The first prediction has to do with the decay length of the LSP \cite{Mukhopadhyaya:1998xj}. This depends on the nature of the LSP (neutralino, stau, \dots), its mass and its couplings. Therefore, it is hard to give a definite prediction. However, some general statements are still possible due to its link with neutrino masses.

On the one hand the amplitude of the LSP decay grows for increasing values of the \rpv couplings whereas, on the other hand, the size of the \rpv couplings determines the absolute scale of neutrino masses. Therefore, one can derive a relation between the amplitude of LSP decay, $\Gamma$, and the absolute scale of neutrino masses, $m_\nu$. In fact, by assuming a neutralino LSP and taking the value $m_{SUSY} \sim 100$ GeV one can estimate

\begin{equation}
\Gamma \simeq 10^{-4} \cdot m_{\nu}
\end{equation}

and then, by using the known formula

\begin{equation}
L = c \tau = \frac{c}{\Gamma}
\end{equation}

one can estimate the decay length $L$ as well. For $m_{\nu} = \sqrt{\Delta m_{atm}^2}$ a distance of the order of cm is obtained. This would be a clear signature at colliders such as the LHC \cite{deCampos:2007bn,DeCampos:2010yu}. Moreover, note that the decay length of the LSP has a lower bound coming from the upper bound on the absolute scale of neutrino masses. Therefore, the prediction of a long decay length is general and holds in most parts of the parameter space.

The second definite prediction that b-\rpv makes is the sharp correlation between some ratios of branching ratios of LSP decays and the neutrino mixing angles \cite{Mukhopadhyaya:1998xj,Choi:1999tq,Romao:1999up,Porod:2000hv}. This tight link allows to use neutrino oscillation data to test the model at colliders. The reason comes again from the fact that the couplings that lead to LSP decay are the same as the ones that give the structure of the neutrino mass matrix.

Let us consider the L violating decay channels $\tilde{\chi}_1^0 \to W^\pm \, l_i^\mp$. If the mass of the LSP is above the $W$ boson mass these channels are open and in fact they typically become dominant in most parts of the parameter space. Note that the breaking of R-parity mixes the charged leptons with the MSSM charginos, in the same way as the neutrinos get mixed with the neutralinos. Therefore, the vertex that is involved in the decay $\tilde{\chi}_1^0 \to W^\pm \, l_i^\mp$ is $\chi_i^0$-$\chi_j^{\mp}$-$W^{\pm}$ and the relevant interaction lagrangian for this process is

\begin{equation}
\begin{split}
\mathcal{L}= & \overline{\chi_i^-} \gamma^{\mu} \big( O_{Lij}^{cnw} P_L + O_{Rij}^{cnw} P_R \big) \chi_j^0 W_{\mu}^- +\\
& \overline{\chi_i^0} \gamma^{\mu} \big( O_{Lij}^{ncw} P_L + O_{Rij}^{ncw} P_R \big) \chi_j^- W_{\mu}^+
\end{split}
\end{equation}

where $\chi_i^0$ represents the neutralino $i$, $\chi_j^\pm$ represents the chargino $j$ and $P_L$ and $P_R$ are the chirality projectors. The couplings $O_{Lij}^{cnw}$, $O_{Rij}^{cnw}$, $O_{Lij}^{ncw}$ and $O_{Rij}^{ncw}$ turn out to be

\begin{equation} \label{o-l1}
O_{Lij}^{cnw} = g S_j^C S_i^N \big( - \mathcal{N}_{j2}^{\ast} \mathcal{U}_{i1} - \frac{1}{\sqrt{2}} \big( \mathcal{N}_{j3}^{\ast} \mathcal{U}_{i2} + \sum_{k=1}^3 \mathcal{N}_{j,4+k}^{\ast} \mathcal{U}_{i,2+k} \big) \big)
\end{equation}

\begin{equation} \label{o-r1}
O_{Rij}^{cnw} = g \big( -\mathcal{N}_{j2} \mathcal{V}_{i1}^{\ast} + \frac{1}{\sqrt{2}} \mathcal{N}_{j4} \mathcal{V}_{i2}^{\ast} \big)
\end{equation}

\begin{equation} \label{o-l2}
O_{Lij}^{ncw} = \big( O_{Lij}^{cnw} \big)^{\ast}
\end{equation}

\begin{equation} \label{o-r2}
O_{Rij}^{ncw} = \big( O_{Rij}^{cnw} \big)^{\ast}
\end{equation}

where $\mathcal{N}$ is the rotation matrix that diagonalizes the $7 \times 7$ neutral fermion mass matrix ($4$ MSSM neutralinos $+ 3$ neutrinos), $\mathcal{U}$ and $\mathcal{V}$ diagonalize the $5 \times 5$ charged fermion mass matrix ($2$ MSSM charginos $+ 3$ charged leptons) and $S_i^C$ and $S_i^N$ are sign parameters defined as

\begin{equation}
S_i^C = \frac{m_{\chi_i^-}}{|m_{\chi_i^-}|} \quad , \quad S_i^N = \frac{m_{\chi_i^0}}{|m_{\chi_i^0}|}
\end{equation}

For the decay of a neutralino LSP one must focus on the couplings $O_{Li1}^{cnw}$ and $O_{Ri1}^{cnw}$. It is easy to find approximated formulas for the rotation matrices $\mathcal{N}$, $\mathcal{U}$ and $\mathcal{V}$ taking advantage of the small mixing between the MSSM and lepton sectors \cite{Hirsch:1998kc}. After applying them the final result is

\begin{equation} \label{o-l1-ap}
\begin{split}
O_{Li1}^{cnw} \simeq & \frac{g \Lambda_i}{\sqrt{2}} \bigg[ - \frac{g' M_2 \mu}{2 \textnormal{Det}_0} N_{11} + g \bigg( \frac{1}{\textnormal{Det}_+} + \frac{M_1 \mu}{2 \textnormal{Det}_0} \bigg) N_{12}\\
& - \frac{v_u}{2} \bigg( \frac{g^2 M_1 + g^{\prime 2} M_2}{2 \textnormal{Det}_0} + \frac{g^2}{\mu \textnormal{Det}_+} \bigg) N_{13} + \frac{v_d(g^2 M_1 + g^{\prime 2} M_2)}{4 \textnormal{Det}_0} N_{14} \bigg]
\end{split}
\end{equation}

\begin{equation} \label{o-r1-ap}
O_{Ri1}^{cnw} \simeq \frac{g Y_e^{ii} v_d}{2 \textnormal{Det}_+} \bigg[ \frac{g v_d N_{12} + M_2 N_{14}}{\mu} \epsilon_i + g \frac{(2\mu^2 + g^2 v_d v_u)N_{12} + (\mu + M_2)g v_u N_{14}}{2 \mu \textnormal{Det}_+} \Lambda_i \bigg]
\end{equation}

Here $\textnormal{Det}_0$ and $\textnormal{Det}_+$ are the determinants of the MSSM neutralino and chargino mass matrices, whereas $N$ is the matrix that diagonalizes the MSSM neutralino mass matrix.

Now it is easy to understand why there are correlations between these decays and the neutrino mixing angles measured at oscillation experiments. For a pure bino LSP ($N_{11} = 1$, $N_{1i} = 0$ for $i \neq 1$) the couplings in equations \eqref{o-l1-ap} and \eqref{o-r1-ap} become

\begin{eqnarray}
O_{Li1}^{cnw} & \simeq & - \frac{g g' M_2 \mu}{2 \sqrt{2} \textnormal{Det}_0} \Lambda_i \\
O_{Ri1}^{cnw} & \simeq & 0
\end{eqnarray}

and then one obtains

\begin{equation} \label{brpv-pred}
\frac{Br(\tilde{\chi}_1^0 \to W \mu)}{Br(\tilde{\chi}_1^0 \to W \tau)} \simeq \left( \frac{\Lambda_\mu}{\Lambda_\tau} \right)^2 = \tan^2 \theta_{23}
\end{equation}

where equation \eqref{brpv-angle23} has been used. As shown in equation \eqref{brpv-pred}, b-\rpv makes a sharp prediction for the ratio between $Br(\tilde{\chi}_1^0 \to W \mu)$ and $Br(\tilde{\chi}_1^0 \to W \tau)$, predicted to be given by $\tan^2 \theta_{23} \simeq 1$, as measured in atmospheric neutrino experiments. A departure from this value would clearly rule out the model.

Only the case of a neutralino LSP has been discussed here. However other possibilities with interesting phenomenological consequences have been considered in the literature as well, see references \cite{Hirsch:2002ys,Hirsch:2003fe}.

The two phenomenological issues briefly discussed here, the long LSP decay length and the correlations with the neutrino mixing angles, are common to all $\epsilon$-type \rpv models. The b-\rpv case, being the simplest one, will be used in the following as the basic reference to compare with.

\subsection{Motivation for extended models}

It has already been shown how b-\rpv is able to accommodate the observed pattern of neutrino masses and mixing angles and the resulting predictions for collider experiments. Such a predictive model would be easily ruled out at the LHC if it does not provide the explanation for neutrino masses. Therefore, it is an interesting alternative to the usual high scale seesaw mechanism.

However, there are some theoretical issues without explanation in b-\rpv. One of these open questions is why some of the \rpv couplings in the superpotential are zero while others, $\epsilon_i$ in particular, are not. This problem has been addressed in the literature by different means. For example, one can break R-parity spontaneously when a scalar field, charged under R-parity, acquires a VEV \cite{Aulakh:1982yn,Masiero:1990uj}. If this is done properly, only some of the \rpv couplings are generated at tree-level. One can also suppose that some couplings are zero at the GUT scale and use the RGEs to show that their running down to the SUSY scale is negligible \cite{Escudero:2008jg}. Finally, there are also approaches that rely on high-energy constructions \cite{Casas:1987us}.

The second open question in b-\rpv is the size of the $\epsilon$ couplings. Since these are SUSY conserving couplings with dimension of mass one would expect that they are generated at some high scale, like the GUT or Planck scales, and thus it is hard to understand why they are so small compared to the electroweak scale. This problem, that is analogous to the $\mu$-problem in the MSSM, can be easily solved if the $\epsilon$ couplings are generated by physics at the SUSY scale. In fact, one can use a solution similar to the one provided in the NMSSM for the $\mu$-problem \cite{Barbieri:1982eh,Nilles:1982dy}. The idea is to add an additional gauge singlet superfield, that will be called $\widehat S$ in this section, whose scalar component acquires a VEV generating an effective $\epsilon$ term

\begin{equation}
\lambda_i S L_i H_u \: \to \: \lambda_i \langle S \rangle L_i H_u = \epsilon_i L_i H_u
\end{equation}

If this VEV is generated at the electroweak scale one can easily explain the smallness of the $\epsilon_i$ parameters by using a small coupling $\lambda_i \sim 10^{-7}-10^{-6}$. This idea can be applied to models that break R-parity spontaneously \cite{Masiero:1990uj} or explicitly \cite{LopezFogliani:2005yw}. In fact, there are also examples in which the field that gets a VEV is not a gauge singlet \cite{Aulakh:1982yn,Ellis:1984gi,Ross:1984yg}, although they are nowadays ruled out due to the strong experimental constraints in the electroweak sector.

In the following chapters these ideas to address the problems of the minimal b-\rpv will be discussed in detail for two extended models: s-\rpv and $\mu \nu$SSM.

\chapter{Spontaneous R-parity violation}
\label{chap:srpv}

Spontaneous R-parity violation (s-\rpv) is a well motivated alternative to the simple explicit violation in b-\rpv. In this type of extended models the original theory conserves R-parity and it is only after symmetry breaking that the \rpv terms are generated. This way one can explain (a) why some \rpv couplings are not generated at tree-level and (b) why the bilinear coupling is much smaller than the electroweak scale.

\section{Introduction}

One of the most important consequences of the spontaneous breaking of R-parity is the presence of a Goldstone boson, the majoron (J) \cite{Chikashige:1980ui,Gelmini:1980re}. It is a well known fact that the breaking of a global continuous symmetry, $U(1)_{B-L}$ in this case, implies the existence of a massless particle \cite{Goldstone:1961eq}. This additional state in the spectrum introduces many changes in the phenomenology and might lead to a completely different experimental picture.

In fact, the nature of the majoron, which is determined by the way R-parity is broken, is extremely relevant for the subsequent phenomenology. Depending on this point, one can classify the s-\rpv models into two types:

\begin{itemize}

\item Models that break R-parity with a gauge non-singlet

\end{itemize}

This was the original type of s-\rpv models. In \cite{Aulakh:1982yn} the breaking of R-parity in the MSSM by a left-handed sneutrino VEV $v_L$ was studied. The resulting phenomenology was discussed in several works \cite{Ellis:1984gi,Ross:1984yg} but a after a few years it was realized that having a doublet majoron, which inherits the nature of the particle that caused its appearance, leads to conflict with LEP bounds and astrophysical data \cite{Raffelt:1996wa,Amsler:2008zzb}.

A doublet majoron would contribute to highly constrained processes. The first example is the cooling of red giant stars. The rate of energy loss by the dangerous process $\gamma e \to J e$ would be clearly enhanced above the known limits \cite{Georgi:1981pg,Fukugita:1982ep} due to the weak couplings of the majorons to matter. Once produced, the majorons would escape from the star contributing to its cooling. The astrophysical bounds can be translated into a limit on the left-handed sneutrino VEV, what will require a strong fine-tuning of the parameters of the theory. Similarly, strong bounds can be also obtained from majoron-emitting decays of light neutrinos in supernovae \cite{Kachelriess:2000qc}. The second example is the contribution to the $Z$ boson invisible decay width. It turns out that the majoron, which is a pseudoscalar, is accompanied by a light scalar $\rho$ with a mass of the order $v_L \ll m_W$, opening a new decay mode for the $Z$ boson, $Z \to J \rho$. Since both products of the decay are very weakly interacting particles that escape detection, this decay mode would increase the $Z$ boson invisible decay width beyond the limits set by the LEP experiment \cite{GonzalezGarcia:1989zh,Romao:1989yh}.

For these reasons this type of models is nowadays ruled out.

\begin{itemize}

\item Models that break R-parity with a gauge singlet

\end{itemize}

By using a gauge singlet one avoids the main problems present in the previous models. Reference \cite{Masiero:1990uj} showed how to break R-parity in a phenomenologically acceptable way by extending the spectrum to include gauge singlets whose scalar components acquire VEVs at the electroweak scale. This way the majoron has a singlet nature, which reduces its coupling to the $Z$ boson and the rate of red giant star cooling.

This is the type of models that will be discussed in this chapter.

Although the presence of a massless majoron is allowed by the experimental contraints, it dramatically changes the phenomenology both at collider and low-energy experiments \cite{Hirsch:2008ur,Hirsch:2009ee}.

On the one hand, whether R-parity is conserved or not can, in principle, be easily 
decided at colliders in the case of explicit R-parity violation since (a) neutrino 
physics implies that the lightest supersymmetric particle (LSP) will 
decay inside a typical detector of existing and future high energy
experiments \cite{Porod:2000hv,Hirsch:2003fe} and (b) the branching 
fraction for (completely) invisible LSP decays is at most ${\cal O}(10 \%)$ 
and typically smaller\cite{Porod:2000hv,Hirsch:2006di}. {\em Spontaneous} 
violation of R-parity (s-\rpv)
\cite{Aulakh:1982yn,Masiero:1990uj}, on the other hand, implies the 
existence of a Goldstone boson, the majoron. In s-\rpv 
the lightest neutralino can then decay according to $\tilde{\chi}_1^0
\rightarrow J + \nu$, i.e.~completely invisible. It has been shown 
\cite{Hirsch:2006di} that this decay mode can in fact be the dominant 
one, with branching ratios close to 100 \%, despite the smallness of 
neutrino masses, in case the scale of R-parity breaking is relatively 
low. In this limit, the accelerator phenomenology of models with 
spontaneous violation of R-parity can resemble the MSSM with conserved 
R-parity and large statistics might be necessary before it can be 
established that R-parity indeed is broken.

In addition, measurements of the LSP branching ratios can lead to tests of the model 
as the origin of the neutrino mass, in case sufficient statistics 
for the final states with charged leptons can be obtained. This result is completely analogous
to what is found in b-\rpv, see section \ref{sec:brpv-pheno}. However, the additional possibility of a singlino-like LSP, 
studied in \cite{Hirsch:2008ur} for the first time, allows interesting cross-checks with respect to neutrino physics 
different from the usual bino LSP.

On the other hand, the phenomenology at low-energy experiments is also expected to change.
The search for majorons in charged lepton decays with majoron emission has attracted little 
attention. Indeed, the limits on $l_i\to l_j J$ quoted by the Particle 
Data Group \cite{Amsler:2008zzb} are all based on experimental data which 
is now more than 20 years old. Probably this apparent lack of interest 
from the experimental side is due to the fact that both, the triplet 
\cite{Gelmini:1980re} and the doublet majoron \cite{Aulakh:1982yn}, are 
ruled out by LEP data, while the (classical) singlet majoron model 
\cite{Chikashige:1980ui} predicts majoron-neutrino and majoron-charged-lepton 
couplings which are unmeasurably small. Nevertheless, in the model studied in this thesis 
the majoron can play an important role phenomenologically. In 
\cite{Romao:1991tp} $l_i\to l_j J$ was calculated for a tau neutrino 
mass of $m_{\nu_\tau} \simeq $ MeV. Below we show that (a) despite 
the fact that current neutrino mass bounds are of the order of eV 
or less, theoretically $\mu\to e J$ can be (nearly) arbitrarily large 
in s-\rpv, and (b) $\mu\to e J$ is large in the same part of SUSY 
parameter space where the invisible neutralino decay is large, 
making the discovery of R-parity violation at the LHC difficult. 
$Br(\mu\to e J)$ thus gives complementary information to accelerator 
experiments.

Moreover, the MEG experiment \cite{meg} has started taking 
data. MEG is optimised to search for $Br(\mu\to e \gamma)$ with a 
sensitivity of $Br(\mu\to e \gamma)$ $\sim$ (few) $10^{-14}$. 
While the impressive statistics of the experiment should allow, 
in principle, to improve the existing bound on $Br(\mu\to e J)$ 
\cite{Amsler:2008zzb} by a considerable margin, the experimental 
triggers and cuts make it necessary to resort to a search for the 
radiative majoron emission mode, $Br(\mu\to e J \gamma)$, if one 
wants to limit (or measure) the majoron-charged-lepton coupling. 
Therefore the $Br(\mu\to e J \gamma)$ is also to be considered.

\section{The model}

Spontaneous breaking of a global symmetry leads to a Goldstone boson,  
in case of lepton number breaking usually called the majoron. Spontaneous 
breaking of R-parity through left sneutrinos \cite{Aulakh:1982yn}, 
produces a doublet majoron, which is ruled out by LEP and astrophysical 
data \cite{Raffelt:1996wa,Amsler:2008zzb}. To construct a phenomenologically 
consistent version of s-\rpv it is therefore necessary to extend the 
particle content of the MSSM by at least one singlet field, $\widehat\nu^c$, 
which carries lepton number. For reasons to be explained in more 
detail below, the model under consideration \cite{Masiero:1990uj} contains three 
additional singlet superfields, namely, $\widehat\nu^c$, $\widehat S$ and 
$\widehat\Phi$, with lepton number assignments of $L=-1,1,0$ 
respectively. 

The superpotential can be written as \cite{Masiero:1990uj}
\begin{eqnarray} %
{\cal W} &=& Y_u^{ij} \widehat{Q}_i \widehat{u}_j^c \widehat{H}_u + Y_d^{ij} \widehat{Q}_i \widehat{d}_j \widehat{H}_d + Y_e^{ij} \widehat{L}_i \widehat{e}_j^c \widehat{H}_d \nonumber
\\
        & + & Y_\nu^i\widehat L_i\widehat \nu^c\widehat H_u
          - h_0 \widehat H_d \widehat H_u \widehat\Phi
          + h \widehat\Phi \widehat\nu^c\widehat S +
          \frac{\lambda}{3!} \widehat\Phi^3 .
\label{eq:Wsuppot}
\end{eqnarray}
The basic guiding principle in the construction of equation \eqref{eq:Wsuppot}
is that lepton number is conserved at the level of the superpotential.
The first three terms are the usual MSSM Yukawa terms. The terms
coupling the lepton doublets to $\widehat\nu^c$ fix lepton number.
The coupling of the field $\widehat\Phi$ with the Higgs doublets
generates an effective $\mu$-term \`a la Next to Minimal Supersymmetric
Standard Model \cite{Barbieri:1982eh}. The last two terms, involving
only singlet fields, give mass to $\widehat\nu^c$, $\widehat S$ and
$\widehat\Phi$, once $\Phi$ develops a vacuum expectation value.

For simplicity we consider only one generation of $\widehat\nu^c$ and
$\widehat S$. Adding more generations of $\widehat\nu^c$ or $\widehat S$
does not add any qualitatively new features to the model. Note also,
that the superpotential, equation \eqref{eq:Wsuppot}, does not contain any
terms with dimension of mass, thus potentially offering a solution
to the $\mu$-problem of supersymmetry. The symmetries of the model, 
as defined in \cite{Masiero:1990uj}, allow also to add bilinear terms 
with dimension of mass to eq. \eqref{eq:Wsuppot}. However, we omit such 
terms here for the sake of keeping the number of free parameters of the 
model at the minimum. One could justify the absence of such bilinears - 
in the same way as is usually done in the NMSSM - by introducing a discrete 
$Z_3$ symmetry. This is known to lead to cosmological problems due to the generation
of domain walls during the electroweak phase transition when the $Z_3$ symmetry
gets broken \cite{Abel:1995wk,Abel:1996cr,Panagiotakopoulos:1998yw}.
Nevertheless, several solutions to this problem have been proposed,
like late inflation after the formation of the domain walls,
the introduction of a small explicit breaking or embedding the discrete symmetry
into a gauge one. We note, however, that our numerical results on 
the charged lepton decays are not affected by the presence or 
absence of these terms.

Finally, the inclusion of $\widehat S$ allows 
to generate a ``Dirac''-like mass term for $\widehat\nu^c$, once 
$\widehat\Phi$ gets a VEV. The soft supersymmetry breaking
potential along neutral directions is given by \cite{Hirsch:2004rw}

\begin{eqnarray}
V_{soft}^{s\mbox{-}\rpvm}  &=& \Big[ T_h \Phi \tilde{\nu^c} \tilde{S} - T_{h_0} \Phi H_u H_d + T_\nu^i  \tilde{\nu}_i H_u \tilde{\nu^c} \nonumber \\
&& - B_\mu H_u H_d + \frac{1}{3!} T_{\lambda} \Phi^3 + h.c. \Big] + \sum_{\alpha} \tilde{m}_{\alpha}^2 |z_{\alpha}|^2
\end{eqnarray}

where $z_{\alpha}$ denotes any neutral scalar field in the theory. In this expression the notation for the soft trilinear couplings introduced in \cite{Skands:2003cj,Allanach:2008qq} is used.

At low energy, i.e.~after electroweak symmetry breaking, various
fields acquire VEVs. Besides the usual MSSM Higgs boson VEVs $v_d$ and
$v_u$, these are $\langle \Phi \rangle = v_{\Phi}/\sqrt{2}$, $\langle
{\tilde \nu}^c \rangle = v_R/\sqrt{2}$, $\langle {\tilde S} \rangle =
v_S/\sqrt{2}$ and $\langle {\tilde \nu}_i \rangle = v_i/\sqrt{2}$.
Note, that $v_R \ne 0$ generates effective bilinear terms $\epsilon_i
= Y_\nu^i v_R/\sqrt{2}$ and that $v_R$, $v_S$ and $v_i$ violate lepton
number as well as R-parity. Although other solutions to the tadpole
equations exist, we will focus on minima that break the electroweak
symmetry and R-parity simultaneously. In that sense, EWSB and R-parity
violation are related and cannot be viewed independently.

\section{Neutral fermion mass matrix}

\noindent
In the basis 
\begin{equation}
(-i\tilde{B}^0,-i\tilde{W}_3^0,\widetilde{H}_d^0,\widetilde{H}_u^0,\nu_e,\nu_{\mu},\nu_{\tau},
\nu^c,S,\tilde{\Phi})
\label{eq:defbasis}
\end{equation}
the mass matrix of the neutral fermions following from eq. \eqref{eq:Wsuppot} 
can be written as \cite{Hirsch:2004rw,Hirsch:2005wd} 

\begin{equation}
M_N=
\left(\begin{array}{lllll}
{\cal M}_{\chi^0} & m_{\chi^0\nu}& m_{\chi^0\nu^c}& 0& m_{\chi^0\Phi} \\
\\
m^T_{\chi^0\nu} & 0 & m_{D} & 0 & 0 \\
\\
m^T_{\chi^0\nu^c}&m^T_{D} & 0 & M_{\nu^c S} & M_{\nu^c\Phi} \\
\\
0&0 & M_{\nu^c S} &0 &M_{S\Phi} \\
\\
m^T_{\chi^0\Phi} & 0 & M_{\nu^c\Phi} & M_{S\Phi} & M_{\Phi}
\end{array} \right).
\label{srpv-massF0}
\end{equation}
\noindent
Eq. \eqref{srpv-massF0} can be diagonalized in the standard way,
\begin{equation}
\widehat M_N= {\cal N}^* M_N{\cal N}^{-1}.
\label{eq:diag mass}
\end{equation}
We have chosen the basis in eq. \eqref{eq:defbasis}, such that 
${\cal N}$ reduces to the MSSM neutralino rotation matrix in 
the limit where (a) R-parity is conserved and (b) the field $\Phi$ 
is decoupled.
The various sub-blocks in eq. \eqref{srpv-massF0} are defined as 
follows. The matrix ${\cal M}_{\chi^0}$ is the standard MSSM 
neutralino mass matrix:
\begin{equation}
{\cal M}_{\chi^0} =
\left(\begin{array}{llll}
M_1 & 0    & - \frac{1}{2} g' v_d & + \frac{1}{2} g' v_u \\ \vb{12}
0   & M_ 2 & + \frac{1}{2} g v_d &  - \frac{1}{2} g v_u \\ \vb{12}
- \frac{1}{2} g' v_d & + \frac{1}{2} g v_d &    0 & -\mu \\ \vb{12}
+ \frac{1}{2} g' v_u & - \frac{1}{2} g v_u & -\mu & 0
\end{array} \right).
\label{eq:mntrl}
\end{equation}
Here, $\mu =h_0v_{\Phi}/\sqrt{2}$. $m_{\chi^0\nu}$ is the 
R-parity violating neutrino-neutralino mixing part, which also appears 
in explicit bilinear R-parity breaking models:
\begin{equation}
m^T_{\chi^0\nu} =
\left(\begin{array}{llll}
-\frac{1}{2}g'v_{e} &\frac{1}{2}gv_{e} & 0 & \epsilon_e \\[2mm]
-\frac{1}{2}g'v_{\mu}& \frac{1}{2}gv_{\mu}& 0& \epsilon_{\mu}\\[2mm]
-\frac{1}{2}g'v_{\tau} & \frac{1}{2}gv_{\tau} & 0& \epsilon_{\tau} 
\end{array} \right),
\label{eq:mrpv}
\end{equation}
where $v_i$ are the VEVs of the left-sneutrinos.

\noindent
$m_{\chi^0\nu^c}$ is given as
\begin{equation}
m^T_{\chi^0\nu^c} =
\left(
0, \hskip2mm 0, \hskip2mm  0, \hskip2mm 
\frac{1}{\sqrt{2}}\sum Y_{\nu}^{i} v_i 
\right).
\label{eq:mchinuc}
\end{equation}
and $m^T_{\chi^0\Phi}$ is 
\begin{equation}
m^T_{\chi^0\Phi}
= ( 0 , 0, - \frac{1}{\sqrt{2}}h_0 v_u , - \frac{1}{\sqrt{2}}h_0 v_d).
\label{eq:mchiphi}
\end{equation}
The ``Dirac'' mass matrix is defined in the usual way:
\begin{equation}
(m_{D})_{i} = \frac{1}{\sqrt{2}}Y_{\nu}^{i}v_u.
\label{eq:mD}
\end{equation}
And, finally, 
\begin{equation}
M_{\nu^c S} = \frac{1}{\sqrt{2}}h v_{\Phi}, \hskip5mm 
M_{\nu^c\Phi} = \frac{1}{\sqrt{2}} h v_S, \hskip5mm 
M_{S\Phi} = \frac{1}{\sqrt{2}} h v_R, \hskip5mm 
M_{\Phi} = \frac{\lambda}{\sqrt{2}} v_{\Phi}.
\label{eq:mp_other}
\end{equation}
The matrix, eq. \eqref{srpv-massF0}, produces ten eigenvalues with 
vastly different masses. First, since ${\rm Det}(M_N)=0$, one 
state is massless at tree-level. Then there are two more very light 
states, together they form to a good approximation the three observed, 
light doublet neutrinos. Their masses and mixing will be discussed in 
detail in the next subsection. 

The remaining seven eigenstates are typically heavy. They can be 
sub-divided into two groups: Mainly doublet and mainly singlet states. 
There are usually four states which are very similar to the well-known 
MSSM neutralinos. Unless $h_0$ is large {\em and} $\lambda v_{\Phi}$ 
small, mixing between the phino and the higgsinos is small  
\footnote{As in the NMSSM, if the field $\Phi$ is light and the coupling 
$h_0$ large, one has five neutralino states.}, and there are three 
singlets. From these singlets, unless $(h-\lambda)\le v_R/v_{\Phi}$, 
$\nu^c$ and $S$ form a quasi-Dirac pair, which we will loosely call 
``the singlino'', 
${\cal S}_{1,2} \simeq \frac{1}{\sqrt{2}} (\nu^c \mp S)$. Note, that
this is a different state compared to the NMSSM singlino \cite{Franke:2001nx}
which corresponds to $\tilde \Phi$ in our notation.

Which of the seven, heavy states is the lightest depends on a number 
of unknown parameters and can not be predicted. In our analysis below 
we will concentrate on two cases: (a) As in mSugra motivated 
scenarios, $M_1$ is the smallest mass parameter and the lightest state 
mainly a bino. We study this case in order to work out the 
differences to (i) the well-studied phenomenology of the MSSM; 
and (ii) to the explicit R-parity violating case studied in 
\cite{Porod:2000hv}. The second case we consider is (b) the 
singlino $\cal S$ being the lightest state. This case is 
interesting, since it is the only part of the parameter space, 
where singlets indeed can be produced and studied at accelerators.

\section{Neutrino masses}
\label{srpv:numass}

The smallness of the \rpv terms imply that equation \eqref{srpv-massF0} has the same structure as the analogous equation \eqref{brpv-massF0} for the bilinear R-parity breaking model. Therefore, one can follow the same approach and find the effective neutrino mass matrix in a 
seesaw--type approximation \cite{Hirsch:2004rw,Hirsch:2005wd}. 
First we define the useful dimensionless expansion parameters $\xi_{ij}$, which 
characterize the mixing between the neutrino sector and the 
seven heavy neutral fermion states, the ``neutralinos'' of the 
model, 
\begin{equation}\label{eq:defxi}
\xi=m_{3\times 7}\cdot M_H^{-1}.
\end{equation}
The sub-matrix describing the seven heavy states of eq. \eqref{srpv-massF0} 
is
\begin{equation}
M_H=
\left(\begin{array}{llll}
{\cal M}_{\chi^0} & 0 & 0& m_{\chi^0\Phi} \\
\\
0& 0 & M_{\nu^c S} & M_{\nu^c\Phi} \\
\\
0 &M_{\nu^c S} &0 &M_{S\Phi} \\
\\
m^T_{\chi^0\Phi} &M^T_{\nu^c\Phi} & M_{S\Phi} & M_{\Phi}
\end{array} \right).
\label{eq:mass77}
\end{equation}
\noindent
and
\begin{equation}
m_{3\times 7}=
\left(\begin{array}{llll}
m^T_{\chi^0\nu} & m_D & 
0 & 0
\end{array} \right).
\label{eq:mass37}
\end{equation}
\noindent
We have neglected $m_{\chi^0\nu^c}$ in eq. \eqref {eq:mass77} 
since it is doubly suppressed. The ``effective'' ($3,3$) neutrino mass 
matrix is then given in seesaw approximation by
\begin{equation}
m_{eff}^{\nu\nu} =
- m_{3\times 7}\cdot M_H^{-1}\cdot m_{3\times 7}^T.
\end{equation}
In the following we will use the symbol $N_{ij}$ with $i,j=1,...7$ as
the matrix which diagonalizes eq. \eqref{eq:mass77}. Our $N$ reduces
to the MSSM neutralino mixing matrix $N$, in the limit where the
singlets decouple, i.e. $h_0\to 0$ or $M_{\Phi}\to \infty$. It is not
to be confused with ${\cal N}$, the matrix that diagonalizes $M_N$,
the complete $10 \times 10$ mass matrix of the neutral fermions. After
some straightforward algebra $\xi_{ij}$ can be written as
\begin{equation}\label{def:xi}
\xi_{ij}=K_\Lambda^j \Lambda_i + K_\epsilon^j \epsilon_i,
\end{equation}
where the effective bilinear R--parity violating parameters 
$\epsilon_{i}$ and $\Lambda_i$ are 
\begin{equation}
  \label{eq:eps}
\epsilon_{i} = Y_{\nu}^{i}\, \frac{v_R}{\sqrt{2}}  
\end{equation}
and
\begin{equation}
\Lambda_i = \epsilon_i v_d + \mu v_i. 
\label{eq:deflam0}
\end{equation}
The coefficients $K$ are given as 
\begin{eqnarray}\label{defK}
K_\Lambda^1 = -\frac{2 g' M_2 \mu}{m_\gamma}a, 
&\hskip5mm & K_\epsilon^1 = -\frac{2 g' M_2 \mu}{m_\gamma}b \\ \nonumber
K_\Lambda^2 = \frac{2 g M_1 \mu}{m_\gamma}a, 
&\hskip5mm & K_\epsilon^2 = \frac{2 g M_1 \mu}{m_\gamma}b \\ \nonumber
K_\Lambda^3 = -v_u a + \frac{v_d b}{2 v_u},
&\hskip5mm & K_\epsilon^3 = -\frac{c}{2 \mu v_u^2} 
\big(\frac{4 {\rm Det}(M_H) a}{h^2 m_\gamma}-v_d v_u \mu \big)-\frac{v^2 b}{2 v_u} 
\\ \nonumber
K_\Lambda^4 = v_d a + \frac{b}{2},
&\hskip5mm & K_\epsilon^4 = \frac{h^2 \mu v_u}{4 {\rm Det}(M_H)}
(4 M_1 M_2 \mu v_u - m_\gamma v_d v^2) \\ \nonumber
K_\Lambda^5 = \frac{v_R b}{2 v_u}, 
&\hskip5mm & K_\epsilon^5 = \frac{v_R c}{2 v_u} \\ \nonumber
K_\Lambda^6 = \frac{v_S b}{2 v_u}, 
&\hskip5mm & K_\epsilon^6 = \frac{c}{2 \sqrt{2} v_u v_R v_\Phi h} 
\big[ \frac{8 {\rm Det}(M_H) a}{h^2 m_\gamma}+\sqrt{2} h v_\Phi v_R v_S \big] \\ \nonumber
& & \hskip9mm -\frac{2 \sqrt{2} {\rm Det}(M_H) b^2}{h^3 m_\gamma v_u v_R v_\Phi} \\ \nonumber
K_\Lambda^7 = -\frac{v_\Phi b}{2 v_u},
&\hskip5mm & K_\epsilon^7 = -\frac{v_\Phi c}{2 v_u}
\end{eqnarray}
The coefficients $a$, $b$ and $c$ are defined as
\begin{eqnarray}\label{def:abc}
a=\frac{m_\gamma h^2 v_\Phi}{4 \sqrt{2} {\rm Det}(M_H)} 
(-h v_R v_S+\frac{1}{2} \lambda v_\Phi^2+h_0 v_d v_u), \\ \nonumber
b=\frac{m_\gamma h^2 \mu}{4 {\rm Det}(M_H)} v_u (v_u^2-v_d^2), \\ \nonumber 
c=\frac{h^2 \mu}{{\rm Det}(M_H)} v_u^2 (2 M_1 M_2 \mu - m_\gamma v_d v_u).
\end{eqnarray}
${\rm Det}(M_H)$ is the determinant of the ($7,7$) matrix of the heavy 
neutral states,
\begin{eqnarray}
{\rm Det}(M_H)=\frac{1}{16} h_0 h^2 v_\Phi^2 & \big[ & 4(2 M_1 M_2 \mu 
- m_\gamma v_d v_u)(-h v_R v_S+\frac{1}{2} 
\lambda v_\Phi^2+h_0 v_d v_u) \nonumber \\
& & - h_0 m_\gamma (v_u^2-v_d^2)^2 \hskip3mm \big] \label{def:detmh}
\end{eqnarray}
and $v^2=v_u^2+v_d^2$. The ``photino'' mass parameter is defined 
as $m_{\gamma} = g^2M_1 +g'^2 M_2$.
Note that the $K_{\Lambda}^i$ and  $K_{\epsilon}^i$ reduce to the 
expressions of the explicit bilinear R-parity breaking model 
\cite{Hirsch:2000ef}, in the limit $M_{\Phi}\rightarrow \infty$ and 
in the limit $h,h_0\rightarrow 0$, i.e.~ $b=c=0$. 

The effective neutrino mass matrix at tree-level can then be cast into 
a very simple form 
\begin{equation}
 -(m_{eff}^{\nu \nu})_{ij} = a \Lambda_i \Lambda_j + 
     b (\epsilon_i \Lambda_j + \epsilon_j \Lambda_i) +
     c \epsilon_i \epsilon_j\,.
\label{srpv:effnu}
\end{equation}
Equation \eqref{srpv:effnu} resembles very closely the corresponding expression 
for the explicit bilinear R-parity breaking model, once the tree-level 
and the dominant 1-loop contributions are taken into account 
\cite{Romao:1999up,Hirsch:2000ef,Diaz:2003as}. Eq. \eqref{srpv:effnu} 
reduces to the tree-level expression of the explicit model 
\footnote{In the definition of the coefficient $a$ given in 
\cite{Hirsch:2004rw} there is a relative sign to the corresponding 
definition for the explicit case \cite{Hirsch:2000ef}.}
\begin{equation}
(m_{eff}^{\nu\nu})_{ij} = \frac{m_{\gamma}}
{4 {\rm Det} ({\cal M}_{\chi^0})} \Lambda_i \Lambda_j 
\label{eq:effexpl}
\end{equation}
in the limit $M_{\Phi}\rightarrow \infty$ and in the limit 
$h,h_0\rightarrow 0$. Different from the explicit model, however, 
the spontaneous model has in general two non-zero neutrino masses at 
tree-level. With the lightest neutrino mass zero at tree-level, the 
s-\rpv model could generate degenerate neutrinos only in regions of 
parameter space where the two tree-level neutrino masses of eq. 
\eqref{srpv:effnu} are highly fine-tuned against the loop corrections. 
We will disregard this possibility in the following. 

Neutrino physics constrains the parameters $\Lambda_i$ and
$\epsilon_i$. The mass matrix in equation \eqref{srpv:effnu} must
reproduce the current data on neutrino masses and mixing angles. In
particular, one must correctly fit the solar and atmospheric mass
scales. However, in the spontaneous model there is no a priori reason
which of the terms gives the dominant contribution to the neutrino
mass matrix, thus two possibilities to fit the neutrino data exist:
\begin{itemize}
\item case (c1) $\vec \Lambda$ generates the atmospheric mass scale, 
          $\vec \epsilon$ the solar mass scale
\item case (c2) $\vec \epsilon$ generates the atmospheric mass scale, 
          $\vec \Lambda$ the solar mass scale
\end{itemize}
The absolute scale of neutrino mass requires both $|\vec\Lambda|/\mu$ 
and $|\vec\epsilon|/\mu$ to be small, the exact numbers depending 
on many unknown parameters. For typical SUSY masses order 
${\cal O}(100\hskip1mm{\rm \text{GeV}})$, $|\vec\Lambda|/\mu^2 \sim 10^{-6}$--
$10^{-5}$. If some of the singlet fields are light, i.e.~have masses 
in the range of ${\cal O}(0.1-{\rm few})$ TeV, also $|\epsilon_i/\mu|$ 
can be as small as $|\vec\epsilon|/\mu \sim 10^{-6}$--$10^{-5}$. On 
the other extreme, independent of the singlet spectrum, $|\vec\epsilon|/\mu$ 
can not be larger than, say, $|\vec\epsilon|/\mu \sim 10^{-3}$, due to 
contributions from sbottom and stau loops to the neutrino mass 
matrix \cite{Romao:1999up,Hirsch:2000ef,Diaz:2003as}. 

The observed mixing angles in the neutrino sector then require certain
ratios for the parameters $\Lambda_i/\Lambda_j$ and $\epsilon_i/\epsilon_j$. 
This can be most easily understood as follows. As first observed in 
\cite{Harrison:2002er}, the so-called tri-bimaximal mixing pattern
\begin{equation}
\label{eq:UHPS}
U^{\rm TBM} =
\left(\begin{array}{cccc}
\sqrt{\frac{2}{3}} & \frac{1}{\sqrt{3}} & 0 \cr
- \frac{1}{\sqrt{6}} &  \frac{1}{\sqrt{3}} & - \frac{1}{\sqrt{2}} \cr
- \frac{1}{\sqrt{6}} &  \frac{1}{\sqrt{3}} & \frac{1}{\sqrt{2}}
\end{array}\right).
\end{equation}
is a good first-order approximation to the observed neutrino angles. In
case of hierarchical neutrinos ${\cal M}_\nu^{diag}=(0,m,M)$, where 
$m$ ($M$) stands for the solar (atmospheric) mass scale, rotating with 
$U^{\rm TBM}$ to the flavor basis leads to the following neutrino mass 
matrix
\begin{equation}
\label{eq:MHPS}
{\cal M}_\nu^{\rm TBM} =
\frac{1}{2}\left(\begin{array}{cccc}
0 &  0 &  0 \cr
0 &  M & -M \cr
0 & -M &  M
\end{array}\right) +
\frac{1}{3}\left(\begin{array}{cccc}
m &  m &  m \cr
m &  m &  m \cr
m &  m &  m
\end{array}\right).
\end{equation}
In case the coefficient $b$ in equation \eqref{def:abc} is exactly zero,
i.e. for $\tan\beta=1$, the model would produce a tri-bimaximal 
mixing pattern for $\Lambda_e=0,\Lambda_\mu=-\Lambda_\tau$ and 
$\epsilon_e=\epsilon_\mu=\epsilon_\tau$, in case (i). For case (ii) the 
conditions on $\Lambda_i$ should be exchanged with the conditions for 
the $\epsilon_i$ and vice versa. 

In reality, since $\tan\beta\ne 1$ in general, neither is $b$ exactly 
zero, nor need the neutrino mixing angles be exactly those of eq. 
\eqref{eq:UHPS}. One then finds certain allowed ranges for ratios of 
the $\Lambda_i$ and $\epsilon_i$. In case (i) one gets approximately 
\begin{eqnarray}\label{fitnucase1}
\Big(\frac{\Lambda_e}{\sqrt{\Lambda_\mu^2+\Lambda_\tau^2}}\Big)^2 
\simeq \tan^2\theta_{13},
\\ \nonumber
\Big(\frac{\Lambda_\mu}{\Lambda_\tau}\Big)^2 \simeq \tan^2\theta_{23},
\\ \nonumber
\Big(\frac{{\tilde\epsilon_e}}{{\tilde\epsilon_\mu}}\Big)^2 
\simeq \tan^2\theta_{12}.
\end{eqnarray}
Here, ${\tilde\epsilon} = U_{\nu}^T\cdot{\vec\epsilon}$ with 
$(U_{\nu})^T$ being the matrix which diagonalizes the ($3,3$) 
effective neutrino mass matrix. In case (i) $(U_{\nu})^T$ is 
(very) approximately given by
\begin{equation}\label{def:tildeeps}
{\tilde\epsilon} = 
\left(\begin{array}{cccc}
\frac{\sqrt{\Lambda_\mu^2 + \Lambda_\tau^2}}{|\vec\Lambda|} &  
-\frac{\Lambda_e\Lambda_\mu}{\sqrt{\Lambda_\mu^2 + \Lambda_\tau^2}|\vec\Lambda|} &  
-\frac{\Lambda_e\Lambda_\tau}{\sqrt{\Lambda_\mu^2 + \Lambda_\tau^2}|\vec\Lambda|} \cr 
 0 &  
\frac{\Lambda_\tau}{\sqrt{\Lambda_\mu^2 + \Lambda_\tau^2}} 
& -\frac{\Lambda_\mu}{\sqrt{\Lambda_\mu^2 + \Lambda_\tau^2}}  \cr 
 \frac{\Lambda_e}{|\vec\Lambda|} & 
 \frac{\Lambda_\mu}{|\vec\Lambda|} & 
 \frac{\Lambda_\tau}{|\vec\Lambda|} \cr
\end{array}\right)\cdot
{\vec \epsilon}
\end{equation}
Note that $ U_{\Lambda}^T$ is the matrix which diagonalizes only 
the part of the effective neutrino mass matrix proportional to 
$\Lambda_i\Lambda_j$. Again, for the case (ii) replace 
$\Lambda_i\leftrightarrow \epsilon_i$ in all expressions.

\section{Scalar sector}

The scalar sector of s-\rpv has many unconventional properties, leading to a collider phenomenology different from what is expected in the MSSM. It is very important to study in detail the possible deviations from the standard phenomenology, in order to be ready for new signatures. In fact, the well established Higgs boson search strategies for the LHC might need to be changed if s-\rpv is realized in nature and we live in a region of parameter space far from the MSSM. For detailed works on the scalar sector in s-\rpv see references \cite{Hirsch:2004rw,Hirsch:2005wd}, where most of the material in this section is taken from.

From a phenomenological point of view the most important difference 
between the scalar sectors of spontaneous and explicit R-parity violating models is the 
appearance of the majoron. As will be shown below, this models naturally leads to a singlet majoron, evading the strong experimental constraints.

We can follow the general procedure described in reference \cite{romao:1992vu} and evaluate the second
derivatives of the scalar potential \cite{Hirsch:2004rw} at the
minimum. In s-\rpv this results in $8 \times 8$ mass matrices for the real
and imaginary parts of the neutral scalars. The pseudo-scalar sector of the model we 
consider has eight different eigenstates. Two of them are Goldstone 
bosons. The standard one is eaten by the $Z^0$ boson, the remaining 
state is identified with the majoron.
In the basis
$A'_0=(H_d^{0 I},H_u^{0 I},\tilde{\nu}^{1 I},\tilde{\nu}^{2 I},
\tilde{\nu}^{3 I},\Phi^I,\tilde{S}^I,\tilde{\nu}^{c I})$ 
these fields are given as,
\begin{eqnarray}
  \label{eq:3a}
  G_0&=&(N_0\, v_d,-N_0\, v_u,N_0\,v_{e},N_0\, v_{\mu},N_0\, v_{\tau},0,0,0)
   \\[+2mm]
  J&=&N_4 (-N_1 v_d,N_1 v_u, N_2 v_{e}, N_2 v_{\mu}, N_2 v_{\tau},0,
  N_3 v_S,-N_3 v_R) \nonumber
\end{eqnarray}
where the normalization constants $N_i$ are given as
\begin{eqnarray}
  \label{eq:4a}
  N_0&=&\frac{1}{\sqrt{v_d^2+v_u^2+v_{e}^2 + v_{\mu}^2 + v_{\tau}^2}}
\nonumber \\[+2mm]
  N_1&=&v_{e}^2 + v_{\mu}^2 + v_{\tau}^2\nonumber \\[+2mm]
  N_2&=&v_d^2 + v_u^2\nonumber \\[+2mm]
  N_3&=&N_1 + N_2 \nonumber \\[+2mm]
  N_4&=&\frac{1}{\sqrt{N_1^2 N_2 + N_2^2 N_1 +N_3^2(v_R^2+v_S^2)}}
\end{eqnarray}
and can easily be checked to be orthogonal, i.~e. they satisfy $G_0
\cdot J=0$ \cite{Hirsch:2004rw}. 

It is useful to have an approximation for the majoron profile. In the limit $v_{i} \ll v_R,v_S$ 
this is given by the simple expression 
\begin{equation}\label{smplstmaj}
R_{Jm}^{P^0} \simeq 
\big(0,0,\frac{v_{k}}{V},0,\frac{v_S}{V},-\frac{v_R}{V}\big). 
\end{equation}
Here, $V=\sqrt{v_R^2+v_S^2}$ and terms of order $\frac{v_L^2}{V v}$, 
where $v_L^2 = \sum_i v_{i}^2$, have been neglected.

From equation \eqref{smplstmaj} one can see that, apart from the negligible components in the left-handed sneutrino directions, the majoron is mainly given as a combination of the CP-odd components of the $\tilde{S}$ and $\tilde{\nu}^{c}$ fields. In conclusion, the s-\rpv majoron is a mixture of gauge singlet states, as needed to evade LEP and astrophysical bounds\footnote{For comparison purposes, let us mention that the majoron profile in the model of reference \cite{Aulakh:1982yn} is given by $J = Im \left( \sum_i \frac{v_i}{v_L} {\tilde{\nu}_L^i} + \frac{v_L}{v^2}(v_d {H_d^0} - v_u {H_u^0}) \right)$, this is, a doublet majoron.}.

In addition to the massless majoron, there are considerable parts of the parameter 
space where one also finds a rather light singlet scalar, called the 
``scalar partner'' of the majoron in \cite{Hirsch:2005wd}, $S_J$. Different from the majoron, however, there is no simple analytical approximation for $R_{S_J}$. This state will be very important when we discuss invisible LSP decays, since it decays to a pair of majorons, $S_J \to J J$, with a branching ratio very close to $100 \%$.

Finally, although it is not our purpose to review the phenomenology of the scalar sector in detail, that was worked out in references \cite{Hirsch:2004rw,Hirsch:2005wd}, we must comment on an important result concerning Higgs boson decays into majorons. This is one of the best examples that clearly show how the phenomenology in the scalar sector might be totally modified.

Let us consider the ratio
\begin{equation}
  \label{eq:ratio}
  R_{Jb}=\frac{\Gamma(h \to JJ)}{\Gamma(h \to b \bar{b})}
\end{equation}
of the Higgs invisible decay to the Standard Model decay into b-jets. This quantity measures the departure from the SM branching ratio. The smaller $R_{Jb}$ is the closer we are to the standard phenomenology.

In order to compute this quantity one has to look separately at the decay widths,
\begin{equation}
  \label{eq:JJ}
  \Gamma(h\to  JJ)=\frac{g_{hJJ}^2}{32\pi m_h}
\end{equation}
and
\begin{equation}
  \label{eq:bb}
  \Gamma(h\to  b \bar{b})=\frac{3 G_F \sqrt{2}}{8\pi\cos^2\beta}\,
  \left(R^S_{11}\right)^2\, m_h\,  m_b^2 \left[ 1-4
  \left(\frac{m_b}{m_h}\right)^2 \right]^{3/2}
\end{equation}

where $g_{hJJ}^2$ is the $h^0-J-J$ coupling and $R^S_{11}$ is the $h^0$ component in the $H_d^0$ direction.

\begin{figure}
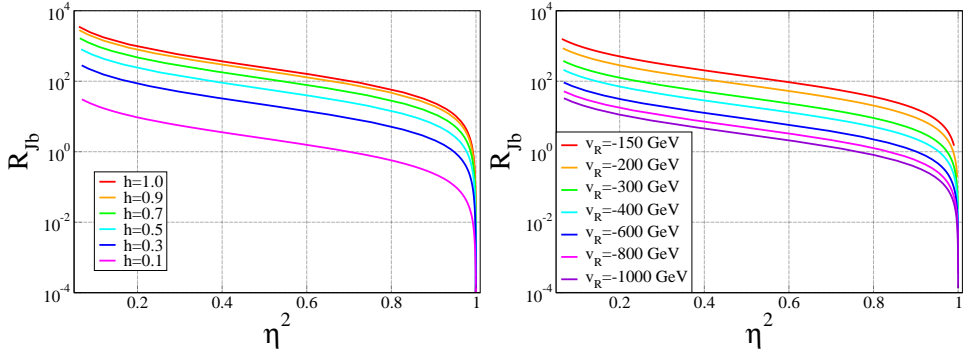

\begin{center}
\vspace{5mm}
\includegraphics[width=0.49\textwidth]{img/higgs1.eps}
\includegraphics[width=0.49\textwidth]{img/higgs2.eps}
\end{center}
\vspace{-5mm}
\caption{Ratio $R_{Jb}$, defined in equation \eqref{eq:ratio}, as a function 
  of $\eta^2$.  a) To the left, for different values of the
  parameter $h$, from top to bottom: $h=1,\, 0.9,\, 0.7,\, 0.5,\, 0.3,\, 0.1$. 
  b) To the right, for different values of the
  parameter $v_R=v_S$: $-v_R= 150,\, 200,\, 300,\, 400,\, 600,\, 800,\, 1000$ GeV. 
  This is the main result of \cite{Hirsch:2004rw}.}
\label{fig:plot54}
\end{figure}

In general, these two decay widths depend on many unknown parameters and one cannot give a definite prediction for their values. In addition, if one wants to enhance the branching ratio for $h \to J J$ by enlarging its singlet component, the Higgs boson production cross section becomes very suppressed. However, it is possible to find large regions in parameters space where (a) there is a large $h^0$ doublet component, called here $\eta$, and (b) the decay into majorons is dominant. This important result is shown in figure \ref{fig:plot54}, taken from reference \cite{Hirsch:2004rw}, which shows explicitly that $R_{Jb}>1$ is possible even for $\eta \simeq 1$. This implies that the 
lightest Higgs boson can decay mainly invisibly into a pair of majorons and, at the same time, have a production cross section essentially equal to the standard (MSSM) doublet Higgs boson production cross section.

\section{Phenomenology at colliders}

In this section we discuss the phenomenology of a neutralino LSP in 
s-\rpv at future colliders \cite{Hirsch:2008ur} \footnote{In order to make the notation simpler, this section will use the simplifications $\chi_i^0 \equiv \tilde{\chi}_i^0$, $\tilde{B} \equiv \tilde{B}^0$ and $\tilde{W} \equiv \tilde{W}^0$.}. We do not attempt to do an exhaustive 
study of the (quite large) parameter space of the model. Instead 
we will focus on the most important qualitative differences between 
s-\rpv, the previously studied case of explicit bilinear \rpv 
\cite{Mukhopadhyaya:1998xj,Choi:1999tq,Porod:2000hv,deCampos:2007bn,DeCampos:2010yu} and the MSSM. All numerical 
results shown below have been obtained using the program package 
SPheno \cite{Porod:2003um}, extended to include the new singlet 
superfields $\widehat\nu^c$, $\widehat S$ and $\widehat\Phi$. 

Unless mentioned otherwise, we have always chosen the \rpv parameters in 
such a way that solar and atmospheric neutrino data \cite{Schwetz:2008er} 
are fitted in the correct way. The numerical procedure to fit neutrino 
masses is the following. Compared to the MSSM we have a number of 
new parameters. For the superpotential of eq. \eqref{eq:Wsuppot} these 
are $h_0$, $h$ and $\lambda$, as well as the neutrino Yukawas $Y_{\nu}^i$. 
In addition, there are in principle also the soft SUSY breaking terms, 
which generate non-zero VEVs, $v_R$, $v_S$, $v_{\Phi}$ and $v_{i}$ 
for ${\tilde\nu}^c$, ${\tilde S}$, $\Phi$ and ${\tilde\nu}_i$, respectively. 
We trade the unknown soft parameters for the VEVs. For any random 
choice of MSSM parameters, we can reproduce the ``correct'' MSSM value 
of $\mu$ for a random value of $v_{\Phi}$, by appropriate choice of 
$h_0$. For any random set of $h$, $\lambda$, $v_S$ and $v_R$, we can 
then calculate those values of $Y_{\nu}^i$ and $v_{i}$, using 
eq. \eqref{srpv:effnu}, such that the corresponding $\epsilon_i$ and 
$\Lambda_i$ give correct neutrino masses and mixing angles. There 
are two options, how neutrino data can be fitted, i.e. the cases (c1) 
and (c2), defined in section \ref{srpv:numass}. We discuss 
the differences between these two possibilities below. 

In the following we will study only two 'limiting' cases, which we 
consider to be the simplest possibilities to realize within the parameter 
space of the model: (a) a bino-like LSP and (b) a singlino LSP. We note, 
however, that theoretically also other possibilities exist at least in 
some limited parts of parameter space. For example, one could also have 
that the phino, ${\tilde\Phi}$, is the lightest $R_p$ odd state. However, 
with the superpotential of eq. \eqref{eq:Wsuppot}, for any given value 
of $\mu$, $v_{\Phi}$ has a minimum value. Since the product 
$\lambda v_{\Phi}$ also determines approximately the phino mass, a very 
light phino requires a certain hierarchy $\lambda \ll h_0,h$, which might 
be considered to be a rather special case. Also in mSugra in the region 
where $m_0$ is large one can find points in which $\mu \sim M_1$ and the 
lightest (MSSM) neutralino has a significant higgsino component.

\subsection{LSP production}

Since neutrino physics requires that the R-parity violating parameters
are small, supersymmetric production cross sections are very similar
to the corresponding MSSM values, see for example
\cite{AguilarSaavedra:2005pw} and references therein. Over most of the
MSSM parameter space one expects that mainly gluinos and squarks are
directly produced the LHC and that the lightest neutralinos appear as
the ``final'' decay products at the end of possibly long decay chains
of sparticles. In addition charginos, neutralinos and sleptons
can be produced directly via Drell-Yan processes provided that they are
relatively light.

Cross sections for direct production of singlinos are 
always negligible. There are essentially two possibilities how singlinos
can be produced in cascade decays.
Firstly, a somewhat exotic chance to produce singlinos occurs if 
at least one of the MSSM Higgsinos is heavier than $\tilde\Phi$ {\em and 
both} $h_0$ and $h$ are large. In this case ${\cal S}_i$ appear in  
decay chains such as $\tilde H_{u,d}\to \tilde\Phi + \,X_1 \to {\cal S}+\,X_2$,
where $X_i$ denotes the additionally produced particles.
Secondly, there is the possiblity that singlinos are the LSPs. 
Squarks and gluinos will then decay fast to the NLSP, which then decays 
to ${\cal S}$. A typical decay chain might be ${\tilde q} \to q 
+ {\tilde B} \to q + {\cal S}_{1,2} + J$. Other NLSPs such as, 
${\tilde\tau}_1$ will decay mainly via ${\tilde\tau}_1 \to {\cal S}_{1,2} 
+ \tau$, i.e. again ending up in singlinos. The total number of singlino 
events therefore will be simply approximately equal to the number of 
SUSY events for singlino LSPs.

\subsection{LSP decays}
\label{sec:LSPdec}

\begin{figure}
\begin{center}
\vspace{5mm}
\includegraphics[width=0.49\textwidth]{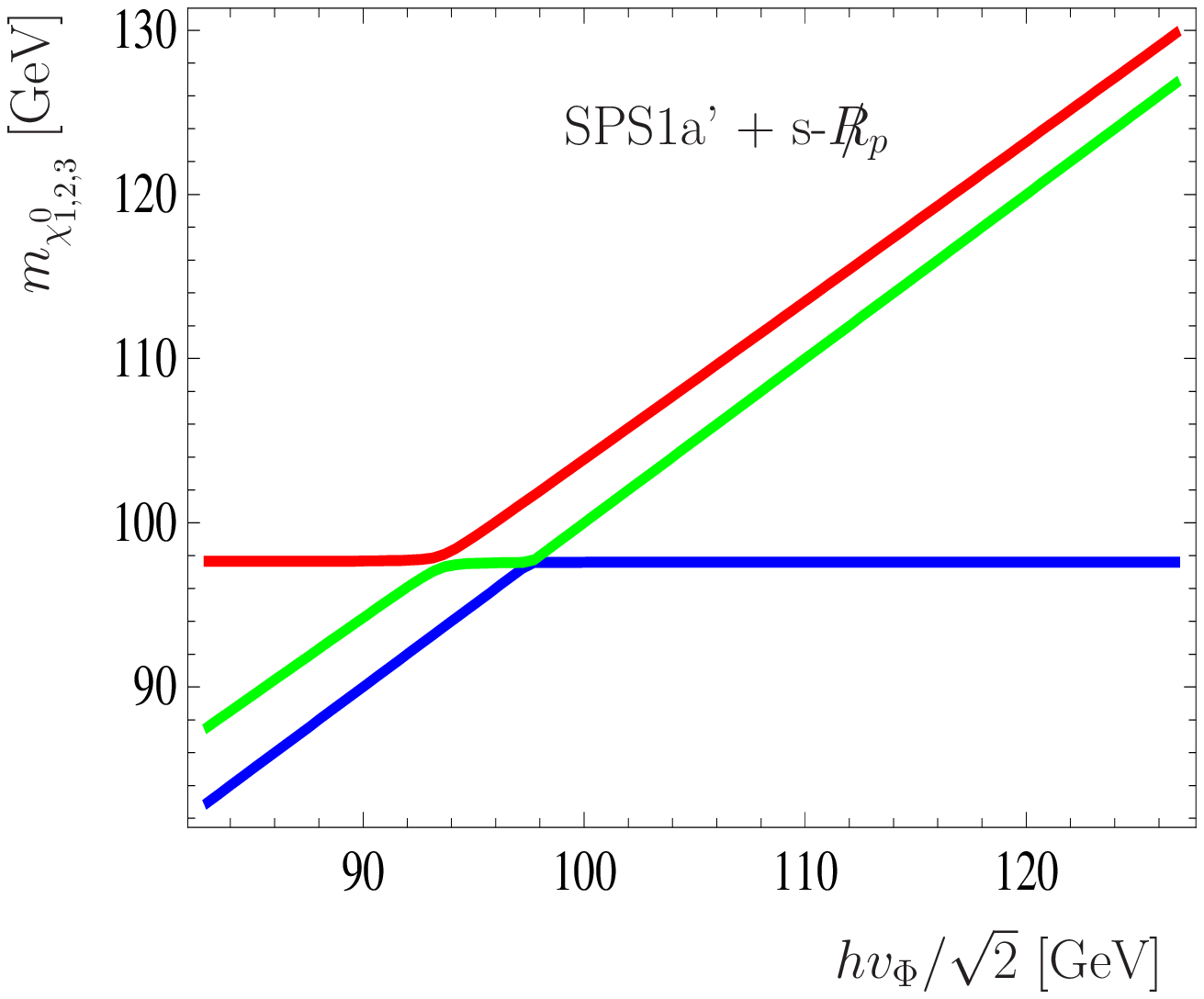}
\includegraphics[width=0.49\textwidth]{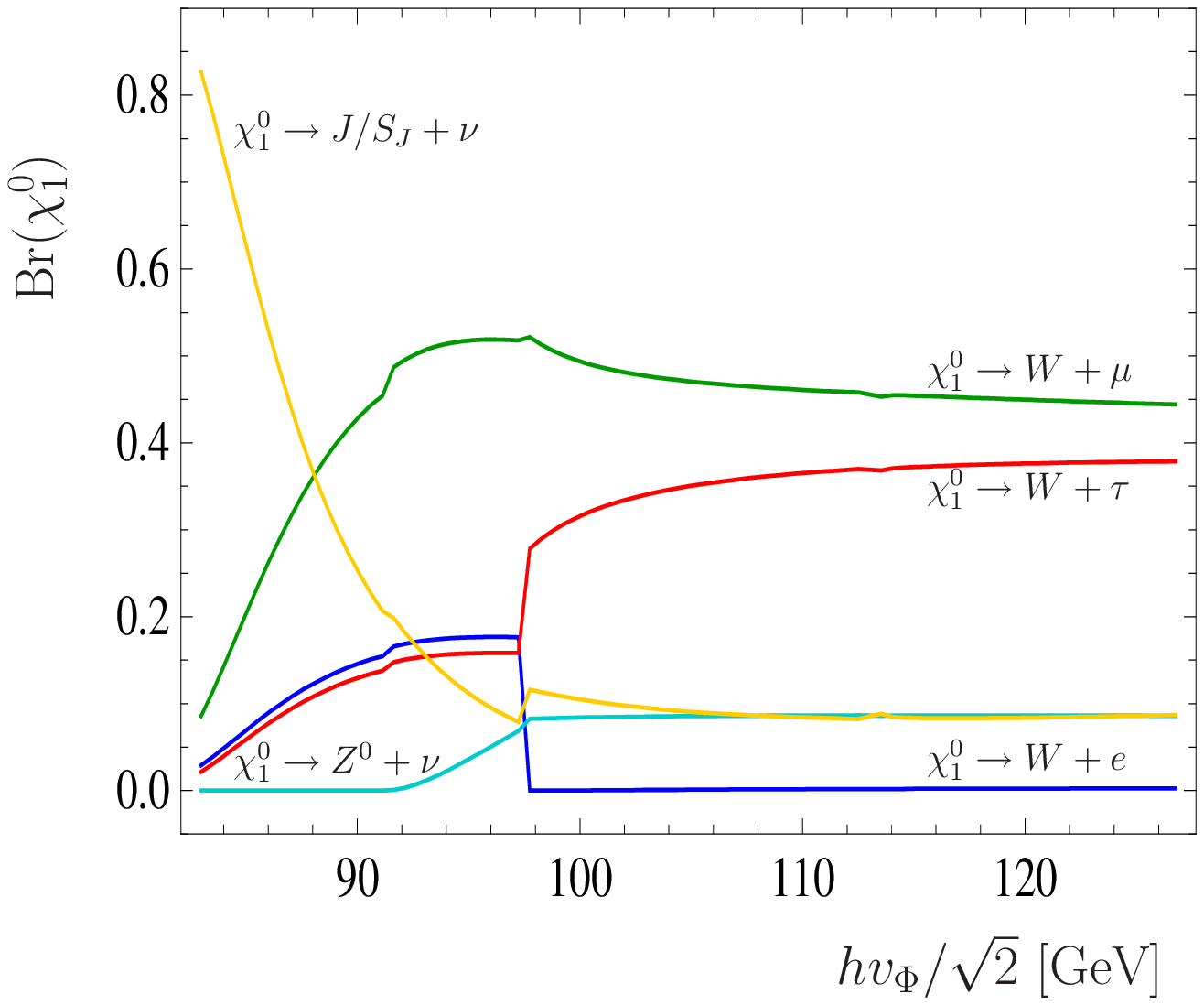}
\end{center}
\caption{Masses of the three lightest neutralinos (left) and branching 
ratios for the most important decay modes of the lightest state (right) 
versus $\frac{1}{\sqrt{2}}hv_{\Phi}$ for a specific, but typical example 
point. The MSSM parameters have been adjusted such that the sparticle 
spectrum of the standard point SPS1a' is approximately reproduced. The 
singlet parameters have been chosen randomly, $v_R=v_S=1$ TeV, 
${\vec\epsilon}$ and $\vec\Lambda$ have been fitted to neutrino data, 
such that $\vec\Lambda$ generates the atmospheric scale and ${\vec\epsilon}$ 
the solar scale. For a detailed discussion see text.}
\label{fig:Exa_mass_br}
\end{figure}

Here we will discuss the main decay modes of bino and singlino LSPs. For a qualitative understanding of the results, which are based on exact numerical computations, the approximate formulas for the neutralino couplings in appendix \ref{sect:app1} are helpful.

We will first discuss the parameter range, where $m_{\chi^0_1} 
\ge m_{W^{\pm}}$, such that two-body decays of $\chi^0_1$ to gauge bosons 
are kinematically allowed.  Figure \ref{fig:Exa_mass_br} shows an
example of the three lightest neutralino mass eigenvalues (left) and
the main decay modes of $\chi^0_1$ (right) as a function of
$\frac{hv_{\Phi}}{\sqrt{2}}$ for fixed values of all other
parameters. This point has been constructed in such a way, that the
MSSM part of the spectrum, all production cross sections and all decay
branching ratios, apart from the lightest neutralino decays, match
very closely the mSugra standard point
SPS1a' \cite{AguilarSaavedra:2005pw}.  Here, $v_R=v_S=1$ TeV has been
chosen as an arbitrary example, but the result can be also obtained
for other values of these two VEVs.

The left part of figure \ref{fig:Exa_mass_br} shows how the quasi-Dirac 
pair ${\cal S}_{1,2}$ evolves as a function of $\frac{hv_{\Phi}}{\sqrt{2}}$. 
For low values (i.e. $\lesssim M_1$) of this parameter combination ${\cal S}_{1}$ 
is the LSP, for large 
values a ${\tilde B}$ is the LSP. The right side of the figure shows 
the final states with the largest branching ratios. For low values of 
the LSP mass, $J/S_J+\nu$ is usually the most important, i.e. there is 
a sizeable decay to invisible final states, even for a relatively high 
$v_R$, see also the discussion for figure \ref{fig:Inv_vr}. Next in 
importance are the final states involving $W^{\pm}$ and charged leptons. 
Note, that the model predicts 
\begin{equation}\label{ZtoW}
\frac{\sum_i Br(\chi_1^0\to Z^0 + \nu_i)}
     {2 \sum_i Br(\chi_1^0\to W^{+} + l_i^{-})} 
     \simeq \frac{g}{4 \cos^2\theta_W}
\end{equation}
with $g$ being a phase space correction factor, with $g \to 1$ 
in the limit $m_{\chi^0_1}\to\infty$ \cite{Hirsch:2005ag}. Equation 
\eqref{ZtoW} can be understood with the help of the approximative couplings 
\eqref{eq:cnwdef} and \eqref{eq:cnndef}. The relative size 
of the branching ratios for the final states $W+e$, $W+\mu$ and $W+\tau$ 
depends on both, (a) the nature of the LSP and (b) the fit to the 
neutrino data. We will discuss this important feature in more detail 
in section \ref{sec:corr}.

Generally, for $m_{\chi^0_1}\ge m_{W^\pm}$ three-body final states of 
the neutralino decay are less important than the two-body decays shown 
in figure \ref{fig:Exa_mass_br}. Especially one expects that the final 
state $\nu b{\bar b}$ has a smaller branching than in the case of 
explicit \rpv \cite{Porod:2000hv}. This is essentially due to the fact, 
that $|\vec\epsilon|/\mu$ is smaller in s-\rpv with a ``light'' singlet 
spectrum than in a model with explicit bilinear \rpv, simply because in s-\rpv
it enters the neutrino mass matrix at tree-level while in b-\rpv it only appears at the 1-loop level. A smaller $|\vec\epsilon|/\mu$ 
leads to smaller couplings between $\chi^0_1-l-{\tilde l}$, 
$\chi^0_1-q-{\tilde q}$ and especially $\chi^0_1-\nu-h^0$, see also 
couplings in \cite{Porod:2000hv}. We have checked numerically, that 
$Br(\chi^0_1\to\nu + h^0)$, if kinematically open, is typically below 
$1 \%$ for singlets in the ${\cal O}({\rm \text{TeV}})$ range.

\begin{figure}
\begin{center}
\vspace{5mm}
\includegraphics[width=0.49\textwidth]{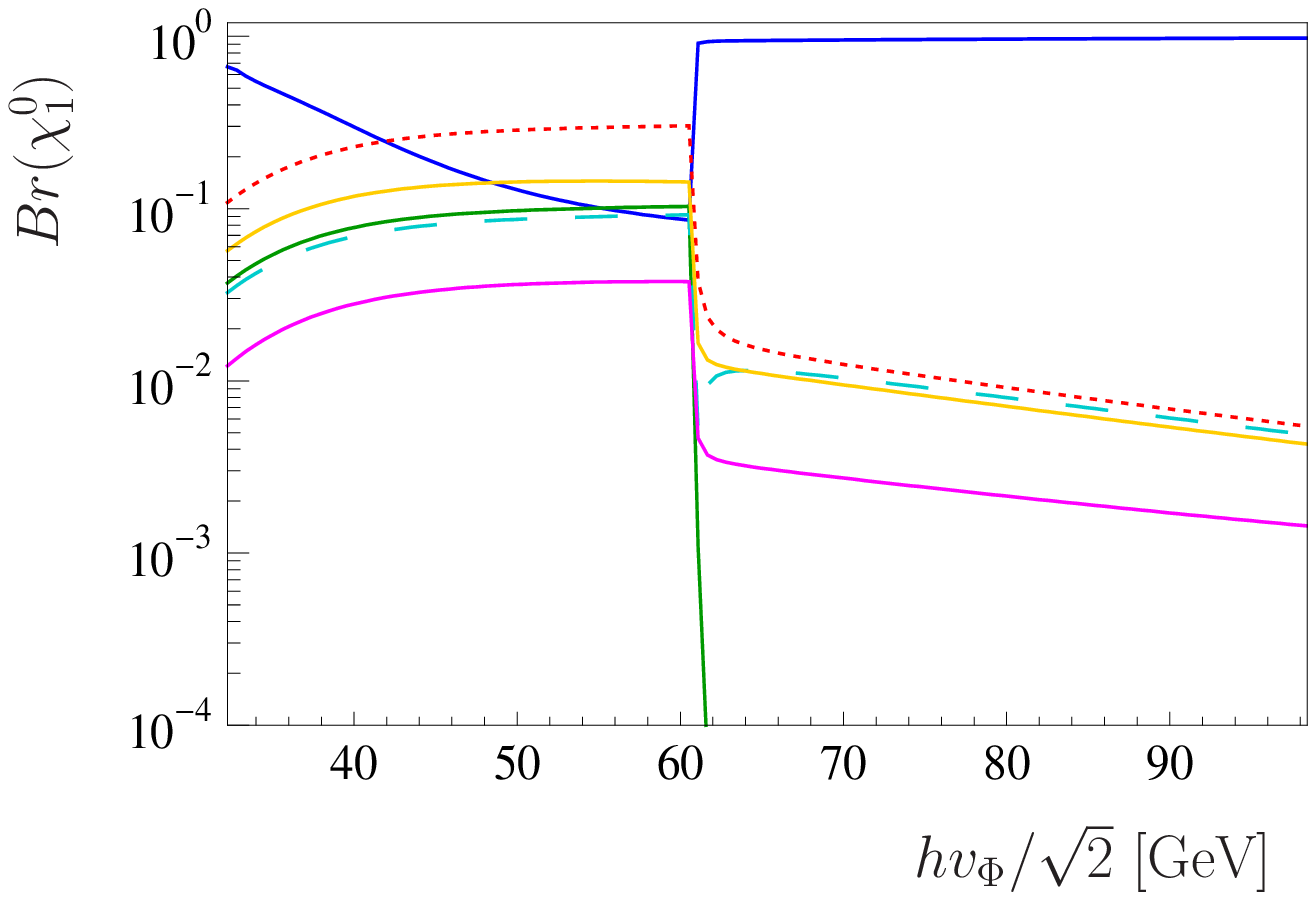}
\includegraphics[width=0.49\textwidth]{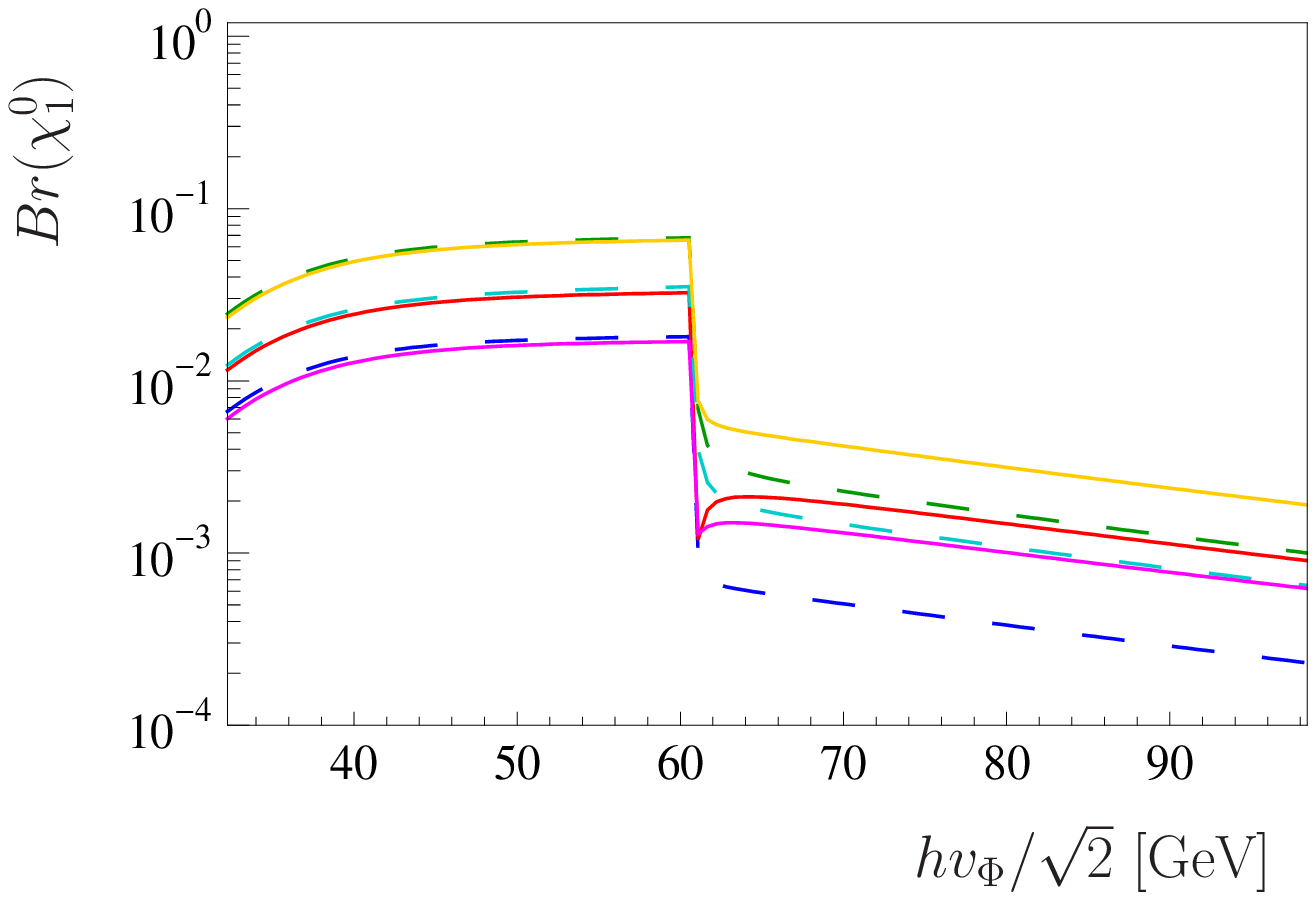}
\end{center}

\caption{Branching ratios for the most important decay modes of the 
lightest neutralino state versus $\frac{1}{\sqrt{2}}hv_{\Phi}$ for a 
specific, but typical example point. The MSSM parameters have been 
adjusted such that the sparticle spectrum of the standard point SU4 
is approximately reproduced. The singlet parameters have been chosen 
randomly, $v_R=v_S=1$ TeV, ${\vec\epsilon}$ and $\vec\Lambda$ have 
been fitted to neutrino data, such that $\vec\Lambda$ generates the 
atmospheric scale and ${\vec\epsilon}$ the solar scale. The different 
final states are as follows. In the left figure, as ordered on the right 
side, from top to bottom the lines are 
$Br(\chi^0_1\to [{\rm invisible}])$ (full line, blue), 
$Br(\chi^0_1\to \mu qq')$ (short-dashed, red), 
$Br(\chi^0_1\to \tau qq')$ (large-dashed, light blue), 
$Br(\chi^0_1\to \nu q{\bar q})$ (full, yellow),  
$Br(\chi^0_1\to \nu b{\bar b})$ (full, pink) and 
$Br(\chi^0_1\to e qq')$ (full, green). In the right 
figure, purely leptonic modes, from top to bottom (on the right 
side): 
$Br(\chi^0_1\to \nu \mu\tau)$ (full, yellow), 
$Br(\chi^0_1\to \nu e \mu)$ (dashed, green), 
$Br(\chi^0_1\to \nu e \tau)$ (full, red), 
$Br(\chi^0_1\to \nu \mu \mu)$ (dashed, light blue),
$Br(\chi^0_1\to \nu \tau \tau)$ (full, pink) and
$Br(\chi^0_1\to \nu e e)$ (dashed, darker blue). 
For a detailed discussion see text.}
\label{fig:Exa_br_low}
\end{figure}

For the case of $m_{\chi^0_1}\le m_{W^\pm}$ figure \ref{fig:Exa_br_low} 
shows an example for the most important final states of the lightest 
neutralino decay as a function of $\frac{hv_{\Phi}}{\sqrt{2}}$. As in 
the figure \ref{fig:Exa_mass_br} to the left the lightest neutralino 
is a singlino, to the right of the ``transition'' region the lightest 
neutralino is a bino. Note that the point SU4 \footnote{Benchmark point defined by the ATLAS collaboration at the Rome meeting, 2004. See \cite{Aad:2009wy}. The values of the SU4 mSUGRA parameters can be also found at https://twiki.cern.ch/twiki/bin/view/Atlas/SusyPublicResults.}
produces a bino mass of approximately $m_{\tilde B} \simeq 60$ GeV, 
thus the only two body decay modes which are kinematically allowed 
are $J+\nu$ and - very often, but not always - $S_J+\nu$. One observes 
that these invisible decay modes have typically a larger branching 
ratio than in the case $m_{\chi^0_1}\ge m_{W^\pm}$ shown in figure 
\ref{fig:Exa_mass_br}. This fact is essentially due to the propagator 
and phase space suppression factors for three body decays. For a bino 
LSP the invisible decay has the largest branching fraction. Semileptonic 
modes are next important with typically $l_i qq'$ being larger than 
$\nu q{\bar q}$. It is interesting to note, that in the purely leptonic 
decays, lepton flavor violating final states such as $\mu\tau$ have 
branching ratios typically as large or larger than the corresponding 
charged lepton flavor diagonal decays ($\mu\mu$ and $\tau\tau$). These 
large flavor off-diagonal decays can be traced to the fact that 
neutrino physics requires two large mixing angles. The branching ratios 
shown in figure \ref{fig:Exa_br_low} should be understood only as 
representative examples - not as firm predictions. Especially for the 
case of a bino LSP, the partial width to the final state 
$J+\nu$, i.e. invisible final state, can vary by several orders of 
magnitude, see the discussion below. The predictions for relative 
ratios of the different (partially or completely) visible final states 
is fixed much tighter, because these final states correlate with neutrino 
physics, as we discuss in section \ref{sec:corr}.

\begin{figure}
\begin{center}
\vspace{5mm}
\includegraphics[width=0.49\textwidth]{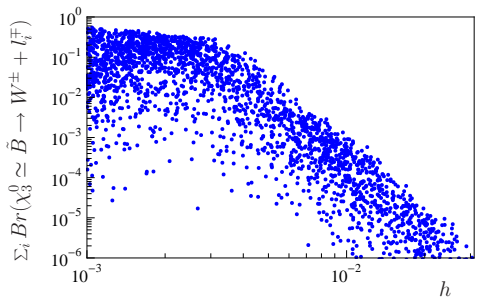}
\end{center}

\caption{Sum over 
$\sum_i Br(\chi^0_3 \simeq {\tilde B} \rightarrow  W^{\pm} + l_i^{\mp})$ 
versus $h$ for MSSM parameters resembling the standard point SPS1a', 
random values of the singlet parameters and with the condition of 
${\cal S}_1$ being the LSP. The dominant decay mode for the ${\tilde B}$ 
in all points is ${\tilde B} \rightarrow {\cal S}_1 + \eslash$, with the 
missing energy due to either $J$ or 2$J$ emission. For low values of 
$h$ one can have visible decays of the ${\tilde B}$ reaching $(20-30) \%$, 
for $h$ larger than, for say, $h=0.05$ ${\tilde B}$ decays to 
${\cal S}_1$ plus missing energy with nearly $100 \%$.}
\label{fig:chi3vh}
\end{figure}

\begin{figure}
\begin{center}
\vspace{5mm}
\includegraphics[width=0.49\textwidth]{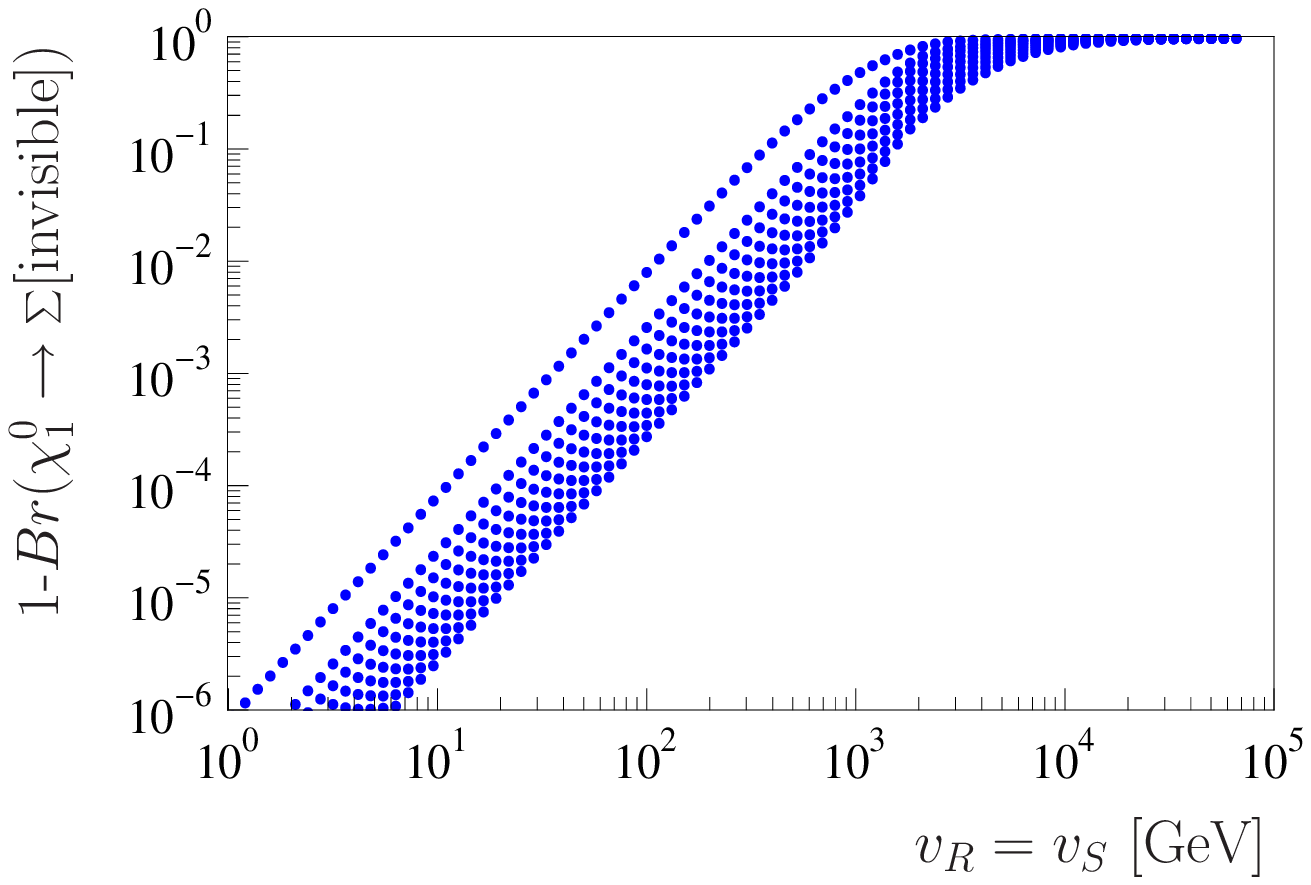}
\includegraphics[width=0.49\textwidth]{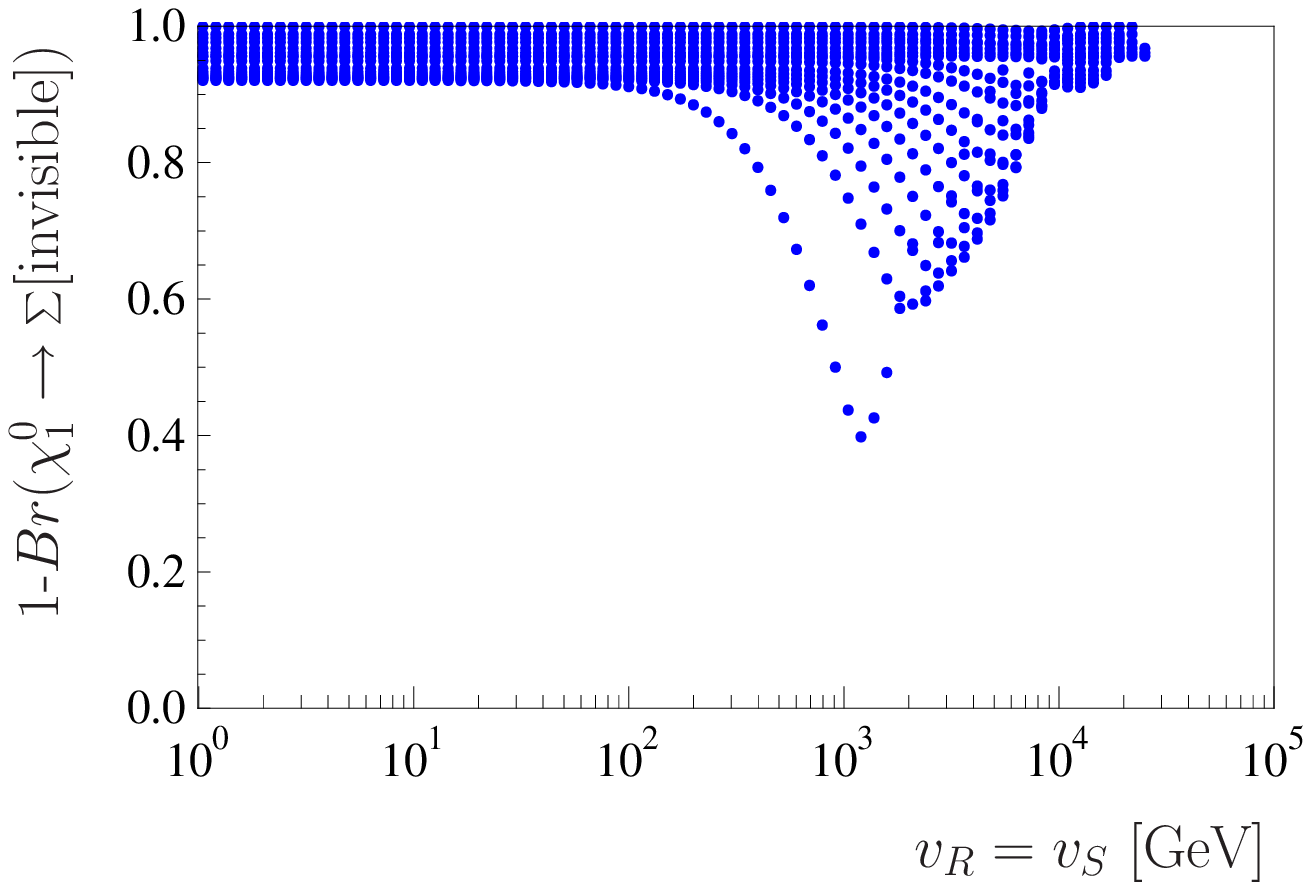}
\end{center}

\caption{Sum over all at least partially visible decay modes 
of the lightest neutralino versus $v_R$ in GeV, for a set of 
$v_{\Phi}$ values $v_{\Phi} = 10$--40 TeV for the mSUGRA parameter 
point $m_0=280$ GeV, $m_{1/2}=250$ GeV, $\tan\beta=10$, $A_0=-500$ GeV 
and $sgn(\mu)=+$. To the left $\chi_1^0 \simeq {\tilde B}$; to the right 
$\chi_1^0 \simeq {\cal S}$. The plot demonstrates that the branching 
ratio into ${\tilde B} \to J + \nu$ does depend strongly on the value 
of $v_R$ and to a minor extend on $v_{\Phi}$. Lowering $v_R$ one can 
get branching ratios for the invisible decay of the ${\tilde B}$ very close 
to 100 \%, thus a very MSSM-like phenomenology. This plot can be also found in figure \ref{fig:vis}, where
the different values for $v_{\Phi}$ are also indicated. The right plot demonstrates 
that such a possibility does not exist in the case of an ${\cal S}$ LSP.}
\label{fig:Inv_vr}
\end{figure}

If the ${\cal S}$ is the LSP, a bino NLSP decays dominantly to the singlino 
plus missing energy, as is shown in figure \ref{fig:chi3vh}. The final 
state can be either ${\cal S}_1 + J$ or ${\cal S}_1 + 2 J$, the latter 
due to the chain ${\tilde B} \to {\cal S}_2 + J \to {\cal S}_1 + 2J$, 
where the 2nd step has always a branching fraction very close to $100 \%$. 
However, a special opportunity arises if $h$ is low. In this case 
$\sum_i Br(\chi^0_3 \simeq {\tilde B} \rightarrow  W^{\pm} + l_i^{\mp})$ 
can easily reach several percent and it becomes possible to test the 
model with the bino decays and the singlino decays {\em at the same time}. 
This would allow a much more detailed study of the model parameters 
than for the more ``standard'' case where only either singlino or 
bino decay visibly. We note that for any fixed value of $h$, 
$\sum_i Br(\chi^0_3 \simeq {\tilde B} \rightarrow  W^{\pm} + l_i^{\mp})$ 
depends mostly on $v_R$ (and to some extend on $v_{\Phi}$). Low values 
if $v_R$ lead to low 
$\sum_i Br(\chi^0_3 \simeq {\tilde B} \rightarrow  W^{\pm} + l_i^{\mp})$ 
as we will discuss next. 

Figure \ref{fig:Inv_vr} shows the sum over all at least partially 
visible decay modes of the lightest neutralino versus $v_R$ in GeV, 
for a set of $v_{\Phi}$ values $v_{\Phi} = 10 - 40$ TeV for the mSUGRA 
parameter point ($m_0=280$~GeV, $m_{1/2}=250$~GeV, $\tan\beta=10$, 
$A_0=-500$~GeV and sgn$(\mu)=+$). This point was constructed to produce 
formally a $\Omega_{\chi^0_1}h^2 \simeq 1$ in case of conserved R-parity, much 
larger than the observed relic DM density \cite{Komatsu:2010fb}. The 
left plot shows the case $\chi_1^0 \simeq {\tilde B}$, the right plot 
$\chi_1^0 \simeq {\cal S}$. For ${\tilde B}$, $Br({\tilde B} \to J + \nu)$ 
very close to 100 \% are found for low values of $v_R$. This feature 
is independent of the mSugra parameters, see the correspoding figure 
in \cite{Hirsch:2006di}. In this case large statistics becomes necessary 
to find the rare visible neutralino decays, which prove that R-parity is 
broken. The inconsistency between the calculated $\Omega_{\chi^0_1}h^2$ 
and the measured $\Omega_{CDM}h^2$ might give a first indication for 
a non-standard SUSY model.

Figure \ref{fig:Inv_vr} to the right shows that the case $\chi_1^0 \simeq 
{\cal S}$ has a very different dependence on $v_R$. We have checked that 
this feature is independent of the mSugra point. For other choices of mSugra 
parameters larger branching ratios for $Br({\cal S} \to J + \nu)$ can be 
obtained, but contrary to the bino LSP case, the sum over the invisible 
decay branching ratios never approaches 100 \%. 

\begin{figure}
\begin{center}
\vspace{5mm}
\includegraphics[width=0.49\textwidth]{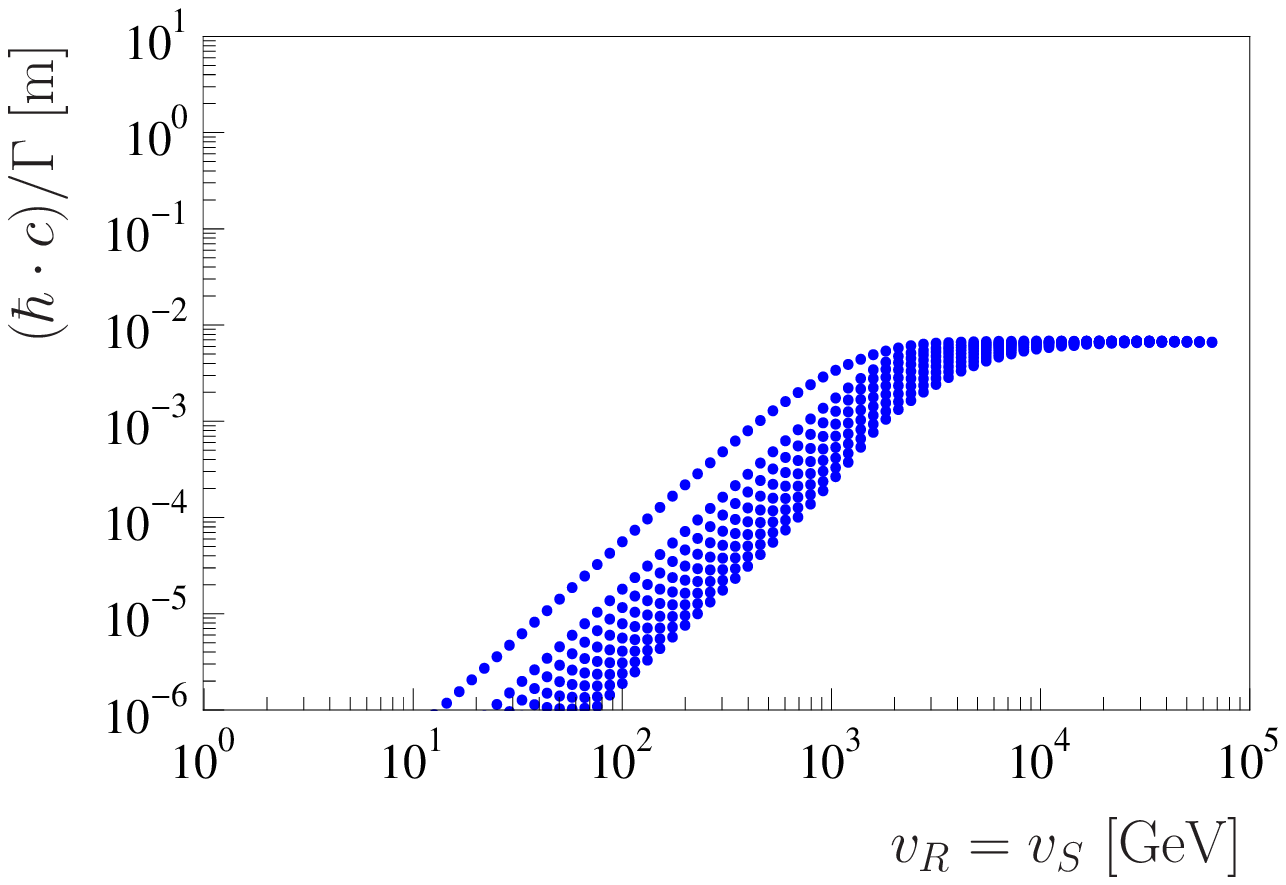}
\includegraphics[width=0.49\textwidth]{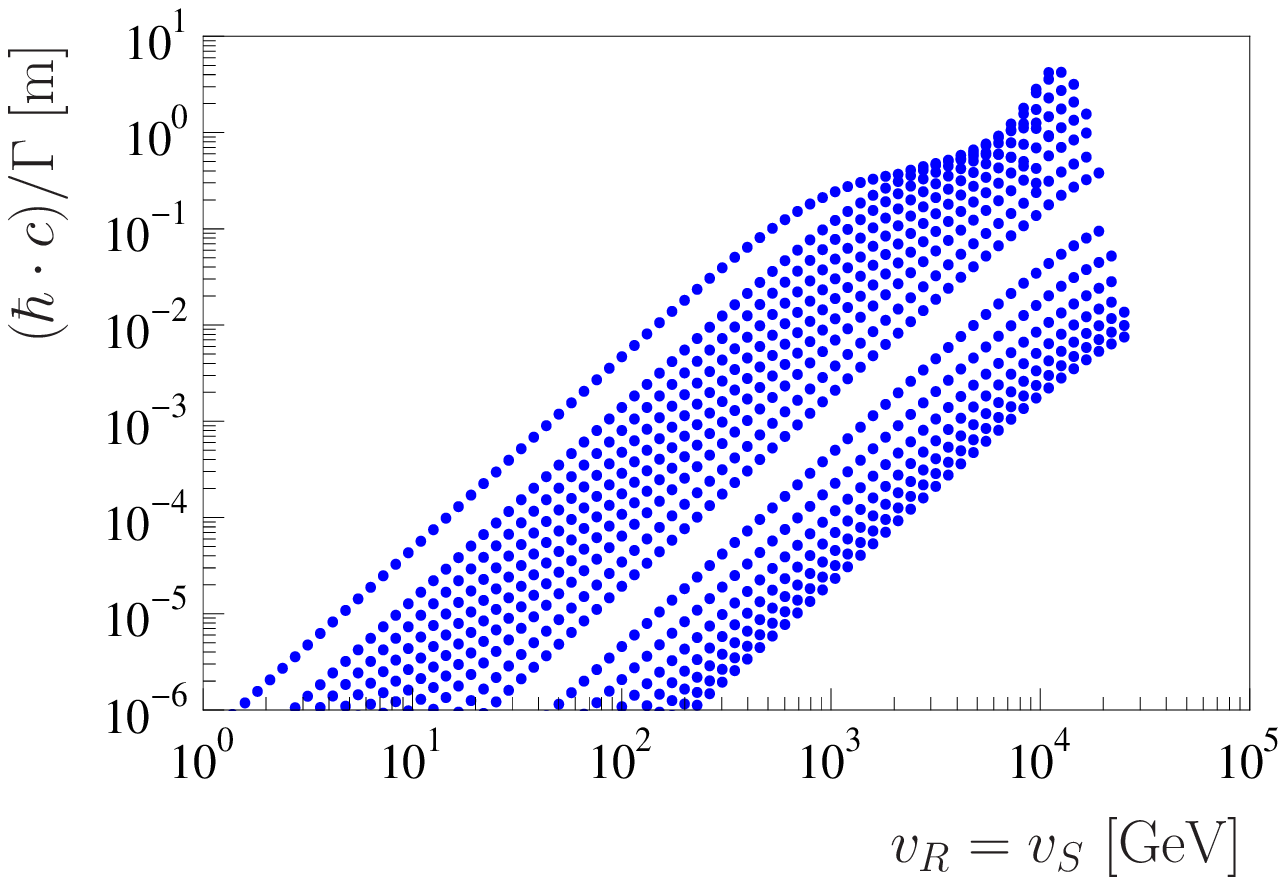}
\end{center}

\caption{Decay length of the lightest neutralino in meter 
versus $v_R$. To the left: Bino LSP; to the right: Singlino 
LSP. All parameters have been chosen as in figure \ref{fig:Inv_vr}.}
\label{fig:L_vr}
\end{figure}

Figure \ref{fig:L_vr} shows the calculated decay lengths for 
the lightest neutralino for the same choice of parameters as 
shown in figure \ref{fig:Inv_vr}. To the left the case $\chi_1^0 \simeq 
{\tilde B}$, to the right $\chi_1^0 \simeq {\cal S}$. Decay 
lengths depend strongly on $v_R$. Singlinos tend to have larger 
decay lengths than binos for the same choice of parameters. 
However, a measurement of the decay length alone is not sufficient 
to decide whether the LSP is a singlino or a bino. If the nature 
of the LSP is known, observing a finite decay length allows a rough 
estimate of the scale $v_R$, or at least to establish a rough lower 
limit on $v_R$. 

Summarizing this discussion, it can be claimed that observing a decay 
branching ratio of the LSP into completely invisible final states larger 
than Br$({\chi_1^0} \to \sum [{\rm invisible}])\ge 0.1$ is an indication 
for s-\rpv. Finding Br$({\chi_1^0} \to \sum [{\rm invisible}]) \simeq 100 \%$ 
shows furthermore that the ${\chi_1^0}$ must be a bino and measuring 
Br$({\chi_1^0} \to \sum [{\rm visible}])$ for a bino LSP gives 
an order-of-magnitude estimate of $v_R$.

\subsection{Possible observables to distinguish between Singlino LSP and
bino LSP}

Since bino and singlino LSP decays have, in principle, the same final 
states, simply observing some visible decay products of the LSP does not 
allow to decide the nature of the LSP. In this subsection we will 
schematically discuss some possible measurements, which would allow to 
check for the LSP nature.

As shown above, if the singlino is the LSP and the bino the NLSP, 
one can have that the bino decays to standard model particles 
competing with the decay to the singlino LSP. If both particles, the 
bino and the singlino LSP have visible decay modes, it is guaranteed 
that the singlino is the LSP. If the bino decays only invisibly to 
the singlino, a different strategy is called for. We discuss two 
examples in the following.

In the following discussion we will replace the neutralino mass
eigenstates by the particles which correspond to their main content
to avoid confusion with indices. At the LHC one will mainly produce
squarks and gluinos which will decay in general in cascades. A typical
example is $\tilde q_L \to q \tilde W$ and the wino decays further to a 
 bino LSP as for example:
\begin{eqnarray}
 \tilde W \to e^+  \tilde e^-
                 \to e^- e^+ \tilde B
                 \to e^- e^+ \mu q \bar{q} \\
\label{eq:bino1}
 \tilde W \to e^+  \tilde e^-
                 \to e^- e^+ \tilde B
                 \to e^- e^+  J \nu
\label{eq:bino2}
\end{eqnarray}
In this case one can measure in principle the neutralino mass from the first
decay chain. In the invariant momentum spectrum of the $e^+ e^-$ pair
the edge must correspond to this mass. 
In the case of the singlino
\begin{eqnarray}
 \tilde W \to e^+  \tilde e^-
                 \to e^- e^+ \tilde B
                 \to e^- e^+  J {\cal S}
                 \to e^- e^+ J \mu q \bar{q} \\
\label{eq:singlino1}
 \tilde W \to e^+  \tilde e^-
                 \to e^- e^+ \tilde B
                 \to e^- e^+  J {\cal S}
                 \to e^- e^+  J J \nu 
\label{eq:singlino2}
\end{eqnarray}
In this case one has on average more missing energy than for a bino LSP. 
However, in both cases one can study spectra combining the jet 
stemming from the squark and the $e^+ e^-$ pair and obtain
information on the masses from the so-called edge variables 
\cite{Allanach:2000kt}. In addition one can use additional variables 
like, for example, $m_{T2}$ \cite{Barr:2003rg,Barr:2007hy,Cho:2007dh,Barr:2010zj}
to obtain information on the LSP mass. Note, that this variable works
also if there are additional massless particles involved, although at 
the expense of available statistics \cite{Barr:private}.  In addition
one can obtain the invariant mass of the LSP from the final state 
$\mu q \bar{q}$. In the case where the LSP has a decay length measurable 
at the LHC, one can separate the latter decay products from the other 
particles in the event and, thus, reduce considerably the combinatorial 
problems associated with the correct assignment of the jets. In the case 
of a bino LSP one would find that all the three different measurements 
yield the same mass for the LSP. In the case of a singlino LSP, on the 
other hand, one would obtain that the LSP mass reconstructed from the 
edge variables does not coincide with the mass reconstructed from the 
$\mu q \bar{q}$ spectrum. This would indicate that there are two different
particles involved. (Such a difference might also be visible in the $m_{T2}$ 
variable.) However, in all cases detailed Monte Carlo studies will be 
necessary to work out the required statistics, etc.

Distinguishing bino and singlino LSPs  will become considerably easier 
at a future international linear collider. In $e^+ e^-$ one can directly 
produce a bino LSP but not a singlino LSP and, thus, the identification of
the correct scenario should be fairly straightforward. 

\subsection{Correlations between LSP decays and neutrino mixing angles}
\label{sec:corr}

Correlations between LSP decays and neutrino mixing angles depend 
on the nature of the LSP. Above we have discussed some possible 
measurements which, at least in principle, allow to distinguish 
bino from singlino LSPs. In this subsection we assume that the 
nature of the LSP is known.

\subsubsection{Bino LSP}

We note that the following discussion is valid also if the bino is the 
NLSP which, as discussed above, decays with some final, but probably 
small percentage to visible final states. 

\begin{figure}
\begin{center}
\vspace{5mm}
\includegraphics[width=0.49\textwidth]{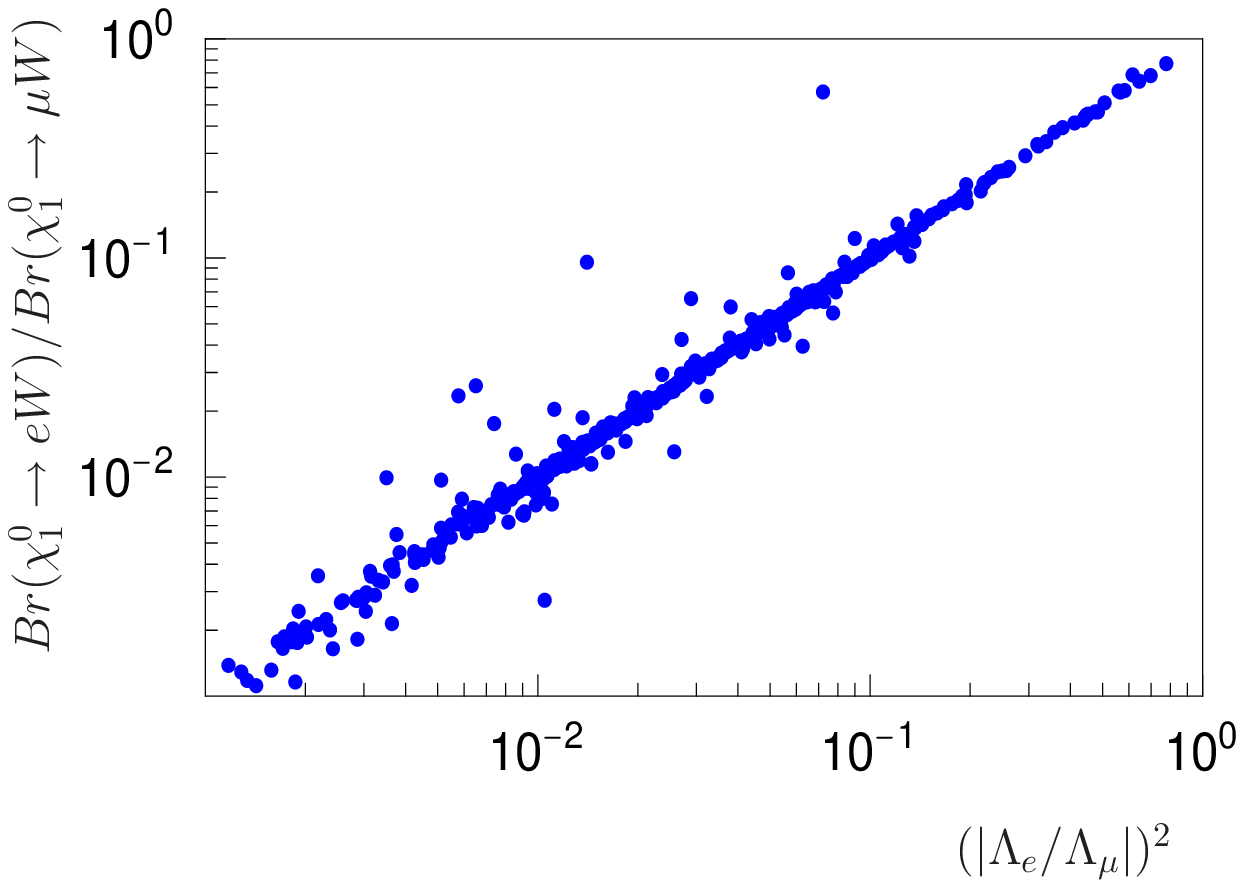}
\includegraphics[width=0.49\textwidth]{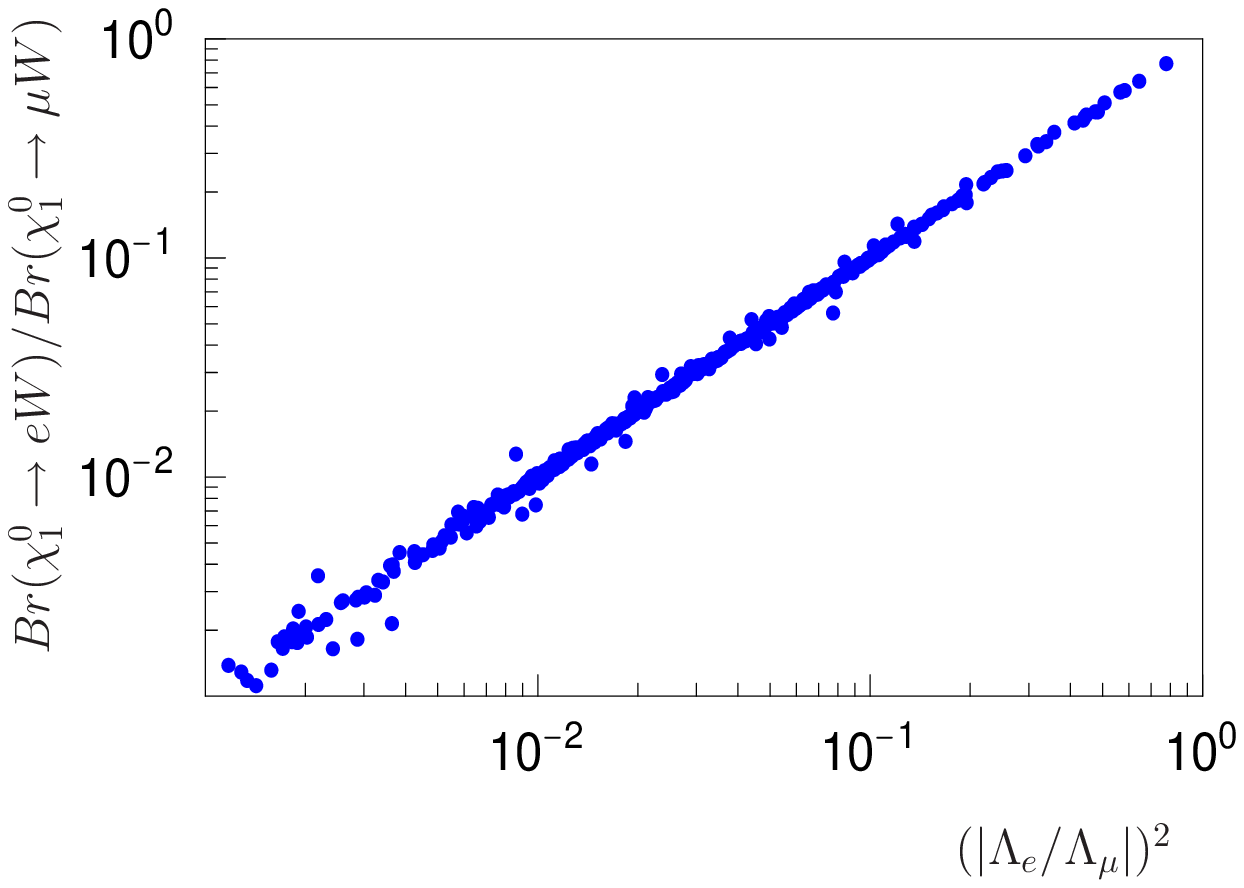}

\end{center}
\vspace{0mm}
\caption{Ratio $\frac{Br(\chi^0_1\rightarrow e W)}
{Br(\chi^0_1\rightarrow \mu W)}$ versus $(\Lambda_{e}/\Lambda_{\mu})^2$ 
for a bino LSP. To the left: ``Bino-purity'' $N_{11}^2 > 0.5$, to the 
right: $N_{11}^2 > 0.9$. All points with $m_{LSP} > m_{W}$.}
\label{fig:BinoeWmuW}
\end{figure}

\begin{figure}
\begin{center}
\vspace{5mm}
\includegraphics[width=0.49\textwidth]{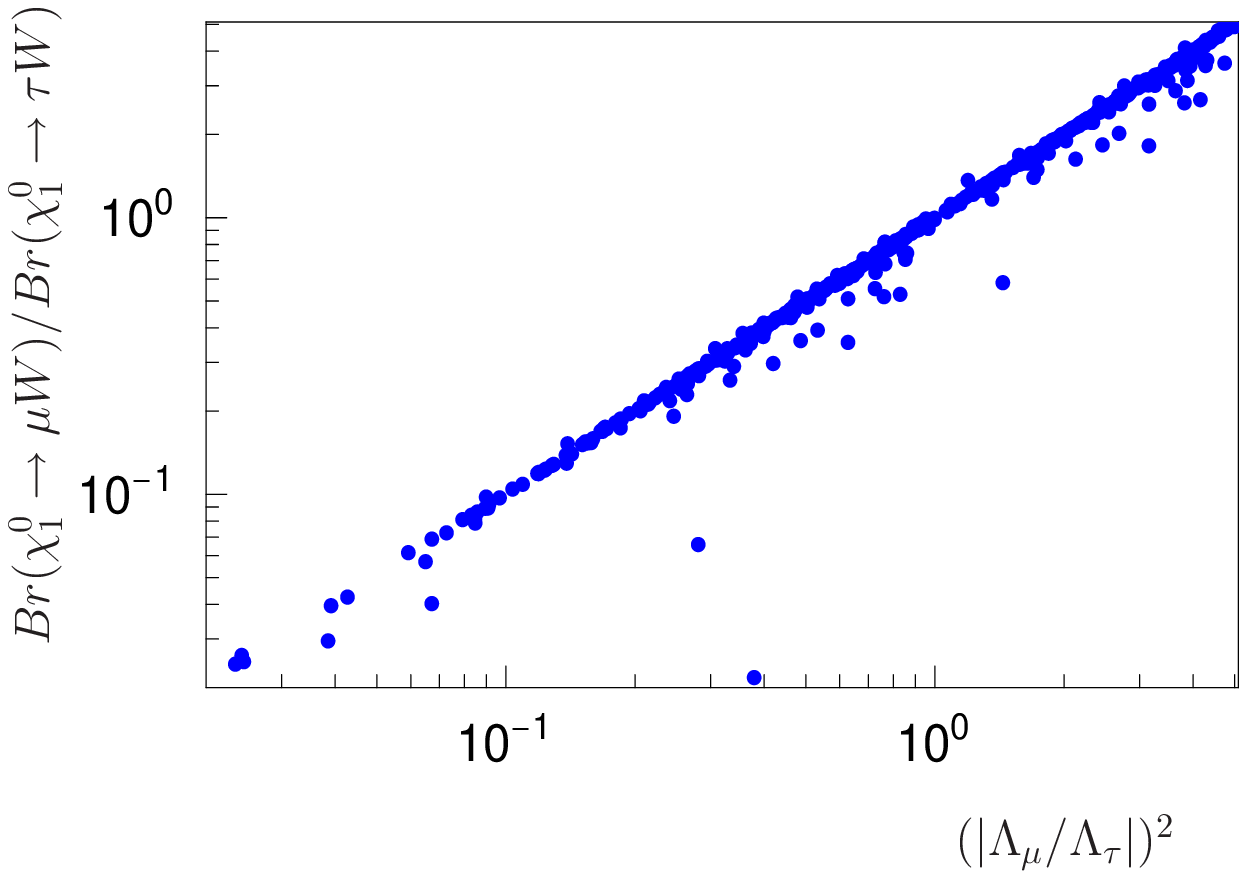}
\includegraphics[width=0.49\textwidth]{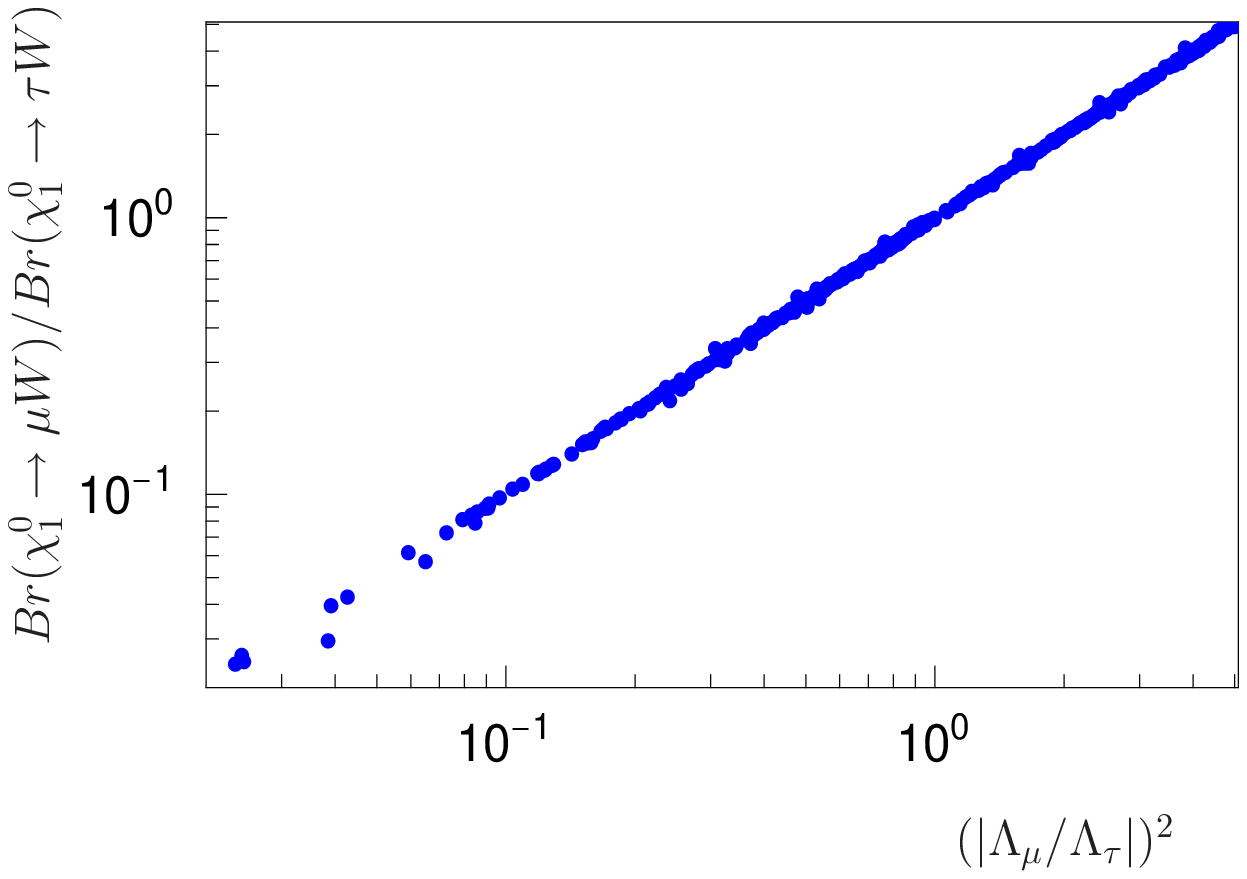}

\end{center}
\vspace{0mm}
\caption{Ratio $\frac{Br(\chi^0_1\rightarrow \mu W)}
{Br(\chi^0_1\rightarrow \tau W)}$ versus $(\Lambda_{\mu}/\Lambda_{\tau})^2$ 
for a bino LSP. To the left: ``Bino-purity'' $N_{11}^2 > 0.5$, to the 
right: $N_{11}^2 > 0.9$.}
\label{fig:BinomuWtauW}
\end{figure}

In explicit bilinear R-parity violation the coupling of the bino component 
of the neutralino to gauge bosons and leptons is completely dominated by 
terms proportional to $\Lambda_i$, as has been shown in \cite{Porod:2000hv}. 
Although the coefficients for the spontaneous model are more complicated, 
see the discussion in appendix \ref{sect:app1}, generation dependence for the 
coefficients for the coupling $\chi_1^0-W-l_i$ appear only in the 
terms $\Lambda_i$ and $\epsilon_i$, i.e. $K_{\Lambda}^i$ and  $K_{\epsilon}^i$ 
are independent of the lepton generation. Numerically one finds than 
that the terms proportional to $\Lambda_i$ dominate the $\chi_1^0-W-l_i$ 
coupling for a bino LSP always. This is demonstrated in figures 
\ref{fig:BinoeWmuW} and \ref{fig:BinomuWtauW}. Here we have 
numerically scanned the mSugra parameter space, with random singlet 
parameters and the additional condition that the LSP is a bino.
For the left (right) figures we have numerically applied the cut 
$N_{11}^2>0.5$ ($N_{11}^2>0.9$). 

Figure \ref{fig:BinoeWmuW} [ \ref{fig:BinomuWtauW}] shows the ratio 
of branching ratios $Br({\tilde B}\to W + e)$/$Br({\tilde B}\to W + \mu)$ 
[$Br({\tilde B}\to W + \mu)$/$Br({\tilde B}\to W + \tau)$] versus 
$(\Lambda_e/\Lambda_{\mu})^2$ [$(\Lambda_{\mu}/\Lambda_{\tau})^2$]. 
To establish a correlation between ratios of $\Lambda_i$ and the 
bino decay branching ratios, a bino purity of $N_{11}^2>0.5$ is 
usually sufficient. The figures demonstrate that the correlations 
get sharper with increasing bino purity.

We have checked that for neutralinos with mass lower than $m_W$ one 
can use ratios of the decays ${\tilde B}\rightarrow l_i qq'$ for the 
different $l_i$ in the same way to perform a measurement of $\Lambda_i$ 
ratios. Plots for this parameter region are rather similar to the ones 
shown for the case ${\tilde B}\rightarrow l_i W$, although with a 
somewhat larger dispersion, and we therefore do not repeat them here. 

\begin{figure}
\begin{center}
\vspace{5mm}
\includegraphics[width=0.49\textwidth]{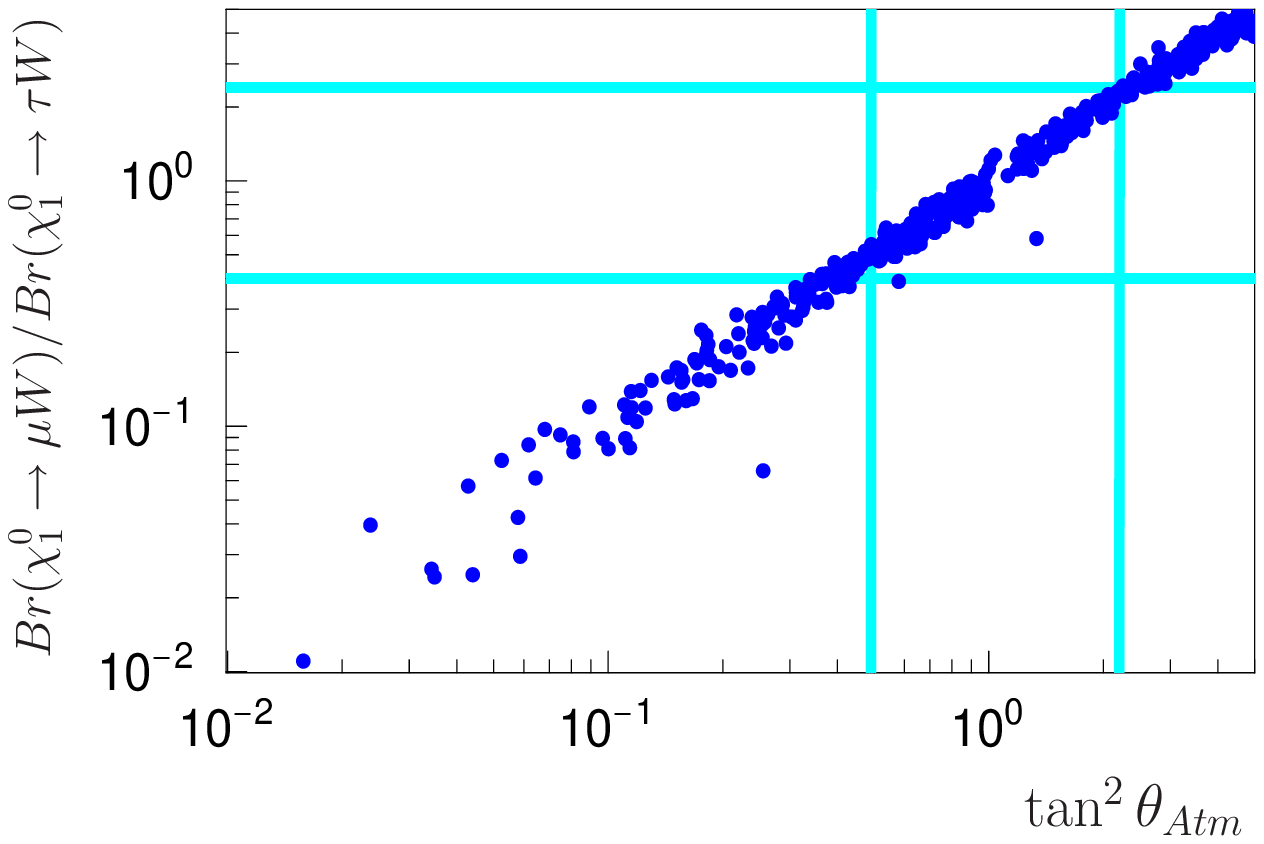}
\includegraphics[width=0.49\textwidth]{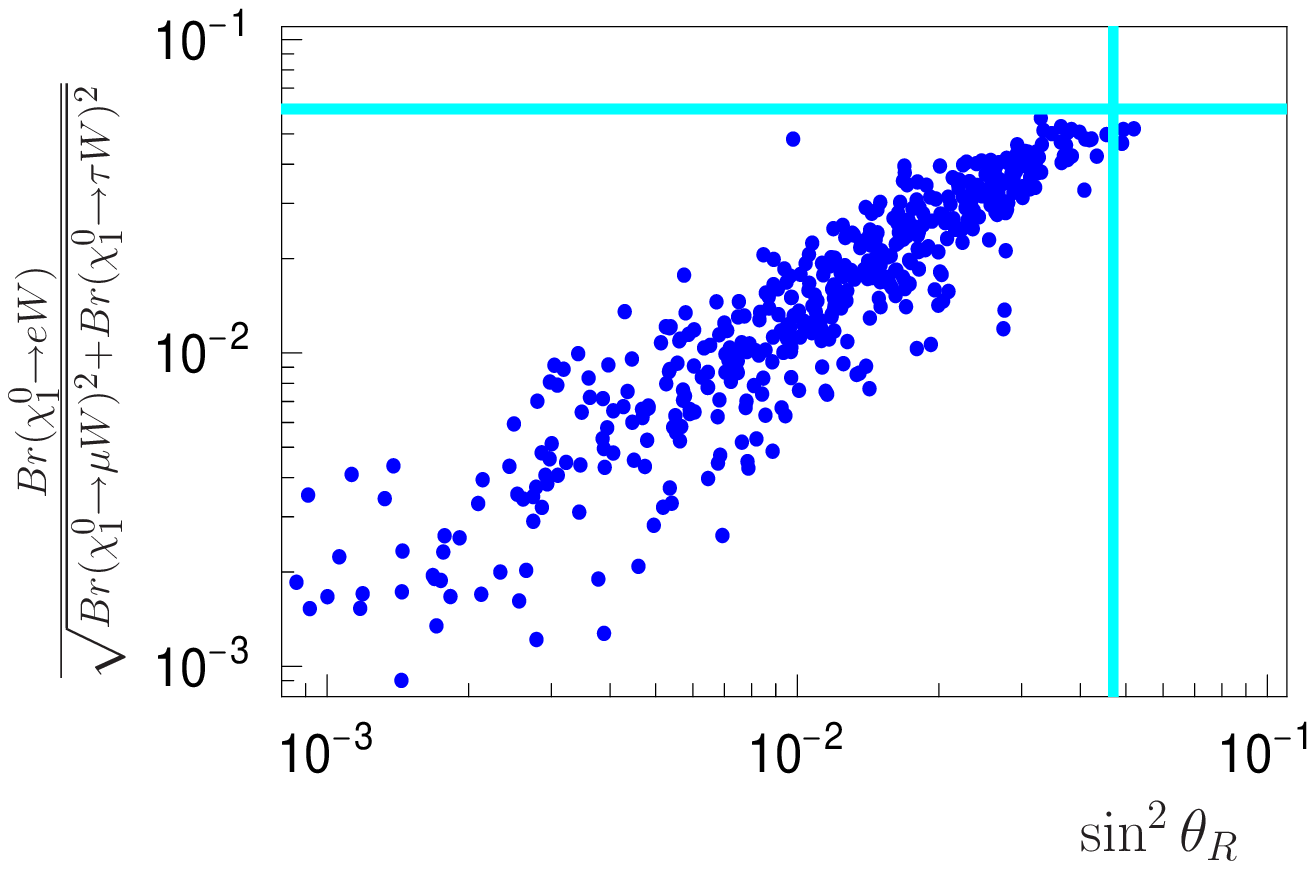}

\end{center}
\vspace{0mm}
\caption{Ratio ${\cal R}_{\mu}= \frac{Br(\chi^0_1\rightarrow \mu W)}
{Br(\chi^0_1\rightarrow \tau W)}$ versus $\tan^2\theta_{Atm} \equiv \tan^2\theta_{23}$ (left) 
and ${\cal R}_e= \frac{Br(\chi^0_1\rightarrow e W)}
{\sqrt{Br(\chi^0_1\rightarrow \mu W)^2+Br(\chi^0_1\rightarrow \tau W)^2}}$ 
versus $\sin^2\theta_R \equiv \sin^2\theta_{13}$ (right) for a bino LSP. ``Bino-purity'' 
$N_{11}^2 > 0.8$. Vertical lines are the $3 \sigma$ c.l. allowed experimental 
ranges, horizontal lines the resulting predictions for the fit (c1), 
see text.} 
\label{fig:BinoNf1pred}
\end{figure}

With the measurement of ratios of branching ratios different consistency 
checks of the model can be performed. In case (c1), i.e. $\vec\Lambda$
explaining the atmospheric scale, the atmospheric and the reactor angle 
are related to $W+l$ final states, as shown in figure \ref{fig:BinoNf1pred}. 
Here we show the ratios ${\cal R}_{\mu} = 
\frac{Br(\chi^0_1\rightarrow \mu W)}{Br(\chi^0_1\rightarrow \tau W)}$ 
versus $\tan^2\theta_{23}$ (left) and ${\cal R}_e = 
\frac{Br(\chi^0_1\rightarrow e W)}
{\sqrt{Br(\chi^0_1\rightarrow \mu W)^2+Br(\chi^0_1\rightarrow \tau W)^2}}$ 
versus $\sin^2\theta_{13}$ (right) for a bino LSP, for an assumed 
bino-purity of $N_{11}^2 > 0.8$. The vertical lines are the 
$3 \sigma$ c.l. allowed experimental ranges (upper bound), horizontal 
lines the resulting predictions for the two different observables 
${\cal R}$. Given the current experimental data, one expects 
$\frac{Br(\chi^0_1\rightarrow \mu W)}{Br(\chi^0_1\rightarrow \tau W)}$ 
in the range $[0.4,2.1]$ and $\frac{Br(\chi^0_1\rightarrow e W)}
{\sqrt{Br(\chi^0_1\rightarrow \mu W)^2+Br(\chi^0_1\rightarrow \tau W)^2}} 
\le 0.06$.

\begin{figure}
\begin{center}
\vspace{5mm}
\includegraphics[width=0.49\textwidth]{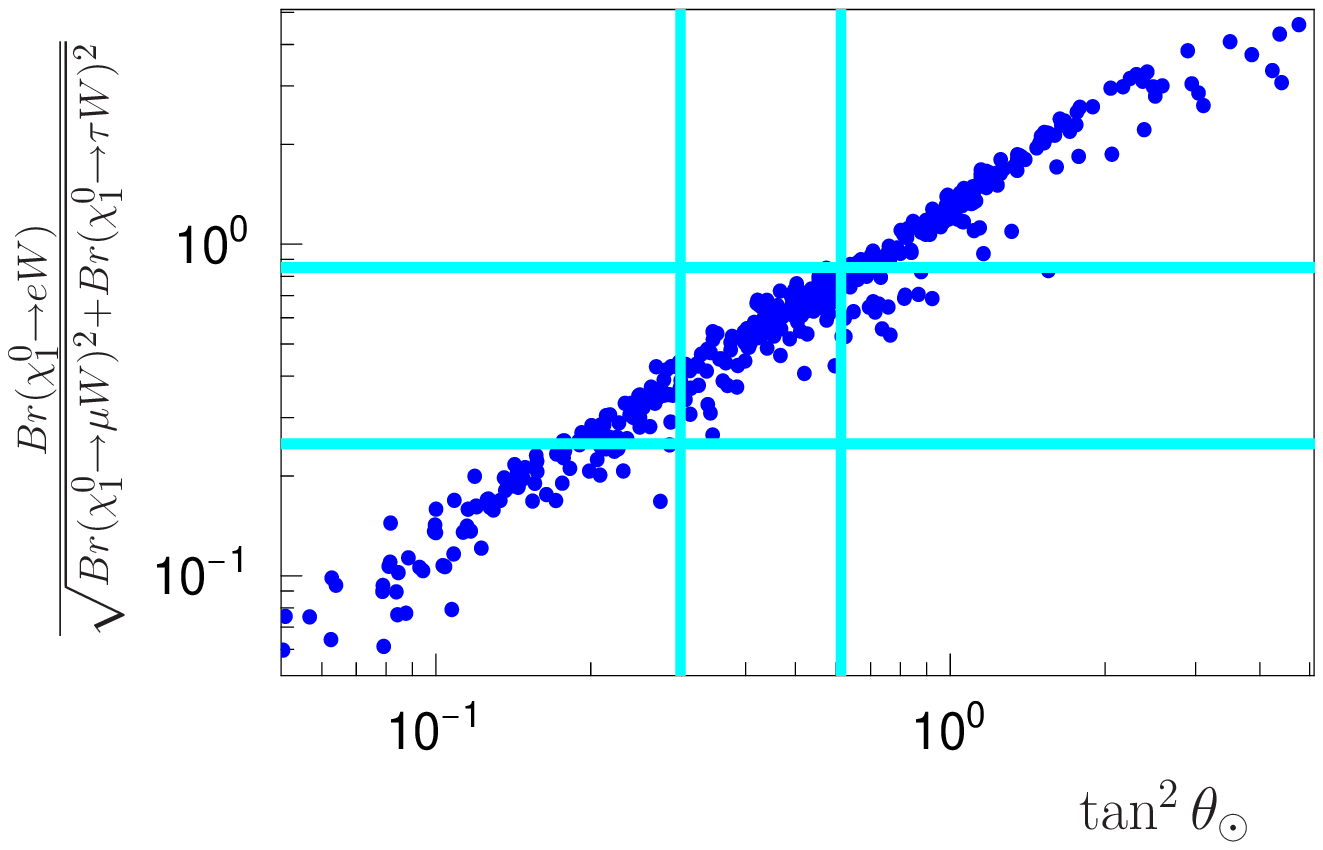}
\end{center}
\vspace{0mm}
\caption{Ratio 
${\cal R}_e= \frac{Br(\chi^0_1\rightarrow e W)}
{\sqrt{Br(\chi^0_1\rightarrow \mu W)^2+Br(\chi^0_1\rightarrow \tau W)^2}}$ 
versus $\tan^2\theta_{\odot} \equiv \tan^2\theta_{12}$ for a bino LSP. ``Bino-purity'' 
$N_{11}^2 > 0.8$. Vertical lines are the $3 \sigma$ c.l. allowed experimental 
ranges, horizontal lines the resulting predictions for the fit 
(c2), see text.}
\label{fig:BinoNf3pred}
\end{figure}

Different from figure \ref{fig:BinoNf1pred}, in case of (c2), i.e.
$\vec\Lambda$ explaining the solar scale, the ratio
$\frac{Br(\chi^0_1\rightarrow e W)} {\sqrt{Br(\chi^0_1\rightarrow \mu
W)^2+Br(\chi^0_1\rightarrow \tau W)^2}}$ correlates with
$\tan^2\theta_{12}$, as shown in figure
\ref{fig:BinoNf3pred}. Therefore, from the $3 \sigma$ c.l. allowed
range of the solar angle as measured by oscillation experiments one
expects to find $\frac{Br(\chi^0_1\rightarrow e W)}
{\sqrt{Br(\chi^0_1\rightarrow \mu W)^2+Br(\chi^0_1\rightarrow \tau
W)^2}}
\simeq [0.25,0.85]$. Finding this ratio experimentally to be larger than 
the one indicated by the solar data, i.e. 
$\frac{Br(\chi^0_1\rightarrow e W)}{\sqrt{Br(\chi^0_1\rightarrow \mu W)^2
+Br(\chi^0_1\rightarrow \tau W)^2}} \gg 1$, rules out the model as the 
origin of the observed neutrino oscillation data. Similarly a low (high) 
experimental value for this ratio indicates (for a bino LSP) that 
case (c1) [(c2)] is the correct explanation for the two observed 
neutrino mass scales.

\subsubsection{Singlino LSP}

\begin{figure}
\begin{center}
\vspace{5mm}
\includegraphics[width=0.49\textwidth]{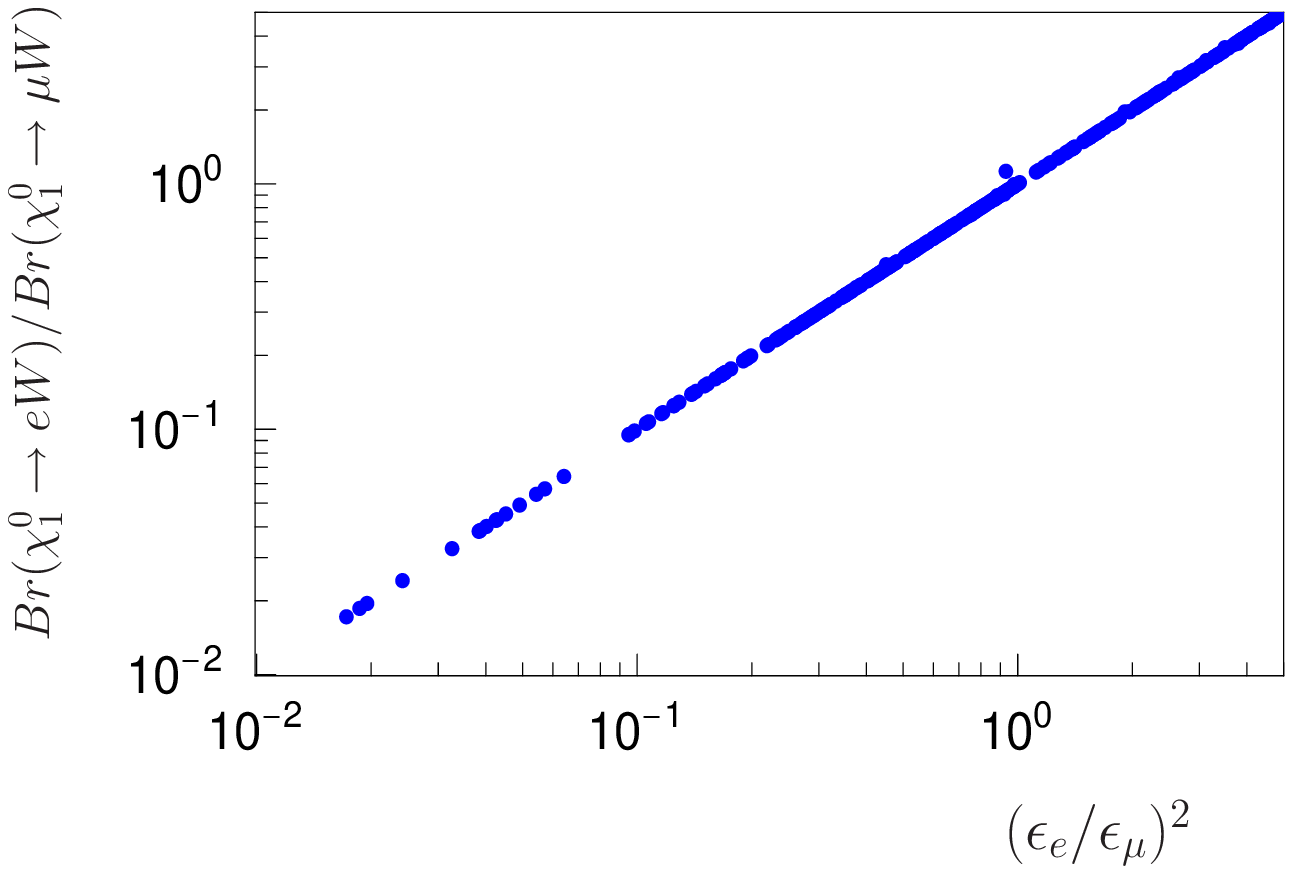}
\includegraphics[width=0.49\textwidth]{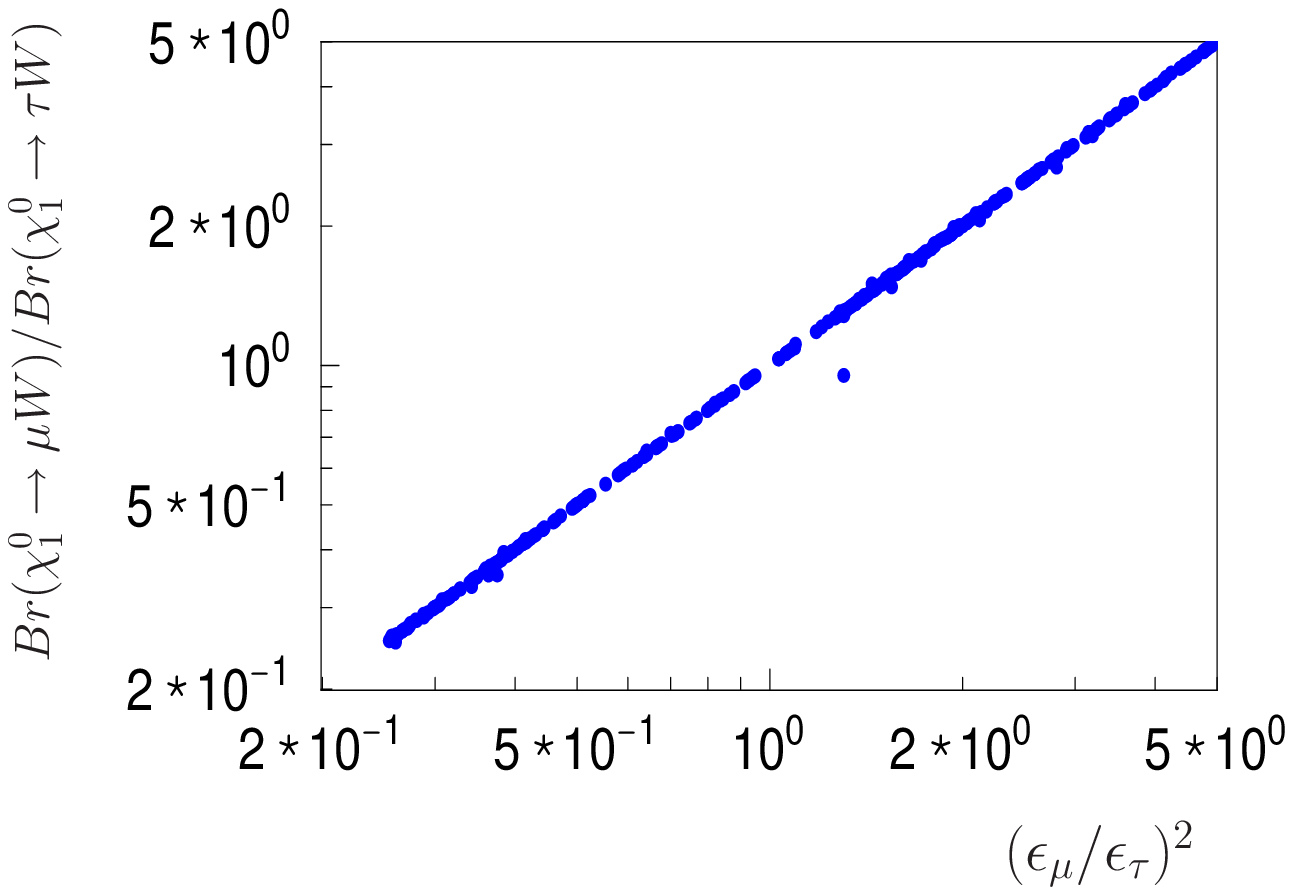}

\end{center}
\vspace{0mm}
\caption{Ratio $\frac{Br(\chi^0_1\rightarrow e W)}
{Br(\chi^0_1\rightarrow \mu W)}$ (left) versus 
$(\epsilon_{e}/\epsilon_{\mu})^2$ and 
$\frac{Br(\chi^0_1\rightarrow \mu W)}
{Br(\chi^0_1\rightarrow \tau W)}$ (right) versus 
$(\epsilon_{\mu}/\epsilon_{\tau})^2$ 
for a ``singlino'' LSP.} 
\label{fig:SngleWmuW}
\end{figure}

Different from the bino LSP case, for singlinos coupling to a lepton 
$l_i$-W pair terms proportional to $\epsilon_i$ dominate by far. 
This is demonstrated in figure \ref{fig:SngleWmuW}, where we show the 
ratios $\frac{Br(\chi^0_1\rightarrow e W)}{Br(\chi^0_1\rightarrow \mu W)}$ 
(left) versus $(\epsilon_{e}/\epsilon_{\mu})^2$ and 
$\frac{Br(\chi^0_1\rightarrow \mu W)}{Br(\chi^0_1\rightarrow \tau W)}$ 
(right) versus $(\epsilon_{\mu}/\epsilon_{\tau})^2$ 
for a singlino LSP. Note that mixing between singlinos and 
the doublet neutralinos of the model is always very small, 
unless the singlino is highly degenerate with the bino. 
Consequently singlinos are usually very ``pure'' singlinos and 
the correlations of the $l_i$-W with the $\epsilon_i$ ratios is 
very sharp.

\begin{figure}
\begin{center}
\vspace{5mm}
\includegraphics[width=0.49\textwidth]{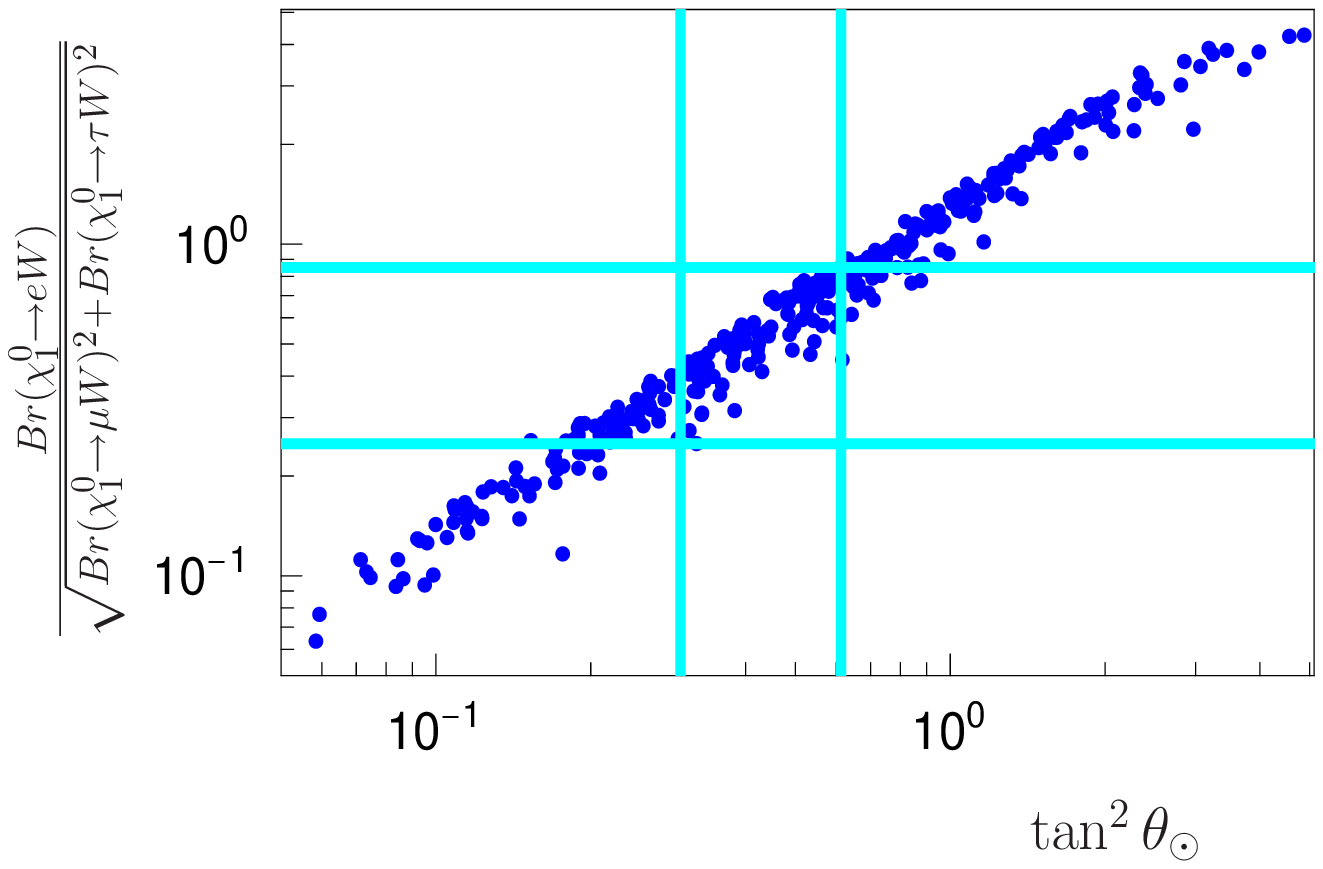}
\end{center}
\vspace{0mm}
\caption{Ratio ${\cal R}_e= \frac{Br(\chi^0_1\rightarrow e W)}
{\sqrt{Br(\chi^0_1\rightarrow \mu W)^2+Br(\chi^0_1\rightarrow \tau W)^2}}$ 
versus $\tan^2\theta_{\odot} \equiv \tan^2 \theta_{12}$ for a singlino LSP. 
Vertical lines are the $3 \sigma$ c.l. allowed experimental 
ranges, horizontal lines the resulting predictions for the fit (c1), 
see text.}
\label{fig:SnglNf1pred}
\end{figure}

\begin{figure}
\begin{center}
\vspace{5mm}
\includegraphics[width=0.49\textwidth]{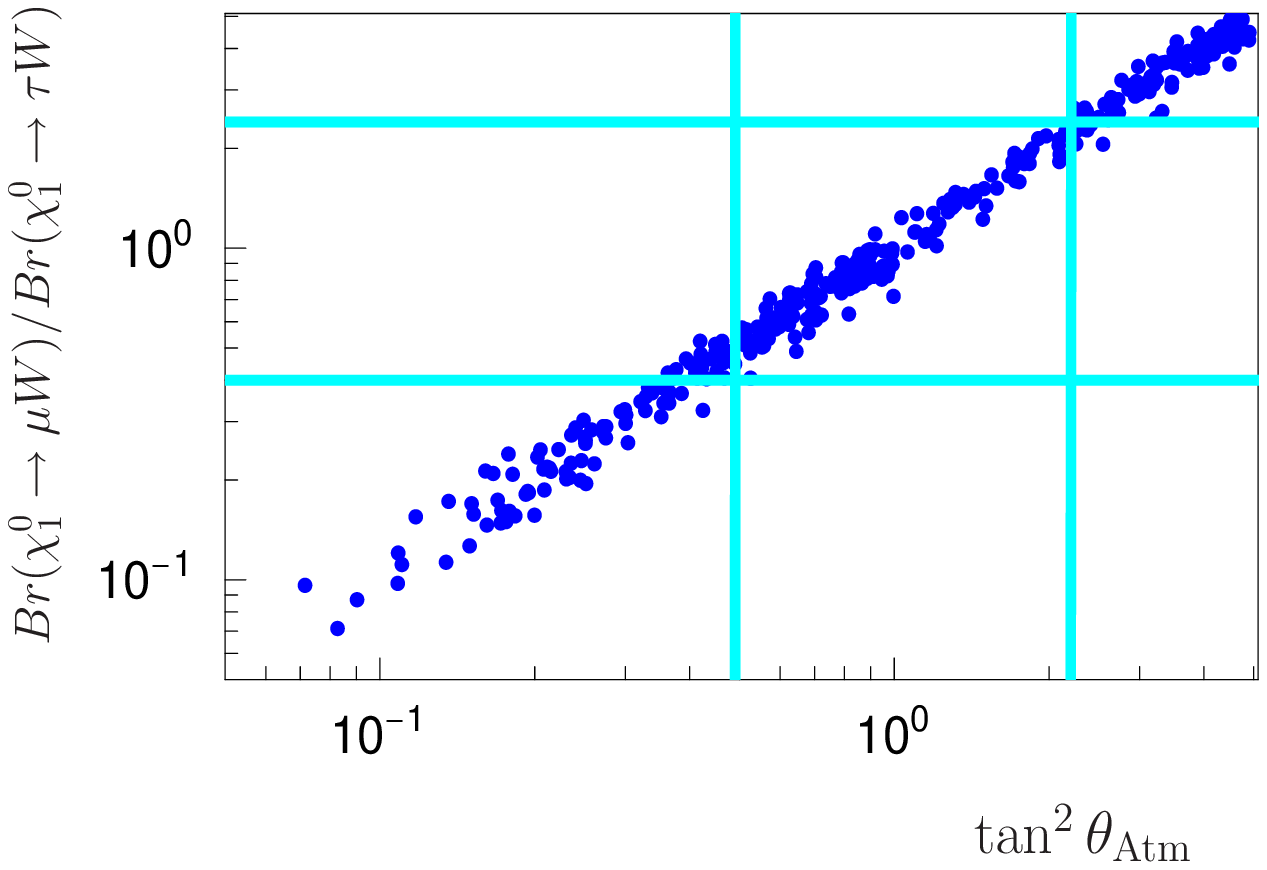}
\includegraphics[width=0.49\textwidth]{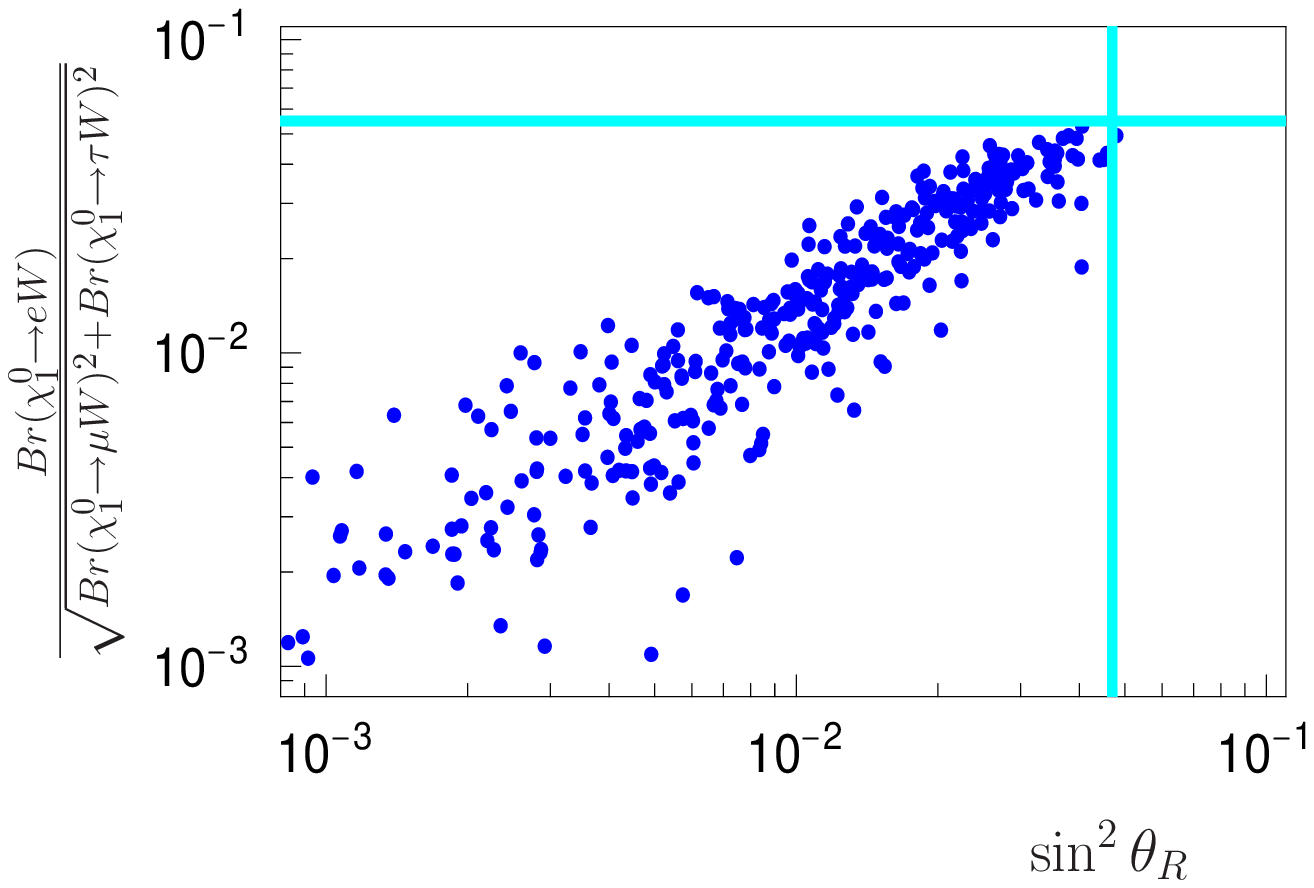}
\end{center}
\vspace{0mm}
\caption{Ratio ${\cal R}_{\mu}= \frac{Br(\chi^0_1\rightarrow \mu W)}
{Br(\chi^0_1\rightarrow \tau W)}$ versus $\tan^2\theta_{Atm} \equiv \tan^2\theta_{23}$ (left) 
and ${\cal R}_e= \frac{Br(\chi^0_1\rightarrow e W)}
{\sqrt{Br(\chi^0_1\rightarrow \mu W)^2+Br(\chi^0_1\rightarrow \tau W)^2}}$ 
versus $\sin^2\theta_R \equiv \sin^2\theta_{13}$ (right) for a singlino LSP. 
Vertical lines are the $3 \sigma$ c.l. allowed experimental 
ranges, horizontal lines the resulting predictions for the fit (c2), 
see text.}
\label{fig:SnglNf3pred}
\end{figure}

Depending on which case, (c1) or (c2), is chosen to fit the 
neutrino data, the corresponding ratios of branching ratios 
are then either sensitive to the atmospheric and reactor or 
the solar angle. This is demonstrated in figures \ref{fig:SnglNf1pred}
and \ref{fig:SnglNf3pred}. Here, figure \ref{fig:SnglNf1pred} 
shows the correlation of ${\cal R}_e= \frac{Br(\chi^0_1\rightarrow e W)}
{\sqrt{Br(\chi^0_1\rightarrow \mu W)^2+Br(\chi^0_1\rightarrow \tau W)^2}}$ 
with $\tan^2\theta_{12}$ for the fit (c1). This result is very 
similar to the one obtained for the fit (c2) and a bino LSP. For 
this reason the nature of the LSP needs to be known, before one 
can decide, whether the measurement of a ratio of branching ratio 
is testing (c1) or (c2).

Figure \ref{fig:SnglNf3pred} shows the dependence of 
${\cal R}_{\mu}= \frac{Br(\chi^0_1\rightarrow \mu W)}
{Br(\chi^0_1\rightarrow \tau W)}$ versus $\tan^2\theta_{23}$ (left) 
and ${\cal R}_e= \frac{Br(\chi^0_1\rightarrow e W)}
{\sqrt{Br(\chi^0_1\rightarrow \mu W)^2+Br(\chi^0_1\rightarrow \tau W)^2}}$ 
versus $\sin^2\theta_{13}$ (right) for a singlino LSP, using the 
neutrino fit (c2). Again one observes that this result is very 
similar to the one obtained for a bino LSP and fit (c1). This 
simply reflects that fact, that neutrino angles can be either fitted 
with ratios of $\epsilon_i$ or with ratios of $\Lambda_i$ and 
singlinos couple mostly proportional to $\epsilon_i$, while binos 
are sensitive to $\Lambda_i$. 

Similar correlations can be found for scenarios with $m_{\chi_1^0} <
m_W$. In this case one must look at three-body decays like
$\chi_1^0 \to q q' l_i$, which are mediated by virtual $W$ bosons.

Let us finally point out that in case the singlino is the
LSP and the bino, as the NLSP, decays with some measurable branching
ratios to $W-l_i$, both $\Lambda_i$ and $\epsilon_i$ ratios could be
reconstructed, which would allow for a much more comprehensive test of
the model.

\section{Phenomenology at low-energy experiments}
\label{sec:lowenergy}

Now we will discuss the phenomenology at low-energy experiments. The presence of a massless particle, the majoron, gives rise to new processes like $\mu \to e J$ and $\mu \to e J \gamma$, providing additional information not present in collider experiments \cite{Hirsch:2009ee}.

Majorons are weakly coupled, thus potentially lead to a decay mode 
for the lightest neutralino which is invisible. As demonstrated in section \ref{sec:LSPdec}, this
new decay mode can be dominant in some regions of parameter space, making the model indistinguishable
from the MSSM with conserved R-parity. Here we extend the argument and look for additional signatures
at low-energy experiments.

Neutralino-Majoron couplings can be calculated from the general coupling ${\chi}_i^0-{\chi}_j^0-P^0_k$, see appendix \ref{sect:app1}. Mixing between the neutralinos and the neutrinos then leads to a coupling $\chi^0_1-\nu_k-J$ which, in the limit $v_R, v_S \gg \epsilon_i, v_{i}$, can be written as \cite{Hirsch:2008ur}
\begin{equation}
\label{eq:majcl-sec}
|O_{\tilde\chi^0_1\nu_kJ}| \simeq  - \frac{{\tilde \epsilon}_k}{V}N_{14} +
\frac{{\tilde v}_{k}}{2 V}(g' N_{11} - g N_{12})
+ \cdots, 
\end{equation}

For the notation see appendix \ref{sect:app1}. In addition to the majoron there is also a rather light singlet 
scalar, called the ``scalar partner'' of the majoron in \cite{Hirsch:2005wd}, 
$S_J$. The lightest neutralino has a coupling $O_{\tilde\chi^0_1\nu_kS_J}$, 
which is of the same order as $O_{\tilde\chi^0_1\nu_kJ}$. Since $S_J$ 
decays to nearly 100 $\%$ to two majorons, this decay mode contributes 
sizeably to the invisble width of the lightest neutralino, for more 
details see \cite{Hirsch:2008ur}. 

The decays $l_i\to l_j J$ can be calculated from the general 
coupling $\chi^+_i-\chi^-_j-P^0_k$. In the limit of small 
R-parity violating parameters the relevant interaction lagrangian 
for the $l_i - J - l_j$ coupling is given by
\begin{equation}
\mathcal{L}= \bar{l}_i 
\big( O_{LijJ}^{ccp} P_L + O_{RijJ}^{ccp} P_R \big) l_j J
\end{equation}
with
\begin{eqnarray}\label{cpl_llJ}
O_{RijJ}^{ccp} &=& - \frac{i (Y_e)^{jj}}{\sqrt{2} V} \big[ 
                    \frac{v_d v_L^2}{v^2} \delta_{ij} 
  + \frac{1}{\mu^2}(C_1 \Lambda_i \Lambda_j + C_2 \epsilon_i \epsilon_j 
+ C_3 \Lambda_i \epsilon_j + C_4 \epsilon_i \Lambda_j) \big]\nonumber \\
O_{LijJ}^{ccp} &=& \big(O_{RjiJ}^{ccp}\big)^*.
\end{eqnarray}
The $C$ coefficients are different combinations of MSSM parameters
\begin{eqnarray}
C_1 &=& \frac{g^2}{2 \textnormal{Det}_+^2}
(- g^2 v_d v_u^2 - v_d \mu^2 + v_u M_2 \mu) \\
C_2 &=& -2 v_d \hskip10mm
C_3 = - \frac{g^2 v_d v_u}{\textnormal{Det}_+} \hskip10mm
C_4 = 1 - \frac{g^2 v_d v_u}{2 \textnormal{Det}_+} \nonumber
\end{eqnarray}
where $\textnormal{Det}_+$ is the determinant of the MSSM chargino mass matrix
$\textnormal{Det}_+ = M_2 \mu - \frac{1}{2} g^2 v_d v_u$. Eq. \eqref{cpl_llJ} 
shows that one expects large partial widths to majorons, if 
$v_R$ is low. 

For a charged lepton $l_i$, with polarization vector $\vec P_i$, the decay
$l_i \rightarrow l_j J$ has a differential decay width given by
\begin{eqnarray}\label{decwid}
\frac{d\Gamma (l_i \rightarrow l_j J)}{d \cos \theta} &=& 
             \frac{m_i^2 - m_j^2}{64 \pi m_i^3} 
   \big[ |O_{LijJ}^{ccp}|^2 \big( m_i^2 + m_j^2 \pm (m_i^2 - m_j^2) 
           P_i \cos \theta \big) \nonumber \\
&& + |O_{RijJ}^{ccp}|^2 \big( m_i^2 + m_j^2 \mp (m_i^2 - m_j^2) 
        P_i \cos \theta \big) \\
&& + 4 m_i m_j Re( {O_{LijJ}^{ccp}}^* O_{RijJ}^{ccp} ) \big] \nonumber
\label{eq:dgampol}
\end{eqnarray}
where $\theta$ is the angle between the polarization vector $\vec P_i$ 
and the momentum $\vec p_j$ of the charged lepton in the final state, 
and $P_i = |\vec P_i|$ is the polarization degree of the decaying 
charged lepton.

\noindent
In the limit $m_j \simeq 0$ the expression \eqref{decwid} simplifies to
\begin{equation}
\frac{d\Gamma (l_i \rightarrow l_j J)}{d \cos \theta} 
= \frac{m_i}{64 \pi} |O_{LijJ}^{ccp}|^2 \big( 1 \pm P_i \cos \theta \big)
\end{equation}
since $|O_{RijJ}^{ccp}|^2 \propto ( Y_e^{jj} )^2 \propto m_j^2$. The 
angular distribution of the majoron emitting lepton decay is thus very 
similar to the standard model muon decay \cite{Amsler:2008zzb}, 
up to corrections of the order $(m_j/m_i)^2$, which are negligible in 
practice. 

\begin{figure}
\begin{center}
\vspace{5mm}
\includegraphics[width=0.49\textwidth]{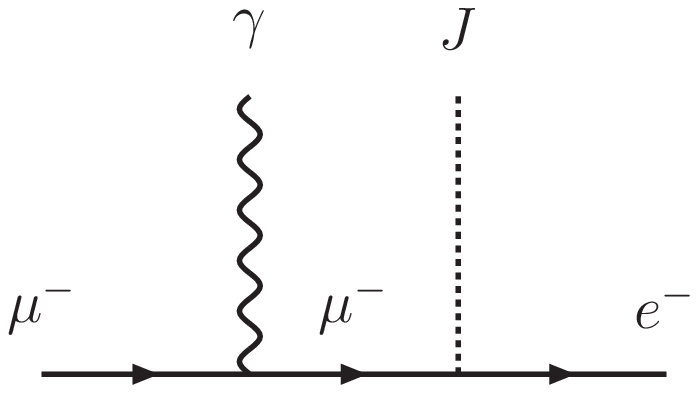}
\includegraphics[width=0.49\textwidth]{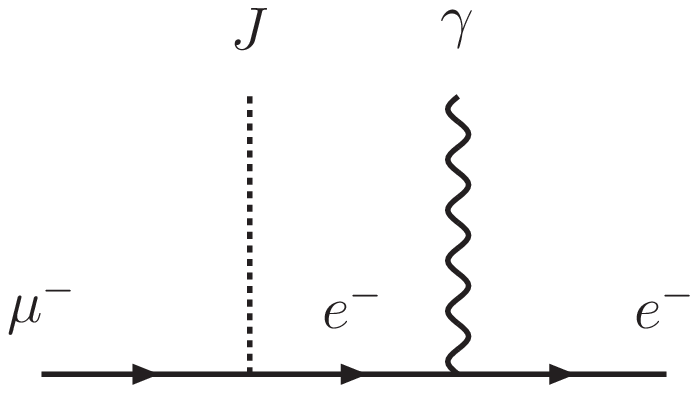}
\end{center}
\caption{Feynman diagrams for the decay $\mu \to e J \gamma$. 
As in the standard model radiative decay $\mu \to e {\bar\nu}\nu\gamma$ 
these diagrams contain an infrared divergence for $m_{\gamma}=0$, see 
text.}
\label{fig:feynG}
\end{figure}

We next consider the decay $\mu \to e J \gamma$ which might be more
interesting due to the existent experiments looking for $\mu \to
e \gamma$\footnote{Formulas for the radiative majoron decays of the
$\tau$ can be found from straightforward replacements.}. It is induced
by the Feynman diagrams shown in figure \ref{fig:feynG}.

In the approximation $m_e \simeq 0$ the partial decay width for the 
process $\mu \to e J \gamma$ can be written as
\begin{equation}
\Gamma(\mu \to e J \gamma) = \frac{\alpha }{64 \pi^2} 
              |O_{L\mu eJ}^{ccp}|^2 m_\mu {\cal I}(x_{min},y_{min})
\end{equation}
where ${\cal I}(x_{min},y_{min})$ is a phase space integral given by
\begin{equation}
{\cal I}(x_{min},y_{min}) = \int dx dy f(x,y) = 
\int dx dy \frac{(x-1)(2-xy-y)}{y^2(1-x-y)},
\end{equation}
the dimensionless parameters $x$, $y$ are defined as usual
\begin{equation}
x = \frac{2 E_e}{m_\mu} \quad , \quad y = \frac{2 E_\gamma}{m_\mu}
\end{equation}
and $x_{min}$ and $y_{min}$ are the minimal electron and photon 
energies measured in a given experiment. 

Note that the integral ${\cal I}(x_{min},y_{min})$ diverges for $y_{min}=0$. 
This infrared divergence is well-known from the standard model 
radiative decay $\mu \to e {\bar\nu}\nu\gamma$
\cite{Behrends:1955mb,Yennie:1961ad,Weinberg:1965nx,Ross:1972yq}, and can be taken 
care off in the standard way by introducing a non-zero photon 
mass $m_{\gamma}$. Note that in the limit $m_e=0$ there also appears 
a colinear divergence, just as in the SM radiative decay. Since in any 
practical experiment there is a minimum measurable photon energy, 
$y_{min}$, as well as a minimum measurable photon-electron angle 
($\theta_{e\gamma}$), neither divergence affects us in practice. 
We simply integrate 
from the minimum value of $y$ up to $y_{max}$ when estimating the 
experimental sensitiviy of $Br(\mu \to e J \gamma)$ on the majoron 
coupling.

In the calculation of the integral ${\cal I}(x_{min},y_{min})$ one 
has to take into account not only the experimental cuts applied 
to the variables $x$ and $y$, but also the experimental cut for 
the angle between the directions of electron and photon. This 
angle is fixed for kinematical reasons to
\begin{equation}
\cos \theta_{e \gamma} = 1 + \frac{2-2(x+y)}{xy}.
\label{eq:ctheta}
\end{equation}
This relation restricts $x_{max}$ to be $x_{max}\le 1$ as a 
function of $y$ (and vice versa) and to $x_{max}< 1$ for 
$\cos \theta_{e \gamma} > -1$. 

Using the formula for $\Gamma(\mu \to e J)$, in the approximation 
$m_e \simeq 0$,
\begin{equation}
\Gamma(\mu \to e J) = \frac{m_\mu}{32 \pi} |O_{L\mu eJ}^{ccp}|^2
\end{equation}
one finds a very simple relation between the two branching ratios
\begin{equation}
Br(\mu \to e J \gamma) = \frac{\alpha}{2 \pi} {\cal I}(x_{min},y_{min}) 
                          Br(\mu \to e J).
\label{eq:relBr}
\end{equation}
We will use eq. \eqref{eq:relBr} in section \ref{sec:exp} when 
we discuss the relative merits of the two different measurements.

\subsection{Numerical results}
\label{sec:nummueJ}

As in the collider phenomenology section, all numerical results shown in this section have been obtained using the 
program package SPheno \cite{Porod:2003um}, with the required extension to include $\widehat\nu^c$, $\widehat S$ and $\widehat\Phi$. 
The \rpv parameters are chosen in such a way that solar and 
atmospheric neutrino data \cite{Schwetz:2008er} are fitted correctly.
In the plots shown below we use $\Lambda_i$ for the atmospheric 
scale and $\epsilon_i$ for the solar scale.

\begin{figure}
\begin{center}
\vspace{5mm}
\includegraphics[width=0.49\textwidth]{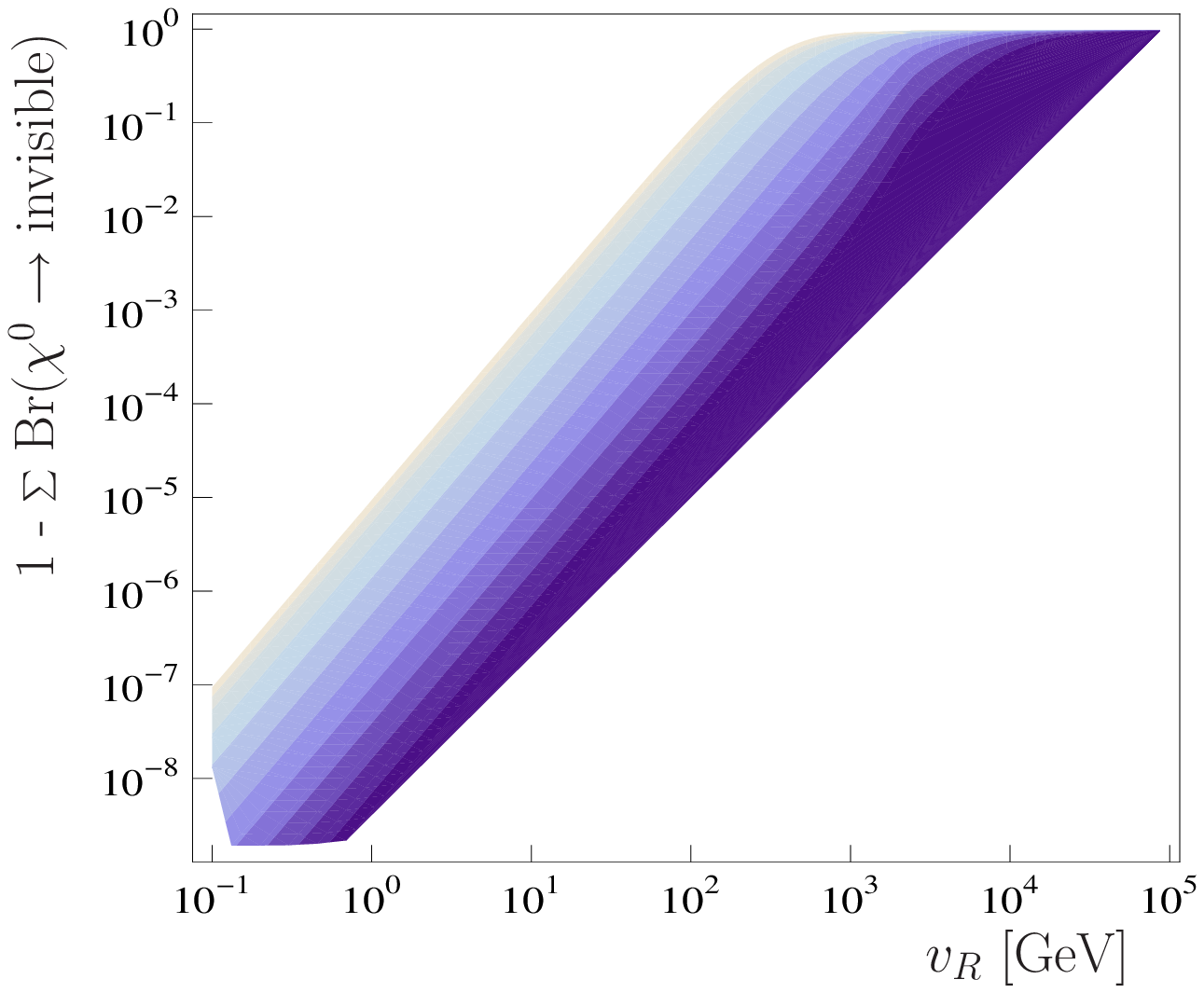}
\includegraphics[width=0.49\textwidth]{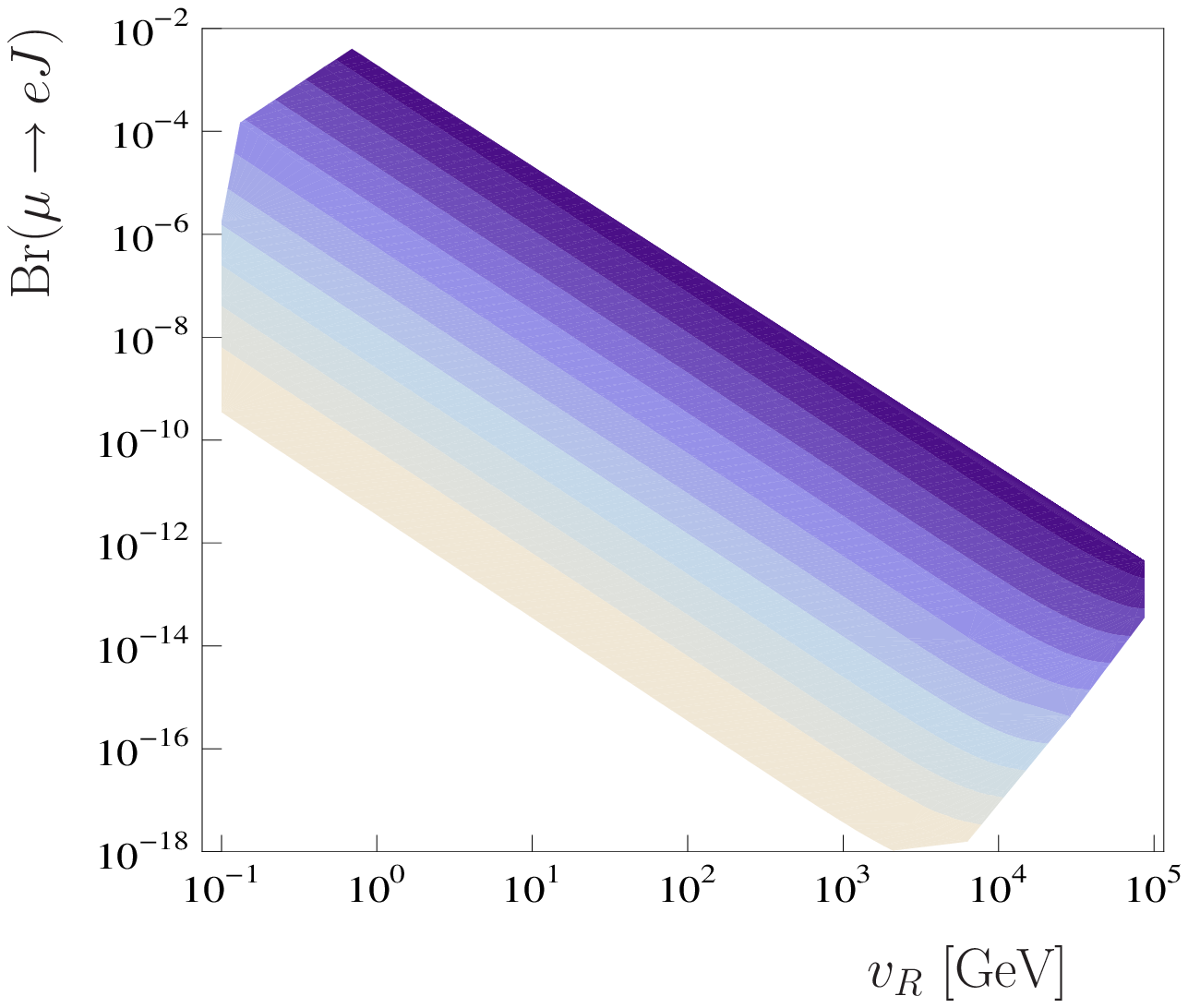}
\end{center}

\caption{Branching ratios for visible lightest neutralino decay (left) 
and branching ratio $Br(\mu\to e J)$ (right) versus $v_R$ in GeV 
for a number of different choice of $v_\Phi$ between [$1,10^2$] TeV 
indicated by the different colors. Darker colors indicate larger 
$v_\Phi$ in a logarithmic scale. mSugra parameters defined in the text. 
There is very little dependence on the actual mSugra parameters, however, 
see discussion and figure \ref{fig:VisvMuEj}.}

\label{fig:vis}
\end{figure}

As shown previously \cite{Hirsch:2008ur} if the lightest neutralino is
mainly a bino, the decay to majoron plus neutrino is dominant if $v_R$
is low. This was shown in figure \ref{fig:Inv_vr} and is demonstrated
again for a bino LSP in figure \ref{fig:vis}, to the left, with the
same sample point (mSugra parameters $m_0=280$~GeV, $m_{1/2}=250$~GeV,
$\tan\beta=10$, $A_0=-500$~GeV and sgn$(\mu)=+$).  We stress that this
result is independent of the choice of mSugra parameters to a large
degree \cite{Hirsch:2006di}.  A scan over $v_\Phi$ has been performed
in this plot, varying $v_\Phi$ in the huge interval [$1,10^2$]
TeV. Large values of $v_\Phi$ lead to small values of the constant $c$
in the neutrino mass matrix, see eqs. \eqref{def:abc}
and \eqref{def:detmh}. Small $c$ require, for constant neutrino
masses, large values of $\epsilon_i$, which in turn lead to a large
invisible width of the neutralino. The largest values of $v_\Phi$
(dark areas) therefore lead to the smallest visible neutralino decay
branching ratios shown in figure \ref{fig:vis}. Let us mention that
this region of parameter space requires small values for $h_0$ in
order to keep $\mu$ at the weak scale.

Figure \ref{fig:vis}, to the right, shows the branching ratio 
$Br(\mu\to e J)$ as a function of $v_R$ for different values of 
$v_\Phi$. All parameters have been fixed to the same values as 
shown in the left figure. As the figure demonstrates, small values 
of $v_R$ (and large values of $v_\Phi$) lead to large values of 
$Br(\mu\to e J)$. This agrees with the analytic expectation, 
compare to equation \eqref{cpl_llJ}. 

\begin{figure}
\begin{center}
\vspace{5mm}
\includegraphics[width=0.49\textwidth]{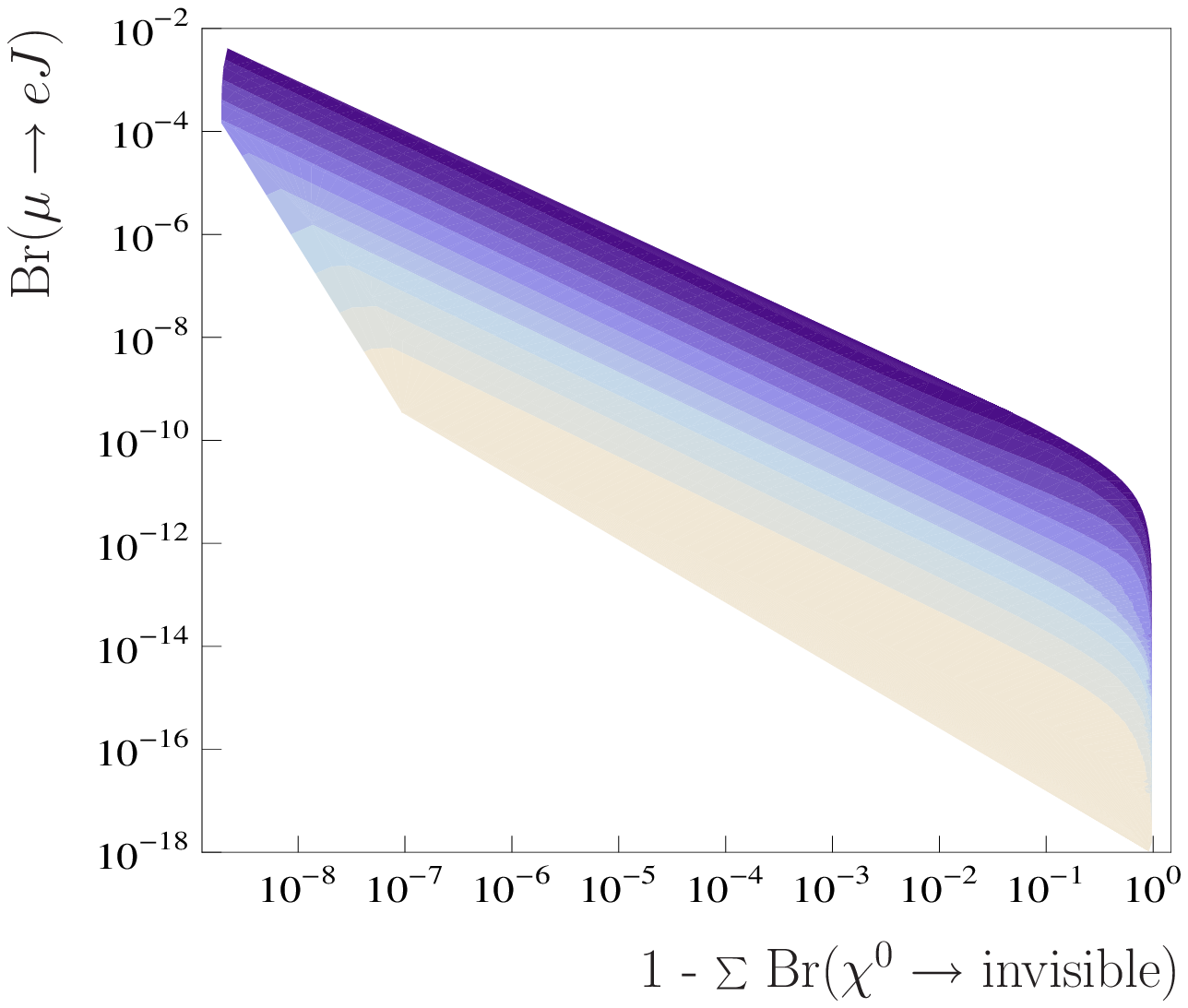}
\includegraphics[width=0.49\textwidth]{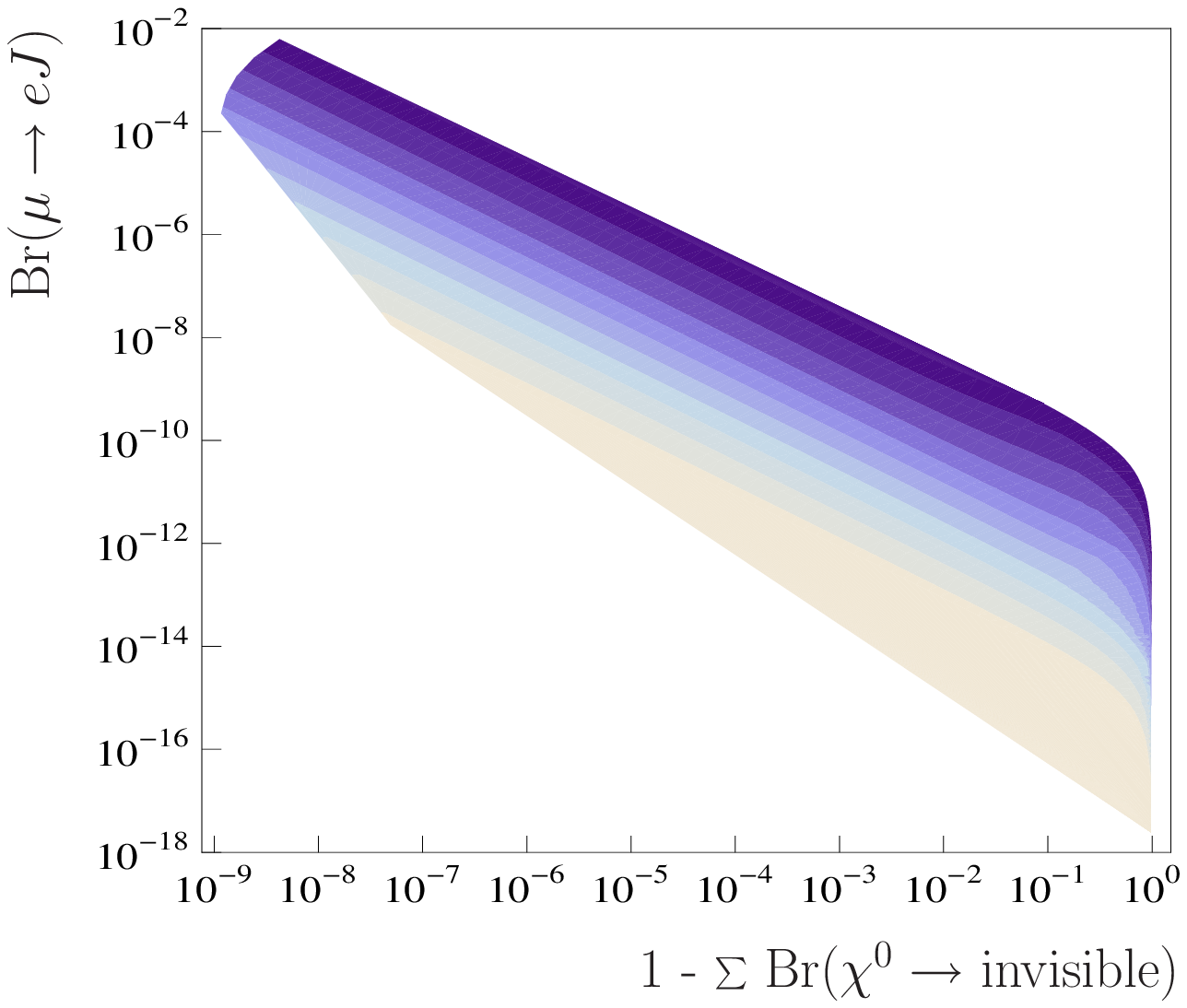}
\end{center}

\caption{Branching ratios for visible lightest neutralino decay versus  
branching ratio $Br(\mu\to e J)$ for two mSugra points, for various 
choices of $v_\Phi$, see figure \ref{fig:vis}. To the left, same 
mSugra parameters as figure \ref{fig:vis}, to the right SPS1a'.}

\label{fig:VisvMuEj}
\end{figure}

Our main result in this section is shown in figure \ref{fig:VisvMuEj}. In this 
figure we show $Br(\mu\to eJ)$ versus the sum of all branching 
ratios of neutralino decays leading to at least one visible particle 
in the final state for two different choices of mSugra parameters.
The similarity of the two plots shows that our result is only 
weakly dependent on the true values of mSugra parameters. We have 
checked this fact also by repeating the calculation for other 
mSugra points, although we do not show plots here.  
As expected $Br(\mu\to e J)$ anticorrelates with the visible 
bino decay branching fraction and thus probes a complementary 
part in the supersymmetric parameter space. An upper bound on 
$Br(\mu\to e J)$ will constrain the maximum branching ratio 
for invisible neutralino decay, thus probing the part of parameter 
space where spontaneous R-parity breaking is most easily confused 
with {\em conserved} R-parity at accelerators.

We have checked the points shown in the plots for various 
phenomenological constraints. LEP bounds are trivially fulfilled 
by $v_{i} < v_R$. Double beta decay bounds on $g_{\nu\nu J}$ 
\footnote{We will use the symbol 
$g$ when discussing experimental bounds, to differentiate from the 
model dependent couplings $O^{ccp}_L$ and $O^{ccp}_R$ defined previously in 
this section.} are of 
the order of $10^{-4}$ \cite{Arnold:2006sd} and, since the coupling 
$g_{\nu\nu J}$ is suppressed by two powers of R-parity violating 
parameters, are easily satisfied in our model. More interesting 
is the astrophysical limit on $g_{eeJ}$. 
Ref. \cite{Raffelt:1990yz} quotes a bound of $g_{eeJ} \le 3 \cdot 10^{-13}$. 
Although this bound is derived from the coupling of the majoron to two 
electrons, thus constraining actually the products $v_{L_e}^2$, 
$\epsilon_e^2$ and $\Lambda_e^2$, whereas $Br(\mu\to e J)$ is proportional 
to $\epsilon_e\epsilon_{\mu}$ and $\Lambda_e\Lambda_{\mu}$, it still 
leads to a (weak) constraint on $Br(\mu\to eJ)$, since neutrino physics 
shows that two leptonic mixing angles are large. This requires that either 
$\epsilon_e\sim \epsilon_{\mu}$ {\em or} $\Lambda_e \sim \Lambda_{\mu}$. 
For the case studied in our plots, where $\epsilon_i$ generate the solar 
scale, $\tan^2\theta_{12} \simeq 1/2$ requires $\epsilon_e\sim 
\epsilon_{\mu}$. Numerically we then find that $g_{eeJ} \le 3 \cdot 10^{-13}$ 
corresponds to an upper bound on $Br(\mu \to e J)$ of very roughly 
$Br(\mu \to e J) \lesssim {\rm (few)} \cdot 10^{-5}$. In case neutrino 
data is fitted with $\vec\epsilon$ for the atmospheric scale, the 
corresponding bound is considerably weaker.

\subsection{Experimental constraints and $Br(\mu\to eJ\gamma)$}
\label{sec:exp}

The Particle Data Group \cite{Amsler:2008zzb} cites \cite{Jodidio:1986mz} 
with an upper limit on the branching ratio of $Br(\mu\to e X^0)$ 
$\le 2.6 \times 10^{-6}$, where $X^0$ is a scalar boson called the 
familon. This constraint does not apply to the majoron we consider 
here, since it is derived from the decay of polarized muons in a 
direction opposite to the direction of polarization. The authors of 
\cite{Jodidio:1986mz} concentrated on this region, since it minimizes 
events from standard model $\beta$-decay. As shown in eq. \eqref{eq:dgampol}, 
the majoron emitting decay has a very similar angular distribution 
as the standard model decay, with the signal approaching zero in the 
data sample analyzed by \cite{Jodidio:1986mz}. Nevertheless, from the 
spin processed data shown in figure (7) of \cite{Jodidio:1986mz}, which 
seems to be in good agreement with the SM prediction, it should 
in principle be possible to extract a limit on $Br(\mu\to e J)$. 
From this figure we estimate very roughly that this limit should 
be about one order of magnitude less stringent than the one for 
familon decay. For a better estimate a re-analysis of this data, 
including systematic errors, would be necessary.

Ref. \cite{Picciotto:1987pp} searched for majorons in the 
decay of $\pi\to e \nu J$, deriving a limit of $Br(\pi\to e \nu J)$ 
$\le 4 \cdot 10^{-6}$. Since the experimental cuts used in this 
paper \cite{Picciotto:1987pp} are designed to reduce the standard 
model background from the decay chain $\pi\to\mu \to e$, the 
contribution from on-shell muons is reduced by about five orders 
of magnitude. The limit then essentially is a limit on the 
majoron-neutrino-neutrino coupling, $g_{\nu\nu J}$, leaving only a
very weak constraint on the coupling $g_{\mu e J}$. 
Also an analysis searching for $Br(\mu\to eJ\gamma)$ has been published 
previously \cite{Goldman:1987hy}. From a total data sample of 
$8.15 \cdot 10^{11}$ stopped muons over the live time of the 
experiment \cite{Goldman:1987hy} derived a limit on $Br(\mu\to eJ\gamma)$ 
of the order of $Br(\mu\to eJ\gamma) \le 1.3 \cdot 10^{-9}$. 
For the cuts used in this analysis, we calculate ${\cal I} \simeq 10^{-3}$. 
Thus, see eq. \eqref{eq:relBr}, this limit translates into only a 
rather weak bound $Br(\mu\to eJ) \le 1.1\cdot 10^{-3}$.

Currently, the MEG experiment \cite{meg} is the most advanced experiment
investigating muon decay. With a muon stopping rate of $(0.3-1)\cdot 10^8$ 
per second, it expects a total of the order of $10^{15}$ muons over 
the expected live time of the experiment. An analysis of {\em electron 
only} events near the endpoint should therefore allow, in principle, 
to improve the existing limits on $Br(\mu\to eJ)$ by an estimated ($2-3$) 
orders of magnitude, if systematic errors can be kept under control. 

However, the MEG experiment, as it is designed to search for 
$Br(\mu\to e\gamma)$, uses a trigger that requires a photon in the 
event with a minimum energy of $E_{\gamma}^{min}\ge 45$ MeV. Data for muon decays without a
photon in the final state are only kept for background measurements and calibration purposes.
Therefore, until these data are released, it is interesting to constrain the majoron-charged-lepton coupling 
via searching for $\Gamma(\mu\to eJ\gamma)$. See appendix \ref{sect:app2} for a comparison between
the constraining power of $\Gamma(\mu\to eJ)$ and $\Gamma(\mu\to eJ\gamma)$.

\begin{figure}
\begin{center}
\vspace{5mm}
\includegraphics[width=0.49\textwidth]{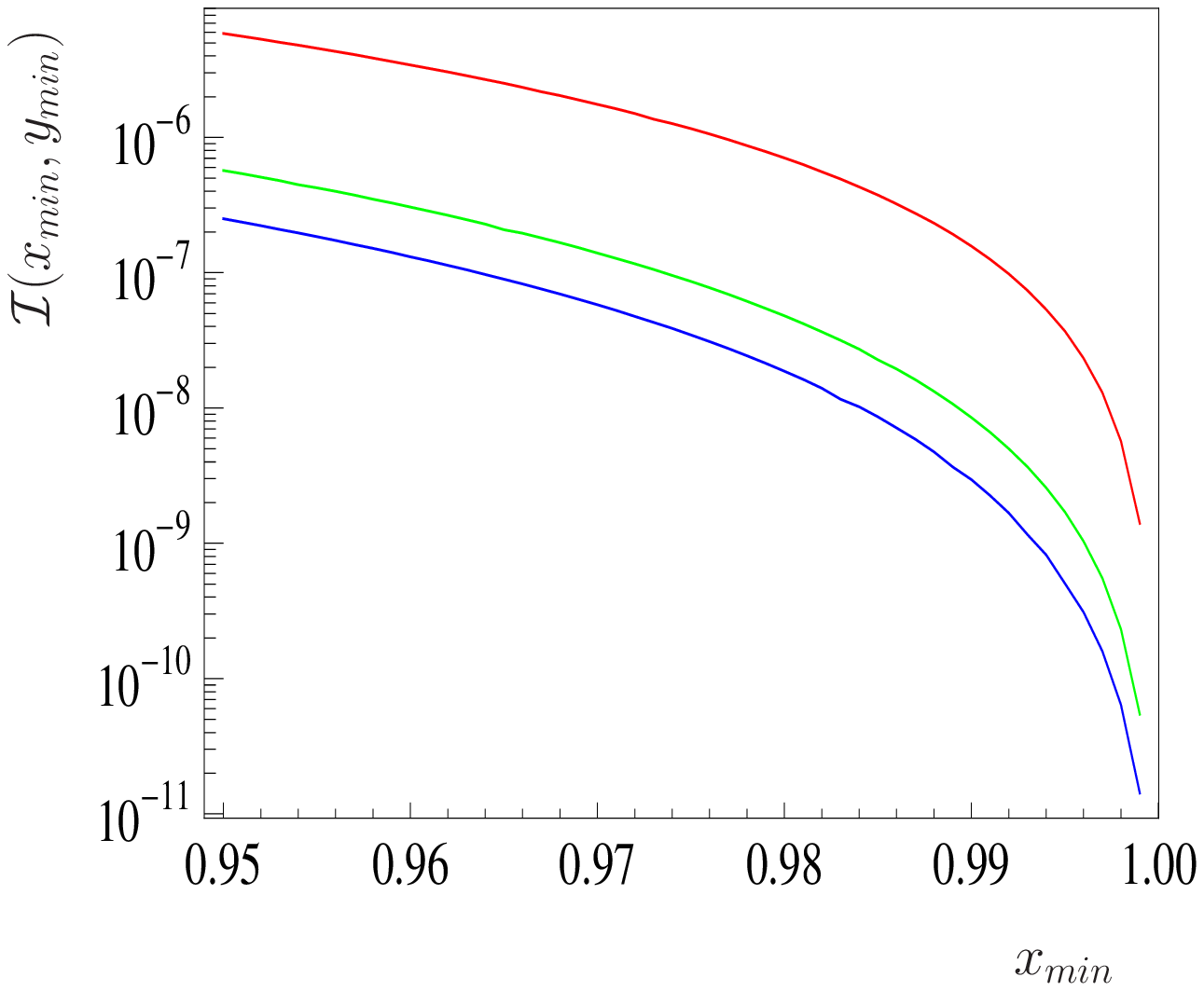}
\includegraphics[width=0.49\textwidth]{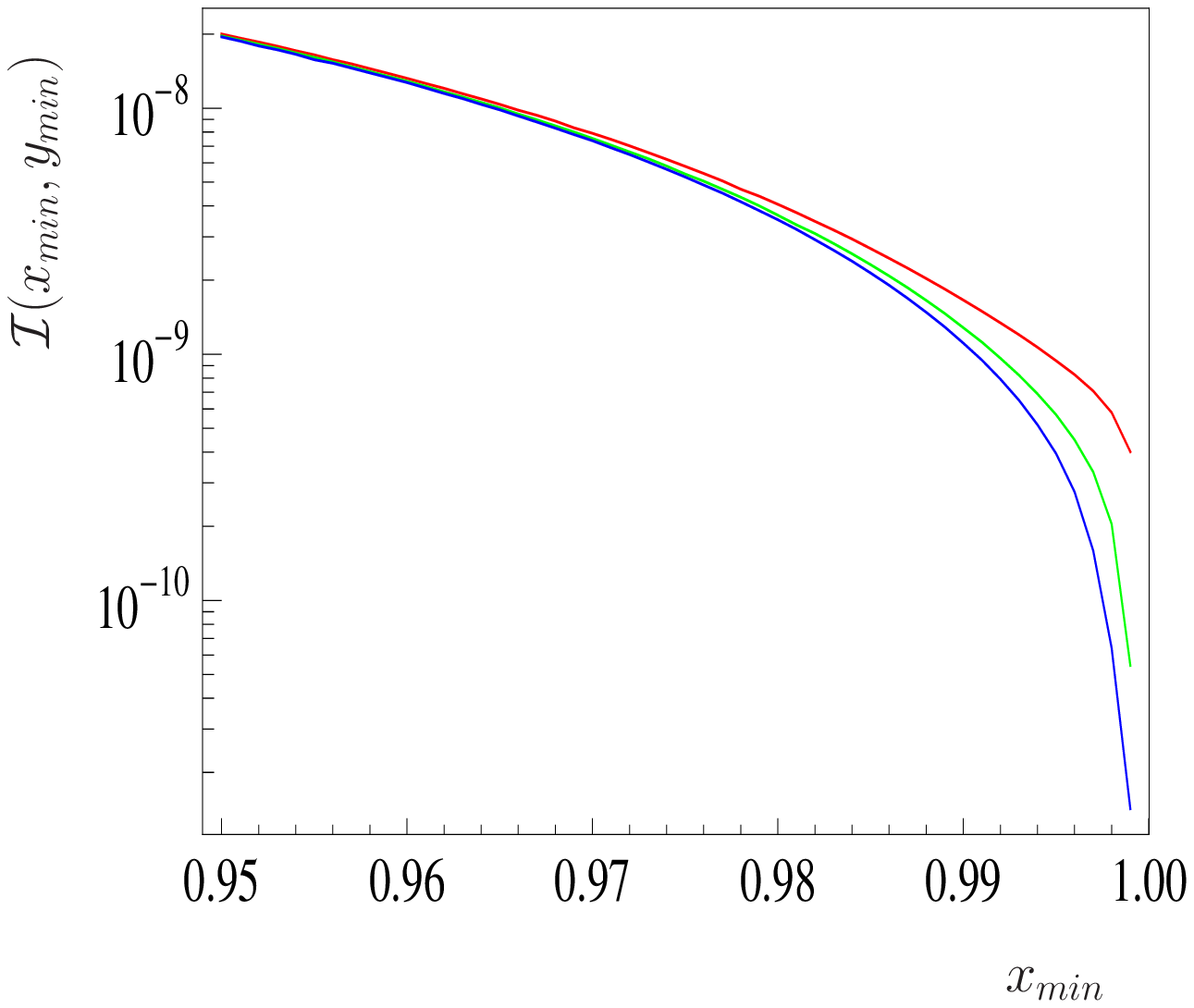}
\end{center}

\caption{The phase space integral for the decay $\mu\to e J \gamma$ 
as a function of $x_{min}$ for three different values of $y_{min}= 
0.95, 0.99, 0.995$ from top to bottom and for two different values of 
$\cos\theta_{e\gamma}$. To the left $\cos\theta_{e\gamma}=-0.99$, 
to the right $\cos\theta_{e\gamma}=-0.99997$.}

\label{fig:int}
\end{figure}

Figure \ref{fig:int} shows the value of the phase space integral 
${\cal I}(x_{min},y_{min})$ as a function of $x_{min}$ for three 
different values of $y_{min}$ and for two choices of $\cos\theta_{e\gamma}$. 
The MEG proposal describes the cuts used in the search for 
$\mu\to e \gamma$ as $x_{min}\ge 0.995$, $y_{min}\ge 0.99 $  
and $|\pi - \theta_{e\gamma}| \le$ 8.4 mrad. For these values we find 
a value of ${\cal I} \simeq 6 \cdot 10^{-10}$. A limit for 
$Br(\mu\to e \gamma)$ of $Br(\mu\to e \gamma)$ $\le 10^{-13}$ then 
translates into a limit of Br$(\mu\to e J) \le 0.14$, obviously 
not competitive. To improve upon this bound, it is necessary to 
relax the cuts. For example, relaxing the cut on the opening angle 
to $\cos\theta_{e\gamma}=-0.99$, the value of the integral increases 
by more than 3 orders of magnitude for $x_{min}=y_{min}\ge 0.95$.

On the other hand, such a change in the analysis is prone to 
induce background events, which the MEG cuts were designed for 
to avoid. The MEG proposal discusses as the two most important 
sources of background: (a) Prompt events from the standard model 
radiative decay $\mu \to e \nu {\bar\nu} \gamma$; and (b) accidental 
background from muon annihilation in flight. For the current 
experimental setup the accidental background is larger than the 
prompt background. Certainly, a better timing resolution of the 
experiment would be required to reduce this background. For the 
prompt background we estimate, using the formulas of \cite{Kuno:1999jp}, 
that for a total of $10^{13}$ muon events, one background event 
from the radiative decay will enter the analysis window for 
$x_{min}=y_{min} \simeq 0.96$ for the current cut on $\cos\theta_{e\gamma}$.

A further relaxation of the cuts can lead, in principle, to much 
larger values for ${\cal I}(x_{min},y_{min})$. However, the search 
for Br$(\mu\to e J\gamma)$ than necessarily is no longer background 
free. Since all the events from $\mu\to e J\gamma$ lie along the line 
of $\cos\theta_{e\gamma}$ defined by eq. \eqref{eq:ctheta}, whereas 
events from the SM radiative mode fill all of the $\cos\theta_{e\gamma}$
space, such a strategy might be advantageous, given a large enough 
data sample. 

Before closing this section, we mention that tau decays with majoron 
emission are less interesting phenomenologically for two reasons. 
First the existing experimental limits are much weaker for taus 
\cite{Baltrusaitis:1985fh} $Br(\tau \to \mu J) \le 2.3$ $\%$  and 
$Br(\tau \to e J) \le 0.73$ $\%$. And, second, although the coupling
$\tau-\mu-J$ is larger than the coupling $\mu-e-J$ by a factor
$m_{\tau}/m_{\mu}$, the total width of the tau is much larger than the
width of the muon, thus the resulting theoretical predictions for tau
branching ratios to majorons are actually smaller than for the muon by
a factor of approximately $10^4$. Finally, we also point out that
$\mu$ -- $e$ conversion in nuclei is another important observable,
intimately linked to those studied here. Experiments like COMET and
PRISM/PRIME \cite{Kuno:2008zz}, still in the research and design
stage, will have sensitivities for conversion rates as low as
$10^{-16}-10^{-18}$, and thus very promising results are expected.

\section{Summary}

We have studied the theory and phenomenology of a supersymmetric 
model in which neutrino oscillation data is explained by spontaneous 
R-parity violation. This setup provides a testable framework that explains
neutrino masses and solves some of the problems of b-\rpv.

The model has many distinctive signatures. From the collider point of view, the LSP decay clearly
distinguishes this model from the standard MSSM, and provides a tool to test the model.

We have concentrated the discussion on the case that 
the LSP is either a bino, like in a typical mSugra point, or a singlino 
state, novel to the current model. We have worked out the most important 
phenomenological signals of the model and how it might be 
distinguished from the well-studied case of the MSSM, as well as from 
a model in which the violation of R-parity is explicit. 

There are regions in parameter space, where $\tilde\chi^0$ decays 
invisibly with branching ratios close to 100 \%, despite the smallness 
of neutrino masses. In this limit, spontaneous violation of R-parity 
can resemble the MSSM with conserved R-parity at the LHC and the 
experimentalists would have to search for the very rare visible 
decay channels to establish the R-parity indeed is broken.

The perhaps most important test of the model as the origin of the observed 
neutrino masses comes from measurements of ratios of branching ratios 
to $W$-boson and charged lepton final states. Ratios of these decays are 
always related to measured neutrino angles. If SUSY has a spectrum 
light enough to be produced at the LHC, the spontaneous model of 
R-parity violation is therefore potentially testable.

However, as stated above, invisible decays might be dominant and do not allow
for these signals to be seen at the LHC. In that case one needs to study additional
observables that are enhanced in the region of parameter space where the branching ratio
for majoron final states approaches 100 \%. These observables are provided by low-energy
experiments.

We have calculated branching ratios for exotic muon and tau 
decays involving majorons in the final state. Branching ratios 
can be measurably large, if the scale of lepton number breaking 
is low. Note that this is the region of parameter space where a bino
LSP decays mainly to invisible states, and thus the combination between
collider and low-energy experiments might allow to distinguish s-\rpv from
the MSSM with conserved R-parity.

This result is independent of the absolute value of the 
neutrino mass. The lowest possible values of $v_R$ (at large 
values of $v_\Phi$) are already explored by the existing limit 
on $Br(\mu\to e J)$. 

We have briefly discussed the status of experimental limits. It 
will not be an easy task to improve the current numbers in future 
experiments. While MEG \cite{meg} certainly has a high number of 
muon events in the detector, a search for $Br(\mu\to e J\gamma)$ 
instead of $Br(\mu\to e J)$ suffers from a small value of the 
available phase space integral, given current MEG cuts. An 
improvement will only be possible, if a dedicated search by the 
experimentalists is carried out. Nevertheless, we believe this 
is a worthwhile undertaking, since measuring a finite value for 
$Br(\mu\to e J)$ will establish that R-parity is broken in a 
region of SUSY parameter space complementary to that probed by 
accelerator searches.

\chapter{$\mu \nu$SSM}
\label{chap:munuSSM}

The $\mu \nu$SSM was recently proposed as an economical way to introduce neutrino masses in the MSSM. This R-parity breaking model introduces just one singlet to address the $\mu$ and $\epsilon$ problems. In that sense, the $\mu \nu$SSM is a simpler scheme than s-\rpv, where three additional singlets are used. The main difference is, however, the explicit breaking of R-parity. This implies a spectrum without majoron.

\section{Motivation}

The superpotential of the MSSM contains a mass term for the Higgs
superfields, $\mu {\widehat H_d}{\widehat H_u}$. For phenomenological
reasons this parameter $\mu$ must be of the order of the electroweak
scale. However, if there is a larger scale in the theory, like the
grand unification or Planck scales, the natural value of $\mu$ lies at
this large scale. This naturalness problem is, in short, the
$\mu$-problem of the MSSM \cite{Kim:1983dt}.  The Next-to-Minimal SSM
(NMSSM) provides a solution \cite{Barbieri:1982eh,Nilles:1982dy} at
the cost of introducing a new singlet field and imposing the scale
invariance of the superpotential by assuming a discrete $Z_3$
symmetry. The VEV of the singlet produces the $\mu$ term, once the
electroweak symmetry is broken (for recent reviews on the NMSSM
see \cite{Maniatis:2009re,Ellwanger:2009dp}, whereas for some recent
papers on the phenomenology of the NMSSM, see for example
\cite{Djouadi:2008uw,Djouadi:2008uj,Ellwanger:2005uu,Ellwanger:2005dv} 
and references therein.).

The $\mu\nu$SSM \cite{LopezFogliani:2005yw} proposes to use the same 
singlet superfield(s) which generate the $\mu$ term to also generate 
Dirac mass terms for the observed left-handed neutrinos. Lepton number 
in this approach is broken explicitly by cubic terms coupling only 
singlets. $R_p$ is broken also and Majorana neutrino masses are generated 
once electroweak symmetry is broken. Two recent papers have studied the 
$\mu\nu$SSM in more detail. In \cite{Escudero:2008jg} the authors analyze 
the parameter space of the $\mu\nu$SSM, putting special emphasis on 
constraints arising from correct electroweak symmetry breaking, avoiding 
tachyonic states and Landau poles in the parameters. The phenomenology 
of the $\mu\nu$SSM has been studied also in \cite{Ghosh:2008yh}. In this 
paper formulas for tree-level neutrino masses are given and decays of 
a neutralino LSP to two-body ($W$-lepton) final states have been 
calculated \cite{Ghosh:2008yh}.

There have been recent works expanding our knowledge on the
$\mu \nu$SSM. The authors of reference \cite{Choi:2009ng} investigate
indirect signals coming from a decaying gravitino in the
$\mu \nu$SSM. This is one of the usual dark matter candidates in \rpv
SUSY. Reference \cite{Fidalgo:2009dm} studies spontaneous CP violation
in the $\mu \nu$SSM, showing that it can indeed happen. This recent
contribution also addresses the seesaw mechanism in the $\mu \nu$SSM,
describing its origin and how it can accommodate the observed neutrino
masses and mixing angles. Finally, reference \cite{Ghosh:2010zi} makes
a detailed computation of the 1-loop corrections to neutrino masses
and mixing angles in the $\mu \nu$SSM.

In the present chapter, we study the collider phenomenology of the
$\mu\nu$SSM \cite{Bartl:2009an}, extending previous works
\cite{LopezFogliani:2005yw,Escudero:2008jg,Ghosh:2008yh}.
In contrast to \cite{Ghosh:2008yh} all kinematically allowed final
states are considered. This does not only cover scenarios where
two-body decays are important, but also those where three-body decays
are dominant. We consider two different variations of the model. In
its simplest form the $\mu\nu$SSM contains only one new singlet. This
version produces one neutrino mass at tree-level, while the remaining
two neutrinos receive mass at the loop-level. This feature is very
similar to bilinear R-parity breaking, although as discussed below,
the relative importance of the various loops is different for the
explicit bilinear model and the $\mu\nu$SSM. As in the explicit
bilinear model neutrino angles restrict the allowed range of \rpv
parameters and correlations between certain ratios of decay branching
ratios of the LSP and neutrino angles appear. In the second version we
allow for $n$ singlets.  Neutrino masses can then be fitted with
tree-level physics only.  However, many of the features of the one
generation model remain at least qualitatively true also in the $n$
singlet variants. LSP decays (for a bino or a singlino LSP) can be
correlated with either the solar or atmospheric angle, thus allowing
to construct explicit tests of the model for the LHC.

Finally, similar proposals have been discussed in the literature. 
\cite{Kitano:1999qb} studied a model in which the NMSSM singlet is 
coupled to (right-handed) singlet neutrino superfields. Effectively 
this leads to a model which is very similar to the NMSSM with explicit 
bilinear terms, as studied for example also in \cite{Abada:2006qn}. In 
\cite{Chemtob:2006ur} the authors propose a model similar to the 
$\mu\nu$SSM, but with only one singlet.

\section{The model}

In this section we introduce the model, work out its most important
properties related to phenomenology and neutrino masses and mixings. 
As explained in the introduction, we will consider the $n$ generations 
case in this section. Approximate formulas are then given for scalar 
masses for the one ($1$) $\hat\nu^c$-model and for neutrino masses 
for the $1$ and $2$ $\hat\nu^c$-model.

\subsection{Superpotential}
\label{subsec:superpot}

The model contains $n$ generations of right-handed neutrino
singlets. The superpotential can be written as
\begin{eqnarray} %
{\cal W} &=& Y_u^{ij} \widehat{Q}_i \widehat{u}_j^c \widehat{H}_u + Y_d^{ij} \widehat{Q}_i \widehat{d}_j \widehat{H}_d + Y_e^{ij} \widehat{L}_i \widehat{e}_j^c \widehat{H}_d \nonumber\\
        & + & Y_{\nu}^{is} \widehat L_i \widehat \nu^c_s \widehat H_u
          - \lambda_s \widehat \nu^c_s \widehat H_d \widehat H_u
 +\frac{1}{3!}\kappa_{stu} \widehat \nu^c_s \widehat \nu^c_t \widehat \nu^c_u\quad.
\label{eq:munuWsuppot}
\end{eqnarray}
The last three terms include the right-handed neutrino superfields,
which additionally play the role of the $\widehat \Phi$ superfield in
the NMSSM \cite{Barbieri:1982eh}, a gauge singlet with respect to the
SM gauge group. The model does not contain any terms with dimensions
of mass, providing a natural solution to the $\mu$-problem of the
MSSM. Like in the NMSSM or s-\rpv, this can be enforced by introducing
a discrete $Z_3$ symmetry. For the associated domain wall problem and its possible solutions we refer the reader to references \cite{Abel:1995wk,Abel:1996cr,Panagiotakopoulos:1998yw}.

Note, that as the number of right-handed neutrino
superfields can be different from $3$ we use the letters $s$, $t$
and $u$ as generation indices for the $\widehat{\nu}^c$ superfields and
reserve the letter $i$, $j$ and $k$ as generation indices for the
usual MSSM matter fields.

The last two terms in \eqref{eq:munuWsuppot} explicitly
break lepton number and thus R-parity giving rise to neutrino
masses. Note that $\kappa_{stu}$ is completely symmetric in all its
indices.  In contrast to other models with R-parity violation, this
model does not need the presence of unnaturally small parameters with
dimensions of mass, like in bilinear R-parity breaking models
\cite{Hirsch:2004he}, and there is no Goldstone boson associated with
the breaking of lepton number
\cite{Chikashige:1980ui,Gelmini:1980re,Aulakh:1982yn}, since 
breaking of $R_p$ is done explicitly.

The absence of the R-parity violating superpotential trilinear couplings can be justified by assuming that they vanish at some high-energy scale. Then, although they are generated at the SUSY scale by RGE running, their values remain negligible \cite{Escudero:2008jg}.

For practical purposes, it is useful to write the superpotential in
the basis where the right-handed neutrinos have a diagonal mass
matrix. Since their masses are induced by the $\kappa$ term in
\eqref{eq:munuWsuppot}, this is equivalent to writing this term including
only diagonal couplings: 

\begin{equation}\label{eq:kappaconv}
\kappa_{stu} \widehat \nu^c_s \widehat \nu^c_t \widehat \nu^c_u 
\qquad\Longrightarrow\qquad \sum_{s=1}^n \kappa_s
(\widehat \nu^c_s)^3
\end{equation}

\subsection{Soft terms}
\label{subsec:soft}

The soft SUSY breaking terms of the model are
\begin{equation}
V_{soft} = V_{soft}^{MSSM - B_\mu} + V_{soft}^{singlets}\quad.
\end{equation}
$V_{soft}^{MSSM - B_\mu}$ contains all the usual soft terms of the MSSM
but the $B_\mu$-term, see equation \eqref{mssm-soft}, and $V_{soft}^{singlets}$ includes the new terms with singlets:

\begin{eqnarray}
V_{soft}^{singlets} &=& (m_{\nu^c}^2)^{st} \tilde{\nu}_s^c
 \tilde{\nu}_t^{c \: \ast} + \epsilon_{ab} \big[ T_\nu^{st}
 \tilde{L}_s^a \tilde{\nu}_t^c H_u^b - T_\lambda^s \tilde{\nu}_s^c
 H_d^a H_u^b+ h.c. \big] \label{eq:softsing} \\
&+& \big[\frac{1}{3!}T_{\kappa}^{stu} \tilde{\nu}_s^c \tilde{\nu}_t^c \nonumber
 \tilde{\nu}_u^c +h.c.\big]
\end{eqnarray}
In these expressions the notation for the soft trilinear couplings
introduced in \cite{Skands:2003cj,Allanach:2008qq} is used.
Note that the rotation made in the superpotential does not necessarily
diagonalize the soft trilinear terms $T_{\kappa}^{stu}$ implying 
in general additional mixing between the right-handed sneutrinos.

\subsection{Scalar potential and its minimization}
\label{subsec:scalartadpole}

Summing up the different contributions, the scalar potential
considering only neutral fields reads
\begin{equation}
V = V_D + V_F + V_{soft}
\end{equation}
with

\begin{eqnarray}\label{eq:dterms}
 V_D & = &  \frac{1}{8} (g^2 + g'^2) \big( |H_u^0|^2 - |H_d^0|^2 -
\sum_{i=1}^3 |\tilde{\nu}_i|^2 \big)^2 \\ V_F &=& | Y_\nu^{is}
\tilde{\nu}_i \tilde{\nu}_s^c - \lambda_s \tilde{\nu}_s^c H_d^0 |^2 +
| \lambda_s \tilde{\nu}_s^c H_u^0 |^2 + \sum_{i=1}^3 |Y_\nu^{is}
\tilde{\nu}_s^c H_u^0 |^2 \label{eq:fterms} \\
& + & \sum_{s=1}^n |Y_\nu^{is}
\tilde{\nu}_i H_u^0 - \lambda_s H_u^0 H_d^0 + \frac{1}{2} \kappa_s
(\tilde{\nu}_s^c)^2 |^2\quad, \nonumber
\end{eqnarray}
where summation over repeated indices is implied. 

This scalar potential determines the structure of the vacuum, inducing
VEVs: 

\begin{equation}
\langle H_d^0 \rangle = \frac{v_d}{\sqrt{2}}, \hskip5mm
\langle H_u^0 \rangle = \frac{v_u}{\sqrt{2}}, \hskip5mm
\langle \tilde{\nu}_s^c \rangle = \frac{v_{Rs}}{\sqrt{2}}, \hskip5mm
\langle \tilde{\nu}_i \rangle = \frac{v_i}{\sqrt{2}}
\label{eq:ewsb}
\end{equation}
In particular, the VEVs for the right-handed sneutrinos generate
effective bilinear couplings:

\begin{eqnarray}
 Y_{\nu}^{is} \widehat L_i \widehat \nu^c_s \widehat H_u &-& \lambda_s
 \widehat \nu^c_s \widehat H_d \widehat H_u \nonumber \\
&\Downarrow& \\
Y_{\nu}^{is} \widehat L_i \frac{v_{Rs}}{\sqrt{2}} \widehat H_u
 - \lambda_s \frac{v_{Rs}}{\sqrt{2}} \widehat H_d \widehat H_u &\equiv&
 \epsilon_i \widehat L_i \widehat H_u - \mu \widehat H_d \widehat
 H_u \nonumber
\end{eqnarray}

Since by electroweak symmetry breaking an effective $\mu$ term is generated, 
it is naturally at the electroweak scale.
Minimizing the scalar potential gives the tadpole equations at tree-level, see appendix \ref{munuapp1}.

As usual in R-parity breaking models with right-handed neutrinos, see
for example the model proposed in \cite{Masiero:1990uj} and discussed
in chapter \ref{chap:srpv}, it is possible to explain the smallness of
the $v_i$ in terms of the smallness of the Yukawa couplings $Y_\nu$,
that generate Dirac masses for the neutrinos. This can be easily seen
from equation \eqref{eq:tadpoleneu}, where both quantities are
proportional. Moreover, as shown in
\cite{LopezFogliani:2005yw}, taking the limit $Y_\nu \to 0$ and,
consequently, $v_i \to 0$, one recovers the tadpole equations of the
NMSSM, ensuring the existence of solutions to this set of equations.

Finally, as in chapter \ref{chap:srpv}, although other vacuum
configurations exist, we will focus on minima that break the
electroweak symmetry and R-parity simultaneously.

\subsection{Masses of the neutral scalars and pseudoscalars}
\label{sec:scalarmass}

In this subsection we work out the main features of the neutral scalar sector 
mainly focusing on singlets. The complete mass matrices are given
in appendix \ref{munuapp2}. We start with the one generation case
which closely resembles the NMSSM, considered, for example, in
\cite{Franke:1995xn,Miller:2003ay}. This already implies an upper
bound on the lightest doublet Higgs mass $m(h^0)$, on which we will focus
on at the end of this subsection.
A correct description of neutrino physics
implies small values for the VEVs $v_i$ of the left sneutrinos and
small Yukawa couplings $h_\nu$ as we will see later. Neglecting mixing
terms proportional to these quantities, the $(6\times 6)$ mass matrix of
the pseudoscalars in the basis $Im (H_d^0, H_u^0, \tilde{\nu}^c, \tilde{\nu}_i)$
given in appendix \ref{munuapp2}, equation \eqref{eq:masspseudoscalars}, can be
decomposed in two $\left(3\times 3\right)$ blocks. By using the
tadpole equations we obtain
\begin{equation}
M_{P^0}^2=\begin{pmatrix}M_{HH}^2&M_{HS}^2&0\\\left(M_{HS}^{2}\right)^T&M_{SS}^2&0\\0&0&M_{\tilde{L}\tilde{L}}^2\end{pmatrix}
\end{equation}
with
{\allowdisplaybreaks
\begin{eqnarray}
M_{HH}^2&=&\begin{pmatrix}\left(\Omega_1+\Omega_2\right)\frac{v_u}{v_d}&\Omega_1+\Omega_2\\\Omega_1+\Omega_2&\left(\Omega_1+\Omega_2\right)\end{pmatrix} \\
M_{HS}^2&=&\begin{pmatrix}\left(-2\Omega_1+\Omega_2\right)\frac{v_u}{v_R}\\\left(-2\Omega_1+\Omega_2\right)\frac{v_d}{v_R}\end{pmatrix} \\
M_{SS}^2&=&\left(4\Omega_1+\Omega_2\right)\frac{v_dv_u}{v_R^2}-3\Omega_3 \\
\left(M_{\tilde{L}\tilde{L}}^2\right)_{ij}&=&\tfrac{1}{2}\left(m_L^2\right)_{ij}+\tfrac{1}{2}\left(m_L^2\right)_{ji}+\delta_{ij}\left[\tfrac{1}{8}\left(g^2+g'^2\right)u^2\right]
\end{eqnarray}
}
where $u^2 = v_d^2 - v_u^2 + v_1^2 + v_2^2 + v_3^2$. The parameters $\Omega_i$ 
are defined as:
{\allowdisplaybreaks
\begin{eqnarray}
\Omega_1 &=& \frac{1}{8}\left(\lambda\kappa^*+\lambda^*\kappa\right)v_R^2 \\
\Omega_2 &=& \frac{1}{2\sqrt{2}}\left(T_\lambda+T_\lambda^*\right)v_R \\
\Omega_3 &=& \frac{1}{4\sqrt{2}}\left(T_\kappa+T_\kappa^*\right)v_R
\end{eqnarray}
}%
The upper $\left(3\times 3\right)$ block contains the mass terms for
$Im(H_d)$, $Im(H_u)$ and $Im(\tilde{\nu}^c)$ and we get analytic
expressions for the eigenvalues:
{\allowdisplaybreaks
\begin{align}
\nonumber m^2(P^0_1)&=0\\
m^2(P^0_2)&=\frac{1}{2}\left(\Omega_1+\Omega_2\right)\left(\frac{v_d}{v_u}+\frac{v_u}{v_d}+\frac{v_dv_u}{v_R^2}\right)-\frac{3}{2}\Omega_3-\sqrt{\Gamma}\\
\nonumber
m^2(P^0_3)&=\frac{1}{2}\left(\Omega_1+\Omega_2\right)\left(\frac{v_d}{v_u}+\frac{v_u}{v_d}+\frac{v_dv_u}{v_R^2}\right)-\frac{3}{2}\Omega_3+\sqrt{\Gamma}\\
\nonumber
\end{align}
}

with

\begin{eqnarray}
\Gamma&=&\left(\frac{1}{2}\left(\Omega_1+\Omega_2\right)\left(\frac{v_d}{v_u}+\frac{v_u}{v_d}+\frac{v_dv_u}{v_R^2}\right)-\frac{3}{2}\Omega_3\right)^2 \\
&+& 3\left(\Omega_1+\Omega_2\right)\Omega_3\left(\frac{v_d}{v_u}+\frac{v_u}{v_d}\right)-9\Omega_1\Omega_2\left(\frac{v_R^2}{v_d^2}+\frac{v_R^2}{v_u^2}\right) \nonumber
\end{eqnarray}
The first eigenvalue corresponds to the Goldstone boson due to spontaneous 
symmetry breaking. To get only positive eigenvalues for the physical 
states, the condition
\begin{equation}
\Omega_3<\frac{v_dv_u}{v_R^2}\frac{3\Omega_1\Omega_2}{\Omega_1+\Omega_2}=:f_1\left(\Omega_2\right) \end{equation}
has to be fulfilled, implying that $T_\kappa$ has in general the
opposite sign of $v_R$. Additional constraints on the parameters are
obtained from the positiveness of the squared masses of the neutral scalars.
Taking the scalar mass matrix from appendix \ref{munuapp2}, equation \eqref{eq:massscalars}, 
in the basis $Re (H_d^0, H_u^0,
\tilde{\nu}^c, \tilde{\nu}_i)$ in the same limit as above we obtain
\begin{equation}
M_{S^0}^2=\begin{pmatrix}M_{HH}^2&M_{HS}^2&0\\\left(M_{HS}^{2}\right)^T&M_{SS}^2&0\\0&0&M_{\tilde{L}\tilde{L}}^2\end{pmatrix}
\end{equation}
with
{\allowdisplaybreaks
\begin{eqnarray}
 M_{HH}^2&=&\begin{pmatrix}\left(\Omega_1+\Omega_2\right)\frac{v_u}{v_d}+\Omega_6\frac{v_d}{v_u}&-\Omega_1-\Omega_2-\Omega_6+\Omega_4\\-\Omega_1-\Omega_2-\Omega_6+\Omega_4&\left(\Omega_1+\Omega_2\right)\frac{v_d}{v_u}+\Omega_6\frac{v_u}{v_d}\end{pmatrix} \\
 M_{HS}^2&=&\begin{pmatrix}\left(-2\Omega_1-\Omega_2\right)\frac{v_u}{v_R}+\Omega_4\frac{v_R}{v_u}\\\left(-2\Omega_1-\Omega_2\right)\frac{v_d}{v_R}+\Omega_4\frac{v_R}{v_d}\end{pmatrix} \\
 M_{SS}^2&=&\Omega_2\frac{v_dv_u}{v_R^2}+\Omega_3+\Omega_5 \\
 \left(M_{\tilde{L}\tilde{L}}^2\right)_{ij}&=&\tfrac{1}{4}\left(g^2+g'^2\right)v_iv_j+\tfrac{1}{2}\left(m_L^2\right)_{ij}+\tfrac{1}{2}\left(m_L^2\right)_{ji} \\
&+& \delta_{ij}\left[\tfrac{1}{8}\left(g^2+g'^2\right)u^2\right] \nonumber
\end{eqnarray}
}%
using the additional parameters
{\allowdisplaybreaks
\begin{eqnarray}
\Omega_4 &=&\lambda\lambda^*v_dv_u>0 \\
\Omega_5 &=&\frac{1}{2}\kappa\kappa^*v_R^2>0 \\
\Omega_6 &=&\frac{1}{4}\left(g^2+g'^2\right)v_dv_u>0
\end{eqnarray}
}%
An analytic determination of the eigenvalues is possible but not very
illuminating.  However, one can use the following theorem: A symmetric
matrix is positive definite, if all eigenvalues are positive and this
is equal to the positiveness of all principal minors (Sylvester
criterion). This results in the following three conditions
\begin{eqnarray}
\nonumber 
0 &<&
\left(\Omega_1+\Omega_2\right)\frac{v_u}{v_d}+\Omega_6\frac{v_d}{v_u}\\\nonumber
0 &<& \left(\Omega_1+\Omega_2\right)\left(\Omega_6\left(\frac{v_d^2}{v_u^2}+\frac{v_u^2}{v_d^2}\right)-2\Omega_6+2\Omega_4\right)+2\Omega_4\Omega_6-\Omega_4^2\\
0 &<& \Omega_3-f_2\left(\Omega_2\right)\quad,
\end{eqnarray}
where $f_2(\Omega_2)$ is given by $f_2\left(\Omega_2\right) = \frac{\Sigma_1}{\Sigma_2}$, with
{\allowdisplaybreaks
\begin{align}
\nonumber&\Sigma_1=\left(\Omega_1+\Omega_2\right)\Omega_5\left(-2\Omega_4+2\Omega_6\right)+\left(\Omega_4^2-2\Omega_4\Omega_6\right)\Omega_5\\
\nonumber&\qquad +\left(\Omega_1+\Omega_2\right)\Omega_4^2v_R^2\left(\frac{v_d}{v_u^3}+\frac{v_u}{v_d^3}\right)+\left(4\Omega_1^2+3\Omega_1\Omega_2\right)\Omega_6\frac{1}{v_R^2}\left(\frac{v_d^3}{v_u}+\frac{v_u^3}{v_d}\right)\\
\nonumber&\qquad -\left(\Omega_1+\Omega_2\right)\Omega_5\Omega_6\left(\frac{v_d^2}{v_u^2}+\frac{v_u^2}{v_d^2}\right)+2\left(\Omega_1+\Omega_2-\Omega_4+2\Omega_6\right)\Omega_4^2\frac{v_R^2}{v_dv_u}\\
\nonumber&\qquad -2\left(2\Omega_1+\Omega_2\right)\left(2\Omega_1+2\Omega_2-\Omega_4+2\Omega_6\right)\Omega_4\left(\frac{v_d}{v_u}+\frac{v_u}{v_d}\right)\\
\nonumber&\qquad +\left[16\Omega_1^3+8\left(4\Omega_2-\Omega_4+\Omega_6\right)\Omega_1^2+10\Omega_1\Omega_2\left(2\Omega_2-\Omega_4+\Omega_6\right)\right.\\
&\left.\qquad\quad +\Omega_2\left(2\Omega_2-\Omega_4\right)\left(2\Omega_2-\Omega_4+2\Omega_6\right)\right]\frac{v_dv_u}{v_R^2}\\
&\nonumber \Sigma_2=\left(\Omega_1+\Omega_2\right)\Omega_6\left(\frac{v_d^2}{v_u^2}+\frac{v_u^2}{v_d^2}\right)+2\left(\Omega_1+\Omega_2\right)\left(\Omega_4-\Omega_6\right)\\
&\qquad +2\Omega_4\Omega_6-\Omega_4^2
\label{formelfuerbeding}
\end{align}
}%
The first two conditions are in general fulfilled, but for special
values of $\tan\beta$ or $\lambda$. Putting all the above together we get
the following conditions:
\begin{equation}
f_2(\Omega_2)<\Omega_3<f_1(\Omega_2)
\end{equation}
It turns out that by taking a negative value of $\Omega_3$ ($\propto
T_\kappa$) near $f_2(\Omega_2)$ one obtains a very light singlet scalar,
whereas for a value of $\Omega_3$ near $f_1(\Omega_2)$ one gets a very
light singlet pseudoscalar. In between one finds a value of
$\Omega_3$, where both particles have the same mass. This discussion is
comparable to formula (37) in \cite{Miller:2003ay} for the
NMSSM. Moreover, a small mass of the singlet scalar and/or
pseudoscalar comes always together with a small mass of the singlet
fermion.

In the $n$ generation case a similar result holds as long as
$T_{\kappa}$ and $m^2_{{\tilde \nu}^c}$ do not have sizeable off-diagonal
entries. Inspecting eqs. \eqref{SSscalar} and
\eqref{SSpseudoscalar} it is possible to show that the singlet scalars
and pseudoscalars can be made heavy by properly choosing values for the
off-diagonal entries of $T_\kappa$ while keeping at the same time the
singlet fermions relatively light, as will be discussed later. As
pointed out in \cite{Escudero:2008jg}, the NMSSM upper bound on the
lightest doublet Higgs mass of about $\sim 150$ GeV, which also applies
in the $\mu\nu$SSM, can be relaxed to ${\cal O}(300)$ GeV, if one
does not require perturbativity up to the GUT scale.

\section{Neutrino masses}
\label{subsec:neutrinomass}

In the basis

\begin{equation}
\big( \psi^0 \big)^T = \big( -i{\tilde B}^0, -i{\tilde W}_3^0, {\tilde H}_d^0, {\tilde H}_u^0, \nu_s^c, \nu_i \big)
\end{equation}
the mass matrix of the neutral fermions, see appendix 
\ref{subsec:neutralinos}, has the structure
\begin{equation}
M_N =
\left(\begin{array}{cc}
M_H & m \\
m^T & 0
\end{array} \right)\quad.
\label{eq:neutralmass}
\end{equation}
Here $M_H$ is the submatrix including the heavy states, which
consists of the usual four neutralinos of the MSSM and $n$ generations of
 right-handed neutrinos. The matrix $m$ mixes the heavy
 states with the left-handed neutrinos and contains the R-parity
 breaking parameters.

The matrix $M_N$ can be diagonalized in the standard way:
\begin{equation}
\widehat{M}_N = {\cal N}^* M_N {\cal N}^{-1}
\label{eq:diagmass}
\end{equation}
As we saw in the previous chapters, the smallness of neutrino masses allows to 
find the effective neutrino mass matrix in a seesaw approximation
\begin{equation}\label{eq:defmnueff}
m_{eff}^{\nu \nu} = - m^T \cdot M_H^{-1} \cdot m
 = - \xi \cdot m\quad,
\end{equation}
where the matrix $\xi$ contains the small expansion parameters which
characterize the mixing between the neutrino sector and the heavy
states.

Since the superpotential explicitly breaks lepton number, at least one
mass for the left-handed neutrinos is generated at tree-level. In the case
of the $1$ $\hat\nu^c$-model the other neutrino masses are generated
at loop-level. With more than one
generation of right-handed neutrinos additional neutrino masses
are generated at tree-level, resulting in different possibilities 
to fit the neutrino oscillation data, see the discussion below.

\subsection{$1$ generation of right-handed neutrinos}
\label{subsec:1genneut}

With only one generation of right-handed neutrinos the matrix $\xi$ is
given by
\begin{equation}\label{xi}
\xi_{ij} = K_{\Lambda}^j \Lambda_i - \frac{1}{\mu} \epsilon_i \delta_{j3}\quad,
\end{equation}
where the $\epsilon_i$ and $\Lambda_i$ parameters are defined as usual

\begin{eqnarray}
\epsilon_i & = & \frac{1}{\sqrt{2}} Y_{\nu}^{i}v_{R} \\
\Lambda_i &=& \mu v_i + \epsilon_i v_d
\end{eqnarray}
and $K^j_{\Lambda}$ as 
\begin{eqnarray}\label{defK1}
K_\Lambda^1 &=& \frac{2 g' M_2 \mu}{m_\gamma}a \nonumber \\
K_\Lambda^2 &=& -\frac{2 g M_1 \mu}{m_\gamma}a \nonumber \\
K_\Lambda^3 &=& \frac{m_\gamma}{8 \mu {\rm Det}(M_H)}(\lambda^2 v_d v^2 + 2 M_R \mu v_u) \nonumber\\
K_\Lambda^4 &=& -\frac{m_\gamma}{8 \mu {\rm Det}(M_H)}(\lambda^2 v_u v^2 + 2 M_R \mu v_d) \nonumber \\
K_\Lambda^5 &=& \frac{\lambda m_\gamma}{4 \sqrt{2} {\rm Det}(M_H)}(v_u^2 - v_d^2)
\end{eqnarray}
The parameters $m_\gamma = g^2M_1 +g'^2 M_2$ and $v^2 = v_d^2 + v_u^2$ have the same definition as in the models studied in the previous chapters, $M_R$ and $a$ are given by

\begin{eqnarray}
M_R &=& \frac{1}{\sqrt{2}} \kappa v_R \\
a &=& \frac{m_\gamma}{4 \mu {\rm Det}(M_H)}(v_d v_u \lambda^2 + M_R \mu)
\end{eqnarray}
and ${\rm Det}(M_H)$ is the determinant of the $(5 \times 5)$ mass matrix of
the heavy states
\begin{equation}
{\rm Det}(M_H) = \frac{1}{8} m_\gamma (\lambda^2 v^4 + 4 M_R \mu v_d v_u) 
- M_1 M_2 \mu(v_d v_u \lambda^2 + M_R \mu)\quad.
\end{equation}
Using these expressions the tree-level effective neutrino mass matrix
takes the form
\begin{equation}\label{eq:effone}
(m_{eff}^{\nu\nu})_{ij} = a \Lambda_i \Lambda_j\quad.
\end{equation}
The projective form of this mass matrix 
implies that only one neutrino gets a tree-level mass, while the other
two remain massless. Therefore, as in models with bilinear R-parity
violation \cite{Romao:1999up,Hirsch:2000ef,Diaz:2003as} 1-loop
corrections are needed in order to correctly explain the oscillation
data, which requires at least one additional massive neutrino.
The absolute scale of neutrino mass constrains the $\vec{\Lambda}$ and 
$\vec{\epsilon}$ parameters, which have to be small. For typical SUSY masses 
order ${\cal O}(100\hskip1mm{\rm \text{GeV}})$, one finds $|\vec\Lambda|/\mu^2 \sim
10^{-7}$--$10^{-6}$ and $|\vec{\epsilon}|/\mu \sim
10^{-5}$--$10^{-4}$. This implies a ratio of
$|\vec{\epsilon}|^2/|\vec{\Lambda}|\sim 10^{-3}$--$10^{-1}$.

\begin{figure}
\begin{center}
\includegraphics[width=0.6\textwidth]{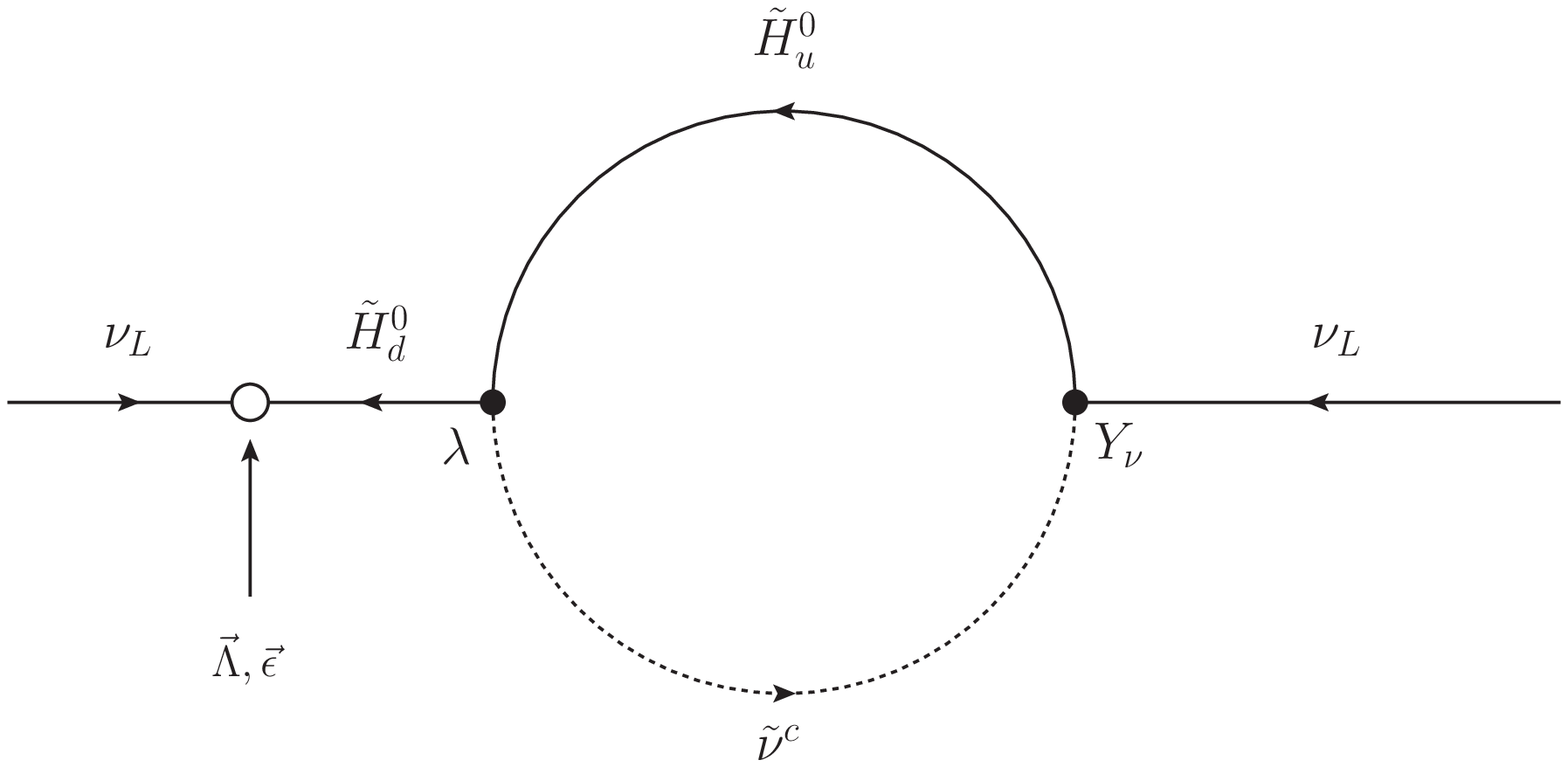}
\caption{Example of one 1-loop correction to the effective neutrino mass matrix
involving the singlet scalar/pseudoscalar.}
\label{loopdiag}
\end{center}
\end{figure}

\begin{figure}
\begin{center}
\includegraphics[width=0.6\textwidth]{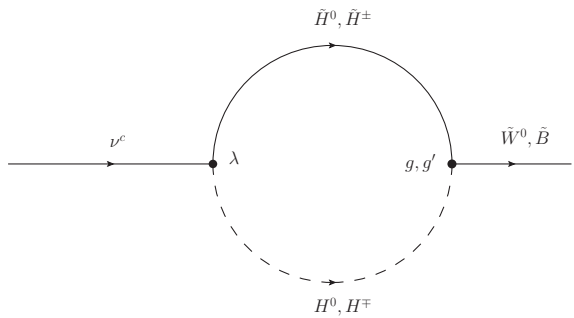}
\caption{1-loop mixing between gauginos and the right-handed neutrinos.}
\label{loopdiagMuRGaugino}
\end{center}
\end{figure}

General formulas for the 1-loop contributions can be found in
\cite{Hirsch:2000ef} and adjusted to the $\mu\nu$SSM with appropriate
changes in the index ranges for neutralinos and scalars. See also reference \cite{Ghosh:2010zi}, which provides a
complete calculation of the 1-loop corrections to the neutrino mass matrix. Important
contributions to the neutrino mass matrix are due to $b-{\tilde b}$
and $\tau-{\tilde\tau}$ loops as in the models with b-\rpv
\cite{Diaz:2003as}. In addition there are two new important 
contributions: (i) loops containing the singlet scalar and singlet
pseudoscalar shown in figure \ref{loopdiag}.  As shown in
\cite{Hirsch:1997vz,Hirsch:1997dm,Grossman:1997is}, the sum of both
contributions is proportional to the squared mass difference
$\Delta_{12} = m_R^2 - m_I^2 \propto \kappa^2 v_R^2$ between the 
singlet scalar and pseudoscalar mass eigenstates. Note that this splitting 
can be much larger than the corresponding one for the left sneutrinos. 
Thus the sum of both loops can be more important than $b
- \tilde b$ and $\tau - \tilde \tau$ loops in the current model. See appendix \ref{munuapp3} for more details.
(ii) At loop-level a direct mixing between the
right-handed neutrinos and the gauginos is possible which is zero at
tree-level, see figure \ref{loopdiagMuRGaugino}. 

\subsection{$n$ generations of right-handed neutrinos}
\label{subsec:ngenneut}

In this class of models with $n>1$ one can explain the neutrino data
using the tree-level neutrino mass matrix only. In general one finds
that the loop corrections are small if the conditions given at the end of
this section are fulfilled.

For the sake of simplicity, let us consider two generations of
right-handed neutrinos. This choice contains all relevant features.  The
matrix $\xi$ in equation \eqref{eq:defmnueff} takes the form
\begin{equation}\label{xi2}
\xi_{ij} = K_{\Lambda}^j \Lambda_i + K_{\alpha}^j \alpha_i 
- \frac{\epsilon_i}{\mu} \delta_{j3}
\end{equation}
with

\begin{eqnarray}
\epsilon_i & = & \frac{1}{\sqrt{2}} Y_{\nu}^{is}v_{Rs} \\
\Lambda_i &=& \mu v_i + \epsilon_i v_d \\
\alpha_i &=& v_u ( \lambda_2 Y_\nu^{i1} -\lambda_1 Y_\nu^{i2})\quad.
\label{defalpha}
\end{eqnarray}
The $K_\Lambda$ and $K_\alpha$ coefficients have complicated dependencies on the parameters of the model. We list them here:
{\allowdisplaybreaks
\begin{eqnarray}\label{defK2}
K_\Lambda^1 &=& \frac{2 g' M_2 \mu}{m_\gamma}a \nonumber \\
K_\Lambda^2 &=& -\frac{2 g M_1 \mu}{m_\gamma}a \nonumber \\
K_\Lambda^3 &=& \frac{v_u}{2}a + \frac{m_\gamma}{8 \mu {\rm Det}(M_H)} \left[ v_d^3 (M_{R1} \lambda_2^2 + M_{R2} \lambda_1^2) + M_{R1} M_{R2} \mu v_u \right] \nonumber \\
K_\Lambda^4 &=& -\frac{v_d}{2}a - \frac{m_\gamma}{8 \mu {\rm Det}(M_H)} \left[ v_u^3 (M_{R1} \lambda_2^2 + M_{R2} \lambda_1^2) + M_{R1} M_{R2} \mu v_d \right] \nonumber \\
K_\Lambda^5 &=& \frac{M_{R2} \lambda_1 m_\gamma}{4 \sqrt{2} {\rm Det}(M_H)} (v_u^2 - v_d^2) \nonumber \\
K_\Lambda^6 &=& \frac{M_{R1} \lambda_2 m_\gamma}{4 \sqrt{2} {\rm Det}(M_H)} (v_u^2 - v_d^2) \\
K_{\alpha}^1 &=& \frac{2 g' M_2 \mu}{m_\gamma}b \nonumber \\ 
K_{\alpha}^2 &=& -\frac{2 g M_1 \mu}{m_\gamma}b \nonumber \\  
K_{\alpha}^3 &=& \frac{b}{m_\gamma (v_u^2 - v_d^2)} (m_\gamma v^2 v_u - 4 M_1 M_2 \mu v_d) \nonumber \\   
K_{\alpha}^4 &=& \frac{b}{m_\gamma (v_u^2 - v_d^2)} (m_\gamma v^2 v_d - 4 M_1 M_2 \mu v_u) \nonumber \\  
K_{\alpha}^5 &=& - \sqrt{2} \lambda_2 c - \frac{4 \textnormal{Det}_0 v_{R1}}{\mu m_\gamma (v_u^2 - v_d^2)}b \nonumber \\ 
K_{\alpha}^6 &=& \sqrt{2} \lambda_1 c - \frac{4 \textnormal{Det}_0 v_{R2}}{\mu m_\gamma (v_u^2 - v_d^2)}b
\end{eqnarray}
}

The effective neutrino mass matrix reads as
\begin{equation}\label{eq:efftwo}
(m_{eff}^{\nu \nu})_{ij} = a \Lambda_i \Lambda_j + b (\Lambda_i \alpha_j + \Lambda_j \alpha_i) + c \alpha_i \alpha_j
\end{equation}
with
\begin{eqnarray}
a &=& \frac{m_\gamma}{4 \mu {\rm Det}(M_H)} (M_{R1} \lambda_2^2 v_u v_d + M_{R2} \lambda_1^2 v_u v_d + M_{R1} M_{R2} \mu)  \\
b &=& \frac{m_\gamma}{8 \sqrt{2} \mu {\rm Det}(M_H)} (v_u^2 - v_d^2)(M_{R1} v_{R1} \lambda_2 - M_{R2} v_{R2} \lambda_1) \\
c &=& - \frac{1}{16 \mu^2 {\rm Det}(M_H)} \big[\mu^2(m_\gamma v^4 - 8 M_1 M_2 \mu v_u v_d) \\
 && \hskip30mm + 4 \textnormal{Det}_0 (M_{R1} v_{R1}^2 + M_{R2} v_{R2}^2) \big] \nonumber
\end{eqnarray}
using $M_{Rs} = \frac{1}{\sqrt{2}} \kappa_s v_{Rs}$. The determinant of the $(6\times6)$ mass matrix of the heavy states is
\begin{equation}
{\rm Det}(M_H) = \frac{1}{8} \big[(M_{R2} \lambda_1^2 + M_{R1} \lambda_2^2)(m_\gamma v^4 - 8 M_1 M_2 \mu v_u v_d) + 8 M_{R1} M_{R2} \textnormal{Det}_0 \big] 
\end{equation}
with $\textnormal{Det}_0$ being the determinant of the usual MSSM neutralino mass matrix
\begin{equation}
 \textnormal{Det}_0 = \frac{1}{2} m_\gamma \mu v_d v_u - M_1 M_2 \mu^2 \quad.
\end{equation}

Note that the $\vec \alpha$ parameters play the role of the $\vec \epsilon$ parameters in s-\rpv, see equation \eqref{srpv:effnu}. Therefore, one can follow an analogous procedure to fit neutrino data.

The mass matrix in equation \eqref{eq:efftwo} has two nonzero
eigenvalues and therefore the loop corrections are not needed to
explain the experimental data. Two different options arise:
\begin{itemize}
\item $\vec \Lambda$ generates the atmospheric mass scale, $\vec
\alpha$ the solar mass scale
\vspace{-1mm}
\item $\vec \alpha$ generates the atmospheric mass scale, $\vec
\Lambda$ the solar mass scale
\end{itemize}

In both cases one obtains in general a hierarchical spectrum.
A strong fine-tuning would be necessary to generate an inverted
hierarchy which is not stable against small variations of the parameters
or radiative corretions.
Moreover the absolute scale of neutrino mass requires both $|\vec
\Lambda|/\mu^2$ and $|\vec \alpha|/\mu$ to be small. For typical SUSY
masses order ${\cal O}(100\hskip1mm{\rm \text{GeV}})$ we find in the first
case $|\vec\Lambda|/\mu^2 \sim 10^{-7}$--$10^{-6}$ and
$|\vec\alpha|/\mu \sim 10^{-9}$--$10^{-8}$. In the second case we find
$|\vec\Lambda|/\mu^2 \sim 10^{-8}$--$10^{-7}$ and $|\vec\alpha|/\mu
\sim 10^{-8}$--$10^{-7}$. The ratios including $\vec{\epsilon}$ or
$\vec{\alpha}$ are much smaller than those in the $1$
$\widehat{\nu}^c$ case.  We find that 1-loop corrections 
to \eqref{eq:efftwo} are negligible if
\begin{eqnarray}
\frac{|\vec \alpha|^2}{|\vec \Lambda|}  \lesssim  10^{-3}  \qquad\,\, \mathrm{and}
\qquad \,\,
\frac{|\vec \epsilon|^2}{|\vec \Lambda|}  \lesssim  10^{-3} \label{seccond}
\end{eqnarray}
are fulfilled.
Note that the mixing of the neutrinos with the higgsinos, given by the
third column in the matrix $\xi$ in equation \eqref{xi2}, depends not
only on $\alpha_i$ but also on $\epsilon_i$.  This leads to 1-loop
corrections to the neutrino mass matrix with pieces proportional to
the $\epsilon_i$ parameters, as also happens in the $1$
$\widehat{\nu}^c$-model. Therefore, both conditions in
equation \eqref{seccond} need to be fulfilled. See ref. \cite{Ghosh:2010zi} for more
details on the 1-loop corrected neutrino mass matrix.
Finally, in models with more generations of right-handed neutrinos
there will be more freedom due to additional contributions to the
neutrino mass matrix. For example, the case of three generations is
discussed in \cite{Ghosh:2008yh}, where the additional freedom is 
also used to generate an inverted hierarchy for the neutrino masses.

\section{Choice of the parameters and experimental constraints}
\label{sec:Strate}

In the subsequent sections we work out collider signatures for various
scenarios. To facilitate the comparison with existing studies we adopt
the following strategy: We take existing study points and augment them
with the additional model parameters breaking R-parity. These points
are SPS1a' \cite{AguilarSaavedra:2005pw}, SPS3, SPS4, SPS9
\cite{Allanach:2002nj} and the ATLAS SU4 point \cite{Aad:2009wy}.
SPS1a' contains a relative light spectrum so that at LHC a high
statistic can be achieved, SPS3 has a somewhat heavier spectrum and in
addition the lightest neutralino and the lighter stau are close in
mass which affects also the R-parity violating decays of the lightest
neutralino. SPS4 is chosen because of the large $\tan\beta$ value and
SPS9 is an AMSB scenario where not only the lightest neutralino but
also the lighter chargino has dominant R-parity violating decay modes.
In all these points the lightest neutralino is so heavy that it can
decay via two-body modes, as long as it's not a light $\nu^c$.
In contrast for the SU4 point all two-body decay modes (at
tree-level) are kinematically forbidden. As the parameters of these
points are given at different scales we use the program {\tt SPheno}
\cite{Porod:2003um} to evaluate them at $Q=m_Z$ where we add the
additional model parameters. Note that we allow $\mu$ to depart from 
their standard SPS values to be consistent with the LEP bounds on 
Higgs masses, discussed below.

The additional model parameters are subject to theoretical and
experimental constraints. In \cite{Escudero:2008jg} the question
of color and charge breaking minima, perturbativity up to the GUT
scale as well as the questions of tachyonic states for the neutral scalar
and pseudoscalars have been investigated. The last issue has already
been addressed in section \ref{sec:scalarmass} where we derived
conditions on the parameters. By choosing the coupling constants
$\lambda, \kappa<0.6$ in the $1$ $\widehat{\nu}^c$-model
and $\lambda_s,\kappa_s<0.5$ in the $2$ $\widehat{\nu}^c$-model, 
perturbativity up to the GUT scale is guaranteed \cite{Escudero:2008jg}.
Note, that choosing
somewhat larger values for $\lambda$ and/or $\kappa$ up to $1$ does not
change any of the results presented below. We also address the
question of color and charge breaking minimas by choosing
$\lambda_{s}>0$, $\kappa_{s}>0$, $T_{\lambda}^{s}>0$, $T_\kappa^{stu}
<0$, whereas the Yukawa couplings $Y_\nu^{is}$ can either be positive
or negative, but those values are small $<\mathcal{O}(10^{-6})$ due to
constraints from neutrino physics. Our $T_\nu^{is}$ are negative,
so the condition (2.8) of \cite{Escudero:2008jg} is easy to fulfill.

Concerning experimental data we take the following constraints into
account:
\begin{itemize}
\item We check that the neutrino data are fulfilled within the $2$-$\sigma$
      range given in Table \ref{tab:nudata} taken from ref.\
      \cite{Schwetz:2008er} if not stated otherwise. These data can
      easily be fitted using the effective neutrino mass matrices
      given in section \ref{subsec:neutrinomass}.
\item Breaking lepton number implies that flavor violating decays of
     the leptons like $\mu\to e \gamma$ are possible, where strong
     experimental bounds exist \cite{Amsler:2008zzb}.  However, in the
     model under study it turns out that these bounds are automatically
     fulfilled once the constraints from neutrino physics are taken
     into account similar to the case of models with bilinear R-parity
     breaking \cite{Carvalho:2002bq}.
\item Bounds on the masses of the Higgs bosons
      \cite{Schael:2006cr,Amsler:2008zzb}. For this purpose we have
      added the dominant 1-loop correction to the (2,2) entry of the
      scalar mass matrix in appendix \ref{subsec:scalars}. Moreover,
      we have checked in the $1$ ${\hat \nu}^c$-model with the help of
      the program NMHDECAY \cite{Ellwanger:2005dv} that in the NMSSM
      limit the experimental constraints are fulfilled.
\item Constraints on the chargino and charged slepton masses given by
the PDG \cite{Amsler:2008zzb}.
\item The bounds on squark and gluino masses from TEVATRON
      \cite{Amsler:2008zzb} are automatically fulfilled by our choices
      of the study points.
\end{itemize}

The smallness of the \rpv parameters guarantees that the direct
production cross sections for the SUSY particles are very similar to
the corresponding MSSM/NMSSM values. Note that for low values of
$\lambda$ the singlet states are decoupled from the rest of the particles, 
leading to low production rates. 

\section{Phenomenology of the $1$ $\widehat{\nu}^c$ model}
\label{sec:phenoone}

In this section we discuss the phenomenology of the $1$
$\widehat{\nu}^c$-model, including mass hierarchies, mixings in the
scalar and fermionic sectors, decays of the scalar and fermionic
states and the correlations between certain branching
ratios and the neutrino mixing angles.

\begin{figure}
\begin{center}
\vspace{0mm}
\includegraphics[width=0.49\textwidth]{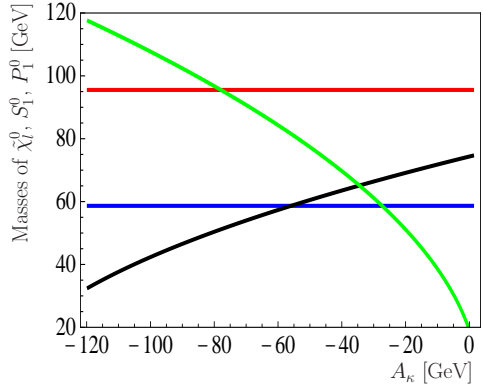} 
\end{center}
\vspace{-5mm}
\caption{Masses of the lightest neutralinos $\tilde{\chi}_l^0$ and
the lightest scalar $S_1^0=Re(\tilde{\nu}^c)$/pseudoscalar
$P_1^0=Im(\tilde{\nu}_1^c)$ as a function of $A_\kappa=T_\kappa/\kappa$ for
$\lambda=0.24$, $\kappa=0.12$, $\mu=150$ GeV and $T_\lambda=360$ GeV for SPS1a'.
The different colors refer to the singlino $\tilde{\chi}_1^0$ (blue),
the bino $\tilde{\chi}_2^0$ (red), the singlet scalar $S_1^0$ (black)
and the singlet pseudoscalar $P_1^0$ (green).}
\label{fig:1NuR_akappatotal}
\end{figure}

In the following discussion we call a neutralino $\tilde{\chi}_l^0$ a bino (singlino) 
if $|\mathcal{N}_{l+3,1}|^2>0.5$ ($|\mathcal{N}_{l+3,5}|^2>0.5$). 
As discussed below, light scalar $S_m^0$ or pseudoscalar states
$P_m^0$ appear, especially in case of the singlino being the lightest 
neutralino. In the following we discuss possible mass
hierarchies and mixings in more detail.

The diagonal entry of the singlet right-handed neutrino in the mass
matrix of the neutral fermions is $M_R=\tfrac{1}{\sqrt{2}}\kappa v_R$,
see appendix \ref{subsec:neutralinos}. A singlino as lightest
neutralino is obtained by choosing small values for $\kappa$ and/or
$v_R$. Since the masses of the four MSSM neutralinos are mainly fixed
by the chosen SPS point, we can either generate a bino-like or a
singlino-like lightest neutralino by varying $\kappa$ and/or $v_R$,
where the latter case means a variation of $\lambda$ due to a fixed
$\mu$-parameter. A light singlet scalar and/or pseudoscalar can be
obtained by appropriate choices of $T_\lambda$ and $T_\kappa$. An
example spectrum is shown in figure \ref{fig:1NuR_akappatotal}. The
MSSM parameters have been chosen according to SPS1a' except for
$\mu=150$ GeV.  The scalar state $S_2^0= h^0$ can easily get too light
to be consistent with current experimental data, although the
production rate $e^+e^-\rightarrow ZS_2^0$ is lowered, since a mixing
with the lighter singlet scalar $S_1^0= \tilde{\nu}^c$ reduces its
mass. By reducing $\mu$ the mixing can be lowered (see mass matrices)
and this problem can be solved.

\begin{figure}
\begin{center}
\vspace{2mm}
\includegraphics[width=0.49\textwidth]{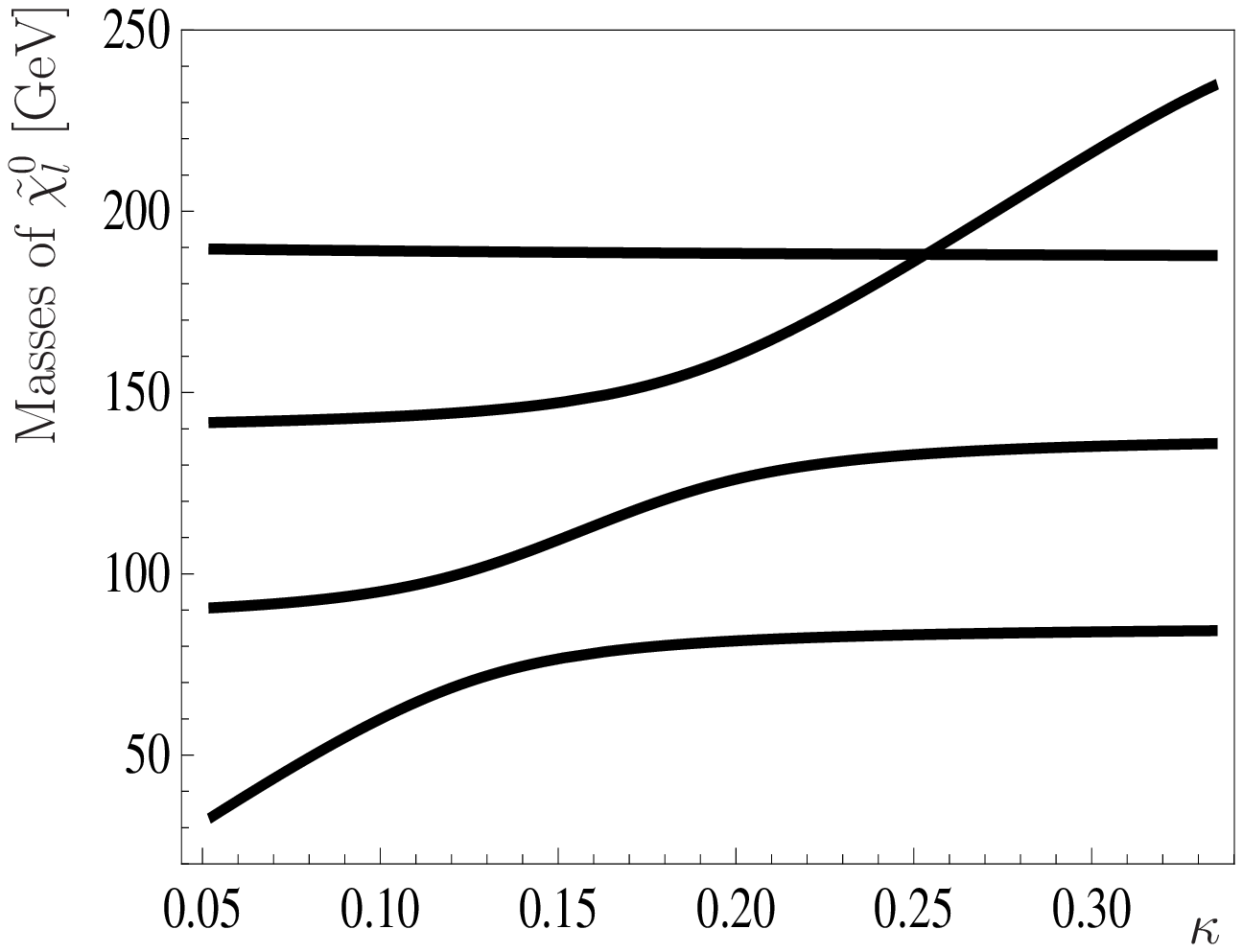}
\includegraphics[width=0.49\textwidth]{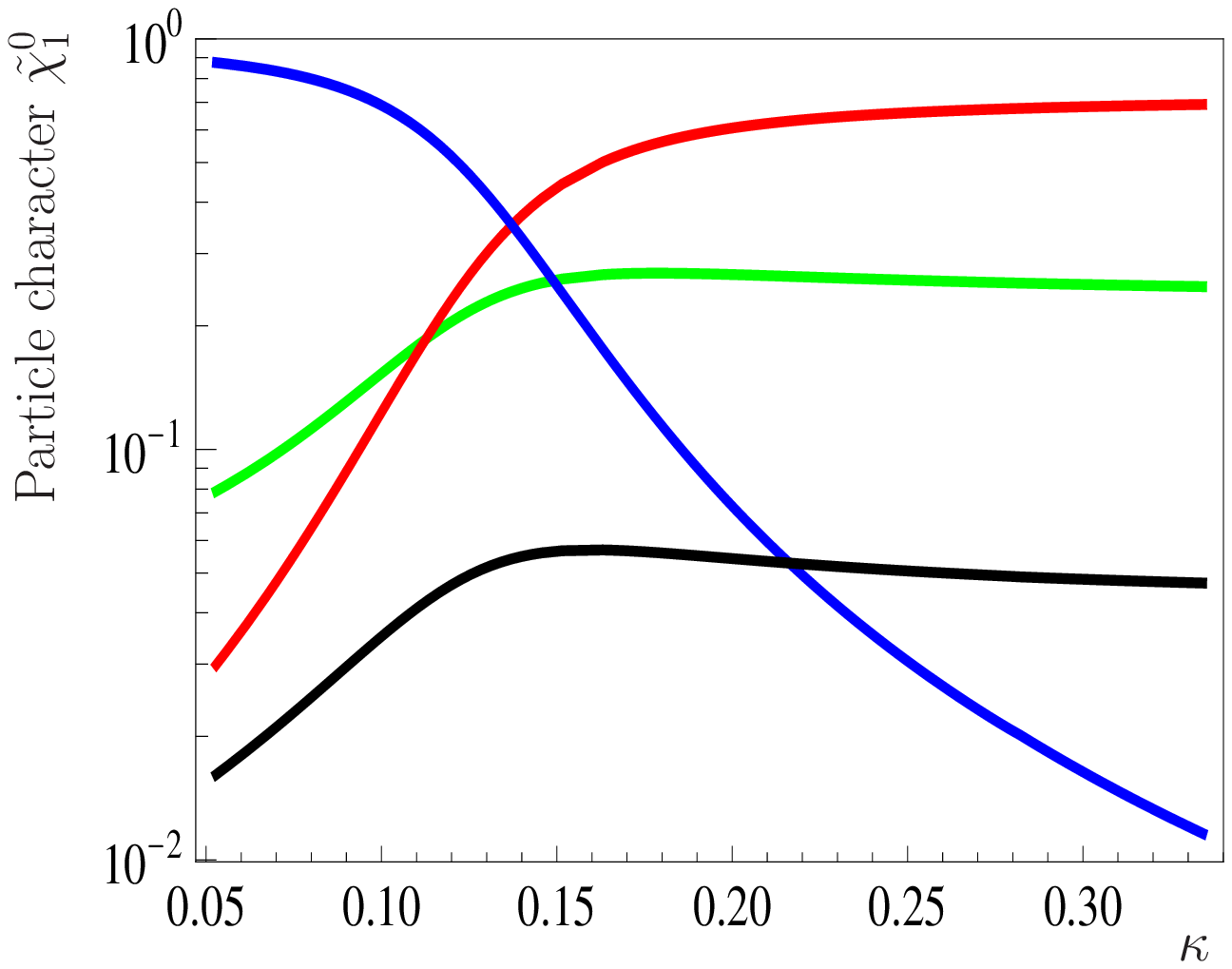} 
\vspace{2mm}
\includegraphics[width=0.49\textwidth]{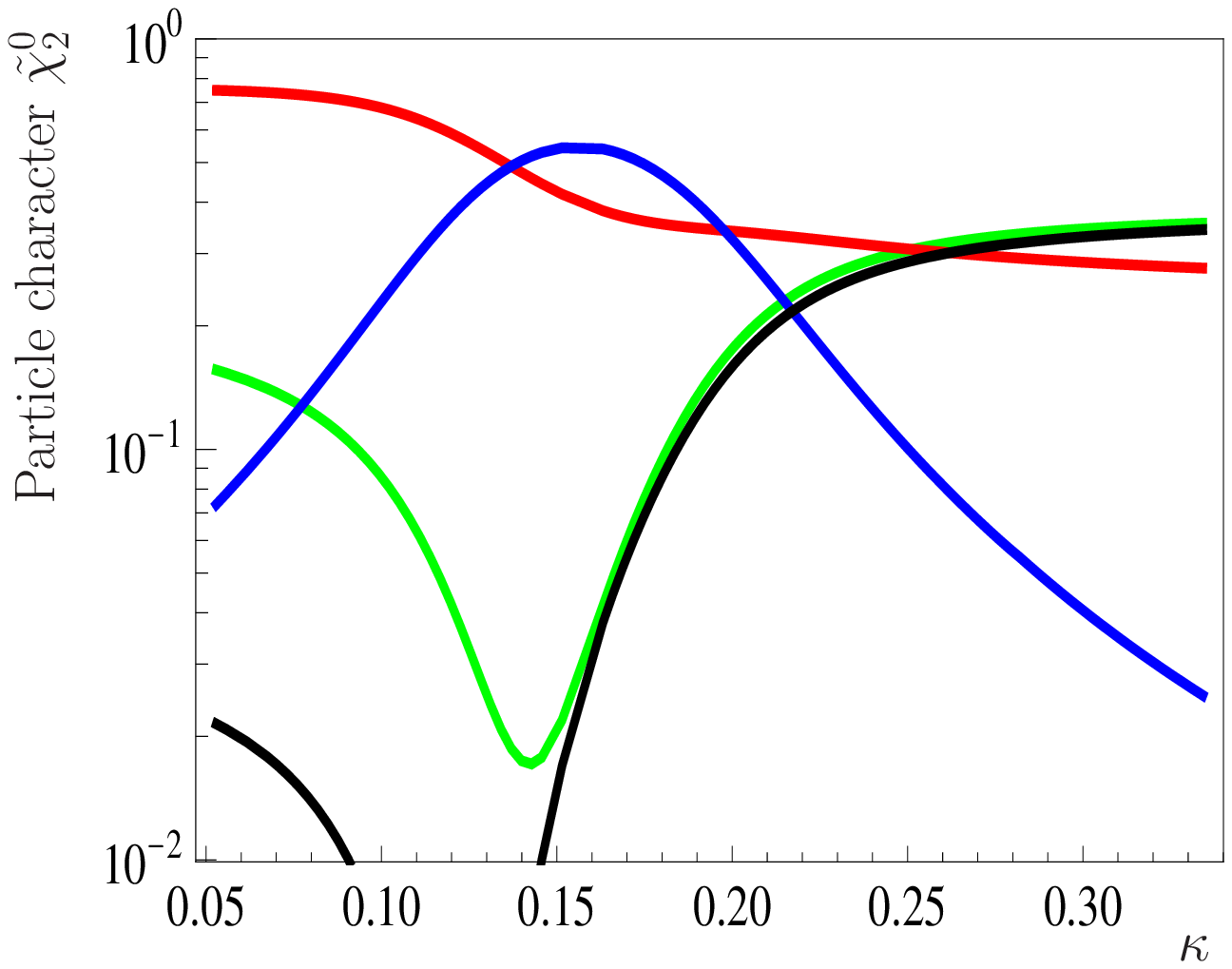} 
\includegraphics[width=0.49\textwidth]{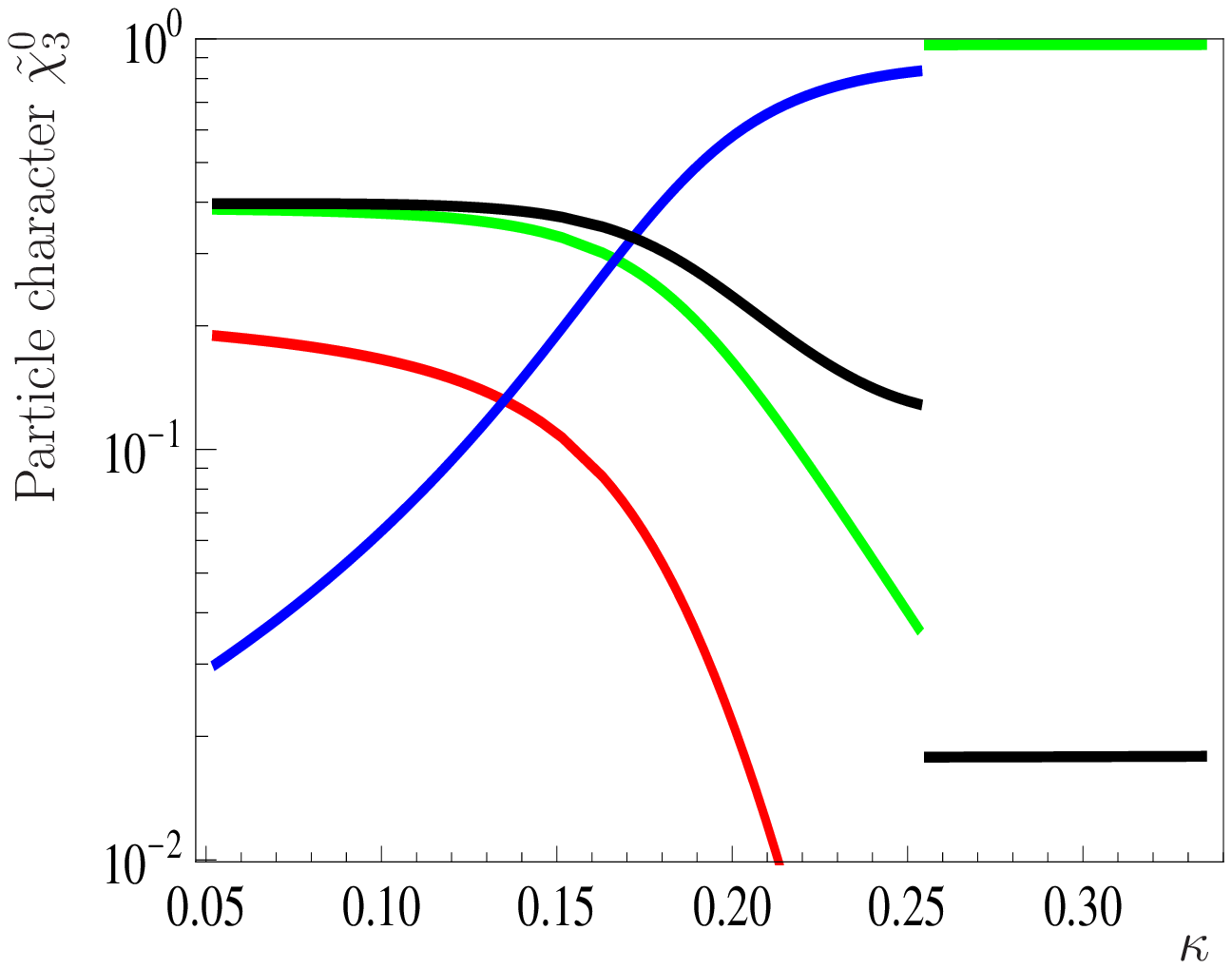} 
\end{center}
\vspace{-2mm}
\caption{Masses and particle characters of the lightest neutralinos
$\tilde{\chi}_l^0$ as a function of $\kappa$ for $\lambda=0.24$,
$\mu=170$ GeV, $T_\lambda=360$ GeV and $T_\kappa=-\kappa\cdot 50$ GeV for SPS1a'.
The different colors refer to
singlino purity $|\mathcal{N}_{l+3,5}|^2$ (blue), bino purity
$|\mathcal{N}_{l+3,1}|^2$ (red), wino purity
$|\mathcal{N}_{l+3,2}|^2$ (black) and higgsino purity
$|\mathcal{N}_{l+3,3}|^2+|\mathcal{N}_{l+3,4}|^2$ (green).}
\label{fig:1NuR_mixingtotal}
\end{figure}

Another example spectrum for neutral fermions is shown in figure 
\ref{fig:1NuR_mixingtotal}. Again SPS1a' parameters have been 
chosen, except $\mu=170$ GeV. As the figure demonstrates for 
this reduced value of $\mu$ the states are usually quite mixed, 
which is important for their decay properties, as discussed 
below. Note that the abrupt change in composition in $\tilde{\chi}^0_3$
is due to the level crossing in the mass eigenstates. 

\begin{figure}
\begin{center}
\vspace{3mm}
\includegraphics[width=0.49\textwidth]{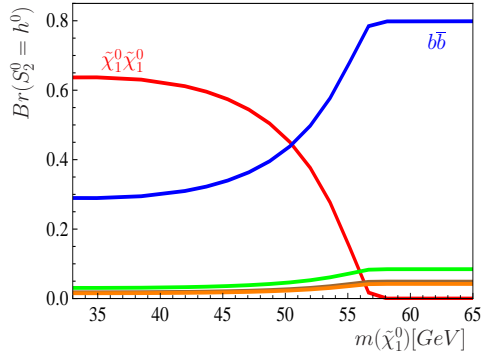}
\end{center}
\vspace{-5mm}
\caption{
Branching ratios
$Br(S_2^0 = h^0)$ as a function of $m(\tilde{\chi}_1^0)$ for
the parameter set of figure \ref{fig:1NuR_mixingtotal} (variation of
$\kappa$). The colors indicate the different final states:
$\tilde{\chi}_1^0\tilde{\chi}_1^0$ (red), $b\overline{b}$ (blue),
$\tau^+\tau^-$ (green), $c\overline{c}$ (orange) and
$Wq\overline{q}$ (brown).}
\label{fig:1NuR_decayscalar}
\end{figure}

The decay properties of the lightest scalars/pseudoscalars are in
general quite similar to those found in the NMSSM \cite{Miller:2003ay,
Ellwanger:2005uu}. The lightest doublet Higgs boson similar to the
$h^0$ decays mainly like in the MSSM, apart from the possible final
states $2 \tilde{\chi}^0_1$ and $2 P_1^0$, if kinematically
possible. The decay to a pair of light pseudoscalars has not been
studied in this thesis, and the interested reader is referred to
earlier works in the NMSSM \cite{Miller:2003ay, Ellwanger:2005uu},
where this possibility is addressed with conserved
R-parity. Concerning the decay to a pair of light neutralinos, an
example is shown in figure
\ref{fig:1NuR_decayscalar}, which display the branching ratios of
$S^0_2 = h^0$ versus $m(\tilde{\chi}_1^0)$. $\tilde{\chi}^0_1$ in this plot
is mainly a singlino (see figure \ref{fig:1NuR_mixingtotal}),
variation of $\kappa$ varies its mass, since $v_R$ is kept fixed
here. In contrast to the NMSSM this does not lead to an invisible
Higgs, since the neutralinos themselves decay. For the range of 
parameters where the decay to $2\tilde{\chi}^0_1$ is large, 
$\tilde{\chi}^0_1$ decays mainly to $\nu b\overline{b}$, leading to 
the final state 4 $b$-jets plus missing energy. Note that the $S_1^0$
which is mainly singlet here decays dominantly to $b{\bar b}$ final
states, followed by $\tau\tau$ final states.

\subsection{Decays of a gaugino-like lightest neutralino}
\label{subsec:decaysfermions}

\begin{figure}
\begin{center}
\includegraphics[width=0.6\textwidth]{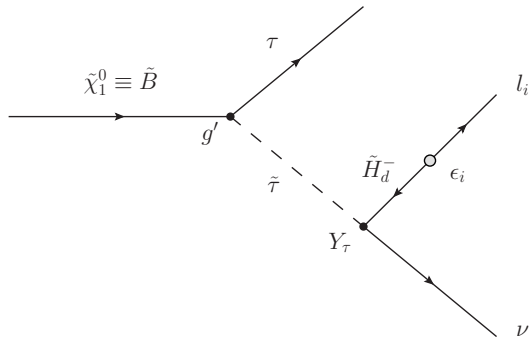}
\caption{Dominant Feynman graph for the decay $\tilde{\chi}_1^0
\rightarrow l_i\tau\nu$ with $l_i=e,\mu$.}
\label{fig:3bodydecay}
\end{center}
\end{figure}

\begin{figure}
\begin{center}
\vspace{2mm}
\includegraphics[width=0.49\textwidth]{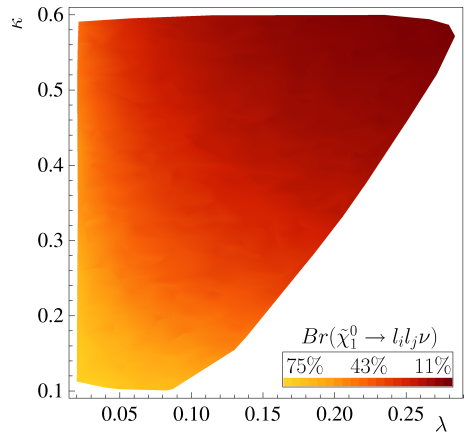}
\end{center}
\vspace{-5mm}
\caption{Dependence of allowed $\kappa(\lambda)$ for values of
$\lambda\in\left[0.02,0.5\right]$ and $\kappa\in\left[0.1,0.6\right]$
and $Br$($\tilde{\chi}_1^0\rightarrow l_il_j\nu$) as function of
$\lambda$ and $\kappa$ exemplary for SPS1a' with $\mu=390$ GeV, 
$T_\lambda=\lambda\cdot 1.5$ TeV and $T_\kappa=-\kappa\cdot 100$ GeV.}
\label{fig:1NuR_lambdakappa}
\end{figure}

We first consider the case of a bino as lightest neutralino. Although
$m(\tilde{\chi}_1^0)>m_W$ in the SPS points we have chosen, two-body
decay modes are not necessarily dominant. The three-body decay
$\tilde{\chi}_1^0\rightarrow l_il_j\nu$ dominated by a virtual
$\tilde{\tau}$ also can have a sizeable branching ratio, see Table
\ref{table:binodecays} and figure \ref{fig:1NuR_lambdakappa}. 
The importance of this final state can be understood from the 
Feynman graph shown in figure \ref{fig:3bodydecay}, giving the dominant
contribution due to $\tilde{H}_d^-$-$l_i$-mixing ($l_i=e,\mu$).

In the case $l_i=\tau$ there's an additional contribution due to
$\tilde{H}_d^0$-$\nu$-mixing. As figure \ref{fig:1NuR_lambdakappa} shows
there exist parameter 
combinations in the $\lambda$-$\kappa$-plane,
where the decay mode $\tilde{\chi}_1^0\rightarrow l_il_j\nu$ is more
important than $\tilde{\chi}_1^0\rightarrow Wl$. 
The strong variation in the branching ratios for SPS1a' is mainly due
to the strong dependence of the partial decay width of
$\tilde{\chi}_1^0\rightarrow l_il_j\nu$, where the decays with $i=j$
and $i\neq j$ both play a role. Other important final states are
$\tilde{\chi}_1^0\rightarrow Z\nu$ and in case of a light scalar with
$m(\tilde{\chi}_1^0)>m(h^0)$ the decay $\tilde{\chi}_1^0\rightarrow
h^0\nu$, as demonstrated in Table \ref{table:binodecays}.

\begin{table}
\begin{center}
{
\renewcommand\arraystretch{1.2} 
\begin{tabular}{|c|c|c|c|}
\hline
 $Br(\tilde{\chi}_1^0)$& SPS1a' & SPS3 & SPS4 \\
\hline \hline
 $Wl$ & $23-80$ & $12-55$ & $68-72$\\
 $l_il_j\nu$ & $11-75$ & $2-31$ & $2.6-3.9$ \\
 $Z\nu$ & $2.2-8.9$ & $5-28$ & $25-28$ \\
 $h^0\nu$ & $-$ & $15-53$ & $<2.0$\\
\hline
Decay length [mm] & $1.6-7.0$ & $0.1-0.5$ & $1.4-1.6$\\\hline
\end{tabular}
}
\caption{Branching ratios (in \%) and total decay length in mm of the
decay of the lightest bino-like neutralino for different values of
$\lambda\in\left[0.02,0.5\right]$ and $\kappa\in\left[0.1,0.6\right]$
with a dependence of allowed $\kappa(\lambda)$ similar to
\cite{Escudero:2008jg} and to figure \ref{fig:1NuR_lambdakappa}
and $T_\lambda=\lambda\cdot 1.5$ TeV and $T_\kappa=-\kappa\cdot 100$ GeV.}
\label{table:binodecays}
\end{center}
\end{table}

\begin{figure}
\begin{center}
\vspace{2mm}
\includegraphics[width=0.49\textwidth]{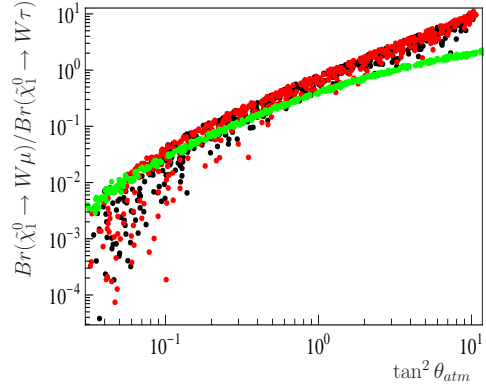} 
\end{center}
\vspace{-5mm}
\caption{Ratio $\frac{Br(\tilde{\chi}_1^0\rightarrow
W\mu)}{Br(\tilde{\chi}_1^0\rightarrow W\tau)}$ versus
$\tan^2\theta_{atm} \equiv \tan^2\theta_{23}$ for different SPS scenarios (SPS1a'
(black), SPS3 (red), SPS4 (green)) and for different values of
$\lambda\in\left[0.02,0.5\right]$ and $\kappa\in\left[0.1,0.6\right]$
with a dependence of allowed $\kappa(\lambda)$ similar to
\cite{Escudero:2008jg} and to figure \ref{fig:1NuR_lambdakappa} 
and $T_\lambda=\lambda\cdot 1.5$ TeV and $T_\kappa=-\kappa\cdot 100$ GeV.}
\label{fig:1NuR_binocorr}
\end{figure}

In the $\mu\nu$SSM one finds correlation between the decays of the
lightest neutralino and the neutrino mixing angles, because neutralino
couplings depend on the same \rpv parameters as the neutrino masses.
figure \ref{fig:1NuR_binocorr} shows the correlation between the
branching ratios of the decay $\tilde{\chi}_1^0\rightarrow Wl$ as a
function of the atmospheric angle. Although a clear correlation is 
visible it is not as pronounced as in the $n$ generation
case, see below and \cite{Ghosh:2008yh}, due to inclusion of 
1-loop effects in the neutrino masses and mixing angles. 

\begin{figure}
\begin{center}
\vspace{2mm}
\includegraphics[width=0.49\textwidth]{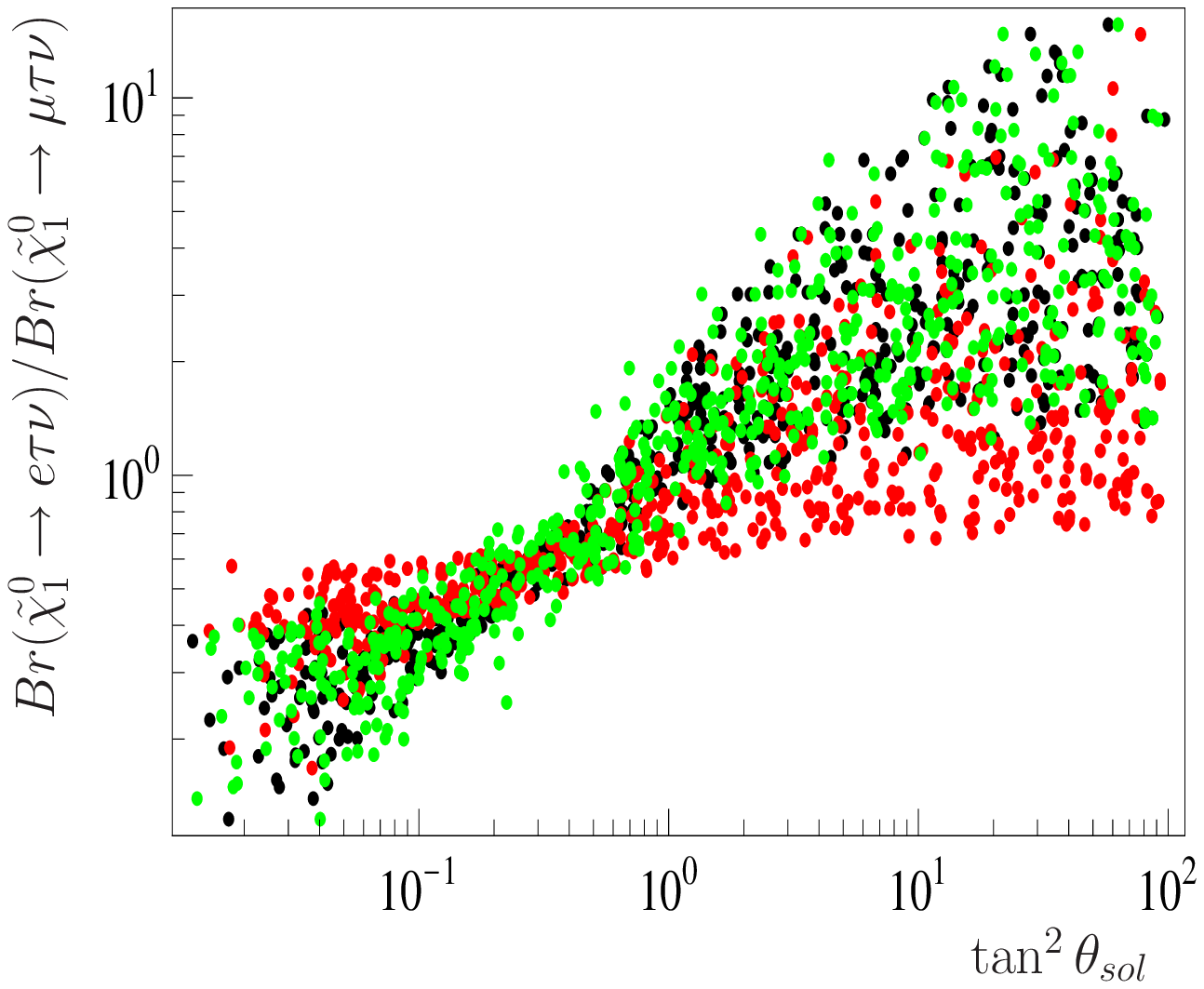}
\includegraphics[width=0.49\textwidth]{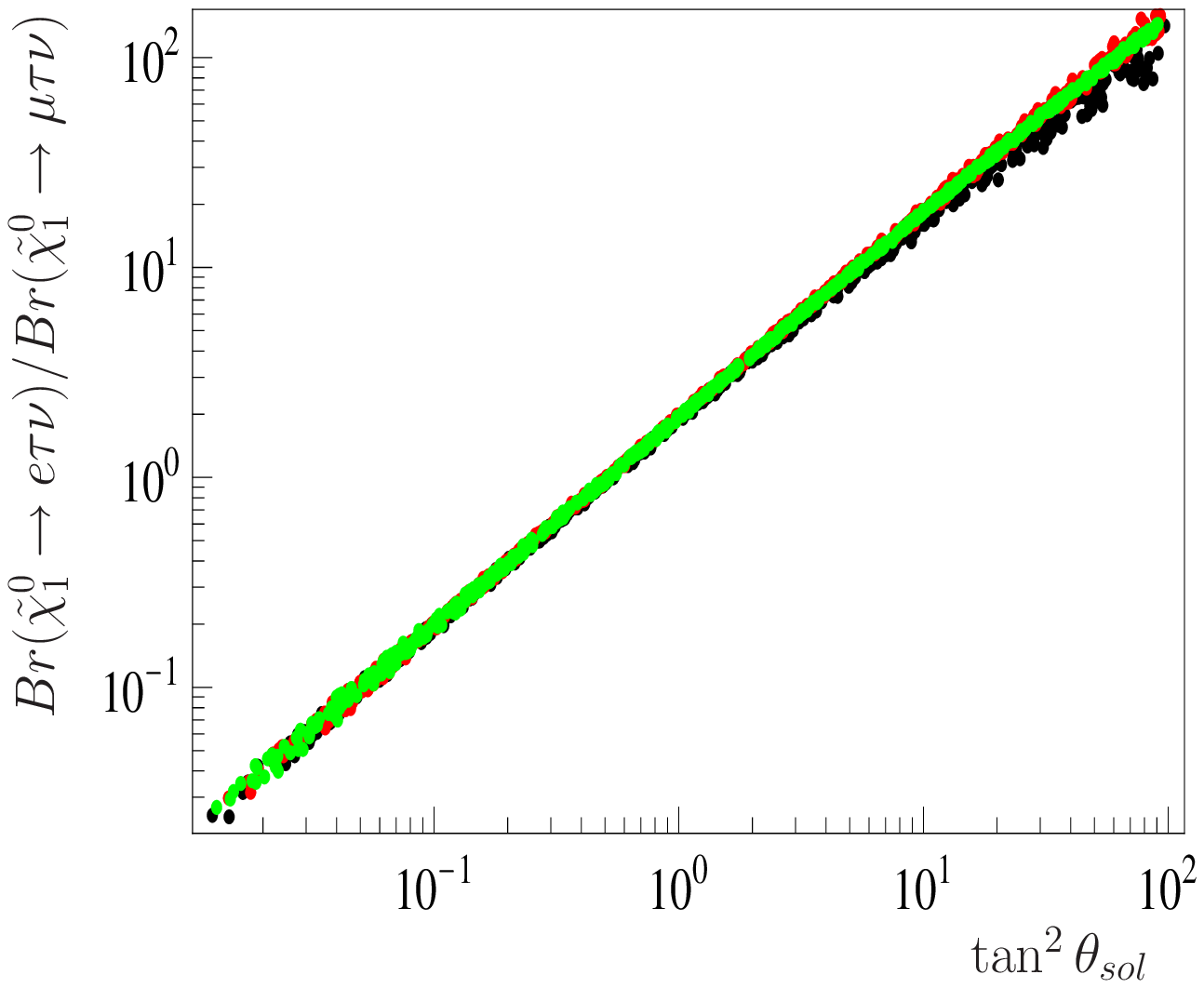} 
\end{center}
\vspace{-5mm}
\caption{Ratio $\frac{Br(\tilde{\chi}_1^0\rightarrow
e\tau\nu)}{Br(\tilde{\chi}_1^0\rightarrow \mu\tau\nu)}$ versus
$\tan^2\theta_{sol} \equiv \tan^2 \theta_{12}$ with same set of parameters as figure 
\ref{fig:1NuR_binocorr}. Bino purity $|\mathcal{N}_{41}|^2>0.97$.
To the left (a) two-body plus three-body contributions, to the right
(b) three-body contributions only. For a discussion see text.}
\label{fig:1NuR_binocorr2}
\end{figure}

Also the three-body decay $\tilde{\chi}_1^0\rightarrow l_il_j\nu$
exemplifies a correlation with neutrino physics. However, this decay 
is connected to the solar angle, see figure \ref{fig:1NuR_binocorr2}. 
There are two main contributions to this final state: $\tilde{\chi}^0_1 
\to W l \to l_il_j\nu$ and $\tilde{\chi}^0_1 \to {\tilde\tau}^* l \to l_il_j\nu$.
While the former is mainly sensitive to $\Lambda_i$, the latter is 
dominated by $\epsilon_i$-type couplings (see figure \ref{fig:3bodydecay}),
causing the connection to 
solar neutrino angle. In case the $W$ is on-shell as in the SPS1a' 
point, one could in principle devise kinematical cuts reducing this 
contribution. Such a cut can significantly improve the quality of 
the correlation.

\begin{figure}
\begin{center}
\vspace{-1mm}
\includegraphics[width=0.49\textwidth]{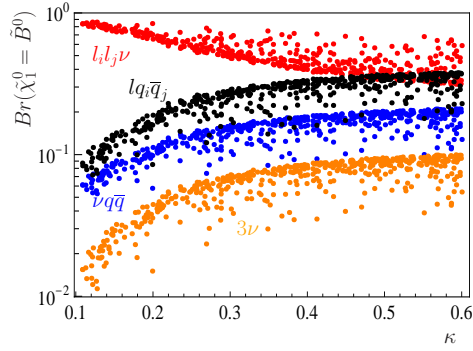}
\end{center}
\vspace{-7mm}
\caption{Decay branching ratios for bino-like lightest neutralino 
as a function of $\kappa$ for $\lambda\in\left[0.02,0.5\right]$, 
$T_\lambda=\lambda\cdot 1.5$ TeV, $T_\kappa=-\kappa \cdot 100$ GeV and for MSSM parameters 
defined by the study point SU4 of the ATLAS collaboration \cite{Aad:2009wy}. The colors
indicate the different final states: $l_il_j\nu$ (red),
$lq_i\overline{q}_j$ (black), $\nu q{\bar q}$ (blue) and $3\nu$ (orange).}
\label{fig:SU4tot}
\end{figure}

\begin{figure}[htbp]
\begin{center}
\vspace{-1mm}
\includegraphics[width=0.49\textwidth]{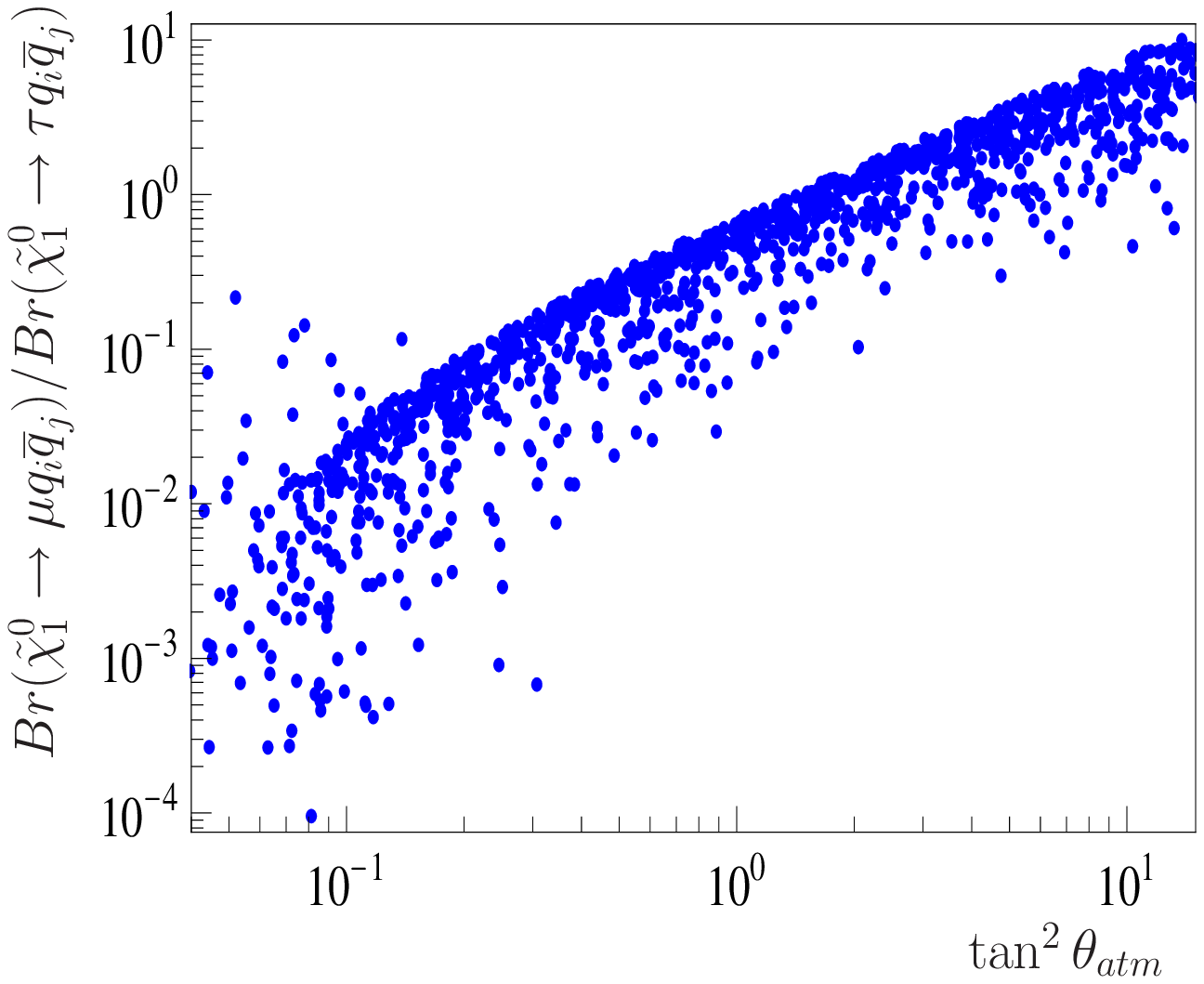} 
\includegraphics[width=0.49\textwidth]{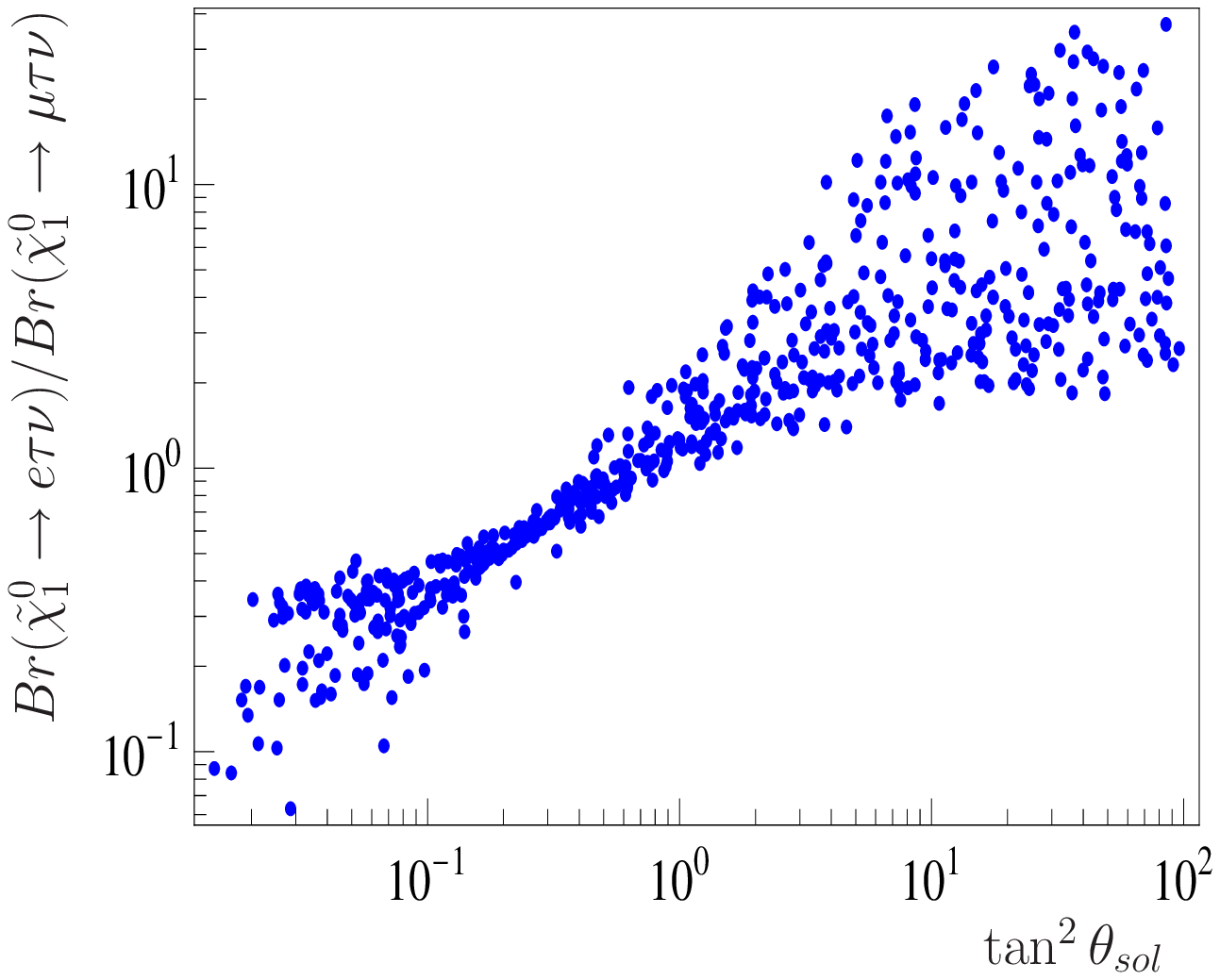} 
\end{center}
\vspace{-7mm}
\caption{To the left (a) ratio $\frac{Br(\tilde{\chi}_1^0\rightarrow
\mu q_i\overline{q}_j)}{Br(\tilde{\chi}_1^0\rightarrow
\tau q_i\overline{q}_j)}$ versus $\tan^2\theta_{atm} \equiv \tan^2\theta_{23}$ for the SU4
scenario of the ATLAS collaboration \cite{Aad:2009wy} and to the right (b) ratio
$\frac{Br(\tilde{\chi}_1^0\rightarrow
e\tau\nu)}{Br(\tilde{\chi}_1^0\rightarrow \mu\tau\nu)}$ versus
$\tan^2\theta_{sol} \equiv \tan^2 \theta_{12}$ with same set of parameters as (a).
Bino purity $|\mathcal{N}_{41}|^2>0.94$.}
\label{fig:1NuR_binoSU4corr}
\end{figure}

The SU4 scenario of the ATLAS collaboration \cite{Aad:2009wy} has 
a very light SUSY spectrum close to the Tevatron bound with a bino-like
neutralino $m(\tilde{\chi}_1^0)\approx 60$ GeV. Thus, for 
SU4 the lightest neutralino has only three-body decay modes. 
Most important branching ratios are shown in figure \ref{fig:SU4tot}.
The lightness of the bino-like neutralino $\tilde{\chi}_1^0$ in this
scenario implies a larger average decay length of $(8-90)$ cm,
depending on the parameter point in the $\lambda$-$\kappa$-plane. Note
that the decay length becomes smaller for smaller values of
$\lambda,\kappa$. In general the decay length scales as 
$L \propto m^{-4}(\tilde{\chi}^0_1)$ for $m(\tilde{\chi}^0_1) < m_W$.
Also for this point a correlation between the branching ratios 
and the neutrino mixing angles is found 
as illustrated in figure \ref{fig:1NuR_binoSU4corr}.

\begin{figure}
\begin{center}
\vspace{-1mm}
\includegraphics[width=0.49\textwidth]{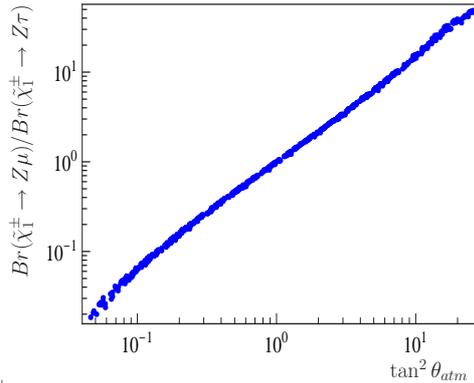}
\end{center}
\vspace{-8mm}
\caption{Ratio $\frac{Br(\tilde{\chi}_1^\pm\rightarrow
Z\mu)}{Br(\tilde{\chi}_1^\pm\rightarrow Z\tau)}$ versus
$\tan^2\theta_{atm} \equiv \tan^2\theta_{23}$ for the AMSB scenario SPS9 and for 
different values of
$\lambda\in\left[0.02,0.5\right]$, $\kappa\in\left[0.1,0.6\right]$, 
$T_\lambda=\lambda\cdot 1.5$ TeV and $T_\kappa=-\kappa \cdot 100$ GeV.}
\label{fig:1NuR_Winoatm}
\end{figure}

In addition to the SUGRA scenarios discussed up to now we have 
also studied SPS9, which is a typical AMSB point. The most 
important difference between this point and the previously discussed 
cases is the near degeneracy between lightest neutralino and 
lightest chargino. This near degeneracy is the reason that the chargino 
decay is dominated by \rpv final states. Varying 
$\lambda$ and $\kappa$ as before we find a total decay length of
$(0.12-0.16)$mm with $Br(\tilde{\chi}_1^\pm\rightarrow
W\nu)=(42-57)$\%, $Br(\tilde{\chi}_1^\pm\rightarrow Zl)=(20-26)$\% and
$Br(\tilde{\chi}_1^\pm\rightarrow h^0l)=(17-40)$\%. This is 
especially interesting since, similar to $Wl$ in case of the gaugino-like
lightest neutralino, the decay to $Zl$ of the chargino is linked 
to the atmospheric angle, see figure \ref{fig:1NuR_Winoatm}.

\subsection{Decays of a singlino-like lightest neutralino}

\begin{figure}
\begin{center}
\vspace{0mm}
\includegraphics[width=0.49\textwidth]{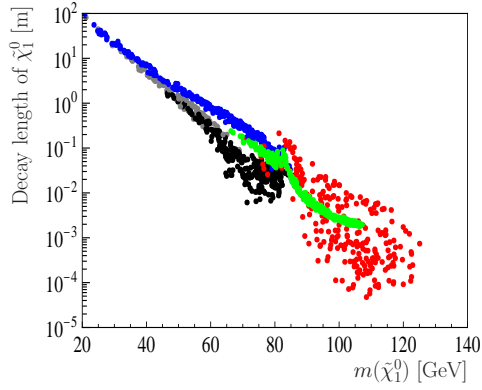}
\end{center}
\vspace{-5mm}
\caption{Decay length of the lightest neutralino $\tilde{\chi}_1^0$ in
m as a function of its mass $m(\tilde{\chi}_1^0)$ in GeV for different
values of $\lambda\in\left[0.2,0.5\right]$,
$\kappa\in\left[0.025,0.2\right]$ and $\mu\in[110,170]$ GeV with a
dependence of allowed $\kappa(\lambda)$ similar to
\cite{Escudero:2008jg} and to figure \ref{fig:1NuR_lambdakappa}
and $T_\lambda=\lambda\cdot 1.5$ TeV, whereas $T_\kappa\in [-20,-0.05]$ GeV is
chosen in such a way, that no lighter scalar or pseudoscalar states
with $\lbrace m(S_1^0),m(P_1^0)\rbrace <m(\tilde{\chi}_1^0)$ appear.
Note that the different colors stand for SPS1a' (real singlino,
$|\mathcal{N}_{45}|^2>0.5$) (gray), SPS1a' (mixture state) (black),
SPS3 (real singlino) (blue), SPS3 (mixture state) (red) and SPS4
(mixture state) (green).}
\label{fig:1NuR_decaylength}
\end{figure}

We now turn to the case of a singlino-like LSP. As already explained,
this scenario is connected to
a light singlet scalar and pseudoscalar. Recall, that the particles in
the fermionic sector are mixed for
$\lambda,\kappa=\mathcal{O}(10^{-1})$ due to the reduced
$\mu$-parameter as can be seen in figure
\ref{fig:1NuR_mixingtotal}. We will first discuss the average decay
length of the lightest neutralino $\tilde{\chi}_1^0$. Figure
\ref{fig:1NuR_decaylength} shows the average decay length in meter for
different SPS scenarios as a function of the mass of the lightest
neutralino $m(\tilde{\chi}_1^0)$.  Composition of the neutralino is
indicated by color code, as given in the caption. $\lambda$,
$\kappa$, $T_\kappa$ and $\mu$ are varied in this plot. Note that by
variation of $T_\kappa$ the parameter points in figure
\ref{fig:1NuR_decaylength} are chosen in such a way, that all scalar
and pseudoscalar states are heavier than the lightest neutralino.
Singlino purity in this plot increases with decreasing mass and for
pure singlinos the decay length is mainly determined by its mass and
the experimentally determined neutrino masses.  For neutralino masses 
below about 50 GeV decay lengths become larger than 1 meter, implying 
that a large fraction of neutralinos will decay outside typical 
collider detectors. Note that if one
allows for lighter scalar states so that at least one of the decays
$\tilde{\chi}_1^0\rightarrow S_1^0(P_1^0)\nu$ appears, the average
decay length can be easily reduced by several orders of magnitude.

\begin{figure}
\begin{center}
\vspace{2mm}
\includegraphics[width=0.49\textwidth]{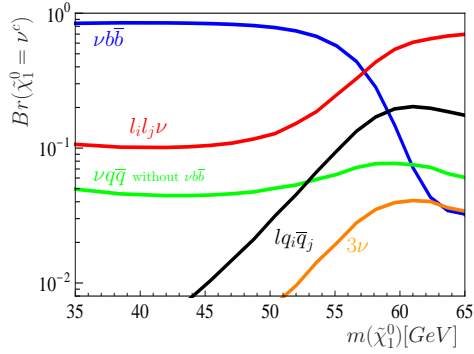}
\end{center}
\vspace{-5mm}
\caption{Singlino decay branching ratios as a function of its mass, 
for the same parameter choices as in figure \ref{fig:1NuR_decayscalar}. 
The colors indicate the different final states: $\nu b{\bar b}$ (blue), $l_il_j\nu$ (red), 
$lq_i\overline{q}_j$ (black), $3\nu$ (orange) and $\nu q{\bar q}$ ($q \ne b$, 
green).}  
\label{fig:1NuR_singlinodecay}
\end{figure}

\begin{figure}
\begin{center}
\vspace{2mm}
\includegraphics[width=0.49\textwidth]{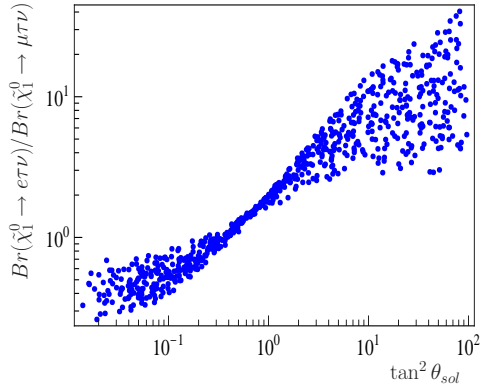} 
\end{center}
\vspace{-5mm}
\caption{Ratio $\frac{Br(\tilde{\chi}_1^0\rightarrow
e\tau\nu)}{Br(\tilde{\chi}_1^0\rightarrow \mu\tau\nu)}$ versus
$\tan^2\theta_{sol} \equiv \tan^2 \theta_{12}$ for the SPS1a' scenario and
$\lambda\in\left[0.2,0.5\right]$, $\mu\in[110,170]$ GeV,
$\kappa=0.035$, $T_\lambda=\lambda\cdot 1.5$ TeV and $T_\kappa=-0.7$ GeV.}
\label{fig:1NuR_singlinocorr}
\end{figure}

Again typical decays are $Wl$, $lq_i\overline{q}_j$, $Z\nu$, $\nu
q\overline{q}$, $l_il_j\nu$ and the invisible decay to $3\nu$.  For
the region of $m(\tilde{\chi}^0_1)$ below the $W$ threshold see figure
\ref{fig:1NuR_singlinodecay}. The dominance of $\nu b\overline{b}$
for smaller values of $m(\tilde{\chi}_1^0)$ is due to the decay chain
$\tilde{\chi}_1^0\rightarrow S_1^0 \nu\rightarrow \nu b\overline{b}$,
whereas for larger values of $m(\tilde{\chi}_1^0)$ we find
$m(S_1^0)>m(\tilde{\chi}_1^0)$. Final state ratios show correlations
with neutrino physics also in this case.  As an example we show
$l_il_j\nu$ branching ratios versus the solar neutrino mixing angle in figure
\ref{fig:1NuR_singlinocorr}. Singlino purity for this plot
$|\mathcal{N}_{45}|^2\in[0.75,0.83]$ and mass
$m(\tilde{\chi}_1^0)\in[22,53]$ GeV. The absolute values for the
branching ratios are comparable to those of the described SU4 scenario
with a bino-like lightest neutralino.
We note that for the parameters in figure \ref{fig:1NuR_singlinocorr} 
the light Higgs $S_2^0= h^0$ decays to $\tilde{\chi}_1^0\tilde{\chi}_1^0$ 
with a branching ratio of $Br(S_2^0= h^0\rightarrow
\tilde{\chi}_1^0\tilde{\chi}_1^0)=(21-91)$\%.

Up to now we have considered values of $\lambda$ and $\kappa$ larger 
than $10^{-2}$. For very small values of these couplings, the singlet 
sector, although very light, effectively decouples.
This implies that R-parity conserving decays of $\tilde{\chi}_2^0$, e.g. 
decays to final states like $\tilde{\chi}_1^0S_1^0$, $\tilde{\chi}_1^0P_1^0$,
$\tilde{\chi}_1^0l^+l^-$ or $\tilde{\chi}_1^0 q\overline{q}$, are strongly
suppressed and the \rpv decay modes dominate, implying decays with correlations
as in the case of the explicit b-\rpv.

\section{Phenomenology of the $n$ $\widehat{\nu}^c$ model}
\label{sec:nNuC}

In the previous section the phenomenology for the one generation case
of the model has been worked out in detail. Most of the signals discussed 
so far are independent of the number of right-handed neutrinos. 
However, the $n$ generation variants also offer some additional 
phenomenology, which we discuss here for the simplified case of $n=2$. 

In the $\mu \nu$SSM with one right-handed neutrino superfield a light singlino
will always imply a light scalar/pseudoscalar. This connection between
the neutral fermion sector and scalar/pseudoscalar sector is a
well-known property of the NMSSM (see again 
\cite{Franke:1995xn,Miller:2003ay}). In the $\mu \nu$SSM with more than one generation of
singlets, the off-diagonal $T_\kappa$ terms in equation
\eqref{eq:softsing} induce mixing between the different generations of
singlet scalars and pseudoscalars.  This opens up the possibility, not
considered in previous publications
\cite{LopezFogliani:2005yw,Escudero:2008jg,Ghosh:2008yh}, to have the
singlet scalars considerably heavier than the singlet fermions.

\begin{figure}
\begin{center}
\vspace{0mm}
\includegraphics[width=0.49\textwidth]{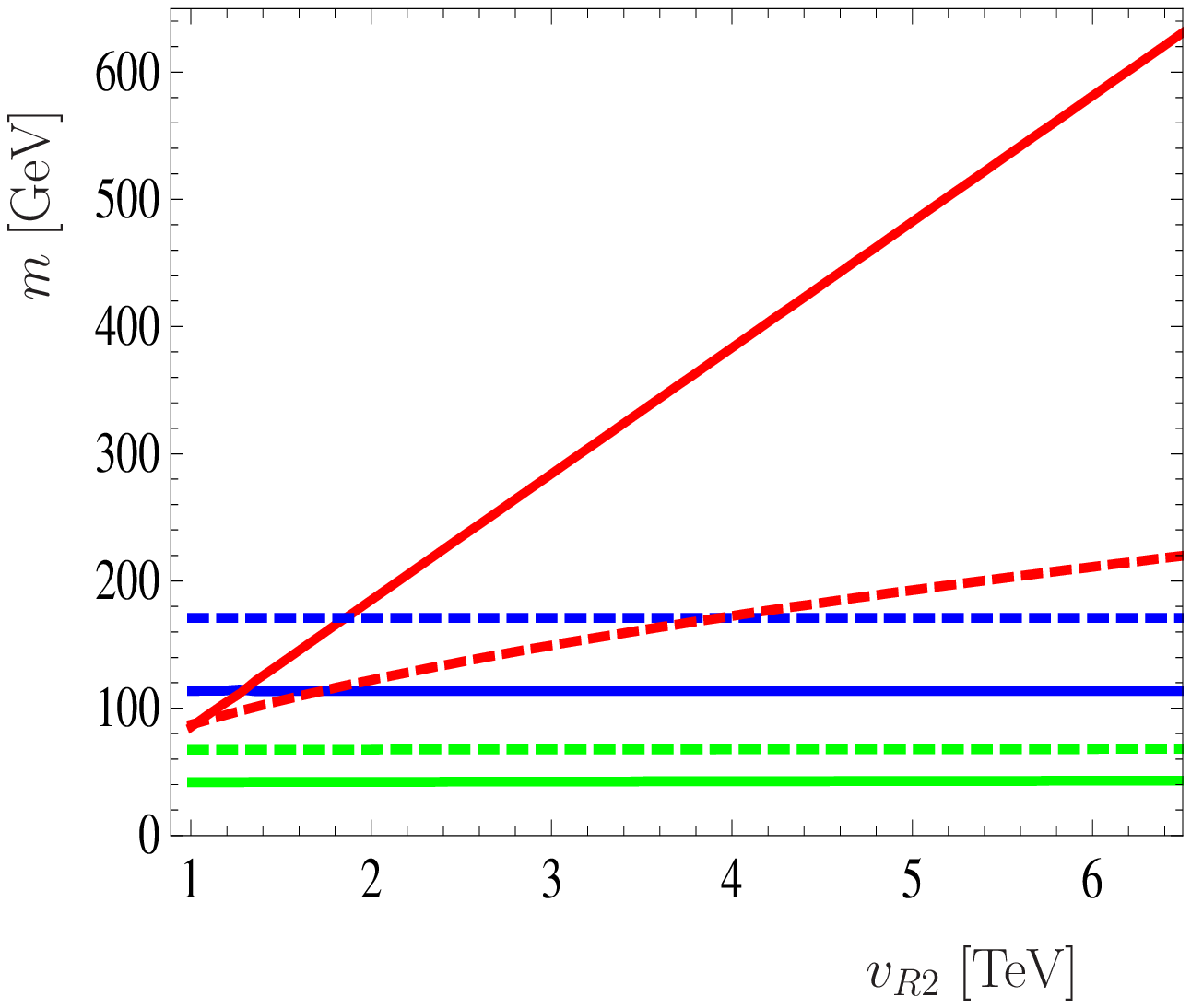}
\includegraphics[width=0.49\textwidth]{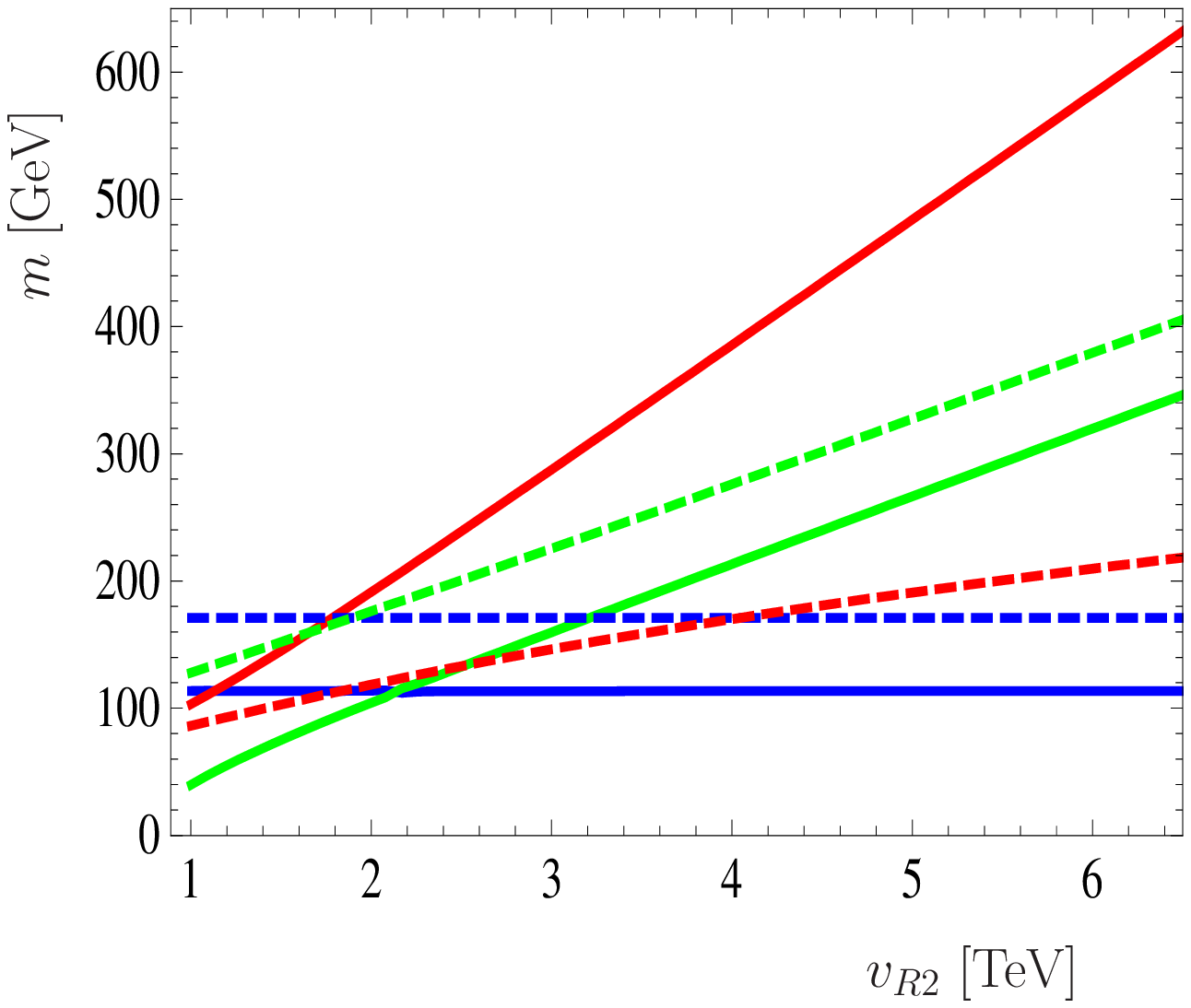}
\end{center}
\vspace{-7mm}
\caption{Masses of the scalar states $Re(\tilde{\nu}_1^c)$ (green),
$Re(\tilde{\nu}_2^c)$ (red) and $h^0$ (blue) and the pseudoscalar
states $Im(\tilde{\nu}_1^c)$ (dashed green), $Im(\tilde{\nu}_2^c)$
(dashed red) and $Im(\tilde{\nu}_1)$ (dashed blue) as a function of
$v_{R2}$ for different values of $T_\kappa^{112}=T_\kappa^{122}$. To
the left (a) $T_\kappa^{112}=T_\kappa^{122}=0$ whereas to the right (b)
$T_\kappa^{112}=T_\kappa^{122}=-2$ GeV. The MSSM parameters have been
taken such that the standard SPS1a' point is reproduced. The light
singlet parameters $\kappa_1 = 0.16$ and $v_{R1} = 500$ GeV ensure
that in all points the lightest neutralino is mostly $\nu_1^c$, with a
mass of $47-48$ GeV. In addition, $T_\lambda^1 = 300$ GeV and
$T_\lambda^2 \in [10,200]$ GeV.}
\label{fig:tkappa}
\end{figure}

Let us illustrate this feature with a simple example. Imagine a light
singlino $\nu_1^c$, and a heavy singlino $\nu_2^c$, in a model with
non-zero trilinear couplings $T_\kappa^{112}$. In that case, the
contributions to the mass of the $\tilde{\nu}_1^c$, scalar or
pseudoscalar, coming from the large value of $v_{R2}$ are proportional
to $T_\kappa^{112}$.  Without these contributions the mass of
$\tilde{\nu}_1^c$ would only depend on the small $v_{R1}$, thus making
it light like the singlino of the same generation. With non-zero
$T_\kappa^{112}$ the mass of both $\tilde{\nu}_s^c$ are dominated by the
larger of the $v_{Rs}$.  This feature is demonstrated in figure
\ref{fig:tkappa}. In the two plots the lightest neutralino is mostly
$\nu_1^c$, with a mass of $\sim 50$ GeV. These plots show the
dependence of the masses of the singlet scalar states
$Re(\tilde{\nu}_1^c)$ and $Re(\tilde{\nu}_2^c)$ and the corresponding
pseudoscalar states $Im(\tilde{\nu}_1^c)$ and $Im(\tilde{\nu}_2^c)$
with $v_{R2}$ for different values of
$T_\kappa^{112}=T_\kappa^{122}$. The masses of the light Higgs boson
$h^0$ and the lightest left-handed sneutrino $Im(\tilde{\nu}_1)$ are
also shown for reference. Note that for
$T_\kappa^{112}=T_\kappa^{122}=0$ the mass of the state
$Re(\tilde{\nu}_1^c)$ does not depend on $v_{R2}$, whereas for
$T_\kappa^{112}=T_\kappa^{122}=-2$ GeV the lightest singlet scalar
becomes heavier for larger values of $v_{R2}$. The same feature is
present in the pseudoscalar sector, where the effect is even 
more pronounced.

\subsection{Correlations with neutrino mixing angles in the 
$n$ $\widehat{\nu}^c$-model}

The connection between decays and neutrino angles is not a
particular property of the $1$ $\widehat{\nu}^c$-model and is also
present in a general $n$ $\widehat{\nu}^c$-model. However,
since the structure of the approximate couplings
$\tilde{\chi}_1^0-W^{\pm}-l^{\mp}_i$ is different, see appendix
\ref{munuapp4}, we encounter additional features for $n=2$. 

As explained in section \ref{subsec:ngenneut}, we have now 
two possibilities to fit neutrino
data. If the dominant contribution to the neutrino mass matrix comes
from the $\Lambda_i \Lambda_j$ term in equation \eqref{eq:efftwo} one
can link it to the atmospheric mass scale, using the $\alpha_i\alpha_j$ term to
fit the solar mass scale. This case will be called option fit1. On the
other hand, if the dominant contribution is given by the $\alpha_i
\alpha_j$ term one has the opposite situation, where the atmospheric
scale is fitted by the $\alpha_i$ parameters and the solar scale is
fitted by the $\Lambda_i$ parameters. This case will be called option
fit2.

\begin{figure}
\begin{center}
\vspace{0mm}
\includegraphics[width=0.49\textwidth]{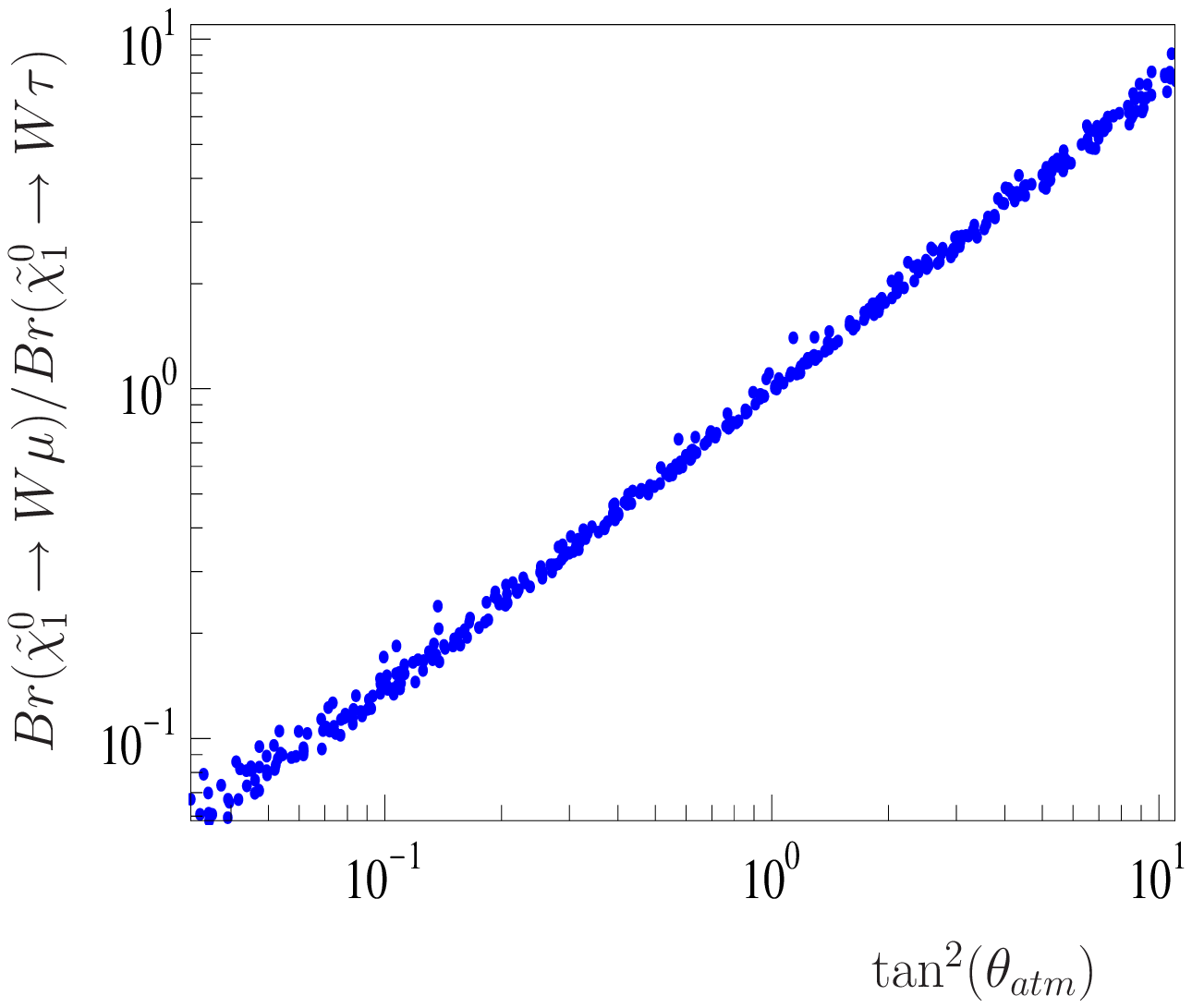}
\includegraphics[width=0.49\textwidth]{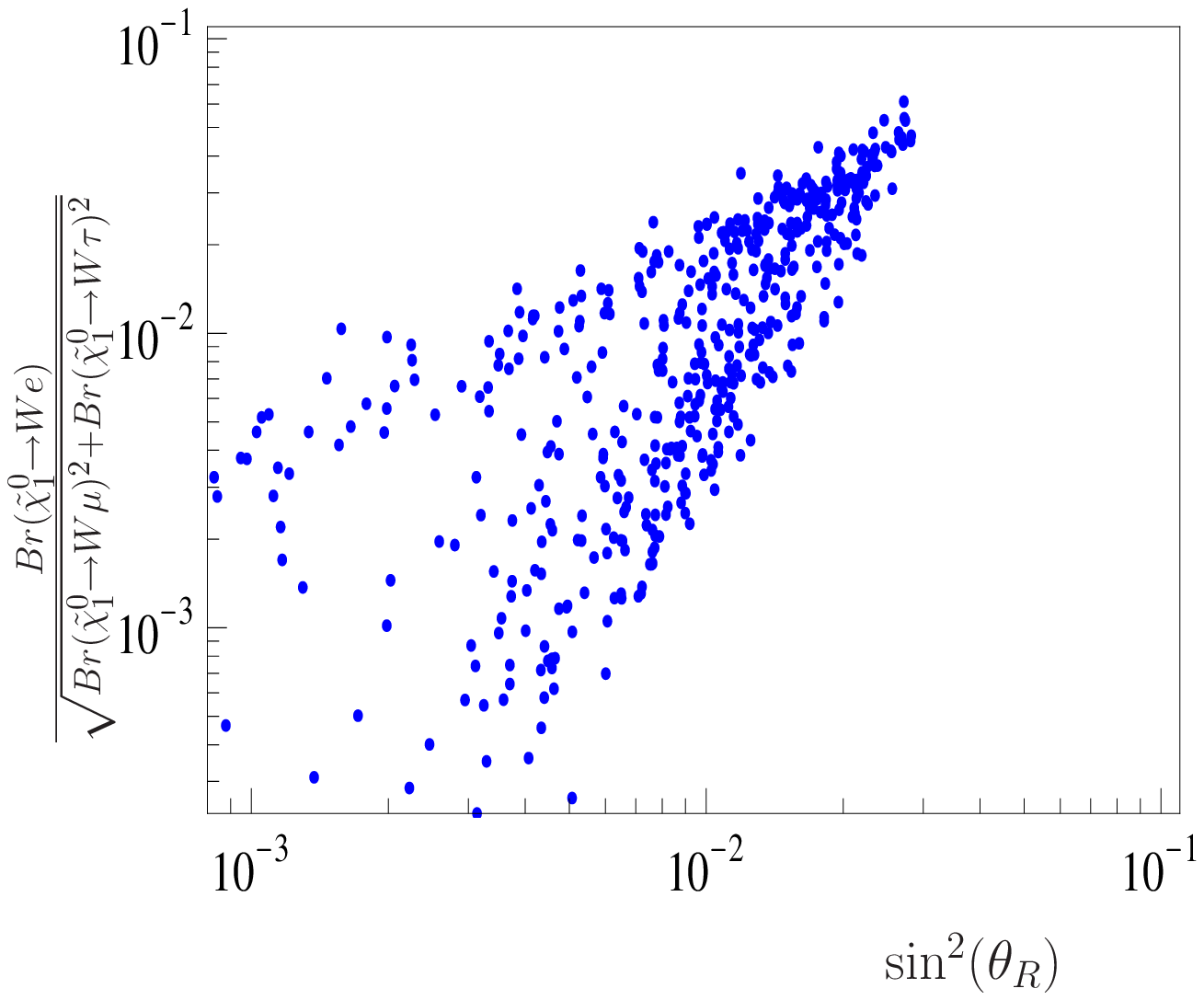}
\end{center}
\vspace{-6mm}
\caption{To the left (a) ratio $\frac{Br(\tilde{\chi}^0_1\to 
W\mu )}{Br(\tilde{\chi}^0_1\to W\tau)}$ versus $\tan^2 \theta_{atm} \equiv \tan^2\theta_{23}$
and to the right (b) ratio $\frac{Br(\tilde{\chi}^0_1\to 
We)}{\sqrt{Br(\tilde{\chi}^0_1\to W\mu )^2+Br(\tilde{\chi}^0_1\to 
W\tau )^2}}$ versus $\sin^2 \theta_R \equiv \sin^2 \theta_{13}$ for a bino LSP. Bino purity
$|\mathcal{N}_{41}|^2 > 0.9$. Neutrino data is fitted using option
fit1.}
\label{fig:binofit1}
\end{figure}

\begin{figure}
\begin{center}
\vspace{0mm}
\includegraphics[width=0.49\textwidth]{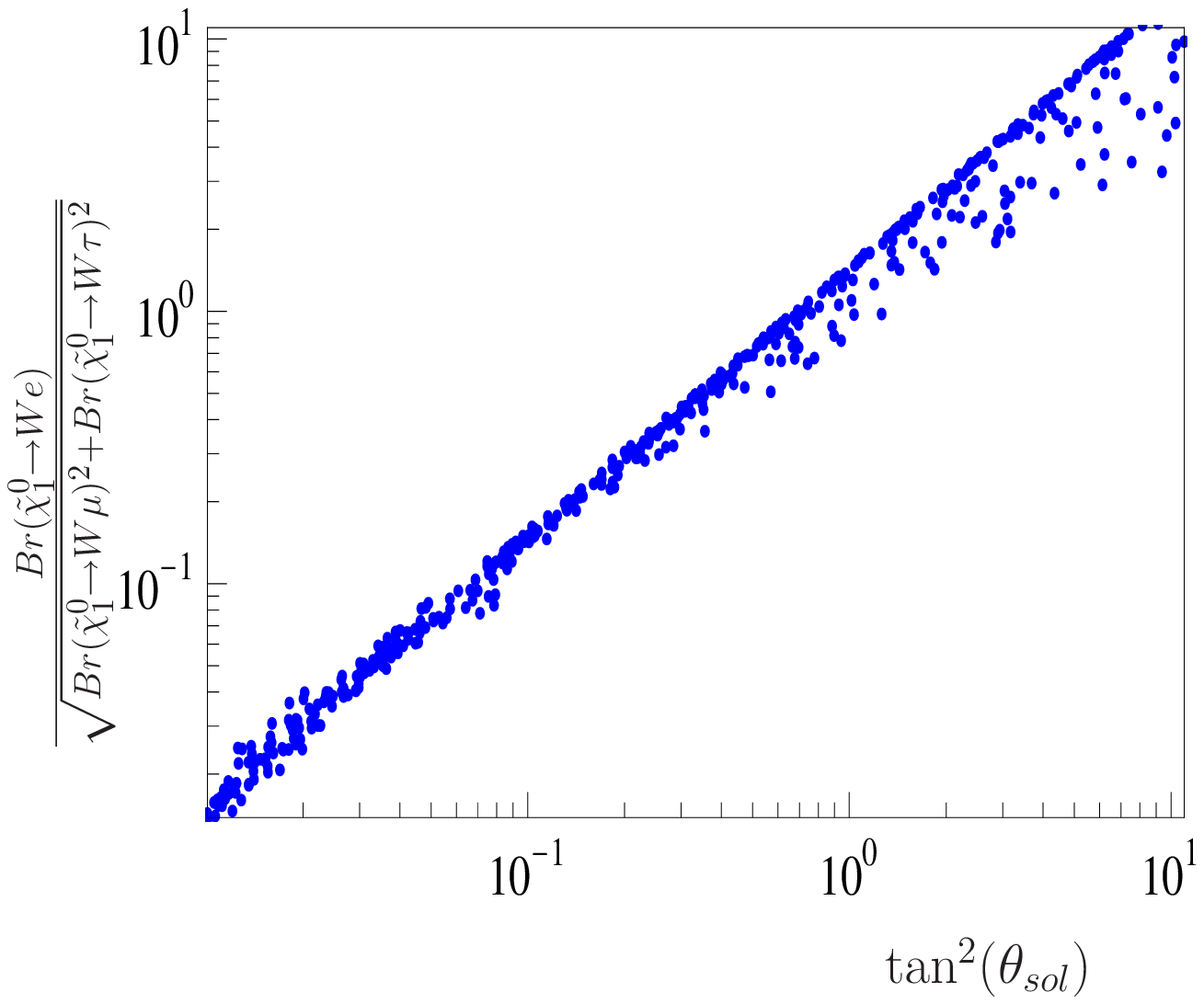}
\end{center}
\vspace{-7mm}
\caption{Ratio $\frac{Br(\tilde{\chi}^0_1\to 
We)}{\sqrt{Br(\tilde{\chi}^0_1\to W\mu )^2+Br(\tilde{\chi}^0_1\to 
W\tau )^2}}$ versus $\tan^2 \theta_{sol} \equiv \tan^2 \theta_{12}$ for a bino LSP. Bino purity
$|\mathcal{N}_{41}|^2 > 0.9$. Neutrino data is fitted using option
fit2.}
\label{fig:binofit2}
\end{figure}

For the case of a bino-like lightest neutralino one can
show that the coupling is proportional to $\Lambda_i$ whereas for the
case of a singlino-like lightest neutralino the dependence is on
$\alpha_i$, as shown in appendix \ref{munuapp4}. 
Figure \ref{fig:binofit1} shows the ratio
$\frac{Br(\tilde{\chi}^0_1\to W\mu)}{Br(\tilde{\chi}^0_1\to W\tau)}$
versus $\tan^2(\theta_{23})$ (left) and $\frac{Br(\tilde{\chi}^0_1\to
We)}{\sqrt{Br(\tilde{\chi}^0_1\to W\mu)^2+Br(\tilde{\chi}^0_1\to
W\tau)^2}}$ versus $\sin^2(\theta_{13})$ (right) for a bino LSP and
option fit1. The correlation with the
atmospheric angle and the upper bound on $\frac{Br(\tilde{\chi}^0_1\to 
We)}{\sqrt{Br(\tilde{\chi}^0_1\to W\mu)^2+Br(\tilde{\chi}^0_1\to 
W\tau)^2}}$ from $\sin^2(\theta_{13})$ is more pronounced than in the $1$ $\hat\nu^c$-model,
because we fit neutrino data with tree-level physics only.
Recall that this implies that the ratio $|\vec\epsilon|^2/|\vec\Lambda|$ is 
much smaller than in the plots shown in the previous section. 
A correlation between $\frac{Br(\tilde{\chi}^0_1\to 
We)}{\sqrt{Br(\tilde{\chi}^0_1\to W\mu)^2+Br(\tilde{\chi}^0_1\to 
W\tau)^2}}$ and $\tan^2 \theta_{12}$ is found instead, if neutrino data
is fitted with option fit2, as figure \ref{fig:binofit2} shows.

\begin{figure}
\begin{center}
\vspace{-3mm}
\includegraphics[width=0.49\textwidth]{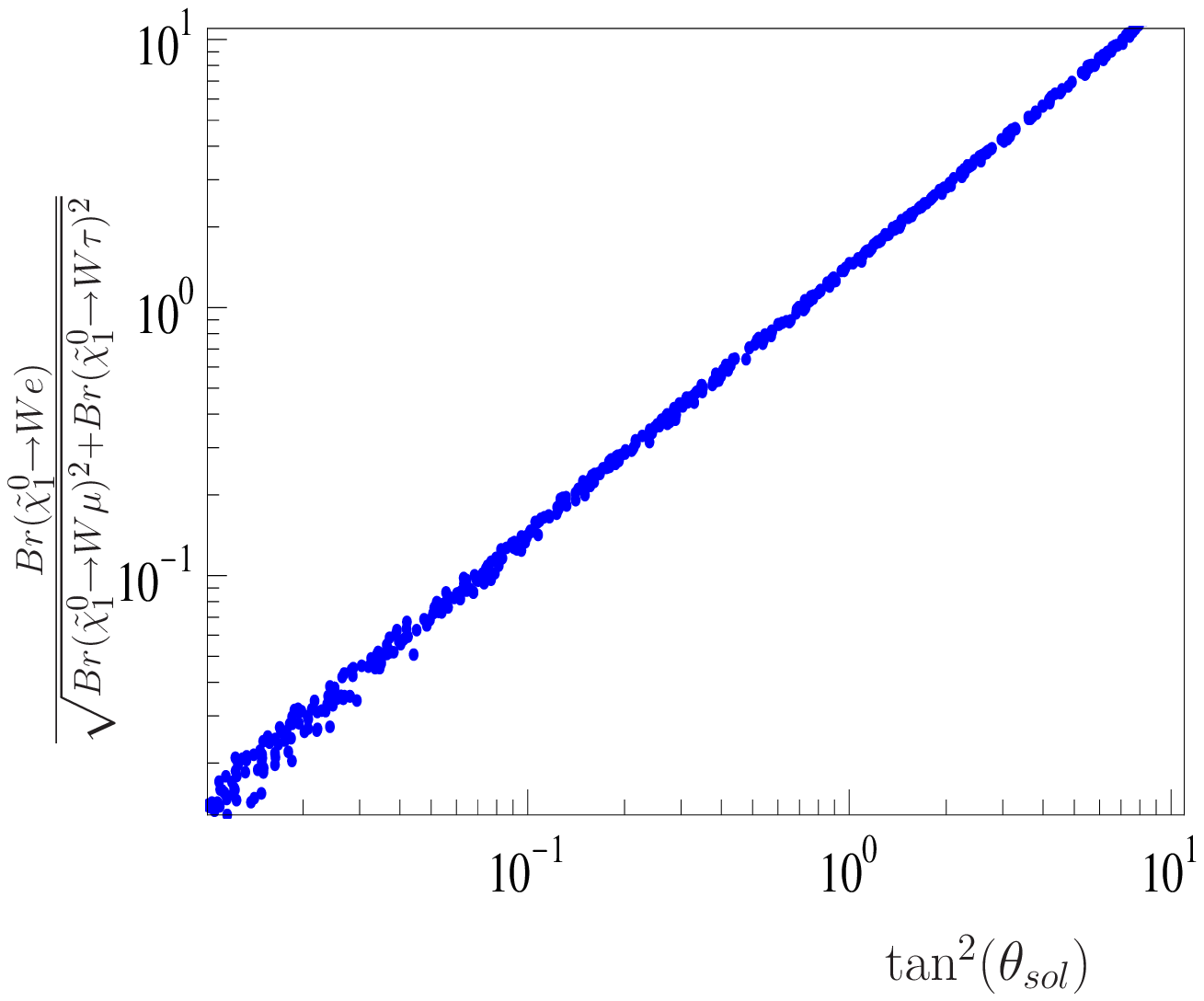}
\end{center}
\vspace{-8mm}
\caption{Ratio $\frac{Br(\tilde{\chi}^0_1\to 
We)}{\sqrt{Br(\tilde{\chi}^0_1\to W\mu )^2+Br(\tilde{\chi}^0_1\to W \tau 
)^2}}$ versus $\tan^2 \theta_{sol} \equiv \tan^2 \theta_{12}$ for a singlino
LSP. Singlino purity $|\mathcal{N}_{45}|^2 > 0.9$. Neutrino data is
fitted using option fit1.}
\label{fig:singlinofit1}
\end{figure}

\begin{figure}
\begin{center}
\vspace{0mm}
\includegraphics[width=0.49\textwidth]{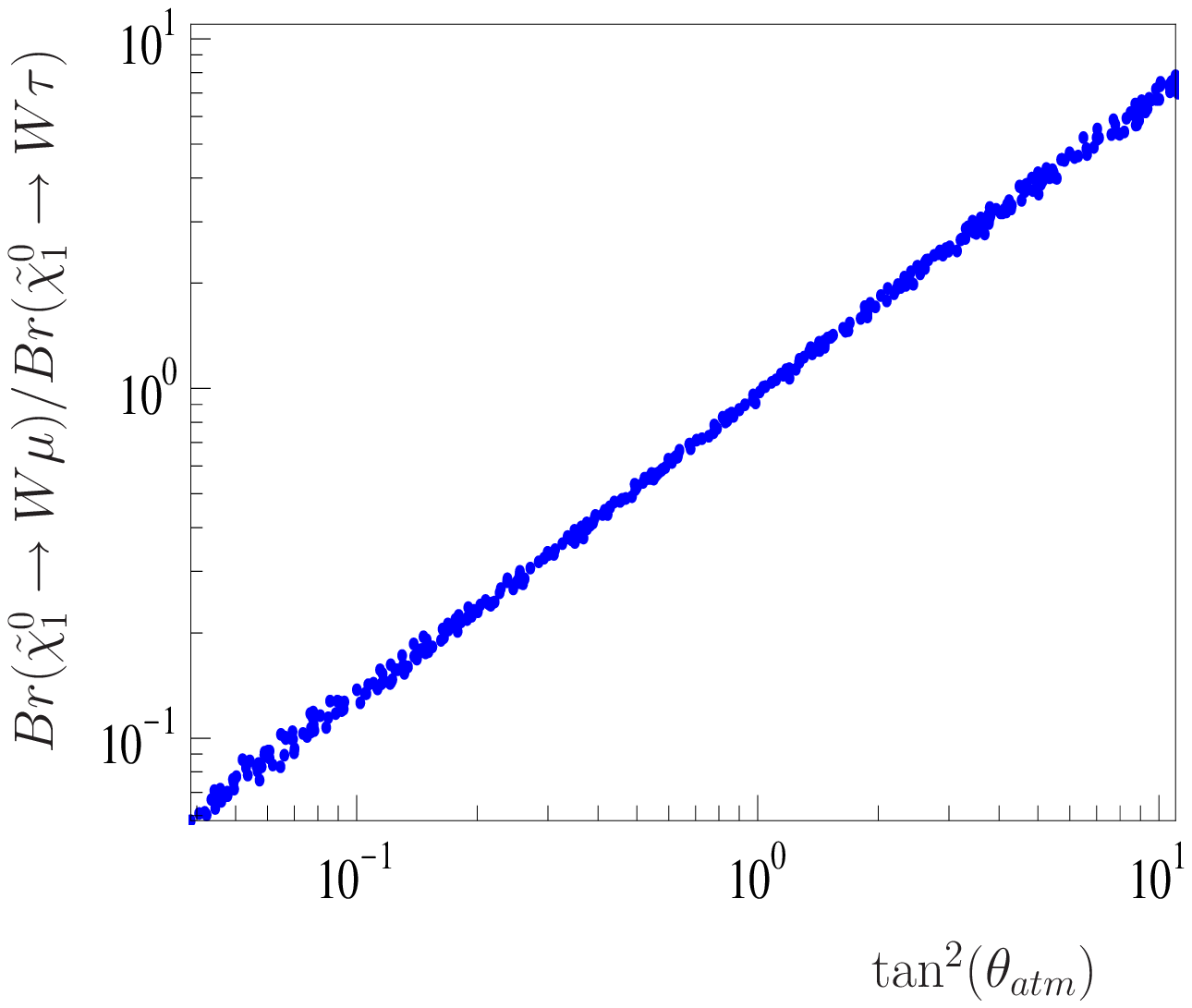}
\includegraphics[width=0.49\textwidth]{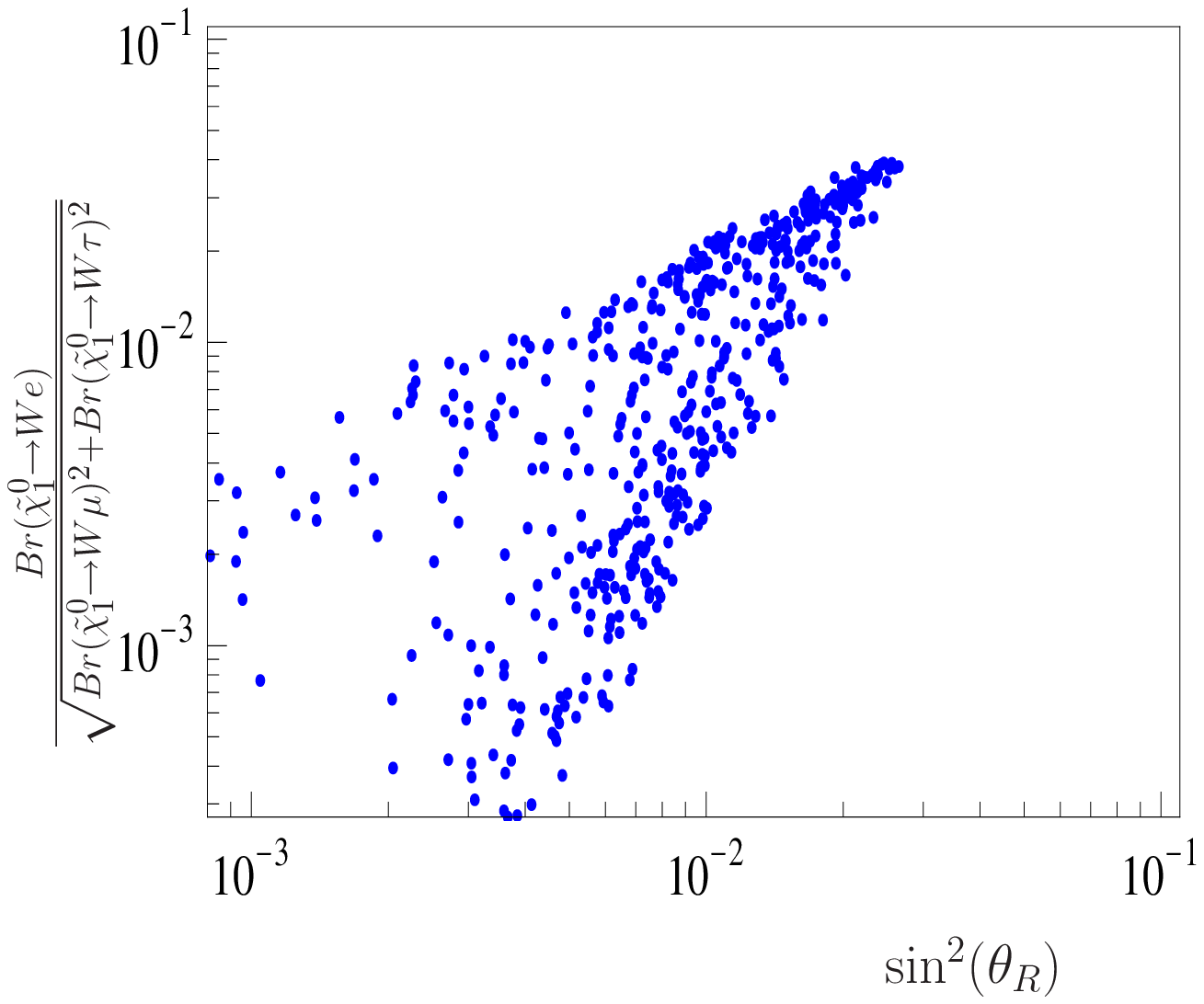}
\end{center}
\vspace{-8mm}
\caption{To the left (a) ratio $\frac{Br(\tilde{\chi}^0_1\to 
W \mu)}{Br(\tilde{\chi}^0_1\to W \tau )}$ versus $\tan^2 \theta_{atm} \equiv \tan^2\theta_{23}$
and to the right (b) ratio $\frac{Br(\tilde{\chi}^0_1\to 
We)}{\sqrt{Br(\tilde{\chi}^0_1\to W\mu )^2+Br(\tilde{\chi}^0_1\to 
W \tau)^2}}$ versus $\sin^2 \theta_R \equiv \sin^2 \theta_{13}$ for a singlino
LSP. Singlino purity $|\mathcal{N}_{45}|^2 > 0.9$. Neutrino data is
fitted using option fit2.}
\label{fig:singlinofit2}
\end{figure}

For the case of a singlino LSP the correlations and types of fit to
neutrino data are swapped with respect to the gaugino case. Since the
couplings $\tilde{\chi}_1^0-W^{\pm}-l^{\mp}_i$ are mainly proportional
to $\alpha_i$, instead of $\Lambda_i$, a scenario with a singlino LSP
and option fit1 (fit2) will be similar to bino LSP and option fit2
(fit1). This similarity is demonstrated in figures
\ref{fig:singlinofit1} and \ref{fig:singlinofit2}.  To decide which
case is realized in nature, one would need to determine the particle
character of the lightest neutralino. This might be difficult at the
LHC, but could be determined by a cross section measurement at the
ILC. We want to note, that in the $2$ $\widehat{\nu}^c$-model
we cannot reproduce all correlations for a singlino LSP presented
for the $3$ $\widehat{\nu}^c$-model in \cite{Ghosh:2008yh}.

The results shown so far in this section were all calculated for 
the SPS1a' scenario. We have checked explicitly that for all the 
other standard points results remain unchanged. 
We have also checked that for a LSP with a mass below $m_W$ the
three-body decays $\tilde{\chi}_1^0 \to l q_i \bar{q}_j$, mediated by
virtual W bosons, show the same correlations.

A final comment is in order. In a $n$ $\widehat{\nu}^c$-model with
$n>2$, the effective neutrino mass matrix will have additional terms
with respect to \eqref{eq:efftwo}, due to the contributions coming
from the new right-handed neutrinos. For this richer structure there
is one additional contribution to $m_{eff}^{\nu \nu}$,
which could be sub-dominant. Therefore, one can imagine a scenario 
in which a third generation of singlets produces a negligible 
contribution to neutrino masses while the corresponding singlino, 
$\nu_3^c$, is the LSP. In such a scenario the correlations 
between the $\nu_3^c$ LSP decays and the neutrino mixing angles 
will be lost. 

\subsection{$\tilde{\chi}_1^0$ decay length and type of fit}
\label{subsec:declengfit}

\begin{figure}
\begin{center}
\vspace{0mm}
\includegraphics[width=0.49\textwidth]{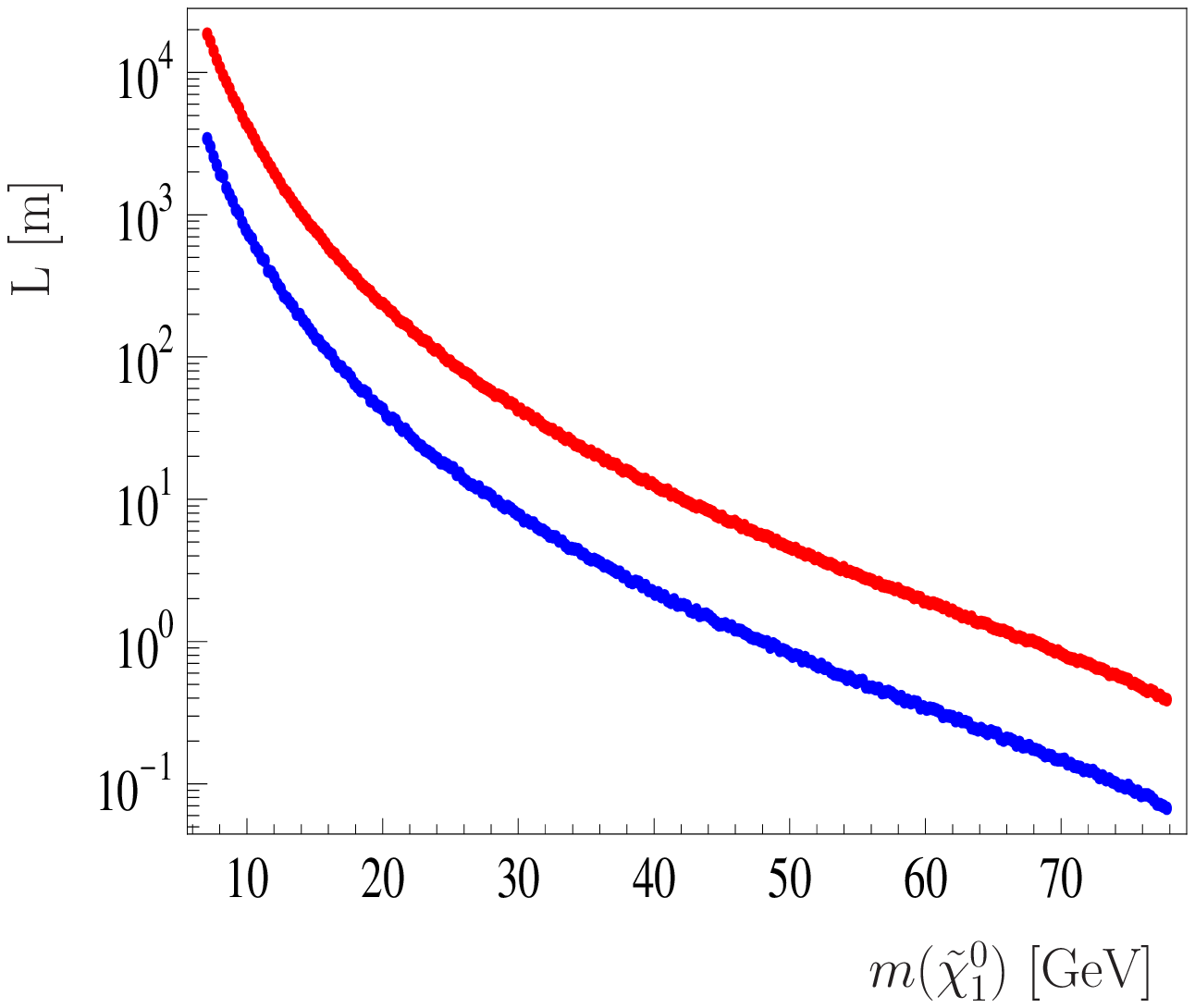}
\includegraphics[width=0.49\textwidth]{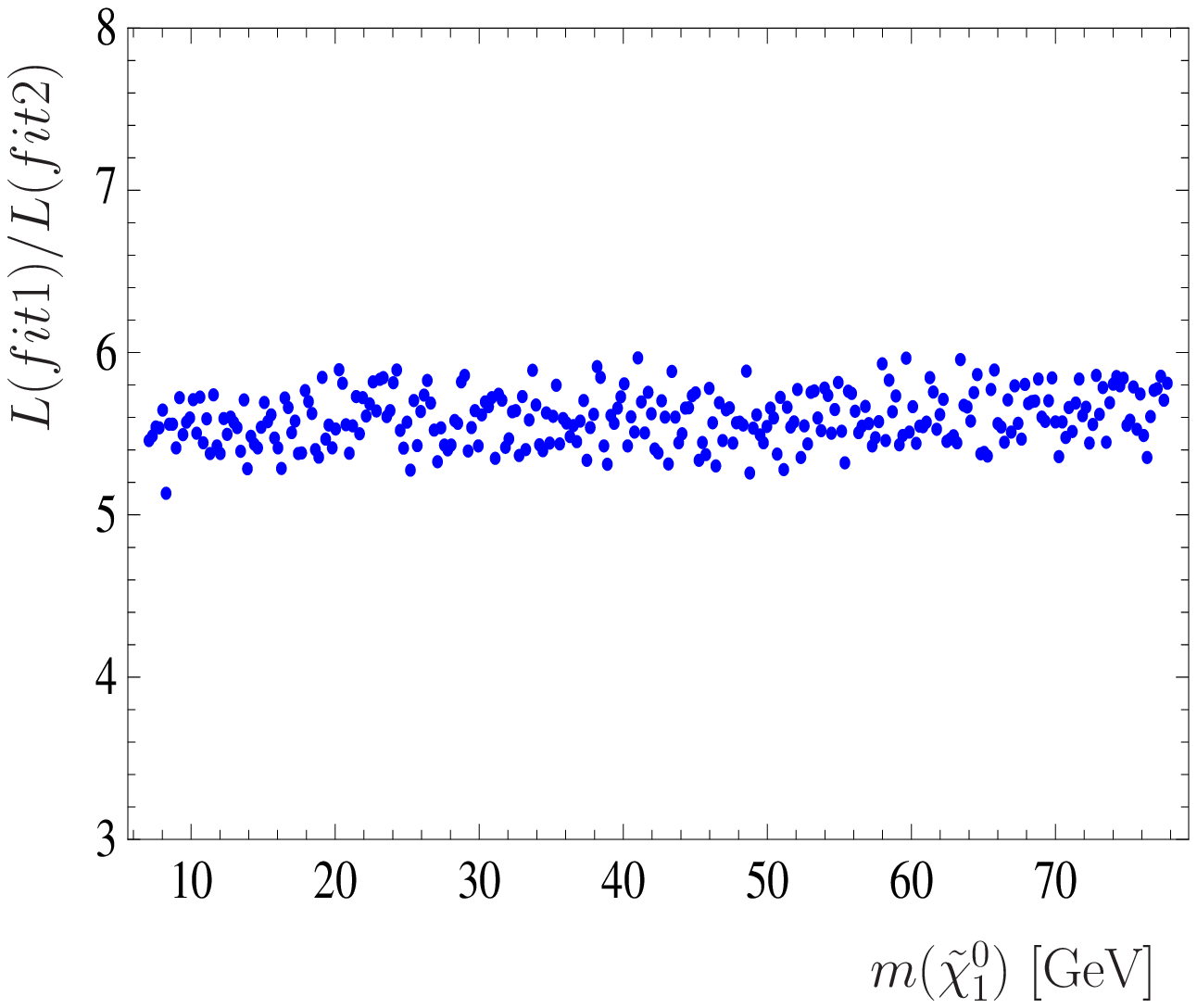}
\end{center}
\vspace{-6mm}
\caption{Decay length of the lightest neutralino and its dependence on
the type of fit to neutrino data. To the left (a) the decay length of the
lightest neutralino versus $m(\tilde{\chi}_1^0)$ for the case fit1
(red) and the case fit2 (blue). To the right (b) the ratio
$L(\text{fit1})/L(\text{fit2})$ versus $m(\tilde{\chi}_1^0)$. The MSSM
parameters have been taken such that the standard SPS1a' point is
reproduced. The light singlet parameter $\kappa$ is varied in the
range $\kappa\in\left[0.01,0.1\right]$. In all the points the lightest
neutralino has a singlino purity higher than $0.99$.}
\label{fig:nfitlength}
\end{figure}

As already discussed we have two different possiblities to fit 
neutrino data: $\vec \Lambda$ generates the atmospheric mass scale and
$\vec \alpha$ the solar mass scale (case fit1), or vice versa (case
fit2). It turns out that the decay length of the lightest neutralino
is sensitive to the type of fit, due to the proportionality between
its couplings with gauge bosons and the \rpv parameters (see appendix
\ref{munuapp4} for exact and approximated formulas of the
couplings $\tilde{\chi}_1^0-W^{\pm}-l^{\mp}_i$ and their simplified
expressions in particular limits). For example, a singlino-like
neutralino couples to the gauge bosons proportionally to the $\alpha_i$
parameters. This implies that its decay length will follow $L \propto
1/|\vec \alpha|^2$ and obeys the approximate relation
\begin{equation}
\frac{L(fit1)}{L(fit2)} \simeq \frac{m_{atm}}{m_{sol}} \simeq 6\quad.
\end{equation}
In figure \ref{fig:nfitlength} the decay length of the lightest
neutralino and its dependence on the type of fit to neutrino data is
shown. Once mass and length are known this dependence can be used to
determine which parameters generate which mass scale.  Note that this
feature is essentially independent of the MSSM parameters. However,
this property is lost if either the lightest neutralino has a sizeable 
gaugino/higgsino component or if there are singlet scalars/pseudoscalars
lighter than the singlino.

\subsection{Several light singlets}
\label{subsec:scenariod}

In scenarios with two (or more) light singlets, the phenomenology has
additional features. The light Higgs boson $h^0$ can decay with
measurable branching ratios to pairs of right-handed neutrinos of
different generations. Similarly, the bino can decay to the different
light right-handed neutrinos. 

In the following, the case of two light singlinos and two light
scalars/pseudo\-scalars will be considered. For the neutral fermion sector
this implies that the mass eigenstates $\tilde{\chi}_1^0$ and
$\tilde{\chi}_2^0$ will always be the singlets $\nu_1^c$ and $\nu_2^c$
and the bino will be the $\tilde{\chi}_3^0$. In the
scalar sector one has two very light mostly singlet states $S_1^0$ 
and $S_2^0$, which are consistent with the LEP bounds. Finally, 
the state $S_3^0$ will be the light doublet Higgs boson
$h^0$. One can also have light singlet pseudoscalars.

The decays of a bino-like $\tilde{\chi}_3^0$ can be very important to
distinguish between the one generation model and models with more than
one generation of singlets. 
In principle, the most important decay channels strongly depend on the
couplings of the bino to the two generations of singlinos and the
configuration of masses of singlinos and scalars. Therefore, a general
list of signals cannot be given. Nevertheless, there are some features
which are always present:

\begin{figure}
\begin{center}
\vspace{0mm}
\includegraphics[width=0.49\textwidth]{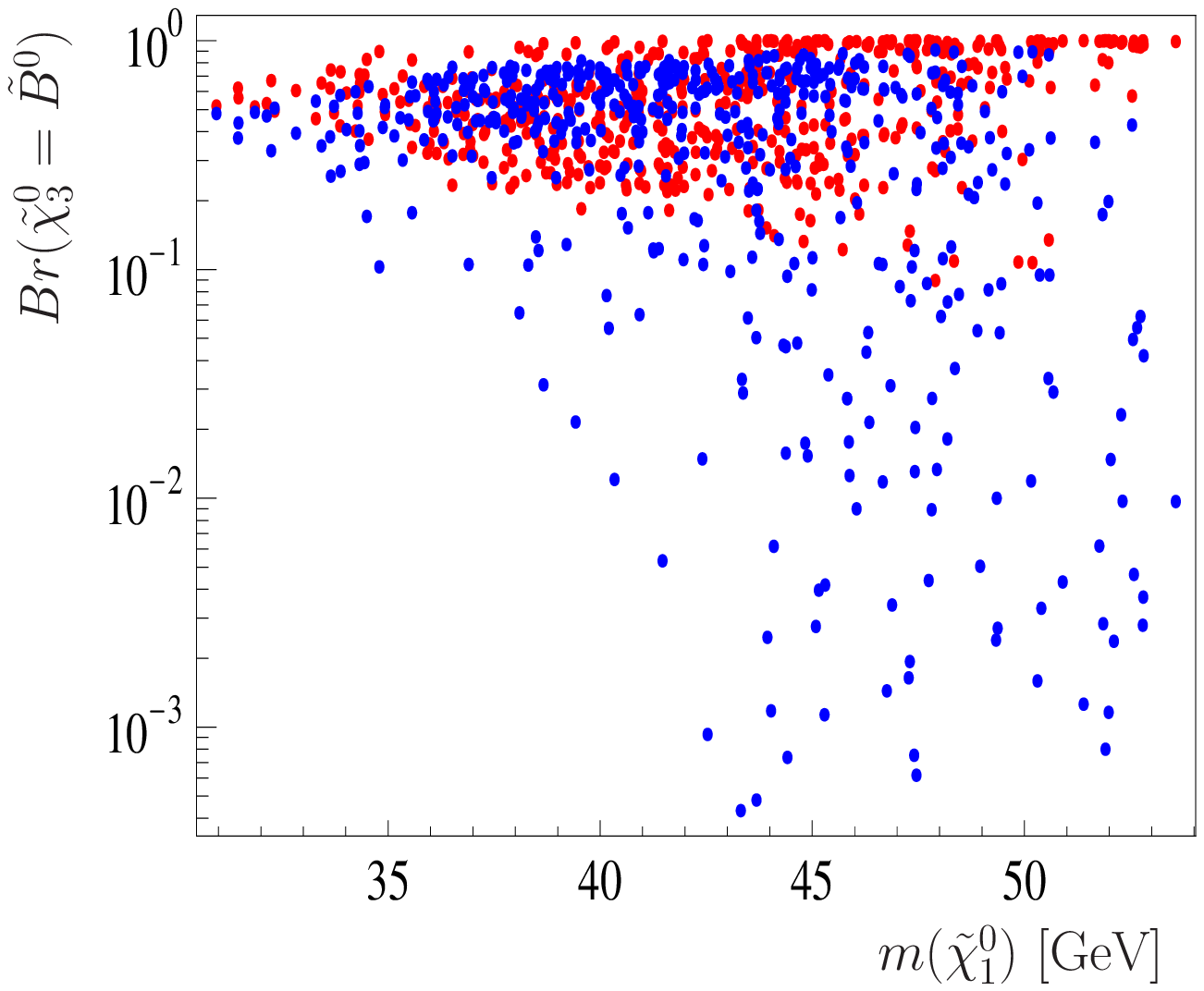}
\end{center}
\vspace{-6mm}
\caption{Branching ratios $Br(\tilde{\chi}_3^0 = \tilde{B}^0 \to
\tilde{\chi}_1^0)$ (red) and $Br(\tilde{\chi}_3^0 = \tilde{B}^0 \to
\tilde{\chi}_2^0)$ (blue) as a function of the mass of the lightest
neutralino for the scenario considered in section
\ref{subsec:scenariod}. The MSSM parameters have been taken such that
the standard SPS1a' point is reproduced, whereas the singlet
parameters are chosen randomly in the ranges $v_{R1},v_{R2} \in
[400,600]$ GeV, $\lambda_1,\lambda_2 \in [0.0,0.4]$, $T_\kappa^{111} =
T_\kappa^{222} \in [-15,-1]$ GeV, $T_\kappa^{112} = T_\kappa^{122} \in
[-1.5,-0.005]$ GeV and $T_\lambda^1,T_\lambda^2 \in [0,600]$ GeV. $\kappa_1 =
\kappa_2 = 0.16$ is fixed to ensure the lightness of the two
singlinos.}
\label{fig:binodecD}
\end{figure}

When kinematically allowed, the decays $\tilde{\chi}_3^0 \to
\tilde{\chi}_{1,2}^0 \: S_1^0 (P_1^0)$ dominate, with the sum of the
branching ratios typically larger than 50 \%.  The relative importance
of the different channels is mainly dictated by kinematics. This
feature is illustrated in figure \ref{fig:binodecD}, where these two
quantities are shown as a function of the mass of the lighest
neutralino. The MSSM parameters are fixed to the standard point
SPS1a', with light singlet parameters taken randomly. One can see
that the relative importance of each singlino cannot be predicted in
general, but both branching ratios are at least of order
$10^{-3}-10^{-4}$, given enough statistics. For very light singlinos
two-body decays including scalars and pseudoscalars are open, and thus
both $Br(\tilde{\chi}_3^0 \to \tilde{\chi}_1^0)$ and
$Br(\tilde{\chi}_3^0 \to \tilde{\chi}_2^0)$ are close to $50\%$,
as expected if the values of the singlet parameters are of the same
order for the two light generations. On the other hand, if the mass of
the lightest neutralino is increased some of the two-body decays are
kinematically forbidden, specially those of the $\tilde{\chi}_2^0$,
which has to be produced through three-body decays, leading to a
suppresion in $Br(\tilde{\chi}_3^0 \to \tilde{\chi}_2^0)$.  Note that
it is also possible to find points where the decay mode
$\tilde{\chi}_3^0 \to \tilde{\chi}_{1,2}^0 \: S_2^0 (P_2^0)$ has a
branching ratio about 10\%-20\%, giving additional information.

The other possible signals are the usual bino 
decays of the NMSSM. Final states with standard model particles, like
$\tilde{\chi}_{1,2}^0 l^+ l^-$ or $\tilde{\chi}_{1,2}^0 q \bar{q}$,
become very important when the decays to scalars and pseudoscalars are
kinematically forbidden.

In addition, the decays of the light Higgs boson $h^0$ can also
play a very important role in the study of the different generations,
provided it can decay to final states including
$\tilde{\chi}_1^0$ or $\tilde{\chi}_2^0$. In this case typically 
the standard Higgs boson decays are reduced to less than 40\%, 
completely changing the usual search strategies.

\begin{figure}
\begin{center}
\vspace{-2mm}
\includegraphics[width=0.49\textwidth]{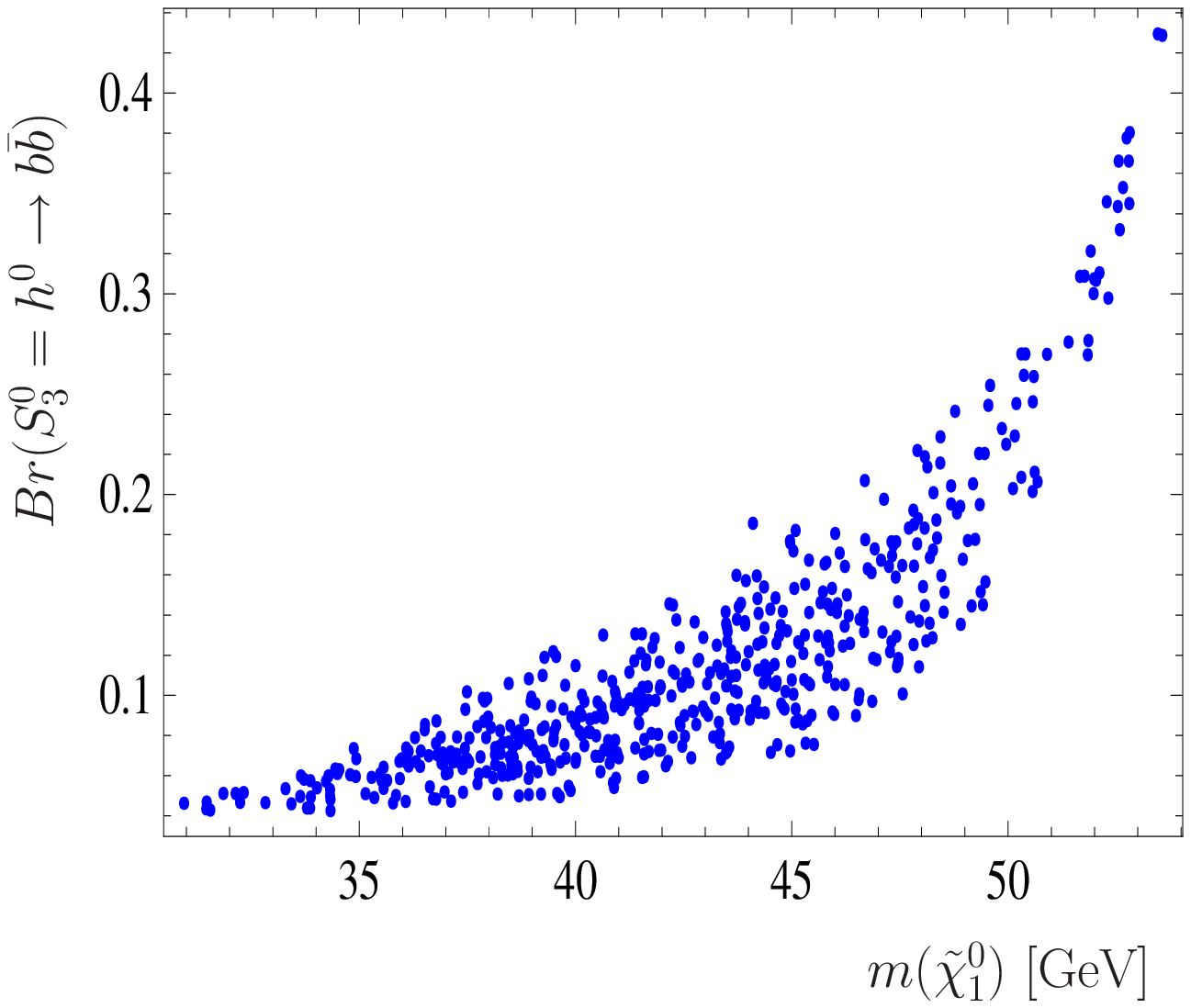}
\includegraphics[width=0.49\textwidth]{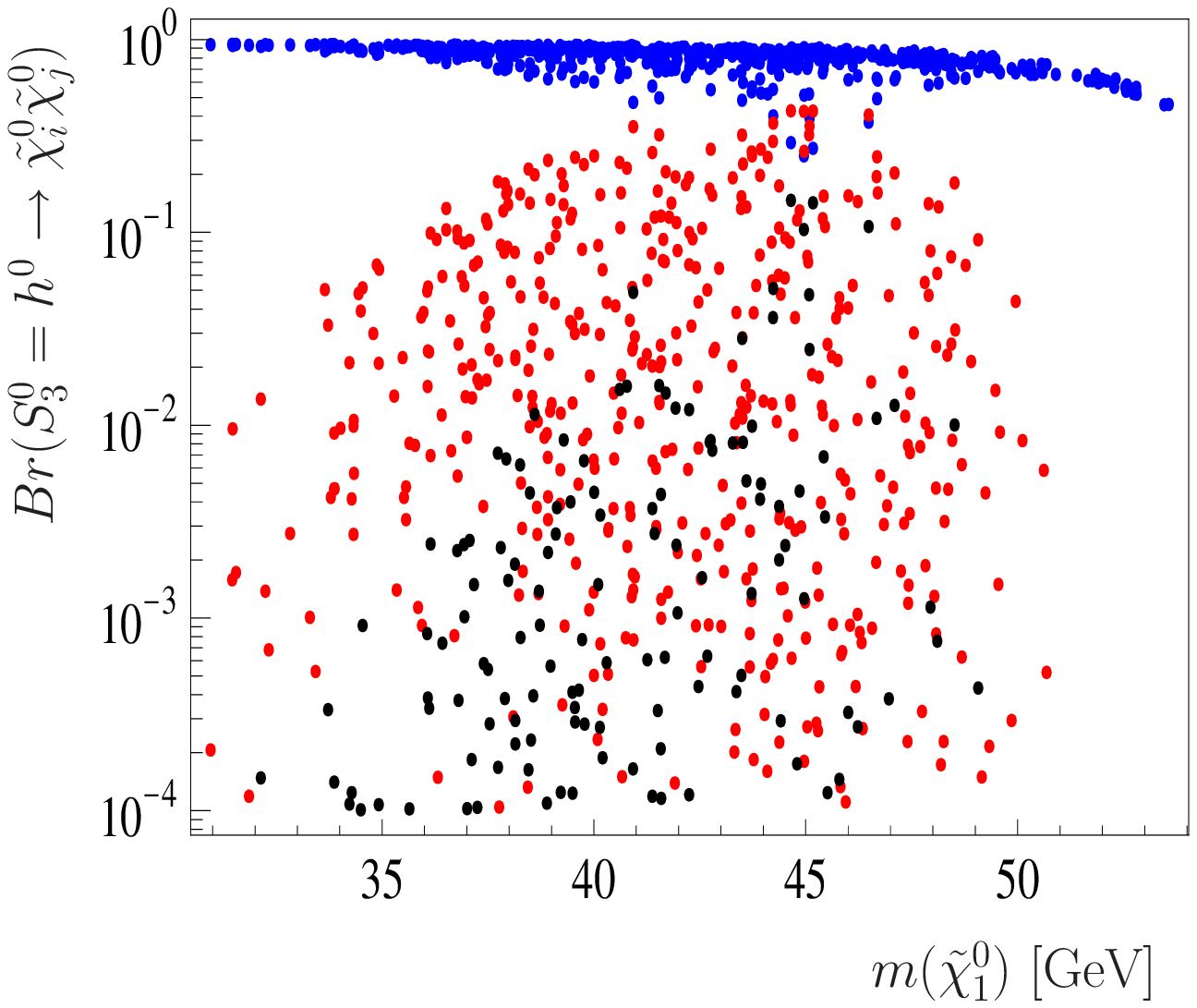}
\end{center}
\vspace{-6mm}
\caption{Higgs boson decays as a function of the mass of the lightest
neutralino for the scenario considered in section
\ref{subsec:scenariod}. To the left (a) the standard decay channel $h^0
\to b \bar{b}$, whereas to the right (b) the exotic decays to pairs of
singlinos $h^0 \to \tilde{\chi}_1^0\tilde{\chi}_1^0$ (red), $h^0 \to
\tilde{\chi}_1^0\tilde{\chi}_2^0$ (blue) and $h^0 \to
\tilde{\chi}_2^0\tilde{\chi}_2^0$ (black). The parameters are 
chosen as in figure \ref{fig:binodecD}.}
\label{fig:higgsdecaysD}
\end{figure}

In figure \ref{fig:higgsdecaysD} the branching ratios of standard and
exotic Higgs boson decay channels are shown. The left plot shows the
suppressed branching ratio of the standard $b \bar{b}$ channel.  The
main decay channel is $\tilde{\chi}_1^0 \: \tilde{\chi}_1^0$, but
there is a sizeable branching ratio to $\tilde{\chi}_1^0 \:
\tilde{\chi}_2^0$. Note that $\tilde{\chi}_2^0$ decays dominantly to
$\tilde{\chi}_1^0$ plus two SM fermions. This feature allows us to
distinguish between the $1$ $\widehat{\nu}^c$-model and models with
more than one generation of singlets. Finally, the branching ratio to
$\tilde{\chi}_2^0 \: \tilde{\chi}_2^0$ is small due to kinematics, but
leads to interesting final states with up to eight b-jets plus missing
energy.

A final comment is in order. In these kind of scenarios with many
light singlets $\tilde{\chi}_1^0$ decays to $\nu b{\bar b}$ can 
be dominant. This will reduce the available statistics in the 
interesting $l_i l_j \nu$ and $l q_i \overline{q}_j$ channels. Moreover, 
the correlations are less pronounced due to mixing effects in 
the singlet sector. 

\section{Summary}

The phenomenology of the $\mu\nu$SSM has been studied in this chapter. This proposal 
solves at the same time the $\mu$-problem of the MSSM and generates 
small neutrino masses, consistent with data from neutrino oscillation 
experiments. Neutrino data put very stringent constraints on the 
parameter space of the model. Both the left-sneutrino vacuum expectation 
values and the effective bilinear parameters have to be small compared 
to MSSM soft SUSY breaking parameters. As a result all SUSY production 
cross sections and all decay chains are very similar to the NMSSM, the 
only, but phenomenologically very important, exceptions being the decay 
of the LSP and NLSP (the latter only in some parts of the parameter space)
plus the decays of the lightest Higgses.

We have discussed in some details two variants of the model. In the 
simplest version with only one generation of singlets 1-loop corrections 
to the neutralino-neutrino mass matrix need to be carefully calculated 
in order to explain neutrino data correctly. The advantage of this minimal 
scheme is that effectively it contains only six new (combinations of) 
\rpv parameters, which can be fixed to a large extent by the requirement 
that oscillation data is correctly explained. This feature of the model 
is very similar to explicit bilinear R-parity breaking, although, as we have 
discussed, the relative importance of the different 1-loop contributions 
is different in the $\mu\nu$SSM and in bilinear \rpv. Certain ratios of 
decay branching ratios depend on the same parameter combinations as 
neutrino angles and are therefore predicted from neutrino physics, to a 
large extent independent of NMSSM parameters. We have also calculated the 
decay length of the LSP, which depends mostly on the LSP mass and the 
(experimentally determined) neutrino masses. Lengths sufficiently large 
to observe displaced vertices are predicted over most parts of the parameter 
space. However, for neutralinos lighter than approximately 30 GeV, decay 
lengths become larger than 10 meters, making the observation of \rpv 
difficult for LHC experiments. However, if there is a singlet scalar or 
pseudoscalar with a mass smaller than the lightest neutralino, 
$\tilde{\chi}^0_1 \to S_m^0(P_m^0) \nu$ is the dominant decay mode and the corresponding 
decay lengths become much smaller, such that the displaced vertex 
signature of \rpv might even be lost in some points of this part of parameter 
space. On the other hand, in case the mass of the lightest scalar is larger 
than twice the singlino mass, the decay $S_m^0 \to 2 \tilde{\chi}^0_1$ becomes 
important, both for $S^0_m \sim {\tilde\nu^c}$ and $S^0_m \sim h^0$. 
If this kinematical situation is realized also the Higgs search at the 
LHC will definitely be affected. 

The more involved $n$ generation variants of the $\mu\nu$SSM can explain 
all neutrino data at tree-level and therefore are {\em calculationally} 
simpler. Depending on the nature of the neutralino, neutralino LSP decays 
show different correlations with either solar or atmospheric neutrino angles. 
This is guaranteed in the two generation version of the model and likely, 
but not always true, for $n$ generations. If the NMSSM coupling $\lambda$ 
is sufficiently small also the NLSP has decays to \rpv final states 
with potentially measurable branching ratios. In this part of parameter 
space it seems possible,  in principle, to test both solar and atmospheric 
neutrino angles. If only the singlino(s) are light, i.e. the singlet 
scalars are heavier than, say, the $h^0$, the decay length of the singlino 
is very sharply predicted as a function of its mass and either the solar or 
atmospheric neutrino mass scale. If both, singlinos and singlet scalars 
(or pseudoscalars) are light, bino NLSP and $h^0$ will decay not only 
to the lightest singlinos/singlets but also to next-to-lightest states. 
This leads to enhanced multiplicities in the final states and the 
possibility to observe multiple displaced vertices.

In conclusion, the $\mu\nu$SSM offers a very rich phenomenology. 
Especially scenarios with light singlets deserve further, much  
more detailed studies.

\chapter[Comparison between \rpv schemes]{Comparison between R-parity breaking schemes}
\label{chap:comprpv}

We now briefly discuss possible differences in collider 
phenomenology of R-parity breaking schemes. 
Different models of R-parity breaking appear clearly distinct at 
the Lagrangian level. However, at accelerator experiments it can 
be very hard to distinguish the different proposals. This can be 
easily understood from the fact that for a heavy singlet sector 
all \rpv models approach necessarily the MSSM with explicit R-parity 
breaking terms. It is therefore an interesting question to ask, what 
- if any - kind of signals could exist, which at least might hint at 
which model is the correct description of \rpv. Given the large variety 
of possibilities and the very limited predictive power of the most 
general cases, any discussion {\em before} the discovery of SUSY must 
be rather qualitative. 

First one should mention that not all \rpv models explaining neutrino 
data show correlations between LSP decay branching ratios and neutrino 
angles. Especially the large number of free parameters in trilinear 
models exclude the possibility to make any definite predictions. 
\rpv models which do show such correlations, on the other hand, lead 
usually to very similar predictions for the corresponding LSP decays. 
For example, fitting the atmospheric data with tree-level \rpv terms, 
a bino LSP in explicit bilinear models and in the $\mu\nu$SSM decay 
with the same ratio of branching ratios into $W l$ (or $lq_i{\bar q}_j$) 
final states. Thus, to distinguish the different proposals other signals 
are needed.

We will briefly discuss the main differences in collider phenomenology 
between the following three proposals: (i) MSSM with explicit bilinear 
terms; (ii) Spontaneous \rpv model and (iii) 
$\mu\nu$SSM. Table \ref{tab:comp} shows a brief summary of this 
comparison. Differences occur in (a) the observability of a displaced 
vertex of the lightest neutralino decay; (b) the upper limit on the 
branching ratio of the lightest neutralino decaying completely invisible 
and (c) standard versus non-standard lightest Higgs decays.

The decay length of the lightest neutralino is fixed in both, the
b-\rpv model and the $\mu\nu$SSM, essentially by the mass of the
lightest neutralino and the experimentally determined neutrino masses.
For $m(\tilde{\chi}^0_1)$ larger than the W-mass decay lengths are
typically of the order of ${\cal O}(mm)$ and proportional to
$m^{-1}(\tilde{\chi}^0_1)$.  For lighter neutralinos, larger decay
lengths are expected, see figures \ref{fig:1NuR_decaylength} and
\ref{fig:nfitlength}, which scale like
$m^{-4}(\tilde{\chi}^0_1)$. Shorter decay lengths are not possible in
b-\rpv and possible in the $\mu\nu$SSM only if at least one
(singlet) scalar or pseudoscalar is lighter than $\tilde{\chi}^0_1$,
when ${\tilde{\chi}^0_1} \to S_m^0(P_m^0) \nu$ dominates. Since in the
$\mu\nu$SSM the singlet scalars decay with a short decay length to
${\bar b}b$, one expects that in the $\mu\nu$SSM short
$\tilde{\chi}^0_1$ decay lengths correlate with the dominance of
${\bar b}b$ + missing energy final states. In the s-\rpv, on the other
hand, the $\tilde{\chi}^0_1$ decay length can be shorter than in the
b-\rpv, due to the new final state $\tilde{\chi}^0_1 \to J + \nu$,
where $J$ is the Majoron. Therefore, different from the $\mu\nu$SSM,
the neutralino decay length in the s-\rpv model anti-correlates with
the branching ratio for the invisible neutralino decay.

Finally, in the b-\rpv one expects that the decay properties of the 
lightest Higgs ($h^0$) are equal to the MSSM expectations, the only 
exception being the case when $h^0 \to 2 \tilde{\chi}^0_1$ is possible 
kinematically, in which the $\tilde{\chi}^0_1$ decays themselves can then 
lead to a non-standard signal in the Higgs sector. This is different 
in s-\rpv, where for a low-scale of spontaneous R-parity breaking, 
the $h^0$ can decay to two Majorons, i.e. large branching ratios 
of Higgs to invisible particles are possible. In the $\mu\nu$SSM 
the $h^0$ decays can be non-standard, if the lightest singlino is 
lighter than $m(h^0)/2$. However, since the singlinos decay, 
this will not lead to an invisible Higgs, unless the mass of the 
singlino is so small, that the decays occur outside the detector.

\begin{table}

\begin{center}
\begin{tabular}{|c|m{0.12\textwidth}|m{0.25\textwidth}|m{0.15\textwidth}|m{0.2\textwidth}|}
\hline
   & \centering Displaced vertex & \centering Comment & \centering $Br(\text{invisible})$ & \begin{center}Higgs decays\end{center} \\
\hline
\hline
b-\rpv & \centering Yes & \centering Visible & \centering $\le 10$ \% & \begin{center}standard\end{center} \\
\hline
s-\rpv & \centering Yes/No & \centering anti-correlates with invisible & \centering any & \begin{center}non-standard 
(invisible)\end{center} \\
\hline
$\mu \nu$SSM & \centering Yes/No & \centering anti-correlates with non-standard Higgs 
& \centering $\le 10$ \% & \begin{center}non-standard\end{center} \\
\hline
\end{tabular}
\end{center}
\caption{Comparison of displaced vertex signals, completely invisible 
final state branching ratios for LSP decays and lightest Higgs decays 
for three different R-parity violating models. For a discussion see text.}
\label{tab:comp}
\end{table}

To summarize this brief discussion, b-\rpv, s-\rpv and $\mu\nu$SSM can, 
in principle, be distinguished experimentally {\em if the singlets are 
light enough to be observed} in case of s-\rpv and $\mu\nu$SSM. We note in 
passing that we have not found any striking differences in collider 
phenomenology of the $\mu\nu$SSM and the NMSSM with explicit bilinear 
terms.

\chapter{Lepton Flavor Violation in SUSY Left-Right models}
\label{chap:susylr}

Left-right symmetric models are well motivated extensions of the
SM. Apart from their original motivation, the restoration of parity at
high energies, they have very interesting properties. In particular,
their particle spectra contain right-handed neutrinos and thus they
can accommodate a seesaw mechanism, generating neutrino masses quite
naturally. Moreover, in the case of supersymmetry, and due to the
inclusion of $U(1)_{B-L}$ in the left-right gauge group, the low
energy theory potentially conserves R-parity. In this chapter the
phenomenology of a SUSY left-right model is studied, with emphasis on
lepton flavor violation at low energy experiments and colliders. The
main novelty with respect to minimal seesaw models is the presence of
signatures in the right slepton sector. These provide a clear hint of
the underlying left-right symmetry of the model.

\section{Introduction}

The most popular explanation for the observed smallness of neutrino
masses is certainly the seesaw mechanism
\cite{Minkowski:1977sc,Mohapatra:1979ia,Schechter:1980gr,Cheng:1980qt}.
Literally hundreds of theoretical papers based on ``the seesaw'' have
been published since the discovery of neutrino oscillations
\cite{Fukuda:1998mi}.  Unfortunately, attractive as this idea might
appear from the theoretical point of view, ``the seesaw'' will never
be {\em directly} tested due to the high scales involved\footnote{Of
  course, we are referring here to the \emph{usual} high energy
  realizations of the seesaw mechanism.}.

This situation might change slightly, if supersymmetry (SUSY) is found 
at the LHC, essentially because scalar leptons provide potentially 
additional information about seesaw parameters. Assuming SUSY gets 
broken at a high energy scale, the seesaw parameters leave their imprint 
on the soft parameters in the RGE running. Then, at least in principle, 
indirect tests of the seesaw become possible. Indeed, this has 
been pointed out already in \cite{Borzumati:1986qx}, where it was shown 
that lepton flavor violating (LFV) off-diagonal mass terms for 
sleptons are automatically generated in seesaw (type-I), even if SUSY 
breaking is completely flavor blind at the GUT scale as in minimal 
supergravity.

Motivated by the above arguments, many authors have then studied LFV 
in SUSY models. For the seesaw type-I, low energy LFV decays such as 
$l_i \to l_j \gamma$ and $l_i \to 3 l_j$ have been calculated in 
\cite{Hisano:1995nq,Hisano:1995cp,Ellis:2002fe,Deppisch:2002vz,Petcov:2003zb,
Arganda:2005ji,Petcov:2005yh,Antusch:2006vw,Deppisch:2004fa,Hirsch:2008dy}; 
$\mu-e$ conversion in nuclei has been studied in 
\cite{Arganda:2007jw,Deppisch:2005zm}. The type-II seesaw has 
received much less attention, although it has actually fewer 
free parameters than type-I. The latter implies that ratios of LFV 
decays of leptons can actually be predicted as a function of neutrino 
angles in mSUGRA, as has been shown in \cite{Rossi:2002zb,Hirsch:2008gh}.
Finally, for completeness we mention that LFV in SUSY seesaw type-III 
has been studied in \cite{Esteves:2010ff}.

Measurements at colliders, once SUSY is discovered, can provide
additional information. LFV decays of left sleptons within mSUGRA have
been studied for type-I in \cite{Hisano:1998wn} and for type-II in
\cite{Hirsch:2008gh,Esteves:2009vg}.  Precise mass measurements might
also show indirect effects of the seesaw
\cite{Blair:2002pg,Freitas:2005et,Deppisch:2007xu}. Most prominently,
type-II and type-III seesaw contain non-singlet superfields, so gauge
couplings run differently from pure MSSM. One then expects that
sparticle spectra show a characteristic ``deformation'' with respect
to mSUGRA predictions. From different combinations of masses one can
form ``invariants'', i.e. numbers which to leading order depend only
on the seesaw scale \cite{Buckley:2006nv}, although there are
important corrections at 2-loop \cite{Hirsch:2008gh,Esteves:2010ff}, which
have to be included before any quantitative analysis can be done.
Experimentally interesting is also that at the LHC the mass splitting
between selectrons and smuons may be constrained down to ${\cal
O}(10^{-4})$ for 30 $fb^{-1}$ of integrated luminosity
\cite{Allanach:2008ib}.  In mSUGRA, one expects this splitting to be
unmeasurably tiny, whereas in mSUGRA plus seesaw significantly
different masses can be generated, as has been shown for type-I in
\cite{Abada:2010kj}.

Interestingly, in pure seesaw models with flavor blind SUSY boundary
conditions all of the effects discussed above show up only in the left
slepton sector. Naturally one expects that in a supersymmetric model
with an intermediate left-right symmetric stage, also the right
sleptons should contain some indirect information about the high
energy parameters. This simple observation forms the main motivation
for the current investigation \cite{Esteves:2010si}. Before entering
in the details of our calculation, let us first briefly discuss
left-right symmetric models.

\subsection{Left-right symmetric models}

Quite a large number of different left-right (LR) symmetric models
have been discussed in the literature. Originally LR models were
introduced to explain the observed left-handedness of the weak
interaction as a consequence of symmetry breaking
\cite{Pati:1974yy,Mohapatra:1974gc,Senjanovic:1975rk}.  However, LR
models offer other advantages as well. First, the particle content of
LR models contains automatically the right-handed neutrino and thus
the ingredients for generating a (type-I) seesaw
mechanism \footnote{Breaking the LR symmetry with triplets can
  generate also a type-II \cite{Mohapatra:1979ia}.}. Second, the gauge
group $SU(3)_c \times SU(2)_L \times SU(2)_R \times U(1)_{B-L}$ is one
of the possible chains through which $SO(10)$
\cite{Georgi:1974my,Fritzsch:1974nn} can be broken to the standard
model gauge group. In addition, it has been shown that they provide
technical solutions to the SUSY CP and strong CP problems
\cite{Mohapatra:1996vg} and they give an understanding of the $U(1)$
charges of the standard model fermions. Interesting only for the
supersymmetric versions of LR models, (B-L) is gauged and thus,
potentially, the low energy theory conserves R-parity
\cite{Mohapatra:1986su,Martin:1992mq}.

This last argument requires possibly some elaboration. R-parity, defined 
as $R_P = (-1)^{3(B-L)+2s}$ (where $B$ and $L$ stand for baryon and 
lepton numbers and $s$ for the spin of the particle), is imposed in 
the MSSM to avoid dangerous baryon and lepton number violating 
operators. However, the origin of $R_P$ is not explained within the 
MSSM. In early LR models $SU(2)_R$ doublets were used to break the gauge
symmetry. The non-supersymmetric model proposed in references
\cite{Mohapatra:1974gc,Senjanovic:1975rk} introduced two additional 
scalar doublets $\chi_L$ and $\chi_R$, where $\chi_L \equiv \chi_L(1,2,1,1)$ 
and $\chi_R \equiv \chi_R(1,1,2,-1)$ under $SU(3)_c \times SU(2)_L \times 
SU(2)_R \times U(1)_{B-L}$. Parity conservation 
implies that both, $\chi_L$ and $\chi_R$, are needed. When the neutral 
component of $\chi_R$ gets a VEV, $\langle \chi_R^0 \rangle \neq 0$, the 
gauge symmetry is broken down to the SM gauge group. However, $\chi_R$ 
is odd under $U(1)_{B-L}$ and thus, in the SUSY versions of this 
setup, $R_P$ is broken at the same time\footnote{This could be solved 
by imposing additional discrete symmetries on the model that forbid 
the dangerous \rpv operators \cite{Malinsky:2005bi}, but this cannot 
be regarded as automatic R-parity conservation.}. A possible solution to 
this problem is to break the gauge symmetry by $SU(2)_R$ fields with 
even charge under $U(1)_{B-L}$, i.e. by triplets. For a SUSY LR model, 
this was in fact proposed in reference \cite{Cvetic:1983su}, where four 
triplets were added to the MSSM spectrum: $\Delta(1,3,1,2)$, 
$\Delta^c(1,1,3,-2)$, $\bar{\Delta}(1,3,1,-2)$ and $\bar{\Delta}^c(1,1,3,2)$. 
Breaking the symmetry by the VEV of $\Delta^c$ produces at the same 
time a right-handed neutrino mass via the operator $L^c \Delta^c L^c$, 
leading to a type-I seesaw mechanism. Depending on whether or not 
$\Delta$ gets a VEV, also a type-II seesaw can be generated 
\cite{Akhmedov:2006de}.

However, whether R-parity is conserved in this setup is not clear.
The reason is that the minimum of the potential might prefer a
solution in which also the right-handed scalar neutrino gets a vev,
thus breaking $R_P$, as has been claimed to be the case in
\cite{Kuchimanchi:1993jg}.  Later \cite{Babu:2008ep} calculated some
1-loop corrections to the scalar potential, concluding that $R_P$
conserving minima can be found.  However, this contradicts the earlier
claim \cite{Kuchimanchi:1993jg} that 1-loop corrections can not
eliminate the dangerous \rpv minima. On the other hand, as first noted
in \cite{Kuchimanchi:1995vk} and later showed by Aulakh and
collaborators \cite{Aulakh:1997ba,Aulakh:1997fq}, by the addition of two
more triplets, $\Omega(1,3,1,0)$ and $\Omega^c(1,1,3,0)$, with zero
lepton number one can achieve LR breaking with conserved $R_P$
guaranteed already at tree-level. Lacking a general proof that the
model \cite{Cvetic:1983su} conserves $R_P$ we will follow
\cite{Aulakh:1997ba,Aulakh:1997fq} as the setup for our numerical
calculations.

Finally, for completeness we mention the existence of left-right models with
R-parity violation. For example, if the left-right symmetry is broken
with the VEVs of right-handed sneutrinos R-parity gets broken as well
and the resulting phenomenology is
totally different, as shown in \cite{Hayashi:1984rd,FileviezPerez:2008sx}.

Compared to the long list of papers about indirect tests of the seesaw, 
surprisingly little work on the ``low-energy'' phenomenology of SUSY LR 
models has been done. One loop RGEs for two left-right SUSY models 
have been calculated in \cite{Setzer:2005hg}. These two models are (with 
one additional singlet): (a) breaking LR by doublets a la 
\cite{Mohapatra:1974gc,Senjanovic:1975rk} and (b) by triplets following 
\cite{Cvetic:1983su}, but no numerical work at all was done in this paper. 
The possibility that right sleptons might have flavor violating decays in 
the left-right symmetric SUSY model of \cite{Cvetic:1983su} was mentioned 
in \cite{Chao:2007ye}. A systematic study of all the possible signals 
discussed above for the seesaw case is lacking and to our knowledge 
there is no publication of any calculation of these signals for the model 
of \cite{Aulakh:1997ba,Aulakh:1997fq}.
(For completeness we would like to mention that in GUTs based on SU(5) one 
can have the situation the LFV occurs {\em only} in the right slepton 
sector, as pointed out in \cite{Barbieri:1994pv}. However, this 
model \cite{Barbieri:1994pv} is in a different class from all the models 
discussed above, since it does not contain non-zero neutrino masses.)\\

The rest of this paper is organized as follows. In the next section we 
define the model \cite{Aulakh:1997ba,Aulakh:1997fq} and discuss its 
particle content and main features at each symmetry breaking scale. 
We have calculated the RGEs for each step complete at the 2-loop level 
following the general description by \cite{Martin:1993zk} using the 
Mathematica package SARAH \cite{Staub:2008uz,Staub:2009bi,Staub:2010jh}. 
A summary is given in the appendix, the complete set of equations 
and the SARAH model files can be found at \cite{SarahWeb}. 
Neutrino masses can be fitted to experimental data via a type-I seesaw 
mechanism and we discuss different ways to implement the fit. We then 
turn to the numerical results. The output of SARAH has been passed to 
the program package SPheno \cite{Porod:2003um} for numerical evaluation. 
We calculate the SUSY spectra and LFV slepton decays, such as 
${\tilde\tau}_{L/R} \to \mu \tilde{\chi}^0_1$ and 
${\tilde\tau}_{L/R} \to e \tilde{\chi}^0_1$ and 
$\tilde{\chi}^0_2 \to e \mu \tilde{\chi}^0_1$, 
as well as low-energy decays $l_i \to l_j \gamma$ for some sample 
points as a function of the 
LR and (B-L) scales. Potentially measurable signals are found in both, 
left and right slepton sectors, if (a) the seesaw scale is above 
(very roughly) $10^{13}$ GeV and (b) if the scale of LR breaking is 
significantly below the GUT scale. Since we find sizeable LFV soft 
masses in both slepton sectors, also the polarization in 
$\mu \to e \gamma$ is different from the pure seesaw expectation. 
We then close with a short summary.

\section{The model}

In this section we define the model, its particle content and give 
a description of the different symmetry breaking steps. The fit 
to neutrino masses and its connection to LFV violation in the 
slepton sector is discussed in some detail, to prepare for 
the numerical results given in the next section. We summarize briefly 
the free parameters of the theory.

The model essentially follows \cite{Aulakh:1997ba,Aulakh:1997fq}. 
We have not attempted to find a GUT completion. We will, however, 
assume that gauge couplings and soft SUSY parameters can be unified, 
i.e. implicitly assume that such a GUT model can indeed be constructed. 

\subsection{Step 1: From GUT scale to $SU(2)_R$ breaking scale}

Just below the GUT scale the gauge group of the model is $SU(3)_c
\times SU(2)_L \times SU(2)_R \times U(1)_{B-L}$. In addition it
is assumed that parity is conserved, see below. The matter content 
of the model is given in table \ref{tab:particles-step1}. Here $Q$, 
$Q^c$, $L$ and $L^c$ are the quark and lepton superfields of the MSSM 
with the addition of (three) right-handed neutrino(s) $\nu^c$.

\begin{table}
\centering
{
\renewcommand\arraystretch{1.3} 
\begin{tabular}{c c c c c c}
\hline
Superfield & generations & $SU(3)_c$ & $SU(2)_L$ & $SU(2)_R$ & $U(1)_{B-L}$ \\
\hline
$Q$ & 3 & 3 & 2 & 1 & $\frac{1}{3}$ \\
$Q^c$ & 3 & $\bar{3}$ & 1 & 2 & $-\frac{1}{3}$ \\
$L$ & 3 & 1 & 2 & 1 & -1 \\
$L^c$ & 3 & 1 & 1 & 2 & 1 \\
$\Phi$ & 2 & 1 & 2 & 2 & 0 \\
$\Delta$ & 1 & 1 & 3 & 1 & 2 \\
$\bar{\Delta}$ & 1 & 1 & 3 & 1 & -2 \\
$\Delta^c$ & 1 & 1 & 1 & 3 & -2 \\
$\bar{\Delta}^c$ & 1 & 1 & 1 & 3 & 2 \\
$\Omega$ & 1 & 1 & 3 & 1 & 0 \\
$\Omega^c$ & 1 & 1 & 1 & 3 & 0 \\
\hline
\end{tabular}
}
\caption{Matter content between the GUT scale and the $SU(2)_R$ breaking scale.}
\label{tab:particles-step1}
\end{table}

Two $\Phi$ superfields, bidoublets under $SU(2)_L \times SU(2)_R$, are
introduced. They contain the standard $H_d$ and $H_u$ MSSM Higgs
doublets. In this model, two copies are needed for a non-trivial 
CKM matrix. Although there are known attempts to build a realistic LR 
model with only one bidoublet generating the quark mixing angles at the 
loop level \cite{Babu:1998tm}, we will not rely on such a mechanism. 
Finally, the rest of the superfields in table \ref{tab:particles-step1} 
are introduced to break the LR symmetry, as explained above.

Table \ref{tab:particles-step1} shows also the gauge charges for the matter
content in the model. In particular, the last column shows the $B-L$
value for the different superfields. However, the following definition
for the electric charge operator will be used throughout this paper
\begin{equation}
Q = I_{3L} + I_{3R} + \frac{B-L}{2}
\end{equation}
and thus the $U(1)_{B-L}$ charge is actually $\frac{B-L}{2}$.

With the representations in table \ref{tab:particles-step1}, the most
general superpotential compatible with the gauge symmetry and parity
is
\begin{eqnarray} \label{eq:Wsuppot1}
{\cal W} &=& Y_Q Q \Phi Q^c 
          +  Y_L L \Phi L^c 
          - \frac{\mu}{2} \Phi \Phi
          +  f L \Delta L
          +  f^* L^c \Delta^c L^c \nonumber \\
         &+& a \Delta \Omega \bar{\Delta}
          +  a^* \Delta^c \Omega^c \bar{\Delta}^c
          + \alpha \Omega \Phi \Phi
          +  \alpha^* \Omega^c \Phi \Phi \nonumber \\
         &+& M_\Delta \Delta \bar{\Delta}
          +  M_\Delta^* \Delta^c \bar{\Delta}^c
          +  M_\Omega \Omega \Omega
          +  M_\Omega^* \Omega^c \Omega^c \thickspace.
\end{eqnarray}
Note that this superpotential is invariant under the parity transformations
$Q  \leftrightarrow  (Q^c)^*$, $L \leftrightarrow (L^c)^*$, 
$\Phi \leftrightarrow  \Phi^\dagger$, $\Delta \leftrightarrow 
(\Delta^c)^*$, $\bar{\Delta} \leftrightarrow (\bar{\Delta}^c)^*$, 
$\Omega \leftrightarrow (\Omega^c)^*$. This discrete symmetry fixes, 
for example, the $L^c \Delta^c L^c$ coupling to be $f^*$, the complex 
conjugate of the $L \Delta L$ coupling, thus reducing the number of free 
parameters of the model. 

Family and gauge indices have been omitted in eq. \eqref{eq:Wsuppot1}, 
more detailed expressions can be found in \cite{Aulakh:1997ba}. 
$Y_Q$ and $Y_L$ are quark and lepton Yukawa couplings. However, with 
two bidoublets there are two copies of them, and thus there are four 
$3 \times 3$ Yukawa matrices. Conservation of parity implies that they 
must be hermitian. $\mu$ is a $2 \times 2$ symmetric matrix, whose entries
have dimensions of mass, $f$ is a $3 \times 3$ (dimensionless) complex symmetric
matrix, and $\alpha$ is a $2 \times 2$ antisymmetric matrix, and thus
it only contains one (dimensionless) complex parameter, $\alpha_{12}$. 
The mass parameters $M_\Omega$ and $M_\Delta$ can be exchanged for
$v_R$ and $v_{BL}$, the vacuum expectation values of the scalar fields
that break the LR symmetry, see below. 

The soft terms of the model are
\begin{eqnarray}
- {\cal L}_{soft} &=& m_Q^2 \tilde{Q}^\dagger \tilde{Q} + m_{Q^c}^2
\tilde{Q^c}^\dagger \tilde{Q^c} + m_L^2 \tilde{L}^\dagger \tilde{L} +
m_{L^c}^2 \tilde{L^c}^\dagger \tilde{L^c} \nonumber \\ &+& m_{\Phi}^2
\Phi^\dagger \Phi + m_{\Delta}^2 \Delta^\dagger \Delta +
m_{\bar{\Delta}}^2 \bar{\Delta}^\dagger \bar{\Delta} + m_{\Delta^c}^2
{\Delta^c}^\dagger \Delta^c \nonumber \\
&+& m_{\bar{\Delta}^c}^2 \bar{\Delta}^{c \:
  \dagger} \bar{\Delta}^c + m_{\Omega}^2 \Omega^\dagger
\Omega + m_{\Omega^c}^2 {\Omega^c}^\dagger \Omega^c \label{eq:soft1} \\
&+& \frac{1}{2}
\big[ M_1 \tilde{B}^0 \tilde{B}^0 + M_2 (\tilde{W_L} \tilde{W_L} +
  \tilde{W_R} \tilde{W_R}) + M_3 \tilde{g} \tilde{g} + h.c. \big]
\nonumber \\ 
&+& \big[ T_Q \tilde{Q} \Phi \tilde{Q^c} + T_L \tilde{L}
  \Phi \tilde{L^c} + T_f \tilde{L} \Delta \tilde{L} + T_f^*
  \tilde{L^c} \Delta^c \tilde{L^c} \nonumber \\ &+& T_a \Delta \Omega
  \bar{\Delta} + T_a^* \Delta^c \Omega^c \bar{\Delta^c} + T_\alpha
  \Omega \Phi \Phi + T_\alpha^* \Omega^c \Phi \Phi + h.c. \big]
\nonumber \\ 
&+& \big[ B_\mu \Phi \Phi + B_{M_\Delta} \Delta
  \bar{\Delta} + {B_{M_\Delta}}^* \Delta^c \bar{\Delta}^c +
  B_{M_\Omega} \Omega \Omega + {B_{M_\Omega}}^* \Omega^c \Omega^c +
  h.c. \big] \nonumber \thickspace.
\end{eqnarray}
Again, family and gauge indices have been omitted for the sake of
simplicity.  The LR model itself does not, of course, fix the values
of the soft SUSY breaking terms.  In the numerical evaluation of the
RGEs we will resort to mSUGRA-like boundary conditions, i.e.  $m_0^2
\: \mathcal{I}_{3 \times 3} = m_Q^2 = m_{Q^c}^2 = m_L^2 = m_{L^c}^2$,
$m_0^2 \: \mathcal{I}_{2 \times 2} = m_\Phi^2$, $m_0^2 = m_\Delta^2 =
m_{\bar{\Delta}}^2 = m_{\Delta^c}^2 = m_{\bar{\Delta}^c}^2 =
m_\Omega^2 = m_{\Omega^c}^2$, $M_{1/2} = M_1 = M_2 = M_3$, $T_Q = A_0
Y_Q, T_L = A_0 Y_L, T_f = A_0 f, T_a = A_0 a, T_\alpha = A_0 \alpha$,
$B_\mu = B_0, B_{M_\Delta} = B_0 M_\Delta, B_{M_\Omega} = B_0
M_\Omega$. Here $\mathcal{I}_{3 \times 3}$ and $\mathcal{I}_{2 \times
  2}$ are the $3 \times 3$ and $2 \times 2$ identity matrices,
respectively. The superpotential couplings $f$, $Y_Q$ and $Y_L$ are
fixed by the low-scale fermion masses and mixing angles. Their values
at the GUT scale are obtained by RGE running. This will be discussed
in more detail in section \ref{sec:yukawas-lr}.

The breaking of the LR gauge group to the MSSM gauge group takes place 
in two steps: $SU(2)_R \times U(1)_{B-L} \rightarrow U(1)_R \times 
U(1)_{B-L} \rightarrow U(1)_Y$. In the first step the neutral 
component of the triplet $\Omega$ takes a VEV:
\begin{equation}
\langle \Omega^{c \: 0} \rangle = \frac{v_R}{\sqrt{2}}
\end{equation}
which breaks $SU(2)_R$. However, since $I_{3R} (\Omega^{c \: 0}) = 0$ 
there is a $U(1)_R$ symmetry left over. Next, the group 
$U(1)_R \times U(1)_{B-L}$ is broken by
\begin{equation}
\langle \Delta^{c \: 0} \rangle = \frac{v_{BL}}{\sqrt{2}} \thickspace, \qquad 
\langle \bar{\Delta}^{c \: 0} \rangle = \frac{\bar{v}_{BL}}{\sqrt{2}} \thickspace.
\end{equation}
The remaining symmetry is now $U(1)_Y$ with hypercharge defined as
$Y = I_{3R} + \frac{B-L}{2}$. 

The tadpole equations do not link $\Omega^c$, $\Delta^c$ and $\bar{\Delta}^c$ 
with their left-handed counterparts, due to supersymmetry. Thus, the left-handed
triplets can have vanishing VEVs \cite{Aulakh:1997ba} and the model produces 
only a type-I seesaw.

Although a ``hierarchy'' between the two breaking scales may exist,
$v_{BL} \ll v_R$, one cannot neglect the effects of the second
breaking stage on the first one, since mass terms of $\Omega$ and 
$\Delta$ enter in both tadpole equations. If we assume
$\bar{v}_{BL} = v_{BL}$ the tadpole equations of the model can 
be written
\begin{eqnarray}
\frac{\partial V}{\partial v_R} &=& 4 |M_\Omega|^2 v_R + 
 \frac{1}{2} |a|^2 v_{BL}^2 v_R - \frac{1}{2} v_{BL}^2 
\left[ a^* (M_\Delta + M_\Omega) + c.c \right] = 0 \label{tadpoleeqs-1} \thickspace, \\
\frac{\partial V}{\partial v_{BL}} &=& |M_{\Delta}|^2 v_{BL} + 
\frac{1}{4} |a|^2 (v_{BL}^2 + v_R^2) v_{BL} \nonumber \\
&& - \frac{1}{2} v_{BL} v_R
\left[ a^* (M_\Delta + M_\Omega) + c.c \right]= 0 \label{tadpoleeqs-2} \thickspace.
\end{eqnarray}
In these equations (small) soft SUSY breaking terms have been
neglected.  Similarly, at this stage there are no electroweak symmetry
breaking VEVs $v_d$ and $v_u$. From equations \eqref{tadpoleeqs-1} and
\eqref{tadpoleeqs-2} one sees that, in fact, there is an inverse
hierarchy between the VEVs and the superpotential masses $M_\Delta$,
$M_\Omega$, given by
\begin{equation} \label{tadpolesol}
v_R = \frac{2 M_\Delta}{a} \thickspace, \qquad v_{BL} = \frac{2}{a} (2 M_\Delta M_\Omega)^{1/2} \thickspace.
\end{equation}
And so, $v_{BL} \ll v_R$ requires $M_\Delta \gg M_\Omega$, as has already 
been discussed in \cite{Aulakh:1997ba}.

\subsection{Step 2: From $SU(2)_R$ breaking scale to $U(1)_{B-L}$ breaking scale}

At this step the gauge group is $SU(3)_c \times SU(2)_L \times U(1)_R
\times U(1)_{B-L}$. The particle content of the model from the
$SU(2)_R$ breaking scale to the $U(1)_{B-L}$ breaking scale is given
in table \ref{tab:particles-step2}.

\begin{table}
\centering
{
\renewcommand\arraystretch{1.3} 
\begin{tabular}{c c c c c c}
\hline
Superfield & generations & $SU(3)_c$ & $SU(2)_L$ & $U(1)_R$ & $U(1)_{B-L}$ \\
\hline
$Q$ & 3 & 3 & 2 & 0 & $\frac{1}{3}$ \\
$d^c$ & 3 & $\bar{3}$ & 1 & $\frac{1}{2}$ & $-\frac{1}{3}$ \\
$u^c$ & 3 & $\bar{3}$ & 1 & $-\frac{1}{2}$ & $-\frac{1}{3}$ \\
$L$ & 3 & 1 & 2 & 0 & $-1$ \\
$e^c$ & 3 & 1 & 1 & $\frac{1}{2}$ & $1$ \\
$\nu^c$ & 3 & 1 & 1 & $-\frac{1}{2}$ & $1$ \\
$H_d$ & 1 & 1 & 2 & $-\frac{1}{2}$ & 0 \\
$H_u$ & 1 & 1 & 2 & $\frac{1}{2}$ & 0 \\
$\Delta^{c \: 0}$ & 1 & 1 & 1 & 1 & -2 \\
$\bar{\Delta}^{c \: 0}$ & 1 & 1 & 1 & -1 & 2 \\
$\Omega$ & 1 & 1 & 3 & 0 & 0 \\
$\Omega^{c \: 0}$ & 1 & 1 & 1 & 0 & 0 \\
\hline
\end{tabular}
}
\caption{Matter content from the $SU(2)_R$ breaking scale to the 
$U(1)_{B-L}$ breaking scale.}
\label{tab:particles-step2}
\end{table}

Some comments might be in order. Despite $M_{\Delta}$ being of the
order of $v_R$ (or larger), see eq. \eqref{tadpolesol}, not all
components of the $\Delta$ superfields receive large masses. The neutral
components of $\Delta^c$ and $\bar{\Delta}^c$ lie at the $v_{BL}$
scale.  One can easily check that the F-term contributions to their
masses vanish in the minimum of the scalar potential eq.~\eqref{tadpolesol}. Moreover, $\Omega^c$ does not generate D-terms
contributions to their masses. Therefore, contrary to the other
components of the $\Delta$ triplets, they only get masses at the
$v_{BL}$ scale. On the other hand, one might guess that all components
in the $\Omega$,$\Omega^c$ superfields should be retained at this
stage, since their superpotential mass $M_\Omega$ is required to be
below $v_{BL}$. However, some of their components get contributions
from $SU(2)_R$ breaking, and thus they become heavy. The charged 
components of $\Omega^c$ do develop large masses, in the case of the 
scalars through D-terms, while in the case of the fermions due to 
their mixing with the charged gauginos $\tilde{W}_R^\pm$, which have 
masses proportional to $v_R$. However, the neutral components of 
$\Omega^c$ do not get $SU(2)_R$ breaking contributions, since they 
have $I_{3R} (\Omega^{c  \: 0}) = 0$, and then they must be
included in this energy regime. See reference \cite{Aulakh:1997fq} for 
a more quantitative discussion.

After $SU(2)_R$ breaking the two bidoublets $\Phi_1$ and $\Phi_2$ get
split into four $SU(2)_L$ doublets. Two of them must remain light,
identified with the two Higgs doublets of the MSSM, responsible for EW
symmetry breaking, while, at the same time, the other two get masses
of the order of $v_R$. This strong hierarchy can be only obtained by
imposing a fine-tuning condition on the parameters involved in the
bidoublet sector.

The superpotential terms mixing the four $SU(2)_L$ doublets can be
rewritten as
\begin{equation}
{\cal W}_M = (H_d^f)^T M_H H_u^f
\end{equation}
where $H_d^f = ( H_d^1, H_d^2)$ and $H_u^f = ( H_u^1, H_u^2)$ are the
\emph{interaction eigenstates}. In this basis reads the matrix
\begin{equation}
M_H = 
\left(
\begin{array}{cc}
\mu_{11} & \mu_{12} + \alpha_{12} M_R \\
\mu_{12} - \alpha_{12} M_R & \mu_{22} 
\end{array}
\right) \thickspace,
\end{equation}
where the relations $\mu_{ij} = \mu_{ji}$ and $\alpha_{ij} = -
\alpha_{ji}$ have been used and $M_R = \frac{v_R}{2}$ has been
defined. In order to get two light doublets we impose the fine-tuning
condition \cite{Aulakh:1997fq}
\begin{equation} \label{fine-tuning}
{\rm Det}(M_H) = \mu_{11} \mu_{22} - (\mu_{12}^2 - \alpha_{12}^2 M_R^2) = 0 \thickspace.
\end{equation}
The result of eq.~\eqref{fine-tuning} is to split the two Higgs bidoublets
into two pairs of doublets $(H_d,H_u)_L$ and $(H_d,H_u)_R$, where
$(H_d,H_u)_L$ is the light pair that appears in table
\ref{tab:particles-step2}, and $(H_d,H_u)_R$ a heavy pair with mass of
order of $v_R$. In practice, equation \eqref{fine-tuning} implies that
one of the superpotential parameters must be chosen in terms of the
others. Since this fine-tuning condition is not protected by any
symmetry, the RGEs do not preserve it, and one must impose it at
the $SU(2)_R$ breaking scale. In our computation we chose to compute
$\mu_{11}$ in terms of the free parameters $\mu_{12}$, $\mu_{22}$,
$\alpha_{12}$ and $v_R$.

In order to compute the resulting couplings for the light Higgs
doublets one must rotate the original fields into their mass
basis. Since $M_H$ is not a symmetric matrix (unless $\alpha_{12} =0$)
one has to rotate independently $H_d^f$ and $H_u^f$, i.e.  $H_d^f =
D H_d^m$, $H_u^f = U H_u^m$, where $D$ and $U$ are orthogonal
matrices and $H_d^m = ( H_d^L, H_d^R)$ and $H_u^m = ( H_u^L, H_u^R)$
are the \emph{mass eigenstates}. This way one finds
\begin{equation}
{\cal W}_M = (H_d^f)^T M_H H_u^f = (H_d^m)^T D^T M_H U H_u^m =
(H_d^m)^T {\hat M}_H H_u^m
\end{equation}
where ${\hat M}_H$ is a diagonal matrix, with eigenvalues
\begin{eqnarray}\nonumber
\hat{M}_{H,1}^2 &=& 0 \thickspace, \\
\hat{M}_{H,2}^2 &=& \frac{1}{\mu_{22}^2}
\left(\alpha_{12}^4 M_R^4 + 2 \alpha_{12}^2 M_R^2 
(\mu_{22}^2-\mu_{12}^2) + (\mu_{22}^2+\mu_{12}^2)^2 \right) \thickspace.
\end{eqnarray}
The $D$ and $U$ rotations are, in general, different. We can compute
them by using the following identities
\begin{eqnarray}\nonumber
{\hat M}_H^2 = {\hat M}_H ({\hat M}_H)^T &=& D^T M_H (M_H)^T D \label{id1} \thickspace, \\ 
{\hat M}_H^2 = ({\hat M}_H)^T {\hat M}_H &=& U^T (M_H)^T M_H U \label{id2} \thickspace,
\end{eqnarray}
where we used $D D^T = D^T D = U U^T = U^T U = I$. If we parametrize the
rotations as
\begin{equation}
D = \left( \begin{array}{c c}
\cos \theta_1 & \sin \theta_1 \\
- \sin \theta_1 & \cos \theta_1
\end{array} \right) \thickspace, \qquad U = \left( \begin{array}{c c}
\cos \theta_2 & \sin \theta_2 \\
- \sin \theta_2 & \cos \theta_2
\end{array} \right)
\end{equation}
one gets
\begin{eqnarray}\nonumber
H_d^1 &=& \cos \theta_1 H_d^L + \sin \theta_1 H_d^R \label{higgs-rot-1} \thickspace, \\
H_d^2 &=& - \sin \theta_1 H_d^L + \cos \theta_1 H_d^R \label{higgs-rot-2} \thickspace,
\end{eqnarray}
and similar for $H_u$.  In general the angles $\theta_1$ and
$\theta_2$ are different. However, they are connected to the same
matrix $M_H$ and can be calculated by diagonalizing $M_H (M_H)^T$ or
$(M_H)^T M_H$. Using eq.~\eqref{id1}  one finds
\begin{eqnarray}
\tan \theta_{1,2} &=& 
\frac{\mu_{12} \pm \alpha_{12} M_R}{\mu_{22}} \label{ang1} \thickspace.
\end{eqnarray}
In these expressions ${\rm Det}(M_H)=0$ has been used to simplify the
result.  Exact ${\rm Det}(M_H)=0$ implies that the $\mu$-term of the MSSM is 
zero, so this condition can only be true up to small corrections, 
see the discussion below.
Note that there are two interesting limits. First, $\mu_{12} \gg
\alpha_{12} M_R$ : this implies $\tan \theta_1 = \tan \theta_2$ and
therefore $D = U$. This is as expected, since that limit makes $M_H$
symmetric. And, second, $\mu_{12} \ll \alpha_{12} M_R$ : this implies
$\tan \theta_1 = - \tan \theta_2$ and therefore $D = U^T$.

The superpotential at this stage is
\begin{eqnarray} \label{eq:Wsuppot2}
{\cal W} &=& Y_u Q H_u u^c + Y_d Q H_d d^c 
          +  Y_e L H_d e^c + Y_\nu L H_u \nu^c 
          +  \mu H_u H_d \nonumber \\
         &+& f_c^1 \nu^c \nu^c  \Delta^{c \: 0}
          +  M_{\Delta^c}^1 \Delta^{c \: 0} \bar{\Delta}^{c \: 0}
          +  a_c^1 \Delta^{c \: 0} \bar{\Delta}^{c \: 0} \Omega^{c \: 0} 
\nonumber \\
         &+& b \Omega H_d H_u
          +  b_c \Omega^{c \: 0} H_d H_u
          +  M_\Omega \Omega \Omega
          +  M_{\Omega^c} \Omega^{c \: 0} \Omega^{c \: 0} .
\end{eqnarray}
Particles belonging to the same $SU(2)_R$ gauge multiplets split due 
to their different $U(1)_R$ charges. At this stage both the LR group, 
that symmetrizes the $SU(2)_L$ and $SU(2)_R$ gauge interactions, and 
the discrete parity symmetry that we imposed on the couplings are broken. 

The soft terms are
\begin{eqnarray} \label{eq:soft2}
- {\cal L}_{soft} &=& m_Q^2 \tilde{Q}^\dagger \tilde{Q} + m_{u^c}^2
\tilde{u^c}^\dagger \tilde{u^c} + m_{d^c}^2 \tilde{d^c}^\dagger
\tilde{d^c} + m_L^2 \tilde{L}^\dagger \tilde{L} + m_{e^c}^2
\tilde{e^c}^\dagger \tilde{e^c} + m_{\nu^c}^2 \tilde{\nu^c}^\dagger
\tilde{\nu^c} \nonumber \\ &+& m_{H_u}^2 H_u^\dagger H_u + m_{H_d}^2
H_d^\dagger H_d + m_{\Delta^{c \: 0}}^2 {\Delta^{c \: 0}}^\dagger
\Delta^{c \: 0} \nonumber \\
&+& m_{\bar{\Delta}^{c \: 0}}^2 \bar{\Delta}^{c \: 0 \:
  \dagger} \bar{\Delta}^{c \: 0} + m_{\Omega}^2
\Omega^\dagger \Omega + m_{\Omega^{c \: 0}}^2 \Omega^{c \: 0 \:
  \dagger} \Omega^{c \: 0} \\
&+& \frac{1}{2} \big[ M_1 \tilde{B}^0
  \tilde{B}^0 + M_L \tilde{W_L} \tilde{W_L} + M_R \tilde{W_R^0}
  \tilde{W_R^0} + M_3 \tilde{g} \tilde{g} + h.c. \big] \nonumber \\
&+& \big[ T_u \tilde{Q} H_u \tilde{u^c} + T_d \tilde{Q} H_d
  \tilde{d^c} + T_e \tilde{L} H_d \tilde{e^c} + T_\nu \tilde{L} H_u
  \tilde{\nu^c} \nonumber \\ &+& T_{f_c}^1 \tilde{\nu^c} \tilde{\nu^c} \Delta^{c
    \: 0} + T_{a_c}^1 \Delta^{c \: 0} \Omega^{c \: 0} \bar{\Delta}^{c
    \: 0} + T_b \Omega H_d H_u + T_{b^c} \Omega^{c \: 0} H_d H_u +
  h.c. \big] \nonumber \\ &+& \big[ B_\mu H_u H_d + B_{M_{\Delta^c}^1}
  \Delta^{c \: 0} \bar{\Delta}^{c \: 0} + B_{M_\Omega} \Omega \Omega +
  B_{M_\Omega^c} \Omega^{c \: 0} \Omega^{c \: 0} + h.c. \big] \thickspace.
\nonumber
\end{eqnarray}
Again we  suppress gauge and family indices.

We must impose matching conditions at the $SU(2)_R$ breaking scale. 
These are for superpotential parameters given by
\begin{eqnarray}\nonumber
Y_d = Y_Q^1 \cos \theta_1 - Y_Q^2 \sin \theta_1 \thickspace,
&\qquad& Y_u = - Y_Q^1 \cos \theta_2 + Y_Q^2 \sin \theta_2  \thickspace, \\ \nonumber
Y_e = Y_L^1 \cos \theta_1 - Y_L^2 \sin \theta_1 \thickspace,
&\qquad& Y_\nu = - Y_L^1 \cos \theta_2 + Y_L^2 \sin \theta_2 \thickspace, \\ \nonumber
f_c^1 = - f^* \thickspace, &\qquad& a_c^1 = - \frac{a^*}{\sqrt{2}}  \thickspace, \\ \nonumber
M_{\Delta^c}^1 = M_\Delta^* \thickspace, &\qquad& M_{\Omega^c} = M_\Omega^* \thickspace, \\
b = 2 \alpha R \thickspace, &\qquad& b_c = \sqrt{2} \alpha^* R \thickspace,
\end{eqnarray}
where $R = \sin (\theta_1 - \theta_2)$. For the soft masses we have 
\begin{eqnarray}
m_{u^c}^2 = m_{d^c}^2 &=& m_{Q^c}^2 \thickspace,\\ \nonumber
m_{e^c}^2 = m_{\nu^c}^2 &=& m_{L^c}^2 \thickspace, \\ \nonumber
m_{\Delta^{c \: 0}}^2 &=& m_{\Delta^c}^2 \thickspace, \\ \nonumber
m_{\bar{\Delta}^{c \: 0}}^2 &=& m_{\bar{\Delta}^c}^2 \thickspace ,\\ \nonumber
m_{\Omega^{c \: 0}}^2 &=& m_{\Omega^c}^2 \thickspace, \\ \nonumber
M_L = M_R &=& M_2 \thickspace.
\end{eqnarray}

The matching of the soft trilinears follows corresponding conditions. 
In addition, one has
\begin{eqnarray}
m_{H_d}^2 &=& \cos^2 \theta_1 (m_\Phi^2)_{11} + \sin^2 \theta_1
(m_\Phi^2)_{22} - \sin \theta_1 \cos \theta_1 \left[ (m_\Phi^2)_{12} +
(m_\Phi^2)_{21} \right] \thickspace, \nonumber \\ 
m_{H_u}^2 &=& \cos^2 \theta_2
(m_\Phi^2)_{11} + \sin^2 \theta_2 (m_\Phi^2)_{22} - \sin \theta_2 \cos
\theta_2 \left[ (m_\Phi^2)_{12} + (m_\Phi^2)_{21} \right] \thickspace, \nonumber
\end{eqnarray}
as obtained when the operator $m_\Phi^2 \Phi^\dagger \Phi$ is
projected into the light Higgs doublets operators $(H_d^L)^\dagger
H_d^L$ and $(H_u^L)^\dagger H_u^L$. Gauge couplings are matched as 
$g_L = g_R = g_2$.

\subsection{Step 3: From $U(1)_{B-L}$ breaking scale to EW/SUSY scale}

We mention this stage only for completeness, since the last regime is 
just the usual MSSM. We need matching conditions in the
gauge sector. Since $U(1)_R \times U(1)_{B-L}$ breaks to $U(1)_Y$, the
MSSM gauge coupling $g_1$ will be a combination of $g_R$ and $g_{BL}$.
The resulting relationship is
\begin{equation}
g_1 = \frac{\sqrt{5} g_R g_{BL}}{\sqrt{2 g_R^2 + 3 g_{BL}^2}} \thickspace .
\end{equation}
Analogously, the following condition holds for gaugino masses
\begin{equation}
M_1({\rm MSSM}) = \frac{2 g_R^2 M_1 + 3 g_{BL}^2 M_R}{2 g_R^2 + 3 g_{BL}^2} \thickspace .
\end{equation}
Note that in the last two equations the gauge couplings are GUT-normalized. 
Electroweak symmetry breaking occurs as in the MSSM. We take the Higgs
doublet VEVs
\begin{equation} \label{higgs-VEVs}
\langle H_d^0 \rangle = \frac{v_d}{\sqrt{2}}  \thickspace,
\qquad \langle H_u^0 \rangle = \frac{v_u}{\sqrt{2}} \thickspace, 
\end{equation}
as free parameters and then solve the tadpole equations to find
$\mu_{\rm MSSM}$ and $B^\mu$. $\mu_{\rm MSSM}$ must be different from
zero, that is ${\rm Det}(M_H)$ can not be exactly zero. Instead the
tuning must be exact up to ${\rm Det}(M_H)={\cal O}(\mu_{\rm
  MSSM}^2)$. This strong fine-tuning is required for a correct EWSB
and is nothing but the $\mu$-problem of the MSSM, which we do not
attempt to solve here. Finally, $\tan \beta = \frac{v_u}{v_d}$ is used
as a free parameter.  Also the sign of $\mu_{\rm MSSM}$ is not
constrained as usual.

\subsection{Neutrino masses, LFV and Yukawa couplings}
\label{sec:yukawas-lr}

Neutrino masses are generated after $U(1)_{B-L}$ breaking through a
type-I seesaw mechanism. The matrix $f_c^1$ leads to Majorana
masses for the right-handed neutrinos once $\Delta^{c \: 0}$ gets a
VEV. We define the seesaw scale as the lightest eigenvalue of 

\begin{equation}
M_S \equiv f_c^1 v_{BL} \thickspace .
\end{equation}

As usual, we can always rotate the fields to a basis where $M_S$ is diagonal. However, this will introduce lepton flavor violating entries in the $Y_{L_i}$ Yukawas, see discussion below. As mentioned above, contrary to non-supersymmetric LR models \cite{Mohapatra:1979ia}, 
there is no type-II contribution to neutrino masses.

Global fits to all available experimental data provide values for the
parameters involved in neutrino oscillations, see table \ref{tab:nudata} 
for updated results. As first observed in \cite{Harrison:2002er}, these 
data imply that the neutrino mass matrix can be diagonalized to a good 
approximation by the so-called tri-bimaximal mixing pattern:
\begin{equation}
\label{eq:UTBM}
U_{TBM} =
\left(\begin{array}{cccc}
\sqrt{\frac{2}{3}} & \sqrt{\frac{1}{3}} & 0 \cr
- \frac{1}{\sqrt{6}} &  \frac{1}{\sqrt{3}} & - \frac{1}{\sqrt{2}} \cr
- \frac{1}{\sqrt{6}} &  \frac{1}{\sqrt{3}} & \frac{1}{\sqrt{2}}
\end{array}\right).
\end{equation}
The matrix product $Y_\nu \cdot (f_c^1)^{-1} \cdot Y_\nu^T$ is constrained 
by this particular structure. LFV entries can be present in both $Y_\nu$ 
and $f_c^1$, see also the discussion about parameter counting in the 
next subsection. However, in the numerical section we will consider only 
two specific kinds of fits:
\begin{itemize}

\item $Y_\nu$-fit: flavor structure in $Y_\nu$ and diagonal $f_c^1$.

\item $f$-fit: flavor structure in $f_c^1$ and diagonal $Y_\nu$.

\end{itemize}
While at first it may seem either way of doing the fit is equivalent, 
$f_c^1$ and $Y_{\nu}$ in our setup can leave different traces in the 
soft slepton mass parameters if $v_{BL} \ll v_R$. This last condition 
is essential to distinguish between both possibilities, because otherwise 
one obtains the straightforward prediction that LFV entries in left and 
right slepton are equal, due to the assumed LR symmetry above $v_R$.

These two types of fit were already discussed in reference \cite{Babu:2002tb},
which investigates low energy LFV signatures in a supersymmetric seesaw model
where the right-handed neutrino mass is generated from a term of the form
$f \Delta^c \nu^c \nu^c$. When the scalar component of $\Delta^c$ acquires a VEV
a type-I seesaw is obtained, generating masses for the light neutrinos.
Therefore, this model has the ingredients to accommodate a $Y_\nu$-fit,
named as \emph{Dirac LFV} in \cite{Babu:2002tb}, or a $f$-fit, named as
\emph{Majorana LFV}. Note, however, that the left-right symmetry, central in
our work, is missing in this reference, thus implying different signatures at
the electroweak scale.

The difference in phenomenology of the two fits can be easily understood 
considering approximated expressions for the RGEs for $m_L^2$ and 
$m_{e^c}^2$. In the first step, from the GUT scale to the $v_R$ 
scale RGEs at 1-loop order can be written in leading-log approximation 
as \cite{Chao:2007ye}
\begin{eqnarray}\nonumber
\Delta m_L^2 &=& - \frac{1}{4 \pi^2} \left( 3 f f^\dagger + Y_L^{(k)} Y_L^{(k) 
\: \dagger} \right) (3 m_0^2 + A_0^2) \ln \left( \frac{m_{GUT}}{v_R} \right) \thickspace,
\label{apprge1} \\
\Delta m_{L^c}^2 &=& - \frac{1}{4 \pi^2} \left( 3 f^\dagger f + Y_L^{(k) 
\: \dagger} Y_L^{(k)} \right) (3 m_0^2 + A_0^2)
\ln \left( \frac{m_{GUT}}{v_R} \right) \thickspace. \label{apprge2}
\end{eqnarray}
Of course, also the $A$ parameters develop LFV off-diagonals in the 
running. We do not give the corresponding approximated equations for 
brevity. After parity breaking at the $v_R$ scale the Yukawa coupling 
$Y_L$ splits into $Y_e$, the charged lepton Yukawa, and $Y_\nu$, the
neutrino Yukawa. The later contributes to LFV entries in the running
down to the $v_{BL}$ scale. Thus,
\begin{eqnarray}
\Delta m_L^2 &\sim & - \frac{1}{8 \pi^2} Y_\nu Y_\nu^\dagger
  \left( 3 m_L^2|_{v_R} + A_e^2|_{v_R}\right)
\ln \left( \frac{v_R}{v_{BL}} \right) \thickspace, \nonumber  \\
\Delta m_{e^c}^2 & \sim & 0  \thickspace, \label{apprge4}
\end{eqnarray}
where $m_L^2|_{v_R}$ is the matrix $m_L^2$ at the scale $v_R$ and 
$A_e^2|_{v_R}$ is defined as $T_e = Y_e A_e$ and also has to be taken 
at $v_R$. 
In order to understand the main difference between the two fits, let
us first consider the $f$-fit.  This assumes that $Y_{\nu}$ is
diagonal at the seesaw scale and thus the observed low energy mismatch
between the neutrino and charged lepton sectors is due to a
non-trivial flavor structure in $f_c^1$. Of course, non-diagonal 
entries in $f$ generate in the running also non-diagonal entries in 
$Y_{\nu}$ and $Y_e$, but these can be neglected in first approximation. 
In this case, equations \eqref{apprge2} and \eqref{apprge4} show that 
the LR symmetry makes $m_L^2$ and $m_{e^c}^2$ run with the same flavor 
structure and the magnitudes of their off-diagonal entries at the SUSY 
scale are similar. If, on the other hand, $Y_{\nu}$ is non-trivial 
($Y_{\nu}$-fit), while $f$ is diagonal, the running from the GUT scale 
to the $v_R$ scale induces again the same off-diagonal entries in $m_L^2$ 
and $m_{L^c}^2$. However, from $v_R$ to $v_{BL}$ the off-diagonals entries 
in $m_L^2$ continue to run, while those in $m_{e^c}^2$ do not. This effect, 
generated by the right-handed neutrinos via the $Y_\nu$ Yukawas, induces 
additional flavor violating effects in the L sector compared to the R sector. 
Seeing LFV in both left- and right slepton sectors thus allows us to 
indirectly learn about the high energy theory. We will study this in 
some detail in the phenomenology section below, where the numerical results
will be presented.

\subsection{Parameter counting}

Let us briefly summarize the free parameters of the model. With the 
assumption of mSUGRA (or better: mSUGRA-like) boundary conditions, 
in the SUSY breaking sector we only have the standard parameters 
$m_0$, $M_{1/2}$, $A_0$, $\tan \beta$, $sign(\mu_{\rm MSSM})$. Thus, 
we count 4+1 parameters in the soft terms. We note in passing that the 
soft terms of the heavy sector, of course, do not have to follow strictly 
the conditions outlined in equation \eqref{eq:soft1}, as long as these 
parameters are small compared to $v_{BL}$ there are no changes compared 
to the above discussion.

In the superpotential we have $a$, $\alpha$, $\mu$, $M_{\Delta}$ and 
$M_{\Omega}$. This leaves, at first sight, 7 parameters free. However, 
we can reduce them to 4+2 parameters as follows. Since $\alpha_{ij} 
= - \alpha_{ji}$, $\alpha$ only contains one free parameter: $\alpha_{12}$. 
The matrix $\mu$ has 3 entries, but one of them, $\mu_{11}$, is fixed 
by the fine-tuning condition ${\rm Det}(M_H)={\cal O}(\mu_{\rm MSSM}^2)$. This 
leaves two free parameters, $\mu_{12}$, $\mu_{22}$. We have traded 
$M_{\Delta}$ and $M_{\Omega}$ for the VEVs $v_R$, $v_{BL}$, since 
$\ln ( \frac{v_R}{v_{BL}})$ and $\ln ( \frac{v_{GUT}}{v_{R}})$ enter 
into the RGEs and thus can, at least in principle, be determined from 
low-energy spectra. There are then in summary 6 parameters, four  
independent of low-energy constraints and two which could be fixed 
from LFV data, see below.

In addition, in the superpotential we have the Yukawa matrices $Y_{Q_i}$, 
$Y_{L_i}$ and $f$. Let's consider the quark sector first. Since we 
can always go to a basis in which one of the $Y_{Q_i}$ is diagonal 
with only real entries, there are 12 parameters. Ten of them are 
fixed by six quark masses, three CKM angles and the CKM phase, leaving 
two phases undetermined. 

In the lepton sector we have the symmetric matrices, $Y_{L_1}$ and 
$Y_{L_2}$. As with the quark sector, a basis change shows that there 
are only 12 free parameters. $f$ is symmetric and thus counts as another 
9 parameters. Going to a basis in which $f$ is diagonal does not reduce 
the number of free parameters, since in this basis we can no longer 
assume one of the $Y_{L_i}$ to be diagonal. In summary there are thus 
free 21 parameters in these three matrices. 

In the simple, pure seesaw type-I with three generations of right-handed 
neutrinos the number of free parameters is 21. Only 12 of them can be 
fixed from low-energy data: three neutrino and three charged lepton masses, 
three leptonic mixing angles and three phases (two Majorana and one 
Dirac phase). However, as pointed out in \cite{Ellis:2002fe}, in principle, 
$m_L^2$ contains 9 observable entries and thus, if the normalization (i.e. 
$m_0$, $A_0$, $\tan\beta$ etc.) is known from other sfermion measurements, 
one could re-construct the type-I seesaw parameters \footnote{Of course, 
this discussion is slightly academic, since at least one of the Majorana 
phases will never be measured in praxis.}.

How does the SUSY LR model compare to this? We have, as discussed above, 
also 21 parameters in the three coupling matrices, but neutrino masses 
depend also on $v_{BL}$. However, in principle, we have 9 more observables 
in $m_{e^c}^2$, assuming again that the soft SUSY breaking terms 
can be extracted 
from other measurements. Since in the RGEs also $v_R$ appears we have 
in total 23 parameters which need to be determined. The number of observables, 
on the other hand is fixed to 30 in total, as we have 12 (low-energy lepton 
sector) plus 9 (left sleptons) plus 9 (right sleptons) possible 
measurements. 

\section{Phenomenology}

\subsection{Procedure for numerics}

All necessary, analytical expressions were calculated with SARAH. For
this purpose, two different model files for the model above the two
threshold scales were created and used to calculate the full set of
2-loop RGEs. SARAH calculates the RGEs using the generic expressions
of \cite{Martin:1993zk} in the most general form respecting the
complete flavor structure. These RGEs were afterwards exported to
Fortran code and implemented in SPheno. As starting point for the RGE
running, the gauge and Yukawa couplings at the electroweak scale are
used. In the calculation of the gauge and Yukawa couplings we follow
closely the procedure described in reference \cite{Porod:2003um}: the
values for the Yukawa couplings giving mass to the SM fermions and the
gauge couplings are determined at the scale \(M_Z\) based on the
measured values for the quark, lepton and vector boson masses as well
as for the gauge couplings.  Here, we have included the 1-loop
corrections to the mass of W- and Z-boson as well as the SUSY
contributions to \(\delta_{VB}\) for calculating the gauge
couplings. Similarly, we have included the complete 1-loop
corrections to the self-energies of SM fermions
\cite{Pierce:1996zz}. Moreover, we have resummed the $\tan\beta$
enhanced terms for the calculation of the Yukawa couplings of the
$b$-quark and the $\tau$-lepton as in \cite{Porod:2003um}. 
 The vacuum expectation values \(v_d\) and
\(v_u\) are calculated with respect to the given value of
\(\tan\beta\) at \(M_Z\). Since we are working with two distinct
threshold scales, not all heavy fields are integrated out at their
mass and the corresponding 1-loop boundary conditions at the
threshold scales are needed. It is known that these particles cause a
finite shift in the gauge couplings and gaugino masses.  The general
expressions are \cite{Hall:1980kf}
\begin{eqnarray}
\label{eq:shift1}
 g_i & \rightarrow & g_i \left( 1\pm \frac{1}{16 \pi^2} g_i^2 I^i_2(r) 
\ln\left(\frac{M^2}{M_T^2}\right)\right)  \thickspace ,\\
\label{eq:shift2}
 M_i & \rightarrow & M_i \left( 1\pm \frac{1}{16 \pi^2} g_i^2 I^i_2(r) 
\ln\left(\frac{M^2}{M_T^2}\right)\right) \thickspace .
\end{eqnarray}
\(I^i_2(r)\) is the Dynkin index of a field transforming as
representation \(r\) with respect to the gauge group belonging to the
gauge coupling \(g_i\), \(M\) is the mass of this particle and \(M_T\)
is the threshold scale. When evaluating the RGEs from the low to the
high scale, the contribution is positive, when running down, it is
negative. The different masses used for calculating the finite shifts
are the eigenvalues of the full tree-level mass matrix of the charged,
heavy particles removed from the spectrum.  The correct mass spectrum
is calculated in an iterative way.

\begin{figure}
\begin{center}
\vspace{5mm}
\includegraphics[width=0.5\textwidth]{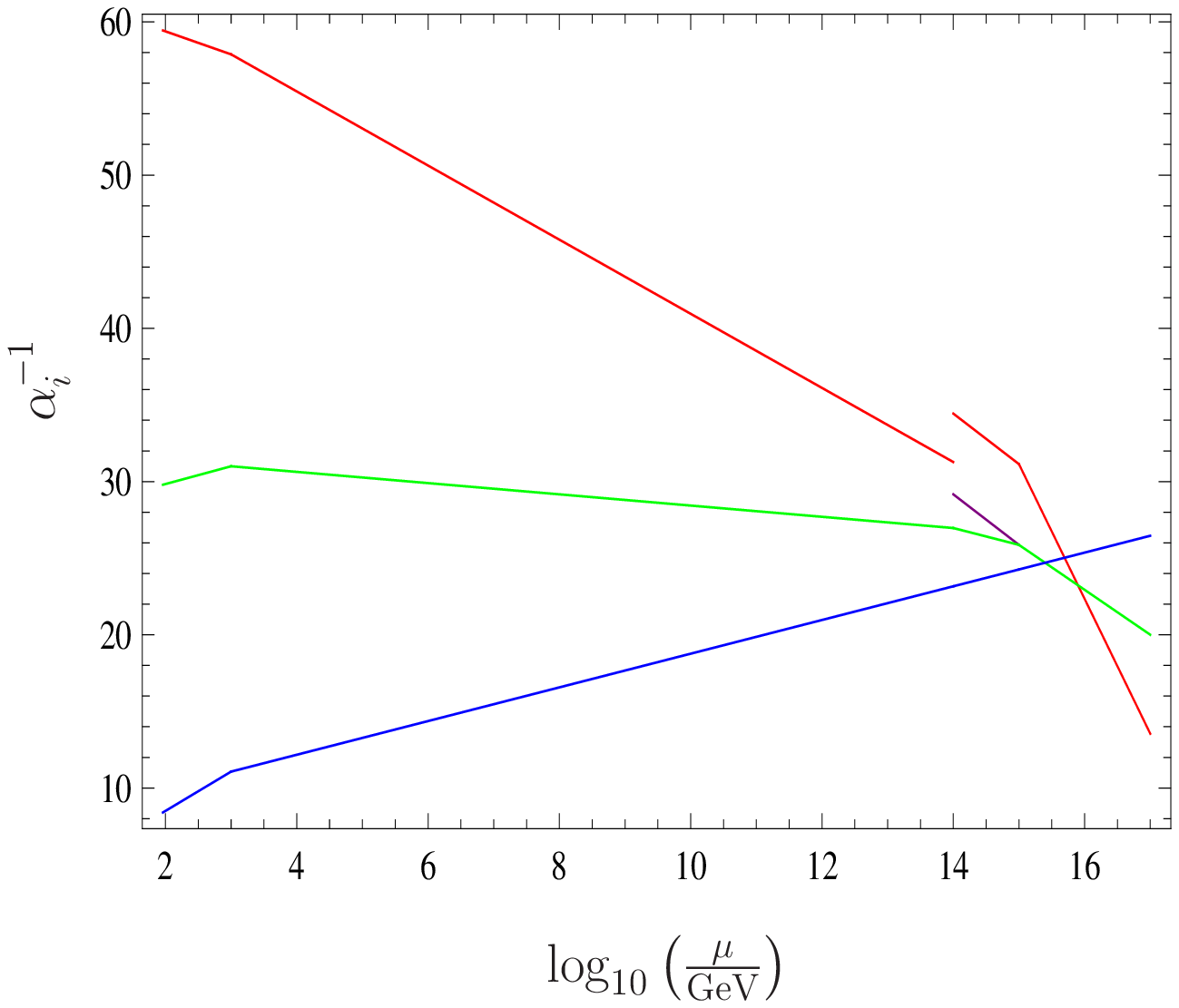}
\end{center}
\vspace{-5mm}
\caption{1-loop running of the gauge couplings for the choice of scales $m_{SUSY} = 1$ TeV, $v_{BL} = 10^{14}$ GeV and $v_R = 10^{15}$ GeV. The dependence of $\alpha_i^{-1}$, where $\alpha_i = \frac{g_i^2}{4 \pi}$, on the energy scale $\mu$ is shown. Different gauge couplings are represented in the different energy regimes. For $\mu \in \left[m_Z,v_{BL}\right]$ one has $\alpha_3^{-1}$ (blue), $\alpha_L^{-1}$ (green) and $\alpha_Y^{-1}$ (red). For $\mu \in \left[v_{BL},v_R\right]$ one has $\alpha_3^{-1}$ (blue), $\alpha_L^{-1}$ (green), $\alpha_R^{-1}$ (purple) and $\alpha_{BL}^{-1}$ (red). For $\mu > v_R$ one has $\alpha_3^{-1}$ (blue), $\alpha_2^{-1} \equiv \alpha_L^{-1} = \alpha_R^{-1}$ (green) and $\alpha_{BL}^{-1}$ (red).}
\label{fig:gcu-original}
\end{figure}

\begin{figure}
\begin{center}
\vspace{5mm}
\includegraphics[width=0.49\textwidth]{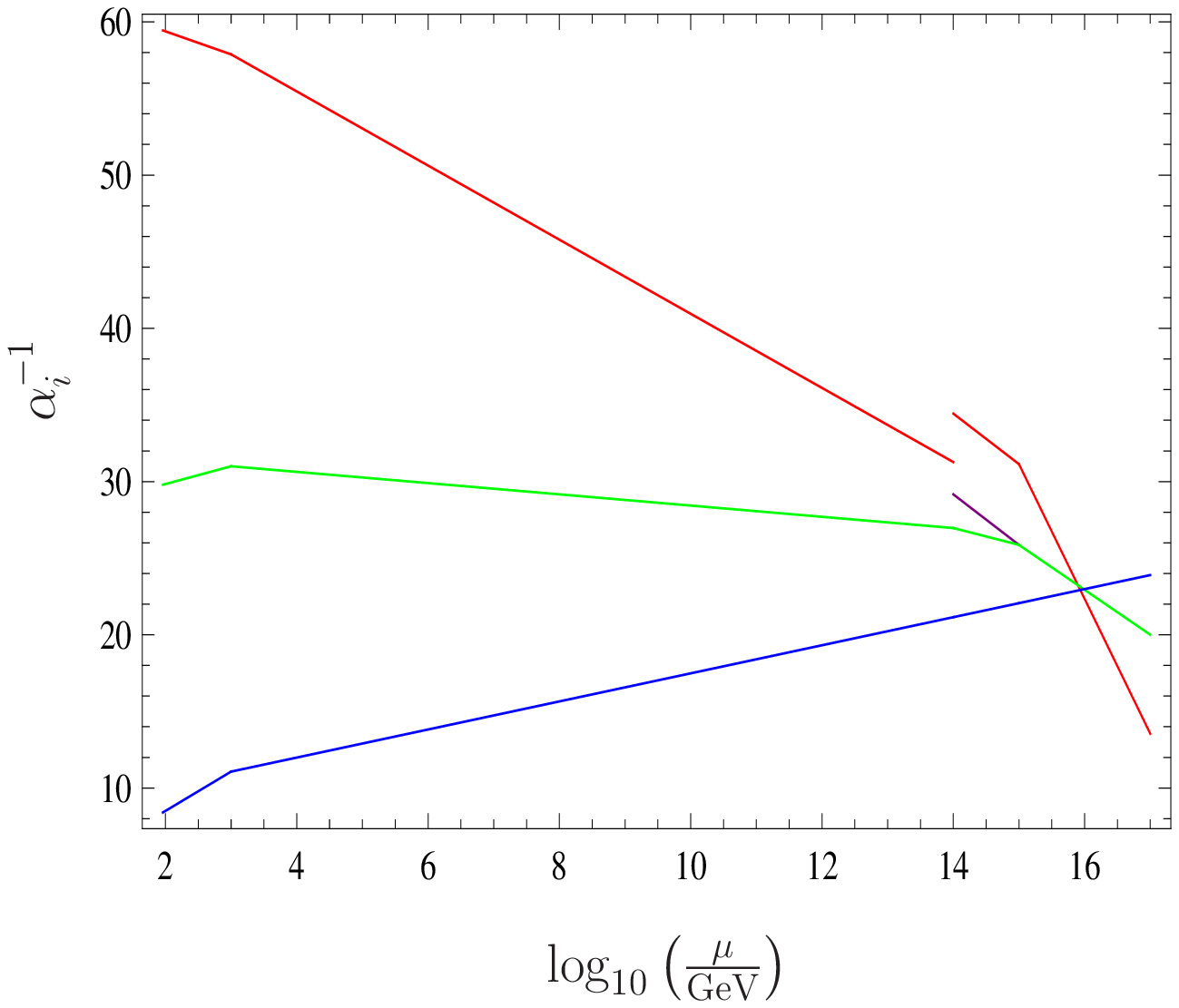}
\includegraphics[width=0.49\textwidth]{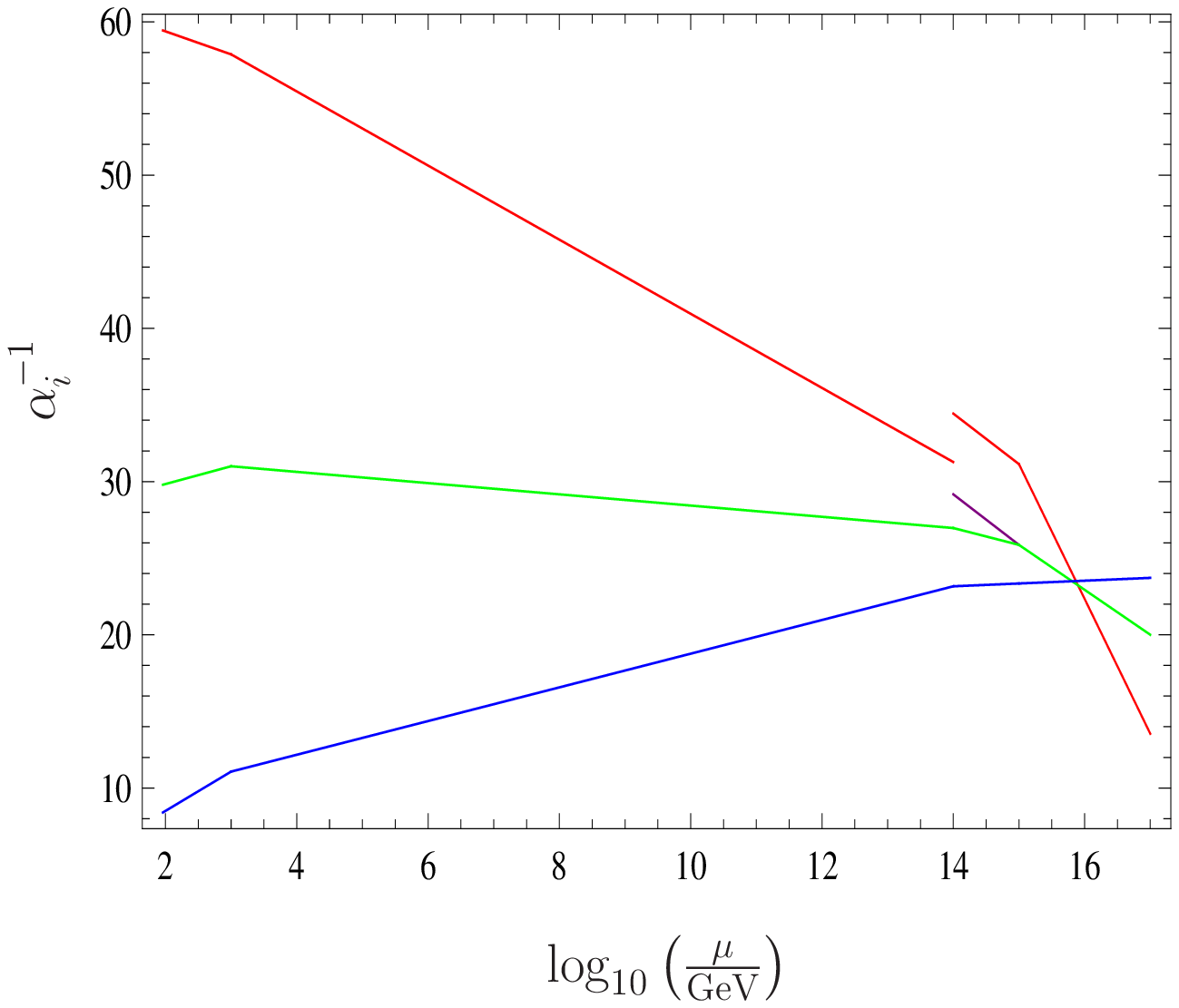}
\end{center}
\vspace{-5mm}
\caption{1-loop running of the gauge couplings for the choice of scales $m_{SUSY} = 1$ TeV, $v_{BL} = 10^{14}$ GeV and $v_R = 10^{15}$ GeV. Contrary to figure \ref{fig:gcu-original}, gauge coupling unification is obtained thanks to additional colored superfields. In the left panel, one triplet under $SU(3)_c$, singlet under the other gauge subgroups, is added at $m_{SUSY}$, whereas in the right panel five generations of the same superfield are added at $v_{BL}$. See figure \ref{fig:gcu-original} for the color code.}
\label{fig:gcu-solved}
\end{figure}

Some comments on the determination of the GUT scale are now in
order. In our numerical procedure the GUT scale is defined as the
scale at which \(g_{BL} = g_2 = g_{GUT}\) holds. Generally, there is
difference with \(g_3\) to \(g_{GUT}\) in the percent range, the
actual numerical mismatch depending on the scales $v_{BL}$ and $v_R$
and being larger for lower values of $v_{BL}$ and $v_R$. Figure
\ref{fig:gcu-original} shows an example for the choice of scales
$m_{SUSY} = 1$ TeV, $v_{BL} = 10^{14}$ GeV and $v_R = 10^{15}$ GeV. It
has been stressed in particular in \cite{Kopp:2009xt} that within
supersymmetric LR models, the LR symmetry breaking scale has to be
close to the GUT scale, otherwise this mismatch will grow too large.

However, several solutions are known. In \cite{Majee:2007uv} it was
pointed out that GUT thresholds - unknown unless the GUT model,
including the complete Higgs sector used to break the GUT symmetry, is
specified - can lead to important corrections, accounting for this
apparent non-unification\footnote{For a discussion of these effects in
  the context of \(SU(5)\) see \cite{Martens:2010nm}.}. Another
possibility is the addition of new particles to the spectrum. As
clearly seen in figure \ref{fig:gcu-original}, unification is not
obtained due to a too fast running of $g_3$. This can be fixed by
adding new superfields charged under $SU(3)_c$ but singlet under the
other gauge subgroups, as pointed out in \cite{Borah:2010kk}. Two
examples are shown in figure \ref{fig:gcu-solved}, where the addition
of triplets of $SU(3)_c$ has been considered. To the left, one
generation is added at $m_{SUSY}$, whereas to the right five
generations are added at $v_{BL}$. In both cases the new contributions
to the running of $g_3$ are sufficient to obtain gauge coupling
unification.

Nevertheless, in our numerical procedure we simply use $g_{BL} = g_2 =
g_{GUT}$ and attribute departures from complete unification to
(unknown) thresholds and/or the existence of additional colored
particles below $m_{GUT}$.

After applying the
GUT scale boundary conditions, the RGEs are evaluated down to the low scale
and the mass spectrum of the MSSM is calculated. The MSSM masses are,
in general, calculated at the 1-loop level in the
\(\overline{\mbox{DR}}\) scheme using on-shell external momenta. For
the Higgs fields also the most important 2-loop contributions are taken into
account. We note that the corresponding Fortran routines are also
written by SARAH but they are equivalent to the routines
included in the public version of SPheno based on
\cite{Pierce:1996zz}.  The iteration stops when the largest change
in the calculation of the SUSY and Higgs boson masses at
\(m_{SUSY}\) is below one per-mille  between two iterations.

\subsection{Mass spectrum}

\begin{figure}
\begin{center}
\vspace{5mm}
\includegraphics[width=0.47\textwidth]{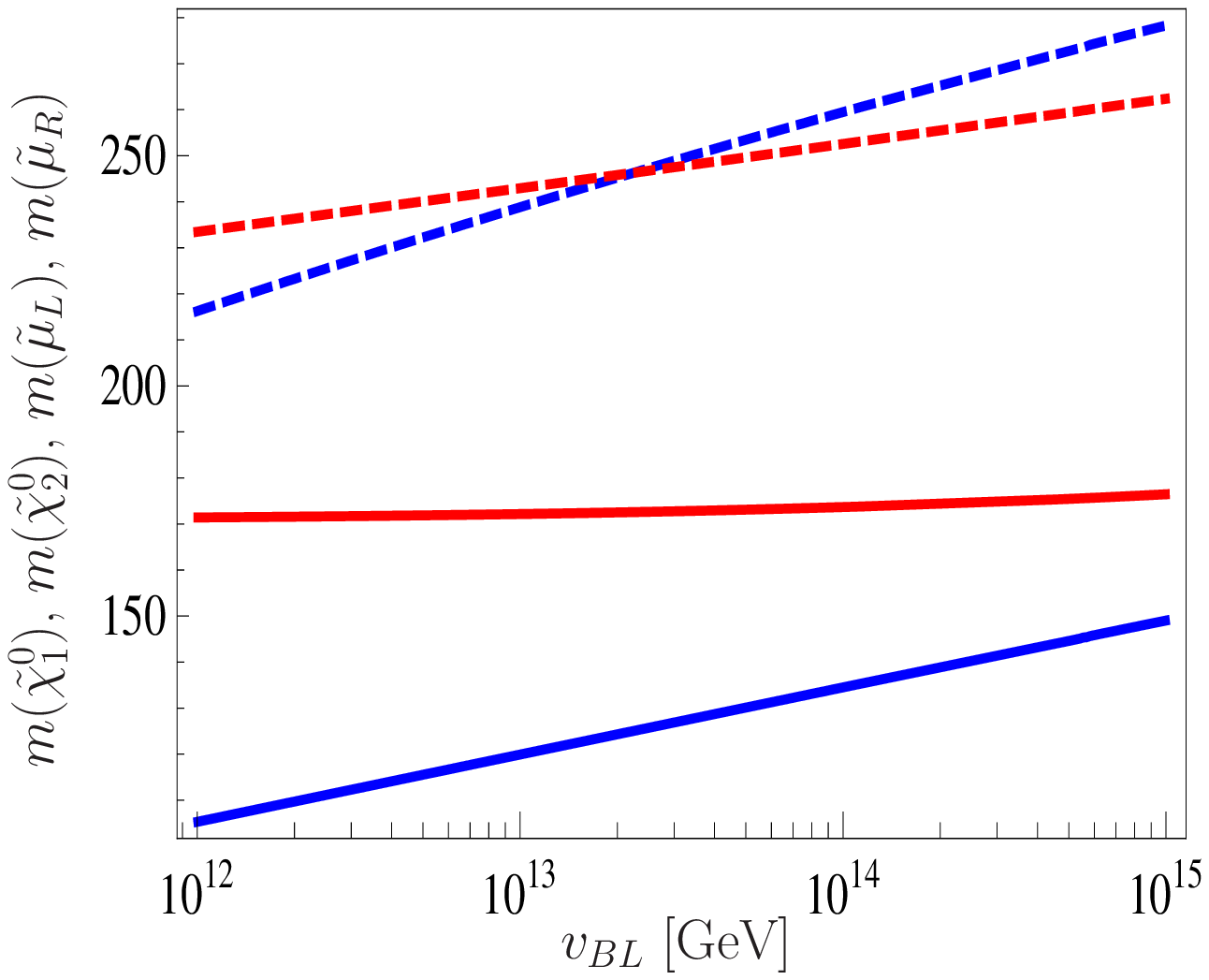}
\hspace{5mm}
\includegraphics[width=0.47\textwidth]{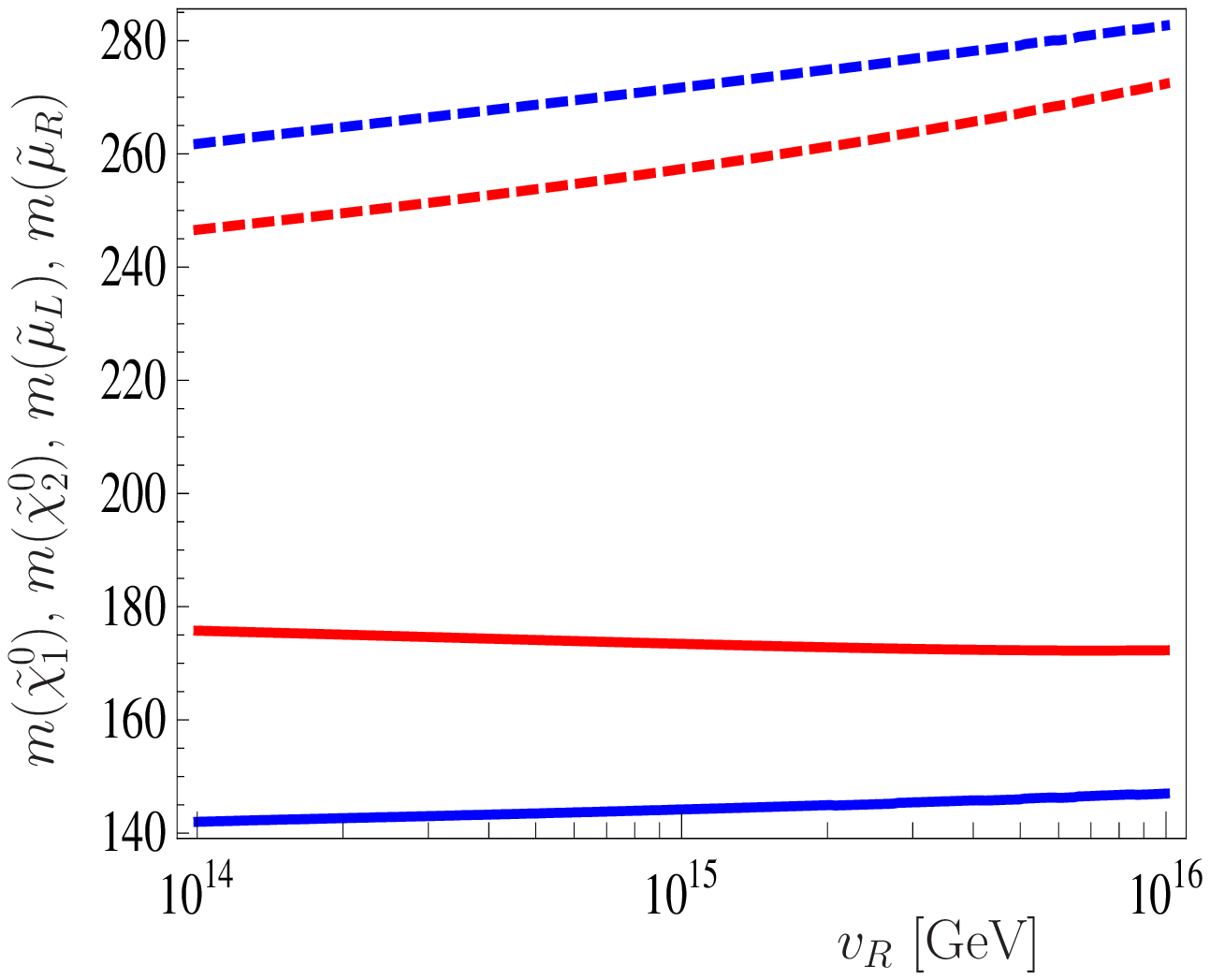}
\end{center}
\vspace{-5mm}
\caption{Example of spectra at the SUSY scale and its dependence on
$v_{BL}$ (left side) and $v_R$ (right side). The masses of four states
are shown: $\tilde{\chi}_1^0$ (blue line), $\tilde{\chi}_2^0$ (blue
dashed line), $\tilde{\mu}_R$ (red line) and $\tilde{\mu}_L$ (red
dashed line). In both panels the mSUGRA parameters have been taken as
in the SPS3 benchmark point.}
\label{fig:masses-scales}
\end{figure}

The appearance of charged particles at scales between the electroweak
scale and the GUT scale leads to changes in the beta functions
of the gauge couplings \cite{Rossi:2002zb,Buckley:2006nv}.  
This does not only change the evolution of the gauge couplings but 
also the evolution of the gaugino and scalar mass parameters 
\cite{Buckley:2006nv,Hirsch:2008gh}. The LR model contains additional 
triplets, and similar to what is observed in the seesaw models 
\cite{Esteves:2010ff} the mass spectrum at low energies is shifted with 
respect to mSUGRA expectations. Two examples of this behaviour are 
shown in figure \ref{fig:masses-scales}. In this figure we show the 
two lightest neutralino masses and the masses of the left and right 
smuons versus $v_{BL}$ (left side) and $v_R$ (right side). We note 
that also all other sfermion and gaugino masses show the same dependence and
in general smaller values are obtained for lower values of $v_{BL}$ and $v_R$. 
One finds that gaugino masses depend stronger on $v_{BL}$ and $v_R$ 
than sfermion masses and that right sleptons are the sfermions for
which the sensitivity to these VEVs is smallest. 

\begin{figure}
\begin{center}
\vspace{5mm}
\includegraphics[width=0.47\textwidth]{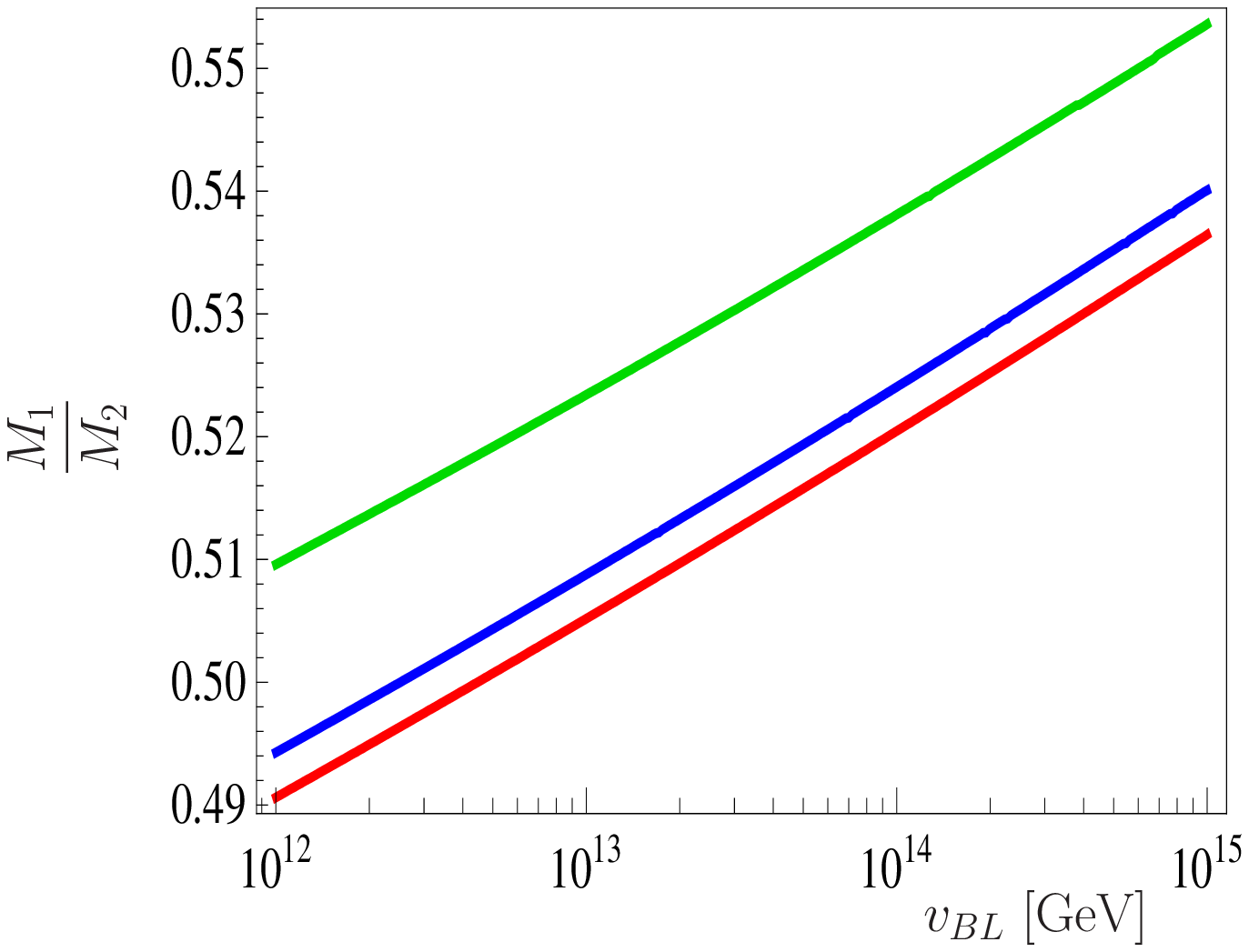}
\hspace{5mm}
\includegraphics[width=0.47\textwidth]{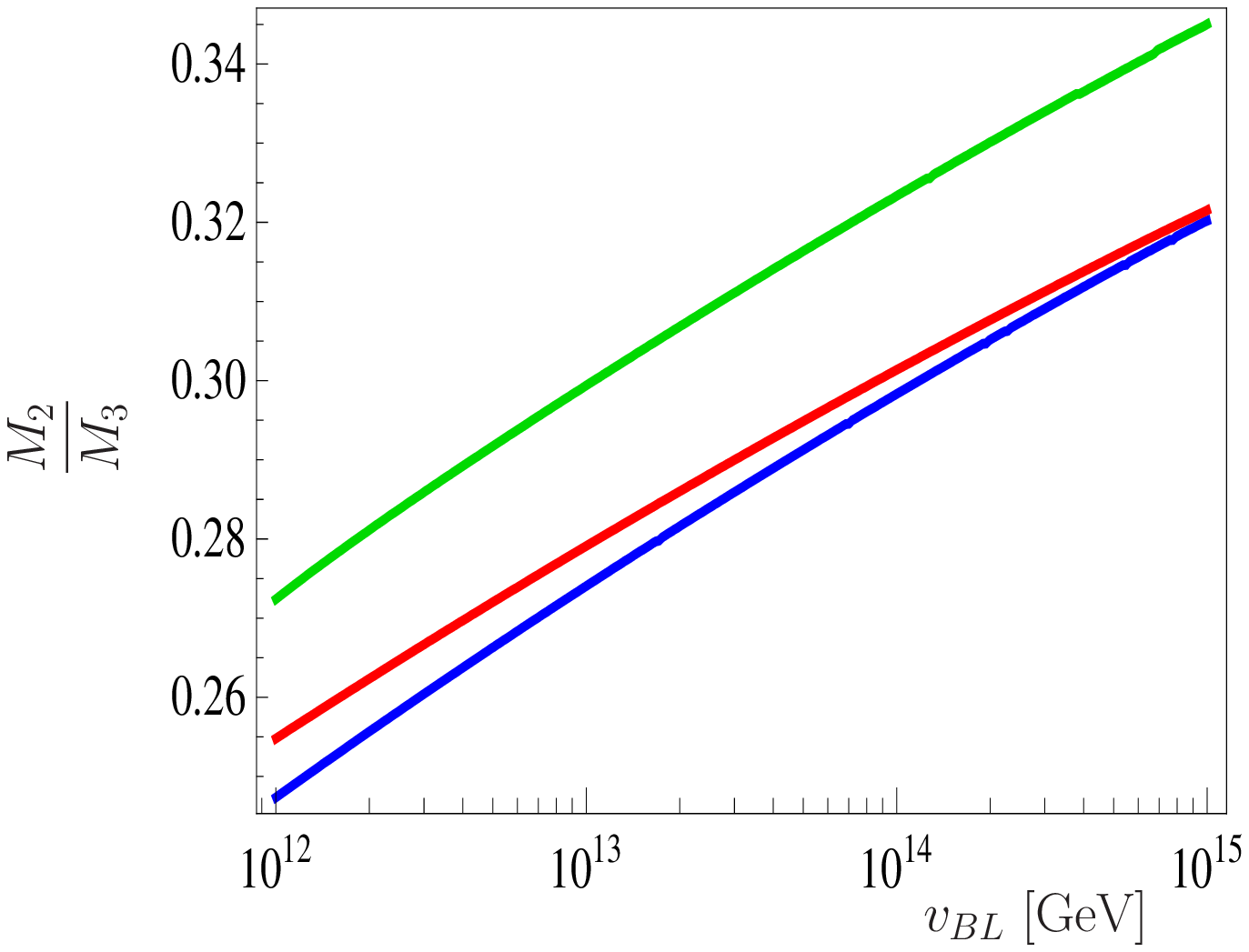}
\end{center}
\vspace{-5mm}
\caption{Gaugino mass ratios as a function of $v_{BL}$ for the fixed
value $v_R = 10^{15}$ GeV. To the left, $M_1 / M_2$, whereas to the
right $M_2 / M_3$. In both figures the three colored lines correspond
to three mSUGRA benchmark points: SPS1a' (blue), SPS3 (green) and SPS5
(red). Note the small variation in the numbers on the Y axis.}
\label{fig:ratiogauginosvBL}
\end{figure}

\begin{figure}
\begin{center}
\vspace{5mm}
\includegraphics[width=0.47\textwidth]{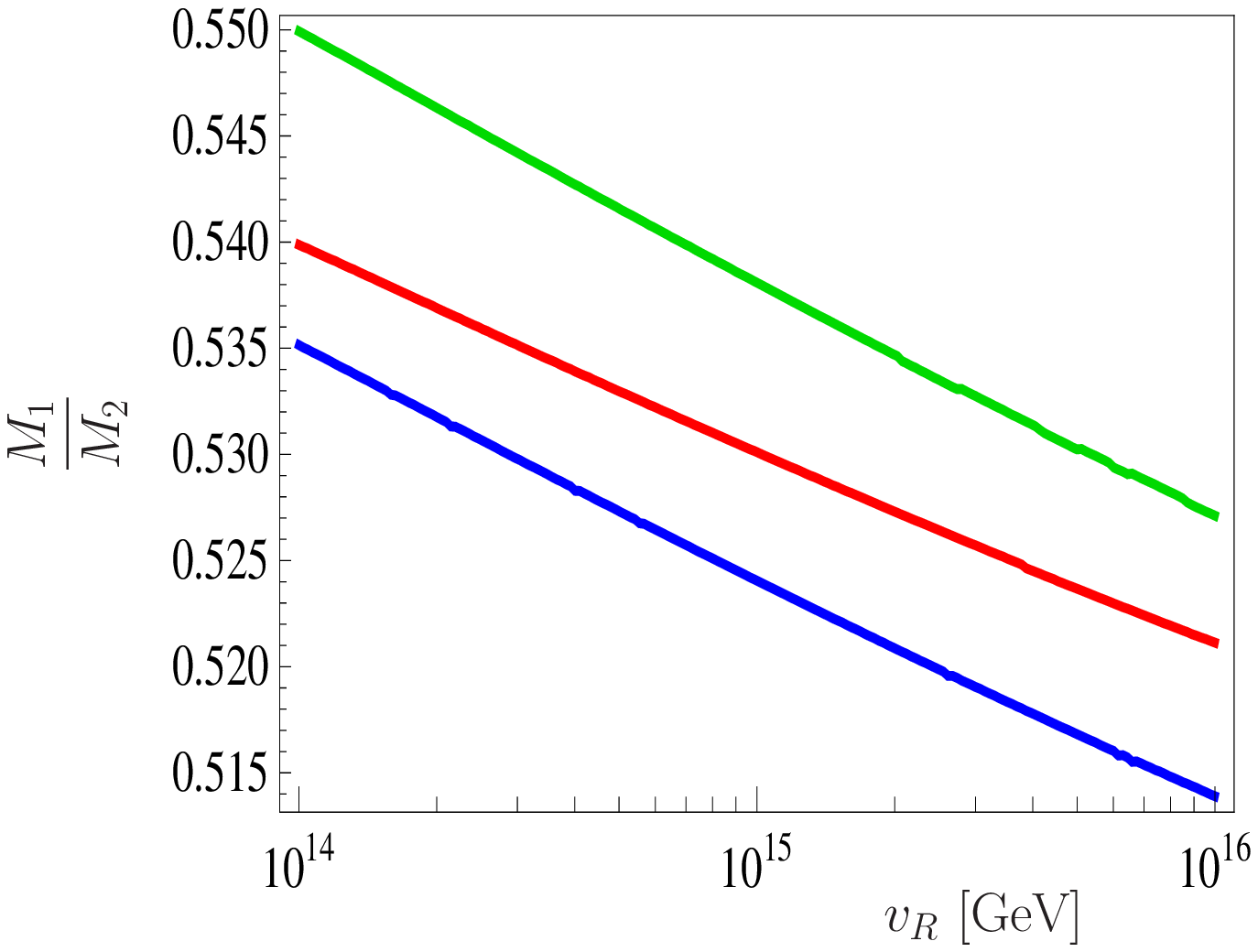}
\hspace{5mm}
\includegraphics[width=0.47\textwidth]{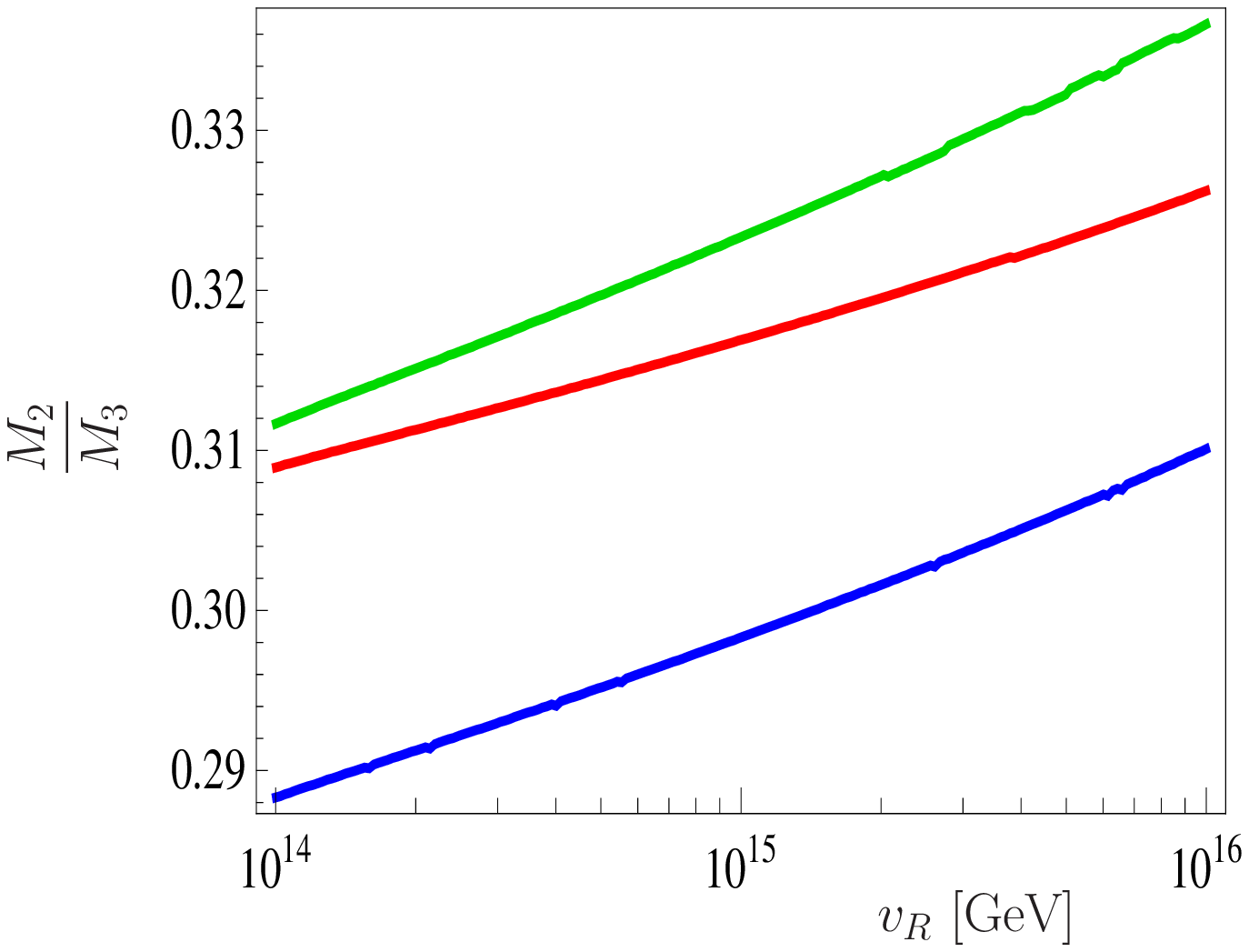}
\end{center}
\vspace{-5mm}
\caption{Gaugino mass ratios as a function of $v_R$ for the fixed
value $v_{BL} = 10^{14}$ GeV. To the left, $M_1 / M_2$, whereas to the
right $M_2 / M_3$. In both figures the three colored lines correspond
to three mSUGRA benchmark points: SPS1a' (blue), SPS3 (green) and SPS5
(red). Note the small variation in the numbers on the Y axis.}
\label{fig:ratiogauginosvR}
\end{figure}

The change in the low energy spectrum, however, maintains to a good 
degree the standard mSUGRA expectation for the ratios of gaugino 
masses, as shown in figures \ref{fig:ratiogauginosvBL} and
\ref{fig:ratiogauginosvR}. Here, figure \ref{fig:ratiogauginosvBL} 
shows the ratios $M_1/M_2$ and $M_2/M_3$ versus $v_{BL}$, while 
figure \ref{fig:ratiogauginosvR} shows the same ratios versus $v_R$. 
Shown are the results for three different SUSY points, which in 
the limit of $v_R,v_{BL} \to m_{GUT}$ approach the standard SPS 
points SPS1a' \cite{AguilarSaavedra:2005pw}, SPS3 and SPS5 
\cite{Allanach:2002nj}. For example, the ratio $M_1/M_2$ is 
expected to be $(5/3)\tan^2\theta_W \simeq 0.5$ at 1-loop order 
in mSUGRA. The exact ratio, however, depends on higher order corrections,
and thus on the SUSY spectrum. The LR model will thus appear rather mSUGRA 
like, if these ratios are measured. Only with very high precision 
on mass measurements, possible only at a linear collider, can one 
hope to find any (indirect) dependence on $v_{BL}$ and $v_R$.

\subsection{Low energy LFV}

Lepton flavor violation in charged lepton decays has attracted a lot
of attention for decades. Processes like $\mu \to e \gamma$ are highly
suppressed in the standard model (plus non-zero neutrino masses) due 
to the GIM mechanism \cite{Glashow:1970gm}, and thus the observation 
of these rare decays would imply new physics. The MEG experiment \cite{meg} 
is currently the most advanced experimental setup in the search for 
$\mu^+ \to e^+ \gamma$. This rare decay will be observed if its branching 
ratio is above the MEG expected sensitivity, around $Br(\mu \to e \gamma) \sim
10^{-13}$. 

LFV decays like $l_i \to l_j \gamma$ are induced by 1-loop diagrams
with the exchange of neutralinos and sleptons. They can be described
by the effective Lagrangian, see for example the review \cite{Kuno:1999jp},
\begin{equation}
\mathcal{L}_{eff} = e \frac{m_i}{2} \bar{l}_i \sigma_{\mu \nu} F^{\mu \nu} 
(A_L^{ij} P_L + A_R^{ij} P_R) l_j + h.c. \thickspace.
\end{equation}

Here $P_{L,R} = \frac{1}{2}(1 \mp \gamma_5)$ are the usual chirality
projectors and therefore the couplings $A_L$ and $A_R$ are generated
by loops with left and right sleptons, respectively. In our 
numerical calculation we use exact expressions for $A_L$ and $A_R$. 
However, for an easier understanding of the numerical results, 
we note that the relation between these couplings and the slepton soft 
masses is very approximately given by 
\begin{equation} \label{A-dependence}
A_L^{ij} \sim \frac{(m_L^2)_{ij}}{m_{SUSY}^4} \quad , \quad A_R^{ij} 
\sim \frac{(m_{e^c}^2)_{ij}}{m_{SUSY}^4} \thickspace, 
\end{equation}
where $m_{SUSY}$ is a typical supersymmetric mass. Here it has been 
assumed that (a) chargino/neutralino masses are similar to slepton 
masses and (b) A-terms mixing left-right transitions are negligible. 
Therefore, due to the negligible off-diagonal entries in 
$m_{e^c}^2$, a pure seesaw model predicts $A_R \simeq 0$.

The branching ratio for $l_i \to l_j \gamma$ can be calculated from
the previous formulas. The result is
\begin{equation} \label{brLLG}
Br(l_i \to l_j \gamma) = \frac{48 \pi^3 \alpha}{G_F^2} 
\left( |A_L^{ij}|^2 + |A_R^{ij}|^2 \right) Br(l_i \to l_j \nu_i \bar{\nu}_j) \thickspace .
\end{equation}

\begin{figure}
\begin{center}
\vspace{5mm}
\includegraphics[width=0.49\textwidth]{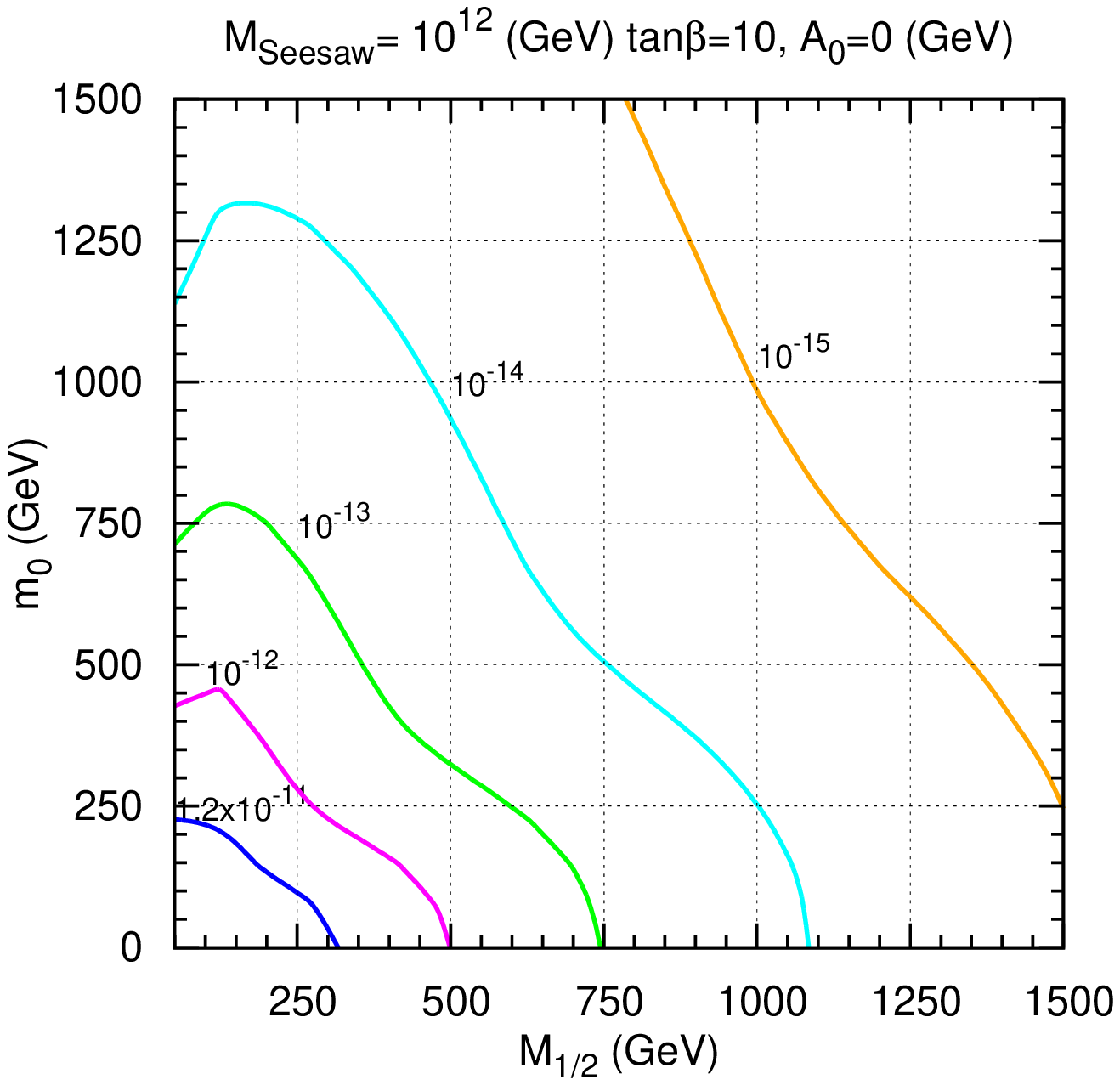}
\includegraphics[width=0.49\textwidth]{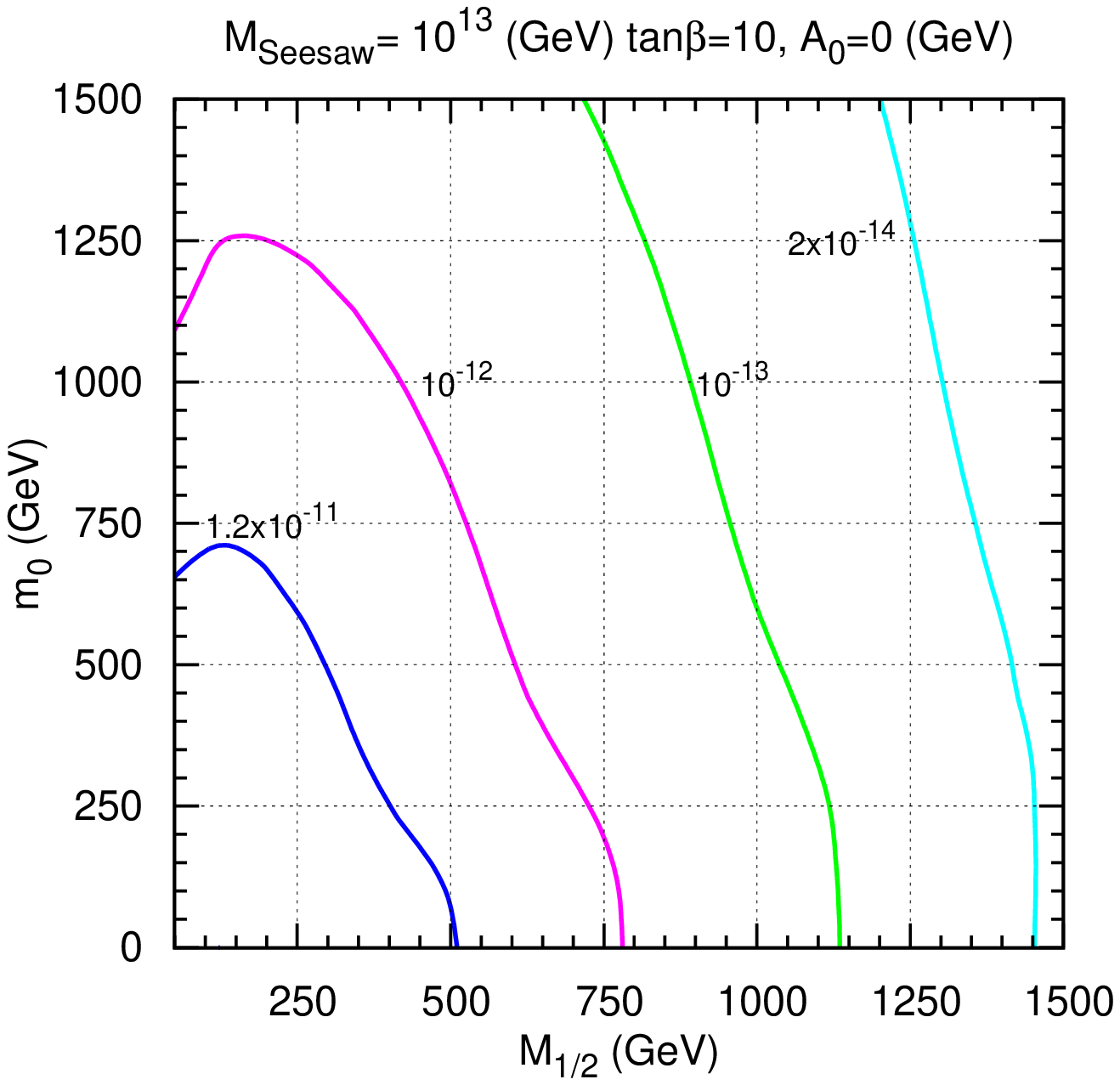}
\end{center}
\vspace{-5mm}
\caption{Contours of $Br(\mu \to e \gamma)$ in the $m_0,M_{1/2}$ plane
for $v_{BL} = 10^{14}$ GeV and $v_R = 10^{15}$ GeV. To the left $M_S =
10^{12}$ GeV, whereas to the right $M_S = 10^{13}$ GeV. Neutrino
oscillation data have been fitted with the $Y_\nu$ fit, assuming
degenerate right-handed neutrinos, $M_{R i} = M_S$.}
\label{fig:contourMuEG}
\end{figure}

Figure \ref{fig:contourMuEG} shows two examples for $Br(\mu \to e
\gamma)$ in the $m_0,M_{1/2}$ plane. Here, we have fixed $v_{BL} = 10^{14}$ 
GeV and $v_R = 10^{15}$ GeV and show to the left $M_S =10^{12}$ GeV, 
whereas to the right $M_S = 10^{13}$ GeV. Once Yukawas are fitted 
to explain the observed neutrino masses, the branching ratio shows 
an approximately quadratic dependence on the seesaw scale, with lower 
$M_S$ giving smaller $Br(\mu \to e\gamma)$. As expected, the branching ratio
also strongly decreases as $m_0$ and/or $M_{1/2}$ increase. This is because
the superparticles in the loops leading to $\mu \to e \gamma$ become
heavier in these directions, suppressing the decay rate. In fact, from
equations \eqref{A-dependence} and \eqref{brLLG} one easily finds the
dependence
\begin{equation}
Br(\mu \to e \gamma) \sim \frac{48 \pi^3 \alpha}{G_F^2} 
\frac{(m_{L,\tilde{e}^c}^2)_{ij}^2}{m_{SUSY}^8} \thickspace,
\end{equation}
which shows that $Br(\mu \to e \gamma)$ decreases as $m_{SUSY}^{-8}$.

It is also remarkable that for a given seesaw scale, $Br(\mu \to e \gamma)$ 
is sizeably larger in the LR model than in a pure seesaw type-I model, 
see for example \cite{Esteves:2009vg}. 
The explanation of this is that right sleptons contribute 
significantly in the LR model to $Br(\mu \to e \gamma)$ and these 
contributions are absent in seesaw models. 

As already discussed, a pure seesaw model predicts simply $A_R \simeq
0$. However, in the LR model we expect a more complicated picture. 
Left-right symmetry implies that, above the parity breaking scale, 
non-negligible flavor violating entries are generated in 
$m_{e^c}^2$. Therefore, $A_R \ne 0$ is obtained at 
low energy. The angular distribution of the outgoing positron at, 
for example, the MEG experiment could be used to discriminate between 
left- and right-handed polarized states \cite{Okada:1999zk,Hisano:2009ae}. 
If MEG is able to measure the positron polarization asymmetry, defined as

\begin{equation}
\mathcal{A}(\mu^+ \to e^+ \gamma) = \frac{|A_L|^2-|A_R|^2}{|A_L|^2+|A_R|^2},
\end{equation}

\noindent there will be an additional observable to distinguish from minimal 
seesaw models. In a pure seesaw model one expects $\mathcal{A} \simeq +1$
to a very
good accuracy. However, the LR model typically leads to significant 
departures from this expectation, giving an interesting signature 
of the high energy restoration of parity.

\begin{figure}
\begin{center}
\vspace{5mm}
\includegraphics[width=0.49\textwidth]{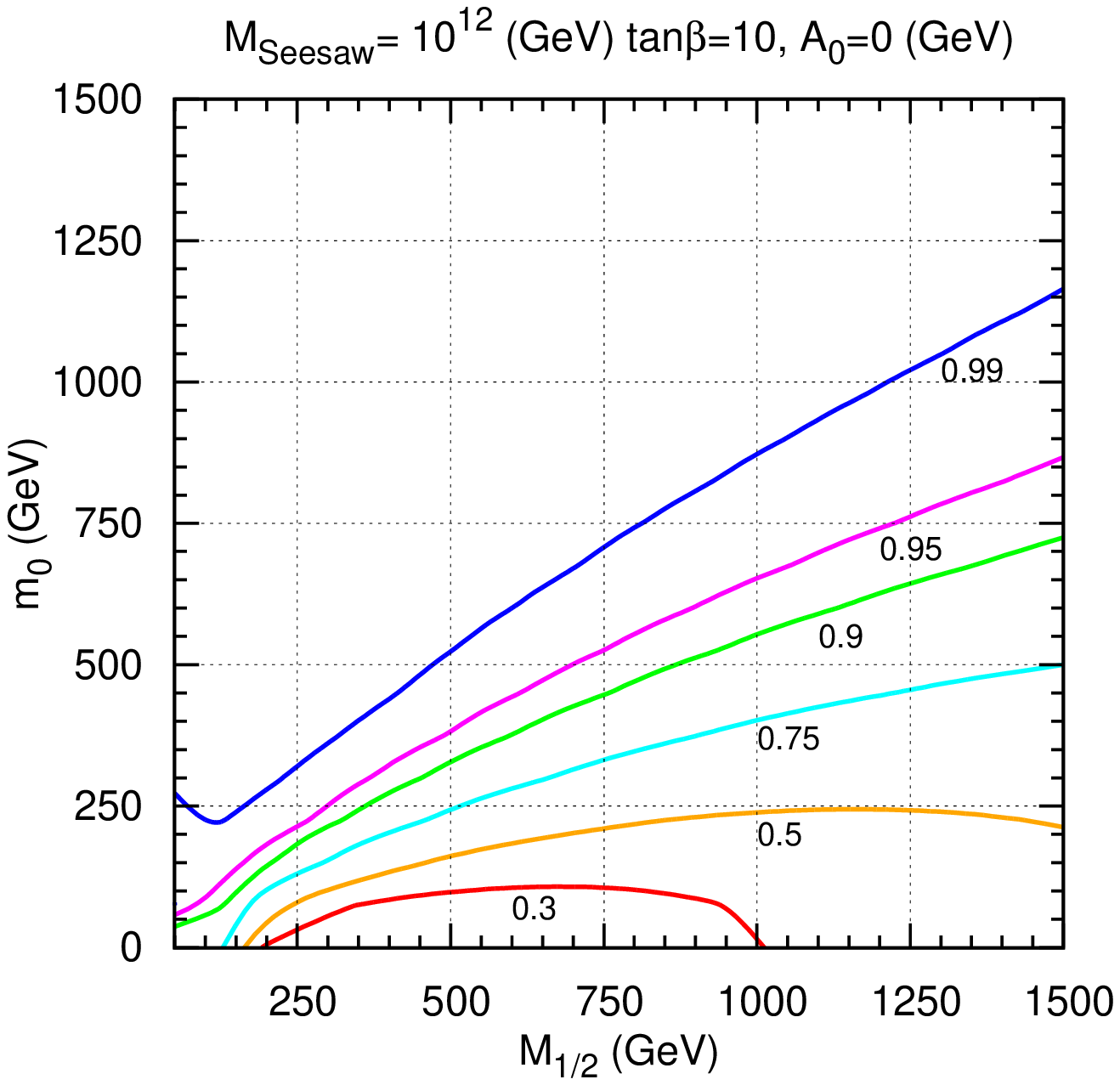}
\includegraphics[width=0.49\textwidth]{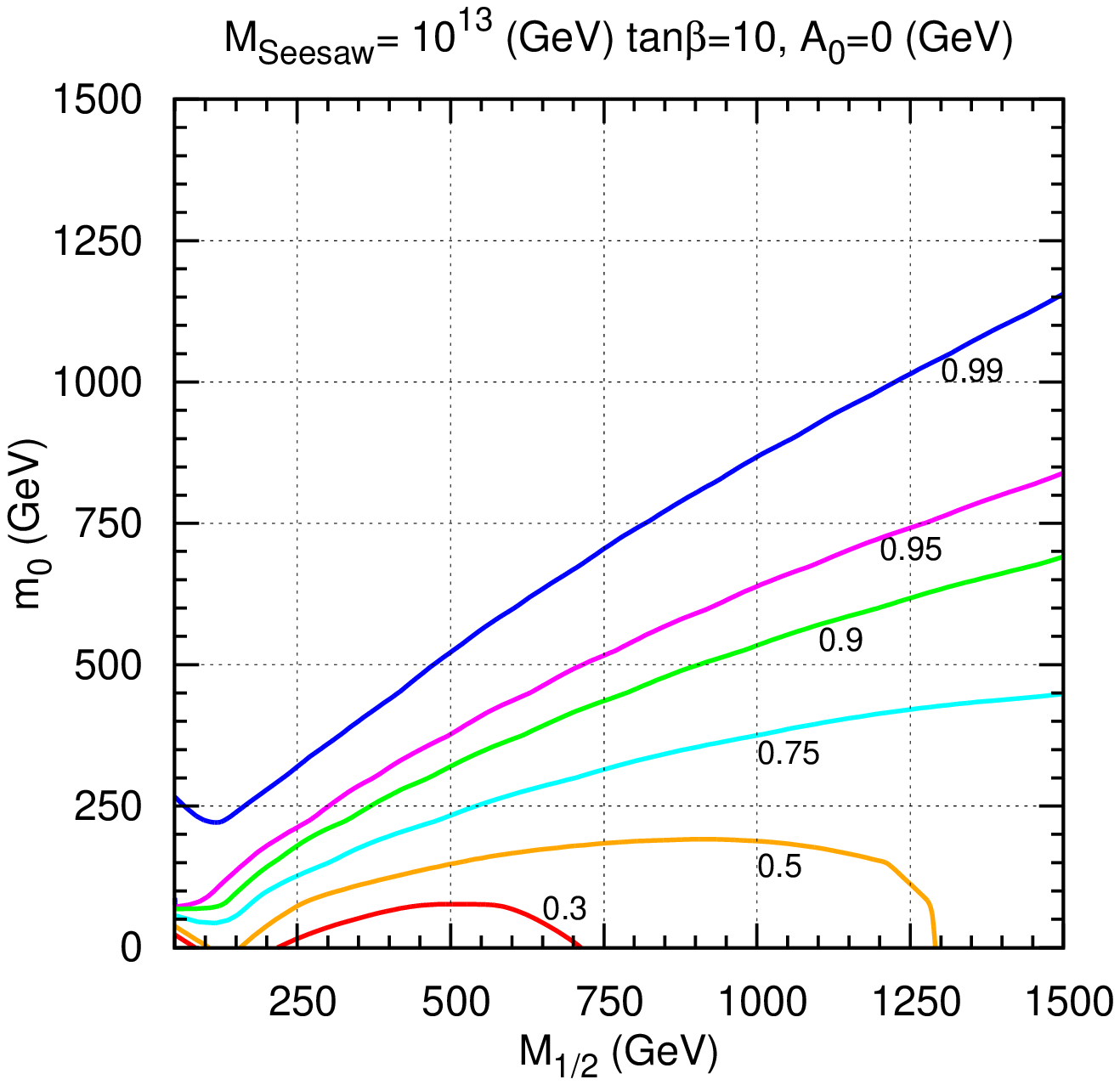}
\end{center}
\vspace{-5mm}
\caption{Contours of $\mathcal{A}(\mu^+ \to e^+ \gamma)$ in the
$m_0,M_{1/2}$ plane. To the left $M_S = 10^{12}$ GeV, whereas to the
right $M_S = 10^{13}$ GeV. The parameters have been chosen as in
figure \ref{fig:contourMuEG}.}
\label{fig:contourALR}
\end{figure}

Figure \ref{fig:contourALR} shows contours for $\mathcal{A}(\mu^+ \to
e^+ \gamma)$ in the $m_0,M_{1/2}$ plane. For the corresponding branching 
ratios see figure \ref{fig:contourMuEG}. Note the rather strong dependence 
on $m_0$. The latter can be understood as follows. Since $v_{BL}$ in 
these examples is one order of magnitude smaller than $v_R$,
and the $Y_\nu$ fit has been used, the LFV 
mixing angles in the left slepton sector are larger than the corresponding 
LFV entries in the right sleptons. At very large values of $m_0$, were the 
masses of right and left sleptons are of comparable magnitude, therefore 
``left'' LFV is more important and the model approaches the pure seesaw 
expectation. At smaller values of $m_0$, right sleptons are lighter 
than left sleptons, and due to the strong dependence of $\mu\to e\gamma$ 
on the sfermion masses entering the loop calculation, see eq. 
\eqref{A-dependence}, $A_R$ and $A_L$ can become comparable, despite 
the smaller LFV entries in right slepton mass matrices. In the limit 
of very small right slepton masses the model then approaches 
$\mathcal{A} \sim 0$. We have not explicitly searched for regions of 
parameter space with $\mathcal{A} < 0$, but one expects that negative 
values for $\mathcal{A}$ are possible if $v_{BL}$ is not much below 
$v_R$ and sleptons are light at the same time, i.e. small values of 
$m_0$ and $M_{1/2}$. Note that, again due to the LR symmetry above to $v_R$,
the model can never approach the limit $\mathcal{A} =-1$ exactly.

\begin{figure}
\begin{center}
\vspace{5mm}
\includegraphics[width=0.49\textwidth]{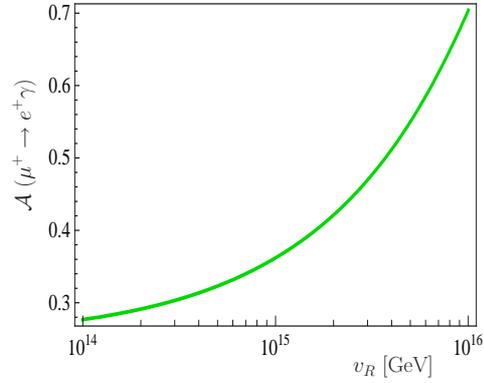}
\end{center}
\vspace{-5mm}
\caption{Positron polarization asymmetry $\mathcal{A}(\mu^+ \to e^+
\gamma)$ as a function of $v_R$ for the parameter choice $M_S =
10^{13}$ GeV and $v_{BL} = 10^{14}$ GeV. The mSUGRA parameters have
been taken as in the SPS3 benchmark point and neutrino oscillation
data have been fitted with the $Y_\nu$ fit, assuming degenerate
right-handed neutrinos, $M_{R i} = M_S$.}
\label{fig:SPS3-megpolvR-Yv}
\end{figure}

The positron polarization asymmetry is very sensitive to the high
energy scales. Figure \ref{fig:SPS3-megpolvR-Yv} shows $\mathcal{A}$
as a function of $v_R$ for $M_S = 10^{13}$ GeV, $v_{BL} = 10^{14}$ GeV
and the mSUGRA parameters as in the SPS3 benchmark point. The plot has
been obtained using the $Y_\nu$ fit. This example shows that as $v_R$
approaches $m_{GUT}$ the positron polarization $\mathcal{A}$
approaches $+1$, which means $A_L$ dominates the calculation. This is 
because, in the $Y_\nu$ fit, the right-handed LFV soft slepton masses, 
and thus the corresponding $A_R$ coupling, only run from $m_{GUT}$ to $v_R$. 

\begin{figure}
\begin{center}
\vspace{5mm}
\includegraphics[width=0.5\textwidth]{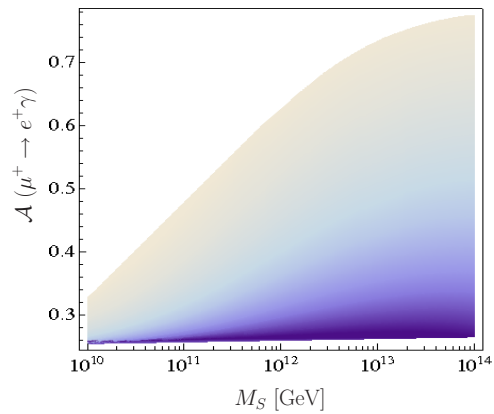}
\end{center}
\vspace{-5mm}
\caption{Positron polarization asymmetry $\mathcal{A}(\mu^+ \to e^+
\gamma)$ as a function of the seesaw scale, defined as the mass of the
lightest right-handed neutrino, for the parameter choice $v_{BL} =
10^{15}$ GeV and $v_R \in [10^{15},10^{16}]$ GeV. Lighter colors mean
higher values of $v_R$. The mSUGRA parameters have been taken as in
the SPS3 benchmark point and neutrino oscillation data have been
fitted with the $Y_\nu$ fit, assuming degenerate right-handed
neutrinos, $M_{R i} = M_S$.}
\label{fig:megpol-MR}
\end{figure}

$\mathcal{A}(\mu^+ \to e^+ \gamma)$ also has an important dependence
on the seesaw scale. This is shown in figure \ref{fig:megpol-MR},
where $\mathcal{A}$ is plotted as a function of the lightest
right-handed neutrino mass. This dependence can be easily understood from
the seesaw formula for neutrino masses. It implies that larger $M_S$ 
requires larger Yukawa parameters in order to fit neutrino masses which, 
in turn, leads to larger flavor violating soft terms due to RGE
running. However, note that, for very small seesaw scales all lepton flavor 
violating effects are negligible and no asymmetry is produced, since $A_L \sim
A_R \sim 0$.

In addition, figure \ref{fig:megpol-MR} shows again the relevance of
$v_R$, which determines the parity breaking scale at which the LFV
entries in the right-handed slepton sector essentially stop running. 
Lighter colors indicate larger $v_R$. As shown already in figure
\ref{fig:SPS3-megpolvR-Yv} for a particular point, the positron
polarization approaches $+1$ as $v_R$ is increased. 

\begin{figure}
\begin{center}
\vspace{5mm}
\includegraphics[width=0.5\textwidth]{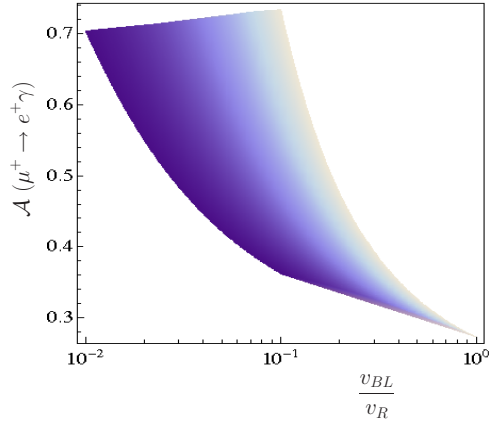}
\end{center}
\vspace{-5mm}
\caption{Positron polarization asymmetry $\mathcal{A}(\mu^+ \to e^+
\gamma)$ as a function of the ratio $v_{BL} / v_R$. The seesaw scale
$M_S$ has been fixed to $10^{13}$ GeV, whereas $v_{BL}$ and $v_R$ take
values in the ranges $v_{BL} \in [10^{14},10^{15}]$ GeV and $v_R \in
[10^{15},10^{16}]$ GeV. Lighter colors indicate larger $v_{BL}$. The
mSUGRA parameters have been taken as in the SPS3 benchmark point and
neutrino oscillation data have been fitted with the $Y_\nu$ fit,
assuming degenerate right-handed neutrinos, $M_{R i} = M_S$.}
\label{fig:megpol-ratio}
\end{figure}

Below the $SU(2)_R$ breaking scale parity is broken and left and
right slepton soft masses evolve differently. The approximate
solutions to the RGEs in equations \eqref{apprge2} and \eqref{apprge4}
show that, if neutrino data is fitted according to the $Y_\nu$ fit,
the left-handed ones keep running from the $SU(2)_R$ breaking scale to
the $U(1)_{B-L}$ scale. In this case one expects larger flavor
violating effects in the left-handed slepton sector and a correlation
with the ratio $v_{BL}/v_R$, which measures the difference between the
breaking scales. This correlation, only present in the $Y_\nu$ fit, 
is shown in figure \ref{fig:megpol-ratio}. On the one hand, one finds 
that as $v_{BL}$ and $v_R$ become very different, $v_{BL}/v_R \ll 1$, 
the positron asymmetry approaches $\mathcal{A}=+1$. On the other hand, 
when the two breaking scales are close, $v_{BL}/v_R \sim 1$, this effect 
disappears and the positron polarization asymmetry approaches
$\mathcal{A}=0$. Note that the $Y_\nu$ fit does not usually produce a 
negative value for $\mathcal{A}$ since the LFV terms in the right
slepton sector never run more than the corresponding terms in the
left-handed sector. The only possible execption to this general rule 
is, as discussed above, in the limit of very small $m_0$ and 
$v_{BL}/v_R \sim 1$. 

\begin{figure}
\begin{center}
\vspace{5mm}
\includegraphics[width=0.5\textwidth]{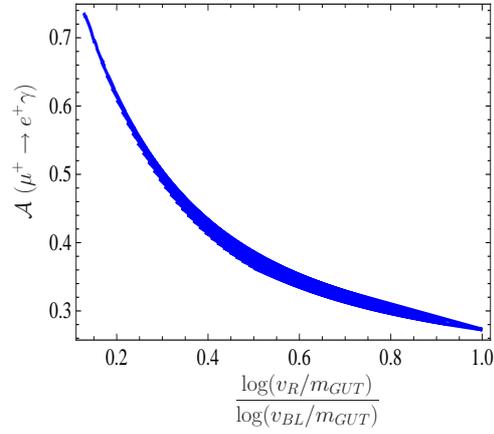}
\end{center}
\vspace{-5mm}
\caption{Positron polarization asymmetry $\mathcal{A}(\mu^+ \to e^+
\gamma)$ as a function of $\log (v_R/m_{GUT}) / \log
(v_{BL}/m_{GUT})$. The parameters have been chosen as in figure
\ref{fig:megpol-ratio}.}
\label{fig:megpol-ratio2}
\end{figure}

The determination of the ratio $v_{BL}/v_R$ from figure
\ref{fig:megpol-ratio} is shown to be very inaccurate. This is due to
the fact that other parameters, most importantly $m_{GUT}$ (which itself 
has an important dependence on the values of $v_{BL}$ and $v_R$), have a
strong impact on the results. Therefore, although it would be possible
to constrain the high energy structure of the theory, a precise
determination of the ratio $v_{BL}/v_R$ will require additional
input. Figure \ref{fig:megpol-ratio2}, on the other hand, shows that
the polarization asymmetry $\mathcal{A}(\mu^+ \to e^+ \gamma)$ is much
better correlated with the quantity $\log (v_R/m_{GUT}) / \log
(v_{BL}/m_{GUT})$. This is as expected from equations \eqref{apprge2} and
\eqref{apprge4} and confirms the validity of this approximation.

\begin{figure}
\begin{center}
\vspace{5mm}
\includegraphics[width=0.49\textwidth]{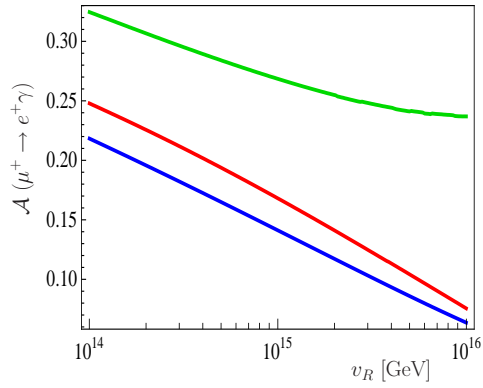}
\end{center}
\vspace{-5mm}
\caption{Positron polarization asymmetry $\mathcal{A}(\mu^+ \to e^+
\gamma)$ as a function of $v_R$ for three different mSUGRA benchmark
points: SPS1a' (blue line), SPS3 (green line) and SPS5 (red
line). In this figure a fixed value $v_{BL} = 10^{14}$ GeV is taken.
Neutrino oscillation data have been fitted with the $f$ fit.}
\label{fig:megpolvR-f}
\end{figure}

We close our discussion on the positron polarization asymmetry with
some comments on the $f$ fit. Since this type of fit leads to $\Delta
m_L^2 \sim \Delta m_{e^c}^2 \sim 0$ in the $v_{BL} - v_R$
energy region, there is little dependence on these symmetry breaking
scales. This is illustrated in figure \ref{fig:megpolvR-f}, where the
asymmetry $\mathcal{A}$ is plotted as a function of $v_R$ for three
different mSUGRA benchmark points: SPS1a' (blue line), SPS3 (green
line) and SPS5 (red line). One clearly sees that the dependence on $v_R$
is quite weak compared to the $Y_\nu$ fit. In fact, the variations in 
this figure are mostly due to the changes in the low energy supersymmetric 
spectrum due to different $v_R$ values. In the case of the $f$-fit one 
then typically finds $\mathcal{A} \in [0.0-0.3]$.

\subsection{LFV at LHC/ILC}

Lepton flavor violation might show up at collider experiments as
well. Although the following discussion is focused on the LHC
discovery potential for LFV signatures, let us emphasize that a future
linear collider will be able to determine the relevant
observables with much higher precision.

\begin{figure}
\begin{center}
\vspace{5mm}
\includegraphics[width=0.47\textwidth]{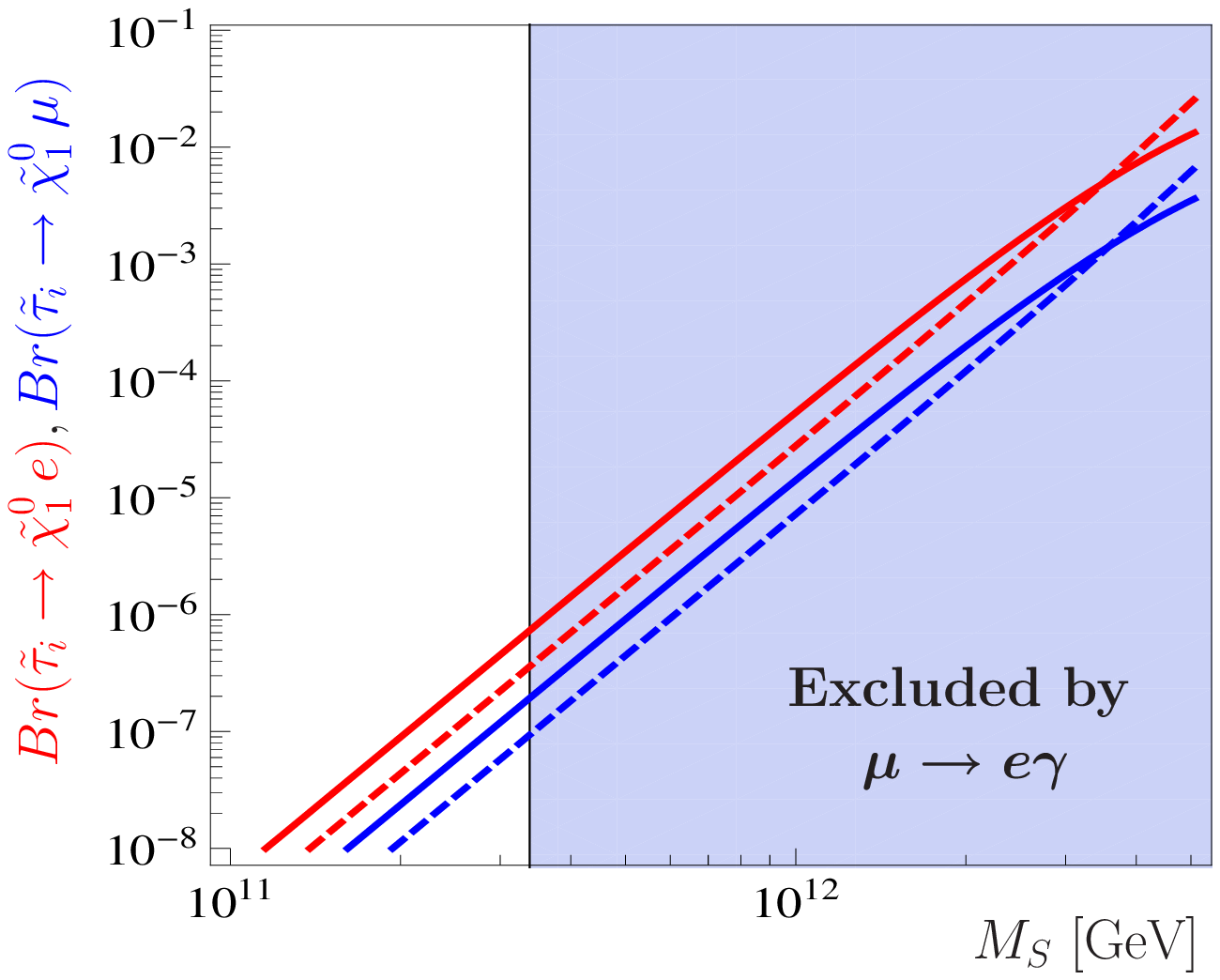}
\hspace{5mm}
\includegraphics[width=0.47\textwidth]{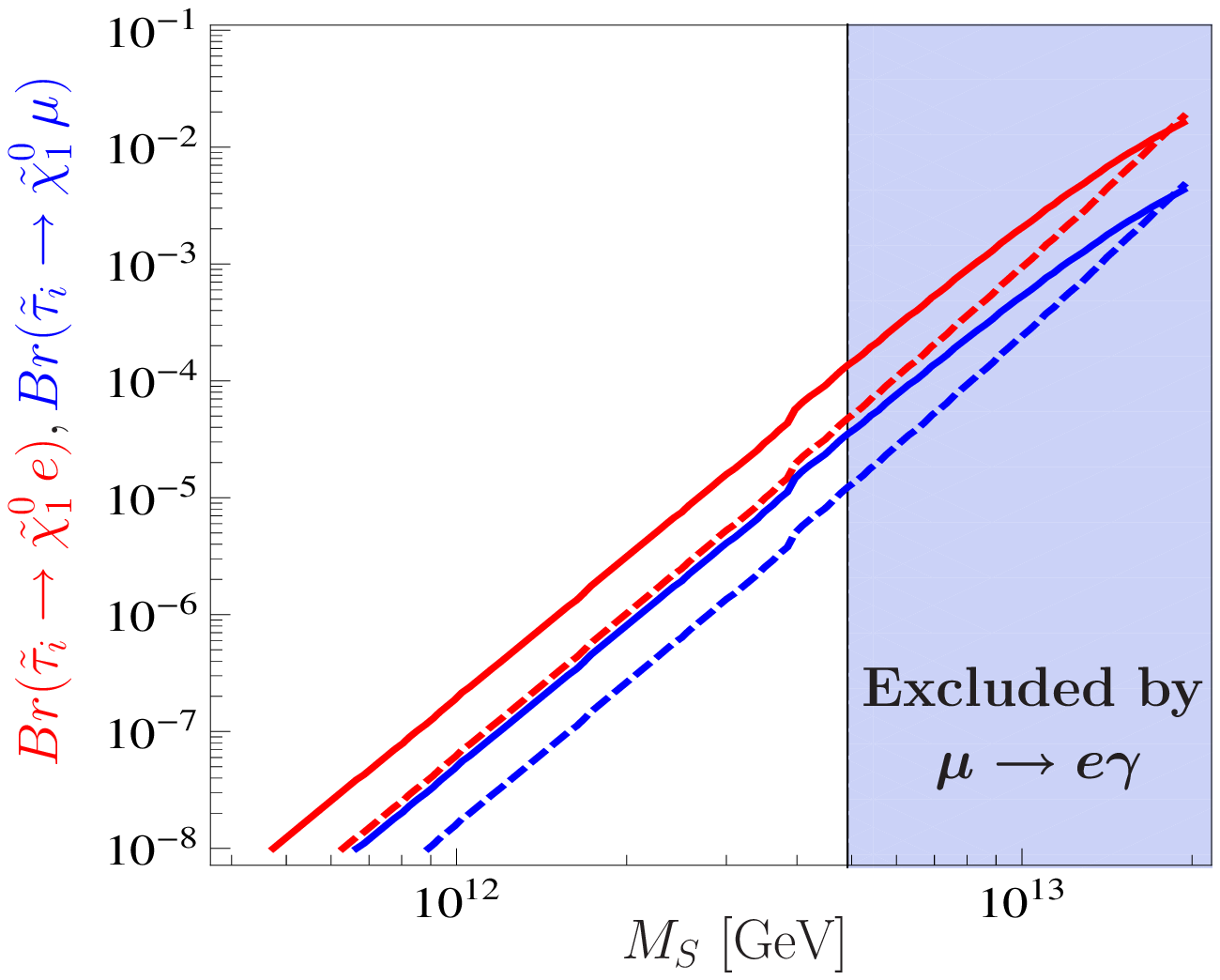}
\end{center}
\vspace{-5mm}
\caption{$Br(\tilde{\tau}_i \to \tilde{\chi}_1^0 \: e)$ and
$Br(\tilde{\tau}_i \to \tilde{\chi}_1^0 \: \mu)$ as a function of the
seesaw scale, defined as the mass of the lightest right-handed
neutrino, for the parameter choice $v_{BL} = 10^{15}$ GeV and $v_R = 5
\cdot 10^{15}$ GeV. The dashed lines correspond to $\tau_1 \simeq
\tau_R$ and the solid ones to $\tau_2 \simeq \tau_L$. To the left, the
mSUGRA parameters have been taken as in the SPS1a' benchmark point,
whereas to the right as in the SPS3 benchmark point. In both figures
neutrino oscillation data have been fitted according to the $f$ fit,
with non-degenerate right-handed neutrinos. The blue shaded regions
are excluded by $\mu \to e \gamma$.}
\label{fig:stauFV}
\end{figure}

Figure \ref{fig:stauFV} shows $Br(\tilde{\tau}_i \to \tilde{\chi}_1^0
\: e)$ and $Br(\tilde{\tau}_i \to \tilde{\chi}_1^0 \: \mu)$ as a
function of the seesaw scale. The dashed lines correspond to $\tau_1 \simeq
\tau_R$ and the solid ones to $\tau_2 \simeq \tau_L$.
As in the case of $\mu\to e \gamma$, 
see figure \ref{fig:contourMuEG}, lower seesaw scales imply less 
flavor violating effects due to smaller Yukawa couplings.
Moreover, figure \ref{fig:stauFV} presents the same results for two
different benchmark points, SPS1a' and SPS3. As already shown in
figure \ref{fig:contourMuEG}, $\mu \to e \gamma$ is strongly dependent
on the SUSY spectrum. For lighter supersymmetric particles, as in the
benchmark point SPS1a', $\mu\to e\gamma$ is large, setting strong limits 
on the seesaw scale and thus on the possibility to observe LFV at 
colliders. In the case of heavier spectrums, as in SPS3, $\mu
\to e \gamma$ is still the most stringent constraint, but larger 
values of the seesaw scale and thus LFV violating branching ratios 
become allowed. Whether decays such as $Br(\tilde{\tau}_i \to \tilde{\chi}_1^0
\: e)$ and $Br(\tilde{\tau}_i \to \tilde{\chi}_1^0 \: \mu)$ are 
observable at the LHC or not, thus depends very sensitively on the 
unknown $m_0$, $M_{1/2}$ and $M_S$. 

Furthermore, the right panel of figure \ref{fig:stauFV} also shows
that right staus can also have LFV decays with sizeable rates.
Of course, as 
emphasized already above, this is the main novelty of the LR model 
compared to pure seesaw models. This is
direct consequence of parity restoration at high energies.

\begin{figure}
\begin{center}
\vspace{5mm}
\includegraphics[width=0.5\textwidth]{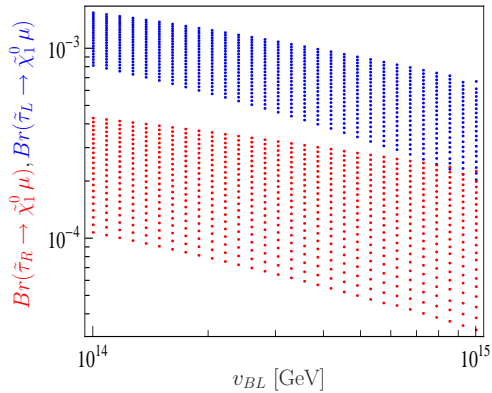}
\end{center}
\vspace{-5mm}
\caption{$Br(\tilde{\tau}_L \to \tilde{\chi}_1^0 \: \mu)$ and
$Br(\tilde{\tau}_R \to \tilde{\chi}_1^0 \: \mu)$ as a function of
$v_{BL}$ for $M_S = 10^{13}$ GeV and $v_R \in [10^{15},5 \cdot
10^{15}]$ GeV. Red dots correspond to $\tau_1 \simeq \tau_R$, whereas
the blue ones correspond to $\tau_2 \simeq \tau_L$. The mSUGRA
parameters have been taken as in the SPS3 benchmark point and neutrino
oscillation data have been fitted with the $Y_\nu$ fit, assuming
degenerate right-handed neutrinos, $M_{R i} = M_S$.}
\label{fig:difLR}
\end{figure}

Moreover, as in our analysis of the positron polarization asymmetry,
one expects to find that if the difference between $v_R$ and $v_{BL}$
is increased, the difference between the LFV entries in the L and R
sectors gets increased as well. This property of the $Y_\nu$ fit is
shown in figure \ref{fig:difLR}, which shows branching ratios for the
LFV decays of the staus as a function of $v_{BL}$ for $v_R \in
[10^{15},5 \cdot 10^{15}]$ GeV. As the figure shows, the theoretical 
expectation is confirmed numerically: the difference between 
$Br(\tilde{\tau}_L)$ and $Br(\tilde{\tau}_R)$ strongly depends on 
the difference between $v_R$ and $v_{BL}$.

\begin{figure}
\begin{center}
\vspace{5mm}
\includegraphics[width=0.5\textwidth]{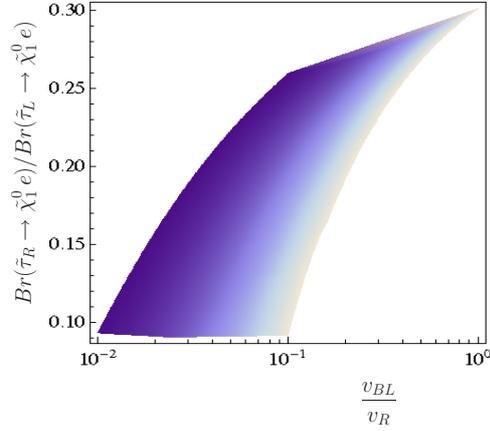}
\end{center}
\vspace{-5mm}
\caption{$Br(\tilde{\tau}_R \to \tilde{\chi}_1^0 \: \mu) /
Br(\tilde{\tau}_L \to \tilde{\chi}_1^0 \: \mu)$ as a function of
$v_{BL} / v_R$. The seesaw scale $M_S$ has been fixed to $10^{13}$
GeV, whereas $v_{BL}$ and $v_R$ take values in the ranges $v_{BL} \in
[10^{14},10^{15}]$ GeV and $v_R \in [10^{15},10^{16}]$ GeV. Lighter
colors indicate larger $v_{BL}$. The rest of the parameters have been
chosen as in figure \ref{fig:difLR}.}
\label{fig:compLR}
\end{figure}

The question arises whether one can determine the ratio $v_{BL}/v_R$
by measuring both $Br(\tilde{\tau}_L)$ and $Br(\tilde{\tau}_R)$ at
colliders. Figure \ref{fig:compLR} attempts to answer this. Here the ratio
$Br(\tilde{\tau}_R \to \tilde{\chi}_1^0 \: e) / Br(\tilde{\tau}_L \to
\tilde{\chi}_1^0 \: e)$ is plotted as a function of $v_{BL} / v_R$. A
measurement of both branching ratios would allow to constrain the
ratio $v_{BL} / v_R$ and increase our knowledge on the high energy
regimes. For the sake of brevity we do not present here the
analogous plots for other LFV slepton decays and/or other lepton final 
states, since they show very similar correlations with
$v_{BL}/v_R$. For example, in principle, one could also use the ratio
$Br(\tilde{\mu}_R \to \tilde{\chi}_1^0 \: \tau) / Br(\tilde{\mu}_L \to
\tilde{\chi}_1^0 \: \tau)$ to determine the ratio between the two high
scales.

\begin{figure}
\begin{center}
\vspace{5mm}
\includegraphics[width=0.5\textwidth]{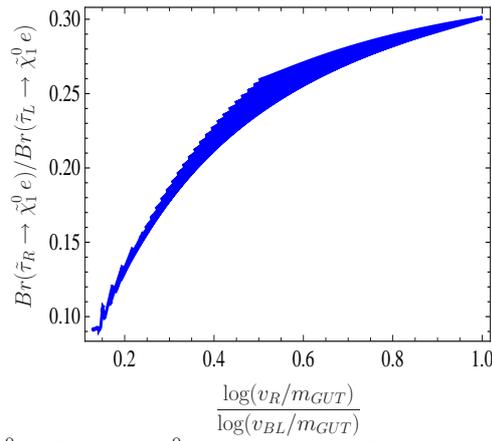}
\end{center}
\vspace{-5mm}
\caption{$Br(\tilde{\tau}_R \to \tilde{\chi}_1^0 \: e) /
Br(\tilde{\tau}_L \to \tilde{\chi}_1^0 \: e)$ as a function of $\log
(v_R/m_{GUT}) / \log (v_{BL}/m_{GUT})$. The parameters have been
chosen as in figure \ref{fig:compLR}.}
\label{fig:compLR2}
\end{figure}

However, as observed also in the polarization asymmetry for $\mu\to e \gamma$ 
there is an important dependence on other parameters of the model, 
especially the exact value of $m_{GUT}$. This implies a theoretical 
uncertainty in the determination of $v_{BL} / v_R$. Again, as for 
$\mathcal{A}$,  a much better correlation with 
$\log (v_R/m_{GUT}) / \log (v_{BL}/m_{GUT})$ is found, 
see figure \ref{fig:compLR2}.

In conclusion, to the determine $v_{BL}$ and $v_R$ individually more 
{\em theoretical} input is needed, such as the GUT scale thresholds, 
which are needed to fix the value of $m_{GUT}$. Recall, that we did 
not specify the exact values of these thresholds in our numerical 
calculation. This leads to a ``floating'' value of $m_{GUT}$ when 
$v_R$ and $v_{BL}$ are varied. Also more experimental data is needed 
to make more definite predictions. Especially SUSY mass spectrum 
measurements, which may or may not be very precise at the LHC, 
depending on the SUSY point realized in nature, will be of great 
importance. Recall that, if in reach of a linear collider, 
slepton mass and branching ratio measurements can be highly precise. 

So far only slepton decays have been discussed. This served to illustrate 
the most interesting signatures of the model, namely, lepton
flavor violation in the right slepton sector. However, LHC 
searches for lepton flavor violation usually concentrate on the decay chain
\cite{Hinchliffe:2000np,Carvalho:2002jg,Carquin:2008gv}
\begin{displaymath}
\tilde{\chi}_2^0 \to \tilde{l}^\pm l^\mp \to \tilde{\chi}_1^0 l^\pm l^\mp \thickspace.
\end{displaymath}
This well known signature has been widely studied due to the accurate
information it can provide about the particle spectrum
\cite{Paige:1996nx,Hinchliffe:1996iu,Bachacou:1999zb,Ball:2007zza,Aad:2009wy}. 
Note that in this decay one assumes usually that the $\tilde{\chi}_2^0$ 
themselves stem from the decay chain ${\tilde q}_L \to q\tilde{\chi}_2^0$. 
If the mass ordering $m_{\tilde{\chi}_2^0} > m_{\tilde{l}} >
m_{\tilde{\chi}_1^0}$ is realized, the dilepton invariant mass
\cite{Bachacou:1999zb,Allanach:2000kt}, defined as $m^2 (l^+ l^-) =
(p_{l^+} + p_{l^-})^2$, has an edge structure with a prominent
kinematical endpoint at
\begin{equation} \label{def-edge}
\left[ m^2 (l^+ l^-) \right]_{max} \equiv m_{ll}^2 = 
\frac{(m_{\tilde{\chi}_2^0}^2-m_{\tilde{l}}^2)
(m_{\tilde{l}}^2-m_{\tilde{\chi}_1^0}^2)}{m_{\tilde{l}}^2}  \thickspace,
\end{equation}
where the masses of the charged leptons have been neglected. The
position of this edge can be measured with impressively high precision 
at the LHC \cite{Paige:1996nx,Hinchliffe:1996iu,Bachacou:1999zb}, 
implying also an accurate determination of the intermediate slepton masses.

In fact, if two different sleptons $\tilde{l}_{1,2}$ have sufficiently
high event rates for $\tilde{\chi}_2^0 \to \tilde{l}_{1,2}^\pm l^\mp_j \to
\tilde{\chi}_1^0 l^\pm_i l^\mp_j$ and their masses allow these chains to 
be on-shell, two different dilepton edge distributions are expected 
\cite{Bartl:2005yy,Allanach:2008ib}. This presents a powerful tool 
to measure slepton mass splittings, which in turn allows to discriminate 
between the standard mSUGRA expectation, with usually negligible mass 
splittings for the first two generations, and extended models with 
additional sources of flavor violation.

The relation between the slepton mass splitting and the variation in
the position of the kinematical is edge is found to be
\cite{Allanach:2008ib}
\begin{equation} \label{edge-split}
\frac{\Delta m_{ll}}{\bar{m}_{ll}} = \frac{\Delta
m_{\tilde{l}}}{\bar{m}_{\tilde{l}}} \frac{m_{\tilde{\chi}_1^0}^2
m_{\tilde{\chi}_2^0}^2 -
\bar{m}_{\tilde{l}}^4}{(\bar{m}_{\tilde{l}}^2-m_{\tilde{\chi}_1^0}^2)
(\bar{m}_{\tilde{l}}^2-m_{\tilde{\chi}_2^0}^2)} \thickspace .
\end{equation}
Here $\Delta m_{ll} (i,j) = m_{l_i l_i} - m_{l_j l_j}$ is the
difference between two edge positions, $\Delta m_{\tilde{l}} =
m_{\tilde{l}_i} - m_{\tilde{l}_j}$ the difference between slepton
masses and $\bar{m}_{ll}$ and $\bar{m}_{\tilde{l}}$ average values of
the corresponding quantities. Note that higher order contributions of
$\frac{\Delta m_{\tilde{l}}}{\bar{m}_{\tilde{l}}}$ have been neglected
in equation \eqref{edge-split}.

A number of studies about the dilepton mass distribution have been performed
\cite{Paige:1996nx,Hinchliffe:1996iu,Bachacou:1999zb}, concluding that
the position of the edges can be measured at the LHC with an accuracy
up to $10^{-3}$. Moreover, as shown in reference
\cite{Allanach:2008ib}, this can be generally translated into a
similar precision for the relative $\tilde{e}-\tilde{\mu}$ mass
splitting, with some regions of parameter space where values as small 
as $10^{-4}$ might be measurable. Since this mass splitting is usually 
negligible in a pure mSUGRA scenario, it is regarded as an interesting 
signature of either lepton flavor violation or non-universality in 
the soft terms. Furthermore, in the context of this paper, it is important 
to emphasize that pure seesaw models can have this signature only in 
the left slepton sector \cite{Abada:2010kj}.

\begin{figure}
\begin{center}
\vspace{5mm}
\includegraphics[width=0.47\textwidth]{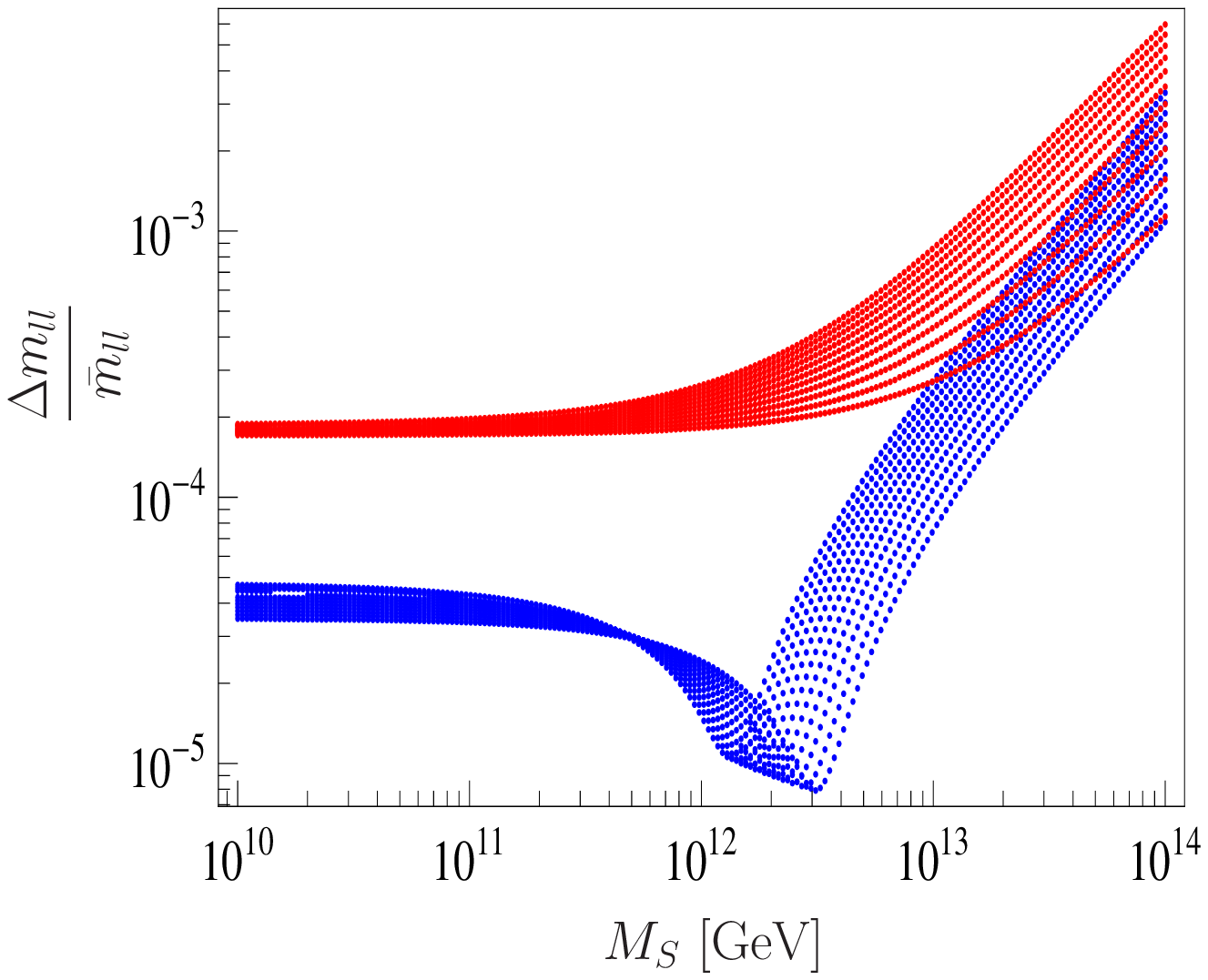}
\hspace{5mm}
\includegraphics[width=0.47\textwidth]{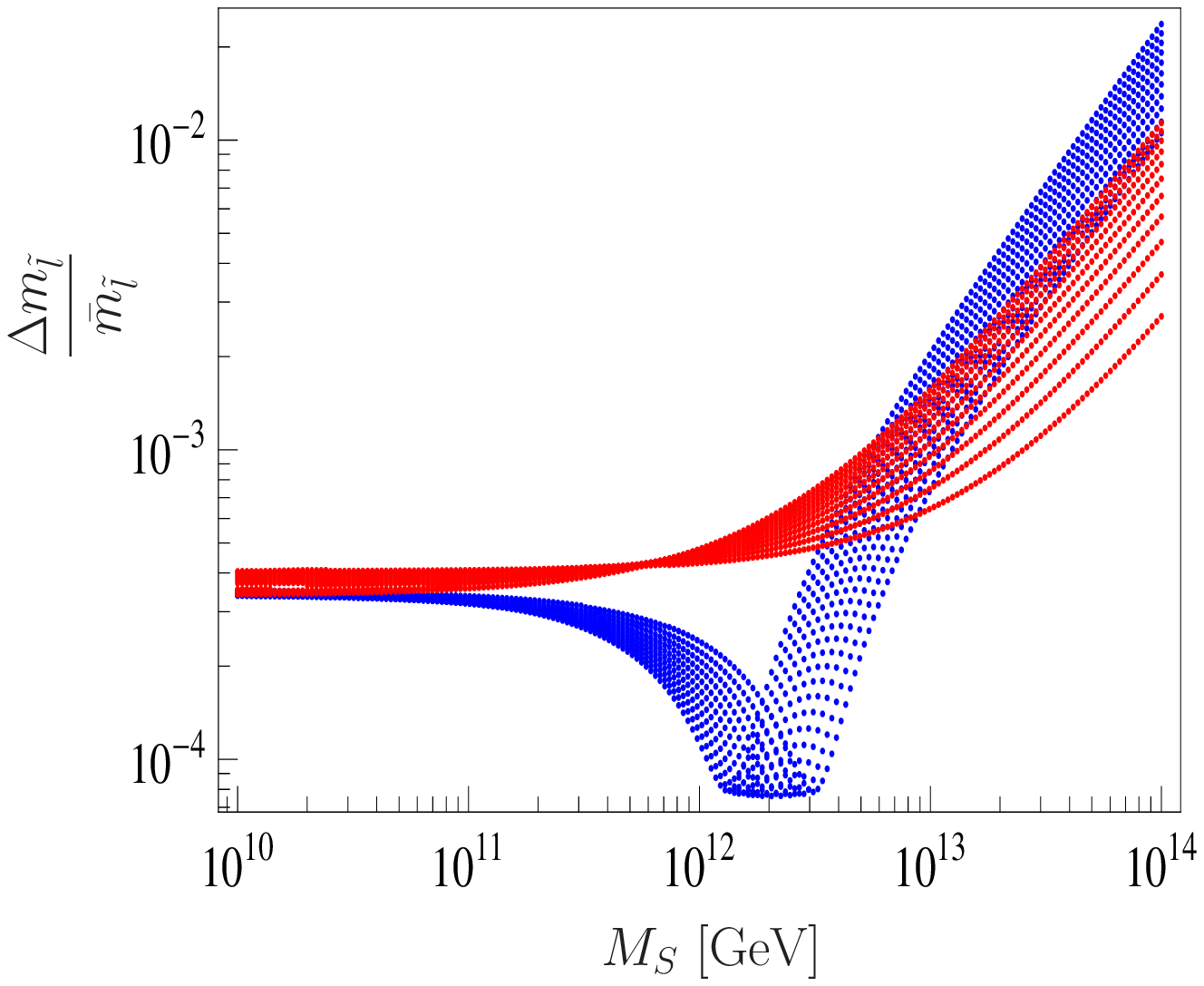}
\end{center}
\vspace{-5mm}
\caption{$\frac{\Delta m_{ll}}{\bar{m}_{ll}}$ (left-hand side) and
$\frac{\Delta m_{\tilde{l}}}{\bar{m}_{\tilde{l}}}$ (right-hand side)
as a function of the seesaw scale, defined as the mass of the lightest
right-handed neutrino, for the parameter choice $v_{BL} = 10^{15}$ GeV
and $v_R \in [10^{15},10^{16}]$ GeV. Blue dots correspond to the mass
distribution generated by intermediate left sleptons whereas
red dots correspond to the mass distribution generated by the
right ones. The mSUGRA parameters have been taken as in the
SPS3 benchmark point and neutrino oscillation data have been fitted
according to the $Y_\nu$ fit, with degenerate right-handed neutrinos.}
\label{fig:edge-split}
\end{figure}

Figure \ref{fig:edge-split} shows our results for the observables
$\frac{\Delta m_{ll}}{\bar{m}_{ll}}$ and $\frac{\Delta
m_{\tilde{l}}}{\bar{m}_{\tilde{l}}}$ as a function of the seesaw
scale. Large values for $M_S$ lead to sizeable deviations from the mSUGRA
expectation, with a distinctive multi-edge structure in the dilepton 
mass distribution. Moreover, this effect is 
found in both left- and right- mediated decays. Observing this 
affect would clearly point towards a non-minimal seesaw model, 
such as the LR model we discuss.

\begin{figure}
\begin{center}
\vspace{5mm}
\includegraphics[width=0.47\textwidth]{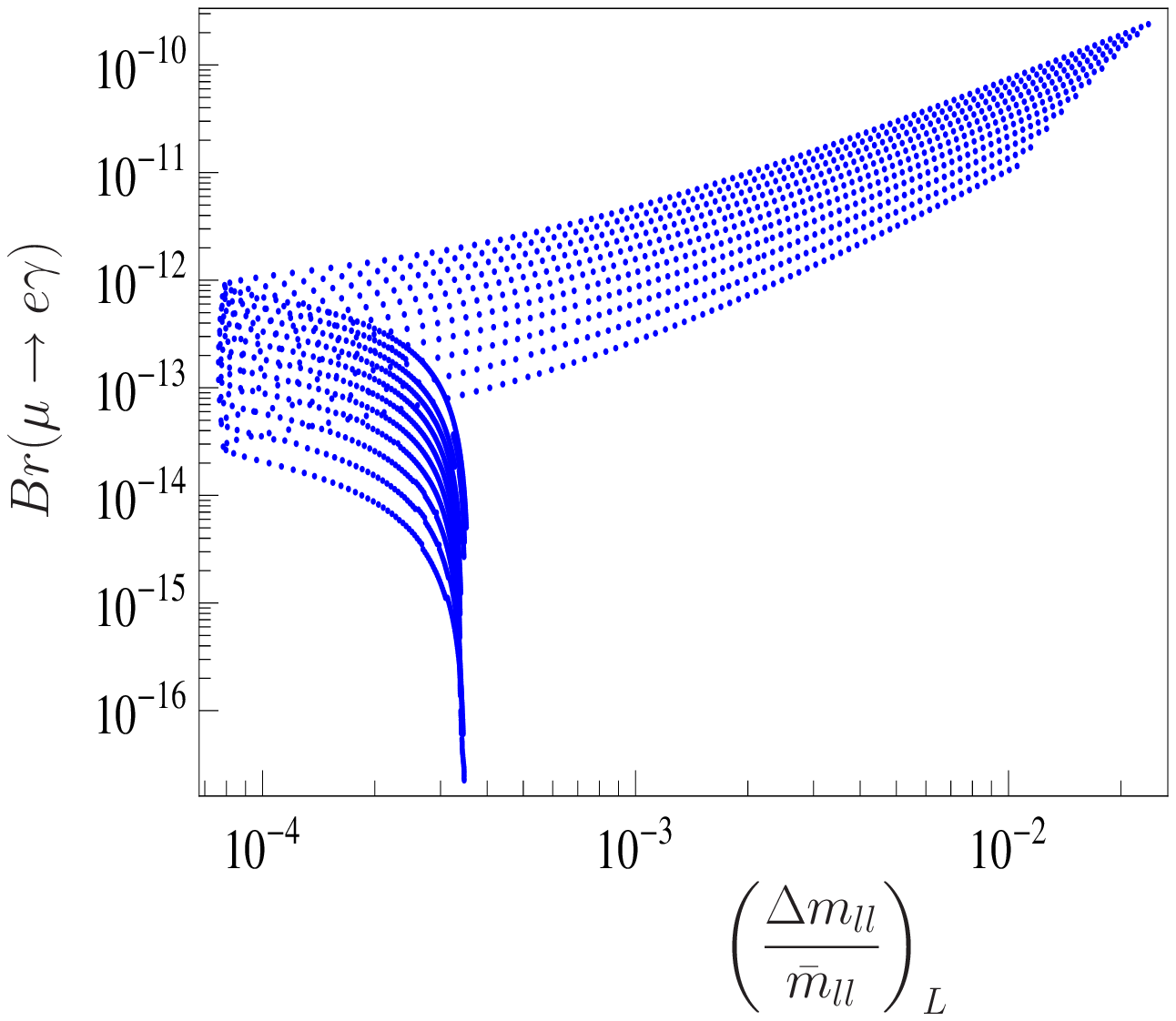}
\hspace{5mm}
\includegraphics[width=0.47\textwidth]{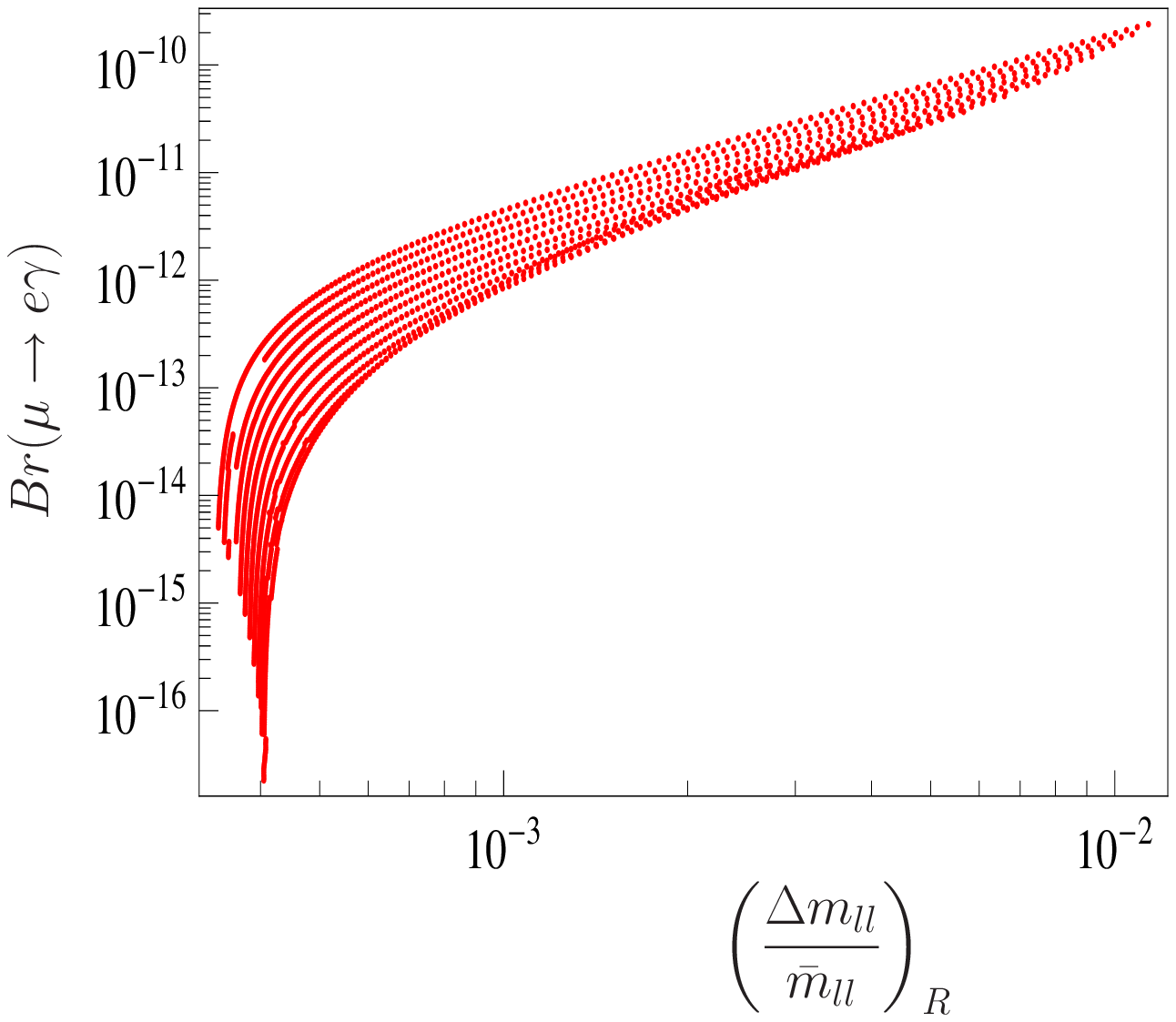}
\end{center}
\vspace{-5mm}
\caption{$Br(\mu \to e \gamma)$ as a function of $\left(\frac{\Delta
m_{ll}}{\bar{m}_{ll}}\right)_L$ (left-hand side) and
$\left(\frac{\Delta m_{ll}}{\bar{m}_{ll}}\right)_R$ (right-hand
side). The parameters are chosen as in figure \ref{fig:edge-split}.}
\label{fig:corredge}
\end{figure}

As expected, these observables are correlated with other LFV signals
\cite{Buras:2009sg,Abada:2010kj}. Figure \ref{fig:corredge} shows
$Br(\mu \to e \gamma)$ as a function of $\left(\frac{\Delta
m_{ll}}{\bar{m}_{ll}}\right)_L$ (mass distribution with intermediate L
sleptons) and $\left(\frac{\Delta m_{ll}}{\bar{m}_{ll}}\right)_R$
(mass distribution with intermediate R sleptons). Again, the main
novelty with respect to the usual seesaw implementations is the
correlation in the right sector, not present in the minimal
case \cite{Abada:2010kj}.

Furthermore, the process $\tilde{\chi}_2^0 \to \tilde{\chi}_1^0 l_i^+
l_j^-$ might provide additional LFV signatures if the rate for decays
with $l_i \neq l_j$ is sufficiently high. Reference
\cite{Andreev:2006sd} has investigated this possibility in great
detail, performing a complete simulation of the CMS detector in the
LHC for the decay $\tilde{\chi}_2^0 \to \tilde{\chi}_1^0 e \mu$. The
result is given in terms of the quantity
\begin{equation}
K_{e \mu} = \frac{Br(\tilde{\chi}_2^0 \to \tilde{\chi}_1^0 e
  \mu)}{Br(\tilde{\chi}_2^0 \to \tilde{\chi}_1^0 e e) +
  Br(\tilde{\chi}_2^0 \to \tilde{\chi}_1^0 \mu \mu)} \thickspace,
\end{equation}
which parametrizes the amount of flavor violation in
$\tilde{\chi}_2^0$ decays. The study, focused on the CMS test point
LM1 ($m_0 = 60$ GeV, $M_{1/2} = 250$ GeV, $A_0 = 0$ GeV, $\tan \beta =
10$, $sign(\mu) = +$) \cite{Ball:2007zza}, concludes that LFV can be
discovered at the LHC at $5 \sigma$ level with an integrated
luminosity of $10 fb^{-1}$ if $K_{e \mu} \ge K_{e \mu}^{min} = 0.04$.

\begin{figure}
\begin{center}
\vspace{5mm}
\includegraphics[width=0.6\textwidth]{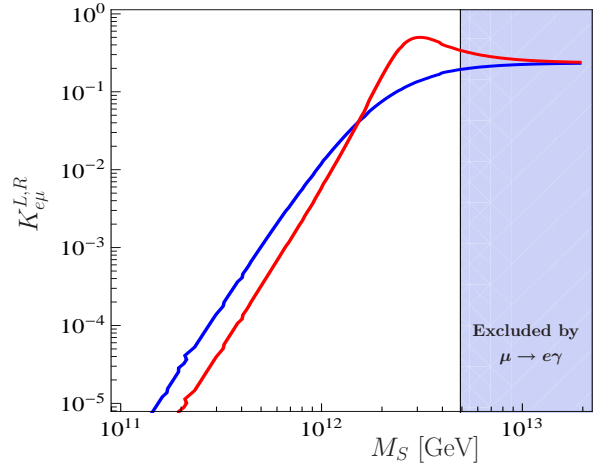}
\end{center}
\vspace{-5mm}
\caption{$K_{e \mu}$ as a function of the lightest right-handed
neutrino mass, for the parameter choice $v_{BL} = 10^{15}$ GeV and
$v_R = 5 \cdot 10^{15}$ GeV. The blue curve corresponds to
contributions from intermediate L sleptons, whereas the red one
corresponds to intermediate R sleptons. The mSUGRA parameters have
been taken as in the SPS3 benchmark point, which satisfies
$m(\tilde{\chi}_2^0) > m(\tilde{l_i})> m(\tilde{\chi}_1^0)$, and thus
the intermediate L and R sleptons can be produced on-shell. Neutrino
oscillation data have been fitted according to the $f$ fit, with
non-degenerate right-handed neutrinos. The blue shaded region is
excluded by $\mu \to e \gamma$.}
\label{fig:Kemu-SPS3}
\end{figure}

Figure \ref{fig:Kemu-SPS3} shows our computation of $K_{e \mu}$ as a
function of the lightest right-handed neutrino mass, for the parameter
choice $v_{BL} = 10^{15}$ GeV and $v_R = 5 \cdot 10^{15}$ GeV. The
results are shown splitting the contributions from intermediate left
(blue) and right (red) sleptons. Although the selected mSUGRA
parameters belong to the SPS3 point, and not to LM1 as in reference
\cite{Andreev:2006sd}, a similar sensitivity for $K_{e \mu}^{min}$ is
expected\footnote{Moreover, the LM1 point, being very similar to
  SPS1a', is strongly constrained by $\mu \to e \gamma$.}. This is because
the reduction in the cross-section due to the slightly heavier
supersymmetric spectrum is possibly partially compensated by the
corresponding reduction in the SM background and thus a limiting value
$K_{e \mu}^{min}$ of a similar order is expected. Moreover,
\cite{Andreev:2006sd} uses 10 $fb^{-1}$ and with larger integrated
luminosities even smaller $K_{e \mu}^{min}$ should become accessible
at the LHC.

The main result in figure \ref{fig:Kemu-SPS3} is that for large $M_S$
values the rates for LFV $\tilde{\chi}_2^0$ decays are measurable for
both left and right intermediate sleptons. In fact, for $M_S
\gtrsim 10^{12}$ GeV the parameter $K_{e \mu}$ is above its minimum
value for the $5 \sigma$ discovery of $\tilde{\chi}_2^0 \to
\tilde{\chi}_1^0 e \mu$. See references
\cite{Andreev:2006sd,delAguila:2008iz} for more details on the LHC
discovery potential in the search for LFV in this channel.

\section{Summary}

This chapter presents a supersymmetric left-right symmetric model. The 
motivation for studying this setup is twofold. First, LR models 
are theoretically attractive, since they contain all the necessary 
ingredients to {\em generate} a seesaw mechanism, instead of adding 
it by hand as is so often done. And, second, in a setup where the 
SUSY LR is supplemented by flavor blind supersymmetry breaking 
boundary conditions, different from all pure seesaw setups, lepton 
flavor violation occurs in both, the left and the right slepton 
sectors. 

We have calculated possible low-energy signals of this SUSY LR model, 
using full 2-loop RGEs for all parameters. We have found that low-energy 
lepton flavor violating decays, such as $\mu\to e \gamma$ are 
(a) expected to be larger than for the corresponding mSUGRA points 
in parameter space of seesaw type-I models and (b) the polarization 
asymmetry $\mathcal{A}$ of the outgoing positron is found to differ 
significantly from the pure seesaw prediction of $\mathcal{A}=+1$ 
in large regions of parameter space. We have also discussed possible 
collider signatures of the SUSY LR model for LHC and a possible ILC. 
Mass splittings between smuons and selectrons and LFV violating 
slepton decays should occur in both the left and the right slepton sector, 
again different from the pure seesaw expectations. 

SUSY LR model is a good example of a
``beyond'' minimal, pure seesaw and offers many interesting novelties.
For example, the impact of the intermediate scales on dark matter
relic density and on certain mass combinations and the influence of
the right-handed neutrino spectrum on low energy observables, are
topics that certainly deserve further studies.

\chapter{Summary}

The subject of this thesis is the phenomenology of neutrino mass models in supersymmetry. Both analytical and numerical tools have been employed in the research this thesis is based on, the results of which have been described in the previous chapters. In the following lines the main conclusions will be summarized and a global picture of the thesis will be presented.

Two different approaches to neutrino masses have been studied in this thesis: (1) R-parity violating models with broken lepton number, and (2) supersymmetric left-right models that conserve R-parity at low energies. In both cases a detailed study of the phenomenology has been performed, obtaining numerical predictions for present and future experiments.

More emphasis has been put on the falsifiability of the models rather than on their verification. In fact, it is impossible to verify a model. In this sense, Popper's idea of science has been applied all along this thesis, looking for clear experimental signatures which, if not observed, disprove the models under investigation.

In the first part of the thesis the phenomenology of two R-parity violating models has been investigated. In this setup neutrinos get masses due to their mixing with the higgsinos in what can be regarded as an electroweak version of the seesaw mechanism. The cleanest prediction in this type of models is the sharp correlation that is found between LSP decays and neutrino physics. This can be used to rule out the model at colliders if a clear deviation is found from the predicted ratios.

When R-parity is broken spontaneously a Goldstone boson appears in the spectrum, the so-called majoron. This leads to many novel signatures, some of which have been studied in this thesis. We found that invisible LSP decays might be dominant if the scale of lepton number breaking is low. This would require large statistics to distinguish s-\rpv from the MSSM with conserved R-parity. However, if low energy experiments are taken into account new perspectives open up. The search for exotic muon decays involving majorons in the final state is of great help to solve the potential confusion, since the branching ratios for these processes are enhanced in the same region of parameter space where the LSP decays mainly to invisible channels.

If, on the other hand, R-parity is broken explicitly, no majoron is generated in the model. This is the case of the $\mu\nu$SSM which, nevertheless, has a very rich phenomenology at colliders. It combines the clear correlations found in bilinear R-parity breaking models with the interesting possibility, also present in the NMSSM, of a light singlet. The phenomenology of the $\mu\nu$SSM with one or two generations of singlet superfields is described in this thesis, together with the characterization of neutrino masses in both cases.

The second part of this thesis is devoted to the study of a non-minimal supersymmetric left-right model that leads to R-parity conservation at low energies. In this case neutrino masses are generated with a type-I seesaw mechanism.

The model is described in detail and its RGEs are computed at 2-loop level including threshold corrections at the intermediate scales. This is numerically implemented with the help of tools like Sarah and SPheno, which have been used extensively in this thesis. The resulting code is able to calculate the soft masses of the sleptons at the SUSY scale and, as usual, the inclusion of the seesaw mechanism leads to the appearance of off-diagonal entries in these matrices. As a consequence of this, lepton flavor violating signatures at the SUSY scale are obtained.

However, in addition to the known phenomenology in the left slepton sector, we have found signatures in the right one, which clearly points to a non-minimal seesaw implementation. Observables like right slepton LFV decays and mass splittings or positron polarization asymmetry in $\mu^+ \to e^+ \gamma$ different from $\mathcal{A}=+1$ would be clear hints towards an underlying left-right symmetry.

Moreover, measurements of ratios like $Br(\tilde{\tau}_R \to \tilde{\chi}_1^0 \mu)/Br(\tilde{\tau}_L \to \tilde{\chi}_1^0 \mu)$ constrain the high energy structure of the model, since they are correlated with the ratio $v_{BL}/v_R$. However, a precise knowledge of the particle spectrum is needed if one wants to set strong and reliable bounds.

Finally, other observables are slightly changed with respect to minimal seesaw models. This is the case of $\mu \to e \gamma$, whose branching ratio gets increased due to the new right-handed slepton contributions.

As a general conclusion from this thesis, neutrino masses make us think that a rich phenomenology will be found at the LHC and other experiments in progress. Only through the careful examination of the new data we will be able to disentangle the different signals and find out which of our models, if any, is the right description of nature.

\appendix

\chapter{s-\rpv : Approximated couplings}
\label{sect:app1}

With broken R-parity the lightest supersymmetric particle decays. 
Here we list the most important couplings of the lightest neutralino in s-\rpv
using the seesaw approximation. In the numerical calculations discussed 
in the thesis, all mass matrices are exactly diagonalized in order to 
obtain the exact couplings. For the understanding of the main 
qualitative features of the LSP decays, however, the approximated 
couplings listed below are very helpful. 

We define the ``rotated'' quantities:
\begin{eqnarray}\label{def:rot}
\tilde{x}_i \equiv \big( U_\nu \big)_{ik}^T x_k, & \hskip5mm & 
\tilde{y}_{ij} \equiv \big( U_\nu \big)_{ik}^T y_{kj}.
\end{eqnarray}

Here $(U_{\nu})^T$ is the matrix 
which diagonalizes either the part of the ($3,3$) effective neutrino mass 
matrix, proportional to $a$ or $c$, depending on which gives the larger 
eigenvalue. 

$\tilde{\chi}_1^0-W^{\pm}-l^{\mp}_i$ couplings are found from the 
general expressions for the $\tilde{\chi}^0-W^{\pm}-{\tilde \chi}^{\mp}$ 
vertices 
\begin{equation}\label{eq:cnw}
\mathcal{L}=\bar{\chi}_i^- \gamma^{\mu} 
\big( O_{Lij}^{cnw} P_L + O_{Rij}^{cnw} P_R \big) \chi_j^0 W_\mu^- 
+ \bar{\chi}_i^0 \gamma^{\mu} \big( O_{Lij}^{ncw} P_L + O_{Rij}^{ncw} 
P_R \big) \chi_j^- W_\mu^+
\end{equation}
as 
\begin{eqnarray}\label{eq:cnwdef}
O_{Li1}^{cnw}& =& \frac{g}{\sqrt{2}} \big[ 
\frac{g N_{12} \Lambda_i}{\textnormal{Det}_+}-\big( \frac{\epsilon_i}{\mu}
+\frac{g^2 v_u \Lambda_i}{2 \mu \textnormal{Det}_+} \big)N_{13}
-\displaystyle\sum_{k=1}^{7} N_{1k} \xi_{ik} \big], \\ \nonumber
O_{Ri1}^{cnw} & = & \frac{1}{2} g (Y_e)^{ii} \frac{v_d}{\textnormal{Det}_+} 
\big[ \frac{g v_d N_{12} + M_2 N_{14}}{\mu} \epsilon_i \\ \nonumber && +
\frac{g(2 \mu^2 +g^2 v_u v_d)N_{12}+g^2 v_u(\mu+M_2)N_{14}}{2 \mu \textnormal{Det}_+} 
\Lambda_i \big].
\end{eqnarray}
$\textnormal{Det}_+$ is the determinant of the MSSM chargino mass matrix and $N$ is 
the matrix which diagonalizes the MSSM neutralino mass matrix. Here, 
\begin{eqnarray}\label{eq:cnwLR}
O_{Li1}^{ncw}& =& \big(O_{Li1}^{cnw}\big)^*, \\
O_{Ri1}^{ncw}& =& \big(O_{Ri1}^{cnw}\big)^*. \nonumber
\end{eqnarray}
The Lagrangian for $\tilde{\chi}_i^0-\tilde{\chi}_j^0-Z$
\begin{equation}
\mathcal{L}=\frac{1}{2} \bar{\chi}_i^0 \gamma^{\mu} \big( O_{Lij}^{nnz} 
P_L + O_{Rij}^{nnz} P_R \big) \chi_j^0 Z_\mu
\end{equation}
gives for $\tilde{\chi}_1^0-\nu_i-Z$
\begin{eqnarray}\label{eq:cnndef}
O_{Li1}^{nnz}& =& -\frac{g}{2 \cos \theta_W} 
\big[\tilde{\xi}_{i1} N_{11}+\tilde{\xi}_{i2} N_{12} 
+ 2 \tilde{\xi}_{i4} N_{14} +\tilde{\xi}_{i5} N_{15}
+\tilde{\xi}_{i6} N_{16}+\tilde{\xi}_{i7} N_{17}\big], \nonumber \\
O_{Ri1}^{nnz}& =& -\big( O_{Li1}^{nnz} \big)^*.
\end{eqnarray}
The most important difference to the explicit R-parity violating 
models comes from the coupling $\tilde{\chi}_i^0-\tilde{\chi}_j^0-P^0_k$
\begin{equation}\label{defnnp}
\mathcal{L}=\frac{1}{2} \bar{\chi}_i^0 
\big( O_{Lij}^{nnp} P_L + O_{Rij}^{nnp} P_R \big) \chi_j^0 P_k^0,
\end{equation}
with
\begin{eqnarray}
O_{Li1J}^{nnp}& =& R_{Jm}^p O_{Li1m}^{' nnp}, \\ \nonumber
O_{Ri1J}^{nnp}& = &\big(O_{Li1J}^{nnp}\big)^*.
\end{eqnarray}
Because the spontaneous breaking of lepton number produces 
a massless pseudo-scalar, eq. \eqref{defnnp} leads to a 
coupling $\tilde{\chi}_1^0-J-\nu_i$, i.e~ a new invisible 
decay channel for the lightest neutralino. In the limit $v_i \ll v_R,v_S$ 
one can find the approximated majoron profile in equation \eqref{smplstmaj} and easily obtain simplified expressions for its couplings.

 The ``unrotated'' couplings $O_{Li1m}^{' nnp}$ 
are
\begin{eqnarray}\label{defOnnpur}
O_{Li1\tilde{L}_k^0}^{' nnp} & =&-\frac{i}{2} \big(U_\nu\big)_{ki}
(g' N_{11}-g N_{12}), \\ \nonumber
O_{Li1\tilde{S}}^{' nnp}& =& \frac{i}{\sqrt{2}} h 
\big( \tilde{\xi}_{i5} N_{17}+\tilde{\xi}_{i7} N_{15} \big), \\ \nonumber
O_{Li1\tilde{\nu}^c}^{' nnp} &= & -i \frac{\tilde{\epsilon}_i}{v_R} N_{14}
+\frac{i}{\sqrt{2}} h \big(\tilde{\xi}_{i6} N_{17}+\tilde{\xi}_{i7} N_{16} 
\big).
\end{eqnarray}
In the limit $v_R, v_S \rightarrow \infty$ one can derive a very simple 
approximation formula for $O_{\tilde\chi^0_1\nu_kJ}$. It s given by
\footnote{We correct here a misprint in \cite{Hirsch:2006di}.}
\begin{equation}
\label{eq:majcl}
|O_{\tilde\chi^0_1\nu_kJ}| \simeq  - \frac{{\tilde \epsilon}_k}{V}N_{14} +
\frac{{\tilde v}_{k}}{2 V}(g' N_{11} - g N_{12})
+ h.O.
\end{equation}
Here, $h.O.$ stands for higher order terms. Eq. \eqref{eq:majcl} serves 
to show that for constant ${\tilde \epsilon}_i$ 
and ${\tilde v}_{i}$, $O_{\tilde\chi^0_1\nu_kJ} \rightarrow 0$ as $v_R$ 
goes to infinity. This is as expected, since for $v_R\rightarrow\infty$ 
the spontaneous model approaches the explicit bilinear model. Note, that 
only the presence of the field $\widehat \nu^c$ is essential for the 
coupling Eq.~ \eqref{eq:majcl}. If $\widehat S$ is absent, replace 
$V \rightarrow v_R$. 

In addition to the Majoron in considerable parts of the parameter 
space one also finds a rather light singlet scalar, called the 
``scalar partner'' of the Majoron in \cite{Hirsch:2005wd}, $S_J$. 
From the Lagrangian
\begin{equation}
\mathcal{L}=\frac{1}{2} \bar{\chi}_i^0 \big( O_{Lij}^{nns} P_L 
+ O_{Rij}^{nns} P_R \big) \chi_j^0 S_k^0,
\end{equation}
one finds the coupling $\tilde{\chi}_1^0-S_J-\nu_i$ as
\begin{eqnarray}
O_{Li1S_J}^{nns}& =& R_{S_Jk}^s O_{Li1k}^{' nns}, \\ \nonumber
O_{Ri1S_J}^{nns}& =& \big(O_{Li1S_J}^{nns}\big)^*.
\end{eqnarray}
Different from the Majoron, however, there is no simple analytical 
approximation for $R_{S_J}$. We write symbolically
\begin{equation}\label{defsj}
R_{S_Jk}^s = \big(R_{S_J H_d},R_{S_J H_u},R_{S_J \tilde{L}_k^0},
R_{S_J \phi},R_{S_J \tilde{S}},R_{S_J \tilde{\nu}^c} \big) ,
\end{equation}
and define unrotated couplings by
\begin{eqnarray}\label{defonnsur}
O_{Li1H_d}^{' nns}& =& \frac{1}{2} \big[ 
(g \tilde{\xi}_{i2}-g' \tilde{\xi}_{i1})N_{13} 
+(g N_{12}-g' N_{11}) \tilde{\xi}_{i3} 
-\sqrt{2} h_0 (\tilde{\xi}_{i7} N_{14}
+\tilde{\xi}_{i4} N_{17})\big], \nonumber \\ \nonumber 
O_{Li1H_u}^{' nns}& =& \frac{1}{2} \big[ (g' \tilde{\xi}_{i1}
-g \tilde{\xi}_{i2})N_{14} +(g' N_{11}-g N_{12}) \tilde{\xi}_{i4} 
-\sqrt{2} h_0 (\tilde{\xi}_{i7} N_{13}+\tilde{\xi}_{i3} N_{17}) \\
&& \nonumber -\sqrt{2} \frac{\tilde{\epsilon}_i}{v_R} N_{15}\big], \\ \nonumber
O_{Li1\tilde{L}_k^0}^{' nns} &=& \frac{1}{2} \big(U_\nu\big)_{ki}
(g' N_{11}-g N_{12}), \\ \nonumber 
O_{Li1\phi}^{' nns} &= & \frac{1}{\sqrt{2}} \big[-h_0 (\tilde{\xi}_{i3} 
N_{14}+\tilde{\xi}_{i4} N_{13})+h(\tilde{\xi}_{i6} N_{15}
+\tilde{\xi}_{i5} N_{16})+\lambda \tilde{\xi}_{i7} N_{i7}\big], \\ \nonumber 
O_{Li1\tilde{S}}^{' nns} &= & \frac{1}{\sqrt{2}} h 
\big( \tilde{\xi}_{i5} N_{17}+\tilde{\xi}_{i7} N_{15} \big), \\
O_{Li1\tilde{\nu}^c}^{' nns} &= & -\frac{\tilde{\epsilon}_i}{v_R} N_{14}
+\frac{1}{\sqrt{2}} h \big(\tilde{\xi}_{i6} N_{17}
+\tilde{\xi}_{i7} N_{16} \big).
\end{eqnarray}
As eqs. \eqref{defonnsur} shows, $\tilde{\chi}_1^0\to S_J + \nu_i$ has a 
partial decay width similar in size to the decay $\tilde{\chi}_1^0\to J +
\nu_i$, as soon as kinematically allowed. Since, on the other hand, $S_J$ 
decays practically always with a branching ratio close to 100 \% 
into two Majorons, $\tilde{\chi}_1^0\to S_J + \nu_i$ gives in general a 
sizeable contribution to the invisible width of the neutralino. 

Finally, we give also the coupling $\tilde{\chi}_i^0-J-\tilde{\chi}_j^0$, 
for the case of two heavy neutralinos. Here, 
\begin{equation}
O_{LijJ}^{nnp}=-\frac{i}{\sqrt{2}} \frac{h}{V} \big[ 
v_S (N_{j7} N_{i5}+N_{i7} N_{j5})-v_R (N_{j7} N_{i6}+N_{i7} N_{j6}) \big].
\end{equation}

\chapter{s-\rpv : $\mu\to eJ$ vs $\mu\to eJ\gamma$}
\label{sect:app2}

Let us compare the constraining power of $\mu \to e J$ and $\mu \to e J \gamma$ in order to know which one is better suited for putting bounds in the s-\rpv parameter space.

\section*{Background}

Both processes have standard model background. Since the majoron escapes detection it is seen experimentally as missing energy, what makes $\mu \to e J$ indistinguishable from the usual $\mu \to e \nu \bar \nu$, and $\mu \to e J \gamma$ indistinguishable from the radiative decay $\mu \to e \nu \bar \nu \gamma$. According to the PDG \cite{Amsler:2008zzb}, these are the branching ratios for these decays:

\begin{eqnarray}
Br(\mu \to e \nu \bar \nu) &\simeq& 100\% \nonumber \\
Br(\mu \to e \nu \bar \nu \gamma) &=& (1.4 \pm 0.4)\%
\end{eqnarray}

In addition, $\mu \to e J \gamma$ has also a very important accidental background due to photons produced in other processes inside the detector. If they happen to be in coincidence with an electron produced in $\mu \to e \nu \bar \nu$, there is no way to distinguish the event from the prompt decay $\mu \to e \nu \bar \nu \gamma$. This will be discussed below.

\section*{Bounds on branching ratios}

The first step to find bounds for the parameters is to put a bound on the corresponding branching ratio.

Let $N$ be the effective number of muons which are used in the experiment (the total number minus the muons which are lost due to non-perfect detector acceptance). Then, for a given process, we can split the possible final states as

\begin{equation}
N = n_s + n_b + n_o
\end{equation}

where $n_s$ is the number of events of the signal we are interested on, $n_b$ is the number of events of its background, and $n_o$ is the number of events involving other final states.

In order to claim detection of a given signal we have to find an excess over the background larger than the background uncertainty. This means

\begin{equation}
n_s > \sqrt{n_b} \: \Rightarrow \: \text{Detection}
\end{equation}

Therefore, if the process is not observed, the bound on the branching ratio is given by

\begin{equation}
Br(signal) < Br(signal)_{bound}
\end{equation}

where

\begin{equation}
Br(signal)_{bound} = \frac{\sqrt{n_b}}{N}
\end{equation}

After this brief introduction, let us estimate expressions for the bounds that $\mu \to e J$ and $\mu \to e J \gamma$ would have if they are not observed.

For the case of $\mu \to e J$ the background has a branching ratio close to 1, and therefore $n_b \simeq N$. This implies

\begin{equation}\label{boundmuej}
Br(\mu \to e J)_{bound} \simeq \frac{1}{\sqrt{N}}
\end{equation}

For the case of $\mu \to e J \gamma$ we need to know the number of events of $\mu \to e \nu \bar \nu \gamma$, what is related to its branching ratio

\begin{equation}\label{boundmuejg}
Br(\mu \to e \nu \bar \nu \gamma) = \frac{\alpha}{8 \pi} I'
\end{equation}

where $I'$ is a four-body phase space integral. Then, since $n = Br \times N$, we get

\begin{equation}
Br(\mu \to e J \gamma)_{bound} = \frac{\sqrt{\frac{\alpha}{8 \pi} I' N}}{N} \simeq 1.7 \cdot 10^{-2} \sqrt{\frac{I'}{N}}
\end{equation}

The integral $I'$ is found to be typically smaller than one, and then the bound on this branching ratio is potentially better. However, since the calculation of parameter bounds using $\mu \to e J \gamma$ also implies small factors, namely another $\alpha$ factor and the three-body phase space integral of the process, it is not clear a priori whether we can get better bounds, with respect to $\mu \to e J$, or not. This is what we want to answer in the following.

\section*{Background supression}

In order to supress as much background as possible we can use kinematical relations involving $E_e$, $E_\gamma$ and $\cos \theta_{e \gamma}$ to distinguish between $\mu \to e J \gamma$ and $\mu \to e \nu \bar \nu \gamma$.

Let us note that in a three-body decay the angle between the directions of two particles can be determined using their energies. This is the case for $\mu \to e J \gamma$, that allows to find a relation

\begin{equation}\label{anglerel}
\theta_{e \gamma} = f(E_e,E_\gamma)
\end{equation}

In contrast, for a four-body decay, like $\mu \to e \nu \bar \nu \gamma$, $\theta_{e \gamma}$ can be taken as another free parameter. Therefore, the relative directions of the additional two particles provide a higher-dimensional phase space. For our calculation this implies that $I'$ has to be calculated integrating over the possible values of $E_e$, $E_\gamma$ and $\cos \theta_{e \gamma}$, while for $\mu \to e J \gamma$ we will get an integral, $I$, over $E_e$ and $E_\gamma$ only.

Moreover, for experimental purposes this means that the prompt decay $\mu \to e \nu \bar \nu \gamma$ can have electron-photon pairs with directions forbidden by the relation \eqref{anglerel}. This can be used to suppress the background, eliminating events which are clearly outside the allowed phase space for $\mu \to e J \gamma$.

In practice one can measure the quantities $E_e$ and $E_\gamma$, with their corresponding errors, and calculate the hypothetical $\theta_{e \gamma}$ that would be predicted with \eqref{anglerel}, also with its corresponding error. The problem is that the errors in the energies of the electron and the photon (order \% for the MEG experiment \cite{meg}) are too large to predict an accurate value for $\theta_{e \gamma}$, giving only an allowed range and implying that many $\mu \to e \nu \bar \nu \gamma$ events can survive this kinematical cut.

For that reason, although the background is supressed and the integral $I'$ has a clearly lower value than it would have without the cut, the result is not optimal, still having a lot of $\mu \to e \nu \bar \nu \gamma$ events which cannot be distinguished from $\mu \to e J \gamma$.

\section*{Comparison}

There is a simple relation between the branching ratios of $\mu \to e J $ and $\mu \to e J \gamma$:

\begin{equation}\label{rel}
Br(\mu \to e J \gamma) = \frac{\alpha}{2 \pi} I Br(\mu \to e J)
\end{equation}

where $I$ is the corresponding three-body phase space integral.

We can use \eqref{rel} to find out which process can give us the best bound. Let us suppose that our experiment is designed to look for photons in the final state. In that case we can get a bound on $Br(\mu \to e J \gamma)$ as explained in the previous section, and use it to obtain an indirect bound on $Br(\mu \to e J)$ thanks to equation \eqref{rel}.

\begin{equation}
Br(\mu \to e J)_{indirect bound} = \frac{2 \pi}{\alpha} \frac{1}{I} Br(\mu \to e J \gamma)_{bound}
\end{equation}

and then, using the expressions \eqref{boundmuej} and \eqref{boundmuejg}, we can get the ratio

\begin{eqnarray}
\frac{Br(\mu \to e J)_{bound}}{Br(\mu \to e J)_{indirect bound}} &=& \frac{\frac{1}{\sqrt{N}}}{\frac{2 \pi}{\alpha} \frac{1}{I} \sqrt{\frac{\alpha}{8 \pi}} \sqrt{\frac{I'}{N}}} \nonumber \\
&\simeq& 0.07 \frac{I}{\sqrt{I'}}
\end{eqnarray}

Numerically it has been found that

\begin{equation}
\frac{I}{\sqrt{I'}} \sim [0.1 - 7]
\end{equation}

depending on the parameters $E_e^{min}$, $E_\gamma^{min}$ and $\theta_{e \gamma}^{min}$, given by the experimental setup.

Therefore

\begin{equation}
\frac{Br(\mu \to e J)_{bound}}{Br(\mu \to e J)_{indirect bound}} < 1
\end{equation}

This means that the bound coming from $\mu \to e J \gamma$ will not be better than a possible bound coming directly from $\mu \to e J$, since the background given by $\mu \to e \nu \bar \nu \gamma$ makes it hard to see a positive signal. This could be different if the measurement of the electron and photon energies was more precise. In that case the uncertainty in the angle $\theta_{e \gamma}$ would be small enough to supress much more background, allowing us to get better bounds on the parameters of the model.

It is also important to remember that the MEG experiment focuses on electrons and photons in the last energy bin (both particles with energies close to $m_\mu/2$), since they are interested in the detection of $\mu \to e \gamma$. The reduction of the accidental background, see below, demands this restriction as well. That way, by taking very small energy windows, the supression of the background $\mu \to e \nu \bar \nu \gamma$ is extremely good, and in practice they work in the regime in which there are no background events from the radiative decay. If we stay in the same phase space region we also have no background, but the integral $I$ decreases strongly. Both things compensate, giving again the same final result for the possible bounds.

\section*{Accidental background}

Finally, let us mention that the process $\mu \to e \gamma$ is also plagued with an important background coming from photons produced in other processes that happen to be accidentally in coincidence with electrons. In fact, this is the strongest limitation in experiments like MEG, which require to focus on very small deviations from {\em coincidence} in order to claim detection.

In principle, one could reduce the accidental background by applying a cut on the relative time $\Delta t_{e\gamma}$, the difference between the detection times of photon and electron. However, this is restricted by current technology to be around 100 ps \cite{meg}. Further developments might provide a better time resolution, what would imply a better reduction of the background from accidental photons.

In conclusion, one has to add the limitation coming from accidental background to the previous discussion. As a consequence, it is not possible to enlarge the measurement window, due to the strong rise of accidental photons.

\chapter{$\mu \nu$SSM: Tadpole equations}
\label{munuapp1}

The tree-level tadpole equations of the $\mu \nu$SSM can be written as

{\allowdisplaybreaks
\begin{eqnarray}
\frac{\partial V}{\partial v_d} &=& \frac{1}{8}(g^2 + g'^2)u^2 v_d +
m_{H_d}^2 v_d + \frac{1}{2} v_d \lambda_s \lambda_t^* v_{Rs} v_{Rt} +
\frac{1}{2}v_d v_u^2 \lambda_s \lambda_s^* \nonumber \\ && -
\frac{1}{8} v_{Rs}^2 v_u (\kappa_s \lambda_s^* + h.c.) -\frac{1}{4}
v_i (\lambda_s^* Y_\nu^{it} v_{Rs} v_{Rt} + h.c.) \nonumber \\ && -\frac{1}{4} v_u^2
v_i (\lambda_s^* Y_\nu^{is} + h.c.) - \frac{1}{2
\sqrt{2}} v_u v_{Rs}(T_\lambda^s + h.c.) = 0 \label{eq:tadpolevd} \\
\frac{\partial V}{\partial v_u} &=& -\frac{1}{8}(g^2 + g'^2)u^2 v_u +
m_{H_u}^2 v_u + \frac{1}{2} v_u \lambda_s \lambda_t^* v_{Rs} v_{Rt} +
\frac{1}{2}v_d^2 v_u \lambda_s \lambda_s^* \nonumber \\ && - \frac{1}{8} v_{Ri}^2 v_d
(\kappa_s \lambda_s^* + h.c.) + \frac{1}{8} v_i
v_{Rs}^2 (\kappa_s^* Y_\nu^{is} + h.c.) \nonumber \\ && -\frac{1}{2} v_d v_u v_i
(\lambda_s^* Y_\nu^{is} + h.c.) + \frac{1}{2} v_u v_i v_j Y_\nu^{is}
(Y_\nu^{js})^* \nonumber \\ && + \frac{1}{2} v_u Y_\nu^{is}
(Y_\nu^{it})^* v_{Rs} v_{Rt} - \frac{1}{2 \sqrt{2}} v_d
v_{Rs}(T_\lambda^s + h.c.) \nonumber \\ && + \frac{1}{2 \sqrt{2}} v_i v_{Rs}
(T_\nu^{is} + h.c.) = 0 \label{eq:tadpolevu} \\
\frac{\partial V}{\partial v_i} &=& \frac{1}{8}(g^2 + g'^2)u^2 v_i + \frac{1}{2}
({m_L^2}_{ij} + {m_L^2}_{ji}) v_j - \frac{1}{4} v_d v_u^2 (
\lambda_s^* Y_\nu^{is} + h.c. ) \nonumber \\ && + \frac{1}{8} v_{Rs}^2
v_u ( \kappa_s^* Y_\nu^{is} + h.c. ) - \frac{1}{4} v_d ( \lambda_s^*
v_{Rs} v_{Rt} Y_\nu^{it} + h.c. ) \nonumber \\ && + \frac{1}{4} v_j ( v_{Rs} v_{Rt}
Y_\nu^{is} (Y_\nu^{jt})^* + h.c. ) + \frac{1}{4} v_u^2
v_j ( Y_\nu^{is} (Y_\nu^{js})^* + h.c. ) \nonumber \\ && + \frac{1}{2 \sqrt{2}} v_u
v_{Rs} (T_\nu^{is} + h.c. ) = 0 \label{eq:tadpoleneu} \\
\frac{\partial V}{\partial v_{Rs}} &=& {m_{\tilde{\nu}^c}^2}_{ss}
v_{Rs} - \frac{1}{4} v_d v_u v_{Rs} (\kappa_s \lambda_s^* + h.c.) +
\frac{1}{4} \kappa_s \kappa_s^* v_{Rs}^3 \nonumber \\ && + \frac{1}{4}
v_u v_{Rs} v_j ( \kappa_s^* Y_\nu^{js} + h.c. ) + \frac{1}{4} (v_u^2 +
v_d^2) ( \lambda_s \lambda_t^* v_{Rt} + h.c. ) \nonumber \\ && + \frac{1}{4} v_u^2 [
Y_\nu^{js} (Y_\nu^{jt})^* v_{Rt} + h.c. ] +
\frac{1}{4} v_m v_n [ (Y_\nu^{ms})^* Y_\nu^{nt} v_{Rt} + h.c. ] \nonumber \\ && -
\frac{1}{4} v_d v_j ( \lambda_t^* Y_\nu^{js} v_{Rt} + \lambda_s^*
v_{Rt} Y_\nu^{jt} + h.c. ) - \frac{1}{2 \sqrt{2}} v_d
v_u (T_\lambda^s + h.c.) \nonumber \\ && + \frac{1}{2 \sqrt{2}} v_u v_j
(T_\nu^{js} + h.c.) + \frac{1}{4\sqrt{2}} v_{Rt} v_{Ru} \big( T_\kappa^{stu} + h.c. \big)=0 \label{eq:tadpoleright}
\end{eqnarray}
}
with
\begin{eqnarray}
u^2 &=& v_d^2 - v_u^2 + v_e^2 + v_\mu^2 + v_\tau^2  \label{eq:deftkappa}
\end{eqnarray}
and there is no sum over the index $s$ in equation \eqref{eq:tadpoleright}.

\chapter{$\mu \nu$SSM: Mass matrices}
\label{munuapp2}

In the scalar mass matrices shown below the tadpole equations 
have not yet been used to reduce the number of free parameters.

\section*{Charged Scalars}
\label{subsec:cscalars}

In the basis
\begin{eqnarray}
\big( {S^+}' \big)^T &=& ((H_d^-)^*,H_u^+,\tilde e_L^*, \tilde
\mu_L^*, \tilde \tau_L^*, \tilde e_R, \tilde \mu_R, \tilde \tau_R)
\nonumber \\ \big( {S^-}' \big)^T &=& (H_d^-,(H_u^+)^*,\tilde e_L,
\tilde \mu_L, \tilde \tau_L, \tilde e_R^*, \tilde \mu_R^*, \tilde
\tau_R^*)
\end{eqnarray}
the scalar potential includes the term
\begin{equation}
V \supset \big( {S^-}' \big)^T M_{S^\pm}^2 {S^+}'\quad,
\end{equation}
where $M_{S^\pm}^2$ is the $(8 \times 8)$ mass matrix of the charged
scalars. In the $\xi = 0$ gauge it can be written as
\begin{equation}
M_{S^\pm}^2 = \left( \begin{array}{c c}
M_{HH}^2 & \big( M_{H \tilde l}^2 \big)^\dagger \\
M_{H \tilde l}^2 & M_{\tilde l \tilde l}^2 \end{array} \right)\quad.
\end{equation}
The $(2 \times 2)$ $M_{HH}^2$ matrix is given by:
 
\begin{eqnarray}
\big( M_{HH}^2 \big)_{11} &=& m_{H_d}^2 + \frac{1}{8} [ (g^2 + g'^2)
v_d^2 + (g^2 - g'^2)(v_u^2 - v_e^2 - v_\mu^2 - v_\tau^2) ] \nonumber \\ &&
+ \frac{1}{2} \lambda_s \lambda_t^* v_{Rs} v_{Rt} + \frac{1}{2} v_i
\big( Y_e Y_e^\dagger \big)_{ij} v_j \nonumber \\ \big( M_{HH}^2
\big)_{12} &=& \frac{1}{4}g^2 v_u v_d - \frac{1}{2} \lambda_s
\lambda_s^* v_u v_d + \frac{1}{4} \lambda_s \kappa_s^* v_{Rs}^2 +
\frac{1}{2}v_u v_i \lambda_s (Y_\nu^{is})^* + \frac{1}{\sqrt{2}}
v_{Rs} T_\lambda^s \nonumber \\ \big( M_{HH}^2 \big)_{21} &=& \big(
M_{HH}^2 \big)_{12}^* \nonumber \\ \big( M_{HH}^2 \big)_{22} &=&
m_{H_u}^2 + \frac{1}{8} [ (g^2 + g'^2) v_u^2 + (g^2 - g'^2)(v_d^2 +
v_e^2 + v_\mu^2 + v_\tau^2) ] \nonumber \\ && + \frac{1}{2} \lambda_s
\lambda_t^* v_{Rs} v_{Rt} + \frac{1}{2} v_{Rs} v_{Rt} Y_\nu^{is}
(Y_\nu^{it})^*
\end{eqnarray}
The $(6 \times 2)$ matrix that mixes the charged Higgs bosons with the
charged sleptons is
\begin{equation}
M_{H \tilde l}^2 = \left( \begin{array}{c}
M_{HL}^2 \\
M_{HR}^2 \end{array} \right)
\end{equation}
with:

\begin{eqnarray}
\big( M_{HL}^2 \big)_{i1} &=& \frac{1}{4} g^2 v_d v_i - \frac{1}{2}
\lambda_s^* Y_\nu^{it} v_{Rs} v_{Rt} - \frac{1}{2}v_d \big( Y_e
Y_e^\dagger \big)_{ij} v_j \nonumber \\ \big( M_{HL}^2 \big)_{i2} &=&
\frac{1}{4} g^2 v_u v_i - \frac{1}{4} \kappa_s^* v_{Rs}^2 Y_\nu^{is} +
\frac{1}{2} v_u v_d \lambda_s^* Y_\nu^{is} -\frac{1}{2} v_u v_j
Y_\nu^{is} (Y_\nu^{js})^* \nonumber \\ && - \frac{1}{\sqrt{2}} v_{Rs} T_{\nu}^{is}
\nonumber \\ \big( M_{HR}^2 \big)_{i1} &=& -\frac{1}{2} v_u v_{Rs}
(Y_e^{ji})^* Y_\nu^{js} - \frac{1}{\sqrt{2}}v_j (T_e^{ji})^* \nonumber
\\ \big( M_{HR}^2 \big)_{i2} &=& -\frac{1}{2} \lambda_s v_{Rs} v_j
(Y_e^{ji})^* - \frac{1}{2} v_d (Y_e^{ji})^* Y_\nu^{js} v_{Rs}
\end{eqnarray}
Finally, the $(6 \times 6)$ mass matrix of the charged sleptons can be
written as
\begin{equation}
M_{\tilde l \tilde l}^2 = \left( \begin{array}{c c}
M_{LL}^2 & M_{LR}^2 \\
M_{RL}^2 & M_{RR}^2 \end{array} \right)
\end{equation}
with:

\begin{eqnarray}
\big( M_{LL}^2 \big)_{ij} &=& \big( m_{L}^2 \big)_{ij} +
\frac{1}{8} (g'^2 - g^2) (v_d^2 - v_u^2 + v_e^2 + v_\mu^2 +
v_\tau^2)\delta_{ij} + \frac{1}{4}g^2 v_i v_j \nonumber \\ && +
\frac{1}{2}v_d^2 \big( Y_e Y_e^\dagger \big)_{ij} + \frac{1}{2} v_{Rs}
v_{Rt} Y_\nu^{is} (Y_\nu^{jt})^* \nonumber \\ M_{LR}^2 &=&
-\frac{1}{2} \lambda_s^* v_{Rs} v_u Y_e + \frac{1}{\sqrt{2}} v_d T_e
\nonumber \\ M_{RL}^2 &=& \big( M_{LR}^2 \big)^\dagger \nonumber \\
\big( M_{RR}^2 \big)_{ij} &=& \big( m_{e^c}^2 \big)_{ij} +
\frac{1}{4} g'^2 (v_u^2 - v_d^2 - v_e^2 - v_\mu^2 - v_\tau^2)\delta_{ij}
\nonumber \\ && + \frac{1}{2}v_d^2 \big( Y_e^\dagger Y_e \big)_{ij} +
\frac{1}{2} v_k v_m (Y_e^{ki})^* Y_e^{mj}
\end{eqnarray}

\section*{Neutral Scalars}
\label{subsec:scalars}

In the basis

\begin{equation}
\big( {S^0}' \big)^T = Re (H_d^0, H_u^0, \tilde{\nu}_s^c, \tilde{\nu}_i)
\end{equation}
the scalar potential includes the term
\begin{equation}
V \supset \big( {S^0}' \big)^T M_{S^0}^2 {S^0}'
\end{equation}
and the $((5+n) \times (5+n))$ neutral scalar mass matrix can be written as
\begin{equation}
M_{S^0}^2 = \left(
\begin{array}{c c c}
M^2_{HH} & M^2_{HS} & M^2_{H\tilde L}\\ \big( M^2_{HS} \big)^T &
M^2_{SS} & M^2_{\tilde L S}\\ \big( M^2_{H\tilde L} \big)^T & \big(
M^2_{\tilde L S} \big)^T & M^2_{\tilde L \tilde L} \end{array} \right)\quad.
\label{eq:massscalars}
\end{equation}
The matrix elements are given as follows:

\begin{eqnarray}
\big( M_{HH}^2 \big)_{11} &=& m_{H_d}^2 + \frac{1}{8} (g^2 + g'^2)
(3v_d^2 - v_u^2 + v_e^2 + v_\mu^2 + v_\tau^2) \nonumber \\ && + \frac{1}{2}
\lambda_s \lambda_t^* v_{Rs} v_{Rt} + \frac{1}{2} v_u^2 \lambda_s
\lambda_s^* \nonumber \\ \big( M_{HH}^2 \big)_{12} &=&
-\frac{1}{4}(g^2 + g'^2)v_d v_u + \lambda_s \lambda_s^* v_d v_u -
\frac{1}{8} v_{Rs}^2 (\lambda_s \kappa_s^* + h.c. ) \nonumber \\ && -
\frac{1}{2} v_u v_i (\lambda_s^* Y_\nu^{is} + h.c.) - \frac{1}{2
\sqrt{2}} v_{Rs} (T_\lambda^s + h.c. ) \nonumber \\ \big( M_{HH}^2
\big)_{21} &=& \big( M_{HH}^2 \big)_{12} \nonumber \\ \big( M_{HH}^2
\big)_{22} &=& m_{H_u}^2 -\frac{1}{8} (g^2 + g'^2) (v_d^2 - 3v_u^2 +
v_e^2 + v_\mu^2 + v_\tau^2) \nonumber \\ && + \frac{1}{2} \lambda_s
\lambda_t^* v_{Rs} v_{Rt} + \frac{1}{2}v_d^2 \lambda_s \lambda_s^* +
\frac{1}{2} v_{Rs} v_{Rt} Y_\nu^{is} (Y_\nu^{it})^* \nonumber \\ &&  + \frac{1}{2} v_i
v_j (Y_\nu^{is})^* Y_\nu^{js} - \frac{1}{2} v_d v_i
(\lambda_s^* Y_\nu^{is} + h.c.)
\end{eqnarray}
\begin{eqnarray}
\big( M^2_{H S} \big)_{1s} &=& -\frac{1}{4} v_u v_{Rs}(\lambda_s^*
\kappa_s + h.c.) + \frac{1}{2} v_d v_{Rt} ( \lambda_s \lambda_t^* +
h.c. ) \nonumber \\ && -\frac{1}{2 \sqrt{2}} v_u (T_\lambda^s + h.c.)
- \frac{1}{4} v_i v_{Rt} ( \lambda_s^* Y_\nu^{it} + \lambda_t^*
Y_\nu^{is} + h.c. ) \nonumber \\ \big( M^2_{H S} \big)_{2s} &=&
-\frac{1}{4} v_d v_{Rs}(\lambda_s^* \kappa_s + h.c.) + \frac{1}{2} v_u
v_{Rt} ( \lambda_s \lambda_t^* + h.c. ) \nonumber \\ && -\frac{1}{2
\sqrt{2}} v_d (T_\lambda^s + h.c.) +\frac{1}{2 \sqrt{2}} v_t
(T_\nu^{ts} + h.c.)  \nonumber \\ && + \frac{1}{4} v_{Rs} v_i ( \kappa_s^*
Y_\nu^{is} + h.c. ) + \frac{1}{2} v_u v_{Rt} [ Y_\nu^{is} (Y_\nu^{it})^* + h.c. ]
\end{eqnarray}
\begin{eqnarray}
\big( M^2_{H \tilde L} \big)_{1i} &=& \frac{1}{4}(g^2 + g'^2)v_d v_i -
\frac{1}{4} v_u^2 ( \lambda_s^* Y_\nu^{is} + h.c. ) - \frac{1}{4}
v_{Rs} v_{Rt}(\lambda_s^* Y_\nu^{it} + h.c.) \nonumber \\ \big( M^2_{H
\tilde L} \big)_{2i} &=& -\frac{1}{4}(g^2 + g'^2)v_u v_i + \frac{1}{8}
v_{Rs}^2 (\kappa_s^* Y_\nu^{is} + h.c.) - \frac{1}{2} v_u v_d
(\lambda_s^* Y_\nu^{is} + h.c.) \nonumber \\ && + \frac{1}{2} v_u v_j
[Y_\nu^{js} (Y_\nu^{is})^* + h.c.] + \frac{1}{2 \sqrt{2}} v_{Rs}
(T_\nu^{is} + h.c.)
\end{eqnarray}
\begin{eqnarray}\label{SSscalar}
\big( M^2_{SS} \big)_{st} &=& \frac{1}{2} [ (m_{\nu^c}^2 )_{st}
+ (m_{\nu^c}^2 )_{ts}] + \frac{1}{4} (\lambda_s \lambda_t^* +
h.c.) (v_d^2 + v_u^2) \nonumber \\ && - \frac{1}{4} v_d v_u (\lambda_s^* \kappa_s +
h.c.) \delta_{st} \nonumber + \frac{3}{4}\kappa_s \kappa_s^*
v_{Rs}^2 \delta_{st} \nonumber \\ &&+ \frac{1}{4} v_u v_i (\kappa_s^* Y_\nu^{is} +
h.c.) \delta_{st} + \frac{1}{4} v_u^2 [(Y_\nu^{is})^* Y_\nu^{it} +
h.c.] \nonumber \\ && + \frac{1}{4} v_i v_j [(Y_\nu^{is})^* Y_\nu^{jt}
+ h.c.] - \frac{1}{4} v_d v_i [\lambda_s^* Y_\nu^{it} + \lambda_t
(Y_\nu^{is})^* + h.c.] \nonumber \\ && + \frac{1}{2 \sqrt{2}} v_{Ru}
(T_\kappa^{stu} + h.c.)
\end{eqnarray}
\begin{eqnarray}
\big( M^2_{\tilde L S} \big)_{si} &=& \frac{1}{4}v_u v_{Rs}(\kappa_s^*
Y_\nu^{is} + h.c.) - \frac{1}{4} v_d v_{Rt} (\lambda_s^* Y_\nu^{it} +
\lambda_t^* Y_\nu^{is} + h.c.) \nonumber \\ && + \frac{1}{2 \sqrt{2}}
v_u (T_\nu^{is} + h.c.) \nonumber \\
&& + \frac{1}{4} v_j v_{Rt} [ Y_\nu^{jt}
(Y_\nu^{is})^* + Y_\nu^{js} (Y_\nu^{it})^* + h.c. ]
\end{eqnarray}
\begin{eqnarray}
\big( M_{\tilde L \tilde L}^2 \big)_{ij} &=& \frac{1}{2} [
(m_L^2)_{ij} + (m_L^2)_{ji}] + \frac{1}{8}(g^2 + g'^2)(v_d^2 - v_u^2 +
v_e^2 + v_\mu^2 + v_\tau^2) \delta_{ij} \nonumber \\ && + \frac{1}{4}(g^2 +
g'^2)v_i v_j + \frac{1}{4} v_u^2 [Y_\nu^{is} (Y_\nu^{js})^* + h.c.] \nonumber \\
&& + \frac{1}{4} v_{Rs} v_{Rt} [ Y_\nu^{is} (Y_\nu^{jt})^* + h.c. ]
\end{eqnarray}

\section*{Pseudoscalars}
\label{subsec:pseudoscalars}

In the basis

\begin{equation}
\big( {P^0}' \big)^T = Im (H_d^0, H_u^0, \tilde{\nu}_s^c, \tilde{\nu}_i)
\end{equation}
the scalar potential includes the term
\begin{equation}
V \supset \big( {P^0}' \big)^T M_{P^0}^2 {P^0}'
\end{equation}
and the $((5+n) \times (5+n))$ pseudoscalar mass matrix can be written as
\begin{equation}
M_{P^0}^2 = \left(
\begin{array}{c c c}
M^2_{HH} & M^2_{HS} & M^2_{H\tilde L}\\ \big( M^2_{HS} \big)^T &
M^2_{SS} & M^2_{\tilde L S}\\ \big( M^2_{H\tilde L} \big)^T & \big(
M^2_{\tilde L S} \big)^T & M^2_{\tilde L \tilde L} \end{array} \right)\quad.
\label{eq:masspseudoscalars}
\end{equation}
The matrix elements are given as follows:
\begin{eqnarray}
\big( M_{HH}^2 \big)_{11} &=& m_{H_d}^2 + \frac{1}{8} (g^2 + g'^2)
(v_d^2 - v_u^2 + v_e^2 + v_\mu^2 + v_\tau^2) \nonumber \\ && + \frac{1}{2}
\lambda_s \lambda_t^* v_{Rs} v_{Rt} + \frac{1}{2} v_u^2 \lambda_s
\lambda_s^* \nonumber \\ \big( M_{HH}^2 \big)_{12} &=& \frac{1}{8}
v_{Rs}^2 (\lambda_s \kappa_s^* + h.c. ) + \frac{1}{2 \sqrt{2}} v_{Rs}
(T_\lambda^s + h.c. ) \nonumber \\ \big( M_{HH}^2 \big)_{21} &=& \big(
M_{HH}^2 \big)_{12} \nonumber \\ \big( M_{HH}^2 \big)_{22} &=&
m_{H_u}^2 -\frac{1}{8} (g^2 + g'^2) (v_d^2 - v_u^2 + v_e^2 + v_\mu^2 +
v_\tau^2) \nonumber \\ && + \frac{1}{2} \lambda_s \lambda_t^* v_{Rs}
v_{Rt} + \frac{1}{2}v_d^2 \lambda_s \lambda_s^* + \frac{1}{2} v_{Rs}
v_{Rt} Y_\nu^{is} (Y_\nu^{it})^* \nonumber \\ && + \frac{1}{2} v_i v_j (Y_\nu^{is})^*
Y_\nu^{js} - \frac{1}{2} v_d v_i (\lambda_s^*
Y_\nu^{is} + h.c.)
\end{eqnarray}

\begin{eqnarray}
\big( M^2_{H S} \big)_{1s} &=& -\frac{1}{4} v_u v_{Rs}(\lambda_s^*
\kappa_s + h.c.) + \frac{1}{4} \sum_{t \neq s} v_i v_{Rt} (
\lambda_s^* Y_\nu^{it} - \lambda_t^* Y_\nu^{is} + h.c. ) \nonumber \\
&& + \frac{1}{2 \sqrt{2}} v_u (T_\lambda^s + h.c.) \nonumber \\ \big(
M^2_{H S} \big)_{2s} &=& - \frac{1}{4} v_d v_{Rs}(\lambda_s^* \kappa_s
+ h.c.) + \frac{1}{4} v_{Rs} v_i ( \kappa_s^* Y_\nu^{is} + h.c. )
\nonumber \\ && + \frac{1}{2 \sqrt{2}} v_d (T_\lambda^s + h.c.) -
\frac{1}{2 \sqrt{2}} v_i (T_\nu^{is} + h.c.)
\end{eqnarray}

\begin{eqnarray}
\big( M^2_{H \tilde L} \big)_{1i} &=& - \frac{1}{4} v_u^2 (
\lambda_s^* Y_\nu^{is} + h.c. ) - \frac{1}{4} v_{Rs}
v_{Rt}(\lambda_s^* Y_\nu^{it} + h.c.) \nonumber \\ \big( M^2_{H \tilde
L} \big)_{2i} &=& - \frac{1}{8} v_{Rs}^2 (\kappa_s^* Y_\nu^{is} +
h.c.) - \frac{1}{2 \sqrt{2}} v_{Rs} (T_\nu^{is} + h.c.)
\end{eqnarray}

\begin{eqnarray}\label{SSpseudoscalar}
\big( M^2_{SS} \big)_{st} &=& \frac{1}{2} [ (m_{\nu^c}^2 )_{st}
+ (m_{\nu^c}^2 )_{ts}] + \frac{1}{4} (\lambda_s \lambda_t^* +
h.c.) (v_d^2 + v_u^2) \nonumber \\
&&+ \frac{1}{4} v_d v_u (\lambda_s^* \kappa_s +
h.c.) \delta_{st} + \frac{1}{4}\kappa_s \kappa_s^*
v_{Rs}^2 \delta_{st} \nonumber \\ &&- \frac{1}{4} v_u v_i (\kappa_s^* Y_\nu^{is} +
h.c.) \delta_{st} + \frac{1}{4} v_u^2 [(Y_\nu^{is})^* Y_\nu^{it} +
h.c.] \nonumber \\ && + \frac{1}{4} v_i v_j [(Y_\nu^{is})^* Y_\nu^{jt}
+ h.c.] - \frac{1}{4} v_d v_i [\lambda_s^* Y_\nu^{it} + \lambda_t
(Y_\nu^{is})^* + h.c.] \nonumber \\ && - \frac{1}{2 \sqrt{2}} v_{Ru}
(T_\kappa^{stu} + h.c.)
\end{eqnarray}

\begin{eqnarray}
\big( M^2_{\tilde L S} \big)_{si} &=& \frac{1}{4} v_u
v_{Rs}(\kappa_s^* Y_\nu^{is} + h.c.) + \frac{1}{4} \sum_{t \neq s} v_d
v_{Rt} (\lambda_t^* Y_\nu^{is} - \lambda_s^* Y_\nu^{it} + h.c.)
\nonumber \\ && + \frac{1}{4} \sum_{t \neq s} v_j v_{Rt} [ Y_\nu^{js}
(Y_\nu^{it})^* - Y_\nu^{jt} (Y_\nu^{is})^* + h.c. ] \nonumber \\ && - \frac{1}{2
\sqrt{2}} v_u (T_\nu^{is} + h.c.)
\end{eqnarray}

\begin{eqnarray}
\big( M_{\tilde L \tilde L}^2 \big)_{ij} &=& \frac{1}{2} [
(m_L^2)_{ij} + (m_L^2)_{ji}] + \frac{1}{8}(g^2 + g'^2)(v_d^2 - v_u^2 +
v_e^2 + v_\mu^2 + v_\tau^2) \delta_{ij} \nonumber \\ && + \frac{1}{4} v_u^2
[Y_\nu^{is} (Y_\nu^{js})^* + h.c.] + \frac{1}{4} v_{Rs} v_{Rt} [
Y_\nu^{is} (Y_\nu^{jt})^* + h.c. ]
\end{eqnarray}

\section*{Neutral Fermions}
\label{subsec:neutralinos}

In the basis

\begin{equation}
\big( \psi^0 \big)^T = \big( -i{\tilde B}^0, -i{\tilde W}_3^0, {\tilde
H}_d^0, {\tilde H}_u^0, \nu_s^c, \nu_i \big)
\end{equation}

the lagrangian of the model includes the term

\begin{equation}
{\cal L} \supset -\frac{1}{2} \big( \psi^0 \big)^T M_N \psi^0 + h.c.
\end{equation}
with the $((7+n) \times (7+n))$ mass matrix of the neutral fermions, which
can be written as:
\begin{equation}
M_N = \left( \begin{array}{c c c}
{\cal M}_{\chi^0} & m_{\tilde{\chi}^0 \nu^c} & m_{\tilde{\chi}^0 \nu} \\
\\
m_{\tilde{\chi}^0 \nu^c}^T & M_R & m_D \\
\\
m_{\tilde{\chi}^0 \nu}^T & m_D^T & 0 \end{array} \right)
\end{equation}
${\cal M}_{\chi^0}$ is the usual mass matrix of the neutralinos in the MSSM
\begin{equation}
M_{\tilde{\chi}^0} = \left( \begin{array}{c c c c}
M_1 & 0 & -\frac{1}{2}g' v_d & \frac{1}{2}g' v_u \\
0 & M_2 & \frac{1}{2}g v_d & -\frac{1}{2}g v_u \\
-\frac{1}{2}g' v_d & \frac{1}{2}g v_d & 0 & -\mu \\
\frac{1}{2}g' v_u & -\frac{1}{2}g v_u & -\mu & 0 \end{array} \right)
\end{equation}
with

\begin{equation}
\mu = \frac{1}{\sqrt{2}} \lambda_s v_{Rs}\quad.
\end{equation}
The mixing between the neutralinos and the singlet $\nu_s^c$ is given by

\begin{equation}
(m_{\tilde{\chi}^0 \nu^c}^T)_s = \left( \begin{array}{c c c c}
0 & 0 & -\frac{1}{\sqrt{2}} \lambda_s v_u & -\frac{1}{\sqrt{2}}\lambda_s v_d + \frac{1}{\sqrt{2}} v_i Y_\nu^{is} \end{array} \right)\quad.
\end{equation}
$m_{\tilde{\chi}^0 \nu}$ is the neutralino-neutrino mixing part
\begin{equation}
m_{\tilde{\chi}^0 \nu}^T = \left( \begin{array}{c c c c}
-\frac{1}{2} g' v_e & \frac{1}{2} g v_e & 0 & \epsilon_e  \\
-\frac{1}{2} g' v_\mu & \frac{1}{2} g v_\mu & 0 &  \epsilon_\mu  \\
-\frac{1}{2} g' v_\tau & \frac{1}{2} g v_\tau & 0 &  \epsilon_\tau \end{array} \right)
\end{equation}
with

\begin{equation}\label{defepstot}
\epsilon_i =  \frac{1}{\sqrt{2}} \sum_{s=1}^n v_{Rs} Y_\nu^{is}\quad.
\end{equation}
The neutrino Dirac term is

\begin{equation}
(m_D)_{is} = \frac{1}{\sqrt{2}} Y_\nu^{is} v_u
\end{equation}
and finally $M_R$ is 

\begin{equation}
(M_R)_{st} = \frac{1}{\sqrt{2}} \kappa_s v_{Rs} \delta_{st}\quad.
\end{equation}

\section*{Charged Fermions}
\label{subsec:charginos}

In the basis
\begin{align}
\nonumber
\left(\psi^-\right)^T&=\left(\tilde{W}^-,\tilde{H}_d^-,e,\mu,\tau\right)\\
\left(\psi^+\right)^T&=\left(\tilde{W}^+,\tilde{H}_u^+,e^c,\mu^c,\tau^c\right),
\end{align}
the $(5 \times 5)$ mass matrix of the charged fermions is given by

\begin{equation}
M_c = \left( \begin{array}{c c c c c}
M_2 & \frac{1}{\sqrt{2}}gv_u & 0 & 0 & 0 \\
\frac{1}{\sqrt{2}}gv_d & \mu & -\frac{1}{\sqrt{2}}Y_e^{i1}v_i & -\frac{1}{\sqrt{2}}Y_e^{i2}v_i & -\frac{1}{\sqrt{2}}Y_e^{i3}v_i \\
\frac{1}{\sqrt{2}}gv_e & -\epsilon_e & \frac{1}{\sqrt{2}}Y_e^{11}v_d & 0 & 0 \\
\frac{1}{\sqrt{2}}gv_\mu & -\epsilon_\mu & 0 & \frac{1}{\sqrt{2}}Y_e^{22}v_d &0 \\
\frac{1}{\sqrt{2}}gv_\tau & -\epsilon_\tau & 0 & 0 & \frac{1}{\sqrt{2}}Y_e^{33}v_d \end{array} \right)\quad.
\end{equation}

\chapter[$\mu \nu$SSM: 1-loop corrections]{$\mu \nu$SSM: Singlet scalar/pseudoscalar 1-loop contributions to the neutrino mass matrix}
\label{munuapp3}

\begin{figure}[t]
\begin{center}
\includegraphics[width=0.6\textwidth]{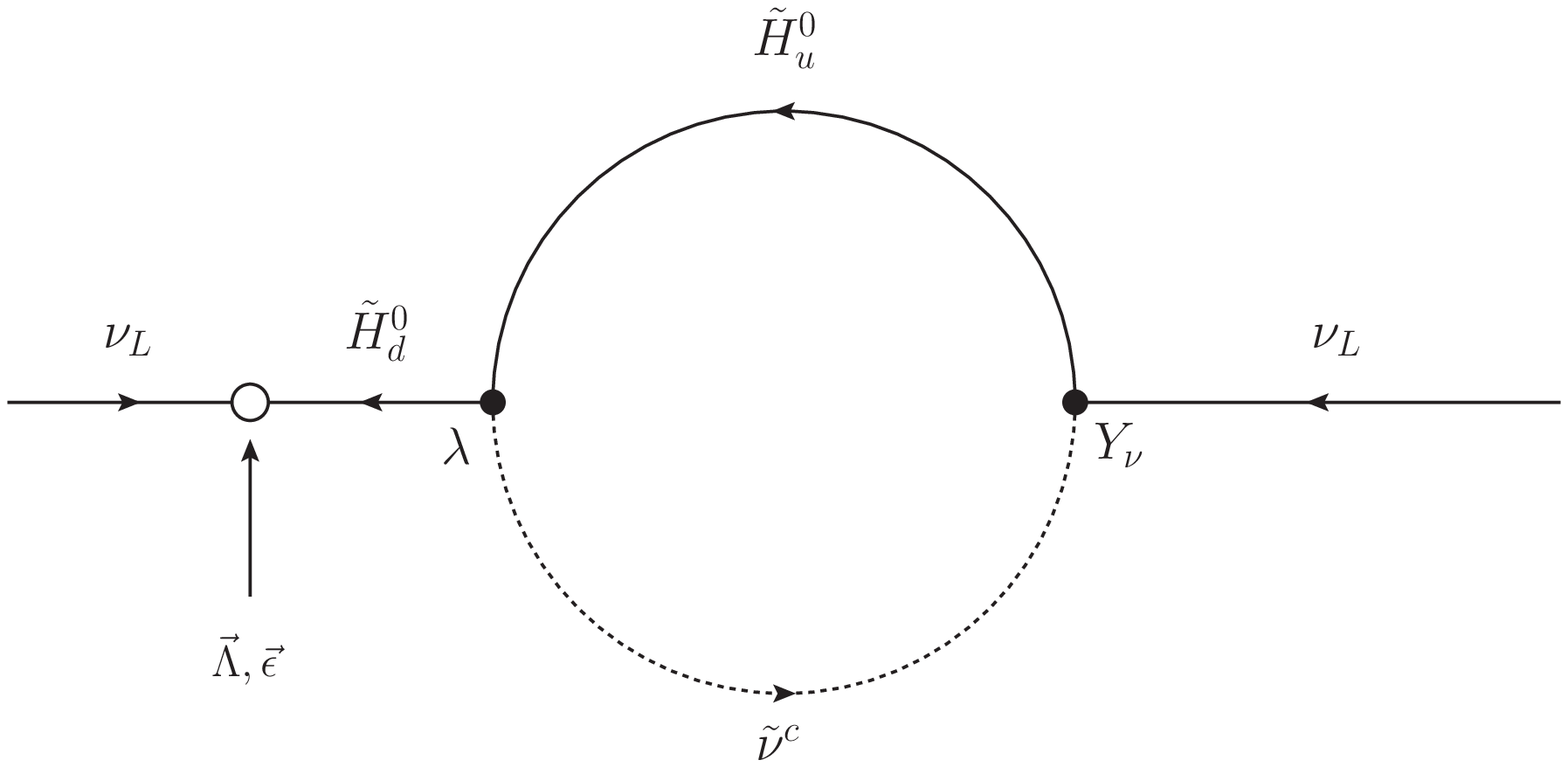}
\caption{1-loop correction to the effective neutrino mass matrix
involving the singlet scalar/pseudoscalar.}
\label{loopdiag-app}
\end{center}
\end{figure}

An estimate of the 1-loop correction to the effective neutrino mass matrix coming from the singlet scalar/pseudoscalar contribution in figure \ref{loopdiag-app} is given in this appendix.

\section*{1 $\widehat{\nu}^c$ model}

The tree-level neutrino mass matrices can be written as

\begin{equation}
m_{ij}^{tree} = a \Lambda_i \Lambda_j
\end{equation}

where $a$ and $\Lambda_i$ are defined in section \ref{subsec:1genneut}. After diagonalizing the tree-level mass matrix and adding (only) the corrections due to loops with right-handed sneutrinos $\tilde{\nu}^c$, we obtain

\begin{eqnarray}
m_{ij}^{1-loop} &=& (m_{ij}^{tree})^{diag} + \Delta M_{ij} \nonumber \\
 &=& (a + a') \tilde{\Lambda}_i \tilde{\Lambda}_j \nonumber
\end{eqnarray}

where

\begin{equation}
a' = - \frac{\Delta_{12}}{32 \pi^2} \sum_{k = \chi} m_k (R_k)^2 C_0(k)
\end{equation}

The $\tilde{\Lambda}_i$ parameters are the alignment parameters $\Lambda_i$ rotated with the matrix $U_\nu$ that diagonalizes the tree-level neutrino mass matrix:

\begin{equation}
\tilde{\Lambda}_i = (U_\nu^T)_{ij} \Lambda_j = (U_\nu)_{ji} \Lambda_j
\end{equation}

$\Delta_{12}$ is the squared mass difference of the two right-handed sneutrino mass eigenstates $\tilde{\nu}^{c R}$ and $\tilde{\nu}^{c I}$ \cite{Hirsch:1997vz,Hirsch:1997dm,Grossman:1997is}:

\begin{eqnarray}
\Delta_{12} &=& m_R^2 - m_I^2 \nonumber \\
 &=& 2 (k^2 v_R^2 - k \lambda v_u v_d + \frac{1}{3} T_k v_R) \nonumber
\end{eqnarray}

$R_k$ is related to the mixing and couplings of the neutralino $k$

\begin{eqnarray}
R_k = \frac{\lambda m_\gamma}{8 \mu {\rm Det}(M_H)} & \big[ & \lambda^2 v^2 (v_d N_{k4} + v_u N_{k3}) + 2 M_R \mu (v_d N_{k3} + v_u N_{k4}) \nonumber \\
 & - & \sqrt{2} k \mu (v_u^2 - v_d^2) N_{k5} \big]
\end{eqnarray}

where $N_{ij}$ is the matrix that diagonalizes the $5 \times 5$ mass matrix of the heavy neutralinos. Finally, $C_0(k)$, is a Passarino-Veltman function \cite{Passarino:1978jh}

\begin{equation}
C_0(k) \equiv C_0 (0,0,0,m_k^2,m_R^2,m_I^2)
\end{equation}

In order to compare this loop correction with the tree-level value, it is interesting to define the ratio

\begin{equation}
R = \frac{a'}{a}
\end{equation}

which can be expressed in good approximation as

\begin{equation}
R \simeq - \frac{\lambda^2 m_\gamma M_R^2}{32 \pi^2 \textnormal{Det}_0} \sum_{k = \chi} m_k (v_d N_{k3} + v_u N_{k4})^2 C_0(k)
\end{equation}

where $\textnormal{Det}_0$ is the determinant of the usual MSSM neutralino mass matrix.

Now, using this expression, it is possible to understand the importance of this loop correction to the neutrino mass matrix. Since

\begin{equation}
\frac{m_\gamma}{\textnormal{Det}_0} \sim \frac{1}{\mu^2 M_{gaugino}}
\end{equation}

the ratio $R$ is proportional to

\begin{equation}
R \propto \frac{\lambda^2 m_\gamma M_R^2}{\textnormal{Det}_0} \propto \frac{\lambda^2 M_R^2}{\mu^2} \propto k^2
\end{equation}

If we stay in the perturbative regime, the superpotential coupling $k$ is at most $O(1)$ and therefore this loop correction cannot be very large. This conclusion has been checked numerically.

\section*{2 $\widehat{\nu}^c$ model}

In this case we have contributions from loops with $\tilde{\nu}_1^c$ and $\tilde{\nu}_2^c$:

\begin{equation}
m_{ij}^{1-loop} = (m_{ij}^{tree})^{diag} + \Delta M_{ij}^{(1)} + \Delta M_{ij}^{(2)}
\end{equation}

with

\begin{equation}
\Delta M_{ij}^{(n)} = - \frac{\Delta_{12}^{(n)}}{32 \pi^2} \sum_{k = \chi} m_k R_{jk}^{(n)} R_{ki}^{(n)} C_0^{(n)}(k)
\end{equation}

The definitions of these parameters are similar to those shown in the previous section, but the expressions are different. In fact, the contributions from both right-handed sneutrinos take the same form, just exchanging the indices. Then, we give only the formulas for the first right-handed sneutrino, omitting the index in the notation.

The squared mass difference of the two right-handed sneutrino mass eigenstates is

\begin{equation}
\Delta_{12} = 2 (k_1^2 v_{R1}^2 - k_1 \lambda^1 v_u v_d + \frac{1}{3} T_k^{111} v_{R1} + T_k^{112} v_{R2})
\end{equation}

The factors $R_{ik}$ can be written as

\begin{equation}
R_{ik} = C_{\epsilon_1}^k \epsilon_1^i + C_{\epsilon_2}^k \epsilon_2^i + C_{\Lambda}^k \Lambda_i
\end{equation}

with

\begin{eqnarray}
C_{\epsilon_1}^k &=& \frac{\lambda^2 v_u}{\mu m_\gamma v_{R1}(v_d^2 - v_u^2)} \big[- \sqrt{2} b \mu \lambda^1 (\Delta_{du} N_{k3} + \Delta_{ud} N_{k4}) + \frac{v_{R2}}{v_u} m_\gamma (v_d^2 - v_u^2) N_{k4} \nonumber \\
 && + 2 \sqrt{2} M_{R1} \big( \frac{\lambda^2}{v_{R1}} c \mu m_\gamma (v_d^2 - v_u^2) - 2 \sqrt{2} b \textnormal{Det}_0 \big) N_{k5} \big] \\
C_{\epsilon_2}^k &=& - \frac{\mu_1}{\mu_2} C_{\epsilon_1}^k \\
C_{\Lambda}^k &=& \frac{\lambda^1}{8 \mu {\rm Det}(M_H)} \big[\big( 4 a \mu {\rm Det}(M_H) + M_{R1} M_{R2} m_\gamma \mu \big)(v_u N_{k4} - v_d N_{k3}) \nonumber \\
 && + m_\gamma \big((\lambda^1)^2 M_{R2} + (\lambda^2)^2 M_{R1} \big)(v_d^3 N_{k4} - v_u^3 N_{k3}) \nonumber \\
 && + \sqrt{2} k_1 \mu m_\gamma M_{R2}(v_d^2 - v_u^2) N_{k5} \big]
\end{eqnarray}

where $\mu_i = \frac{1}{\sqrt{2}} \lambda_i v_{R i}$, the coefficients $a$, $b$ and $c$ are given in section \ref{subsec:ngenneut} and we have defined

\begin{eqnarray}
\Delta_{du} &=& m_\gamma v^2 v_d - 4 M_1 M_2 \mu v_u \\
\Delta_{ud} &=& m_\gamma v^2 v_u - 4 M_1 M_2 \mu v_d
\end{eqnarray}

The connection between $C_{\epsilon_1}^k$ and $C_{\epsilon_2}^k$ allows us to prove that the determinant of the 1-loop mass matrix vanishes, implying that after adding this loop correction there is still one zero eigenvalue. In conclusion, this loop does not generate a new mass scale.

The complicated structure of the resulting expressions does not allow to find an approximated formula to get a clue about the importance of the loop. Numerically it has been shown that for low values of $v_{R1}$ and $v_{R2}$, below 1-5 TeV, the corrections to $m_{\nu_2}$ and $m_{\nu_3}$ are typically of order $\sim 10^{-3} m_{\nu_{2,3}}^{tree}$, with a few points in parameter space where it reaches $\sim 10^{-2} m_{\nu_{2,3}}^{tree}$. For higher values of $v_{R1}$ and $v_{R2}$ the model approaches the explicit limit (b-\rpv model): the corrections for $m_{\nu_3}$ are still small, but the corrections to $m_{\nu_2}$ are very important, since the tree-level value goes to zero in this limit.

\chapter{$\mu \nu$SSM: Coupling $\tilde{\chi}_1^0-W^{\pm}-l^{\mp}_i$}
\label{munuapp4}

Approximate formulas for the coupling
$\tilde{\chi}_1^0-W^{\pm}-l^{\mp}_i$ can be obtained from the general
$\tilde{\chi}_i^0-W^{\pm}-\tilde{\chi}_j^{\mp}$ interaction lagrangian
\begin{equation}
\mathcal{L} \supset \overline{\tilde{\chi}_i^-} \gamma^{\mu} \big( O_{Lij}^{cnw} P_L + O_{Rij}^{cnw} P_R \big) \tilde{\chi}_j^0 W_\mu^- + \overline{\tilde{\chi}_i^0} \gamma^{\mu} \big( O_{Lij}^{ncw} P_L + O_{Rij}^{ncw} P_R \big) \tilde{\chi}_j^- W_\mu^+\quad,
\end{equation}
where
\begin{eqnarray}\label{Oexact}
O_{Li1}^{cnw} &=& g\left[-{\cal U}_{i1}{\cal N}_{12}^*-\frac{1}{\sqrt{2}}\left({\cal U}_{i2}{\cal N}_{13}^*+\sum_{k=1}^3 {\cal U}_{i,2+k}{\cal N}_{1,5+k}^*\right)\right] \nonumber \\
O_{Ri1}^{cnw} &=& g\left(-{\cal V}_{i1}^*{\cal N}_{12}+\frac{1}{\sqrt{2}}{\cal V}_{i2}^*{\cal N}_{14}\right) \nonumber \\
O_{L1j}^{ncw} &=& \left(O_{Lj1}^{cnw}\right)^* \nonumber\\
O_{R1j}^{ncw} &=& \left(O_{Rj1}^{cnw}\right)^* \quad.
\end{eqnarray}
The matrix ${\cal N}$ diagonalizes the neutral fermion mass matrix
while the matrices ${\cal U}$ and ${\cal V}$ diagonalize the charged
fermion mass matrix, see appendix \ref{munuapp2}.

As  was already mentioned for the case of neutral fermions in
section \ref{subsec:neutrinomass}, it is possible to diagonalize the
mass matrices in very good approximation due to the fact that the \rpv
parameters are small. Defining the matrices $\xi$, $\xi_L$ and
$\xi_R$, that will be taken as expansion parameters, one gets the
leading order expressions
\begin{align}
{\cal N}=\begin{pmatrix}N&N\xi^T\\-U_\nu^T\xi&U_\nu^T\end{pmatrix},\qquad {\cal U}=\begin{pmatrix}U_c&U_c\xi_L^T\\-\xi_L&I_3\end{pmatrix},\qquad {\cal V}=\begin{pmatrix}V_c&V_c\xi_R^T\\-\xi_R&I_3\end{pmatrix}\quad,
\end{align}
where $I_3$ is the $(3 \times 3)$ identity matrix. The expansion
matrices $\xi_L$ and $\xi_R$ are 
\begin{eqnarray}
\left(\xi_L\right)_{i1} &=& \frac{g\Lambda_i}{\sqrt{2} \textnormal{Det}_+} \nonumber \\
\left(\xi_L\right)_{i2} &=& -\frac{\epsilon_i}{\mu}-\frac{g^2v_u\Lambda_i}{2\mu \textnormal{Det}_+} \nonumber \\
\left(\xi_R\right)_{i1} &=& \frac{gv_d Y_e^{ii}}{2 \textnormal{Det}_+}\left[\frac{v_u\epsilon_i}{\mu}+\frac{\left(2\mu^2+g^2v_u^2\right)\Lambda_i}{2\mu \textnormal{Det}_+}\right] \nonumber \\
\left(\xi_R\right)_{i2} &=& -\frac{\sqrt{2}v_d Y_e^{ii}}{2 \textnormal{Det}_+}\left[\frac{M_2\epsilon_i}{\mu}+\frac{g^2\left(v_d\mu +M_2v_u\right)\Lambda_i}{2\mu \textnormal{Det}_+}\right]\quad,
\end{eqnarray}
where $\textnormal{Det}_+=-\frac{1}{2}g^2v_dv_u+M_2\mu$ is the determinant of the
MSSM chargino mass matrix, $\mu = \frac{1}{\sqrt{2}}\lambda_s v_{Rs}$
and $\epsilon_i = \frac{1}{\sqrt{2}} v_{Rs} Y_\nu^{is}$.
The expressions for the matrix $\xi$ depend on the number of
singlet generations in the model. Particular cases can be found in
\eqref{xi} and \eqref{xi2}.

Using the previous equations and assuming that all parameters are real
, one gets the approximate formulas
\begin{eqnarray}\label{Oapprox}
O_{Li1}^{cnw} &=& \frac{g}{\sqrt{2}} \big[ \frac{g N_{12} \Lambda_i}{\textnormal{Det}_+}-\big( \frac{\epsilon_i}{\mu}+\frac{g^2 v_u \Lambda_i}{2 \mu \textnormal{Det}_+} \big)N_{13}-\sum_{k=1}^{n} N_{1k} \xi_{ik} \big] \nonumber \\
O_{Ri1}^{cnw} &=& \frac{1}{2} g (Y_e)^{ii} \frac{v_d}{\textnormal{Det}_+} \big[ \frac{g v_u N_{12} - M_2 N_{14}}{\mu} \epsilon_i \nonumber \\
&& \hskip25mm +\frac{g(2 \mu^2 +g^2 v_u^2)N_{12}-g^2 (v_d \mu +M_2v_u)N_{14}}{2 \mu \textnormal{Det}_+} \Lambda_i \big] \nonumber \\
O_{Li1}^{ncw} &=& \big(O_{Li1}^{cnw}\big)^* \nonumber\\
O_{Ri1}^{ncw} &=& \big(O_{Ri1}^{cnw}\big)^*\quad.
\end{eqnarray}
It is important to emphasize that all previous formulas, and the
following simplified versions, are tree-level results. More
simplified formulas are possible if the lightest neutralino has a
large component in one of the gauge eigenstates. These particular
limits are of great interest to understand the phenomenology:

\section*{Bino-like $\tilde{\chi}_1^0$}

This limit is characterized by $N_{11}^2 = 1$ and $N_{1m} = 0$ for $m
\neq 1$. One gets
\begin{eqnarray}
O_{Li1}^{cnw} &=& - \frac{g}{\sqrt{2}} \xi_{i1} \nonumber \\
O_{Ri1}^{cnw} &=& 0\quad.
\end{eqnarray}
For the $1$ $\widehat{\nu}^c$-model this implies that a bino-like
$\tilde{\chi}_1^0$ couples to $W l_i$ proportionally to $\Lambda_i$, see
equation \eqref{xi}, without any dependence on the $\epsilon_i$
parameters.

On the other hand, for the $2$ $\widehat{\nu}^c$-model, the more complicated
structure of the $\xi$ matrix, see equations \eqref{xi2} and
\eqref{defK2}, implies a coupling of a bino-like $\tilde{\chi}_1^0$
with $W l_i$ dependent on two pieces, one proportional to $\Lambda_i$ and
one proportional to $\alpha_i$:
\begin{equation}
\xi_{i1} = \frac{2 g' M_2 \mu}{m_\gamma} (a \Lambda_i + b \alpha_i)
\end{equation}

However, a simple estimate of the relative importance of
these two terms is possible. By assuming that all masses are at the same scale
$m_{SUSY}$, the couplings $\kappa$ and $\lambda$ are of order $0.1$,
and the \rpv terms $Y_\nu^i$ and $v_i$ are of order $h_{\text{\rpv}}$ and
$m_{SUSY} h_{\text{\rpv}}$ respectively, one can show that $a \Lambda_i \sim
200 \: b \alpha_i$. Therefore, one gets a coupling which is
proportional, in very good approximation, to $\Lambda_i$, as confirmed by
the exact numerical results shown in the main part of the paper.
Similar arguments apply for models with more generations of right-handed neutrinos.

In conclusion, for a bino-like neutralino the coupling
$\tilde{\chi}_1^0-W^{\pm}-l^{\mp}_i$ is  proportional to
$\Lambda_i$ to a good approximation.

\section*{Higgsino-like $\tilde{\chi}_1^0$}

This limit is characterized by $N_{13}^2 + N_{14}^2 = 1$ and $N_{1m} =
0$ for $m \neq 3,4$. If the coupling $O_{Ri1}^{cnw}$ is neglected due
to the supression given by the charged lepton Yukawa couplings, one
gets
\begin{eqnarray}\label{Ohiggs}
O_{Li1}^{cnw} &=& - \frac{g}{\sqrt{2}} \left[ \big( \frac{\epsilon_i}{\mu}+\frac{g^2 v_u \Lambda_i}{2 \mu \textnormal{Det}_+} + \xi_{i3} \big)N_{13} + \xi_{i4} N_{i4} \right] \nonumber \\
O_{Ri1}^{cnw} &\simeq& 0\quad.
\end{eqnarray}
Equations \eqref{xi} and \eqref{xi2} show that the $\epsilon_i$ terms
cancel out in the coupling \eqref{Ohiggs}, and therefore one gets
dependence only on $\Lambda_i$ in the $1$ $\widehat{\nu}^c$-model, and
$(\Lambda_i , \alpha_i)$ in the $2$ $\widehat{\nu}^c$-model.
However, this cancellation is not perfect in $O_{Ri1}^{cnw}$ and thus
one still has some dependence on $\epsilon_i$. 

\section*{Singlino-like $\tilde{\chi}_1^0$}

The limit in which the right-handed neutrino $\nu_s^c$ is the lightest
neutralino is characterized by $N_{1m}^2 = 1$ for $m \ge 5$ 
and $N_{1l} = 0$ for $l \neq m$. One gets

\begin{eqnarray}\label{Osinglino}
O_{Li1}^{cnw} &=& - \frac{g}{\sqrt{2}} \xi_{im} \nonumber \\
O_{Ri1}^{cnw} &=& 0\quad.
\end{eqnarray}

For the $1$ $\widehat{\nu}^c$-model this expression implies that a
pure singlino-like $\tilde{\chi}_1^0$ couples to $W l_i$
proportional to $\Lambda_i$, see equation \eqref{xi}, without any
dependence on the $\epsilon_i$ parameters.
This proportionality to $\Lambda_i$ is different to what is found in
spontaneous R-parity violation, where the different structure of the
corresponding $\xi$ matrix \cite{Hirsch:2008ur} implies that the
singlino couples to $W l_i$ proportionally to $\epsilon_i$.

For the $n$ $\widehat{\nu}^c$-model one finds that the coupling
$\tilde{\chi}_1^0-W^{\pm}-l^{\mp}_i$ for a singlino-like neutralino
has little dependence on $\Lambda_i$. For example, in the $2$
$\widehat{\nu}^c$-model one finds that the element $\xi_{i5}$,
corresponding to the right-handed neutrino $\nu_1^c$, is given by
\begin{equation}\label{singcoup}
\xi_{i5} = \frac{M_{R2} \lambda_1 m_\gamma}{4 \sqrt{2} {\rm Det}(M_H)} (v_u^2 - v_d^2) \Lambda_i - \left( \sqrt{2} \lambda_2 c + \frac{4 \textnormal{Det}_0 v_{R1}}{\mu m_\gamma (v_u^2 - v_d^2)}b \right) \alpha_i\quad.
\end{equation}
The coupling has two pieces, one proportional to $\Lambda_i$ and one
proportional to $\alpha_i$. However, the $\alpha_i$ piece gives the
dominant contribution, as can be shown using an estimate completely
analogous to the one done for a bino-like $\tilde{\chi}_1^0$. In this
case, the ratio between the two terms in equation \eqref{singcoup} is
$\alpha_i$-piece $\sim$ $8 \: \Lambda_i$-piece, sufficient to ensure a
very good proportionality to the $\alpha_i$ parameters. This estimate
has been corroborated numerically.

\chapter{SUSYLR: Renormalization Group Equations}
\label{susylrapp}

We present in the following appendices our results for the RGEs of the model above the $U(1)_{B-L}$ breaking scale. We will only show the \(\beta\)-functions for the gauge couplings and the anomalous dimensions of all chiral superfields. We briefly discuss in this appendix how these results were calculated. Furthermore, we show how they can be used to calculate the other \(\beta\)-functions of the models and give as example the 1-loop results for the soft SUSY breaking masses of the sleptons. The complete results are given online on this site
\begin{verbatim}
http://theorie.physik.uni-wuerzburg.de/~fnstaub/supplementary.html
\end{verbatim}
In addition, the corresponding model files for SARAH are also given on this web page.

\section*{Calculation of supersymmetric RGEs}
\label{sec:genericRGEs}
For a  general $N=1$ supersymmetric gauge theory with superpotential  
\begin{equation}
 W (\phi) = \frac{1}{2}{\mu}^{ij}\phi_i\phi_j + \frac{1}{6}Y^{ijk}
\phi_i\phi_j\phi_k
\end{equation}
the  soft SUSY-breaking scalar terms are given by
\begin{equation}
V_{\hbox{soft}} = \left(\frac{1}{2}b^{ij}\phi_i\phi_j
+ \frac{1}{6}h^{ijk}\phi_i\phi_j\phi_k +\hbox{c.c.}\right)
+(m^2)^i{}_j\phi_i\phi_j^* \thickspace.
\end{equation}
The anomalous dimensions are given by \cite{Martin:1993zk}
\begin{align}
 \gamma_i^{(1)j} = & \frac{1}{2} Y_{ipq} Y^{jpq} - 2 \delta_i^j g^2 C_2(i) \thickspace, \\
 \gamma_i^{(2)j}  = &  -\frac{1}{2} Y_{imn} Y^{npq} Y_{pqr} Y^{mrj} + g^2 Y_{ipq} Y^{jpq} [2C_2(p)- C_2(i)] \nonumber \\
 & \; \;  + 2 \delta_i^j g^4 [ C_2(i) S(R)+ 2 C_2(i)^2 - 3 C_2(G) C_2(i)] \thickspace,
\end{align}
and the \(\beta\)-functions for the gauge couplings are given by
\begin{align}
 \beta_g^{(1)}  =  & g^3 \left[S(R) - 3 C_2(G) \right] \thickspace,\\
 \beta_g^{(2)}  =  & g^5 \left\{ - 6[C_2(G)]^2 + 2 C_2(G) S(R) + 4 S(R) C_2(R) \right\}
    - g^3 Y^{ijk} Y_{ijk}C_2(k)/d(G) \thickspace .
\end{align}
Here, \(C_2(i)\) is the quadratic Casimir for a specific superfield and $C_2(R),C_2(G)$ are the quadratic Casimirs for the matter and adjoint  representations, respectively. \(d(G)\) is the dimension of the adjoint representation.  \\
The $\beta$-functions for the superpotential parameters can be obtained by using superfield technique. The obtained expressions are \cite{West:1984dg,Jones:1984cx}. 
\begin{eqnarray}
 \beta_Y^{ijk} &= & Y^{p(ij} {\gamma_p}^{k)} \thickspace, \\
 \beta_{\mu}^{ij} &= & \mu^{p(i} {\gamma_p}^{j)} \thickspace .
\end{eqnarray}
The $(..)$ in the superscripts denote symmetrization. Most of the \(\beta\)-functions of the models can be derived from these results using the procedure given in \cite{Jack:1997eh} based on the spurion formalism \cite{Yamada:1994id}. In the following, we briefly summarize the basic ideas of this calculation for completeness. \\

The  exact results for the soft $\beta$-functions are given by \cite{Jack:1997eh}:
\begin{eqnarray}
\label{eq:betaM}
\beta_M &=& 2{\cal O} \left[\frac{\beta_g}{g}\right] \thickspace, \\
\beta_{h}^{ijk} &=& h{}^{l(jk}\gamma^{i)}{}_l -
2Y^{l(jk}\gamma_1{}^{i)}{}_l \thickspace, \cr
\beta_{b}^{ij} &=&   
b{}^{l(i}\gamma^{j)}{}_l-2\mu{}^{l(i}\gamma_1{}^{j)}{}_l \thickspace,\cr
\left(\beta_{m^2}\right){}^i{}_j &=& \Delta\gamma^i{}_j \thickspace.
\label{eq:betam2}
\end{eqnarray}
where we defined
\begin{eqnarray}
{\cal O}  &=& Mg^2\frac{\partial}{\partial g^2}-h^{lmn}
\frac{\partial}{\partial Y^{lmn}} \thickspace, \\
(\gamma_1)^i{}_j  &=& {\cal O}\gamma^i{}_j \thickspace, \\
\Delta &=& 2{\cal O} {\cal O}^* +2MM^* g^2 \frac{\partial}{\partial g^2}
+\left[{\tilde Y}^{lmn} \frac{\partial}{\partial Y^{lmn}} + \hbox{c.c.}\right]
+X \frac{\partial}{\partial g} \thickspace.
\end{eqnarray}
Here, $M$ is the gaugino mass and  
${\tilde Y}^{ijk} = (m^2)^i{}_lY^{jkl} +  (m^2)^j{}_lY^{ikl} + (m^2)^k{}_lY^{ijl}.$
Eqs. \eqref{eq:betaM} - \eqref{eq:betam2} hold in a class of renormalization schemes that includes the DRED$'$-one \cite{Jack:1994rk}. We take the known contributions of $X$ from \cite{Jack:1998iy}:
\begin{eqnarray}
X^{\mathrm{DRED}'(1)}&=&-2g^3S \thickspace, \\
X^{\mathrm{DRED}'(2)}&=& (2r)^{-1}g^3 \mathrm{tr} [ W C_2(R)]
-4g^5C_2(G)S-2g^5C_2(G)QMM^* \end{eqnarray}
where
\begin{eqnarray}
S &=&  r^{-1} \mathrm{tr}[m^2C_2(R)] -MM^* C_2(G) \thickspace,  \\
W^j{}_i&=&{1\over2}Y_{ipq}Y^{pqn}(m^2)^j{}_n+ \frac{1}{2}Y^{jpq}Y_{pqn}(m^2)^n{}_i
+2Y_{ipq}Y^{jpr}(m^2)^q{}_r 
+h_{ipq}h^{jpq} \nonumber \\
&& -8g^2MM^*C_2(R)^j{}_i \thickspace .
\nonumber \\
\end{eqnarray}
With $Q = T(R) - 3C_2(G)$, and $T(R) = \mathrm{tr} \left[C_2(R)\right]$, $r$ being  the number of group generators. \\

\section*{From GUT scale to $SU(2)_R$ breaking scale}
In the following sections we will use the definitions
\begin{equation}
 Y^{ij}_{Q_k} = Y^{ijk}_Q \thickspace, \hspace{1cm} Y^{ij}_{L_k} = Y^{ijk}_L
\end{equation}
and in the same way \(T^{ij}_{Q_k}\) and \(T^{ij}_{L_k}\). We will also assume summation of repeated indices.

\subsection*{Anomalous Dimensions}
\label{sec:Ana1}
{\allowdisplaybreaks 
\begin{align} 
\gamma_{\hat{Q}}^{(1)} & = 
2 Y_{Q_k}^{*}  Y_{Q_k}^{T} -\frac{1}{12} \Big(18 g_{2}^{2}  + 32 g_{3}^{2}  + g_{BL}^{2}\Big){\bf 1} \\ 
\gamma_{\hat{Q}}^{(2)} & =  
+\frac{1}{144} \Big(-128 g_{3}^{4}  + 2052 g_{2}^{4}  + 289 g_{BL}^{4}  + 36 g_{2}^{2} \Big(32 g_{3}^{2}  + g_{BL}^{2}\Big) + 64 g_{3}^{2} g_{BL}^{2} \Big){\bf 1} \nonumber \\ 
 &+ Y_{Q_m}^{*} \Big( 6 g_2^2 \delta_{mn} + \frac{27}{4} \mbox{Tr}\Big({\alpha  \alpha^*}\Big) \delta_{mn} -2 \mbox{Tr}\Big({Y_{L_n}^{*}  Y_{L_m}^{T}}\Big) - 6 \mbox{Tr}\Big({Y_{Q_n}^{*}  Y_{Q_m}^{T}}\Big) \Big) Y_{Q_n}^{T} \nonumber \\ 
& -32 Y_{Q_m}^{*} Y_{Q_n}^\dagger Y_{Q_n} Y_{Q_m}^{T} \\
\gamma_{\hat{Q}^c}^{(1)} & = 
2 Y_{Q_k}^{\dagger}  Y_{Q_k} -\frac{1}{12} \Big(18 g_{2}^{2}  + 32 g_{3}^{2}  + g_{BL}^{2}\Big){\bf 1} \\ 
\gamma_{\hat{Q}^c}^{(2)} & =  
+\frac{1}{144} \Big(-128 g_{3}^{4}  + 2052 g_{2}^{4}  + 289 g_{BL}^{4}  + 36 g_{2}^{2} \Big(32 g_{3}^{2}  + g_{BL}^{2}\Big) + 64 g_{3}^{2} g_{BL}^{2} \Big){\bf 1} \nonumber \\ 
 &+ Y_{Q_m}^{\dagger} \Big( 6 g_2^2 \delta_{mn} + \frac{27}{4} \mbox{Tr}\Big({\alpha  \alpha^*}\Big) \delta_{mn} -2 \mbox{Tr}\Big({Y_{L_n}^{*}  Y_{L_m}^{T}}\Big) - 6 \mbox{Tr}\Big({Y_{Q_n}^{*}  Y_{Q_m}^{T}}\Big) \Big) Y_{Q_n} \nonumber \\ 
& -32 Y_{Q_m}^{\dagger} Y_{Q_n} Y_{Q_m}^T Y_{Q_n}^{*} \\
\gamma_{\hat{L}}^{(1)} & =  
2 \Big(3 {f^\dagger  f}  + {Y_{L_k}^{*}  Y_{L_k}^{T}}\Big) -\frac{3}{4} \Big(2 g_{2}^{2}  + g_{BL}^{2}\Big){\bf 1} \\ 
\gamma_{\hat{L}}^{(2)} & =  
\frac{3}{16} \Big(12 g_{2}^{2} g_{BL}^{2}  + 76 g_{2}^{4}  + 99 g_{BL}^{4} \Big){\bf 1} \nonumber \\ &+3 {f^\dagger  f} \Big(-3 |a|^2  -4 \mbox{Tr}\Big({f  f^\dagger}\Big)  + 6 g_{BL}^{2}  + 8 g_{2}^{2} \Big)\nonumber \\ 
&+ Y_{L_m}^{*} \Big( 6 g_2^2 \delta_{mn} + \frac{27}{4} \mbox{Tr}\Big({\alpha  \alpha^*}\Big) \delta_{mn} -2 \mbox{Tr}\Big({Y_{L_n}^{*}  Y_{L_m}^{T}}\Big) - 6 \mbox{Tr}\Big({Y_{Q_n}^{*}  Y_{Q_m}^{T}}\Big) \nonumber \\
& - 11 f^\dagger f \delta_{mn} \Big) Y_{L_n}^{T} -4 Y_{L_m}^{*} Y_{L_n}^\dagger Y_{L_n} Y_{L_m}^{T} - 2 f^\dagger \Big( 17 f f^\dagger + 3 Y_{L_k} Y_{L_k}^\dagger \Big) f \nonumber \\
& - 6 f^\dagger Y_{L_k} f Y_{L_k}^* \\
\gamma_{\hat{L}^c}^{(1)} & =  
2 \Big(3 {f f^\dagger}  + {Y_{L_k}^{\dagger}  Y_{L_k}}\Big) -\frac{3}{4} \Big(2 g_{2}^{2}  + g_{BL}^{2}\Big){\bf 1} \\ 
\gamma_{\hat{L}^c}^{(2)} & =  
\frac{3}{16} \Big(12 g_{2}^{2} g_{BL}^{2}  + 76 g_{2}^{4}  + 99 g_{BL}^{4} \Big){\bf 1} \nonumber \\
&+3 {f f^\dagger} \Big(-3 |a|^2  -4 \mbox{Tr}\Big({f  f^\dagger}\Big)  + 6 g_{BL}^{2}  + 8 g_{2}^{2} \Big)\nonumber \\ 
&+ Y_{L_m}^{\dagger} \Big( 6 g_2^2 \delta_{mn} + \frac{27}{4} \mbox{Tr}\Big({\alpha  \alpha^*}\Big) \delta_{mn} -2 \mbox{Tr}\Big({Y_{L_n}^{*}  Y_{L_m}^{T}}\Big) - 6 \mbox{Tr}\Big({Y_{Q_n}^{*}  Y_{Q_m}^{T}}\Big) \nonumber \\
& - 11 f f^\dagger \delta_{mn} \Big) Y_{L_n} -4 Y_{L_m}^{\dagger} Y_{L_n} Y_{L_m}^T Y_{L_n}^{*} - 2 f \Big( 17 f^\dagger f + 3 Y_{L_k}^T Y_{L_k}^* \Big) f^\dagger \nonumber \\
& - 6 Y_{L_k}^\dagger Y_{L_k} f f^\dagger \\
(\gamma_{\hat{\Phi}}^{(1)})_{ij} & =  
-3 g_{2}^{2} {\bf 1} -\frac{3}{2} \Big({\alpha  \alpha^*} + {\alpha^*  \alpha}\Big) +\delta_{im} \delta_{jn} \Big(3 \mbox{Tr}\Big({Y_{Q_m}^{*}  Y_{Q_n}^{T}}\Big)  + \mbox{Tr}\Big({Y_{L_m}^{*}  Y_{L_n}^{T}}\Big)\Big) \\ 
(\gamma_{\hat{\Phi}}^{(2)})_{ij} & = 
33 g_{2}^{4} {\bf 1} -9 \Big( 2 (\alpha \alpha \alpha^* \alpha^* + \alpha^* \alpha^* \alpha \alpha) + 3 (\alpha \alpha^* \alpha \alpha^* + \alpha^* \alpha \alpha^* \alpha) \Big) \nonumber \\
& -24 (\alpha \alpha^* + \alpha^* \alpha) \Big( 2 g_2^2 + 2 \mbox{Tr}\Big( \alpha \alpha^* \Big) -|a|^2 \Big) \nonumber \\
& - \frac{3}{2} \Big( \alpha_{jm} \alpha_{in}^* + \alpha_{jm}^* \alpha_{in} \Big) \Big( 3 \mbox{Tr}\Big({Y_{Q_m}^{*}  Y_{Q_n}^{T}}\Big) + \mbox{Tr}\Big({Y_{L_m}^{*}  Y_{L_n}^{T}}\Big) \Big) \nonumber \\
& - \frac{1}{2} \delta_{im} \delta_{in} \Big( -3 g_{BL}^2 \mbox{Tr}\Big({Y_{L_m}^{*}  Y_{L_n}^{T}}\Big) - (32 g_3^2 + g_{BL}^2) \mbox{Tr}\Big({Y_{Q_m}^{*}  Y_{Q_n}^{T}}\Big) \nonumber \\
& +2 \Big( 5 \mbox{Tr}\Big({ f f^\dagger Y_{L_n} Y_{L_m}^\dagger }\Big) + \mbox{Tr}\Big({ f Y_{L_n}^* Y_{L_m}^T f^\dagger }\Big)+ 6 \mbox{Tr}\Big({ f Y_{L_m}^* Y_{L_n}^T f^\dagger }\Big) \Big) \nonumber \\
& +4 \Big( 2 \mbox{Tr}\Big({ Y_{L_m}^\dagger Y_{L_n} Y_{L_n}^T Y_{L_n}^* }\Big) + \mbox{Tr}\Big({ Y_{L_m}^\dagger Y_{L_m} Y_{L_n}^T Y_{L_m}^* }\Big) \nonumber \\
& + \mbox{Tr}\Big({ Y_{L_m}^\dagger Y_{L_n} Y_{L_m}^T Y_{L_m}^* }\Big) \Big) \nonumber \\
& +12 \Big( 2 \mbox{Tr}\Big({ Y_{Q_m}^\dagger Y_{Q_n} Y_{Q_n}^T Y_{Q_n}^* }\Big) + \mbox{Tr}\Big({ Y_{Q_m}^\dagger Y_{Q_m} Y_{Q_n}^T Y_{Q_m}^* }\Big) \nonumber \\
& + \mbox{Tr}\Big({ Y_{Q_m}^\dagger Y_{Q_n} Y_{Q_m}^T Y_{Q_m}^* }\Big) \Big) \Big) \\
\gamma_{\hat{\Delta}}^{(1)} & =  
2 \mbox{Tr}\Big({f  f^\dagger}\Big)  -3 g_{BL}^{2}  -4 g_{2}^{2}  + \frac{3}{2} |a|^2 \\ 
\gamma_{\hat{\Delta}}^{(2)} & =  
48 g_{2}^{4} +24 g_{2}^{2} g_{BL}^{2} +81 g_{BL}^{4} \nonumber \\
& +\frac{3}{2} |a|^2 \Big( 4 g_2^2 + \mbox{Tr}\Big( \alpha \alpha^* \Big) - \frac{7}{2} |a|^2 \Big) - \Big(2 g_{2}^{2}  + 3 g_{BL}^{2} \Big)\mbox{Tr}\Big({f  f^\dagger}\Big) \nonumber \\ 
 &-24 \mbox{Tr}\Big({f  f^\dagger  f  f^\dagger}\Big) -6 \mbox{Tr}\Big({f  f^\dagger  Y_{L_k}  Y_{L_k}^{\dagger}}\Big) -2 \mbox{Tr}\Big({f  Y_{L_k}^{*}  Y_{L_k}^{T}  f^\dagger}\Big) \\ 
\gamma_{\hat{\bar{\Delta}}}^{(1)} & =  
-3 g_{BL}^{2}  -4 g_{2}^{2}  + \frac{3}{2} |a|^2 \\ 
\gamma_{\hat{\bar{\Delta}}}^{(2)} & =  
\frac{3}{4} \Big(4 \Big(16 g_{2}^{4}  + 27 g_{BL}^{4}  + 8 g_{2}^{2} g_{BL}^{2} \Big) \nonumber \\
& + |a|^2 \Big(2 \mbox{Tr}\Big({\alpha  \alpha^*}\Big)  -3 \mbox{Tr}\Big({f  f^\dagger}\Big)  -7 |a|^2  + 8 g_{2}^{2} \Big)\Big)\\ 
\gamma_{\hat{\Delta}^c}^{(1)} & =  
2 \mbox{Tr}\Big({f  f^\dagger}\Big)  -3 g_{BL}^{2}  -4 g_{2}^{2}  + \frac{3}{2} |a|^2 \\ 
\gamma_{\hat{\Delta}^c}^{(2)} & =  
48 g_{2}^{4} +24 g_{2}^{2} g_{BL}^{2} +81 g_{BL}^{4} \nonumber \\
& +\frac{3}{2} |a|^2 \Big( 4 g_2^2 + \mbox{Tr}\Big( \alpha \alpha^* \Big) - \frac{7}{2} |a|^2 \Big) - \Big(2 g_{2}^{2}  + 3 g_{BL}^{2} \Big)\mbox{Tr}\Big({f  f^\dagger}\Big) \nonumber \\ 
 &-24 \mbox{Tr}\Big({f  f^\dagger  f  f^\dagger}\Big) -8 \mbox{Tr}\Big({f  Y_{L_k}^T Y_{L_k}^* f^\dagger}\Big) \\ 
\gamma_{\hat{\bar{\Delta}}^c}^{(1)} & =  
-3 g_{BL}^{2}  -4 g_{2}^{2}  + \frac{3}{2} |a|^2 \\ 
\gamma_{\hat{\bar{\Delta}}^c}^{(2)} & =  
\frac{3}{4} \Big(4 \Big(16 g_{2}^{4}  + 27 g_{BL}^{4}  + 8 g_{2}^{2} g_{BL}^{2} \Big) \nonumber \\
& + |a|^2 \Big(2 \mbox{Tr}\Big({\alpha  \alpha^*}\Big)  -3 \mbox{Tr}\Big({f  f^\dagger}\Big)  -7 |a|^2  + 8 g_{2}^{2} \Big)\Big)\\ 
\gamma_{\hat{\Omega}}^{(1)} & =  
2 |a|^2  -4 g_{2}^{2} \\ 
\gamma_{\hat{\Omega}}^{(2)} & =  
3 \mbox{Tr}\Big({\alpha  (\alpha  \alpha^* -\alpha^* \alpha) \alpha^*}\Big)  + 48 g_{2}^{4}  + |a|^2 \Big(12 g_{BL}^{2}  -3 \mbox{Tr}\Big({f  f^\dagger}\Big)  -6 |a|^2  + 8 g_{2}^{2} \Big) \\
\gamma_{\hat{\Omega}^c}^{(1)} & =  
2 |a|^2  -4 g_{2}^{2} \\ 
\gamma_{\hat{\Omega}^c}^{(2)} & =  
3 \mbox{Tr}\Big({\alpha  (\alpha  \alpha^* -\alpha^* \alpha) \alpha^*}\Big)  + 48 g_{2}^{4}  + |a|^2 \Big(12 g_{BL}^{2}  -3 \mbox{Tr}\Big({f  f^\dagger}\Big)  -6 |a|^2  + 8 g_{2}^{2} \Big)
\end{align} }

Note that the previous formulas are totally general and can be applied with any number of bidoublets. Nevertheless, if two bidoublets are considered $\alpha \alpha^* = \alpha^* \alpha$ and further simplifications are possible.

\subsection*{Beta functions for soft breaking masses of sleptons}

Using the procedure explained in section \ref{sec:genericRGEs}, we can calculate the soft breaking masses for the sleptons. The results are

\begin{eqnarray}
16 \pi^2 \frac{d}{dt} m_L^2 &=& 6 f f^\dagger m_L^2 + 12 f m_L^2 f^\dagger + 6 m_L^2 f f^\dagger + 12 m_\Delta^2 f f^\dagger \nonumber \\
&& + 2 Y_{L_k} Y_{L_k}^\dagger m_L^2 + 2 m_L^2 Y_{L_k} Y_{L_k}^\dagger + 4 Y_{L_k} m_{L^c}^2 Y_{L_k}^\dagger \nonumber \\
&& + 4 (m_\Phi^2)_{mn} Y_L^{(m)} Y_L^{(n) \: \dagger} + 12 T_f T_f^\dagger + 4 T_{L_k} T_{L_k}^\dagger \nonumber \\
&& - (3 g_{BL}^2 |M_1|^2 + 6 g_2^2 |M_2|^2 + \frac{3}{2} g_{BL}^2 S_1) {\bf 1} \\
16 \pi^2 \frac{d}{dt} m_{L^c}^2 &=& 6 f^\dagger f m_{L^c}^2 + 12 f^\dagger m_{L^c}^2 f + 6 m_{L^c}^2 f^\dagger f + 12 m_{\Delta^c}^2 f^\dagger f \nonumber \\
&& + 2 Y_{L_k}^\dagger Y_{L_k} m_{L^c}^2 + 2 m_{L^c}^2 Y_{L_k}^\dagger Y_{L_k} + 4 Y_{L_k}^\dagger m_L^2 Y_{L_k} \nonumber \\
&& + 4 (m_\Phi^2)_{mn} Y_L^{(m) \: \dagger} Y_L^{(n)} + 12 T_f^\dagger T_f + 4 T_{L_k}^\dagger T_{L_k} \nonumber \\
&& - (3 g_{BL}^2 |M_1|^2 + 6 g_2^2 |M_2|^2 - \frac{3}{2} g_{BL}^2 S_1) {\bf 1}
\end{eqnarray}

where
\begin{eqnarray}
S_1 &=& 3 (m_\Delta^2 - m_{\bar{\Delta}}^2 - m_{\Delta^c}^2 + m_{\bar{\Delta}^c}^2) \nonumber \\
&& + \sum_{m,n} \left[ (m_Q^2)_{mn} - (m_{Q^c}^2)_{mn} - (m_L^2)_{mn} + (m_{L^c}^2)_{mn} \right]
\end{eqnarray}

\subsection*{Beta functions for gauge couplings}
\label{sec:Beta1}
{\allowdisplaybreaks  \begin{align} 
\beta_{g_{BL}}^{(1)} & =  
24 g_{BL}^{3} \\ 
\beta_{g_{BL}}^{(2)} & =  
\frac{1}{2} g_{BL}^{3} \Big(-192 |a|^2 -93 \mbox{Tr}\Big({f  f^\dagger}\Big) +2 \Big(115 g_{BL}^{2}  + 162 g_{2}^{2}  \nonumber \\ 
 &-2 \mbox{Tr}\Big({Y_{Q_k}^*  Y_{Q_k}^{T}}\Big) -6 \mbox{Tr}\Big({Y_{L_k}^*  Y_{L_k}^{T}}\Big) + 8 g_{3}^{2} \Big)\Big)\\ 
\beta_{g_2}^{(1)} & =  
8 g_{2}^{3} \\ 
\beta_{g_2}^{(2)} & =  
\frac{1}{6} g_{2}^{3} \Big(660 g_{2}^{2} +144 g_{3}^{2} +162 g_{BL}^{2} -192 |a|^2 +108 \mbox{Tr}\Big({\alpha  \alpha^*}\Big) -73 \mbox{Tr}\Big({f  f^\dagger}\Big) \nonumber \\ 
 &-24 \mbox{Tr}\Big({Y_{L_k}^*  Y_{L_k}^{T}}\Big) -72 \mbox{Tr}\Big({Y_{Q_k}^*  Y_{Q_k}^{T}}\Big) \Big)\\ 
\beta_{g_3}^{(1)} & =  
-3 g_{3}^{3} \\ 
\beta_{g_3}^{(2)} & =  
g_{3}^{3} \Big(14 g_{3}^{2}  + 18 g_{2}^{2}  -8 \mbox{Tr}\Big({Y_{Q_k}^*  Y_{Q_k}^{T}}\Big) + g_{BL}^{2}\Big)
\end{align}}

\section*{From $SU(2)_R$ breaking scale to $U(1)_{B-L}$ breaking scale}
\subsection*{Anomalous Dimensions}
\label{sec:Ana2}
{\allowdisplaybreaks \begin{align} 
\gamma_{\hat{Q}}^{(1)} & =  
-\frac{1}{12} \Big(18 g_{L}^{2}  + 32 g_{3}^{2}  + g_{BL}^{2}\Big){\bf 1}  + {Y_d^*  Y_{d}^{T}} + {Y_u^*  Y_{u}^{T}}\\ 
\gamma_{\hat{Q}}^{(2)} & =  
+\frac{1}{144} \Big(109 g_{BL}^{4}  -128 g_{3}^{4}  + 36 g_{BL}^{2} g_{L}^{2}  + 64 g_{3}^{2} \Big(18 g_{L}^{2}  + g_{BL}^{2}\Big) + 972 g_{L}^{4} \Big){\bf 1} \nonumber \\ 
 &-2 \Big({Y_d^*  Y_{d}^{T}  Y_d^*  Y_{d}^{T}} + {Y_u^*  Y_{u}^{T}  Y_u^*  Y_{u}^{T}}\Big)\nonumber \\ 
 &+{Y_d^*  Y_{d}^{T}} \Big(-3 \mbox{Tr}\Big({Y_d  Y_{d}^{\dagger}}\Big)  - |b_c|^2  -\frac{3}{2} |b|^2  - \mbox{Tr}\Big({Y_e  Y_{e}^{\dagger}}\Big)  + g_{R}^{2}\Big)\nonumber \\ 
 &+{Y_u^*  Y_{u}^{T}} \Big(-3 \mbox{Tr}\Big({Y_u  Y_{u}^{\dagger}}\Big)  - |b_c|^2  -\frac{3}{2} |b|^2  - \mbox{Tr}\Big({Y_\nu  Y_{\nu}^{\dagger}}\Big)  + g_{R}^{2}\Big)\\ 
\gamma_{\hat{d}^c}^{(1)} & =  
2 {Y_{d}^{\dagger}  Y_d}  -\frac{1}{12} \Big(32 g_{3}^{2}  + 6 g_{R}^{2}  + g_{BL}^{2}\Big){\bf 1} \\ 
\gamma_{\hat{d}^c}^{(2)} & =  
-\frac{1}{144} \Big(-109 g_{BL}^{4}  + 128 g_{3}^{4}  -12 g_{BL}^{2} g_{R}^{2}  -64 g_{3}^{2} \Big(6 g_{R}^{2}  + g_{BL}^{2}\Big) -684 g_{R}^{4} \Big){\bf 1} \nonumber \\ 
 &-2 \Big({Y_{d}^{\dagger}  Y_d  Y_{d}^{\dagger}  Y_d} + {Y_{d}^{\dagger}  Y_u  Y_{u}^{\dagger}  Y_d}\Big)\nonumber \\ 
 &- {Y_{d}^{\dagger}  Y_d} \Big(2 |b_c|^2  + 2 \mbox{Tr}\Big({Y_e  Y_{e}^{\dagger}}\Big)  + 3 |b|^2  -6 g_{L}^{2}  + 6 \mbox{Tr}\Big({Y_d  Y_{d}^{\dagger}}\Big) \Big)\\ 
\gamma_{\hat{u}^c}^{(1)} & =  
2 {Y_{u}^{\dagger}  Y_u}  -\frac{1}{12} \Big(32 g_{3}^{2}  + 6 g_{R}^{2}  + g_{BL}^{2}\Big){\bf 1} \\ 
\gamma_{\hat{u}^c}^{(2)} & =  
-\frac{1}{144} \Big(-109 g_{BL}^{4}  + 128 g_{3}^{4}  -12 g_{BL}^{2} g_{R}^{2}  -64 g_{3}^{2} \Big(6 g_{R}^{2}  + g_{BL}^{2}\Big) -684 g_{R}^{4} \Big){\bf 1} \nonumber \\ 
 &-2 \Big({Y_{u}^{\dagger}  Y_d  Y_{d}^{\dagger}  Y_u} + {Y_{u}^{\dagger}  Y_u  Y_{u}^{\dagger}  Y_u}\Big)\nonumber \\ 
 &- {Y_{u}^{\dagger}  Y_u} \Big(2 |b_c|^2  + 2 \mbox{Tr}\Big({Y_\nu  Y_{\nu}^{\dagger}}\Big)  + 3 |b|^2  -6 g_{L}^{2}  + 6 \mbox{Tr}\Big({Y_u  Y_{u}^{\dagger}}\Big) \Big)\\ 
\gamma_{\hat{L}}^{(1)} & =  
-\frac{3}{4} \Big(2 g_{L}^{2}  + g_{BL}^{2}\Big){\bf 1}  + {Y_e^*  Y_{e}^{T}} + {Y_\nu^*  Y_{\nu}^{T}}\\ 
\gamma_{\hat{L}}^{(2)} & =  
+\frac{9}{16} \Big(12 g_{L}^{4}  + 13 g_{BL}^{4}  + 4 g_{BL}^{2} g_{L}^{2} \Big){\bf 1} -2 \Big({Y_e^*  Y_{e}^{T}  Y_e^*  Y_{e}^{T}} + {Y_\nu^*  Y_{\nu}^{T}  Y_\nu^*  Y_{\nu}^{T}} \Big) \nonumber \\ 
 &- {Y_\nu^*  F_c  Y_{\nu}^{T}}  \nonumber \\ 
 &+{Y_e^*  Y_{e}^{T}} \Big(-3 \mbox{Tr}\Big({Y_d  Y_{d}^{\dagger}}\Big)  - |b_c|^2  -\frac{3}{2} |b|^2  - \mbox{Tr}\Big({Y_e  Y_{e}^{\dagger}}\Big)  + g_{R}^{2}\Big)\nonumber \\ 
 &+{Y_\nu^*  Y_{\nu}^{T}} \Big(-3 \mbox{Tr}\Big({Y_u  Y_{u}^{\dagger}}\Big)  - |b_c|^2  -\frac{3}{2} |b|^2  - \mbox{Tr}\Big({Y_\nu  Y_{\nu}^{\dagger}}\Big)  + g_{R}^{2}\Big)\\ 
\gamma_{\hat{e}^c}^{(1)} & =  
2 {Y_{e}^{\dagger}  Y_e}  -\frac{1}{4} \Big(2 g_{R}^{2}  + 3 g_{BL}^{2} \Big){\bf 1} \\ 
\gamma_{\hat{e}^c}^{(2)} & =  
+\frac{1}{16} \Big(117 g_{BL}^{4}  + 12 g_{BL}^{2} g_{R}^{2}  + 76 g_{R}^{4} \Big){\bf 1} -2 \Big({Y_{e}^{\dagger}  Y_e  Y_{e}^{\dagger}  Y_e} + {Y_{e}^{\dagger}  Y_\nu  Y_{\nu}^{\dagger}  Y_e}\Big)\nonumber \\ 
 &- {Y_{e}^{\dagger}  Y_e} \Big(2 |b_c|^2  + 2 \mbox{Tr}\Big({Y_e  Y_{e}^{\dagger}}\Big)  + 3 |b|^2  -6 g_{L}^{2}  + 6 \mbox{Tr}\Big({Y_d  Y_{d}^{\dagger}}\Big) \Big)\\ 
\gamma_{\hat{\nu}^c}^{(1)} & =  
2 {Y_{\nu}^{\dagger}  Y_\nu}  -\frac{1}{4} \Big(2 g_{R}^{2}  + 3 g_{BL}^{2} \Big){\bf 1}  + F_c^* \\ 
\gamma_{\hat{\nu}^c}^{(2)} & =  
\frac{1}{16} \Big(117 g_{BL}^{4}  + 12 g_{BL}^{2} g_{R}^{2}  + 76 g_{R}^{4} \Big){\bf 1} - (f_c^{1 \dagger} + f_c^{1 *}) F_c (f_c^{1} + f_c^{1 T}) \nonumber \\
& -2 Y_{\nu}^{\dagger}  \Big( Y_e  Y_{e}^{\dagger}  +  Y_\nu  Y_{\nu}^{\dagger}  \Big) Y_\nu -2 (f_c^{1 \dagger} + f_c^{1 *}) Y_{\nu}^{T}  Y_\nu^*  (f_c^{1} + f_c^{1 T}) \nonumber \\ 
&- {Y_{\nu}^{\dagger}  Y_\nu} \Big(2 \Big(-3 g_{L}^{2}  + 3 \mbox{Tr}\Big({Y_u  Y_{u}^{\dagger}}\Big)  + |b_c|^2 + \mbox{Tr}\Big({Y_\nu  Y_{\nu}^{\dagger}}\Big)\Big) + 3 |b|^2 \Big)\nonumber \\ 
 &+ F_c^* \Big(2 g_{R}^{2}  + 3 g_{BL}^{2}  - |a^1_c|^2  - \mbox{Tr}\Big({f^1_c  f_{c}^{1 \dagger}}\Big)  - \mbox{Tr}\Big({f_{c}^{1 \dagger}  f_{c}^{1 T}}\Big) \Big)\\ 
\gamma_{\hat{H}_d}^{(1)} & =  
3 \mbox{Tr}\Big({Y_d  Y_{d}^{\dagger}}\Big)  -\frac{1}{2} g_{R}^{2}  + \frac{3}{2} |b|^2  -\frac{3}{2} g_{L}^{2}  + |b_c|^2 + \mbox{Tr}\Big({Y_e  Y_{e}^{\dagger}}\Big)\\ 
\gamma_{\hat{H}_d}^{(2)} & =  
\frac{1}{4} \Big(27 g_{L}^{4} +6 g_{L}^{2} g_{R}^{2} +19 g_{R}^{4} +2 \Big(32 g_{3}^{2}  + g_{BL}^{2}\Big)\mbox{Tr}\Big({Y_d  Y_{d}^{\dagger}}\Big) +6 g_{BL}^{2} \mbox{Tr}\Big({Y_e  Y_{e}^{\dagger}}\Big) \nonumber \\ 
 &+|b|^2 \Big(-18 \mbox{Tr}\Big({Y_u  Y_{u}^{\dagger}}\Big) -15 |b|^2 -19 |b_c|^2  + 24 g_{L}^{2}  -6 \mbox{Tr}\Big({Y_\nu  Y_{\nu}^{\dagger}}\Big) \Big)\nonumber \\ 
 &-4 |b_c|^2 \Big(3 |b_c|^2  + 3 \mbox{Tr}\Big({Y_u  Y_{u}^{\dagger}}\Big)  + |a^1_c|^2  + \mbox{Tr}\Big({Y_\nu  Y_{\nu}^{\dagger}}\Big)\Big)-36 \mbox{Tr}\Big({Y_d  Y_{d}^{\dagger}  Y_d  Y_{d}^{\dagger}}\Big) \nonumber \\
& -12 \mbox{Tr}\Big({Y_d  Y_{d}^{\dagger}  Y_u  Y_{u}^{\dagger}}\Big) -12 \mbox{Tr}\Big({Y_e  Y_{e}^{\dagger}  Y_e  Y_{e}^{\dagger}}\Big) -4 \mbox{Tr}\Big({Y_e  Y_{e}^{\dagger}  Y_\nu  Y_{\nu}^{\dagger}}\Big) \Big)\\ 
\gamma_{\hat{H}_u}^{(1)} & =  
3 \mbox{Tr}\Big({Y_u  Y_{u}^{\dagger}}\Big)  -\frac{1}{2} g_{R}^{2}  + \frac{3}{2} |b|^2  -\frac{3}{2} g_{L}^{2}  + |b_c|^2 + \mbox{Tr}\Big({Y_\nu  Y_{\nu}^{\dagger}}\Big)\\ 
\gamma_{\hat{H}_u}^{(2)} & =  
\frac{1}{4} \Big(27 g_{L}^{4} +6 g_{L}^{2} g_{R}^{2} +19 g_{R}^{4} \nonumber \\
& + |b|^2 \Big(-18 \mbox{Tr}\Big({Y_d  Y_{d}^{\dagger}}\Big)  -15 |b|^2 -19 |b_c|^2  + 24 g_{L}^{2}  -6 \mbox{Tr}\Big({Y_e  Y_{e}^{\dagger}}\Big) \Big)\nonumber \\ 
 &-4 |b_c|^2 \Big(3 |b_c|^2  + 3 \mbox{Tr}\Big({Y_d  Y_{d}^{\dagger}}\Big)  + |a^1_c|^2   + \mbox{Tr}\Big({Y_e  Y_{e}^{\dagger}}\Big)\Big) \nonumber \\ 
 &+2 \Big(32 g_{3}^{2}  + g_{BL}^{2}\Big)\mbox{Tr}\Big({Y_u  Y_{u}^{\dagger}}\Big) +6 g_{BL}^{2} \mbox{Tr}\Big({Y_\nu  Y_{\nu}^{\dagger}}\Big) -4 \mbox{Tr}\Big({f^1_c  f_{c}^{1 \dagger}  Y_{\nu}^{T}  Y_\nu^*}\Big) \nonumber \\ 
 &-4 \mbox{Tr}\Big({f^1_c  Y_{\nu}^{\dagger}  Y_\nu  f_{c}^{1 \dagger}}\Big) -12 \mbox{Tr}\Big({Y_d  Y_{d}^{\dagger}  Y_u  Y_{u}^{\dagger}}\Big)-4 \mbox{Tr}\Big({Y_e  Y_{e}^{\dagger}  Y_\nu  Y_{\nu}^{\dagger}}\Big) \nonumber \\ 
 &-36 \mbox{Tr}\Big({Y_u  Y_{u}^{\dagger}  Y_u  Y_{u}^{\dagger}}\Big) -4 \mbox{Tr}\Big({Y_\nu  f_{c}^{1 \dagger}  f_{c}^{1 T}  Y_{\nu}^{\dagger}}\Big) -12 \mbox{Tr}\Big({Y_\nu  Y_{\nu}^{\dagger}  Y_\nu  Y_{\nu}^{\dagger}}\Big) \nonumber \\
&-4 \mbox{Tr}\Big({f_{c}^{1 \dagger}  Y_{\nu}^{T}  Y_\nu^*  f_{c}^{1 T}}\Big) \Big)\\ 
\gamma_{\hat{\Delta}^{c 0}}^{(1)} & =  
-2 g_{R}^{2}  -3 g_{BL}^{2}  + |a^1_c|^2 + \mbox{Tr}\Big({f^1_c  f_{c}^{1 \dagger}}\Big) + \mbox{Tr}\Big({f_{c}^{1 \dagger}  f_{c}^{1 T}}\Big)\\ 
\gamma_{\hat{\Delta}^{c 0}}^{(2)} & =  
\frac{1}{2} \Big(72 g_{BL}^{4} +24 g_{BL}^{2} g_{R}^{2} +44 g_{R}^{4} -4 |a^1_c|^2 \Big(|a^1_c|^2  +| b_c|^2 \Big)\nonumber \\ 
 &- \Big(2 g_{R}^{2}  + 3 g_{BL}^{2} \Big)\Big(\mbox{Tr}\Big({f^1_c  f_{c}^{1 \dagger}}\Big) + \mbox{Tr}\Big({f_{c}^{1 \dagger}  f_{c}^{1 T}}\Big)\Big)-4 \mbox{Tr}\Big({f^1_c  f_{c}^{1 \dagger}  f^1_c  f_{c}^{1 \dagger}}\Big) \nonumber \\
& -8 \mbox{Tr}\Big({f^1_c  f_{c}^{1 \dagger}  f_{c}^{1 T}  f_{c}^{1 \dagger}}\Big) -8 \mbox{Tr}\Big({f^1_c  f_{c}^{1 \dagger}  f_{c}^{1 T}  f^{1 *}_c}\Big) -4 \mbox{Tr}\Big({f^1_c  f_{c}^{1 \dagger}  Y_{\nu}^{T}  Y_\nu^*}\Big) \nonumber \\
& -4 \mbox{Tr}\Big({f^1_c  Y_{\nu}^{\dagger}  Y_\nu  f_{c}^{1 \dagger}}\Big) -4 \mbox{Tr}\Big({Y_\nu  f_{c}^{1 \dagger}  f_{c}^{1 T}  Y_{\nu}^{\dagger}}\Big) -4 \mbox{Tr}\Big({f_{c}^{1 \dagger}  f_{c}^{1 T}  f_{c}^{1 \dagger}  f_{c}^{1 T}}\Big) \nonumber \\
&-8 \mbox{Tr}\Big({f_{c}^{1 \dagger}  f_{c}^{1 T}  f^{1 *}_c  f_{c}^{1 T}}\Big) -4 \mbox{Tr}\Big({f_{c}^{1 \dagger}  Y_{\nu}^{T}  Y_\nu^*  f_{c}^{1 T}}\Big) \Big)\\ 
\gamma_{\hat{\bar{\Delta}}^{c 0}}^{(1)} & =  
-2 g_{R}^{2}  -3 g_{BL}^{2}  + |a^1_c|^2\\ 
\gamma_{\hat{\bar{\Delta}}^{c 0}}^{(2)} & =  
12 g_{BL}^{2} g_{R}^{2}  + 22 g_{R}^{4}  + 36 g_{BL}^{4}  \nonumber \\
& - |a^1_c|^2 \Big(2 |a^1_c|^2  + 2 |b_c|^2  + \mbox{Tr}\Big({f^1_c  f_{c}^{1 \dagger}}\Big) + \mbox{Tr}\Big({f_{c}^{1 \dagger}  f_{c}^{1 T}}\Big)\Big)\\ 
\gamma_{\hat{\Omega}}^{(1)} & =  
-4 g_{L}^{2}  + |b|^2\\ 
\gamma_{\hat{\Omega}}^{(2)} & =  
28 g_{L}^{4}  - |b|^2 \Big(2 |b_c|^2  + 3 |b|^2  + 3 \mbox{Tr}\Big({Y_d  Y_{d}^{\dagger}}\Big)  + 3 \mbox{Tr}\Big({Y_u  Y_{u}^{\dagger}}\Big)  \nonumber \\
& - g_{R}^{2}  + g_{L}^{2} + \mbox{Tr}\Big({Y_e  Y_{e}^{\dagger}}\Big) + \mbox{Tr}\Big({Y_\nu  Y_{\nu}^{\dagger}}\Big)\Big)\\ 
\gamma_{\hat{\Omega}^{c 0}}^{(1)} & =  
2 |b_c|^2  + |a^1_c|^2\\ 
\gamma_{\hat{\Omega}^{c 0}}^{(2)} & =  
-2 |a_{c}^{1}|^4 - |a^1_c|^2 \Big(-4 g_{R}^{2}  -6 g_{BL}^{2}  + \mbox{Tr}\Big({f^1_c  f_{c}^{1 \dagger}}\Big) + \mbox{Tr}\Big({f_{c}^{1 \dagger}  f_{c}^{1 T}}\Big)\Big) \nonumber \\ 
 &-2 |b_c|^2 \Big(2 |b_c|^2  + 3 |b|^2  -3 g_{L}^{2}  + 3 \mbox{Tr}\Big({Y_d  Y_{d}^{\dagger}}\Big)  + 3 \mbox{Tr}\Big({Y_u  Y_{u}^{\dagger}}\Big)  \nonumber \\
& - g_{R}^{2}  + \mbox{Tr}\Big({Y_e  Y_{e}^{\dagger}}\Big) + \mbox{Tr}\Big({Y_\nu  Y_{\nu}^{\dagger}}\Big)\Big)
\end{align} } 

In these expressions we have defined

\begin{equation}
F_c = f_{c}^{1}  f_{c}^{1 \dagger}  +  f_{c}^{1}  f^{1 *}_c  + f_{c}^{1 T}  f_{c}^{1 \dagger}  +  f_{c}^{1 T}  f^{1 *}_c 
\end{equation}

\subsection*{Beta functions for soft breaking masses of sleptons}
Again, the results for the slepton soft SUSY breaking masses at 1-loop are shown. The beta functions read
\begin{eqnarray}
16 \pi^2 \frac{d}{dt} m_L^2 &=& 2 Y_e m_{e^c}^2 Y_e^\dagger + 2 m_{H_d}^2 Y_e Y_e^\dagger + 2 m_{H_u}^2 Y_\nu Y_\nu^\dagger + m_L^2 Y_e Y_e^\dagger \\
&& + Y_e Y_e^\dagger m_L^2 + m_L^2 Y_\nu Y_\nu^\dagger + Y_\nu Y_\nu^\dagger m_L^2 + 2 Y_\nu m_{\nu^c}^2 Y_\nu^\dagger \nonumber \\
&& + 2 T_e T_e^\dagger + 2 T_\nu T_\nu^\dagger - (3 g_{BL}^2 |M_1|^2 + 6 g_L^2 |M_L|^2 + \frac{3}{4} g_{BL}^2 S_2)  {\bf 1} \nonumber \\
16 \pi^2 \frac{d}{dt} m_{e^c}^2 &=& 2 Y_e^\dagger Y_e m_{e^c}^2 + 2 m_{e^c}^2 Y_e^\dagger Y_e + 4 m_{H_d}^2 Y_e^\dagger Y_e + 4 Y_e^\dagger m_L^2 Y_e \\
&& + 4 T_e^\dagger T_e - (3 g_{BL}^2 |M_1|^2 + 2 g_R^2 |M_R|^2 - \frac{3}{4} g_{BL}^2 S_2 - \frac{1}{2} g_R^2 S_3 ) {\bf 1} \nonumber
\end{eqnarray}
where
\begin{eqnarray}
S_2 &=& 2 ( m_{\Delta^{c \: 0}}^2 - m_{\bar{\Delta}^{c \: 0}}^2 ) + Tr \left[ m_{d^c}^2 - m_{e^c}^2 + m_{u^c}^2 - m_{\nu^c}^2 + 2 m_L^2 - 2 m_Q^2 \right] \nonumber \\
S_3 &=& 2 ( m_{\Delta^{c \: 0}}^2 - m_{\bar{\Delta}^{c \: 0}}^2 - m_{H_d}^2 + m_{H_u}^2 ) + Tr \left[ 3 m_{d^c}^2 + m_{e^c}^2 - 3 m_{u^c}^2 - m_{\nu^c}^2 \right] \nonumber
\end{eqnarray}

\subsection*{Beta functions for gauge couplings}
\label{sec:beta2}
{\allowdisplaybreaks  \begin{align} 
\beta_{g_{BL}}^{(1)} & =  
9 g_{BL}^{3} \\ 
\beta_{g_{BL}}^{(2)} & =  
\frac{1}{2} g_{BL}^{3} \Big(16 g_{3}^{2} +50 g_{BL}^{2} +18 g_{L}^{2} +30 g_{R}^{2} -12 |{a_c^1}|^2 -9 \mbox{Tr}\Big({{f_c^1}  {f_c^1}^{\dagger}}\Big) \\ 
 & -2 \mbox{Tr}\Big({Y_d  Y_d^{\dagger}}\Big) -6 \mbox{Tr}\Big({Y_e  Y_e^{\dagger}}\Big) -2 \mbox{Tr}\Big({Y_u  Y_u^{\dagger}}\Big) -6 \mbox{Tr}\Big({Y_v  Y_v^{\dagger}}\Big) -9 \mbox{Tr}\Big({{f_c^1}^{\dagger}  {f_c^1}^{T}}\Big) \Big) \nonumber \\ 
\beta_{g_L}^{(1)} & =  
3 g_{L}^{3} \\ 
\beta_{g_L}^{(2)} & =  
\frac{1}{3} g_{L}^{3} \Big(-28 |b|^2  + 3 \Big(24 g_{3}^{2}  -2 |b_c|^2  -2 \mbox{Tr}\Big({Y_e  Y_e^{\dagger}}\Big)  -2 \mbox{Tr}\Big({Y_v  Y_v^{\dagger}}\Big)  + 3 g_{BL}^{2} \nonumber \\ 
 & + 49 g_{L}^{2} -6 \mbox{Tr}\Big({Y_d  Y_d^{\dagger}}\Big)  -6 \mbox{Tr}\Big({Y_u  Y_u^{\dagger}}\Big)  + g_{R}^{2}\Big)\Big)\\ 
\beta_{g_R}^{(1)} & =  
9 g_{R}^{3} \\ 
\beta_{g_R}^{(2)} & =  
g_{R}^{3} \Big(24 g_{3}^{2} +15 g_{BL}^{2} +3 g_{L}^{2} +15 g_{R}^{2} -4 |{a_c^1}|^2 -4 |b|^2 -2 |b_c|^2 -3 \mbox{Tr}\Big({{f_c^1}  {f_c^1}^{\dagger}}\Big) \nonumber \\
& -6 \mbox{Tr}\Big({Y_d  Y_d^{\dagger}}\Big) -2 \mbox{Tr}\Big({Y_e  Y_e^{\dagger}}\Big) -6 \mbox{Tr}\Big({Y_u  Y_u^{\dagger}}\Big) -2 \mbox{Tr}\Big({Y_v  Y_v^{\dagger}}\Big) -3 \mbox{Tr}\Big({{f_c^1}^{\dagger}  {f_c^1}^{T}}\Big) \Big)\\ 
\beta_{g_3}^{(1)} & =  
-3 g_{3}^{3} \\ 
\beta_{g_3}^{(2)} & =  
g_{3}^{3} \Big(14 g_{3}^{2}  + 3 g_{R}^{2}  -4 \mbox{Tr}\Big({Y_d  Y_d^{\dagger}}\Big)  -4 \mbox{Tr}\Big({Y_u  Y_u^{\dagger}}\Big)  + 9 g_{L}^{2}  + g_{BL}^{2}\Big)
\end{align}}

\bibliographystyle{prsty} 

\end{document}